\documentclass{amsbook}

\usepackage{amsmath,amssymb,amsthm,leftidx}
\usepackage[all,color]{xy}
\usepackage[pagebackref]{hyperref} 
\usepackage{MnSymbol}
\usepackage{bbm}
\usepackage[mathscr]{eucal}
\usepackage{xcolor}

\numberwithin{section}{chapter}
\numberwithin{subsection}{section}
\numberwithin{equation}{section}
\theoremstyle{plain}
        \newtheorem{theorem}[equation]{Theorem}
        \newtheorem{proposition}[equation]{Proposition}
        \newtheorem{lemma}[equation]{Lemma}
        
        \newtheorem{corollary}[equation]{Corollary}

\theoremstyle{definition}
        \newtheorem{assumption}[equation]{Assumption}
        
        \newtheorem{definition}[equation]{Definition}
        
        \newtheorem{remark}[equation]{Remark}
        \newtheorem{example}[equation]{Example}
        \newtheorem{interpretation}[equation]{Interpretation}
        \newtheorem{convention}[equation]{Convention}
        \newtheorem{notation}[equation]{Notation}
        \newtheorem{motivation}[equation]{Motivation}
        \newtheorem{recollection}[equation]{Recollection}
        
\usepackage{tikz}
\usetikzlibrary{arrows,decorations.pathmorphing}
\usetikzlibrary{backgrounds,positioning}
\usetikzlibrary{fit,petri,shapes.misc}
\usetikzlibrary{shapes.geometric}

\tikzset{auto}

\tikzset{empty/.style={circle,inner sep=0pt,minimum size=6mm}}
\tikzset{emptyvt/.style={circle,inner sep=0pt,minimum size=0mm}}

\tikzset{plain/.style={circle,draw,very thick,
inner sep=0pt,minimum size=6mm}}

\tikzset{xplain/.style={circle,draw,very thick,
inner sep=0pt,minimum size=8mm}}

\tikzset{dottedplain/.style={circle,draw,very thick, densely dotted,
inner sep=0pt,minimum size=6mm}}

\tikzset{smallplain/.style={circle,draw,very thick,
inner sep=0pt,minimum size=4mm}}

\tikzset{smalldotted/.style={circle,draw,very thick, densely dotted,
inner sep=0pt,minimum size=3mm}}

\tikzset{xsplain/.style={circle,draw, thick,
inner sep=0pt,minimum size=1.5mm}}

\tikzset{tinyplain/.style={circle,draw, thick,
inner sep=0pt,minimum size=1mm}}

\tikzset{rectplain/.style={rectangle,draw,very thick,minimum size=6mm}}
\tikzset{bigplain/.style={rectangle,draw,very thick,minimum size=1cm}}

\tikzset{triangular/.style={regular polygon, regular polygon sides=3, draw,very thick,
inner sep=0pt,minimum size=1.2cm}}

\tikzset{arrow/.style={->,thick}}
\tikzset{dashedarrow/.style={->,dashed,thick}}
\tikzset{dottedarrow/.style={->,dotted,thick}}
\tikzset{mapto/.style={|->,thick}}

\tikzset{implies/.style={thick,double,double equal sign distance,-implies}} 

\tikzset{line/.style={thick}}
\tikzset{dottedline/.style={dotted,thick}}
\tikzset{dashedline/.style={dashed,thick}}

\tikzset{inputleg/.style={<-,thick}}
\tikzset{outputleg/.style={->,thick}}
\tikzset{dottedinput/.style={<-,dotted,thick}}

\newcommand{\nicearrow}{\SelectTips{cm}{10}}
\newcommand{\nicexy}{\nicearrow\xymatrix@C+5pt}
\newcommand{\narrowxy}{\nicearrow\xymatrix}
\newcommand{\adjoint}{\nicearrow\xymatrix{ \ar@<2pt>[r] & \ar@<2pt>[l]}}

\renewcommand{\to}{\hspace{-.1cm}\nicearrow\xymatrix@C-.3cm{\ar[r]&}\hspace{-.1cm}}

\newcommand{\shortto}{\hspace{-.1cm}\nicearrow\xymatrix@C-.6cm{\ar[r]&}\hspace{-.1cm}}
\newcommand{\iso}{\hspace{-.1cm}\nicearrow\xymatrix@C-.2cm{\ar[r]^-{\cong}&}\hspace{-.1cm}}
\newcommand{\weq}{\hspace{-.1cm}\nicearrow\xymatrix@C-.2cm{\ar[r]^-{\sim}&}\hspace{-.1cm}}
\renewcommand{\mapsto}{\hspace{-.1cm}\nicexy@C-.2cm{\ar@{|->}[r]&}\hspace{-.1cm}}

\newcommand{\pushoutcat}{\narrowxy{1 & 0 \ar[l] \ar[r] & 2}}

\newcommand{\edge}{\vert}

\newcommand{\fieldk}{\mathbbm{K}}
\newcommand{\fieldr}{\mathbbm{R}}

\newcommand{\fraka}{\mathfrak{A}}
\newcommand{\frakb}{\mathfrak{B}}

\newcommand{\cala}{\mathcal{A}}
\newcommand{\calc}{\mathcal{C}}
\newcommand{\calf}{\mathcal{F}}
\newcommand{\cali}{\mathcal{I}}
\newcommand{\calj}{\mathcal{J}}
\newcommand{\call}{\mathcal{L}}
\newcommand{\calr}{\mathcal{R}}
\newcommand{\calw}{\mathcal{W}}

\newcommand{\scF}{\mathscr{F}}

\newcommand{\colorb}{\mathfrak{B}}
\newcommand{\colorc}{\mathfrak{C}}

\newcommand{\colord}{\mathfrak{D}}
\newcommand{\Cor}{\mathrm{Cor}} 
\newcommand{\Corucd}{\Cor_{(\uc;d)}} 
\newcommand{\Ed}{\mathsf{Ed}}
\newcommand{\Flag}{\mathsf{Flag}}
\newcommand{\graft}{\mathsf{Graft}}
\newcommand{\Lin}{\mathrm{Lin}}
\newcommand{\lin}{\mathrm{lin}}
\newcommand{\Leg}{\mathsf{Leg}}

\newcommand{\Ord}{\mathsf{Ord}}
\newcommand{\Prof}{\mathsf{Prof}}
\newcommand{\Profb}{\Prof(\colorb)}
\newcommand{\Profbb}{\Profb \times \colorb}
\newcommand{\Profc}{\Prof(\colorc)}
\newcommand{\Profcc}{\Profc \times \colorc}

\newcommand{\profofg}{\Prof(G)}
\newcommand{\profofh}{\Prof(H)}
\newcommand{\profofhv}{\Prof(H_v)}
\newcommand{\profofk}{\Prof(K)}
\newcommand{\profl}{\Prof(L)}
\newcommand{\profoft}{\Prof(T)}

\newcommand{\profofu}{\Prof(u)}
\newcommand{\profofv}{\Prof(v)}
\newcommand{\profofw}{\Prof(w)}
\newcommand{\profofz}{\Prof(z)}
\newcommand{\Tun}{\mathsf{Tun}}
\newcommand{\Vt}{\mathsf{Vt}}

\newcommand{\zeroc}{\{0,c\}}
\newcommand{\zerod}{\{0,d\}}
\newcommand{\zerocd}{\{0,c,d\}}

\newcommand{\diag}{\mathsf{diag}}
\newcommand{\fact}{\mathsf{fact}}
\newcommand{\op}{\mathsf{op}}
\newcommand{\A}{\mathsf{A}}

\newcommand{\C}{\mathsf{C}}
\newcommand{\Cop}{\C^{\op}}
\newcommand{\D}{\mathsf{D}}
\newcommand{\Dop}{\D^{\op}}
\newcommand{\Doft}{\D(T)}
\newcommand{\Deltaop}{\Delta^{\op}}
\newcommand{\E}{\mathsf{E}}
\newcommand{\F}{\mathsf{F}}
\newcommand{\I}{\mathsf{I}}
\newcommand{\J}{\mathsf{J}}
\newcommand{\Jminus}{\J^{-}}
\newcommand{\Lder}{\mathsf{L}}
\newcommand{\M}{\mathsf{M}}
\newcommand{\Mbar}{\overline{\M}}
\newcommand{\N}{\mathsf{N}}
\renewcommand{\O}{\mathsf{O}}
\newcommand{\Ominus}{\O^{-}}
\newcommand{\jominus}{(\J\otimes\O)^{-}}
\newcommand{\Otom}{\O^{\M}}
\newcommand{\Oton}{\O^{\N}}
\renewcommand{\P}{\mathsf{P}}

\newcommand{\R}{\mathsf{R}}
\newcommand{\Sing}{\mathsf{Sing}}
\newcommand{\W}{\mathsf{W}}

\newcommand{\cod}{\mathsf{codomain}}
\newcommand{\coequal}{\mathsf{coequal}}
\newcommand{\colim}{\mathsf{colim}}
\newcommand{\ellm}{\ell^{\M}}
\newcommand{\wellm}{\W\ellm}
\newcommand{\ellmstar}{(\ellm)^*}
\newcommand{\ellmst}{\ellm_!}
\newcommand{\wellmst}{\wellm_!}
\newcommand{\wellmstar}{(\wellm)^*}
\newcommand{\End}{\mathsf{End}}
\newcommand{\etao}{\eta^{\O}}
\newcommand{\etaostar}{(\etao)^*}
\newcommand{\etap}{\eta^{\P}}
\newcommand{\etapstar}{(\etap)^*}
\newcommand{\Ho}{\mathsf{Ho}}
\newcommand{\Hom}{\mathsf{Hom}}
\newcommand{\Homm}{\Hom_{\M}}
\newcommand{\Id}{\mathrm{Id}}
\newcommand{\id}{\mathrm{id}}

\newcommand{\Lan}{\mathsf{Lan}}
\newcommand{\limit}{\mathsf{lim}}

\newcommand{\Mor}{\mathsf{Mor}}
\newcommand{\Morc}{\Mor(\C)}

\newcommand{\Ob}{\mathsf{Ob}}
\newcommand{\Obc}{\Ob(\C)}
\newcommand{\Obcsinv}{\Ob(\Csinv)}
\newcommand{\Obd}{\Ob(\D)}
\newcommand{\Opc}{\mathsf{Op}^{\colorc}}
\newcommand{\Res}{\mathrm{Res}}
\newcommand{\spinzero}{\mathrm{Spin}_0}

\newcommand{\operadunit}{\mathsf{1}}

\newcommand{\tensorunit}{\mathbbm{1}}

\newcommand{\comp}{\circ}
\newcommand{\compi}{\circ_i}

\newcommand{\colimover}[1]{\underset{#1}{\mathsf{colim}}}

\newcommand{\coprodover}[1]{\underset{#1}{\coprod}}
\newcommand{\prodover}[1]{\underset{#1}{\prod}}
\newcommand{\tensorover}[1]{\underset{#1}{\otimes}}
\newcommand{\bigtensorover}[1]{\underset{#1}{\bigotimes}}

\newcommand{\deltam}{\delta^{\M}}
\newcommand{\deltamin}{\delta_{\mathsf{min}}}
\newcommand{\deltamax}{\delta_{\mathsf{max}}}
\newcommand{\deltaminm}{\deltamin^{\M}}
\newcommand{\deltamaxm}{\deltamax^{\M}}
\newcommand{\deltaminmst}{(\deltaminm)_!}
\newcommand{\deltaminmstar}{(\deltaminm)^*}

\newcommand{\deltamaxmstar}{(\deltamaxm)^*}

\newcommand{\dqed}{\hfill$\diamond$}
\newcommand{\inv}[1]{{#1}^{-1}}
\newcommand{\sigmainv}{\inv{\sigma}}
\newcommand{\finverse}{\inv{f}}

\newcommand{\fbar}{\overline{f}}
\newcommand{\fstar}{f^*}
\newcommand{\fstaro}{\fstar\O}
\newcommand{\Fstar}{F_*}
\newcommand{\Fstaro}{\Fstar\O}
\newcommand{\gammao}{\gamma^{\O}}
\newcommand{\gammafstaro}{\gamma^{\fstar\O}}
\newcommand{\gammap}{\gamma^{\P}}

\newcommand{\Sinv}{\inv{S}}
\newcommand{\sinv}{\inv{s}}
\newcommand{\Szero}{S_0}
\newcommand{\xinv}{\inv{x}}
\newcommand{\yinv}{\inv{y}}

\newcommand{\Config}{\triangle} 

\newcommand{\Configc}{\Config^{\!\C}}
\newcommand{\Configczero}{\Configc_{0}}
\newcommand{\Configcmax}{\Configc_{\mathsf{max}}}
\newcommand{\Configcmin}{\Configc_{\mathsf{min}}}
\newcommand{\Configd}{\Config^{\!\D}}
\newcommand{\Configl}{\Config^{\! L}}

\newcommand{\Configx}{\Config^{\! X}}
\newcommand{\Configxg}{\Configx_G}

\newcommand{\perpen}{~\perp}
\newcommand{\perpmax}{\perp_{\mathsf{max}}}
\newcommand{\perpenc}{\perpen^{\C}}
\newcommand{\perpend}{\perpen^{\D}}
\newcommand{\perpc}{\perp^{\C}}
\newcommand{\perpd}{\perp^{\D}}

\newcommand{\Cbar}{\overline{\C}}
\newcommand{\Cbarmin}{\overline{\C_{\mathsf{min}}}}
\newcommand{\Cbarmax}{\overline{\C_{\mathsf{max}}}}

\newcommand{\Chat}{\widehat{\C}}
\newcommand{\Chatmin}{\widehat{\C_{\mathsf{min}}}}
\newcommand{\Chatmax}{\widehat{\C_{\mathsf{max}}}}
\newcommand{\Csinv}{\C[\inv{S}]}
\newcommand{\Csinvbar}{\overline{\Csinv}}
\newcommand{\Csinvdiag}{\Csinv^{\diag}}

\newcommand{\Lbar}{\overline{L}}
\newcommand{\Lhat}{\widehat{L}}
\newcommand{\Osubc}{\O_{\C}}
\newcommand{\Ocm}{\O_{\C}^{\M}}
\newcommand{\Ocbar}{\O_{\Cbar}}
\newcommand{\Ocbarmin}{\O_{\Cbarmin}}
\newcommand{\Ocbarmax}{\O_{\Cbarmax}}
\newcommand{\Ocsinvbar}{\O_{\Csinvbar}}
\newcommand{\Ochat}{\O_{\Chat}}
\newcommand{\Ochatsinv}{\Ochat[\inv{S}]}
\newcommand{\Ochatszeroinv}{\Ochat[\inv{\Szero}]}
\newcommand{\Ocbarm}{\Ocbar^{\M}}
\newcommand{\Ocbarminm}{\Ocbarmin^{\M}}
\newcommand{\Ocbarmaxm}{\Ocbarmax^{\M}}
\newcommand{\Ocsinvbarm}{\Ocsinvbar^{\M}}
\newcommand{\Ochatm}{\Ochat^{\M}}
\newcommand{\Ochatminm}{\Otom_{\Chatmin}}
\newcommand{\Ochatmaxm}{\Otom_{\Chatmax}}
\newcommand{\Ochatsinvm}{\Ochatsinv^{\M}}
\newcommand{\Ochatszeroinvm}{\Ochatszeroinv^{\M}}
\newcommand{\Olhat}{\O_{\Lhat}}

\newcommand{\Osinv}{\O[\Sinv]}
\newcommand{\Osinvtom}{\Osinv^{\M}}
\newcommand{\Osinvm}{\Osinvtom}

\newcommand{\Dbar}{\overline{\D}}

\newcommand{\Dhat}{\widehat{\D}}
\newcommand{\Odbar}{\O_{\Dbar}}
\newcommand{\Odhat}{\O_{\Dhat}}
\newcommand{\Odbarm}{\Odbar^{\M}}
\newcommand{\Odhatm}{\Odhat^{\M}}
\newcommand{\Odhatsinv}{\Odhat[\Sinv]}
\newcommand{\Odhatsinvm}{\Odhatsinv^{\M}}

\newcommand{\Ab}{\mathsf{Ab}}
\newcommand{\As}{\mathsf{As}}
\newcommand{\Was}{\W\As}

\newcommand{\Bgloc}{\Locd_G}
\newcommand{\Bglocsginv}{\Bgloc[\inv{S_G}]}
\newcommand{\Bglocsginvbar}{\overline{\Bglocsginv}}
\newcommand{\Bglocbar}{\overline{\Bgloc}}

\newcommand{\Bgconloc}{\Locd_{G,\mathsf{con}}}
\newcommand{\Bgconlocbar}{\overline{\Bgconloc}}

\newcommand{\Bgconlocsginv}{\Bgconloc[\inv{S_G}]}
\newcommand{\Bgconlocsginvbar}{\overline{\Bgconlocsginv}}

\newcommand{\Cdiag}{\C^{\mathsf{diag}}}
\newcommand{\Wcdiag}{\W\Cdiag}
\newcommand{\Wcsinvdiag}{\W\Csinvdiag}
\newcommand{\Cat}{\mathsf{Cat}}
\newcommand{\Chain}{\mathsf{Chain}}
\newcommand{\Chaink}{\mathsf{Chain}_{\fieldk}}
\newcommand{\Chainz}{\mathsf{Chain}_{\mathbb{Z}}}
\newcommand{\Com}{\mathsf{Com}}
\newcommand{\Comc}{\Com^{\C}}
\newcommand{\Comcset}{\Comc_{\Set}}
\newcommand{\Comm}{\Com(\M)}
\newcommand{\Commc}{\Comm^{\C}}
\newcommand{\Wcom}{\W\Com}
\newcommand{\Wcomc}{\W\Comc}
\newcommand{\Configcat}{\mathsf{ConfCat}}

\newcommand{\Ddiag}{\D^{\mathsf{diag}}}

\newcommand{\Disc}{\mathsf{Disc}}

\newcommand{\Discd}{\Disc^d}
\newcommand{\Discdbar}{\overline{\Discd}}
\newcommand{\Discdhat}{\widehat{\Discd}}

\renewcommand{\emptyset}{\varnothing}
\newcommand{\Emb}{\mathsf{Emb}}
\newcommand{\Embn}{\Emb^n}
\newcommand{\Embnbarmax}{\overline{\Embn_{\mathsf{max}}}}
\newcommand{\Embnhatmax}{\widehat{\Embn_{\mathsf{max}}}}
\newcommand{\Fdiag}{F^{\mathsf{diag}}}
\newcommand{\Fdiagstar}{(F^{\mathsf{diag}})^*}
\newcommand{\Fun}{\mathsf{Fun}}
\newcommand{\Gh}{\mathsf{Gh}}
\newcommand{\Ghx}{\Gh(X)}

\newcommand{\Ghxbar}{\overline{\Ghx}}
\newcommand{\Ghxsinv}{\Ghx[\Sinv]}
\newcommand{\Ghxsinvbar}{\overline{\Ghx[\Sinv]}}
\newcommand{\Ghxhat}{\widehat{\Ghx}}

\newcommand{\Grp}{\mathsf{Grp}}

\newcommand{\Hol}{\mathsf{Hol}}
\newcommand{\Holn}{\Hol^n}
\newcommand{\Holnbarmax}{\overline{\Holn_{\mathsf{max}}}}
\newcommand{\Holnhatmax}{\widehat{\Holn_{\mathsf{max}}}}
\newcommand{\Loc}{\mathsf{Loc}}

\newcommand{\Locd}{\Loc^d}
\newcommand{\Locdbar}{\overline{\Locd}}
\newcommand{\Locdhat}{\widehat{\Locd}}
\newcommand{\Locdsinv}{\Locd[\Sinv]}
\newcommand{\Locdsinvbar}{\overline{\Locdsinv}}

\newcommand{\Man}{\mathsf{Man}}

\newcommand{\Mand}{\Man^d}
\newcommand{\Mandbar}{\overline{\Mand}}
\newcommand{\Mandhat}{\widehat{\Mand}}
\newcommand{\MFun}{\mathsf{MFun}}
\newcommand{\Mod}{\mathsf{Mod}}
\newcommand{\Mon}{\mathsf{Mon}}
\newcommand{\Monm}{\Mon(\M)}
\newcommand{\Monmc}{\Monm^{\C}}

\newcommand{\Open}{\mathsf{Open}}
\newcommand{\Openx}{\Open(X)}
\newcommand{\Openr}{\Open(\fieldr)}
\newcommand{\Openxbar}{\overline{\Openx}}
\newcommand{\Openxhat}{\widehat{\Openx}}
\newcommand{\Openrbar}{\overline{\Openr}}
\newcommand{\Openrhat}{\widehat{\Openr}}
\newcommand{\Opensonebar}{\overline{\Open(S^1)}}
\newcommand{\Openxg}{\Openx_G}
\newcommand{\Openxgbar}{\overline{\Openxg}}
\newcommand{\Openxghat}{\widehat{\Openxg}}

\newcommand{\Operad}{\mathsf{Operad}}
\newcommand{\Operadc}{\Operad^{\colorc}}
\newcommand{\Operadcset}{\Operadc(\Set)}
\newcommand{\Operadcm}{\Operadc(\M)}
\newcommand{\Operadm}{\Operad(\M)}
\newcommand{\Operadcn}{\Operadc(\N)}
\newcommand{\Orthcat}{\mathsf{OrthCat}}
\newcommand{\PFA}{\mathsf{PFA}}
\newcommand{\HPFA}{\mathsf{HPFA}}
\newcommand{\QFT}{\mathsf{QFT}}
\newcommand{\HQFT}{\mathsf{HQFT}}

\newcommand{\Reg}{\mathsf{Reg}}
\newcommand{\Regx}{\Reg(X)}
\newcommand{\Regxbar}{\overline{\Regx}}
\newcommand{\Regxhat}{\widehat{\Regx}}
\newcommand{\Regxsinv}{\Regx[\inv{S_X}]}
\newcommand{\Regxsinvbar}{\overline{\Regxsinv}}

\newcommand{\Regxzero}{\Reg(X_0)}
\newcommand{\Regxzerobar}{\overline{\Regxzero}}
\newcommand{\Regxzerosinv}{\Regxzero[\inv{S_{X_0}}]}
\newcommand{\Regxzerosinvbar}{\overline{\Regxzerosinv}}

\newcommand{\Riem}{\mathsf{Riem}}

\newcommand{\Riemd}{\Riem^d}
\newcommand{\Riemdbar}{\overline{\Riemd}}
\newcommand{\Riemdhat}{\widehat{\Riemd}}

\newcommand{\Sloc}{\mathsf{SLoc}}
\newcommand{\Slocd}{\Sloc^d}
\newcommand{\Slocdbar}{\overline{\Slocd}}

\newcommand{\Slocdsinv}{\Slocd[\inv{S_{\pi}}]}
\newcommand{\Slocdsinvbar}{\overline{\Slocdsinv}}

\newcommand{\Set}{\mathsf{Set}}
\newcommand{\Sset}{\mathsf{SSet}}
\newcommand{\SMFun}{\mathsf{SMFun}}
\newcommand{\Str}{\mathsf{Str}}
\newcommand{\Strbar}{\overline{\Str}}
\newcommand{\Strhat}{\widehat{\Str}}
\newcommand{\Strsinv}{\Str[\inv{S_{\pi}}]}
\newcommand{\Strsinvbar}{\overline{\Strsinv}}

\newcommand{\Top}{\mathsf{Top}}
\newcommand{\Linear}{\mathsf{Linear}}
\newcommand{\uLinear}{\underline{\Linear}}
\newcommand{\Linearc}{\Linear^{\colorc}}
\newcommand{\uLinearc}{\uLinear^{\colorc}}
\newcommand{\Tree}{\mathsf{Tree}}
\newcommand{\uTree}{\underline{\Tree}}

\newcommand{\uTreeb}{\uTree^{\colorb}}
\newcommand{\Treec}{\Tree^{\colorc}}
\newcommand{\uTreec}{\uTree^{\colorc}}
\newcommand{\uTreecn}{\uTreec_n}
\newcommand{\uTreeceqn}{\uTreec_{=n}}
\newcommand{\uTreecduc}{\uTreec\duc}
\newcommand{\uTreecnduc}{\uTreecn\duc}
\newcommand{\uTreeceqnduc}{\uTreeceqn\duc}
\newcommand{\uTreecducop}{\uTreecduc^{\op}}
\newcommand{\uTreecnducop}{\uTreecn\duc^{\op}}
\newcommand{\uTreeceqnducop}{\uTreeceqn\duc^{\op}}

\newcommand{\Treeopc}{\mathsf{TreeOp}^{\colorc}}

\newcommand{\uTreed}{\uTree^{\colord}}

\newcommand{\uTreezero}{\uTree^{\{0\}}}
\newcommand{\uTreezeroc}{\uTree^{\zeroc}}
\newcommand{\uTreezerod}{\uTree^{\zerod}}
\newcommand{\uTreezerocd}{\uTree^{\zerocd}}
\newcommand{\Vectk}{\mathsf{Vect}_{\fieldk}}

\newcommand{\wf}{\W f}
\newcommand{\wo}{\W\O}
\newcommand{\wno}{\W_n\O}
\newcommand{\weqno}{\W_{=n}\O}
\newcommand{\wom}{\W\Otom}
\newcommand{\wocm}{\W\Ocm}

\newcommand{\wochat}{\W\Ochat}
\newcommand{\wocbarminm}{\W\Ocbarminm}
\newcommand{\wocbarmaxm}{\W\Ocbarmaxm}

\newcommand{\wocbarm}{\W\Ocbarm}
\newcommand{\wocsinvbarm}{\W\Ocsinvbarm}
\newcommand{\wochatm}{\W\Ochatm}
\newcommand{\wochatminm}{\W\Ochatminm}
\newcommand{\wochatmaxm}{\W\Ochatmaxm}

\newcommand{\wochatsinv}{\wochat[\Sinv]}
\newcommand{\wochatsinvm}{\wochatsinv^{\M}}
\newcommand{\wodbarm}{\W\Odbarm}
\newcommand{\wodhatm}{\W\Odhatm}
\newcommand{\wofp}{\W\P}

\newcommand{\wolhatm}{\wom_{\Lhat}}

\newcommand{\Sigmaop}{\Sigma^{\op}}

\newcommand{\Sigmac}{\Sigma_{\colorc}}
\newcommand{\Sigmacop}{\Sigmaop_{\colorc}}
\newcommand{\Sigmacopc}{\Sigmacop\times\colorc}
\newcommand{\Mtoc}{\M^{\colorc}}
\newcommand{\Mcstar}{\M^{\C}_*}
\newcommand{\Mtod}{\M^{\colord}}

\newcommand{\symseq}{\mathsf{SymSeq}}
\newcommand{\symseqcm}{\symseq^{\colorc}(\M)}

\newcommand{\alg}{\mathsf{Alg}}
\newcommand{\algc}{\alg_{\C}}
\newcommand{\algct}{\algc(T)}

\newcommand{\algm}{\alg_{\M}}
\newcommand{\algmas}{\algm(\As)}
\newcommand{\algmcdiag}{\algm(\Cdiag)}
\newcommand{\algmcom}{\algm(\Com)}
\newcommand{\algmcomc}{\algm\bigl(\Comc\bigr)}
\newcommand{\algmwcom}{\algm(\Wcom)}
\newcommand{\algmwcomc}{\algm\bigl(\Wcomc\bigr)}

\newcommand{\algmo}{\algm(\O)}
\newcommand{\algmotom}{\algm(\Otom)}
\newcommand{\algmocm}{\algm(\Ocm)}

\newcommand{\algmosinvtom}{\algm\bigl(\Osinvtom\bigr)}

\newcommand{\algmocbarm}{\algm\bigl(\Ocbarm\bigr)}
\newcommand{\algmocsinvbarm}{\algm\bigl(\Ocsinvbarm\bigr)}

\newcommand{\algmochatm}{\algm\bigl(\Ochat^{\M}\bigr)}
\newcommand{\algmochatsinvm}{\algm\bigl(\Ochatsinvm\bigr)}
\newcommand{\algmochatszeroinvm}{\algm\bigl(\Ochatszeroinvm\bigr)}
\newcommand{\algmodhatsinvm}{\algm\bigl(\Odhatsinvm\bigr)}

\newcommand{\algmwas}{\algm(\Was)}
\newcommand{\algmwcdiag}{\algm(\Wcdiag)}
\newcommand{\algmwo}{\algm(\wo)}
\newcommand{\algmwocm}{\algm(\wocm)}

\newcommand{\algmwocbarm}{\algm\bigl(\wocbarm\bigr)}
\newcommand{\algmwocsinvbarm}{\algm\bigl(\wocsinvbarm\bigr)}
\newcommand{\algmwochatm}{\algm\bigl(\wochatm\bigr)}
\newcommand{\algmwocbarminm}{\algm\bigl(\wocbarminm\bigr)}
\newcommand{\algmwocbarmaxm}{\algm\bigl(\wocbarmaxm\bigr)}
\newcommand{\algmwochatsinvm}{\algm\bigl(\wochatsinvm\bigr)}
\newcommand{\algmwochatminm}{\algm\bigl(\wochatminm\bigr)}
\newcommand{\algmwochatmaxm}{\algm\bigl(\wochatmaxm\bigr)}

\newcommand{\algmodbarm}{\algm\bigl(\Odbarm\bigr)}
\newcommand{\algmodhatm}{\algm\bigl(\Odhatm\bigr)}
\newcommand{\algmwodbarm}{\algm\bigl(\wodbarm\bigr)}
\newcommand{\algmwodhatm}{\algm\bigl(\wodhatm\bigr)}

\newcommand{\algmp}{\algm(\P)}

\newcommand{\algmwp}{\algm(\wofp)}

\newcommand{\ua}{\underline a}
\newcommand{\ub}{\underline b}
\newcommand{\uc}{\underline c}
\newcommand{\ud}{\underline d}
\newcommand{\ue}{\underline e}
\newcommand{\uf}{\underline f}
\newcommand{\ug}{\underline g}
\newcommand{\us}{\underline s}
\newcommand{\uU}{\underline U}

\newcommand{\smallprof}[1]
{\raisebox{.05cm}{\scalebox{0.8}{#1}}}

\newcommand{\sbinom}[2]{\raisebox{.05cm}{\scalebox{0.8}{$\binom{#1}{#2}$}}}
\newcommand{\inout}[1]{\raisebox{.05cm}{\scalebox{0.8}{$\binom{\out(#1)}{\inp(#1)}$}}}
\newcommand{\inoutv}{\inout{v}}
\newcommand{\finoutv}{\sbinom{f\out(v)}{f\inp(v)}}
\newcommand{\bua}{\smallprof{$\binom{b}{\ua}$}}
\newcommand{\baoneam}{\smallprof{$\binom{b}{a_1,\ldots,a_m}$}}
\newcommand{\bjiuaji}{\smallprof{$\binom{b^j_i}{\ua^j_i}$}}

\newcommand{\ca}{\smallprof{$\binom{c}{a}$}}
\newcommand{\cab}{\smallprof{$\binom{c}{a,b}$}}
\newcommand{\cempty}{\smallprof{$\binom{c}{\varnothing}$}}

\newcommand{\cjuaj}{\smallprof{$\binom{c_j}{\ua_j}$}}
\newcommand{\cb}{\smallprof{$\binom{c}{b}$}}
\newcommand{\ciub}{\smallprof{$\binom{c_i}{\ub}$}}
\newcommand{\ciubi}{\smallprof{$\binom{c_i}{\ub_i}$}}
\newcommand{\cjub}{\smallprof{$\binom{c_j}{\ub}$}}
\newcommand{\cjubj}{\smallprof{$\binom{c_j}{\ub_j}$}}
\newcommand{\cjubjtauj}{\smallprof{$\binom{c_j}{\ub_j\tau_j}$}}
\newcommand{\csigmajubsigmaj}{\smallprof{$\binom{c_{\sigma(j)}}{\ub_{\sigma(j)}}$}}
\newcommand{\cc}{\smallprof{$\binom{c}{c}$}}

\newcommand{\cici}{\smallprof{$\binom{c_i}{c_i}$}}
\newcommand{\cjcj}{\smallprof{$\binom{c_j}{c_j}$}}
\newcommand{\cd}{\smallprof{$\binom{c}{d}$}}
\newcommand{\cjudj}{\smallprof{$\binom{c_j}{\ud_j}$}}

\newcommand{\ccc}{\smallprof{$\binom{c}{c,\ldots,c}$}}

\newcommand{\czerozeroc}{\smallprof{$\binom{c}{0,\ldots,0,c}$}}

\newcommand{\dempty}{\smallprof{$\binom{d}{\varnothing}$}}

\newcommand{\dua}{\smallprof{$\binom{d}{\ua}$}}
\newcommand{\dub}{\smallprof{$\binom{d}{\ub}$}}

\newcommand{\duboneubn}{\smallprof{$\binom{d}{\ub_1,\ldots,\ub_n}$}}
\newcommand{\dubsigmaoneubsigman}{\smallprof{$\binom{d}{\ub_{\sigma(1)},\ldots,\ub_{\sigma(n)}}$}}
\newcommand{\dubonetauoneubntaun}{\smallprof{$\binom{d}{\ub_1\tau_1,\ldots,\ub_n\tau_n}$}}
\newcommand{\dc}{\smallprof{$\binom{d}{c}$}}
\newcommand{\dcc}{\smallprof{$\binom{d}{c,\ldots,c}$}}
\newcommand{\ddd}{\smallprof{$\binom{d}{d,\ldots,d}$}}
\newcommand{\FdFc}{\smallprof{$\binom{Fd}{Fc}$}}
\newcommand{\duc}{\smallprof{$\binom{d}{\uc}$}}
\newcommand{\ducsigma}{\smallprof{$\binom{d}{\uc\sigma}$}}
\newcommand{\ductau}{\smallprof{$\binom{d}{\uc\tau}$}}

\newcommand{\dconecn}{\smallprof{$\binom{d}{c_1,\ldots,c_n}$}}
\newcommand{\dconecm}{\smallprof{$\binom{d}{c_1,\ldots,c_m}$}}
\newcommand{\dud}{\smallprof{$\binom{d}{\ud}$}}

\newcommand{\fdufc}{\smallprof{$\binom{fd}{f\uc}$}}
\newcommand{\fdfuc}{\fdufc}
\newcommand{\fzerodufzeroc}{\smallprof{$\binom{f_0d}{f_0\uc}$}}
\newcommand{\dd}{\smallprof{$\binom{d}{d}$}}

\newcommand{\dzerozeroc}{\smallprof{$\binom{d}{0,\ldots,0,c}$}}
\newcommand{\dzerozerod}{\smallprof{$\binom{d}{0,\ldots,0,d}$}}
\newcommand{\starnstar}{\smallprof{$\binom{*}{*,\ldots,*}$}}
\newcommand{\tus}{\smallprof{$\binom{t}{\us}$}}
\newcommand{\VuU}{\smallprof{$\binom{V}{\uU}$}}
\newcommand{\vuoneum}{\smallprof{$\binom{V}{U_1,\ldots,U_m}$}}

\newcommand{\zerozerozero}{\smallprof{$\binom{0}{0,\ldots,0}$}}

\newcommand{\inp}{\mathsf{in}}
\newcommand{\out}{\mathsf{out}}

\newcommand{\andspace}{\quad\text{and}\quad}
\newcommand{\byspace}{\quad\text{by}\quad}
\newcommand{\ifspace}{\quad\text{if}\quad}
\newcommand{\iffspace}{\quad\text{if and only if}\quad}
\newcommand{\inspace}{\quad\text{in}\quad}
\newcommand{\forspace}{\quad\text{for}\quad}
\newcommand{\forsomespace}{\quad\text{for some}\quad}

\newcommand{\stspace}{\quad\text{such that}\quad}

\newcommand{\withspace}{\quad\text{with}\quad}

\makeindex

\begin{document}
\title{Homotopical Quantum Field Theory}

\author{Donald Yau}
\address{Department of Mathematics\\
	 The Ohio State University at Newark\\
	 1179 University Drive\\ 
	 Newark, OH 43055, USA}
\email{yau.22@osu.edu}


\begin{abstract}
Algebraic quantum field theory and prefactorization algebra are two mathematical approaches to quantum field theory.  In this monograph, using a new coend definition of the Boardman-Vogt construction of a colored operad, we define homotopy algebraic quantum field theories and homotopy prefactorization algebras and investigate their homotopy coherent structures.  Homotopy coherent diagrams, homotopy inverses, $A_\infty$-algebras, $E_\infty$-algebras, and $E_\infty$-modules arise naturally in this context.  In particular, each homotopy algebraic quantum field theory has the structure of a homotopy coherent diagram of $A_\infty$-algebras and satisfies a homotopy coherent version of the causality axiom.  When the time-slice axiom is defined for algebraic quantum field theory, a homotopy coherent version of the time-slice axiom is satisfied by each homotopy algebraic quantum field theory.  Over each topological space, every homotopy prefactorization algebra has the structure of a homotopy coherent diagram of $E_\infty$-modules over an $E_\infty$-algebra.  To compare the two approaches, we construct a comparison morphism from the colored operad for (homotopy) prefactorization algebras to the colored operad for (homotopy) algebraic quantum field theories and study the induced adjunctions on algebras.
\end{abstract}

\subjclass[2000]{18A25, 18A40, 18D10, 18D50, 18G55, 81T05}
\keywords{Algebraic quantum field theory, prefactorization algebra, causality axiom, time-slice axiom, operad, Boardman-Vogt construction, homotopy coherent diagram, $A_\infty$-algebra, $E_\infty$-algebra, $E_\infty$-module.}

\date{\today}

\maketitle

\cleardoublepage
\thispagestyle{empty}
\vspace*{13.5pc}
\begin{center}
To Eun Soo and Jacqueline
\end{center}
\cleardoublepage

\setcounter{page}{6}

\tableofcontents

\chapter{Introduction}\label{ch:introduction}

Algebraic quantum field theory and prefactorization algebra are two mathematical approaches to quantum field theory.  One of the main aims of this book is to provide robust definitions of homotopy algebraic quantum field theories and homotopy prefactorization algebras using a new definition of the Boardman-Vogt construction of a colored operad.  To compare the two mathematical approaches to quantum field theory as well as their homotopy coherent analogues, we work within the framework of operads.  This approach allows us to employ the powerful machinery from operad theory to quantum field theory.  In the rest of this introduction, we briefly introduce each of these topics without going into too much details.

\section{Algebraic Quantum Field Theory}\label{sec:intro-aqft}

Algebraic quantum field theory as introduced by Haag and Kastler \cite{hk} provides one mathematical approach to quantum field theory that takes into account both quantum features and the theory of relativity.  An algebraic quantum field theory $\fraka$ assigns to each suitable spacetime region $U$ in a fixed Lorentzian spacetime $X$ an algebra $\fraka(U)$.  To each inclusion $i^V_U : U \subset V$, it assigns an algebra morphism \[\fraka(i^V_U) : \fraka(U) \to \fraka(V)\] in a functorial way.  In other words, the morphism assigned to $i^U_U : U = U$ is the identity morphism of $\fraka(U)$, and if $U \subset V \subset W$ then there is an equality
\begin{equation}\label{aqft-functoriality}
\fraka(i^W_V) \circ \fraka(i^V_U) = \fraka(i^W_U).
\end{equation}
This is just another way of saying that $\fraka$ is a functor from the category of spacetime regions in $X$ to the category of algebras.  Physically $\fraka(U)$ is the algebra of quantum observables in the region $U$.  Each algebra $\fraka(U)$ is only required to be associative, not commutative as in the classical case. The morphism $\fraka(i^V_U)$ sends observables in $U$ to observables in $V$.  

An algebraic quantum field theory is more than just a functor from the category of spacetime regions in $X$ to algebras.  It is required to satisfy Einstein's \emph{causality axiom}.  It states that if $U$ and $V$ are causally disjoint regions in $W \subset X$, then the images of $\fraka(U)$ and $\fraka(V)$ in $\fraka(W)$ commute.  The causality axiom, also known as causal locality\index{causal locality} or just \index{locality}locality, is a precise way of saying that physical influences cannot propagate faster than the speed of light.  So causally disjoint regions are independent systems.  An algebraic quantum field theory is also required to satisfy the \emph{time-slice axiom}.  It states that if $U \subset V$ contains a Cauchy surface of $V$, then the morphism \[\fraka(i^V_U) : \fraka(U) \iso \fraka(V)\] is an isomorphism of algebras.  Physically this means that all the observables in a spacetime region $V$ are already determined by observables in a small time interval.

Traditionally one also asks that $\fraka$ satisfy the \index{isotony axiom}\emph{isotony axiom}, which states that each $\fraka(i^V_U)$ is an injective morphism of algebras.  However, various models of  quantum gauge theories do not satisfy the isotony axiom.  Therefore, recent literature on algebraic quantum field theory does not always include the isotony axiom.  We follow this practice and only ask that each $\fraka(i^V_U)$ be a morphism of algebras.

The Haag-Kastler framework is flexible in the sense that one can replace the domain category of spacetime regions in a fixed spacetime by another category $\C$ of spacetimes to obtain other versions of quantum field theories.  One example is the category of all oriented, time-oriented, and globally hyperbolic Lorentzian manifolds of a fixed dimension.  The resulting algebraic quantum field theories are locally covariant quantum field theories \cite{bfv,fewster,fv}.  Similarly, to obtain chiral conformal quantum field theories \cite{bdh} and Euclidean quantum field theories \cite{schlingemann}, one uses the domain category of oriented manifolds and oriented Riemannian manifolds of a fixed dimension.   One can also consider the category of spacetimes with extra geometric structures, such as principal bundles, connections, and spin structure \cite{bs17}, and the category of spacetimes with timelike boundaries \cite{bds}.  

To implement the causality axiom, one asks for a small category $\C$ that has a chosen subset $\perp$ of pairs of morphisms $\bigl\{\nicexy@C-.5cm{U_1 \ar[r] & V & U_2 \ar[l]}\bigr\}$ with a common codomain.  Such a pair formalizes the idea that $U_1$ and $U_2$ are disjoint in $V$.  The pair $(\C,\perp)$ is called an \emph{orthogonal category} \cite{bsw}.  To implement the time-slice axiom, one chooses a subset $S$ of morphisms in $\C$, which in the Lorentzian case is the set of Cauchy morphisms.  An algebraic quantum field theory satisfies the time-slice axiom if the structure morphisms corresponding to morphisms in $S$ are isomorphisms.

Furthermore, the target category of algebras over a field $\fieldk$ can also be replaced by other categories of algebras.  For example, instead of $\fieldk$-algebras, which are monoids in the category $\Vectk$ of $\fieldk$-vector spaces, one can use differential graded $\fieldk$-algebras, which are monoids in the category $\Chaink$ of chain complexes of $\fieldk$-vector spaces.  In fact, conceptually it is easier to consider the category $\Monm$ of monoids in a symmetric monoidal category $\M$ and then specify $\M$ as $\Vectk$, $\Chaink$, or whatever setting one wishes to work in, later if necessary.

In short, an algebraic quantum field theory on an orthogonal category $(\C,\perp)$ is a functor \[\fraka : \C \to \Monm\] that satisfies the causality axiom and, if a set $S$ of morphisms in $\C$ is chosen, the time-slice axiom with respect to $S$.  This definition of an algebraic quantum field theory was introduced in \cite{bsw}, and we adopt it in this book.

\section{Homotopy Algebraic Quantum Field Theory}\label{sec:intro-haqft}

Homotopy theory enters the picture with (i) recent toy examples of \index{quantum gauge theory}quantum gauge theories in  \cite{bs17} that are algebraic quantum field theories up to homotopy and (ii) the beginning of a program in \cite{bss} to study \index{quantum Yang-Mills theory}quantum Yang-Mills theory using the homotopy theory of \index{stack}stacks \cite{hola,holb,holc}.  At the most elementary level, this means that the strict equalities in  algebraic quantum field theories are replaced by homotopies.  For example, the equality \eqref{aqft-functoriality} that expresses functoriality is replaced by the homotopy relation \[\fraka(i^W_V) \circ \fraka(i^V_U) \sim \fraka(i^W_U).\] In other words, the composition on the left is chain homotopic to the morphism on the right.  The causality axiom is replaced by a similar homotopical analogue that expresses commutativity up to chain homotopy.  The homotopical version of the time-slice axiom says that, if $s \in S$ is one of the chosen morphisms, then $\fraka(s)$ is a chain homotopy equivalence.

An important lesson from homotopy theory is that homotopies are only the first layer of a much richer homotopy coherent structure.  This means that we do not simply ask for two things to be homotopic.  Instead we ask for specific homotopies as part of the algebraic structure itself, and these homotopies satisfy higher homotopy relations via further structure morphisms, and so forth.  For instance, for monoids and commutative monoids, the higher homotopical analogues are called\index{ainfinityalgebra@$A_\infty$-algebra} $A_\infty$-algebras and\index{einfinityalgebra@$E_\infty$-algebra} $E_\infty$-algebras, respectively.  Set theoretically, in an $A_\infty$-algebra, instead of strict associativity $(ab)c = a(bc)$, it has a specific structure morphism that is a homotopy $(ab)c \sim a(bc)$.  Instead of a strict two-sided unit $1$, it has specific structure morphisms that are homotopies $1a \sim a \sim a1$.  There is an infinite family of higher structure morphisms that relate these homotopies.

Operad theory is a powerful framework originating from homotopy theory that allows one to keep track of the enormous amount of data in higher homotopical structures in a manageable way.  An operad $\O$ has a set of objects $\colorc$, like a small category, but the domain of a morphism \[\nicexy{(c_1,\ldots,c_n) \ar[r]^-{f} & d}\] is a finite, possibly empty, sequence of objects.  Just like in a category, one can compose these morphisms.  Moreover, the domain objects can be permuted.  These morphisms should be thought of as models of $n$-ary operations, and they satisfy some reasonable unity, equivariance, and associativity axioms.  To emphasize the set $\colorc$ of objects, we call it a $\colorc$-colored operad.

Similar to an algebra or a monoid, an operad $\O$ can act on objects, called $\O$-algebras.  For example, there is an associative operad $\As$ whose algebras are monoids, and there is a commutative operad $\Com$ whose algebras are commutative monoids.  As a general rule, if there is an operad for a certain type of structure, then there is a colored operad for $\C$-diagrams of such structure.  So there is a colored operad whose algebras are $\C$-diagrams of monoids in $\M$, i.e., functors $\C \to \Monm$.  With more work, one can even write down a colored operad $\Ocbarm$ whose algebras are algebraic quantum field theories on an orthogonal category $\Cbar=(\C,\perp)$.  In other words, it is possible to incorporate the causality axiom, which is a kind of commutativity, and the time-slice axiom, which is a kind of invertibility, into the colored operad itself.  The construction of the colored operad $\Ocbarm$ for algebraic quantum field theories was made explicit in \cite{bsw}.

To capture homotopy algebraic quantum field theories with all of the higher homotopical structure, we once again follow an established principle in homotopy theory.  If $\O$ is an operad for a certain kind of algebras, then the homotopy coherent versions of these algebras are obtained as algebras over a suitable \index{resolution}resolution $\P \weq \O$ of $\O$.  This is analogous to replacing a module by a \index{projective resolution}projective resolution in homological algebra, so we want $\P$ to be nice in some way.  In the terminology of model category theory, we ask $\P$ to be a cofibrant resolution of $\O$ in the model category of operads.  In homological algebra we learned that projective resolutions of a given module are not unique, and which projective resolution to use depends on one's intended applications.  For instance, corresponding to monoids and commutative monoids, there are different versions of $A_\infty$-algebras and $E_\infty$-algebras, depending on which resolutions one chooses for the associative operad and the commutative operad.  A good choice of a resolution of the colored operad $\Ocbarm$ for algebraic quantum field theories is its Boardman-Vogt resolution, which was originally defined for topological operads in \cite{boardman-vogt} to study homotopy invariant algebraic structures.

In \cite{berger-moerdijk-bv,berger-moerdijk-resolution} the Boardman-Vogt construction of a colored operad was extended from the category of topological spaces to a general symmetric monoidal category equipped with a segment, which provides a concept of length.  For example, for topological spaces, a segment is given by the unit interval $[0,1]$.  For chain complexes over $\fieldk$, a segment is given by the two-stage complex $\nicexy{\fieldk \ar[r]^-{(+,-)} & \fieldk \oplus \fieldk}$ concentrated in degrees $1$ and $0$.  In \cite{berger-moerdijk-bv} the Boardman-Vogt construction of an operad $\O$ is entrywise defined inductively as a sequential colimit, with each morphism in the sequence defined as a pushout that takes input from the previous inductive stage.  To effectively apply the machinery to quantum field theory, we need a more direct construction.  So we will introduce a new definition of the Boardman-Vogt construction of a colored operad that is entrywise defined in one step as a coend indexed by a category of trees, called a \emph{substitution category}.  

Given a flavor of spacetimes, i.e., a choice of an orthogonal category $\Cbar = (\C,\perp)$, we will define homotopy algebraic quantum field theories on $\Cbar$ as algebras over the Boardman-Vogt construction $\wocbarm$ of the colored operad $\Ocbarm$.  All of the higher homotopy relations in homotopy algebraic quantum field theories are parametrized by the substitution categories in the coends.  An algebraic quantum field theory is an $\Ocbarm$-algebra, which in turn is a $\C$-diagram of monoids $\C \to \Monm$ that satisfies the causality axiom and possibly the time-slice axiom if a set $S$ of morphisms in $\C$ is given.  Replacing everything by their higher homotopical analogues, we will show that every homotopy algebraic quantum field theory on $\Cbar$, i.e., $\wocbarm$-algebra, has the structure of a homotopy coherent $\C$-diagram of $A_\infty$-algebras that satisfies a homotopy coherent version of the causality axiom.  Furthermore, if a set $S$ of morphisms in $\C$ is given, then it also satisfies a homotopy coherent version of the time-slice axiom.  

An important point here is that all of the higher homotopies, such as the ones expressing homotopy functoriality, homotopy causality, and homotopy time-slice, are specific structure morphisms of a $\wocbarm$-algebra.  In other words, all of the higher homotopies are already encoded in the Boardman-Vogt construction $\wocbarm$ itself.  Our coend definition of the Boardman-Vogt construction plays a crucial role in our understanding of the structure in homotopy algebraic quantum field theories.

\section{Homotopy Prefactorization Algebra}\label{sec:intro-pfa}

Prefactorization algebras were introduced in \cite{cg} to provide another mathematical framework for quantum field theory that is analogous to the deformation quantization approach to quantum mechanics.  For a given topological space $X$, to each open subset $U \subset X$, a prefactorization algebra $\scF$ on $X$ assigns a chain complex $\scF(U)$.  To each finite sequence $U_1,\ldots,U_n$ of pairwise disjoint open subsets in $V \subset X$, $\scF$ assigns a chain map \[\nicexy@C+.8cm{\scF(U_1) \otimes \cdots \otimes \scF(U_n) \ar[r]^-{\scF^V_{U_1,\ldots,U_n}} & \scF(V)}.\]  In particular, for an inclusion $U \subset V$ of open subsets in $X$, \[\scF^V_U : \scF(U) \to \scF(V)\] is a chain map.  This data is required to satisfy some reasonable unity, equivariance, and associativity conditions.  For example, for open subsets $U \subset V \subset W$ in $X$, a part of the associativity condition is the equality \[\scF^W_V \circ \scF^V_U = \scF^W_U : \scF(U) \to \scF(W).\]  In particular, a prefactorization algebra $\scF$ on $X$ has the structure of a functor \[\Openx \to \Chaink\] from the category of open subsets in $X$ with inclusions as morphisms.  There is also a time-slice axiom in this setting, called\index{local constancy} local constancy in \cite{cg}.  If $S$ is a chosen set of morphisms in $\Openx$, then one asks that each structure morphism $\scF^V_U$ with $(U \subset V) \in S$ be an isomorphism.

Physically $\scF(U)$ is the collection of quantum observables in $U$.  The chain map $\scF^V_{U_1,\ldots,U_n}$ means that if the $U_i$'s are pairwise disjoint in $V$, then their observables can be multiplied in $\scF(V)$.  This is the main difference between a prefactorization algebra and an algebraic quantum field theory.  In an algebraic quantum field theory, every object $\fraka(U)$ is a monoid, so observables in a spacetime region $U$ can always be multiplied.  On the other hand, in a prefactorization algebra $\scF$, only observables from pairwise disjoint regions can be multiplied.  Furthermore, part of the equivariance condition says that $\scF(\varnothing_X)$, where $\varnothing_X \subset X$ denotes the empty subset, is a commutative differential graded algebra.   For each open subset $V \subset X$, since $\varnothing_X$ is disjoint from $V$, there is a structure morphism \[\nicexy@C+1.3cm{\overbrace{\scF(\varnothing_X) \otimes \cdots \otimes \scF(V) \otimes \cdots \otimes \scF(\varnothing_X)}^{\mathrm{only~ one}~\scF(V)} \ar[r]^-{\scF^V_{\varnothing_X,\ldots,V,\ldots,\varnothing_X}} & \scF(V)}\] that gives each $\scF(V)$ the structure of an $\scF(\varnothing_X)$-module.  These objectwise $\scF(\varnothing_X)$-modules are compatible with the structure morphisms $\scF^V_U$.

To facilitate the comparison between the above two mathematical approaches to quantum field theory, we will take a slightly more abstract approach to prefactorization algebras.  To define a prefactorization algebra, what one really needs is a small category $\C$, whose objects are thought of as spacetime regions, that has a suitable notion of pairwise disjointedness.  In other words, one chooses a set $\Config$ of finite sequences of morphisms $\{f_i : U_i \shortto V\}_{i=1}^n$ in $\C$ with a common codomain.  Such a finite sequence, called a \emph{configuration}, formalizes the idea that the $U_i$'s are pairwise disjoint in $V$.  These configurations are required to satisfy some natural axioms, such as closure under composition and permutation.  The pair $\Chat = (\C,\Config)$ is called a \emph{configured category}.  

As in the case of algebraic quantum field theory, we allow the base category to be a general symmetric monoidal category $\M$ instead of just $\Chaink$.  A prefactorization algebra on a configured category $\Chat$ is defined as an algebra over the colored operad $\Ochatm$ whose entries are coproducts $\coprod_{\Config'} \tensorunit$, with $\Config'$ a suitable subset of configurations and $\tensorunit$ the monoidal unit in $\M$.  To implement the time-slice axiom with respect to a set $S$ of morphisms in $\C$, we replace the colored operad $\Ochatm$ by a suitable localization.

Proceeding as in the story of algebraic quantum field theory, we define homotopy prefactorization algebras on a configured category $\Chat$ as algebras over the Boardman-Vogt construction $\wochatm$ of the colored operad $\Ochatm$.    Once again due to our one-step coend definition of the Boardman-Vogt construction, we are able to make explicit the structure of homotopy prefactorization algebras.  Let us take as an example the configured category associated to the category $\Openx$ of open subsets of a topological space $X$ with configurations defined by pairwise disjointedness.  In this setting, we will show that a homotopy prefactorization algebra $Y$ has, first of all, an $E_\infty$-algebra structure in the entry $Y_{\varnothing_X}$ corresponding to the empty subset of $X$.  It also has the structure of a homotopy coherent $\Openx$-diagram and satisfies a homotopy coherent version of the time-slice axiom if a set $S$ of open subset inclusions is given.  Furthermore, for each open subset $V \subset X$, the entry $Y_V$ admits the structure of an $E_\infty$-module over the $E_\infty$-algebra $Y_{\varnothing_X}$.  These objectwise $E_\infty$-modules are homotopy coherently compatible with the homotopy coherent $\Openx$-diagram structure of $Y$.  Once again, all of the higher homotopical structure is already encoded in the Boardman-Vogt construction $\wochatm$ itself.

\section{Comparison}\label{sec:intro-comparison}

Given that both algebraic quantum field theory and prefactorization algebra are mathematical approaches to quantum field theory, a natural question is how they are related.  The two approaches certainly have something in common.  In both settings, we consider functors from some category $\C$, whose objects are thought of as spacetime regions, to some target category, such as $\Vectk$ or $\Chaink$.  Moreover, in each setting there is a time-slice axiom that says that some chosen structure morphisms are invertible.  Our comparison of the two approaches happens at two levels.  

We first compare orthogonal categories, on which (homotopy) algebraic quantum field theories are defined, and configured categories, on which (homotopy) prefactorization algebras are defined.  Informally, every orthogonal category generates a configured category, in which a configuration is a finite sequence of pairwise orthogonal morphisms.  Conversely, every configured category restricts to an orthogonal category, in which the orthogonal pairs are the binary configurations.  The precise version says that the category of orthogonal categories embeds as a full reflective subcategory in the category of configured categories.  This is not an adjoint equivalence, so the two categories are genuinely different.

Next we compare prefactorization algebras on a configured category $\Chat$ and algebraic quantum field theories on the associated orthogonal category $\Cbar$.  We construct a comparison morphism \[\Ochatm \to \Ocbarm\] from the colored operad $\Ochatm$ defining prefactorization algebras on $\Chat$ to the colored operad $\Ocbarm$ defining algebraic quantum field theories on $\Cbar$.  Since our Boardman-Vogt construction is natural, there is an induced comparison morphism \[\wochatm \to \wocbarm\] from the colored operad $\wochatm$ defining homotopy prefactorization algebras to the colored operad $\wocbarm$  defining homotopy algebraic quantum field theories.  These comparison morphisms induce various comparison adjunctions between (homotopy) prefactorization algebras and (homotopy) algebraic quantum field theories, with or without the time-slice axiom.  

Although prefactorization algebras and algebraic quantum field theories are different in general, there is one important case when they are equal.  This situation corresponds to the maximal configured category and the maximal orthogonal category for a given small category $\C$.  In this case, both the category of prefactorization algebras and the category of algebraic quantum field theories are isomorphic to the category of $\C$-diagrams of commutative monoids.  We interpret this situation as saying that the two mathematical approaches to quantum field theory both reduce to the \index{classical case}classical case, where observables form commutative algebras.  Furthermore, since $E_\infty$-algebras are homotopy coherent versions of commutative algebras, in this case both the category of homotopy prefactorization algebras and the category of homotopy algebraic quantum field theories are isomorphic to the category of homotopy coherent $\C$-diagrams of $E_\infty$-algebras.

\section{Organization}\label{sec:intro-organization}

This book is divided into two parts.  The first part is about operads, with special emphasize on our version of the Boardman-Vogt construction in terms of coends.  The second part is the application of the machinery in the first part to algebraic quantum field theory, prefactorization algebra, and their homotopy coherent analogues.  Each chapter has its own introduction.  A brief description of each chapter follows.

To keep this book relatively self-contained, Part 1 begins with Chapter \ref{ch:categories} in which we review basic concepts of category theory, including colimits, coends, adjoint functors, monoidal categories, monads, and localization.  Our coend definition of the Boardman-Vogt construction uses the language of trees.  In Chapter \ref{ch:tree} we review the basic combinatorics of trees and their composition, called tree substitution.

Colored operads are defined in Chapter \ref{ch:operads}.  We give four equivalent definitions of colored operads.  We first define colored operads as monoids with respect to the colored circle product.  Then we give three more equivalent descriptions in terms of generating operations, partial compositions, and trees.  As soon as we start discussing colored operads in this chapter, we will work over a general symmetric monoidal category $\M$.  The reader who is interested in a specific base category, such as $\Chaink$, should feel free to take $\M$ as this category throughout.

In Chapter \ref{ch:operad-construction} we discuss further properties of operads, including change-of-operad adjunctions and change-of-category functors.  We briefly discuss the model category structure on the category of algebras over a colored operad.  This chapter ends with the discussion of a localization of a colored operad, which is analogous to the localization of a category.  A localized colored operad is a colored operad in which some unary elements have been inverted.  We need localized colored operads when we discuss the time-slice axiom in (homotopy) prefactorization algebras.

In Chapter \ref{ch:bv} we define the Boardman-Vogt construction of a colored operad using a coend indexed by a category of trees and discuss its naturality properties.  Each colored operad $\O$ has a Boardman-Vogt construction $\wo$ together with an augmentation $\eta : \wo \to \O$.  In favorable situations, such as when the underlying category is $\Chaink$ with $\fieldk$ a field of characteristic zero, the augmentation is a weak equivalence, and the induced adjunction between the categories of algebras is a Quillen equivalence.  However, to understand the structure of homotopy algebraic quantum field theories and homotopy prefactorization algebras, we only need the Boardman-Vogt construction itself, not its homotopical properties.

In Chapter \ref{ch:w-algebras} we study the Boardman-Vogt construction of various colored operads of interest.  Due to our one-step coend definition of the Boardman-Vogt construction, we are able to write down a coherence theorem for their algebras.  As examples, we discuss in details homotopy coherent diagrams, homotopy inverses in homotopy coherent diagrams, specific models of $A_\infty$-algebras and $E_\infty$-algebras, and homotopy coherent diagrams of $A_\infty$-algebras and of $E_\infty$-algebras.  All of these homotopy coherent algebraic structures are relevant in our study of homotopy algebraic quantum field theories and homotopy prefactorization algebras. This finishes Part 1.

Part 2 begins with Chapter \ref{ch:aqft} in which we discuss the colored operad for algebraic quantum field theories following \cite{bsw}.  Examples include diagrams of (commutative) monoids, quantum field theories on (equivariant) topological spaces, chiral conformal quantum field theories, Euclidean quantum field theories, locally covariant quantum field theories, and quantum field theories on structured spacetimes and on spacetimes with timelike boundary.

In Chapter \ref{ch:haqft} we define homotopy algebraic quantum field theories as algebras over the Boardman-Vogt construction $\wocbarm$ applied to the colored operad $\Ocbarm$ for algebraic quantum field theories.  We record a coherence theorem for homotopy algebraic quantum field theories.  Each homotopy algebraic quantum field theory is shown to have the structure of a homotopy coherent diagram of $A_\infty$-algebras and to satisfy a homotopy coherent version of the causality axiom.  When a set of morphisms in $\C$ is given, each homotopy algebraic quantum field theory also satisfies a homotopy coherent version of the time-slice axiom.

In Chapter \ref{ch:pfa} we define configured categories and prefactorization algebras on them.  We will see that commutative monoids and their modules feature prominently in prefactorization algebras.  In Chapter \ref{ch:hpa} we define homotopy prefactorization algebras on a configured category $\Chat$ as algebras over the Boardman-Vogt construction $\wochatm$ of the colored operad $\Ochatm$ for prefactorization algebras.  We record a coherence theorem for homotopy prefactorization algebras.  In addition to a homotopy coherent diagram structure, we will see that $E_\infty$-algebras and their $E_\infty$-modules play prominent roles in homotopy prefactorization algebras.

In Chapter \ref{ch:comparing} we compare the two mathematical approaches to quantum field theory featured in this book.  We show that the category of orthogonal categories embeds in the category of configured categories as a full reflective subcategory.  Then we construct a comparison morphism $\Ochat \to \Ocbar$ that we use to compare (homotopy) prefactorization algebras and (homotopy) algebraic quantum field theories.  We discuss examples of prefactorization algebras that come from algebraic quantum field theories and those that do not.  This concludes Part 2.

\subsection*{Audience}
This book is intended for graduate students, mathematicians, and physicists.  Throughout this book, we include many examples and a lot of motivation and interpretation of results both mathematically and physically.  Since we actually review the basics of categories and operads, an ambitious advanced undergraduate should be able to follow this book.

\part{Operads}\label{part:operads}

\chapter{Category Theory}\label{ch:categories}

In this chapter, we recall some basic concepts of category theory and some relevant examples.  The reader who is familiar with basic category theory can just read the examples in Section \ref{sec:example-categories} and skip the rest of this chapter.  Our references for category theory are \cite{bor1,bor2,maclane}.  Categories were originally introduced by \index{Eilenberg@Eilenberg, S.}Eilenberg and \index{Maclane@Mac Lane, S.}Mac Lane \cite{eilenberg-maclane}.

In Section \ref{sec:categories} we review categories, functors, natural transformations, and equivalences.  A long list of examples of categories that will be used in later chapters are given in Section \ref{sec:example-categories}.  In Section \ref{sec:limits} we review limits, colimits, and coends, which will play a crucial role in our definition of the Boardman-Vogt construction of a colored operad in Chapter \ref{ch:bv}.  In Section \ref{sec:adjoint} we discuss adjoint functors.  In Section \ref{sec:smc} we review symmetric monoidal categories, which are the most natural setting to discuss colored operads.  In Section \ref{sec:monoids} and Section \ref{sec:monads} we review monoids and monads, which are important because algebras over a colored operad are defined as algebras over the associated monad.  In Section \ref{sec:localization} we review localization of categories, which will be needed to discuss the time-slice axiom in algebraic quantum field theories in Chapter \ref{ch:aqft}.

\section{Basics of Categories}\label{sec:categories}

\begin{definition}\label{def:category}
A \index{category}\emph{category} $\C$ consists of the following data:
\begin{itemize}
\item a class\label{notation:objects-category} $\Obc$ of \index{object}\emph{objects};
\item for any two objects $a,b \in \Obc$, a set\label{notation:morphism-set} $\C(a,b)$ of \index{morphism}\emph{morphisms}  with \index{domain}\emph{domain} $a$ and \index{codomain}\emph{codomain} $b$;
\item for each object $a \in \Obc$, an \index{identity morphism}\index{morphism!identity}\emph{identity morphism}\label{notation:identity-morphism} $\Id_a \in \C(a,a)$;
\item for any objects $a,b,c \in \Obc$, a function called the  \index{composition}\emph{composition}\label{notation:composition-category}
\[\nicexy{\C(b,c) \times \C(a,b) \ar[r]^-{\comp} & \C(a,c)}\]
sending $(g,f)$ to $g \comp f=gf$, called the \emph{composition} of $g$ and $f$.
\end{itemize}
The above data is required to satisfy the following two axioms.
\begin{description}
\item[Associativity]
Suppose $(h,g,f) \in \C(c,d) \times \C(b,c) \times \C(a,b)$.  Then there is an equality\index{associativity!category}
\[h \comp (g \comp f) = (h \comp g) \comp f \inspace \C(a,d).\]
\item[Unity]
For any objects $a,b \in \Obc$ and morphism $f \in \C(a,b)$, there are equalities\index{unity!category}
\[f \comp \Id_a = f = \Id_b \comp f \inspace \C(a,b).\]
\end{description}
The collection of all morphisms in $\C$ is written as $\Mor(\C)$.\label{notation:morphism-C}
\end{definition}

\begin{definition}\label{def:subcat}
Suppose $\C$ is a category.  
\begin{enumerate}
\item The \emph{opposite category}\index{opposite category}\index{category!opposite} $\Cop$ is the category with the same objects as $\C$ and with morphism sets $\Cop(a,b) = \C(b,a)$.\label{notation:opposite-category}  Its identity morphisms and composition are defined by those in $\C$.
\item A \index{subcategory}\emph{subcategory} of $\C$ is a category $\D$ such that:
\begin{enumerate}
\item There is an inclusion $\Obd \subseteq \Obc$ on objects.
\item For any objects $a,b \in \Obd$, there is a subset inclusion $\D(a,b) \subseteq \C(a,b)$ on morphisms.
\item For each object $a \in \Obd$, the identity morphism $\Id_a \in \D(a,a)$ is the identity morphism of $a \in \Obc$.
\item Suppose $(g,f) \in \D(b,c) \times \D(a,b)$.  Then their composition $g \comp f \in \D(a,c)$ in $\D$ is equal to the composition $g \comp f \in \C(a,c)$ in $\C$.
\end{enumerate}
\item A subcategory $\D$ of $\C$ is a \index{full subcategory}\index{subcategory!full}\emph{full subcategory} if, for any objects $a,b \in \Obd$, there is an equality $\D(a,b) = \C(a,b)$ of morphism sets.
\item $\C$ is called a \index{small category}\index{category!small}\emph{small category} if $\Obc$ is a set.
\item An \index{isomorphism}\emph{isomorphism} $f \in \C(a,b)$ is a morphism such that there exists an \emph{inverse} $f^{-1} \in \C(b,a)$ satisfying
\[f \comp f^{-1} = \Id_b \andspace f^{-1} \comp f = \Id_a.\]
An inverse is unique if it exists.  An isomorphism is also denoted by $\cong$.\label{notation:iso}
\item A \index{groupoid}\emph{groupoid} is a category in which all morphisms are isomorphisms.  
\item A \emph{discrete category}\index{discrete category}\index{category!discrete} is a category whose only morphisms are the identity morphisms.
\end{enumerate}
\end{definition}

\begin{definition}\label{def:functor}
A \index{functor}\emph{functor} $F : \C \to \D$ from a category $\C$ to a category $\D$ consists of
\begin{itemize}\item an \emph{assignment on objects}
\[\nicexy{\Obc \ar[r] & \Obd, & a \ar@{|->}[r] & Fa;}\]
\item for any objects $a, b \in \Obc$, a \emph{function on morphism sets}
\begin{equation}\label{functor-on-maps}
\nicexy{\C(a,b) \ar[r] & \D(Fa,Fb), & f \ar@{|->}[r] & Ff.}
\end{equation}
\end{itemize}
The above data is required to satisfy the following two axioms.
\begin{description}
\item[Preservation of Identity]
For each object $a \in \Obc$, there is an equality
\[F(\Id_a) = \Id_{Fa} \inspace \D(Fa,Fa).\]
\item[Preservation of Composition]
For any morphisms $(g,f) \in \C(b,c) \times \C(a,b)$, there is an equality
\[F(g \comp f) = Fg \comp Ff \inspace \D(Fa,Fc).\]
\end{description}
If $\C$ is a small category, we also call a functor $\C \to \D$ a \index{diagram}\emph{$\C$-diagram in $\D$}.  The \emph{identity functor}\index{identity functor} $\Id_{\C} : \C \to \C$ is the functor given by the identity functions on both objects and morphisms.  
\end{definition}

\begin{definition}\label{def:full-functor}
A functor $F : \C \to \D$ is called:
\begin{enumerate}
\item \emph{full} (resp., \emph{faithful})\index{full functor}\index{faithful functor} if for any objects $a,b \in \C$, the function on morphism sets in \eqref{functor-on-maps} is surjective (resp., injective);
\item \emph{essentially surjective}\index{essentially surjective} if for each object $d \in \D$, there exist an object $c \in \C$ and an isomorphism $Fc \iso d$.
\end{enumerate}
An object $d \in \D$ is in the \emph{essential image of $F$}\index{essential image} if there exist an object $c \in \C$ and an isomorphism $Fc \iso d$.
\end{definition}

\begin{definition}\label{def:functor-comp}
Suppose $F : \C \to \D$ and $G : \D \to \E$ are functors.  
\begin{enumerate}
\item The \index{composition!functor}\emph{composition} of functors 
\[GF = G \comp F : \C \to \E\]
is defined by composing the assignments on objects and the functions on morphism sets.
\item We call $F$ an \emph{isomorphism}\index{isomorphism!functor} if there exists a functor $\inv{F} : \D \to \C$ such that
\[F \inv{F} = \Id_{\D} \andspace \inv{F} F = \Id_{\C}.\]
An inverse $\inv{F}$ is unique if it exists.
\end{enumerate}
\end{definition}

\begin{definition}\label{def:natural-transformation}
Suppose $F,G,H : \C \to \D$ are functors from $\C$ to $\D$.
\begin{enumerate}
\item A \index{natural transformation}\emph{natural transformation} $\theta : F\to G$ consists of a \emph{structure morphism} $\theta_a \in \D(Fa, Ga)$ for each $a \in \Obc$ such that, if $f \in \C(a,b)$ is a morphism for some object $b \in \C$, then 
\[Gf \comp \theta_a = \theta_b \comp Ff \inspace \D(Fa,Gb).\]
\item If $\theta : F \to G$ and $\eta : G \to H$ are natural transformations, their \emph{composition}\index{composition!natural transformation} is the natural transformation $\eta\theta : F \to H$ with structure morphisms $(\eta\theta)_a = \eta_a \comp \theta_a$ for $a \in \Obc$.
\item A \index{natural isomorphism}\emph{natural isomorphism} is a natural transformation in which every structure morphism is an isomorphism.
\end{enumerate}
\end{definition}

\begin{definition}\label{def:equivalence-categories}
An \index{equivalence}\emph{equivalence} between categories $\C$ and $\D$ consists of a pair of functors $F : \C \to \D$ and $G : \D \to C$ and a pair of natural isomorphisms
\[\nicexy{\Id_{\C} \ar[r]^-{\cong} & GF} \andspace \nicexy{\Id_{\D} \ar[r]^-{\cong} & FG}.\]
In this setting, we say that $F$ is an \emph{equivalence of categories} and that the categories $\C$ and $\D$ are \emph{equivalent} via the functors $F$ and $G$.  A category is said to be \emph{essentially small}\index{essentially small} if it is equivalent to a small category.
\end{definition}

\begin{notation}
The following notations and conventions will be used.
\begin{itemize}
\item If $x$ is an object or a morphism in a category $\C$, we will often write $x \in \C$ instead of $x \in \Obc$ or $x \in \Morc$. 
\item A morphism $f \in \C(a,b)$ is also written as\label{notation:morphism} $f : a \to b$ or $\narrowxy{a \ar[r]^-{f} & b}$.
\item A functor $F : \C \to \D$ is also written as $\narrowxy{\C \ar[r]^-{F} & \D}$.
\item A natural transformation $\theta : F\to G$ is also written as $\narrowxy{F \ar[r]^-{\theta} & G}$.
\end{itemize}
\end{notation}

\section{Examples of Categories}\label{sec:example-categories}

In this section, we list some relevant examples of categories that we will use later.

\begin{example}\label{ex:empty-cat}
The \index{empty category}\index{category!empty}\emph{empty category}, with no objects and no morphisms, is denoted by $\emptyset$.\dqed
\end{example}

\begin{example}[Functor categories]\label{ex:functor-cat}
Given any categories $\C$ and $\D$, there is a \emph{functor category}\index{functor category} $\Fun(\C,\D)$ with functors $\C \to \D$ as objects and natural transformations between them as morphisms.  If $\C$ is a small category, we also call the functor category $\Fun(\C,\D)$ a \emph{diagram category}\index{diagram category}\index{category!diagram} and denote it by $\D^{\C}$.\dqed
\end{example}

\begin{example}[Product categories]\label{ex:product-cat}
Given any categories $\C$ and $\D$, there is a \emph{product category}\index{product category} $\C \times \D$ with objects $\Obc \times \Obd$ and morphism sets 
\[(\C \times \D)\bigl((c,d),(c',d')\bigr) = \C(c,c') \times \D(d,d')\]
for $c,c' \in \C$ and $d,d' \in \D$.\dqed
\end{example}

\begin{example}[Under categories]\label{ex:undercat}
Suppose $a$ is an object in a category $\C$.  The \emph{under category}\index{under category} $a \downarrow \C$ is the category whose objects are morphisms in $\C$ of the form $a \to b$.  A morphism $f : (a \shortto b) \to (a \shortto c)$ in the under category is a morphism $f : b \to c$ in $\C$ such that the triangle \[\nicexy{a \ar[r] \ar[dr] & b \ar[d]^-{f}\\ & c}\] in $\C$ is commutative.  The identity morphisms and composition are defined by those in $\C$.\dqed
\end{example}

\begin{example}[Sets]\label{ex:set}
There is a category $\Set$ with sets\index{category of sets} as objects and functions as morphisms.\dqed
\end{example}

\begin{example}[Vector spaces]\label{ex:vectk}
For a field\label{notation:fieldk}\index{field} $\fieldk$, there is a category $\Vectk$ with $\fieldk$-vector spaces\index{vector space} as objects and linear maps as morphisms.\dqed
\end{example}

\begin{example}[Chain complexes]\label{ex:chaink}
There is a category $\Chaink$ with chain complexes\index{chain complex} of $\fieldk$-vector spaces as objects and chain maps as morphisms.  Via the reindexing $X_n \mapsto X^{-n}$, one can also regard $\Chaink$ as the category of cochain complexes\index{cochain complex} of $\fieldk$-vector spaces.  With this in mind, everything below about chain complexes also holds for cochain complexes.\dqed
\end{example}

\begin{example}[Topological spaces]\label{ex:top}
There is a category $\Top$ whose objects are compactly generated weak Hausdorff spaces\index{topological spaces} and whose morphisms are continuous maps.\dqed
\end{example}

\begin{example}[Simplex category]\label{ex:simplex-cat}
The \emph{simplex category}\index{simplex category} $\Delta$ has objects the finite totally ordered sets
\[[n] = \{0 < 1 < \cdots < n\}\]
for $n \geq 0$.  A morphism is a weakly order-preserving\index{weakly order-preserving map} map, i.e., $f(i) \leq f(j)$ if $i < j$.\dqed
\end{example}

\begin{example}[Simplicial sets]\label{ex:simplicial-cat}
For a category $\C$, the diagram category \[\C^{\Deltaop} = \Fun(\Deltaop,\C)\] is called the category of \index{simplicial objects}\emph{simplicial objects in $\C$}.  If $\C$ is the category of sets, then $\Set^{\Deltaop}$ is also written as $\Sset$, and its objects are called \index{simplicial set}simplicial sets.\dqed
\end{example}

\begin{example}[Small categories]\label{ex:cat}
There is a category $\Cat$ whose objects are small categories\index{small category} and whose morphisms are functors.\dqed
\end{example}

\begin{example}[Partially ordered sets and lattices]\label{ex:lattice}
Each partially ordered set\index{partially ordered set} $(S,\leq)$ becomes a small category with object set $S$, and there is a morphism $a \to b$ if and only if $a \leq b$.  We will denote this category by $S$.  For $a,b\in S$, the morphism set $S(a,b)$ is either empty or a one-element set.  A \emph{lattice}\index{lattice} is a partially ordered set such that every pair of distinct elements $\{a,b\}$ has both a \index{least upper bound}least upper bound\label{notation:lub} $a \vee b$ and a \index{greatest lower bound}greatest lower bound\label{notation:glb} $a \wedge b$.  A \emph{bounded lattice}\index{bounded lattice} is a lattice with a least element\index{least element} $0$ and a greatest element\index{greatest element} $1$.\dqed
\end{example}

\begin{example}[Open subsets of a topological space]\label{ex:openx}
For each topological space $X$, there is a partially ordered set $(\Openx,\subset)$ consisting of open subsets\index{open subset} of $X$ in which $U \subset V$ if and only if $U$ is a subset of $V$.  By Example \ref{ex:lattice} we will also consider $\Openx$ as a category with open subsets of $X$ as objects and subset inclusions as morphisms.  Note that $\Openx$ is a bounded lattice.  For open subsets $U,V \subset X$, their least upper bound is the union $U \cup V$, and their greatest lower bound is the intersection $U \cap V$.  The least element is the empty subset of $X$, and the greatest element is $X$.\dqed
\end{example}

\begin{example}[Equivariant topological spaces]\label{ex:eq-space}
Suppose $G$ is a group, and $X$ is a topological space\index{equivariant topological space} in which $G$ acts on the left by homeomorphisms.  Suppose $\Openxg$ is the category obtained from $\Openx$ in Example \ref{ex:openx} by adjoining the isomorphisms \[g : U \iso gU\] for each open subset $U \subset X$ and each $g \in G$, subject to the following three relations:
\begin{enumerate}
\item $e : U \to eU = U$ is $\Id_U$, where $e$ is the multiplicative unit in $G$.
\item The composition of $g : U \to gU$ and $h : gU \to hgU$ is $hg : U \to hgU$.
\item The diagram \[\nicexy@C+.4cm{U \ar[d]_-{g} \ar[r]^-{\text{inclusion}} & V \ar[d]^-{g}\\
gU \ar[r]^-{\text{inclusion}} & gV}\] is commutative for all open subsets $U \subset V$ in $X$ and $g \in G$.
\end{enumerate}
Each morphism in $\Openxg$ decomposes as \[\nicexy@C+.4cm{U\ar[r]^-{g}_-{\cong} & gU \ar[r]^-{\text{inclusion}} & gV}\] for some $g \in G$.  If $G$ is the trivial group, then $\Openxg$ is the category $\Openx$ in Example \ref{ex:openx}.\dqed
\end{example}

\begin{example}[Oriented manifolds]\label{ex:man-cat}
For each integer $d \geq 1$, there is a category $\Mand$ with $d$-dimensional oriented manifolds\index{oriented manifold} as objects and orientation-preserving open embeddings as morphisms.  The reader may consult \cite{oneill} for discussion of manifolds.  We always assume that a manifold is \index{Hausdorff}Hausdorff and \index{second-countable}second-countable.  By \index{Whitney Embedding Theorem}Whitney Embedding Theorem, $\Mand$ is \index{essentially small}essentially small.  In what follows, we will tacitly replace $\Mand$ by an equivalent small category.\dqed
\end{example}

\begin{example}[Discs]\label{ex:disc-cat}
There is a full subcategory \[i : \Discd \to \Mand\] whose objects are oriented manifolds diffeomorphic\index{diffeomorphism} to $\fieldr^d$, where\label{notation:fieldr} $\fieldr$ is the field of \index{real numbers}real numbers.\dqed
\end{example}

\begin{example}[Oriented Riemannian manifolds]\label{ex:riem-cat}
For each integer $d \geq 1$, there is a category $\Riemd$ with $d$-dimensional oriented Riemannian manifolds\index{Riemannian manifold} as objects and orientation-preserving isometric open embeddings as morphisms.  As in the Example \ref{ex:man-cat}, $\Riemd$ is essentially small, and we will tacitly replace it by an equivalent small category.\dqed
\end{example}

\begin{example}[Lorentzian manifolds]\label{ex:loc-cat}
The reader is referred to \cite{bgp,bee,oneill} for detailed discussion of Lorentzian geometry.   A \emph{Lorentzian manifold}\index{Lorentzian manifold} is a manifold $X$ equipped with a pseudo-Riemannian metric\index{pseudo-Riemannian metric} $g$ of signature\index{signature} $(+,-,\ldots,-)$.  A tangent vector\index{tangent vector} $v$ in a Lorentzian manifold $(X,g)$ is \emph{timelike} (resp., \emph{causal})\index{timelike}\index{causal} if $g(v,v)>0$ (resp., $g(v,v) \geq 0$).  A smooth curve $f : [0,1]\to X$ is a \emph{timelike/causal curve} if its tangent vectors are all timelike/causal.  A \emph{time-orientation}\index{time-orientation} $t$ on an oriented Lorentzian manifold $(X,g,o)$ is a smooth vector field $t$ on $X$ such that the vector $t_x$ is timelike at each point $x \in X$.  

Suppose $(X,g,o,t)$ is an oriented and time-oriented Lorentzian manifold.   A causal curve $f$ is \emph{future-directed}\index{future-directed} if $g(t_x,\dot{f}_x) > 0$ and \emph{past-directed}\index{past-directed} if $g(t_x,\dot{f}_x)<0$ at each $x \in X$, where $\dot{f}_x$ is the tangent vector of $f$ at $x$.  The \emph{causal future/past}\index{causal future}\index{causal past} of a point $x \in X$ is the set $J^+_X(x)$ (resp., $J^-_X(x)$) consisting of $x$ and points in $X$ that can be reached from $x$ by a future/past-directed causal curve.  A subset $A \subset X$ is \emph{causally compatible}\index{causally compatible} if for each $a \in A$, $J^{\pm}_X(a) \cap A = J^{\pm}_A(a)$.  Two subsets $A$ and $B$ in $X$ are \emph{causally disjoint}\index{causally disjoint} if for each point $a \in \overline{A}$, $J^{\pm}_X(a) \cap \overline{B}=\varnothing$.  A \emph{Cauchy surface}\index{Cauchy surface} in $(X,g,o,t)$ is a smooth hypersurface that intersects every inextensible timelike curve exactly once.  We call $(X,g,o,t)$ \emph{globally hyperbolic}\index{globally hyperbolic} if it contains a Cauchy surface.  

For each integer $d \geq 1$, there is a category $\Locd$ with $d$-dimensional oriented, time-oriented, and globally hyperbolic Lorentzian manifolds as objects.  A morphism in $\Locd$ is an isometric embedding that preserves the orientations and time-orientations whose image is causally compatible and open.   As in Example \ref{ex:man-cat}, $\Locd$ is essentially small, and we will tacitly replace it by an equivalent small category.\dqed
\end{example}

\begin{example}[Globally hyperbolic open subsets]\label{ex:gh-cat}
Similar to Example \ref{ex:openx}, for a fixed Lorentzian manifold $X \in \Locd$, there is a category $\Ghx$ with globally hyperbolic open subsets\index{globally hyperbolic open subset} of $X$ as objects and subset inclusions as morphisms.  There is a functor \[i : \Ghx \to \Locd\] given by restricting the structures of $X$ to globally hyperbolic open subsets.\dqed
\end{example}

\begin{example}[Lorentzian manifolds with bundles]\label{ex:bgloc-cat}
Suppose $G$ is a Lie group.  There is a category $\Bgloc$ in which an object is a pair $(X,P)$ with $X \in \Locd$ and $P$ a principal $G$-bundle\index{principal bundle} over $X$.  A morphism $f : (X,P) \to (Y,Q)$ in $\Bgloc$ is a principal $G$-bundle morphism $f : P \to Q$ covering a morphism $f' : X \to Y$.  There is a forgetful functor \[\pi : \Bgloc \to \Locd\] that forgets about the bundle.\dqed
\end{example}

\begin{example}[Lorentzian manifolds with bundles and connections]\label{ex:bgconloc-cat}
Suppose $G$ is a Lie group.  There is a category $\Bgconloc$ in which an object is a triple $(X,P,C)$ with $(X,P) \in \Bgloc$ and $C$ a connection\index{connection} on $P$.  A morphism  in $\Bgconloc$ is a morphism in $\Bgloc$ that preserves the connections.  There is a forgetful functor \[p : \Bgconloc \to \Bgloc\] that forgets about the connection.  Composing with the forgetful functor in Example \ref{ex:bgloc-cat}, there is a forgetful functor \[\pi p : \Bgconloc \to \Locd\] that forgets about both the bundle and the connection.\dqed
\end{example}

\begin{example}[Lorentzian manifolds with spin structures]\label{ex:sloc-cat}
Suppose $d \geq 4$.  There is a category $\Slocd$ with $d$-dimensional oriented, time-oriented, and globally hyperbolic Lorentzian spin manifolds\index{spin manifold} as objects.  To be more precise, an object is a triple $(X,P,\psi)$ with $X \in \Locd$, $P$ a principle $\spinzero(1,d-1)$-bundle over $X$, and $\psi : P \to FX$ a $\spinzero(1,d-1)$-equivariant bundle map over $\Id_X$ to the pseudo-orthonormal oriented and time-oriented frame bundle $FX$ over $X$.  A morphism $f : (X,P,\psi) \to (Y,Q,\phi)$ in $\Slocd$ is a principal $\spinzero(1,d-1)$-bundle morphism $f : P \to Q$ covering a morphism $f' : X \to Y$ such that $\phi f=f'_* \psi$.  Here $f'_* : FX \to FY$ is the pseudo-orthonormal oriented and time-oriented frame bundle morphism induced by $f'$.  

There is a forgetful functor \[\pi : \Slocd \to \Locd\] that forgets the spin structure such that the fiber $\inv{\pi}(X)$ is a groupoid for each $X \in \Loc^d$.  Here $\inv{\pi}(X)$ is the subcategory of $\Sloc^d$ whose objects are sent to $X$ and whose morphisms are sent to $\Id_X$ by $\pi$.\dqed
\end{example}

\begin{example}[Regions in spacetime with timelike boundary]\label{ex:regions}
Following \cite{bds} we define a \index{spacetime!with timelike boundary}\emph{spacetime with timelike boundary}\index{timelike boundary} as an oriented and time-oriented Lorentzian manifold $X$ with boundary \cite{lee} such that the pullback of the Lorentzian metric along the boundary inclusion $\partial X \to X$ defines a Lorentzian metric on the boundary $\partial X$.  There is a \emph{category of regions}\index{category of regions} in $X$, denoted $\Regx$, in which an object is a causally convex\index{causally convex} open subset in $X$ with inclusions as morphisms.  It contains a full subcategory $\Regxzero$ whose objects are causally convex open subsets contained in the interior $X_0$ of $X$.\dqed
\end{example}

\section{Limits and Colimits}\label{sec:limits}

\begin{definition}\label{def:limit}
Suppose $F : \D \to \C$ is a functor
\begin{enumerate}
\item For an object $c \in \C$, the \emph{constant functor}\index{constant functor}\index{functor!constant} $\Delta_c : \D \to \C$ is the functor that sends every object in $\D$ to $c$ and every morphism in $\D$ to $\Id_c$.
\item A \emph{limit of $F$}\index{limit} is a pair\label{notation:limit} $(\limit F,\theta)$ consisting of 
\begin{itemize}\item an object $\limit F \in \C$ and 
\item a natural transformation $\theta : \Delta_{\limit F} \to F$ 
\end{itemize}
that satisfies the following universal property: If $(y,\phi)$ is another such pair, then there exists a unique morphism \[f : y \to \limit F\in \C \stspace \phi = \theta \comp \Delta_f,\] where $\Delta_f : \Delta_y \to \Delta_{\limit F}$ is the obvious natural transformation induced by $f$.  Omitting $\Delta$ we may represent a limit of $F$ as follows.
\[\nicexy{& \limit F \ar[d]^-{\theta}\\ y \ar[r]^-{\forall\, \phi} \ar@{.>}[ur]^-{\exists !\, f} & F}\]
If a limit of $F$ exists, then it is unique up to a unique isomorphism in $\C$.
\item A \emph{colimit of $F$}\index{colimit} is a pair\label{notation:colim} $(\colim F,\theta)$ consisting of 
\begin{itemize}\item an object $\colim F \in \C$ and 
\item a natural transformation $\theta : F \to \Delta_{\colim F}$ 
\end{itemize}
that satisfies the following universal property: If $(z,\psi)$ is another such pair, then there exists a unique morphism \[f : \colim F \to z\in \C \stspace \psi = \Delta_f \comp  \theta.\]  We may represent a colimit of $F$ as follows.
\[\nicexy{F \ar[r]^-{\theta} \ar[dr]_-{\forall\, \psi} & \colim F \ar@{.>}[d]^-{\exists !\,  f}\\ & z}\]
If a colimit of $F$ exists, then it is unique up to a unique isomorphism in $\C$.  We will often write a colimit of $F$ as either $\colim F$ or $\colim_{\D} F$ if we wish to emphasize its domain category.
\item $\C$ is \emph{(co)complete}\index{complete}\index{cocomplete} if every functor from a small category to $\C$ has a (co)limit.
\end{enumerate}
\end{definition}

\begin{example}
The categories $\Set$, $\Top$, $\Sset$, $\Cat$, $\Vectk$, and $\Chaink$ are complete and cocomplete.\dqed\end{example}

\begin{example}[Initial and Terminal Objects]\label{ex:initial-object}
Taking $\D$ to be the empty category $\emptyset$ and $F : \emptyset \to \C$ the trivial functor, a limit of $F$ (i.e., of the empty diagram) is called a \emph{terminal object}\index{terminal object} in $\C$.  More explicitly, a terminal object $*$ in $\C$ is an object such that, for each object $c \in \C$, there exists a unique morphism $c \to *$.  Dually, a colimit of the empty diagram is called an \emph{initial object}\index{initial object} in $\C$.  An initial object $i \in \C$ is characterized by the universal property that for each object $c \in \C$, there exists a unique morphism $i \to c$.  In what follows, we will often use the symbol\label{notation:initialobject} $\emptyset$ to denote an initial object in $\C$, and the reader should not confuse it with the empty category.

For instance:
\begin{enumerate}
\item In the category $\Set$, the empty set is an initial object, and a terminal object is exactly a one-element set.  
\item In $\Vectk$ and $\Chaink$, the $0$ vector space (or chain complex) is both an initial object and a terminal object.
\item For a topological space $X$, the category $\Openx$ has the empty subset of $X$ as an initial object and $X$ as a terminal object.\dqed
\end{enumerate}
\end{example}

\begin{example}[Coproducts and Products]\label{ex:coproduct}
Taking $\D$ to be a small discrete category, a functor $F : \D \to \C$ is determined by the set of objects $\{Fd : d \in \D\}$.  A (co)limit of $F$ is called a \emph{(co)product}\index{product}\index{coproduct} of the set of objects $\{Fd : d \in \D\}$, denoted by\label{notation:product} $\prod_{d \in \D} Fd$ (resp., $\coprod_{d \in \D} Fd$).  For instance, in $\Vectk$ and $\Chaink$, coproducts and finite products are both given by direct sums.\dqed
\end{example}

\begin{example}[Pushouts and Pullbacks]\label{ex:pushout}
Suppose $\D$ is the category \[\pushoutcat\] with three objects and two non-identity morphisms as indicated.  A colimit of $F : \D \to \C$ is called a \emph{pushout}\index{pushout} of the diagram $\narrowxy{F1 & F0 \ar[l] \ar[r] & F2}$.  A limit of $F : \Dop \to \C$ is called a \emph{pullback}\index{pullback} of the diagram $\narrowxy{F1 \ar[r] & F0 & F2 \ar[l]}$.\dqed
\end{example}

\begin{example}[Coequalizers and Equalizers]\label{ex:coequalizer}
A \emph{coequalizer}\index{coequalizer} of a pair of parallel morphisms $f,g : a \to b$ in $\C$ is a pair $(c,u)$ consisting of an object $c \in \C$ and a morphism $u : b \to c$ such that $uf=ug$ and that is initial among such pairs.  In other words, for every other such pair $(d,v)$, there exists a unique morphism $v' : c \to d$ such that $v = v'u$.  
\[\nicexy{a \ar@<2pt>[r]^-{f} \ar@<-2pt>[r]_-{g} & b \ar[r]^-{u} \ar[d]_-{\forall\, v} & c \ar@{.>}[dl]^-{\exists !\, v'} & uf=ug\\ & d && vf=vg}\]
A \emph{reflexive pair}\index{reflexive pair} is a pair of morphisms $f,g : a \to b$ with a common section $s : b \to a$ in the sense that $fs=gs=\Id_b$.  A \emph{reflexive coequalizer}\index{reflexive coequalizer} is a coequalizer of a reflexive pair.  Note that a coequalizer of $f$ and $g$ is the same as a pushout of the diagram
\[\nicexy{a \coprod b \ar[d]_-{(f,\Id_b)} \ar[r]^-{(g,\Id_b)} & b\\ b &}\]
and hence is a particular kind of colimit.  The dual concept is called an \emph{equalizer}\index{equalizer} of $f$ and $g$.\dqed
\end{example}

\begin{definition}[Coends]\label{def:coend}
Suppose $F : \Cop \times \C \to \M$ is a functor.
\begin{enumerate}
\item A \index{wedge}\emph{wedge} of $F$ is a pair $(X,\zeta)$ consisting of 
\begin{itemize}\item an object $X \in \M$ and 
\item morphisms $\zeta_c : F(c,c) \to X$ for $c \in \C$ 
\end{itemize}
such that the diagram \[\nicexy{F(d,c) \ar[d]_-{F(g,c)} \ar[r]^-{F(d,g)} & F(d,d) \ar[d]^-{\zeta_d}\\
F(c,c) \ar[r]^-{\zeta_c} & X}\] is commutative for each morphism $g : c\to d \in \C$.
\item A \index{coend}\emph{coend} of $F$ is an initial wedge $\left(\int^{c\in \C} F(c,c), \omega\right)$.
\end{enumerate}\end{definition}

In other words, a coend of $F$ is a wedge of $F$ such that given any wedge $(X,\zeta)$ of $F$, there exists a unique morphism \[h : \int^{c\in \C} F(c,c) \to X \in \M\] such that the diagram
\[\nicexy{F(c,c) \ar[r]^-{\omega_c} \ar[dr]_-{\zeta_c} & \int^{c\in \C} F(c,c) \ar[d]^-{h}\\ & X}\]
is commutative for each object $c \in \C$.  The dual concept of a coend is called an end, which is originally due to Yoneda\index{Yoneda@Yoneda, N.} \cite{yoneda}.  We will not need to use ends in this book.

The proof of the following result is a simple exercise in checking the definitions of a coend and of a coequalizer.

\begin{proposition}\label{coend-is-coequal}
Suppose given a functor $F : \Cop \times \C \to \M$ with $\C$ a small category and $\M$ a cocomplete category.  Then a coend of $F$ exists and is given by a \index{coequalizer}coequalizer \[\int^{c\in \C} F(c,c) = \coequal\biggl(\nicexy@C+.5cm{\coprodover{g\in\Mor(\C)} F(d,c) \ar@<2pt>[r]^-{i_d \comp F(d,g)} \ar@<-2pt>[r]_-{i_c \comp F(g,c)} & \coprodover{c\in\C} F(c,c)}\biggr)\]
in which $g : c \to d$ runs through all the morphisms in $\C$ and \[i_c : F(c,c) \to \coprodover{c\in\C} F(c,c)\] is the natural inclusion.  The natural morphism $\omega_c$ is the composition \[\nicexy@C-.4cm{F(c,c) \ar[d]_-{i_c} \ar[r]^-{\omega_c} & \dint^{c\in \C} F(c,c)\\ \coprodover{c\in\C} F(c,c) \ar[ur]_-{\mathrm{natural}} &}\] for each object $c \in \C$.
\end{proposition}
In particular, in the above setting, a coend is a particular kind of colimit.

\section{Adjoint Functors}\label{sec:adjoint}

Adjoint functors will provide us with ways to compare (i) algebraic quantum field theories of various flavors, (ii) prefactorization algebras of various flavors, and (iii) algebraic quantum field theories with prefactorization algebras.  The concept of an adjunction is due to \index{Kan@Kan, D.}Kan \cite{kan}.

\begin{definition}\label{def:adjoint}
Suppose $F : \C \to \D$ and $G : \D \to \C$ are functors.  We call the pair $(F,G)$ an \index{adjoint pair}\emph{adjoint pair}, or an \index{adjunction}\emph{adjunction}, if for each object $c \in \C$ and each object $d \in \D$, there exist a bijection
\[\theta_{c,d} : \D(Fc,d) \cong \C(c,Gd)\]
that is natural in both $c$ and $d$.  In this case:
\begin{enumerate}
\item We call $F$ a \emph{left adjoint of $G$}\index{left adjoint} and $G$ a \emph{right adjoint of $F$}\index{right adjoint} and write $F \dashv G$.\label{notation:adjoints}
\item For each $c \in \C$, the morphism $\eta_c : c \to GFc$ corresponding under $\theta_{c,Fc}$ to $\Id_{Fc}$ is called the \index{unit of adjunction}\emph{unit of $c$}.
\item For each $d \in \D$, the morphism $\epsilon_d : FGd \to d$ corresponding under $\theta_{Gd,d}$ to $\Id_{Gd}$ is called the \index{counit of adjunction}\emph{counit of $d$}.
\end{enumerate}
\end{definition}

\begin{convention} We will always write the left adjoint on top (if displayed horizontally) or on the left (if displayed vertically).\end{convention}

\begin{definition}\label{def:reflective-subcat}
Suppose $i : \D \to \C$ is the inclusion functor of a full subcategory.  Then $\D$ is called a \emph{reflective subcategory}\index{reflective subcategory} of $\C$ if $i$ admits a left adjoint.
\end{definition}

The unit and the counit actually characterize an adjoint pair; the proof of the following result can be found in \cite{bor1} Section 3.1.

\begin{theorem}
Suppose $F : \C \to \D$ and $G : \D \to \C$ are functors.  The following statements are equivalent.
\begin{enumerate}
\item $(F,G)$ is an adjoint pair.
\item There exist natural transformations $\eta : \Id_{\C} \to GF$, called the unit, and $\epsilon : FG \to \Id_{\D}$, called the counit, such that the diagrams
\begin{equation}\label{triangle-identities}
\nicexy{F \ar[r]^-{F\eta} \ar[dr]_-{\Id} & FGF \ar[d]^-{\epsilon F}\\ & F}\qquad
\nicexy{G \ar[dr]_-{\Id} \ar[r]^-{\eta G} & GFG \ar[d]^-{G \epsilon}\\ & G}
\end{equation}
are commutative.
\item There exists a natural transformation $\eta : \Id_{\C} \to GF$  such that, given any morphism $f \in \C(c,Gd)$ with $c \in \C$ and $d \in \D$, there exists a unique morphism $\fbar \in \D(Fc,d)$ such that the diagram
\[\nicexy{& GFc \ar[d]^-{G\fbar}\\ c \ar[r]^-{f} \ar[ur]^-{\eta_c} & Gd}\]
is commutative.
\end{enumerate}
\end{theorem}

The two commutative diagrams in \eqref{triangle-identities} are called the \index{triangle identities}\emph{triangle identities}.  Adjoint functors are unique up to isomorphisms.

\begin{example}
The full subcategory inclusion from the category of abelian groups\index{abelian group} $\Ab$ to the category of groups\index{group} $\Grp$ admits a left adjoint, namely, the abelianization\index{abelianization} functor that sends a group $G$ to the quotient $G/[G,G]$.  So $\Ab$ is a full reflective subcategory of $\Grp$.\dqed  
\end{example}

\begin{example}
In the context of Examples \ref{ex:set} and \ref{ex:vectk}, there is a free-forgetful adjunction\index{free-forgetful adjunction}
\[\nicexy{\Set \ar@<2pt>[r]^-{F} & \Vectk \ar@<2pt>[l]^-{U}}\]
in which the right adjoint $U$ sends a vector space to its underlying set.  The left adjoint $F$ sends a set $X$ to the vector space $\bigoplus_X \fieldk$ freely generated by $X$.\dqed
\end{example}

Recall the concept of an equivalence in Definition \ref{def:equivalence-categories}.

\begin{definition}\label{def:adjoint-equivalence}
An adjoint pair $F : \C \adjoint \D : G$ is an \emph{adjoint equivalence}\index{adjoint equivalence} if the categories $\C$ and $\D$ are equivalent via the functors $F$ and $G$.
\end{definition}

The following characterizations of an adjoint equivalence is \cite{maclane} IV.4 Theorem 1.

\begin{theorem}\label{thm:equivalence-categories}
The following properties of a functor $G : \D \to \C$ are equivalent.
\begin{enumerate} \item $G$ is an equivalence of categories.
\item $G$ admits a left adjoint $F : \C \to \D$ such that $F\dashv G$ is an adjoint equivalence.
\item $G$ is full, faithful, and essentially surjective.
\end{enumerate}
\end{theorem}

An important example of an adjunction is Kan extension.

\begin{definition}\label{def:left-kan}
Suppose $F : \C \to \D$ is a functor, and $\M$ is a category.  If the induced functor
\[F^* = \Fun(F,\M) : \Fun(\D,\M) \to \Fun(\C,\M),\quad F^*(G)=GF\]
on functor categories admits a left adjoint $F_!$, then for a functor $H \in \Fun(\C,\M)$, the image $F_! H \in \Fun(\D,\M)$ is called a \emph{left Kan extension}\index{Kan extension} of $H$ along $F$ and is written as $\Lan_F H$ or $\Lan H$.
\end{definition}

The following existence result is the dual of \cite{maclane} (p.239 Corollary 2).  If $\D$ is also small, then the following result can be obtained as a special case of the change-of-operad Theorem \ref{thm:change-operad}; see Example \ref {ex:change-of-diagram}.

\begin{theorem}\label{thm:left-kan-exists}
Suppose $F : \C \to \D$ is a functor with $\C$ a small category, and $\M$ is a cocomplete category.  Then the induced functor $\Fun(F,\M)$ admits a left adjoint.  In particular, every functor $H : \C \to \M$ admits a left Kan extension along $F$.   
\end{theorem}

\begin{example}[Left Kan Extensions as Coends]\label{ex:kan-as-coend}
In the setting of Theorem \ref{thm:left-kan-exists}, a left Kan extension of $H : \C \to \M$ along $F : \C \to \D$ is given objectwise by the coend (Definition \ref{def:coend})
\begin{equation}\label{kan-coend}
\bigl(\Lan_F H\bigr)(d) = \int^{c \in \C} \D(Fc,d) \cdot Hc
\end{equation}
for each object $d \in \D$.  In this coend formula, the integrand is the \emph{copower}\index{copower} defined by
\begin{equation}\label{copower}
S \cdot X = \coprod_{s\in S} X
\end{equation}
for a set $S$ and an object $X \in \M$.  For a proof that the coend formula \eqref{kan-coend} actually yields a left Kan extension, see \cite{maclane} (p.240 Theorem 1) or \cite{loregian} (p.23).\dqed
\end{example}

An important property of a general left adjoint is that it preserves colimits.  Similarly, a right adjoint preserves limits.  For a proof of the following two results, see \cite{bor1} Section 3.2.

\begin{theorem}[Left Adjoints Preserve Colimits]\label{thm:lapc}
Suppose $F : \C \to \D$ admits a right adjoint, and $H : \E \to \C$ has a colimit $(\colim H, \theta : H \to \Delta_{\colim H})$.  Then the pair\index{left adjoints preserve colimits}
\[\Bigl(F\colim H, F\theta : FH \to F\Delta_{\colim H} = \Delta_{F \colim H}\Bigr)\]
is a colimit of $FH : \E \to \D$.
\end{theorem}

\begin{theorem}[Right Adjoints Preserve Limits]\label{thm:rapl}
Suppose $G : \D \to \C$ admits a left adjoint, and $H : \E \to \D$ has a limit $(\limit H, \theta : \Delta_{\limit H} \to H)$.  Then the pair\index{right adjoints preserve limits}
\[\Bigl(G\lim H, G\theta : G\Delta_{\limit H} = \Delta_{G \limit H} \to GH\Bigr)\]
is a limit of $GH : \E \to \C$.
\end{theorem}

\section{Symmetric Monoidal Categories}\label{sec:smc}

A symmetric monoidal category is the most natural setting to discuss operads and their algebras.

\begin{definition}\label{def:monoidal-category}
A \index{monoidal category}\index{category!monoidal}\emph{monoidal category} is a tuple
\[(\M, \otimes, \tensorunit, \alpha, \lambda, \rho)\]
consisting of the following data.
\begin{itemize}
\item $\M$ is a category.
\item $\otimes : \M \times \M \to \M$\label{notation:monoidal-product} is a functor, called the \index{monoidal product}\emph{monoidal product}.
\item $\tensorunit$ is an object in $\M$, called the \index{monoidal unit}\emph{monoidal unit}.
\item $\alpha$ is a natural isomorphism
\begin{equation}\label{mon-cat-alpha}
\nicexy{(X \otimes Y) \otimes Z \ar[r]^-{\alpha}_-{\cong} &  X\otimes (Y \otimes Z)}
\end{equation}
for all objects $X,Y,Z \in \M$, called the \index{associativity isomorphism}\emph{associativity isomorphism}.
\item $\lambda$ and $\rho$ are natural isomorphisms
\begin{equation}\label{mon-cat-lambda}
\nicexy{\tensorunit \otimes X \ar[r]^-{\lambda}_-{\cong} & X} \andspace \nicexy{X \otimes \tensorunit \ar[r]^-{\rho}_-{\cong} & X}
\end{equation}
for all objects $X \in \M$, called the \emph{left unit}\index{left unit} and the \index{right unit}\emph{right unit}, respectively.
\end{itemize}
This data is required to satisfy the following two axioms.
\begin{description}
\item[Unit Axioms]
The diagram
\begin{equation}\label{monoidal-unit}
\nicexy{(X \otimes \tensorunit) \otimes Y \ar[d]_-{\rho \otimes \Id} \ar[r]^-{\alpha}_-{\cong} 
& X \otimes (\tensorunit \otimes Y) \ar[d]^-{\Id \otimes \lambda}\\ X \otimes Y \ar[r]^-{=} & X \otimes Y}
\end{equation}
is commutative for all objects $X,Y \in \M$, and
\begin{equation}\label{monoidal-unit-two}
\nicexy{\lambda = \rho : \tensorunit \otimes \tensorunit \ar[r]^-{\cong} & \tensorunit.}
\end{equation}
\item[Pentagon Axiom]
The pentagon\index{Pentagon Axiom}
\[\nicexy@C-2cm{& (W \otimes X) \otimes (Y \otimes Z) \ar[dr]^-{\alpha} & \\
\bigl((W \otimes X) \otimes Y\bigr) \otimes Z \ar[ur]^-{\alpha} \ar[d]_-{\alpha \otimes \Id}
&& W \otimes \bigl(X \otimes (Y \otimes Z)\bigr) \\
\bigl(W \otimes (X \otimes Y)\bigr) \otimes Z \ar[rr]^-{\alpha} & \hspace{1cm} & W \otimes \bigl((X \otimes Y) \otimes Z\bigr)\ar[u]_-{\Id \otimes \alpha}}\]
is commutative for all objects $W,X,Y,Z \in \M$.
\end{description}
A \index{strict monoidal category}\index{monoidal category!strict}\emph{strict monoidal category} is a monoidal category in which the natural  isomorphisms $\alpha$, $\lambda$, and $\rho$ are all identity morphisms.
\end{definition}

So in a strict monoidal category, an iterated monoidal product $a_1 \otimes \cdots \otimes a_n$ without any parentheses has an unambiguous meaning.

\begin{convention}\label{conv:empty-tensor}
In a monoidal category, an \index{empty tensor product}\emph{empty tensor product}, written as\label{notation:empty-tensor} $X^{\otimes 0}$ or $X^{\otimes \varnothing}$, means the monoidal unit $\tensorunit$. 
\end{convention}

\begin{definition}\label{def:symmetric-monoidal-category}
A \index{symmetric monoidal category}\index{monoidal category!symmetric}\emph{symmetric monoidal category} is a pair $\left(\M, \xi\right)$ in which:
\begin{itemize}
\item $\M = (\M,\otimes,\tensorunit,\alpha,\lambda,\rho)$ is a monoidal category.
\item $\xi$ is a natural isomorphism\label{notation:symmetry-iso}
\begin{equation}\label{symmetry-isomorphism}
\nicexy{X \otimes Y \ar[r]^-{\xi_{X,Y}}_-{\cong} & Y \otimes X}
\end{equation}
for objects $X,Y \in \M$, called the \index{symmetry isomorphism}\emph{symmetry isomorphism}.
\end{itemize}
This data is required to satisfy the following three axioms.
\begin{description}
\item[Symmetry Axiom]
The diagram
\begin{equation}\label{monoidal-symmetry-axiom}
\nicexy{X \otimes Y \ar[r]^-{\xi_{X,Y}} \ar[dr]_-{=} & Y \otimes X \ar[d]^-{\xi_{Y,X}}
\\ & X \otimes Y}
\end{equation}
is commutative for all objects $X,Y \in \M$.
\item[Compatibility with Units]
The diagram
\[\nicexy{X \otimes \tensorunit \ar[d]_-{\rho} \ar[r]^-{\xi_{X,\tensorunit}}
& \tensorunit \otimes X \ar[d]^-{\lambda}\\ X \ar[r]^-{=} & X}\]
is commutative for all objects $X \in \M$.
\item[Hexagon Axiom]
The diagram\index{Hexagon Axiom}
\begin{equation}
\label{hexagon-axiom}
\nicexy@C-40pt{& X \otimes (Z \otimes Y) \ar[rr]^-{\Id \otimes \xi_{Z,Y}} & \hspace{1.5cm}
& X \otimes (Y \otimes Z) \ar[dr]^-{\alpha^{-1}}&\\ (X \otimes Z) \otimes Y 
\ar[ur]^-{\alpha} \ar[dr]_-{\xi_{X\otimes Z,Y}} &&&& (X \otimes Y) \otimes Z\\
& Y \otimes (X \otimes Z) \ar[rr]^-{\alpha^{-1}} & \hspace{1.5cm}
& (Y \otimes X) \otimes Z \ar[ur]_-{\xi_{Y,X} \otimes \Id} &}
\end{equation}
is commutative for all objects $X,Y, Z \in \M$.
\end{description}
\end{definition}

\begin{definition}\label{def:closed-category}
A \index{symmetric monoidal closed category}\emph{symmetric monoidal closed category} is a symmetric monoidal category $\M$ in which for each object $Y$, the functor
\[- \otimes Y : \M \to \M,\]
admits a right adjoint \index{internal hom}\label{notation:internal-hom}
\[\Homm(Y,-) : \M \to \M,\]
called the \emph{internal hom}.  In other words, for any objects $X,Y,Z \in \M$, there is a specified bijection, called the \emph{$\otimes$-$\Homm$ adjunction},
\begin{equation}\label{tensor-hom-adjunction}
\nicexy{\M\bigl(X \otimes Y, Z\bigr) \ar[r]^-{\phi}_-{\cong} & \M\bigl(X,\Homm(Y,Z)\bigr)}
\end{equation}
that is natural in $X$, $Y$, and $Z$.
\end{definition}

\begin{example}\label{ex:monoidal-categories}
The categories $\Set$, $\Top$, and $\Sset$ are symmetric monoidal closed categories via the Cartesian product.  The category $\Vectk$ is a symmetric monoidal closed category with the usual tensor product of vector spaces.  The category $\Chaink$ is a symmetric monoidal closed category via the monoidal product $X \otimes Y$ with
\[(X \otimes Y)_n = \bigoplus_{k \in \mathbb{Z}} X_k \otimes_{\fieldk} Y_{n-k}\]
and differential
\[d(x \otimes y) = (dx) \otimes y + (-1)^{|x|} x \otimes (dy).\]
For the internal homs, the reader is referred to \cite{hovey} Chapters 2 and 3.\dqed
\end{example}

\begin{definition}\label{def:monoidal-functor}
Suppose $\M$ and $\N$ are monoidal categories.  A \index{monoidal functor}\index{functor!monoidal}\emph{monoidal functor} \[(F,F_2,F_0) : \M \to \N\] consists of the following data:
\begin{itemize}
\item a functor $F : \M \to \N$;
\item a natural transformation
\begin{equation}\label{monoidal-f2}
\nicexy{F(X) \otimes F(Y) \ar[r]^-{F_2} & F(X \otimes Y) \in \N,}
\end{equation}
where $X$ and $Y$ are objects in $\M$;
\item a morphism
\begin{equation}\label{monoidal-f0}
\nicexy{\tensorunit_{\N} \ar[r]^-{F_0} & F(\tensorunit_{\M}) \in \N,}
\end{equation}
where $\tensorunit_{\N}$ and $\tensorunit_\M$ are the monoidal units in $\N$ and $\M$, respectively.
\end{itemize} 
This data is required to satisfy the following three axioms.
\begin{description}
\item[Compatibility with the Associativity Isomorphisms]
The diagram
\begin{equation}\label{f2}
\nicexy{\bigl(F(X) \otimes F(Y)\bigr) \otimes F(Z) \ar[r]^-{\alpha_{\N}}_{\cong} \ar[d]_-{F_2 \otimes \Id} & F(X) \otimes \bigl(F(Y) \otimes F(Z)\bigr) \ar[d]^-{\Id \otimes F_2}\\
F(X \otimes Y) \otimes F(Z) \ar[d]_-{F_2} & F(X) \otimes F(Y \otimes Z) \ar[d]^-{F_2}\\
F\bigl((X \otimes Y) \otimes Z\bigr) \ar[r]^-{F(\alpha_{\M})}_-{\cong} &
F\bigl(X \otimes (Y \otimes Z)\bigr)}
\end{equation}
is commutative for all objects $X,Y,Z \in \M$.
\item[Compatibility with the Left Units]
The diagram
\begin{equation}\label{f0-left}
\nicexy{\tensorunit_{\N} \otimes F(X) \ar[d]_-{F_0 \otimes \Id} \ar[r]^-{\lambda_{\N}}_-{\cong}
& F(X) \\ F(\tensorunit_{\M}) \otimes F(X) \ar[r]^-{F_2} & F(\tensorunit_{\M} \otimes X)
\ar[u]_-{F(\lambda_{\M})}^-{\cong}}
\end{equation}
is commutative for all objects $X \in \M$.
\item[Compatibility with the Right Units]
The diagram
\begin{equation}\label{f0-right}
\nicexy{F(X) \otimes \tensorunit_{\N} \ar[d]_-{\Id \otimes F_0} \ar[r]^-{\rho_{\N}}_-{\cong}
& F(X) \\ F(X) \otimes F(\tensorunit_{\M}) \ar[r]^-{F_2} & F(X \otimes \tensorunit_{\M})
\ar[u]_-{F(\rho_{\M})}^-{\cong}}
\end{equation}
is commutative for all objects $X \in \M$.
\end{description}
A \index{strong monoidal functor}\index{monoidal functor!strong}\emph{strong monoidal functor} is a monoidal functor in which the morphisms $F_0$ and $F_2$ are all isomorphisms. 
\end{definition}

\begin{definition}\label{def:monoidal-nat-transf}
A \emph{monoidal natural transformation}\index{monoidal natural transformation}\index{natural transformation!monoidal} \[\theta : (F,F_2,F_0) \to (G,G_2,G_0)\] between monoidal functors $F,G : \M \to \N$ is a natural transformation of the underlying functors $\theta : F \to G$ that is compatible with the structure morphisms in the sense that the diagrams
\[\nicexy{F(X) \otimes F(Y) \ar[d]_-{F_2} \ar[r]^-{(\theta_X, \theta_Y)} & G(X) \otimes G(Y) \ar[d]^-{G_2}\\ F(X \otimes Y) \ar[r]^-{\theta_{X\otimes Y}} & G(X \otimes Y)} \quad
\nicexy{\tensorunit_{\N} \ar[r]^-{F_0} \ar[dr]_-{G_0} & F(\tensorunit_{M}) \ar[d]^-{\theta_{\tensorunit_{\M}}}\\ & G(\tensorunit_{\M})}\]
are commutative for all objects $X,Y \in \M$.
\end{definition}

The proof of the following result can be found in \cite{maclane} (XI.3). 
 
\begin{theorem}[Mac Lane's Coherence Theorem]\label{maclane-thm}
Suppose\index{Mac Lane's Coherence Theorem}\index{monoidal category!Mac Lane's Coherence Theorem} $\M$ is a monoidal category.  Then there exist a strict monoidal category $\Mbar$ and an adjoint equivalence
\[\nicexy{\Mbar \ar@<2pt>[r]^-{F} & \M \ar@<2pt>[l]^-{G}}\]
such that both $F$ and $G$ are strong monoidal functors.
\end{theorem}

\begin{convention}\label{rk:maclane-theorem}
Following common practice, using Mac Lane's Coherence Theorem \ref{maclane-thm}, we will omit parentheses for monoidal products of multiple objects in a monoidal category, replacing it by an adjoint equivalent strict monoidal category, via strong monoidal functors, if necessary.  In the rest of this book, Mac Lane's Coherence Theorem will be used without further comment. 
\end{convention}

\begin{definition}\label{def:symmetric-monoidal-functor}
Suppose $\M$ and $\N$ are symmetric monoidal categories.  A \index{functor!symmetric monoidal}\index{symmetric monoidal functor}\index{monoidal functor!symmetric}\emph{symmetric monoidal functor} $(F,F_2,F_0) : \M \to \N$
is a monoidal functor that is compatible with the symmetry isomorphisms, in the sense that the diagram
\begin{equation}\label{monoidal-functor-symmetry}
\nicexy@C+.5cm{F(X) \otimes F(Y) \ar[d]_-{F_2} \ar[r]^-{\xi_{FX,FY}}_-{\cong} & F(Y) \otimes F(X) \ar[d]^-{F_2} \\ F(X \otimes Y) \ar[r]^-{F\xi_{X,Y}}_-{\cong} & F(Y \otimes X)}
\end{equation}
is commutative for all objects $X,Y \in \M$.
\end{definition}

\begin{example}
Suppose $(\M,\otimes,\tensorunit)$ is a symmetric monoidal category with all set-indexed coproducts.  Then the functor
\[\nicexy{\Set \ar[r]^-{F} & \M},\quad FX = \coprodover{x \in X} \tensorunit\]
is a strong symmetric monoidal functor.\dqed
\end{example}

\begin{example}
The singular chain functor\index{singular chain functor}\index{functor!singular chain} $C: \Top \to \Chainz$ is a symmetric monoidal functor \cite{massey} (XI.3).\dqed
\end{example}

\begin{example}\label{ex:monoidal-functor-cat}
Given two monoidal categories $\M$ and $\N$, there is a category \[\MFun(\M,\N)\] whose objects are monoidal functors $\M \to \N$ and whose morphisms are monoidal natural transformations between such monoidal functors.  If $\M$ and $\N$ are furthermore symmetric monoidal categories, then there is a category \[\SMFun(\M,\N)\] whose objects are symmetric monoidal functors $\M \to \N$ and whose morphisms are monoidal natural transformations between such symmetric monoidal functors.\dqed
\end{example}

\begin{example}
Suppose $*$ is a category with one object and only the identity morphism.  It has an obvious symmetric strict monoidal structure.\dqed
\end{example}

\begin{example}
For each category $\C$, $\bigl(\Fun(\C,\C),\circ,\Id_{\C}\bigr)$ is a strict monoidal category, where $\circ$ is composition of functors.\dqed
\end{example}

An important property of a symmetric monoidal closed category is that its monoidal product preserves colimits in each variable.  Indeed, a left adjoint preserves colimits (Theorem \ref{thm:lapc}).  So by symmetry each side of a symmetric monoidal product preserves colimits.  The following observation is a special case of the dual of \cite{maclane} (p.231 Corollary)

\begin{theorem}\label{thm:colim-tensor}
Suppose $\M$ is a symmetric monoidal closed category, and $F : \C \to \M$ and $G : \D \to \M$ are functors with $\C$ and $\D$ small categories that admit colimits.  Then there is a canonical isomorphism\index{monoidal product!preserves colimits}\index{colimit!preserved by monoidal product}
\[\nicexy{\colimover{\C\times\D} \, F\otimes G \ar[r]^-{\cong} & \Bigl(\colimover{\C}\, F\Bigr) \otimes \Bigl(\colimover{\D}\, G\Bigr)}\]
in which $F\otimes G$ is the composition of the functors
\[\nicexy{\C \times \D \ar[r]^-{(F,G)} & \M \times \M \ar[r]^-{\otimes} & \M}.\]
\end{theorem}

\begin{example}\label{ex:coproduct-tensor}
Suppose $\M$ is a symmetric monoidal closed category with all set-indexed coproducts.  Then there is a canonical isomorphism\index{coproduct!preserved by monoidal product}
\[\Bigl(\coprod_{a \in A} X_a\Bigr) \otimes \Bigl(\coprod_{b \in B} Y_b\Bigr) \cong
\coprod_{(a,b) \in A \times B} X_a \otimes Y_b\]
for any sets $A$ and $B$ with $X_a, Y_b \in \M$.\dqed
\end{example}

\section{Monoids}\label{sec:monoids}

Below $*$ denotes the category with one object and only the identity morphism.

\begin{definition}[Monoids]\label{def:monoid}
Suppose $\M$ is a monoidal category.  
\begin{enumerate}
\item Define the category\label{notation:monm} \[\Monm = \MFun(*,\M)\] of monoidal functors $* \to \M$, whose objects are called \index{monoid}\emph{monoids in $\M$}.    
\item Suppose $\M$ is also symmetric.  Define the category\label{notation:comm} \[\Comm = \SMFun(*,\M)\] of symmetric monoidal functors $* \to \M$, whose objects are called \index{commutative monoids}\emph{commutative monoids in $\M$}.
\end{enumerate}
\end{definition}

A simple exercise in unwrapping the definitions yields the following more explicit description of a (commutative) monoid.

\begin{proposition}\label{prop:monoid}
Suppose $\M$ is a  monoidal category.
\begin{enumerate}
\item A monoid in $\M$ is exactly a triple $(A,\mu,\varepsilon)$ consisting of 
\begin{itemize}\item an object $A \in \M$, 
\item a multiplication morphism $\mu : A \otimes A \to A$, and 
\item a unit $\varepsilon : \tensorunit \to A$ 
\end{itemize} 
such that the associativity and unity diagrams
\[\nicexy{A \otimes A \otimes A \ar[d]_-{(\mu,\Id_A)} \ar[r]^-{(\Id_A,\mu)} & A \otimes A \ar[d]^-{\mu}\\ A \otimes A \ar[r]^-{\mu} & A}\qquad
\nicexy{\tensorunit \otimes A \ar[dr]_-{\cong} \ar[r]^-{(\varepsilon, \Id_A)} & A \otimes A \ar[d]^-{\mu} & A \otimes \tensorunit \ar[l]_-{(\Id_A,\varepsilon)} \ar[dl]^-{\cong}\\ & A&}\]
are commutative.  A morphism of monoids is a morphism of the underlying objects that is compatible with the multiplications and the units.
\item Suppose $\M$ is a symmetric monoidal category.  Then a commutative monoid in $\M$ is exactly a monoid whose multiplication is commutative in the sense that the diagram
\[\nicexy@C+.4cm{A \otimes A \ar[d]_-{\mu} \ar[r]^-{\mathrm{permute}} & A \otimes A \ar[d]^-{\mu}\\
A \ar[r]^-{\Id_A} & A}\]
is commutative.  A morphism of commutative monoids is a morphism of the underlying objects that is compatible with the multiplications and the units.
\end{enumerate}\end{proposition}

\begin{example}
A monoid in $\Set$ is a monoid in the usual sense.  A monoid in $\Top$ is a \index{topological monoid}topological monoid.\dqed
\end{example}

\begin{example}
In $\Vectk$ a (commutative) monoid is exactly a (commutative) \index{algebra}$\fieldk$-algebra.  In $\Chaink$ a (commutative) monoid is exactly a (commutative) differential graded\index{differential graded algebra}\index{commutative differential graded algebra} $\fieldk$-algebra.\dqed
\end{example}

The following result is a slight extension of Definition \ref{def:monoid} of (commutative) monoids as (symmetric) monoidal functors.

\begin{proposition}\label{prop:finite-coprod}
Suppose $\C$ is a small category with all finite coproducts, regarded as a symmetric monoidal category $(\C,\amalg,\varnothing_{\C})$ under coproducts.  Suppose $\M$ is a monoidal category.  
\begin{enumerate}\item Then there is a canonical isomorphism\index{diagram of monoids}\index{monoid!diagram of} \[\Monm^{\C} \iso \MFun(\C,\M)\] between the category of $\C$-diagrams of monoids in $\M$ and the category of monoidal functors from $\C$ to $\M$.
\item Suppose $\M$ is a symmetric monoidal category. Then there is a canonical isomorphism\index{diagram of commutative monoids}\index{commutative monoid!diagram of} \[\Comm^{\C} \iso \SMFun(\C,\M)\] between the category of $\C$-diagrams of commutative monoids in $\M$ and the category of symmetric monoidal functors from $\C$ to $\M$.
\end{enumerate}
\end{proposition}

\begin{proof}
A (symmetric) monoidal functor $F : \C \to \M$ is equipped with a morphism \eqref{monoidal-f0} \[\nicexy{\tensorunit \ar[r]^-{F_0} & F(\varnothing_{\C})} \in \M\] with $\varnothing_{\C}$ an initial object in $\C$, which exists by our assumption on $\C$.  For each object $c \in \C$, the unique morphism $0_c : \varnothing_{\C} \to c$ and $F_0$ yield the composition \[\nicexy{\tensorunit \ar[d]_-{F_0} \ar[r]^-{\operadunit_c} & F(c)\\ F(\varnothing_{\C}) \ar[r]^-{F(0_c)} & F(c) \ar@{=}[u]} \in \M.\]  The monoidal functor $F$ is also equipped with a morphism \eqref{monoidal-f2} \[\nicexy{F(c) \otimes F(d) \ar[r]^-{F_2} & F(c\amalg d)} \in \M\] that is natural in $c,d \in\C$.   The morphism \[(\Id_c,\Id_c) : c \amalg c \to c \in \C\] and $F_2$ yield the composition \[\nicexy@C+.6cm{F(c) \otimes F(c) \ar[d]_-{F_2} \ar[r]^-{\mu_c} & F(c)\\ F(c\amalg c) \ar[r]^-{F(\Id_c,\Id_c)} & F(c) \ar@{=}[u]} \in \M\] for each object $c \in \C$.

Now one checks that the associativity diagram \eqref{f2} corresponds to the associativity of the morphism $\mu_c$, while the unity diagrams \eqref{f0-left} and \eqref{f0-right} correspond to the property that $\operadunit_c$ is a two-sided unit of $\mu_c$ as in Proposition \ref{prop:monoid}.  Furthermore, the symmetry diagram \eqref{monoidal-functor-symmetry} corresponds to the commutativity of $\mu_c$.  Therefore, $(F(c),\mu_c,\operadunit_c)$ is a (commutative) monoid for each $c\in \C$.  That we have a $\C$-diagram of (commutative) monoids in $\M$ corresponds to the functoriality of $F$ and $F_2$.

Conversely, given $F \in \Monmc$, we will write $(F(c),\mu_c,\operadunit_c) \in \Monmc$ for its value at $c \in \C$.  The monoidal structure on $F$ is defined as follows.  The monoid unit for $F(\varnothing_{\C})$ is a morphism $F_0 : \tensorunit \to F(\varnothing_{\C})$.  The composition \[\nicexy{F(c) \otimes F(d) \ar[d]_-{(F\iota_c,F\iota_d)} \ar[r]^-{F_2} & F(c\amalg d)\\ F(c\amalg d) \otimes F(c\amalg d) \ar[r]^-{\mu_{c \amalg d}} & F(c\amalg d) \ar@{=}[u]}\] is natural in $c,d\in \C$, where \[\iota_c : c \to c\amalg d \in \C\] is the natural morphism.  To simplify the typography below, we will write coproducts in $\C$ as concatenation, so $ab$ means $a \amalg b$.  The desired associativity diagram \eqref{f2} of $F$ is the outermost diagram in
\[\nicexy@C+.6cm@R+.5cm{F(a) \otimes F(b) \otimes F(c) \ar[d]_-{(F\iota_a,F\iota_b,\Id)} \ar[dr]|(.4){(F\iota_a,F\iota_b,F\iota_c)} \ar[r]^-{(\Id,F\iota_b,F\iota_c)} & F(a) \otimes F(bc)^{\otimes 2} \ar[d]^-{(F\iota_a,(F\iota_{bc})^{\otimes 2})} \ar[r]^-{(\Id,\mu_{bc})} & F(a) \otimes F(bc) \ar[d]^-{(F\iota_a,F\iota_{bc})}\\
F(ab)^{\otimes 2} \otimes F(c) \ar@{}[ur]_(.4){(1)} \ar@{}[ur]_(.7){(2)} \ar[d]_-{(\mu_{ab},\Id)} \ar[r]_-{((F\iota_{ab})^{\otimes 2}, F\iota_c)} & F(abc)^{\otimes 3} \ar[d]^-{(\mu_{abc},\Id)} \ar[r]^-{(\Id,\mu_{abc})} & F(abc)^{\otimes 2} \ar[d]^-{\mu_{abc}}\\
F(ab) \otimes F(c) \ar[r]^-{(F\iota_{ab},F\iota_c)} & F(abc)^{\otimes 2} \ar[r]^-{\mu_{abc}} & F(abc)}\] for $a,b,c\in \C$. The sub-diagrams (1) and (2) are commutative by the functoriality of $F$.  The lower left and upper right rectangles are commutative because $F\iota_{ab}$ and $F\iota_{bc}$ are morphisms of monoids, hence compatible with the multiplication.  The lower right rectangle is commutative by the associativity of the monoid multiplication $\mu_{abc}$.

The compatibility with the left unit \eqref{f0-left} is the outer diagram in \[\nicexy@C+.4cm@R+.3cm{\tensorunit \otimes F(c) \ar[r]^-{\cong} \ar[d]_-{(\operadunit_{\varnothing_{\C}},\Id)} \ar[dr]|(.4){(\operadunit_c,\Id)} & F(c)\\ F(\varnothing_{\C}) \otimes F(c) \ar[r]^-{(F0_c,\Id)} & F(c)\otimes F(c) \ar[u]_-{\mu_c}}\] for $c\in \C$.  The lower left triangle is commutative because $F0_c$ preserves the monoid units.  The upper right triangle is commutative by part of the unity condition of the monoid $F(c)$ in Proposition \ref{prop:monoid}.  The compatibility with the right unit \eqref{f0-right} is proved similarly.

Finally, one can check that, under the above correspondence, natural transformations in $\Monmc$ and $\Commc$ correspond to monoidal natural transformations in $\MFun(\C,\M)$ and $\SMFun(\C,\M)$, respectively.
\end{proof}

As one would expect, monoids can act on objects.

\begin{definition}[Modules over a Monoid]\label{def:module-monoid}
Suppose $(A,\mu,\varepsilon)$ is a monoid in a monoidal category $\M$.
\begin{enumerate}
\item A \emph{left $A$-module}\index{module}\index{monoid!module} is a pair $(X,m)$ consisting of 
\begin{itemize}\item an object $X \in \M$ and 
\item a left $A$-action $m : A \otimes X \to X \in \M$ 
\end{itemize}
such that the associativity and unity diagrams \[\nicexy@C+.4cm{A \otimes A \otimes X \ar[d]_-{(\mu,\Id_X)} \ar[r]^-{(\Id_A,m)} & A \otimes X \ar[d]^-{m}\\ A \otimes X\ar[r]^-{m} & X}\qquad
\nicexy{\tensorunit \otimes X \ar[dr]_-{\cong} \ar[r]^-{(\varepsilon, \Id_X)} & A \otimes X \ar[d]^-{m}\\ & X}\]
are commutative.  A morphism of left $A$-modules is a morphism of the underlying objects that is compatible with the left $A$-actions in the obvious sense.
\item The category of left $A$-modules is denoted by $\Mod(A)$.
\end{enumerate}
\end{definition}

\begin{example}
In $\Vectk$ and $\Chaink$, this concept of a left module coincides with the usual one.\dqed
\end{example}

\section{Monads}\label{sec:monads}

\begin{definition}\label{def:monad}
For a category $\C$, a \emph{monad on $\C$}\index{monad} is defined as a monoid in the strict monoidal category $\bigl(\Fun(\C,\C),\circ,\Id_{\C}\bigr)$, where $\circ$ is composition of functors.
\end{definition}

Unwrapping this definition using Proposition \ref{prop:monoid}, a monad can be described more explicitly as follows.

\begin{proposition}\label{prop:monad-explicit}
Given a category $\C$, a monad on $\C$ is exactly a triple $(T,\mu,\varepsilon)$ consisting of 
\begin{itemize}\item a functor $T : \C \to \C$, 
\item a natural transformation $\mu : TT \to T$ called the \emph{multiplication}, and
\item a natural transformation $\varepsilon : \Id_{\C} \to T$ called the \emph{unit}, 
\end{itemize}
such that the following associativity and unity diagrams are commutative.
\[\nicexy{TTT \ar[d]_{\mu T} \ar[r]^-{T\mu} & TT \ar[d]^-{\mu}\\ TT \ar[r]^-{\mu} & T}
\qquad \nicexy{T \ar[r]^-{\varepsilon T} \ar[dr]_-{\Id} & TT \ar[d]^-{\mu} & T \ar[l]_-{T\varepsilon} \ar[dl]^-{\Id}\\ &T&}\]
\end{proposition}

\begin{example}\label{ex:adjunction-monad}
Suppose $F : \C \adjoint \D : G$ is an adjunction.  Then \[T=GF : \C \to \C\] is the functor of a monad on $\C$ whose unit is the unit of the adjunction $\eta : \Id_{\C} \to GF$.  The multiplication is \[\mu=G\epsilon F : TT = GFGF \to GF=T,\] where $\epsilon : FG \to \Id_{\D}$ is the counit of the adjunction.\dqed
\end{example}

We defined monads as monoids in the functor category $\Fun(\C,\C)$.  Conversely, the next example shows that each monoid in a monoidal category yields a monad.

\begin{example}\label{ex:monoid-monad}
Suppose $(\M,\otimes,\tensorunit)$ is a monoidal category, and $(A,\mu,\varepsilon)$ is a monoid in $\M$ as in Proposition \ref{prop:monoid}.  Then there is a monad\index{monoid!induced monad} on $\M$ with the functor $T = A \otimes -$, whose multiplication and unit are induced by those of $A$.  The monadic associativity and unity diagrams are exactly those of the monoid $A$ in Proposition \ref{prop:monoid}.\dqed
\end{example}

\begin{definition}\label{def:monad-algebra}
Suppose $(T,\mu,\varepsilon)$ is a monad on a category $\C$.  A \emph{$T$-algebra}\index{algebra!of a monad}\index{monad!algebra} is a pair $(X,\lambda)$ consisting of 
\begin{itemize}\item an object $X \in \C$ and 
\item a structure morphism $\lambda : TX \to X$ 
\end{itemize}
such that the following associativity and unity diagrams are commutative.
\[\nicexy{TTX \ar[d]_-{\mu_X} \ar[r]^-{T\lambda} & TX \ar[d]^-{\lambda}\\ TX \ar[r]^-{\lambda} & X}\qquad \nicexy{X \ar[dr]_-{\Id} \ar[r]^{\varepsilon_X} & TX \ar[d]^-{\lambda}\\ & X}\]
A morphism of $T$-algebras $f : (X,\lambda) \to (Y,\pi)$ is a morphism $f : X \to Y$ in $\M$ such that the diagram
\[\nicexy{TX \ar[d]_-{\lambda} \ar[r]^-{Tf} & TY \ar[d]^-{\pi}\\ X \ar[r]^-{f} & Y}\]
is commutative.  The category of $T$-algebras and their morphisms is denoted by $\algct$.
\end{definition}

Example \ref{ex:adjunction-monad} says that each adjunction yields a monad on the domain category of the left adjoint.  The next example is the converse.

\begin{example}\label{ex:monad-free-algebra}
Suppose $(T,\mu,\varepsilon)$ is a monad on a category $\C$.  For each object $X \in \C$, the pair \[\bigl(TX,\mu_X : TTX \to TX\bigr)\] is a $T$-algebra, called the \index{monad!free algebra}\index{free algebra}\emph{free $T$-algebra of $X$}.  There is a free-forgetful adjunction
\begin{equation}\label{eilenberg-moore}
\nicexy{\C \ar@<2pt>[r]^-{T} & \algct \ar@<2pt>[l]^-{U}}
\end{equation}
in which the right adjoint $U$ forgets about the $T$-algebra structure and remembers only the underlying object.  The left adjoint sends an object to its free $T$-algebra.  This adjunction is known as the \index{Eilenberg-Moore adjunction}\index{adjunction!Eilenberg-Moore}\emph{Eilenberg-Moore adjunction}.\dqed
\end{example}

\begin{example}\label{ex:monoid-module}
In the setting of Example \ref{ex:monoid-monad}, a $T$-algebra is a pair\index{module!as monad algebra} $(X,\lambda)$ consisting of an object $X \in \M$ and a structure morphism $\lambda : A \otimes X \to X$ such that the following associativity and unity diagrams are commutative.
\[\nicexy{A \otimes A \otimes X \ar[d]_-{(\mu,\Id_X)} \ar[r]^-{(A,\lambda)} & A \otimes X \ar[d]^-{\lambda} \\ A \otimes X \ar[r]^-{\lambda} & X} \qquad
\nicexy{X \cong \tensorunit \otimes X\ar[dr]_-{\Id} \ar[r]^-{(\varepsilon,X)} & A \otimes X \ar[d]^-{\lambda}\\ & X}\]
This is exactly a left $A$-module.\dqed
\end{example}

The following coequalizer characterization of an algebra over a monad is \cite{bor2} Lemma 4.3.3.

\begin{proposition}\label{prop:algebra-coequalizer}
Suppose $(T,\mu,\varepsilon)$ is a monad on a category $\C$, and $(X,\lambda)$ is a $T$-algebra.  Then the diagram
\[\nicexy{\bigl(TTX,\mu_{TX}\bigr) \ar@<2pt>[r]^-{\mu_X} \ar@<-2pt>[r]_-{T\lambda} & \bigl(TX,\mu_X\bigr) \ar[r]^-{\lambda} & (X,\lambda)}\]
is a coequalizer in $\algct$.
\end{proposition}

\begin{interpretation}Proposition \ref{prop:algebra-coequalizer} says that every algebra over a monad $T$ is a quotient of the free $T$-algebra on its underlying object, with relations given by the monad multiplication and the $T$-algebra structure morphism.\dqed\end{interpretation}

\section{Localization}\label{sec:localization}

Localization of categories will play an important role in encoding the time-slice axiom in (homotopy) algebraic quantum field theory.  Here we recall its definition and construction.  The idea of localization is to formally invert some morphisms and make them into isomorphisms.  The process is similar to the construction of the rational numbers from the integers.  Later we will also need the operad version of localization.

\begin{definition}\label{def:localization-cat}
Suppose $\C$ is a category, and $S \subseteq \Morc$.  An \emph{$S$-localization of $\C$}, if it exists, is a pair\index{category!localization}\index{localization of a category} $(\Csinv,\ell)$ consisting of
\begin{itemize}\item a category $\Csinv$ and
\item a functor $\ell : \C \to \Csinv$ 
\end{itemize}
that satisfies the following two conditions:
\begin{enumerate}
\item $\ell(f)$ is an isomorphism for each $f \in S$.
\item $(\Csinv,\ell)$ is initial with respect to the previous property.  In other words, if $F : \C \to \D$ is a functor such that $F(f)$ is an isomorphism for each $f \in S$, then there exists a unique functor \[F' : \Csinv \to \D \stspace F= F'\ell.\]
\begin{equation}\label{localization-triangle}
\nicexy{& \C \ar[r]^-{\ell} \ar[d]_-{\forall\, F} & \Csinv \ar@{.>}[dl]^-{\exists !\, F'}\\ 
F(S) ~\mathrm{iso} & \D &}
\end{equation}
\end{enumerate}
In this setting, $\ell$ is called the\index{localization functor}\index{functor!localization} \emph{$S$-localization functor}.
\end{definition}

By the universal property of an $S$-localization, $\Csinv$ is unique up to a unique isomorphism if it exists.  The following observation says that when $S$ is small enough, the localization always exists.

\begin{theorem}\label{thm:localization-cat}
Suppose $\C$ is a category, and $S$ is a set of morphisms in $\C$.  Then the $S$-localization $\Csinv$ exists such that $\Obc = \Obcsinv$ and that the localization functor $\ell$ is the identity function on objects. 
\end{theorem}

\begin{proof}
The proof can be found in \cite{bor1} Section 5.2.  Since we will need the operad version later,   we provide a sketch of the proof here.  Without loss of generality, we may assume that $S$ is closed under composition.  For each $f \in S$, suppose $\finverse$ is a symbol such that the sets $S$ and $\Sinv=\{\finverse : f \in S\}$ are disjoint.  Define a category $\Csinv$ by setting $\Obc = \Obcsinv$.  For objects $a,b \in \C$, the morphisms in $\Csinv(a,b)$ are the equivalence classes of finite alternating sequences
\[\varphi=\bigl(g_{n+1},\finverse_n, g_n, \cdots, \finverse_2, g_2, \finverse_1, g_1\bigr)\]
with each $g_i \in \Morc$ and each $f_i \in S$, which we visualize as follows.  
\[\narrowxy{a \ar[r]^-{g_1} & \bullet \ar[r]^-{\finverse_1} & \bullet \ar[r]^-{g_2} \ar@/^/[l]^-{f_1} & \cdots \ar[r]^-{g_n} & \bullet \ar[r]^-{\finverse_n} & \bullet \ar@/^/[l]^-{f_n} \ar[r]^-{g_{n+1}} & b}\]
Such a sequence is not required to start or end with some $g_i$.  For $1 \leq i \leq n$, the domain of $f_i$ is the domain of $g_{i+1}$ (if it exists in the sequence), and the codomain of $f_i$ is the codomain of $g_i$ (again if it exists in the sequence).  If $g_1$ is part of $\varphi$, then its domain is $a$.  Otherwise, the codomain of $f_1$ is $a$.  If $g_{n+1}$ is part of $\varphi$, then its codomain is $b$.  Otherwise, the domain of $f_n$ is $b$.

The equivalence relation is generated by the following three identifications:
\begin{enumerate}
\item If $g_i$ is the identity morphism, then $\varphi$ is identified with the sequence $\phi$ obtained by replacing the subsequence $(\finverse_i,g_i,\finverse_{i-1})$ with the entry $\inv{(f_{i-1}f_i)}$.  If this $g_i$ happens to be the first or the last entry of $\varphi$, then it is omitted in $\phi$.
\item If $f_i$ is the identity morphism, then $\varphi$ is identified with the sequence $\phi$  obtained by replacing the subsequence $(g_{i+1},\finverse_i,g_i)$ with the entry $g_{i+1}g_i$.  If this $f_i$ happens to be the first or the last entry of $\varphi$, then it is omitted in $\phi$.
\item If $g_i=f_i$ (resp., $f_i=g_{i+1}$), then the sequence $\varphi$ is identified with the subsequence in which $g_i$ and $\finverse_i$ (resp., $\finverse_i$ and $g_{i+1}$) are omitted.
\end{enumerate}
The assumption that $S$ be a set implies that $\Csinv(a,b)$ is a set.  For each object $a \in \Csinv$, its identity morphism is the equivalence class of the empty sequence.  Composition in $\Csinv$ is induced by concatenation of sequences and composition in $\C$, with $(\inv{h}_1,\finverse_n)$ identified with $\inv{(f_n h_1)}$ if one sequence ends with $\finverse_n$ and the next sequence starts with $\inv{h}_1$ for $f_n,h_1 \in S$.  One checks that this composition is well-defined (i.e., respects the three identifications above) and that $\Csinv$ is indeed a category.

The localization functor $\ell : \C \to \Csinv$ is defined as the identity function on objects.  For a morphism $g \in \C(a,b)$, we define $\ell(g) \in \Csinv(a,b)$ to be the sequence $(g)$.  This defines a functor $\ell$ that sends each morphism $f \in S$ to an isomorphism in $\Csinv$.  Suppose $F : \C \to \D$ is a functor such that $F(f)$ is an isomorphism for each $f \in S$.  The requirement that $F=F'\ell$ \eqref{localization-triangle} forces us to define the functor $F' : \Csinv \to \D$ by defining it to be the same as $F$ on objects and 
\[F'(\varphi) = (Fg_{n+1}) \inv{(Ff_n)} (Fg_n) \cdots \inv{(Ff_1)} (Fg_1)\]
on morphisms.  One checks that this $F'$ is well-defined (i.e., respects the three identifications above).  So $(\Csinv,\ell)$ has the required universal property of the $S$-localization.
\end{proof}

\chapter{Trees}\label{ch:tree}

One of the important descriptions of operads uses the language of trees, which we discuss in this chapter.  The definitions of the Boardman-Vogt construction of a colored operad, homotopy algebraic quantum field theories, and homotopy prefactorization algebras also use trees.  The following material on graphs and trees are adapted from \cite{bluemonster} Part 1, where much more details and many more examples can be found.  In practice, since we mostly work with isomorphism classes of trees, it is sufficient to work pictorially as in the examples below.

\section{Graphs}\label{sec:graphs} 

An \index{involution}\emph{involution} is a self-map $\tau$ such that $\tau^2 = \Id$; it is \emph{free} if it has no fixed points.

\begin{definition}\label{def:graph} Fix an infinite set $\mathfrak{F}$ once and for all.  A \emph{graph}\index{graph} is a tuple\label{notation:flag} \[G = \bigl(\Flag(G),\lambda_G,\iota_G,\pi_G\bigr)\] consisting of:
\begin{itemize}\item a finite set $\Flag(G) \subset \mathfrak{F}$ of \index{flag}\emph{flags};
\item a partition \index{partition} $\lambda_G$ of $\Flag(G)$ into finitely many possibly empty subsets, called \index{cell}\emph{cells}, together with a distinguished cell $G_0$, called the \index{exceptional cell}\emph{exceptional cell};
\item an involution $\iota_G$ on $\Flag(G)$ such that $\iota_G(G_0)=G_0$;
\item a free involution $\pi_G$ on the set of $\iota_G$-fixed points in $G_0$.
\end{itemize}
An \index{isomorphism of graphs}\emph{isomorphism} of graphs is a bijection on flags that preserves the partition and both involutions.  For graphs with any further structure as we will introduce later, an isomorphism is required to preserve that structure as well.
\end{definition}

\begin{definition}\label{def:graph-terminology} Suppose $G$ is a graph.
\begin{itemize}
\item Flags not in the exceptional cell $G_0$ are called \index{ordinary flag}\emph{ordinary flags}.  Flags in $G_0$ are called \index{exceptional flag}\emph{exceptional flags}.
\item $G$ is said to be an \index{ordinary graph}\emph{ordinary graph} if the exceptional cell is empty.
\item A \index{vertex}\emph{vertex} is a cell that is not the exceptional cell.  A flag in a vertex $v$ is said to be \index{adjacent}\emph{adjacent to $v$}.  An \index{isolated vertex}\emph{isolated vertex} is a vertex that is empty.  The cardinality of a vertex $v$ is denoted by $|v|$.\label{notation:vertex}  The set of vertices is denoted by $\Vt(G)$.
\item Two distinct vertices $u$ and $v$ are \emph{adjacent} if there exist flags $a \in u$ and $b \in v$ such that $\iota_G(a) = b$.
\item The fixed points of $\iota_G$ are celled \index{leg}\emph{legs}.  The set of legs is denoted by $\Leg(G)$.\label{notation:leg}  A leg in a vertex is called an \index{ordinary leg}\emph{ordinary leg}.  A leg in the exceptional cell is called an \index{exceptional leg}\emph{exceptional leg}.
\item The orbits of $\pi_G$ and of $\iota_G$ away from its fixed points in $G_0$ are called \index{edge}\emph{edges}.  The set of edges is denoted by $\Ed(G)$.\label{notation:edge}
\item The non-trivial orbits of $\iota_G$ are called \index{internal edge}\emph{internal edges}.  The set of internal edges in $G$ is denoted by $|G|$.  The non-trivial orbits of $\iota_G$ within the vertices are called \index{ordinary internal edge}\emph{ordinary internal edges}.  Those within the exceptional cell are called \index{exceptional loop}\emph{exceptional loops} and denoted by \label{notation:exceptional-loop}$\bigcircle$. 
\item An orbit of $\pi_G$ is called an \index{exceptional edge}\emph{exceptional edge} and denoted by $\edge$.
\item If $f=\{f_{\pm}\}$ is an ordinary internal edge with $f_+ \in u$ and $f_- \in v$, then we say that $f$ is \emph{adjacent} to $u$ and $v$.
\end{itemize}\end{definition}

\begin{example}\label{ex:basic-empty} The \index{graph!empty}\index{empty graph}\emph{empty graph} $\varnothing$ \label{notation:empty-graph} has an empty set of flags, hence an empty exceptional cell, and no vertices.\dqed\end{example}

\begin{example}\label{ex:basic-isolated-vertex} The graph $\bullet$ with an empty set of flags, hence an empty exceptional cell, and a single empty vertex is an \index{isolated vertex}isolated vertex.\dqed\end{example}

\begin{example}\label{ex:basic-exceptional-edge} The graph with no vertices and with only two exceptional legs $f_{\pm}$, which must be paired by the involution $\pi$, is the \index{exceptional edge}exceptional edge $\edge$.\dqed\end{example}

\begin{example}\label{ex:basic-exceptional-loop} The graph with no vertices and with only two exceptional flags $e_{\pm}$ paired by $\iota$ is the \index{exceptional loop}exceptional loop $\bigcircle$.\dqed\end{example}

\begin{definition}\label{def:path} Suppose $G$ is a graph.
\begin{enumerate}\item A \index{path}\emph{path} of length $r \geq 0$ is a pair \[P = \bigl(\{e^i\}_{i=1}^r, \{v_i\}_{i=0}^r\bigr)\] in which:
\begin{itemize} \item the $v_i$'s are distinct vertices, except possibly for $v_0$ and $v_r$;
\item each $e^i$ is an ordinary internal edge adjacent to both $v_{i-1}$ and $v_i$;
\end{itemize}
\item A \index{cycle}\emph{cycle} is a path of length $r \geq 1$ with $v_0=v_r$.
\end{enumerate}\end{definition}

\begin{definition}\label{def:connectivity} A non-empty graph $G$ is:
\begin{enumerate}
\item \index{graph!connected}\index{connected graph}\emph{connected} if it satisfies one of the following two conditions:
\begin{enumerate} \item It is an isolated vertex $\bullet$, the exceptional edge $\edge$, or the exceptional loop $\bigcircle$.
\item It is an ordinary graph that has no isolated vertices such that, for each pair of distinct flags $\{f_1,f_2\}$, there exists a path $P=\bigl(\{e^i\}, \{v_i\}\bigr)$ with $f_1$ adjacent to some $v_k$ and $f_2$ adjacent to some $v_l$.
\end{enumerate}
\item \index{graph!simply-connected}\index{simply-connected}\emph{simply-connected} if it (i) is connected; (ii) is not the exceptional loop; (iii) contains no cycles.
\end{enumerate}\end{definition}

We will need to consider the following extra structures on graphs. 

\begin{definition}\label{def:edge-coloring} For a non-empty set $\colorc$, whose elements are called colors, a  \index{graph!coloring}\index{coloring}\emph{$\colorc$-coloring} of a graph $G$ is a function\label{notation:coloring} \[\kappa : \Flag(G) \to \colorc\] that is constant on each orbit of the involutions $\iota_G$ and $\pi_G$.  In other words, a $\colorc$-coloring assigns to each edge a color.\end{definition}

\begin{definition}\label{def:graph-direction} A \index{graph!direction}\index{direction}\emph{direction} of a graph $G$ is a function\label{notation:direction} \[\delta : \Flag(G) \to \{1,-1\}\] such that:
\begin{itemize} \item If $(f,\iota_G(f))$ is an internal edge, then $\delta(\iota_G(f)) = -\delta(f)$.
\item If $(f,\pi_G(f))$ is an exceptional edge, then $\delta(\pi_G(f)) = -\delta(f)$.\end{itemize}\end{definition}

\begin{definition}\label{def:directed-path} 
Suppose $G$ is a graph equipped with a direction $\delta$.
\begin{itemize} \item A leg $f$ with $\delta(f) = 1$ is an \index{input}\emph{input} of $G$.
\item A leg $f$ with $\delta(f) = -1$ is an \index{output} \index{output}\emph{output} of $G$.
\item If $v$ is a vertex with $f \in v$ and $\delta(f) = 1$, then $f$ is an \index{vertex!input}\emph{input} of $v$.
\item If $v$ is a vertex with $f \in v$ and $\delta(f) = -1$, then $f$ is an \index{vertex!output}\emph{output} of $v$.
\item For $z \in \{G\} \sqcup \Vt(G)$, the set of inputs of $z$ is denoted by $\inp(z)$, and the set of outputs of $z$ is denoted by $\out(z)$.
\item An internal edge is regarded as oriented from the flag with $\delta=-1$ to the flag with $\delta=1$.
\item For an ordinary internal edge $f=\{f_{\pm}\}$ with $\delta(f_{\pm})=\pm 1$, the vertex containing $f_-$ (resp., $f_+$) is the \index{initial vertex}\emph{initial vertex} (resp., \emph{terminal vertex})\index{terminal vertex} of $f$.
\item A \emph{directed path} is a path $P$ as in Definition \ref{def:path} such that each $e^i$ has initial vertex $v_{i-1}$ and terminal vertex $v_i$.  We call $v_0$ (resp., $v_r$) the \emph{initial vertex} (resp., \emph{terminal vertex}) of $P$.
\end{itemize}\end{definition}

\begin{definition}\label{def:unordered-tree}
An \index{unordered tree}\emph{unordered tree} is a pair $(T,\delta)$ consisting of
\begin{itemize}\item a simply-connected graph $T$ and 
\item a direction $\delta$ 
\end{itemize}
such that $|\out(v)|=1$ for each $v \in \Vt(T)$.  In an unordered tree $T$ that is not isomorphic to an exceptional edge, the unique vertex containing the output of $T$ is called the \index{root vertex}\index{vertex!root}\emph{root vertex}.
\end{definition}

\begin{definition}\label{def:directed-graph-listing}
Suppose $(T,\delta)$ is an unordered tree.
\begin{enumerate} \item An \index{ordering!at a vertex}\index{vertex!ordering}\emph{ordering at a vertex $v$} is a bijection\label{notation:ordering} \[\zeta_v : \bigl\{1,\ldots,|\inp(v)|\bigr\} \iso \inp(v).\]
\item An \index{ordering!of an unordered tree}\index{tree!ordering}\emph{ordering of $T$} is a bijection \[\zeta_T : \bigl\{1,\ldots,|\inp(T)|\bigr\} \iso \inp(T).\]\end{enumerate}
A \index{listing}\emph{listing} of $(T,\delta)$ is a choice of an ordering for each $z \in \{T\} \sqcup \Vt(T)$.  Given a listing, we will regard each $\inp(z)$ as an ordered set.
\end{definition}

\begin{definition}\label{def:profofc}
Suppose $\colorc$ is a non-empty set, whose elements are called \index{color}\emph{colors}.  A \index{profile}\emph{$\colorc$-profile} is a finite sequence of elements in $\colorc$.  
\begin{itemize}\item If $\colorc$ is clear from the context, then we simply say \emph{profile}.
\item The empty $\colorc$-profile is denoted by $\emptyset$.  
\item We write $|\uc|=m$ for the \index{length}\emph{length} of a profile\label{notation:profile} $\uc=(c_1,\ldots,c_m)$.  
\item The set of $\colorc$-profiles is denoted by $\Profc$.\label{notation:profc}
\item An element in $\Profcc$ is written either horizontally as\label{notation:duc} $(\uc;d)$ or vertically as $\duc$.
\end{itemize}
\end{definition}

\begin{definition}\label{def:tree}
A \emph{$\colorc$-colored tree}\index{colored tree}\index{tree!colored} is a tuple $(T,\delta,\kappa,\zeta)$ consisting of an unordered tree $(T,\delta)$, a $\colorc$-coloring $\kappa$, and a listing $\zeta$.  Given such a $\colorc$-colored tree, using the $\colorc$-coloring $\kappa$:
\begin{enumerate}
\item For $z \in \{T\} \sqcup \Vt(T)$, we regard the ordered set $\inp(z)$ as a $\colorc$-profile, called the \index{input profile}\index{profile!input}\emph{input profile of $z$}, whose $j$th entry is denoted by $\inp(z)_j$.  Similarly, we regard the element $\out(z) \in \colorc$, called the \index{output color}\index{color!output}\emph{output color of $z$}.
\item For $z \in \{T\} \sqcup \Vt(T)$, the \index{tree!profile}\index{vertex!profile}\emph{profile of $z$} is the pair\label{notation:profz} \[\profofz = \bigl(\inp(z);\out(z)\bigr) = \inout{z} \in \Profcc.\]
\end{enumerate}
The set of isomorphism classes of $\colorc$-colored trees with profile $(\uc;d) \in \Profcc$ is denoted by\label{notation:treecduc} $\Treec(\uc;d)$ or $\Treec\duc$.  We will omit mentioning $\colorc$ if it is clear from the context.
\end{definition}

\begin{convention} For a vertex $v$ in a $\colorc$-colored tree, to simplify the typography, we will often abbreviate $\profofv$ to just $(v)$.  From now on, the single word \emph{tree} will mean a $\colorc$-colored tree, unless otherwise specified.
\end{convention}

\begin{definition}\label{def:linear-graph}
A \emph{$\colorc$-colored linear graph}\index{linear graph} is a $\colorc$-colored tree $T$ such that $|\inp(v)| = 1$ for each $v \in \Vt(T)$.  The set of isomorphism classes of $\colorc$-colored linear graphs with profile $(c;d) \in \colorc^{\times 2}$ is denoted by $\Linearc(c;d)$ or  $\Linearc\dc$.
\end{definition}

\begin{example}[Exceptional edges]\label{ex:colored-exedge} 
The exceptional edge $\edge$ in Example \ref{ex:basic-exceptional-edge} can be given a direction $\delta$ with $\delta(f_{\pm})=\pm 1$.  For each color $c \in \colorc$, it becomes a $c$-colored linear graph $\uparrow_c$ with profile $(c;c)$, called the \index{exceptional edge}\emph{$\colorc$-colored exceptional edge}, in which the bottom (resp., top) flag is $f_+$ (resp., $f_{-}$), with coloring $\kappa(f_{\pm}) = c$ and with a trivial listing.  These colored exceptional edges are the only colored trees with exceptional flags.  As we will see later, the $c$-colored exceptional edge corresponds to the $c$-colored unit of a $\colorc$-colored operad.\dqed
\end{example}

\begin{example}[Linear graphs]\label{ex:linear-graph}
For each $\colorc$-profile $\uc=(c=c_0,\ldots,c_n=d)$ with $n \geq 0$, there is a $\colorc$-colored \emph{linear graph} \[\Lin_{\uc} \in \Linearc\dc\] defined as follows.
\begin{itemize}
\item $\Flag\bigl(\Lin_{\uc}\bigr) = \bigl\{i,e^1_{\pm},\ldots,e^{n-1}_{\pm},o\bigr\}$, all of which are ordinary flags.  Note that the flags $e^j_{\pm}$ are only in $\Flag\bigl(\Lin_{\uc}\bigr)$ if $n\geq 2$.
\item The involution $\iota$ fixes $i$ and $o$, and $\iota(e^j_{\pm})=e^j_{\mp}$ for $1\leq j \leq n-1$.
\item There are $n$ vertices $v_j = \{e^{j-1}_+,e^j_-\}$ for $1\leq j \leq n$, where $e^0_+=i$ and $e^n_-=o$.
\item $\kappa(i)=c_0$, $\kappa(e^j_{\pm}) = c_j$ for $1 \leq j \leq n-1$, and $\kappa(o)=c_n$.
\item $\delta(i)=1$, $\delta(e^j_{\pm})=\pm 1$ for $1 \leq j \leq n-1$, and $\delta(o)=-1$.
\end{itemize}
If $n=0$, then $\Lin_{\uc}$ is the $c$-colored exceptional edge $\uparrow_c$ in Example \ref{ex:colored-exedge}.  If $n \geq 1$, then we depict the linear graph $\Lin_{\uc}$ as follows.
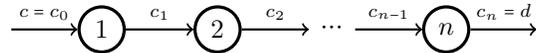
\begin{center}\begin{tikzpicture}
\matrix[row sep=.5cm, column sep=.9cm]{\node [plain] (1) {$1$}; & \node [plain] (2) {$2$}; & \node [empty] (3) {$\cdots$}; & \node [plain] (n) {$n$};\\};
\draw [inputleg] (1) to node[swap]{\scriptsize{$c=c_0$}} +(-1.2cm,0);
\draw [arrow] (1) to node{\scriptsize{$c_1$}} (2);
\draw [arrow] (2) to node{\scriptsize{$c_2$}} (3);
\draw [arrow] (3) to node{\scriptsize{$c_{n-1}$}} (n);
\draw [outputleg] (n) to node{\scriptsize{$c_n=d$}} +(1.2cm,0);
\end{tikzpicture}\end{center}
It has $n-1$ internal edges $e^j=\{e^j_{\pm}\}$ for $1 \leq j \leq n-1$.  The initial vertex of $e^j$ is $v_j$, and its terminal vertex is $v_{j+1}$.  The input flag of $\Lin_{\uc}$ is $i$, and its output flag is $o$.\dqed
\end{example}

\begin{example}[Truncated linear graphs]\label{ex:truncated-linear-graph}
For each $\colorc$-profile $(c_1,\ldots,c_n)$ with $n \geq 1$, the \index{truncated linear graph}\emph{truncated linear graph} \[\lin_{(c_1,\ldots,c_n)}\] with profile $(\varnothing;c_n)$ has the same definition as the linear graph in Example \ref{ex:linear-graph} but without the input flag $i$.  We visualize the truncated linear graph $\lin_{(c_1,\ldots,c_n)}$ as follows.
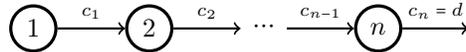
\begin{center}\begin{tikzpicture}
\matrix[row sep=.5cm, column sep=.9cm]{\node [plain] (1) {$1$}; & \node [plain] (2) {$2$}; & \node [empty] (3) {$\cdots$}; & \node [plain] (n) {$n$};\\};
\draw [arrow] (1) to node{\scriptsize{$c_1$}} (2);
\draw [arrow] (2) to node{\scriptsize{$c_2$}} (3);
\draw [arrow] (3) to node{\scriptsize{$c_{n-1}$}} (n);
\draw [outputleg] (n) to node{\scriptsize{$c_n=d$}} +(1.2cm,0);
\end{tikzpicture}\end{center}
Note that it does not have any inputs.\dqed
\end{example}

\begin{example}[Corollas]\label{ex:cd-corolla}
For each pair $\bigl(\uc=(c_1,\ldots,c_m); d\bigr) \in \Profcc$, there is a $\colorc$-colored tree \[\Cor_{(\uc;d)},\] called the \index{corolla}\emph{$(\uc;d)$-corolla}\label{notation:cdcorolla}, with profile $(\uc;d)$.  It is the ordinary $\colorc$-colored tree defined as follows.
\begin{itemize}
\item $\Flag\bigl(\Cor_{(\uc;d)}\bigr) = \{i_1,\ldots,i_{m},o\}$, all of which are ordinary legs at a unique vertex $v$.
\item $\kappa(i_p)=c_p$ for $1 \leq p \leq m$ and $\kappa(o)=d$.
\item $\delta(i_p)=1$ for $1 \leq p \leq m$ and $\delta(o)=-1$.
\item $\zeta_z(p)=i_p$ for $z \in \bigl\{v, \Cor_{(\uc;d)}\bigr\}$ and $1 \leq p \leq m$.
\end{itemize}
The corolla $\Cor_{(\uc;d)}$ is a linear graph if and only if $|\uc|=1$.  We depict the $(\uc;d)$-corolla as
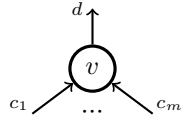
\begin{center}\begin{tikzpicture}
\node[plain] (v) {$v$}; \node[below=.1cm of v] () {$\cdots$};
\draw[outputleg] (v) to node[at end]{\scriptsize{$d$}} +(0cm,.8cm);
\draw[inputleg] (v) to node[swap, at end, outer sep=-2pt]{\scriptsize{$c_1$}} +(-.8cm,-.6cm);
\draw[inputleg] (v) to node[at end, outer sep=-2pt]{\scriptsize{$c_m$}} +(.8cm,-.6cm);
\end{tikzpicture}\end{center}
in which the input legs are drawn from left to right according to their ordering.  The $j$th input leg is $i_j$, and the output is $o$.\dqed
\end{example}

\begin{example}[Permuted corollas]\label{ex:cd-permuted-corolla}
With the same setting as in Example \ref{ex:cd-corolla}, suppose given a permutation $\tau \in \Sigma_m$.  There is a $\colorc$-colored tree \[\Cor_{(\uc;d)}\tau,\] called the \index{permuted corolla}\index{corolla!permuted}\emph{permuted corolla}\label{notation:permuted-corolla}, with profile $(\uc\tau;d)$.  It is defined just like the corolla $\Cor_{(\uc;d)}$, except for the ordering of the whole graph:
\[\zeta_{\Cor_{(\uc;d)}\tau}(p) = i_{\tau(p)} \forspace 1 \leq p \leq m.\]
Note that $\profofv = (\uc;d)$ for its unique vertex $v$, while $\Prof(\Cor_{(\uc;d)}\tau)=(\uc\tau;d)$.  For example, if $\uc = (c_1,c_2)$ and $\tau = (1~2) \in \Sigma_2$, then we may visualize the permuted corolla $\Cor_{(\uc;d)}\tau$ as:
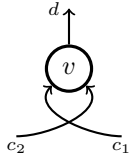
\begin{center}\begin{tikzpicture}
\node[plain] (v) {$v$};
\draw[inputleg, out=-45, in=0] (v) to node[outer sep=-1pt, at end]{\scriptsize{$c_2$}} +(-.7cm,-.9cm);
\draw[inputleg, out=225, in=180] (v) to node[outer sep=-1pt, swap, at end]{\scriptsize{$c_1$}} +(.7cm,-.9cm);
\draw[outputleg] (v) to node[at end]{\scriptsize{$d$}} +(0cm,.8cm);\end{tikzpicture}\end{center}
As we will see below, permuted corollas provide operads with their equivariant structure.\dqed \end{example}

\begin{example}[$2$-level trees]\label{ex:twolevel-tree}
Suppose $d \in \colorc$, $\uc=(c_1,\ldots,c_m) \in \Profc$ with $m\geq 1$, $\ub_j=\bigl(b_{j,1},\ldots,b_{j,k_j}\bigr) \in \Profc$ for $1 \leq j \leq m$ with $|\ub_j|=k_j$, and $\ub=(\ub_1,\ldots,\ub_m)$.  There is a $\colorc$-colored tree \[T\left(\{\ub_j\};\uc;d\right)\] with profile $(\ub;d)$, called a \index{tree!two-level}\emph{$2$-level tree}, that can be pictorially represented as:
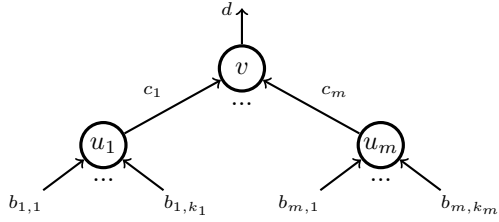
\[\begin{tikzpicture}
\matrix[row sep=.1cm, column sep=1.2cm]{
& \node [plain, label=below:$...$] (v) {$v$}; &\\
\node [plain, label=below:$...$] (u1) {$u_1$}; &&
\node [plain, label=below:$...$] (um) {$u_m$};\\};
\draw [outputleg] (v) to node[at end]{\scriptsize{$d$}} +(0,.8cm);
\draw [arrow] (u1) to node{\scriptsize{$c_1$}} (v);
\draw [arrow] (um) to node[swap]{\scriptsize{$c_m$}} (v);
\draw [inputleg] (u1) to node[below left=.2cm]{\scriptsize{$b_{1,1}$}} +(-.8cm,-.6cm);
\draw [inputleg] (u1) to node[below right=.2cm]{\scriptsize{$b_{1,k_1}$}} +(.8cm,-.6cm);
\draw [inputleg] (um) to node[below left=.2cm]{\scriptsize{$b_{m,1}$}} +(-.8cm,-.6cm);
\draw [inputleg] (um) to node[below right=.2cm]{\scriptsize{$b_{m,k_m}$}} +(.8cm,-.6cm);
\end{tikzpicture}\]
Formally $T=T\left(\{\ub_j\};\uc;d\right)$ is defined as follows.
\begin{itemize}
\item $\Flag(T) = \Bigl\{o,\{f^j_{\pm}\}_{1\leq j \leq m}, \{g_{j,i}\}_{1\leq j \leq m}^{1 \leq i \leq k_j}\Bigr\}$, all of which are ordinary flags.
\item $\iota(f^j_{\pm})=f^j_{\mp}$, and $\iota$ fixes all other flags.
\item There are $m+1$ vertices:
\[v=\bigl\{o,f^1_+,\ldots,f^m_+\bigr\} \andspace 
u_j=\bigl\{f^j_-,g_{j,1},\ldots,g_{j,k_j}\bigr\} \forspace 1 \leq j \leq m\]
\item $\kappa(o)=d$, $\kappa\left(f^j_{\pm} \right)=c_j$, and $\kappa\left(g_{j,i}\right) = b_{j,i}$.
\item $\delta(o) = -1 = \delta\left(f^j_{-}\right)$, and $\delta\left(g_{j,i}\right) = 1 = \delta\left(f^j_+\right)$.
\item $\zeta_v\left(f^j_+\right) = j$, $\zeta_{u_j}\left(g_{j,i}\right) = i$, and $\zeta_T\left(g_{j,i}\right) = i + k_1 + \cdots + k_{j-1}$.
\end{itemize}
There are $m$ internal edges $f^j=\{f^j_{\pm}\}$.  The unique output is the flag $o$, and the flags $g_{j,i}$ are the inputs of $T$.  As we will see below, these $\colorc$-colored trees correspond to the operadic composition $\gamma$.\dqed
\end{example}

In what follows, we will often draw a colored tree without writing down its detailed definition.  The reader can fill in the details using the examples above as a guide.

\section{Tree Substitution}\label{sec:tree-sub}

The main reason for considering trees is the operation called tree substitution.  Suppose $\colorc$ is a non-empty set.  All the trees below are $\colorc$-colored trees.

\begin{definition}[Tree Substitution at a Vertex]\label{def:tree-sub-vertex}
Suppose $T$ is a tree, and $v$ is a vertex in $T$.  Suppose $H$ is a tree such that $\profofh = \profofv$.  Define the tree $T(H)$ with $\Prof\bigl(T(H)\bigr) = \profoft$, called the \index{tree substitution!at a vertex}\emph{tree substitution at $v$}, as follows.
\begin{enumerate}
\item If $H$ is not an exceptional edge, then we identify (i) the ordered sets $\inp(v)$ and $\inp(H)$ and (ii) the flags $\out(v)$ and $\out(H)$.  We define 
\[\begin{split}\Flag\bigl(T(H)\bigr) &= \bigl(\Flag(T) \setminus v\bigr) \coprod \Flag(H),\\
\Vt\bigl(T(H)\bigr) &= \bigl[\Vt(T) \setminus \{v\}\bigr] \coprod \Vt(H),\\
\iota_{T(H)}(f) &= \begin{cases} \iota_H(f) & \text{ if $f \in \Flag(H) \setminus \Leg(H)$},\\ \iota_T(f) & \text{ otherwise},\end{cases}
\end{split}\] with an empty exceptional cell.  Its coloring, direction, and listing are induced from those of $T$ and $H$.
\item If $H$ is the exceptional edge $\uparrow_c$ and if $T$ is the corollary $\Cor_{(c;c)}$ with $v$ its unique vertex, then $T(H) =\, \uparrow_c$.
\item If $H$ is the exceptional edge $\uparrow_c$ and if $T$ is not the corollary $\Cor_{(c;c)}$, then we define \[\Flag\bigl(T(H)\bigr)=\Flag(T) \setminus v \andspace 
\Vt\bigl(T(H)\bigr)=\Vt(T) \setminus \{v\}\] with an empty exceptional cell.  Furthermore: 
\begin{enumerate}
\item If $\out(v)=\out(T)$, then we define \[\iota_{T(H)}(f) =\begin{cases} f & \text{ if $f=\iota_T(\inp(v))$},\\ \iota_T(f) & \text{ otherwise}.\end{cases}\]
\item If $\inp(v)$ is an input of $T$, then we define \[\iota_{T(H)}(f)=\begin{cases} f & \text{ if $f=\iota_T(\out(v))$},\\ \iota_T(f) & \text{ otherwise}. \end{cases}\]
\item If $\out(v) \not= \out(T)$ and if $\inp(v)$ is not an input of $T$, then we define \[\iota_{T(H)}(f)=\begin{cases}\iota_T(\inp(v)) & \text{ if $f=\iota_T(\out(v))$},\\ \iota_T(\out(v)) & \text{ if $f=\iota_T(\inp(v))$},\\ \iota_T(f) & \text{ otherwise}.\end{cases}\]
\end{enumerate}
Its coloring, direction, and listing are induced from those of $T$.
\end{enumerate}\end{definition}

The following properties of tree substitution are proved by directly checking the definitions, so we omit the proofs here.  The reader may consult \cite{bluemonster} Chapter 5 for proofs.  

\begin{proposition}\label{prop:tree-sub-properties}
Consider the setting of Definition \ref{def:tree-sub-vertex}.
\begin{enumerate}
\item If there are isomorphisms $T \cong T'$ and $H \cong H'$, then there is an isomorphism $T(H) \cong T'(H')$.
\item $T(H)$ is a linear graph if and only if $T$ and $H$ are both linear graphs.
\item Internal edges in $H$ yield internal edges in $T(H)$.
\item Suppose $u$ is a vertex in $H$, and $K$ is a tree such that $\profofk = \profofu$.  Then there is a canonical isomorphism
\begin{equation}\label{treesubass1}
\bigl[T(H)\bigr](K) \cong T\bigl(H(K)\bigr).
\end{equation}
\item Suppose $w\not= v$ is another vertex in $T$, and $G$ is a tree such that $\profofg = \profofw$.  Then there is a canonical isomorphism 
\begin{equation}\label{treesubass2}
\bigl[T(H)\bigr](G) \cong \bigl[T(G)\bigr](H).
\end{equation}
\item There are canonical isomorphisms
\begin{equation}\label{treesubunity}
T(\Cor_v) \cong T \andspace \Cor_T(H) \cong H,
\end{equation}
where $\Cor_z$ is the $\bigl(\inp(z);\out(z)\bigr)$-corolla for $z \in \{v,T\}$.
\item If $T$ is a permuted corolla $\Cor_{(\uc;d)}\tau$ as in Example \ref{ex:cd-permuted-corolla} for some permutation $\tau \in \Sigma_{|\inp(T)|}$, then $\bigl(\Cor_{(\uc;d)}\tau\bigr)(H)$ is canonically isomorphic to $H$ except that its input profile is $\inp(H)\tau$.
\item If $H$ is a permuted corolla $\Cor_{v'}\tau$ for some permutation $\tau$, then $T\bigl(\Cor_{v'}\tau\bigr)$ is canonically isomorphic to $T$ except that the input profile at $v'$ is \[\Prof(v') = \bigl(\inp(v)\inv{\tau}; \out(v)\bigr).\]
\end{enumerate}\end{proposition}

\begin{definition}[Tree Substitution]\label{def:tree-substitution}
Suppose $T$ is a tree, and $H_v$ is a tree with $\profofhv=\profofv$ for each vertex $v$ in $T$.  Suppose $\{v_1,\ldots,v_n\}$ is an ordering of the set $\Vt(T)$.  Define the tree $T(H_v)_{v\in \Vt(T)}$ with the same profile as $T$, called the \index{tree substitution}\emph{tree substitution}, by \[T(H_v)_{v\in \Vt(T)} = \Bigl(\cdots\bigl(T(H_{v_1})\bigr)(H_{v_2}) \cdots\Bigr)(H_{v_n}).\]  
\end{definition}

\begin{notation}
To simplify the notation, we will often write $v\in T$ to mean $v \in \Vt(T)$.  Furthermore, we will sometimes abbreviate $T(H_v)_{v\in\Vt(T)}$ to $T(H_v)$.  We will say that $H_v$ is \emph{substituted into $v$}.
\end{notation}

The following properties of tree substitution are consequences of Proposition \ref{prop:tree-sub-properties}.

\begin{corollary}\label{cor:treesub-assunity}
Consider the setting of Definition \ref{def:tree-sub-vertex}.
\begin{enumerate} \item The isomorphism class of the tree substitution $T(H_v)_{v\in T}$ is independent of the choices of (i) an ordering of $\Vt(T)$, (ii) a representative in the isomorphism class of $T$, and (iii) a representative in the isomorphism class of each $H_v$.
\item There is a decomposition 
\begin{equation}\label{treesub-vt}
\Vt\bigl(T(H_v)_{v\in T}\bigr)=\coprodover{v\in T} \Vt(H_v).
\end{equation}
\item Up to isomorphisms, tree substitution is associative in the sense that, if $I_u$ is a tree with $\Prof(I_u) = \profofu$ for each $u \in \Vt(H_v)$ and each $v \in \Vt(T)$, then there is an isomorphism
\[\Bigl[T(H_v)_{v\in T}\Bigr](I_u)_{u\in T(H_v)_{v\in T}} \cong T\Bigl(H_v(I_u)_{u\in H_v}\Bigr)_{v\in T}.\]
\item Up to isomorphisms, tree substitution is unital in the sense that there are isomorphisms
\[T\bigl(\Cor_v\bigr)_{v\in T} \cong T \andspace \Cor_T(T) \cong T.\]
\end{enumerate}
\end{corollary}

\begin{proof}
The first assertion follows from the first assertion in Proposition \ref{prop:tree-sub-properties} and \eqref{treesubass2}.  The decomposition \ref{treesub-vt} on vertex set follows from Definition \ref{def:tree-sub-vertex}.  The associativity isomorphism follows from \eqref{treesubass1} and \eqref{treesubass2}.  The unity isomorphisms follow from \eqref{treesubunity}.
\end{proof}

\begin{convention}\label{conv:tree-isoclass}
To simplify the presentation, in what follows we will minimize the distinction between a tree and its isomorphism class and will use the same symbol to denote both.
\end{convention}

\begin{definition}[Substitution Category]\label{def:treesub-category}
Define the \index{substitution category}\emph{substitution category} $\uTreec$ as the small category with:
\begin{itemize}\item $\colorc$-colored trees as objects;
\item $\uTreec\bigl(K,T\bigr)$ the set of finite sets $(H_v)_{v\in T}$ such that $K=T(H_v)_{v\in T}$;
\item $\bigl(\Cor_v\bigr)_{v\in T}$ as the identity morphism of $T$;
\item composition given by tree substitution.
\end{itemize}  
For a pair $(\uc;d) \in \Profcc$, denote by $\uTreec\duc$ or $\uTreec(\uc;d)$, called the \emph{substitution category} with profile $(\uc;d)$, the full subcategory of $\uTreec$ consisting of trees with profile $(\uc;d)$.  For a vertex $v$ in a $\colorc$-colored tree, we will also write $\uTreec(v)$ for $\uTreec\inoutv$.  We similarly define the substitution categories $\uLinearc$ and $\uLinearc\dc$ \label{notation:ulinear}using linear graphs instead of trees.
\end{definition}

\begin{remark}\label{rk:subcat-composition}
Suppose \[(H_v)_{v\in T} : K \to T \andspace (G_u)_{u\in K} : E \to K\] are morphisms in $\uTreec$.  Then \[E=K(G_u)_{u\in K} = \bigl(T(H_v)_{v\in T}\bigr)(G_u)_{u\in H_v,\, v \in T} = T\Bigl(H_v(G_u)_{u\in H_v}\Bigr)_{v \in T}.\]  This defines the composition
\[\nicexy@C+.4cm{E \ar[r]^-{(G_u)_{u\in K}} \ar`u[rr]`[rr]^-{\bigl(H_v(G_u)_{u\in H_v}\bigr)_{v\in T}}[rr] & K \ar[r]^-{(H_v)_{v\in T}} & T}\] of $(G_u)_{u\in K}$ and $(H_v)_{v\in T}$ in the substitution category $\uTreec$.\dqed\end{remark}

\begin{example}\label{ex:treesub}
Consider the morphism \[(H_u,H_v,H_w) : K \to T\] indicated by the following picture.
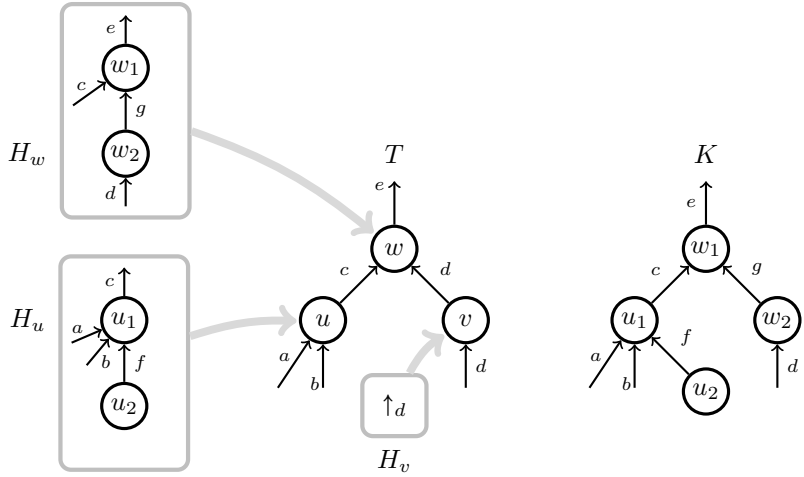
\begin{center}\begin{tikzpicture}
\node[plain] (w) {$w$}; \node[above=.7cm of w] (T) {$T$};
\node[below right=.7cm of w, plain] (v) {$v$}; \node[below left=.7cm of w, plain] (u) {$u$};
\draw[outputleg] (w) to node[pos=.9]{\scriptsize{$e$}} +(0,.9cm);
\draw[arrow] (u) to node{\scriptsize{$c$}} (w);
\draw[arrow] (v) to node[swap]{\scriptsize{$d$}} (w);
\draw[inputleg] (u) to node[swap, outer sep=-2pt]{\scriptsize{$a$}} +(-.6cm,-.9cm);
\draw[inputleg] (u) to node[pos=.9, outer sep=-2pt, swap]{\scriptsize{$b$}} +(0cm,-.9cm);
\draw[inputleg] (v) to node{\scriptsize{$d$}} +(0cm,-.9cm);
\node[left=3cm of T,plain] (w2) {$w_2$}; \node[above=.5cm of w2,plain] (w1) {$w_1$};
\node[left=.6cm of w2] (Hw) {$H_w$};
\draw[outputleg] (w1) to node{\scriptsize{$e$}} +(0,.7cm);
\draw[inputleg] (w1) to node[swap, outer sep=-2pt]{\scriptsize{$c$}} +(-.7cm,-.5cm);
\draw[arrow] (w2) to node[swap]{\scriptsize{$g$}} (w1);
\draw[inputleg] (w2) to node[swap]{\scriptsize{$d$}} +(0,-.7cm);
\node[draw=lightgray,inner sep=15pt,ultra thick,rounded corners,
fit=(w1) (w2)] (Hwbox) {};
\draw [gray!30, ->, line width=3pt, shorten >=0.03cm, bend left=10] (Hwbox) to (w);
\node[left=2cm of u,plain] (u1) {$u_1$}; \node[below=.5cm of u1,plain] (u2) {$u_2$};
\node[left=.6cm of u1] {$H_u$};
\draw[outputleg] (u1) to node{\scriptsize{$c$}} +(0,.7cm);
\draw[inputleg] (u1) to node[swap, outer sep=-2pt]{\scriptsize{$a$}} +(-.7cm,-.3cm);
\draw[inputleg] (u1) to node[outer sep=-2pt]{\scriptsize{$b$}} +(-.5cm,-.6cm);
\draw[arrow] (u2) to node[swap]{\scriptsize{$f$}} (u1);
\node[draw=lightgray,inner sep=15pt,ultra thick,rounded corners,
fit=(u1) (u2)] (Hubox) {};
\draw [gray!30, ->, line width=3pt, shorten >=0.03cm, bend left=10] (Hubox) to (u);
\node[below=1.5cm of w] (Hv) {$\uparrow_d$}; \node[below=.2cm of Hv] () {$H_v$}; 
\node[draw=lightgray,inner sep=4pt,ultra thick,rounded corners,
fit=(Hv)] (Hvbox) {};
\draw [gray!30, ->, line width=3pt, shorten >=0.03cm, bend left=15] (Hvbox) to (v);
\node[right=3.5cm of w,plain] (w1') {$w_1$}; \node[above=.7cm of w1'] {$K$};
\node[below right=.7cm of w1', plain] (w2') {$w_2$}; 
\node[below left=.7cm of w1', plain] (u1') {$u_1$};
\node[below right=.7cm of u1',plain] (u2') {$u_2$};
\draw[outputleg] (w1') to node{\scriptsize{$e$}} +(0,.9cm);
\draw[arrow] (w2') to node[swap]{\scriptsize{$g$}} (w1');
\draw[arrow] (u1') to node{\scriptsize{$c$}} (w1');
\draw[arrow] (u2') to node[swap]{\scriptsize{$f$}} (u1');
\draw[inputleg] (u1') to node[swap, outer sep=-2pt]{\scriptsize{$a$}} +(-.6cm,-.9cm);
\draw[inputleg] (u1') to node[pos=.9, outer sep=-2pt, swap]{\scriptsize{$b$}} +(0cm,-.9cm);\draw[inputleg] (w2') to node{\scriptsize{$d$}} +(0,-.9cm);
\end{tikzpicture}
\end{center}
For simplicity, in each tree, all the orderings at the vertices and for the whole tree are from left to right as displayed.  Each gray arrow indicates substituting the tree inside the originating gray box into the corresponding vertex in $T$.  Observe that 
\[\begin{split}\profofu&=\sbinom{c}{a,b}=\Prof(H_u),\quad \profofv=\dd=\profofhv,\\
\profofw&=\sbinom{e}{c,d}=\Prof(H_w),\andspace \profoft=\sbinom{e}{a,b,d}=\profofk.\end{split}\]
Internal edges in the $H$'s become internal edges in the tree substitution $K$.  The $d$-colored internal edge in $T$ is no longer an internal edge in $K$ because $H_v$ is the $d$-colored exceptional edge.\dqed
\end{example}

\section{Grafting}\label{sec:grafting}

Grafting is a special kind of tree substitution, which can be pictorially interpreted as gluing the outputs of a finite family of trees with the inputs of another tree.  In the next definition, we will use the $2$-level tree in Example \ref{ex:twolevel-tree}.

\begin{definition}[Grafting of Trees]\label{def:grafting}
Suppose 
\begin{itemize}\item $\duc=\dconecm \in \Profcc$ with $m\geq 1$.
\item $\ub_j \in \Profc$ for $1 \leq j \leq m$ with $|\ub_j|=k_j$, and $\ub=(\ub_1,\ldots,\ub_m)$. 
\item $G$ is a $\colorc$-colored tree with profile $(\uc;d)$.
\item $H_j$ is a $\colorc$-colored tree with profile $(\ub_j;c_j)$ for each $1 \leq j \leq m$.  
\end{itemize}
The \index{grafting}\emph{grafting of $G$ with $H_1,\ldots,H_m$} is defined as the tree substitution\label{notation:grafting}
\[\graft\bigl(G;H_1,\ldots,H_m\bigr) = \bigl[T\left(\{\ub_j\};\uc;d\right)\bigr]\bigl(G,H_1,\ldots,H_m\bigr),\]
where $T\left(\{\ub_j\};\uc;d\right)$ is the $2$-level tree with profile $(\ub;d)$ in Example \ref{ex:twolevel-tree}, with $G$ substituted into $v$ and $H_j$ substituted into $u_j$.
\end{definition}

Note that the profile of the grafting $\graft(G;\{H_j\})$ is $(\ub;d)$, which is the same as the profile of the $2$-level tree $T\left(\{\ub_j\};\uc;d\right)$.

\begin{example}\label{ex:graft-exedge}
With $G$ as in Definition \ref{def:grafting}, we have \[\graft\bigl(G;\uparrow_{c_1},\ldots,\uparrow_{c_m}\bigr) = G = \graft\bigl(\uparrow_d;G\bigr).\] That is, grafting with an exceptional edge has no effect.\dqed
\end{example}

\begin{example}\label{ex:grafting}
Suppose $T$ is the tree in Example \ref{ex:treesub} with profile $\sbinom{e}{a,b,d}$.  With the trees $H_1$, $H_2$, and $H_3$ as drawn below, the grafting \[G=\graft\bigl(T;H_1,H_2,H_3\bigr) = \Bigl[T\bigl(\{(f),(b),\varnothing\};(a,b,d);e\bigr)\Bigr]\bigl(T,H_1,H_2,H_3\bigr)\] 
is the tree on the right.
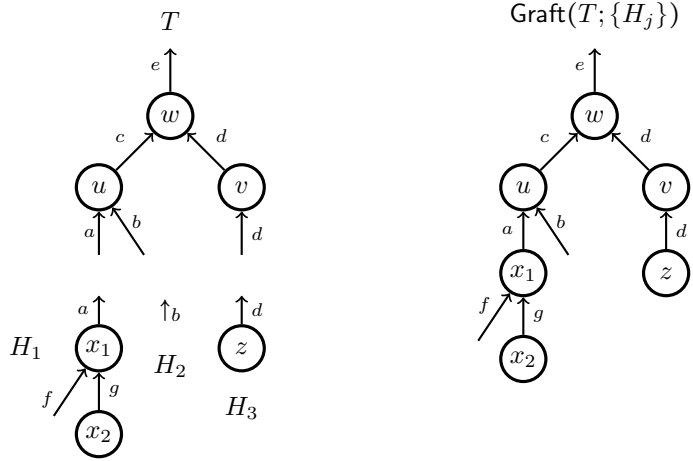
\begin{center}\begin{tikzpicture}
\node[plain] (w) {$w$}; \node[above=.7cm of w] (T) {$T$};
\node[below right=.7cm of w, plain] (v) {$v$}; \node[below left=.7cm of w, plain] (u) {$u$};
\draw[outputleg] (w) to node[pos=.6]{\scriptsize{$e$}} +(0,.9cm);
\draw[arrow] (u) to node{\scriptsize{$c$}} (w);
\draw[arrow] (v) to node[swap]{\scriptsize{$d$}} (w);
\draw[inputleg] (u) to node[swap, outer sep=-2pt]{\scriptsize{$a$}} +(0cm,-.9cm);
\draw[inputleg] (u) to node[outer sep=-2pt]{\scriptsize{$b$}} +(.6cm,-.9cm);
\draw[inputleg] (v) to node{\scriptsize{$d$}} +(0cm,-.9cm);
\node[below=1.5cm of u,plain] (x1) {$x_1$}; \node[below=.5cm of x1,plain] (x2) {$x_2$};
\node[left=.3cm of x1] (H1) {$H_1$};
\draw[outputleg] (x1) to node{\scriptsize{$a$}} +(0,.7cm);
\draw[inputleg] (x1) to node[swap,pos=.9,outer sep=-2pt]{\scriptsize{$f$}} +(-.6cm,-.9cm);
\draw[arrow] (x2) to node[swap]{\scriptsize{$g$}} (x1);
\node[below=2cm of w] (H2) {$\uparrow_b$}; \node[below=.2cm of H2] () {$H_2$}; 
\node[below=1.5cm of v,plain] (z) {$z$}; \node[below=.2cm of z] {$H_3$};
\draw[outputleg] (z) to node[swap]{\scriptsize{$d$}} +(0,.7cm);
\node[right=5cm of w,plain] (w') {$w$}; \node[above=.7cm of w'] {$\graft(T;\{H_j\})$};
\node[below right=.7cm of w', plain] (v') {$v$}; \node[below left=.7cm of w',plain] (u') {$u$};
\node[below=.5cm of u',plain] (x1') {$x_1$}; \node[below=.5cm of x1',plain] (x2') {$x_2$}; \node[below=.5cm of v',plain] (z') {$z$};
\draw[outputleg] (w') to node[pos=.6]{\scriptsize{$e$}} +(0,.9cm);
\draw[arrow] (u') to node{\scriptsize{$c$}} (w');
\draw[arrow] (v') to node[swap]{\scriptsize{$d$}} (w');
\draw[arrow] (x2') to node[swap]{\scriptsize{$g$}} (x1');
\draw[arrow] (x1') to node{\scriptsize{$a$}} (u');
\draw[arrow] (z') to node[swap]{\scriptsize{$d$}} (v');
\draw[inputleg] (x1') to node[swap, outer sep=-2pt]{\scriptsize{$f$}} +(-.6cm,-.9cm);
\draw[inputleg] (u') to node[outer sep=-2pt]{\scriptsize{$b$}} +(.6cm,-.9cm);
\end{tikzpicture}\end{center}
Observe that grafting with an exceptional edge, as with $H_2 =~ \uparrow_b$ above, has no effect.  Internal edges in $T$ and in the $H$'s remain internal edges in the grafting $\graft(T;\{H_j\})$.  New internal edges are created by the grafting, unless it involves an exceptional edge.\dqed
\end{example}

Grafting allows us to construct bigger trees from smaller ones, as illustrated in Example \ref{ex:grafting}.  The following observation says that corollas are the building blocks of trees with respect to grafting.  With slightly different terminology, the following result is \cite{yau-operad} Theorem 5.7.3.

\begin{theorem}\label{thm:grafting-generate}
Suppose $T$ is a $\colorc$-colored tree \index{tree!generation of}with at least one vertex.  Up to a reordering of its inputs, $T$ is an iterated grafting of corollas and exceptional edges.
\end{theorem}

\begin{proof}
If $T$ has only one vertex, then it is a permuted corolla as in Example \ref{ex:cd-permuted-corolla}, which is a corolla with its inputs reordered.  Inductively,  suppose $T$ has $n>1$ vertices, and $v$ is the root vertex in $T$.  Suppose $u_1,\ldots,u_n$ are the vertices adjacent to $v$, with an internal edge $e^j=\{e^j_{\pm}\}$ adjacent to $u_j$ and $v$ for $1 \leq j \leq n$.  For each $j$, suppose $T_j$ is the tree consisting of the largest subset of flags in $T$ with root vertex $u_j$, output $e^j_-$, and the induced direction, $\colorc$-coloring, and listing.  Suppose the inputs of the root vertex $v$ are the flags $\{f_1,\ldots,f_r\}$, which contain $\{e^1_+,\ldots,e^n_+\}$.   Up to a reordering of its inputs, $T$ is the grafting
\[T = \graft\bigl(\Cor_v;H_1,\ldots,H_r\bigr)\] in which:
\begin{itemize}
\item $\Cor_v$ is the corolla with the same profile as $v$.
\item $H_i$ is:
\begin{itemize}\item the $\kappa(f_i)$-colored exceptional edge $\uparrow_{\kappa(f_i)}$ if $f_i \not\in\{e^1_+,\ldots,e^n_+\}$.
\item the tree $T_j$ if $f_i=e^j_+$.
\end{itemize}\end{itemize} 
We finish the proof by observing that the induction hypothesis applies to each $T_j$, since it has strictly fewer vertices than $T$.
\end{proof}

\begin{example}\label{ex2:grafting}
In the setting of Example \ref{ex:grafting}, we have \[T = \graft\bigl(\Cor_w; \Cor_u,\Cor_v\bigr) \andspace H_1=\graft\bigl(\Cor_{x_1}; \uparrow_f,\Cor_{x_2}\bigr).\] Moreover, $H_2=~\uparrow_b$ is an exceptional edge, and $H_3=\Cor_{(\varnothing;d)}$ is a corolla.\dqed
\end{example}

\chapter{Colored Operads}\label{ch:operads}

In this chapter, we define colored operads and their algebras.  A colored operad is a generalization of a category in which the domain of each morphism is a finite sequence of objects.  Colored operads provide an efficient way to encode operations with multiple inputs and one output.  This efficient bookkeeping aspect of operad theory is especially important when we discuss homotopy algebraic quantum field theories and homotopy prefactorization algebras, in which the desired structures are too complicated to encode without colored operads.

Historically, colored non-symmetric operads in $\Set$ were defined by \index{Lambek@Lambek, J.}Lambek \cite{lambek}, who called them \index{multicategory}\emph{multicategories}.  May\index{May@May, J.P.} \cite{may} defined a one-colored \index{operad!topological}\index{topological operad}topological operad and coined the term \emph{operad}.  In \cite{kelly} \index{Kelly@Kelly, G.M.}Kelly gave a more categorical construction of one-colored operads in symmetric monoidal categories in terms of coends and \index{Day convolution}Day convolutions.  For an introduction to colored operads, the reader may also consult \cite{bsw,white-yau,yau-operad}.

In Section \ref{sec:operad-monoid} we define a colored operad as a monoid in certain monoidal category.  In Section \ref{sec:operad-generating} to Section \ref{sec:operad-tree}, we provide three equivalent and more explicit descriptions of a colored operad.  In particular, the definition in Section \ref{sec:operad-generating} in terms of generating structure morphisms and axioms and the definition in Section \ref{sec:operad-tree} in terms of trees will be used throughout the rest of this book.  In Section\ref{sec:algebra-operad}, we define algebras over a colored operad and discuss some key examples.  

Throughout this chapter, $(\M,\otimes,\tensorunit)$ is a cocomplete symmetric monoidal closed category with an initial object $\varnothing$ and an internal hom $\Homm$.

\section{Operads as Monoids}\label{sec:operad-monoid}

As we will see later, each operad yields a monad (Definition \ref{def:monad}).  We defined a monad as a monoid in the strict monoidal category of functors from a category to itself.  In this section, we define operads analogously as monoids in a suitable monoidal category.  We first recall from \cite{bluemonster} some notations regarding colors and profiles.  The symmetric group on $n$ letters is denoted by\label{notation:sigman} $\Sigma_n$, whose unit is $\id_n$.  Recall from Definition \ref{def:profofc} that a $\colorc$-profile is a finite sequence of elements in $\colorc$, and $\Profc$ denotes the set of all $\colorc$-profiles.

\begin{definition}\label{def:profiles}
Fix a non-empty set $\colorc$, whose elements are called \emph{colors}.
\begin{enumerate}
\item If $\ua = (a_1,\ldots,a_m)$ and $\ub$ are $\colorc$-profiles, then a \index{left permutation}\emph{left permutation} $\sigma : \ua \to \ub$ is a permutation $\sigma \in \Sigma_{|\ua|}$ such that\label{notation:left-permutation}
\[\sigma\ua = (a_{\sigma^{-1}(1)}, \ldots , a_{\sigma^{-1}(m)}) = \ub\]
\item The \index{groupoid}\emph{groupoid of $\colorc$-profiles}, with left permutations as the isomorphisms, is denoted by $\Sigmac$.\label{notation:sigmac}  The \index{opposite groupoid}opposite groupoid $\Sigmacop$ is regarded as the groupoid of $\colorc$-profiles with \label{notation:right-permutation}\emph{right permutations}
\[\ua\sigma = (a_{\sigma(1)}, \ldots , a_{\sigma(m)})\]
as isomorphisms.
\item The objects of the diagram category\label{notation:symseqcm} \[\symseqcm = \M^{\Sigmacopc}\] are called \index{symmetric sequence}\emph{$\colorc$-colored symmetric sequences} in $\M$.  For an object $X$ in $\M^{\Profcc}$ or $\symseqcm$, we write \[X(\uc;d) = X\duc \in \M\] for the value of $X$ at $(\uc;d) \in \Profcc$ and call it an \emph{$m$-ary entry of $X$} if $|\uc|=m$.  We call $\uc$ the \index{input profile}\emph{input profile}, $c_i$ the \emph{$i$th input color}, and $d$ the \index{output color}\emph{output color}.
\item An object in the product category $\prod_{\colorc} \M = \M^{\colorc}$\label{notation:mtoc} is called a \index{colored object}\emph{$\colorc$-colored object in $\M$}, and similarly for a morphism of $\colorc$-colored objects.  A $\colorc$-colored object $X$ is also written as $\{X_c\}$ with $X_c \in \M$ for each color $c \in \colorc$.
\item A $\colorc$-colored object $\{X_c\}_{c\in\colorc}$ is also regarded as a $\colorc$-colored symmetric sequence concentrated in $0$-ary entries:
\begin{equation}\label{colored-object-sm}
X\duc= \begin{cases}X_d & \text{ if $\uc=\varnothing$},\\ \varnothing & \text{ if $\uc\not=\varnothing$}.\end{cases}
\end{equation}
\end{enumerate}\end{definition}

\begin{definition}\label{def:colored-circle-product}
Suppose $X,Y  \in \symseqcm$.
\begin{enumerate}
\item For each $\uc = (c_1,\ldots,c_m) \in \Profc$, define the object $Y^{\uc} \in \M^{\Sigmac}$ entrywise as the coend
\begin{equation}\label{ytensorc}
Y^{\uc}(\ub) = \int^{\{\ua_j\}\in\prod_{j=1}^m \Sigmacop} \Sigmacop\bigl(\ua_1,\ldots,\ua_m;\ub\bigr) \cdot \left[\bigotimes_{j=1}^m Y\cjuaj\right] \in \M
\end{equation}
for $\ub \in \Profc$, in which $(\ua_1,\ldots,\ua_m)$ is the concatenation.	 Note that $Y^{\uc}$ is natural in $\uc \in \Profc$ via left permutations of the tensor factors in $\bigotimes_{j=1}^m Y\cjuaj$.
\item The \index{circle product}\emph{$\colorc$-colored circle product} \[X \circ Y \in \symseqcm\] is defined entrywise as the coend
\begin{equation}\label{circle-product}
(X \circ Y)\dub = \int^{\uc \in \Sigmac} X\duc \otimes Y^{\uc}(\ub) 
\end{equation}
for $(\ub;d) \in \Profcc$.
\item Define the object $\I \in \symseqcm$ by
\begin{equation}\label{unit-operad}
\I\duc = \begin{cases} \tensorunit & \text{ if $\uc=d$},\\
\emptyset & \text{ otherwise}\end{cases}
\end{equation}
for $(\uc;d) \in \Profcc$.
\end{enumerate}
\end{definition}

The one-colored case of the following result is in \cite{kelly}.  The general colored case is proved in \cite{white-yau}, in which the colored circle product was written in terms of a left Kan extension.  When the colored circle product is written as a coend as in \eqref{circle-product}, the proof below can be found in \cite{bsw}.

\begin{proposition}\label{circle-product-monoidal}
$\bigl(\symseqcm, \circ, \I\bigr)$ is a \index{monoidal category!of symmetric sequences}\index{symmetric sequence!monoidal category of}monoidal category.
\end{proposition}

\begin{proof}
Suppose $X,Y,Z \in \symseqcm$.  We will exhibit the associativity isomorphism.  First note that, for $\uc=(c_1,\ldots,c_m), \ub \in \Sigmacop$, there exist canonical isomorphisms:
\begin{equation}\label{ycircz}
\begin{split}(Y \circ Z)^{\uc}(\ub) 
&= \int^{\ua_1,\ldots,\ua_m} \Sigmacop\bigl(\ua_1,\ldots,\ua_m;\ub\bigr) \cdot \left[\bigotimes_{j=1}^m (Y\circ Z)\cjuaj\right]\\
&\cong \int^{\ua_1,\ldots,\ua_m} \int^{\ud_1,\ldots,\ud_m} \Sigmacop\bigl(\ua;\ub\bigr) \cdot \bigotimes_{j=1}^m \Bigl[Y\cjudj \otimes Z^{\ud_j}(\ua_j)\Bigr]\\
&\cong \int^{\ud_1,\ldots,\ud_m} \left[\bigotimes_{j=1}^m Y\cjudj\right] \otimes Z^{\ud}(\ub)\\
&\cong \int^{\ud_1,\ldots,\ud_m} \int^{\ue} \Sigmacop(\ud;\ue) \cdot \left[\bigotimes_{j=1}^m Y\cjudj\right] \otimes Z^{\ue}(\ub)\\
&\cong \int^{\ue} Y^{\uc}(\ue) \otimes Z^{\ue}(\ub).\end{split}
\end{equation}
In the above calculation, we wrote $\ua=(\ua_1,\ldots,\ua_m)$ and $\ud=(\ud_1,\ldots,\ud_m)$ for the concatenations.  Now we have the equalities and canonical isomorphism
\[\begin{split}
\bigl((X \circ Y) \circ Z\bigr)\dua 
&= \int^{\uc} (X \circ Y)\duc \otimes Z^{\uc}(\ua)\\
&= \int^{\uc} \int^{\ub} X\dub \otimes Y^{\ub}(\uc) \otimes Z^{\uc}(\ua)\\
&\cong \int^{\ub} X\dub \otimes (Y \circ Z)^{\ub}(\ua)\\
&= \bigl(X \circ (Y \circ Z)\bigr)\dua\end{split}\]
for $(\ua;d) \in \Profcc$, in which the isomorphism uses \eqref{ycircz}.  The rest of the axioms of a monoidal category are straightforward to check.
\end{proof}

Recall from Definition \ref{def:monoid} that each monoidal category has a category of monoids.

\begin{definition}\label{def:operad}
Define the category
\[\Operadcm = \Mon\bigl(\symseqcm\bigr)\]
of \index{colored operad}\emph{$\colorc$-colored operads in $\M$} as the category of monoids in $\bigl(\symseqcm,\circ,\I\bigr)$.  If $\colorc$ has $n<\infty$ elements, we also refer to objects in $\Operadcm$ as \emph{$n$-colored operads}.
\end{definition}

Using Proposition \ref{prop:monoid} we may express a $\colorc$-colored operad as follows.

\begin{corollary}\label{operad-monoid}
A $\colorc$-colored operad in $\M$ is exactly a triple $(\O,\mu,\varepsilon)$ consisting of
\begin{itemize}\item an object $\O$, 
\item a multiplication morphism $\mu : \O \circ \O \to \O$, and 
\item a unit $\varepsilon : \I \to \O$, 
\end{itemize}
all in $\symseqcm$, such that the associativity and unity diagrams
\begin{equation}\label{operad-monoid-axioms}
\nicexy{\O \circ \O \circ \O \ar[d]_-{(\mu,\Id_{\O})} \ar[r]^-{(\Id_{\O},\mu)} & \O \circ \O \ar[d]^-{\mu}\\ \O \circ \O \ar[r]^-{\mu} & \O}\qquad
\nicexy{\I \circ \O \ar[dr]_-{\cong} \ar[r]^-{(\varepsilon, \Id_{\O})} & \O \circ \O \ar[d]^-{\mu} & \O \circ \I \ar[l]_-{(\Id_{\O},\varepsilon)} \ar[dl]^-{\cong}\\ & \O &}
\end{equation}
are commutative.  A morphism of $\colorc$-colored operads is a morphism of the underlying $\colorc$-colored symmetric sequences that is compatible with the multiplications and the units.
\end{corollary}

\begin{notation}
If $\O$ is a $1$-colored operad with color set $\{*\}$, then we write\label{notation:oofn} \[\O(n) = \O\starnstar\] for $\bigl(*,\ldots,*; *\bigr) \in \Sigmaop_{\{*\}} \times \{*\}$ in which the input profile has length $n$.  
\end{notation}

\section{Operads in Terms of Generating Operations}\label{sec:operad-generating}

In Definition \ref{def:operad} above we defined a colored operad as a monoid with respect to the colored circle product, which is defined in terms of coends (Definition \ref{def:colored-circle-product}).  We can unpack the colored circle product to express a colored operad in terms of a few generating operations from \cite{yau-operad} (Section 11.2) and \cite{bluemonster} (Definition 11.14).  In the one-colored topological case, the definition below is due to May \cite{may}.

\begin{definition}\label{def:operad-generating}
A \index{operad!colored}\index{colored operad}\emph{$\colorc$-colored operad} in $(\M,\otimes,\tensorunit)$ is a triple $\bigl(\O, \gamma, \operadunit\bigr)$
consisting of the following data.
\begin{itemize}
\item $\O \in \symseqcm$.
\item For $\bigl(\uc = (c_1, \ldots , c_n); d\bigr) \in \Profcc$ with $n \geq 1$,  $\ub_j \in \Profc$ for $1 \leq j \leq n$, and $\ub = (\ub_1,\ldots,\ub_n)$ their concatenation, 
it is equipped with an \index{operadic composition}\emph{operadic composition} \label{notation:operadic-composition}
\begin{equation}\label{operadic-composition}
\nicexy{\O\duc \otimes \bigotimes\limits_{j=1}^n \O\cjubj \ar[r]^-{\gamma} & \O\dub \in \M.}
\end{equation}
\item For each $c \in \colorc$, it is equipped with a \index{colored operad!unit}\emph{$c$-colored unit}\label{notation:colored-unit}
\begin{equation}\label{c-colored-unit}
\nicexy{\tensorunit \ar[r]^-{\operadunit_c} & \O\cc \in \M.}
\end{equation}
\end{itemize}
This data is required to satisfy the following associativity, unity, and equivariance axioms.
\begin{description}
\item[Associativity]
Suppose that:
\begin{itemize}
\item in \eqref{operadic-composition} $\ub_j = \left(b^j_1, \ldots , b^j_{k_j}\right) \in \Profc$ with at least one $k_j > 0$;
\item $\ua^j_i \in \Profc$ for each $1 \leq j \leq n$ and $1 \leq i \leq k_j$;
\item for each $1 \leq j \leq n$, 
\[\ua_j = \begin{cases}\left(\ua^j_1, \ldots , \ua^j_{k_j}\right) & \text{if $k_j > 0$},\\
\varnothing & \text{if $k_j = 0$}\end{cases}\]
with $\ua = (\ua_1,\ldots , \ua_n)$ their concatenation.
\end{itemize}
Then the \index{associativity!colored operad}\emph{associativity diagram}
\begin{equation}\label{operad-associativity}
\nicexy{\O\duc \otimes 
\left[\bigotimes\limits_{j=1}^n \O\cjubj\right] \otimes \bigotimes\limits_{j=1}^n 
\bigotimes\limits_{i=1}^{k_j} \O\bjiuaji \ar[r]^-{(\gamma, \Id)} \ar[d]_{\text{permute}}^-{\cong} & \O\dub \otimes \bigotimes\limits_{j=1}^{n}\bigotimes\limits_{i=1}^{k_j} \O\bjiuaji \ar[dd]^{\gamma} \\ \O\duc \otimes \bigotimes\limits_{j=1}^n \left[\O\cjubj 
\otimes \bigotimes\limits_{i=1}^{k_j} \O\bjiuaji\right] \ar[d]_{(\Id, \otimes_j \gamma)} &\\
\O\duc \otimes \bigotimes\limits_{j=1}^n \O\cjuaj \ar[r]^-{\gamma} & \O\dua}
\end{equation}
in $\M$ is commutative.
\item[Unity]
Suppose $d \in \colorc$.
\begin{enumerate}
\item For each $\uc = (c_1,\ldots,c_n) \in \Profc$ with $n \geq 1$, the \index{unity!colored operad}\emph{right unity diagram}
\begin{equation}\label{right-unity}
\nicexy{\O\duc \otimes \tensorunit^{\otimes n}\ar[d]_-{(\Id, \otimes \operadunit_{c_j})} \ar[r]^-{\cong} & \O\duc \ar[d]^-{=} \\
\O\duc \otimes \bigotimes\limits_{j=1}^n \O\cjcj\ar[r]^-{\gamma}&\O\duc}
\end{equation}
in $\M$ is commutative.
\item For each $\ub \in \Profc$ the \emph{left unity diagram}
\begin{equation}\label{left-unity}
\nicexy{\tensorunit \otimes \O\dub \ar[d]_-{(\operadunit_d, \Id)} \ar[r]^-{\cong}& \O\dub \ar[d]^-{=}\\ \O\dd \otimes \O\dub\ar[r]^-{\gamma} & \O\dub}
\end{equation}
in $\M$ is commutative.
\end{enumerate}
\item[Equivariance]
Suppose that in \eqref{operadic-composition} $|\ub_j| = k_j \geq 0$.
\begin{enumerate}
\item For each permutation $\sigma \in \Sigma_n$, the \index{colored operad!equivariance}\emph{top equivariance diagram} 
\begin{equation}\label{operadic-eq-1}
\nicexy@C+.3cm{\O\duc \otimes \bigotimes\limits_{j=1}^n \O\cjubj 
\ar[d]_-{\gamma} \ar[r]^-{(\sigma, \sigma^{-1})}& \O\ducsigma \otimes 
\bigotimes\limits_{j=1}^n \O\csigmajubsigmaj \ar[d]^-{\gamma}\\
\O\duboneubn \ar[r]^-{\sigma\langle k_1, \ldots , k_n\rangle}& \O\dubsigmaoneubsigman}
\end{equation}
in $\M$ is commutative.  The bottom horizontal morphism is the equivariant structure morphism of $\O$ corresponding to the block permutation in $\Sigma_{k_1+\cdots+k_n}$ induced by $\sigma$ that permutes $n$ consecutive blocks of lengths $k_1,\ldots,k_n$.  In the top horizontal morphism, $\sigma$ is the equivariant structure morphism of $\O$ corresponding to $\sigma$, and $\inv{\sigma}$ is the left permutation of the $n$ tensor factors.
\item Given permutations $\tau_j \in \Sigma_{k_j}$ for $1 \leq j \leq n$, the \emph{bottom equivariance diagram}
\begin{equation}\label{operadic-eq-2}
\nicexy@C+.3cm{\O\duc \otimes \bigotimes\limits_{j=1}^n \O\cjubj\ar[d]_-{\gamma} \ar[r]^-{(\Id, \otimes \tau_j)}& \O\duc \otimes \bigotimes\limits_{j=1}^n \O\cjubjtauj\ar[d]^-{\gamma} \\
\O\duboneubn \ar[r]^-{\tau_1 \oplus \cdots \oplus \tau_n}& \O\dubonetauoneubntaun}
\end{equation}
in $\M$ is commutative.  In the top horizontal morphism, each $\tau_j$ is the equivariant structure morphism of $\O$ corresponding to $\tau_j \in \Sigma_{k_j}$.  The bottom horizontal morphism is the equivariant structure morphism of $\O$ corresponding to the block sum $\tau_1 \oplus \cdots \oplus \tau_n \in \Sigma_{k_1+\cdots+k_n}$ induced by the $\tau_j$'s.
\end{enumerate}
\end{description}
A morphism of $\colorc$-colored operads is a morphism of the underlying $\colorc$-colored symmetric sequences that is compatible with the operadic compositions and the colored units in the obvious sense.
\end{definition}

\begin{proposition}\label{prop:operad-same-def}
The definition of a $\colorc$-colored operad in Definition \ref{def:operad} and in Definition \ref{def:operad-generating} are equivalent.
\end{proposition}

\begin{proof}
In Corollary \ref{operad-monoid} the domain of the multiplication $\mu$ is $\O \circ \O$, which has entries
\begin{equation}\label{ocompo}\begin{split}
(\O\circ\O)\dub &= \int^{\uc} \O\duc \otimes \O^{\uc}(\ub)\\
&= \int^{\uc, \ua_1, \ldots, \ua_m} \Sigmacop(\ua;\ub) \cdot \O\duc \otimes \bigotimes_{j=1}^m \O\cjuaj\end{split}
\end{equation}
for $(\ub;d) \in \Profcc$, where $\ua=(\ua_1,\ldots,\ua_m)$ is the concatenation.  Observe that $\Sigmacop(\ua;\ub)$ is empty unless $\ub=\ua\sigma$ for some permutation $\sigma$, i.e., the concatenation of the $\ua_j$'s is $\ub$ up to a permutation.  So the multiplication $\mu : \O\circ\O \to \O$ yields the entrywise operadic composition $\gamma$ \eqref{operadic-composition}.  The associativity diagram in \eqref{operad-monoid-axioms} corresponds to the associativity diagram \eqref{operad-associativity}.  The top equivariance diagram \eqref{operadic-eq-1} corresponds to the $\uc$-variable in the coend \eqref{ocompo} and the fact that $\mu$ is a morphism of $\colorc$-colored symmetric sequences.  Similarly, the bottom equivariance diagram \eqref{operadic-eq-2} corresponds to the $\ua_j$ variables in the coend \eqref{ocompo} and the fact that $\mu$ is a morphism of $\colorc$-colored symmetric sequences.

For each $c \in \colorc$, the $c$-colored unit $\operadunit_c$ in \eqref{c-colored-unit} corresponds to the $(c;c)$-entry of the unit morphism $\varepsilon : \I \to \O$ in Corollary \ref{operad-monoid}.  The unity diagram in \eqref{operad-monoid-axioms} corresponds to the right unity diagram \eqref{right-unity} and the left unity diagram \eqref{left-unity}.
\end{proof}

\begin{remark} The reader is cautioned that the definition of a $1$-colored operad in \cite{mss} (p.41, Definition 1.4) is missing the bottom equivariance axiom \eqref{operadic-eq-2}.\dqed
\end{remark}

\section{Operads in Terms of Partial Compositions}\label{sec:compi}

Instead of the operadic composition $\gamma$, it is also possible to express a colored operad in terms of binary operations.

\begin{definition}\label{def:compi}
Suppose $\bigl(\uc = (c_1, \ldots , c_n); d\bigr) \in \Profcc$ with $n \geq 1$, $\ub \in \Profc$, and $1 \leq i \leq n$.  Define the\index{profile!partial composition} $\colorc$-profile
\[\uc \compi \ub = \bigl(\underbrace{c_1,\ldots,c_{i-1}}_{\varnothing~\mathrm{if}~i=1},\ub,\underbrace{c_{i+1},\ldots,c_n}_{\varnothing~\mathrm{if}~i=n}\bigr).\]
\end{definition}

The following is \cite{yau-operad} Definition 16.2.1.

\begin{definition}\label{def:operad-compi}
A \emph{$\colorc$-colored operad} in $(\M,\otimes,\tensorunit)$ is a triple $(\O,\comp,\operadunit)$ consisting of the following data.
\begin{itemize}
\item $\O\in \symseqcm$.
\item For $\bigl(\uc = (c_1, \ldots , c_n);d\bigr) \in \Profcc$ with $n \geq 1$, $1 \leq i \leq n$, and $\ub \in \Profc$, it is equipped with a\index{partial composition} morphism
\begin{equation}\label{operadic-compi}
\nicexy{\O\duc \otimes \O\ciub\ar[r]^-{\compi} & \O\sbinom{d}{\uc\compi\ub} \in \M}
\end{equation}
called the \label{notation:compi-operad}\emph{$\compi$-composition}. 
\item For each color $c \in \colorc$, it is equipped with a $c$-colored unit
\[\nicexy{\tensorunit \ar[r]^-{\operadunit_c} & \O\cc\in \M.}\]
\end{itemize}
This data is required to satisfy the following associativity, unity, and equivariance axioms.  Suppose $d \in \colorc$, $\uc = (c_1, \ldots , c_n) \in \Profc$, $\ub \in \Profc$ with length $|\ub| = m$, and $\ua \in \Profc$ with length $|\ua| = l$.
\begin{description}
\item[Associativity]
There are two associativity axioms.
\begin{enumerate}
\item Suppose $n \geq 2$ and $1 \leq i < j \leq n$.  Then the \index{colored operad!horizontal associativity}\emph{horizontal associativity diagram} in $\M$
\begin{equation}\label{compi-associativity}
\nicexy{\O\duc \otimes \O\sbinom{c_i}{\ua} \otimes \O\cjub \ar[d]_-{\mathrm{permute}}^-{\cong} 
\ar[r]^-{(\compi, \Id)} & \O\sbinom{d}{\uc\compi\ua} \otimes \O\cjub\ar[dd]^-{\comp_{j-1+l}}\\
\O\duc \otimes \O\cjub \otimes \O\sbinom{c_i}{\ua}\ar[d]_-{(\comp_j,\Id)} &\\
\O\sbinom{d}{\uc\comp_j\ub}\otimes \O\sbinom{c_i}{\ua}\ar[r]^-{\compi} & \O\sbinom{d}{(\uc\comp_j\ub)\compi\ua} = \O\sbinom{d}{(\uc\compi\ua)\comp_{j-1+l}\,\ub}}
\end{equation}
is commutative.
\item Suppose $n,m \geq 1$, $1 \leq i \leq n$, and $1 \leq j \leq m$.  Then the \index{colored operad!vertical associativity}\emph{vertical associativity diagram} in $\M$
\begin{equation}\label{compi-associativity-two}
\nicexy{\O\duc \otimes \O\ciub \otimes \O\sbinom{b_j}{\ua} \ar[d]_-{(\compi,\Id)} \ar[r]^-{(\Id,\comp_j)} &
\O\duc \otimes \O\sbinom{c_i}{\ub\comp_j\ua} \ar[d]^-{\compi}\\
\O\sbinom{d}{\uc\compi\ub} \otimes \O\sbinom{b_j}{\ua} \ar[r]^-{\comp_{i-1+j}} & 
\O\sbinom{d}{(\uc\compi\ub)\comp_{i-1+j}\,\ua} = \O\sbinom{d}{\uc\compi(\ub\comp_j\ua)}}
\end{equation}
is commutative.
\end{enumerate}
\item[Unity]
There are two unity axioms.
\begin{enumerate}
\item The \emph{left unity diagram} in $\M$\index{colored operad!unity}
\begin{equation}\label{compi-left-unity}
\nicexy{\tensorunit \otimes \O\duc \ar[dr]_-{\cong} \ar[r]^-{(\operadunit_d,\Id)}&
\O\dd \otimes \O\duc \ar[d]^-{\comp_1}\\ & \O\duc}
\end{equation}
is commutative.
\item If $n \geq 1$ and $1 \leq i \leq n$, then the \emph{right unity diagram} in $\M$
\begin{equation}\label{compi-right-unity}
\nicexy{\O\duc \otimes \tensorunit \ar[dr]_-{\cong} \ar[r]^-{(\Id,\operadunit_{c_i})}& 
\O\duc \otimes \O\cici\ar[d]^-{\compi}\\ & \O\duc}
\end{equation}
is commutative.
\end{enumerate}
\item[Equivariance]
Suppose $|\uc| = n \geq 1$, $1 \leq i \leq n$, $\sigma \in \Sigma_n$, and $\tau \in \Sigma_m$.  Then the \emph{equivariance diagram} in $\M$\index{colored operad!equivariance}
\begin{equation}\label{compi-eq}
\nicexy{\O\duc \otimes \O\sbinom{c_{\sigma(i)}}{\ub}\ar[d]_-{(\sigma,\tau)} \ar[r]^-{\comp_{\sigma(i)}}& 
\O\sbinom{d}{\uc\comp_{\sigma(i)}\ub} \ar[d]^-{\sigma\compi\tau}\\
\O\ducsigma \otimes \O\sbinom{c_{\sigma(i)}}{\ub\tau}\ar[r]^-{\compi} & 
\O\sbinom{d}{(\uc\sigma)\compi(\ub\tau)}=\O\sbinom{d}{(\uc\comp_{\sigma(i)}\ub)(\sigma\compi\tau)}}
\end{equation}
is commutative, where
\[\sigma \compi \tau = \sigma\langle \underbrace{1,\ldots,1}_{i-1}, m,\underbrace{1,\ldots,1}_{n-i}\rangle \comp \bigl(\underbrace{\id \oplus \cdots \oplus \id}_{i-1} \oplus \tau \oplus \underbrace{\id \oplus \cdots \oplus \id}_{n-i}\bigr)\in \Sigma_{n+m-1}\]
is the composition of a block sum induced by $\tau$ with a block permutation induced by $\sigma$ that permutes consecutive blocks of the indicated lengths.
\end{description}
A morphism of $\colorc$-colored operads is a morphism of the underlying $\colorc$-colored symmetric sequences that is compatible with the $\compi$-compositions and the colored units in the obvious sense.
\end{definition}

\begin{proposition}\label{prop:operad-defs-equivalent}
The definition of a $\colorc$-colored operad in Definition \ref{def:operad-generating} and in Definition \ref{def:operad-compi} are equivalent.
\end{proposition}

\begin{proof}
The proof can be found in \cite{yau-operad} Section 16.4.  Let us indicate the correspondence of structures.  Given a $\colorc$-colored operad $(\O,\gamma,\operadunit)$ in the sense of Definition \ref{def:operad-generating}, the associated $\compi$-composition is the composition
\begin{equation}\label{compi-def}
\nicexy{\O\duc\otimes\O\ciub \ar[d]_-{\cong} \ar[r]^-{\compi} & \O\sbinom{d}{\uc\compi\ub}\\
\O\duc \otimes \tensorunit^{\otimes i-1} \otimes \O\ciub \otimes \tensorunit^{\otimes n-i} \ar[r]^-{\{\operadunit_{c_j}\}} &
\O\duc \otimes \Bigl[\bigotimes\limits_{j=1}^{i-1}\O\cjcj\Bigr] \otimes \O\ciub \otimes \Bigl[\bigotimes\limits_{j=i+1}^{n}\O\cjcj\Bigr] \ar[u]_-{\gamma}}
\end{equation}
in which the bottom horizontal morphism is the monoidal product of the colored units $\operadunit_{c_j}$ for $1 \leq j\not=i \leq n$ with the identity morphisms of $\O\duc$ and $\O\ciub$.

Conversely, given a $\colorc$-colored operad $(\O,\comp,\operadunit)$ in the sense of Definition \ref{def:operad-compi}, the operadic composition $\gamma$ is recovered as the composition
\begin{equation}\label{compi-to-gamma}
\nicexy@C+.7cm{\O\duc \otimes\bigotimes\limits_{j=1}^n \O\cjubj \ar[r]^-{\gamma} \ar[d]_-{(\comp_1,\Id)} & \O\dub\\
\O\sbinom{d}{\uc\comp_1\ub_1}\otimes\bigotimes\limits_{j=2}^n\O\cjubj \ar[d]_-{(\comp_{k_1+1},\Id)} 
& \O\sbinom{d}{((\uc\comp_1\ub_1)\cdots)\comp_{k_1+\cdots+k_{n-1}+1}\ub_n} \ar[u]_-{\Id}\\
\cdots \ar[r]^-{(\comp_{k_1+\cdots+k_{n-2}+1},\Id)} & \O\sbinom{d}{((\uc\comp_1\ub_1)\cdots)\comp_{k_1+\cdots+k_{n-2}+1}\ub_{n-1}} \otimes \O\sbinom{c_n}{\ub_n} \ar[u]_-{\comp_{k_1+\cdots+k_{n-1}+1}}}
\end{equation}
in which $k_j = |\ub_j|$.  
\end{proof}

\section{Operads in Terms of Trees}\label{sec:operad-tree}

In Proposition \ref{prop:operad-same-def} and Proposition \ref{prop:operad-defs-equivalent} we observed that there are three equivalent definitions of a colored operad.  In this section, we discuss another equivalent description of a colored operad in terms of trees that we will need later to discuss the Boardman-Vogt construction of a colored operad.  We will need the following concept about monoidal product \cite{mss} (p.64, Definition 1.58).

\begin{definition}[Unordered Monoidal Product]\label{def:unordered-tensor}
Suppose $X$ is a set with $n \geq 1$ elements.
\begin{enumerate}
\item An \index{ordering!of a set}\emph{ordering} of $X$ is a bijection \[\sigma : \{1,\ldots,n\} \iso X.\] The set of all orderings of $X$ is denoted by $\Ord(X)$.
\item Suppose $A_x \in \M$ is an object for each $x \in X$.  
\begin{enumerate}
\item For each ordering $\sigma$ of $X$, define the \emph{ordered monoidal product}\index{monoidal product!ordered} as
\[\bigotimes_{\sigma} A_x\label{note:ordtensor} = 	A_{\sigma(1)} \otimes \cdots \otimes A_{\sigma(n)} \in \M.\]
For each $\tau \in \Sigma_n$, the symmetry isomorphism in $\M$ determines an isomorphism
\[\tau : \bigotimes_{\sigma} A_x \iso \bigotimes_{\sigma\tau} A_x,\]
which defines a $\Sigma_n$-action on the coproduct
$\coprod_{\sigma \in \Ord(X)} \bigotimes_{\sigma} A_x$.
\item Define the \emph{unordered monoidal product}\index{monoidal product!unordered} 
as the colimit
\begin{equation}\label{unordered-tensor}
\bigotimes_{x \in X} A_x= \colimover{\tau \in \Sigma_n} \Bigl(\nicexy{\coprodover{\sigma \in \Ord(X)} \bigtensorover{\sigma} A_x \ar[r]^-{\tau} & \coprodover{\sigma \in \Ord(X)} \bigtensorover{\sigma} A_x}\Bigr).
\end{equation}
\end{enumerate}
\end{enumerate}
\end{definition}

\begin{remark} For each ordering $\sigma$ of $X$, the natural morphism \[\bigotimes_{\sigma} A_x \to \bigotimes_{x\in X} A_x\] from the ordered monoidal product to the unordered monoidal product is an isomorphism.   The point of the unordered monoidal product is that we can talk about the iterated monoidal product of the $A_x$'s without first choosing an ordering of the indexing set $X$.\dqed\end{remark}

As before $\colorc$ is a fixed non-empty set.  All the trees below are $\colorc$-colored trees as in Definition \ref{def:tree}.  Recall that $\Profc$ denotes the set of $\colorc$-profiles.

\begin{definition}[Vertex Decorations]\label{def:tree-decoration}
Suppose $A \in \M^{\Profcc}$, and $T$ is a tree.  Define the \index{decoration}\emph{$A$-decoration of $T$} as the unordered monoidal product\label{notation:vertex-dec} \[A[T] = \bigtensorover{v\in T} A\bigl(\profofv\bigr) = \bigtensorover{v\in T} A\inoutv,\] where $v \in T$ means $v\in \Vt(T)$.
\end{definition}

\begin{proposition}\label{prop:vertexdec-treesub}
Suppose $T$ is a tree, and $H_v$ is a tree with $\profofhv=\profofv$ for each vertex $v$ in $T$.  Then for each $A \in \M^{\Profcc}$, there is an isomorphism
\[A\bigl[T(H_v)_{v\in T}\bigr] \cong \bigtensorover{v\in T} A[H_v],\] where $T(H_v)_{v\in T}$ is the tree substitution in Definition \ref{def:tree-substitution}.  In particular, in the context of Definition \ref{def:grafting}, there is an isomorphism \[A\bigl[\graft(G;\{H_j\})\bigr]\cong A[G]\otimes A[H_1] \otimes \cdots \otimes A[H_m].\]
\end{proposition}

\begin{proof} The first isomorphism follows from the decomposition \[\Vt\bigl(T(H_v)_{v\in T}\bigr) = \coprod_{v\in T} \Vt(H_v).\] The second isomorphism follows from the definition of the grafting as a tree substitution.
\end{proof}

\begin{notation}\label{notation:a-of-v}
In the setting of Definition \ref{def:tree-decoration}, we will sometimes abbreviate $A\bigl(\profofv\bigr) = A\inoutv$ to $A(v)$.\end{notation}

\begin{example}\label{ex:vertexdec-treeesub}
In the context of Example \ref{ex:treesub}, recall that $K$ is the tree substitution $T(H_u,H_v,H_w)$.  There are isomorphisms
\[\begin{split} A[K] &\cong A[H_u] \otimes A[H_v] \otimes A[H_w]\\
&\cong A(u_1) \otimes A(u_2) \otimes \tensorunit \otimes A(w_1) \otimes A(w_2)\\
&\cong A\sbinom{c}{a,b,f} \otimes A\sbinom{f}{\varnothing} \otimes A\sbinom{e}{c,g} \otimes A\sbinom{g}{d}\end{split}\] for each $A \in \M^{\Profcc}$.\dqed
\end{example}

\begin{example}\label{ex:vertexdec-grafting}
In the context of Example \ref{ex:grafting}, recall that $G$ is the grafting $\graft\bigl(T;H_1,H_2,H_3\bigr)$.  There are isomorphisms
\[\begin{split} A[G] &\cong A[T] \otimes A[H_1] \otimes A[H_2] \otimes A[H_3]\\
&\cong A(w) \otimes A(u) \otimes A(v) \otimes A(x_1) \otimes A(x_2) \otimes \tensorunit \otimes A(z)\\
&\cong A\sbinom{e}{c,d} \otimes A\sbinom{c}{a,b} \otimes A\dd \otimes A\sbinom{a}{f,g} \otimes A\sbinom{g}{\varnothing} \otimes A\sbinom{d}{\varnothing}\end{split}\]
for each $A \in \M^{\Profcc}$.\dqed
\end{example}

In the next definition of a colored operad, notice (i) the use of the product category $\M^{\Profcc}$ instead of the category $\symseqcm$ of symmetric sequences and (ii) the apparent absence of an equivariance axiom.

\begin{definition}\label{def:operad-tree}
A \emph{$\colorc$-colored operad} in $\M$ is a pair $(\O,\gamma)$ consisting of 
\begin{itemize}\item an object $\O \in \M^{\Profcc}$ and
\item an \emph{operadic structure morphism}\index{operadic structure morphism} 
\begin{equation}\label{operadic-structure-map}
\nicexy{\O[T] \ar[r]^-{\gamma_T} & \O\bigl(\profoft\bigr) \in \M}
\end{equation}
for each $T \in \uTreec$
\end{itemize}
that satisfies the following unity and associativity axioms.
\begin{description}
\item[Unity] $\gamma_{\Cor_{(\uc;d)}}$ is the \index{colored operad!unity}identity morphism of $\O\duc$ for each $(\uc;d) \in \Profcc$, where $\Cor_{(\uc;d)}$ is the $(\uc;d)$-corolla in Example \ref{ex:cd-corolla}.
\item[Associativity] For each tree substitution $T(H_v)_{v\in T}$, the diagram\index{colored operad!associativity}
\begin{equation}\label{operad-tree-ass}
\nicexy{\O\bigl[T(H_v)_{v\in T}\bigr] \ar[d]_-{\gamma_{T(H_v)_{v\in T}}} \ar[r]^-{\cong} & \bigtensorover{v\in T} \O[H_v] \ar[r]^-{\bigtensorover{v} \gamma_{H_v}} & \bigtensorover{v\in T} \O(v) = \O[T] \ar[d]^-{\gamma_T}\\ \O\bigl(\Prof(T(H_v)_{v\in T})\bigr) \ar[rr]^-{\Id} && \O\bigl(\profoft\bigr)}
\end{equation}
is commutative. 
\end{description}
A morphism $f : (\O,\gammao) \to (\P,\gammap)$ of $\colorc$-colored operads is a morphism $f : \O \to \P \in \M^{\Profcc}$ such that the diagram
\begin{equation}\label{operad-map-tree}
\nicexy{\O[T] \ar[d]_-{\gammao_T} \ar[r]^-{\bigtensorover{v} f} & \P[T] \ar[d]^-{\gammap_T}\\ \O\bigl(\profoft\bigr) \ar[r]^-{f} & \P\bigl(\profoft\bigr)}
\end{equation}
is commutative for each $T \in \uTreec$.
\end{definition}

\begin{theorem}\label{thm:operad-def2=def3}
The definitions of a $\colorc$-colored operad in Definition \ref{def:operad-generating} and in Definition \ref{def:operad-tree} are equivalent.
\end{theorem}

\begin{proof}
This equivalence is \cite{bluemonster} Corollary 11.16.  Let us describe the correspondence of structures.  Suppose $(\O,\gamma)$ is a $\colorc$-colored operad in the sense of Definition \ref{def:operad-tree}.
\begin{enumerate}
\item For a pair $(\uc;d) \in \Profcc$ and a permutation $\tau \in \Sigma_{|\uc|}$, the operadic structure morphism \[\nicexy@C+.5cm{\O\duc=\O\bigl[\Cor_{(\uc;d)}\tau\bigr] \ar[r]^-{\gamma_{\Cor_{(\uc;d)}\tau}} & \O\bigl[\Prof(\Cor_{(\uc;d)}\tau)\bigr]=\O\ductau},\]
where $\Cor_{(\uc;d)}\tau$ is the permuted corolla in Example \ref{ex:cd-permuted-corolla}, corresponds to the $\colorc$-colored symmetric sequence structure in Definition \ref{def:operad-generating}.
\item For $\bigl(\uc = (c_1, \ldots , c_n); d\bigr) \in \Profcc$ with $n \geq 1$,  $\ub_j \in \Profc$ for $1 \leq j \leq n$, $\ub = (\ub_1,\ldots,\ub_n)$ their concatenation, and $T = T\bigl(\{\ub_j\};\uc;d\bigr)$ the $2$-level tree in Example \ref{ex:twolevel-tree}, the operadic structure morphism \[\nicexy{\O\duc \otimes \bigotimes\limits_{j=1}^n \O\cjubj \cong \O[T] \ar[r]^-{\gamma_T} & \O[\profoft] = \O\dub}\] corresponds to the operadic composition $\gamma$ in \eqref{operadic-composition}.
\item For each color $c \in \colorc$, the operadic structure morphism \[\nicexy{\tensorunit = \O[\uparrow_c] \ar[r]^-{\gamma_{\uparrow_c}} & \O[\Prof(\uparrow_c)]=\O\cc,}\] where $\uparrow_c$ is the $c$-colored exceptional edge in Example \ref{ex:colored-exedge}, corresponds to the $c$-colored unit $\operadunit_c$ in \eqref{c-colored-unit}.
\end{enumerate}
The associativity, unity, and equivariance axioms in Definition \ref{def:operad-generating} are now consequences of the associativity and unity of $(\O,\gamma)$.

Conversely, suppose $(\O,\gamma,\operadunit)$ is a $\colorc$-colored operad in the sense of Definition \ref{def:operad-generating}.  Reusing the previous paragraph, we first define the operadic structure morphisms \[\nicexy{\O[G] \ar[r]^-{\gamma_G} & \O[\Prof(G)]} \forspace G \in \Bigl\{\Cor_{(\uc;d)}\tau, \uparrow_c, T\bigl(\{\ub_j\};\uc;d\bigr)\Bigr\}\] as the equivariant structure, the $c$-colored units $\operadunit_c$, and the operadic composition $\gamma$.  For a general tree $T$, a key observation is that it can always be written non-uniquely as an iterated tree substitution involving only permuted corollas, exceptional edges, and $2$-level trees.  We then use any such tree substitution decomposition of $T$ and the associativity diagram \eqref{operad-tree-ass} to define the operadic structure morphism \[\nicexy{\O[T] \ar[r]^-{\gamma_T} & \O[\Prof(T)]}\] as an iterated composition of monoidal products of the already defined operadic structure morphisms $\gamma_G$.  That such a morphism $\gamma_T$ is well-defined is a consequence of the axioms in Definition \ref{def:operad-generating}.  
\end{proof}

\begin{remark}There are two more equivalent descriptions of an operad that we will not need in this book, so we only briefly mention them here.  
\begin{enumerate}
\item There is a $\Profcc$-colored operad $\Opc$ whose category of algebras is precisely the category of\index{colored operad!for colored operads} $\colorc$-colored operads in $\M$.  This is a special case of \cite{bluemonster} Lemma 14.4.  Each entry $\Opc\tus$, with $t$ and each $s_j$ in $\Profcc$, is a coproduct $\coprod \tensorunit$ indexed by pairs $(T,\sigma)$ with
\begin{itemize}\item $T$ a $\colorc$-colored tree with profile $t$ and 
\item $\sigma$ an ordering of the set $\Vt(T)$ such that $s_j = \Prof(\sigma(j))$.  
\end{itemize}
Its equivariant structure comes from reordering of the set $\Vt(T)$, and its colored units correspond to corollas.  Its operadic composition $\gamma$ is induced by tree substitution with the induced lexicographical ordering on vertices.  One checks that $\Opc$-algebras are equivalent to $\colorc$-colored operads in Definition \ref{def:operad-tree}.
\item 
The category of $\colorc$-colored operads in $\M$ is also canonically isomorphic to the category of $\M$-enriched multicategorical functors\index{multicategorical functor} from $\Opc$ to $\M$.  This is a special case of Theorem 14.12 in \cite{bluemonster}, where the reader is referred for the meaning of an enriched multicategorical functor.  One checks that such enriched multicategorical functors are also equivalent to $\colorc$-colored operads in Definition \ref{def:operad-tree}.\dqed
\end{enumerate}\end{remark}

\begin{corollary}\label{cor:operad-functor-subcat}
Suppose $\O$ is a $\colorc$-colored operad in $\M$, and $(\uc;d) \in \Profcc$.  Then $\O$ defines a functor \[\O : \uTreec\duc \to \M\] as follows:
\begin{itemize}\item Each $T \in \uTreec\duc$ is sent to $\O[T]$.
\item Each morphism \[(H_v)_{v\in T} : T(H_v) \to T \in \uTreec\duc\] is sent to the morphism \[\nicexy{\O\bigl[T(H_v)\bigr] = \bigtensorover{v\in T} \O[H_v] \ar[r]^-{\bigtensorover{v} \gamma_{H_v}} & \bigtensorover{v\in T} \O(v) = \O[T]}\]
in which $\gamma_{H_v}$ is the operadic structure morphism \eqref{operadic-structure-map} for $H_v$.
\end{itemize}
\end{corollary}

\begin{proof}
An identity morphism in $\uTreec\duc$ is of the form \[(\Cor_v)_{v\in T} : T \to T.\]  Since $\gamma_{\Cor}$ is the identity morphism for each corolla,  the assignment $\O$ preserves identity morphisms.

Suppose \[(H_v)_{v\in T} : K \to T \andspace (G_u)_{u\in K} : E \to K\] are morphisms in $\uTreec\duc$ as in Remark \ref{rk:subcat-composition}.  Their composition is \[\bigl(H_v(G_u)_{u\in H_v}\bigr)_{v\in T} : E \to T.\]  To see that the assignment $\O$ preserves compositions, observe that
\[\begin{split} \O\bigl(H_v(G_u)_{u\in H_v}\bigr)_{v\in T} &= \bigtensorover{v\in T} \gamma_{H_v(G_u)_{u\in H_v}}\\
&= \bigtensorover{v\in T} \Bigl(\gamma_{H_v} \circ \bigtensorover{u\in H_v}\gamma_{G_u}\Bigr)\\
&= \Bigl(\bigtensorover{v\in T} \gamma_{H_v}\Bigr) \circ \Bigl(\bigtensorover{v\in T}\,\bigtensorover{u\in H_v} \gamma_{G_u}\Bigr)\\
&= \O(H_v)_{v\in T} \circ \O(G_u)_{u\in K}.\end{split}\]
The first and the last equalities are the definitions of the assignment $\O$ on a morphism.  The second equality holds by the associativity axiom \eqref{operad-tree-ass} of the $\colorc$-colored operad $\O$.
\end{proof}

\section{Algebras over Operads}\label{sec:algebra-operad}

In this section, we discuss algebras over colored operads and some relevant examples.  Just as monads are important because of their algebras, operads are important mainly because of their algebras.

\begin{notation}\label{not:x-sub-c}
For a $\colorc$-colored object $X=\{X_c\}_{c\in\colorc}$ in $\M$ and $\uc=(c_1,\ldots,c_m) \in \Profc$, we will write \[X_{\uc} = X_{c_1} \otimes \cdots \otimes X_{c_m},\] which is the initial object $\varnothing$ if $\uc$ is the empty profile.
\end{notation}

\begin{lemma}\label{lem:operad-monad}
Suppose $\O$ is a $\colorc$-colored operad in $\M$.  Then it induces a \index{colored operad!induced monad}\index{monad!induced by an operad}monad whose functor is \[\O \circ - : \Mtoc \to \Mtoc\] and whose multiplication and unit are induced by those of $\O$ as in Corollary \ref{operad-monoid}
\end{lemma}

\begin{proof}
Suppose $Y=\{Y_c\}_{c\in\colorc}$ is a $\colorc$-colored object in $\M$, regarded as a $\colorc$-colored symmetric sequence as in \eqref{colored-object-sm}.  Observe that in \eqref{ytensorc} we have 
\[Y^{\uc}(\ub)= \begin{cases} Y_{\uc} & \text{ if $\ub=\varnothing$},\\
\varnothing & \text{ if $\ub\not=\varnothing$}\end{cases}\] for $\ub,\uc\in\Profc$.  Putting this into the definition \eqref{circle-product} of the $\colorc$-colored circle product, we obtain
\begin{equation}\label{ocompyentries}
(\O\circ Y)\dub = \begin{cases} \dint^{\uc\in\Sigmac} \O\duc \otimes Y_{\uc}& \text{ if $\ub=\varnothing$},\\ \varnothing & \text{ if $\ub\not=\varnothing$.}\end{cases}
\end{equation}
So the restriction of the functor \[\O \circ - : \symseqcm \to \symseqcm\] to the full subcategory $\Mtoc$ yields a functor $\Mtoc \to \Mtoc$.  Corollary \ref{operad-monoid} now shows that $\O\circ -$ is a monad in $\Mtoc$.
\end{proof}

\begin{definition}\label{def:operad-algebra}
Suppose $\O$ is a $\colorc$-colored operad in $\M$.  The category $\algmo$ \index{colored operad!algebra}\index{algebra!of a colored operad}of \emph{$\O$-algebras} is defined as the category of $(\O\circ -)$-algebras for the monad $\O\circ -$ in $\Mtoc$.
\end{definition}

We can describe $\O$-algebras more explicitly by unwrapping this definition.  The detailed colored operad algebra axioms below are from \cite{yau-operad} (Section 13.2) and \cite{bluemonster} (Corollary 13.37).  In the one-colored topological case, the definition below is due to May \cite{may}.  We will use Definition \ref{def:operad-generating} of a $\colorc$-colored operad.

\begin{definition}\label{def:operad-algebra-generating}
Suppose $(\O,\gamma,\operadunit)$ is a $\colorc$-colored operad in $\M$.  An \emph{$\O$-algebra} is a pair $(X,\lambda)$ consisting of 
\begin{itemize}\item a $\colorc$-colored object $X=\{X_c\}_{c\in\colorc}$ and 
\item an \emph{$\O$-action structure morphism}\index{structure morphism!for colored operadic algebras}
\begin{equation}\label{operad-algebra-action}
\nicexy{\O \duc \otimes X_{\uc} \ar[r]^-{\lambda}& X_d \in \M}
\end{equation}
for each $(\uc;d) \in \Profcc$.  
\end{itemize}
It is required that the following associativity, unity, and equivariance axioms hold.
\begin{description}
\item[Associativity]
For $\bigl(\uc = (c_1, \ldots , c_n);d\bigr) \in \Profcc$ with $n \geq 1$, $\ub_j \in \Profc$ for $1 \leq j \leq n$, and $\ub = (\ub_1,\ldots,\ub_n)$ their concatenation, the associativity diagram\index{associativity!colored operadic algebra}
\begin{equation}
\label{operad-algebra-associativity}
\nicexy@C+15pt{\O\duc \otimes \left[\bigotimes\limits_{j=1}^n \O\cjubj\right]\otimes X_{\ub} 
\ar[r]^-{(\gamma, \Id)} \ar[d]_-{\mathrm{permute}}^-{\cong}& \O\dub \otimes X_{\ub} \ar[dd]^-{\lambda}\\ \O\duc \otimes \bigotimes\limits_{j=1}^n \left[\O \cjubj \otimes X_{\ub_j}\right] \ar[d]_-{(\Id,\otimes_j \lambda)} &\\ \O \duc \otimes X_{\uc} \ar[r]^-{\lambda}& X_d}
\end{equation}
in $\M$ is commutative.
\item[Unity] For each $c \in \colorc$, the unity diagram\index{unity!colored operadic algebra}
\begin{equation}\label{operad-algebra-unity}
\nicexy{\tensorunit \otimes X_c \ar[d]_-{(\operadunit_c, \Id)} \ar[r]^-{\cong}& X_c \ar[d]^-{=}\\
\O\cc \otimes X_c \ar[r]^-{\lambda}& X_c}
\end{equation}
in $\M$ is commutative.
\item[Equivariance]
For each $(\uc;d) \in \Profcc$ and each permutation $\sigma \in \Sigma_{|\uc|}$, the equivariance diagram\index{equivariance!colored operadic algebra}
\begin{equation}\label{operad-algebra-eq}
\nicexy{\O\duc \otimes X_{\uc}\ar[d]_-{\lambda} \ar[r]^-{(\sigma,\inv{\sigma})}& \O \ducsigma \otimes X_{\uc\sigma}\ar[d]^-{\lambda}\\ X_d \ar[r]^-{=}& X_d}
\end{equation}
in $\M$ is commutative.   In the top horizontal morphism, $\inv{\sigma}$ is the left permutation on the factors in $X_{\uc}$ induced by $\inv{\sigma} \in \Sigma_{|\uc|}$.
\end{description}
A \index{morphism!colored operadic algebras}\emph{morphism of $\O$-algebras} $f : (X,\lambda) \to (Y,\xi)$ is a morphism $f : X \to Y$ of $\colorc$-colored objects in $\M$ such that the diagram
\begin{equation}\label{algebra-map-compatibility}
\nicexy{\O\duc \otimes X_{\uc}\ar[d]_-{\lambda} \ar[r]^-{(\Id, \otimes f)}& \O\duc \otimes Y_{\uc} \ar[d]^-{\xi}\\ X_d \ar[r]^-{f} & Y_d}
\end{equation}
in $\M$ is commutative for all $(\uc;d) \in \Profcc$.
\end{definition}

\begin{proposition}\label{prop:operad-algebra-defs}
Suppose $\O$ is a $\colorc$-colored operad in $\M$.  Then the definitions of an $\O$-algebra in Definition \ref{def:operad-algebra} and in Definition \ref{def:operad-algebra-generating} are equivalent.
\end{proposition}

\begin{proof}
This is essentially the same as the proof of Proposition \ref{prop:operad-same-def}.  The key is \eqref{ocompyentries}: For a $\colorc$-colored object $X = \{X_c\}_{c\in\colorc}$ in $\M$, the $\colorc$-colored object $\O \circ X$ has entries \[(\O\circ X)_d = \int^{\uc\in\Sigmac} \O\duc \otimes X_{\uc}\] for $d \in \colorc$.  So an $(\O\circ -)$-algebra has $\O$-action structure morphisms as in \eqref{operad-algebra-action}.  The equivariance axiom \eqref{operad-algebra-eq} corresponds to the $\uc$-variable in the coend formula for $(\O\circ X)_d$.  The associativity axiom \eqref{operad-algebra-associativity} and the unity axiom \eqref{operad-algebra-unity} correspond to those of an $(\O\circ -)$-algebra.
\end{proof}

The following result is \cite{white-yau} Proposition 4.2.1.

\begin{proposition}\label{prop:algebra-bicomplete}
Suppose $\O$ is a $\colorc$-colored operad in $\M$.  Then there is a free-forgetful adjunction \[\nicexy{\Mtoc \ar@<2pt>[r]^-{\O\comp -} & \algmo \ar@<2pt>[l]^-{U}}\] in which the right adjoint $U$ forgets about the $\O$-algebra structure.  Moreover, the category $\algmo$ is \index{cocomplete!colored operadic algebra}\index{complete!colored operadic algebra}cocomplete (resp., complete), provided $\M$ is cocomplete (resp., complete).
\end{proposition}

\begin{example}[Colored objects as algebras]\label{ex:ialgebra}
For the unit $\colorc$-colored operad $\I$ in \eqref{unit-operad}, there is an equality \[\algm(\I) = \Mtoc,\] and both functors $U$ and $\I\circ-$ are the identity functors.\end{example}

\begin{example}[Colored endomorphism operads]\label{ex:endomorphism-operad}
For each $\colorc$-colored object $X=\{X_c\}_{c\in\colorc}$ in $\M$, there is a \emph{$\colorc$-colored endomorphism operad}\index{endomorphism operad}\index{colored operad!endomorphism} $\End(X)$ with entries
\[\End(X)\duc = \Homm(X_{\uc},X_d)\] for $(\uc;d) \in \Profcc$.  Its equivariant structure is induced by permutations of the factors in $X_{\uc}$.  Its $d$-colored unit \eqref{c-colored-unit} \[\tensorunit \to \Homm(X_d,X_d)\] is adjoint to the identity morphism of $X_d$.  Its operadic composition $\gamma$ \eqref{operadic-composition} is induced by the $\otimes$-$\Homm$-adjunction.  Another exercise involving the $\otimes$-$\Homm$-adjunction shows that an $\O$-algebra structure $(X,\theta)$ is equivalent to a morphism \[\theta' : \O \to \End(X)\] of $\colorc$-colored operads.  See \cite{yau-operad} Sections 13.8 and 13.9 for details.\dqed
\end{example}

\begin{example}[Tree operad]\label{ex:tree-operad}
There is a \index{tree operad}\index{colored operad!of trees}\emph{$\colorc$-colored tree operad} $\Treeopc$ in $\Set$ in which each entry $\Treeopc\duc$ is the set of $\colorc$-colored trees with profile $(\uc;d)$.  
\begin{itemize}\item The $c$-colored unit is the $c$-colored exceptional edge $\uparrow_c$ in Example \ref{ex:colored-exedge}.  
\item The equivariant structure is given by reordering: If $T \in \Treeopc\duc$ and if $\sigma \in \Sigma_{|\uc|}$, then $T\sigma \in \Treeopc\ducsigma$ is the same as $T$ except that its ordering is $\zeta\sigma$, where $\zeta$ is the ordering of $T$.  
\item The operadic composition $\gamma$ is given by grafting of trees in Definition \ref{def:grafting}.  
\end{itemize}
For each tree $T \in \uTreec\duc$, the operadic structure morphism \[\nicexy{\Treeopc[T] = \prodover{v\in T}\Treeopc(v) \ar[r]^-{\gamma_T} & \Treeopc\duc}\] is given by tree substitution in Definition \ref{def:tree-substitution}, \[\gamma_T\bigl\{H_v\bigr\}_{v\in T} = T\bigl(H_v\bigr)_{v\in T},\]  where each $H_v \in \Treeopc(v)$.\dqed
\end{example}

\begin{example}[Monoids as operads]\label{ex:monoid-unary-operad}
Suppose $(A,\mu,\varepsilon)$ is a monoid in $\M$.  Then it yields a $1$-colored \index{monoid!as an operad}operad $\A$ with entries \[\A(n) = \begin{cases} A & \text{ if $n=1$},\\ \varnothing & \text{ if $n\not= 1$.}\end{cases}\]  Its equivariant structure is trivial.  The operadic composition $\gamma$  and the unit are those of the monoid $A$.  In other words, monoids are $1$-colored operads concentrated in unary entries.\dqed
\end{example}

\begin{example}[Associative operad]\label{ex:operad-as}
There is a $1$-colored operad $\As$ in $\M$, called the \index{associative operad}\emph{associative operad}, with entries \[\As(n) = \coprod_{\sigma \in \Sigma_n} \tensorunit\] for $n \geq 0$ and unit $\operadunit : \tensorunit \to \As(1)$ the identity morphism.  Its operadic composition $\gamma$ is induced by the map
\[\Sigma_n \times \Sigma_{k_1} \times \cdots \times \Sigma_{k_n} \to \Sigma_{k_1+\cdots+k_n}\] that sends $(\sigma; \sigma_1,\ldots,\sigma_n)$ to the composition
\begin{equation}\label{as-comp}
\sigma(\sigma_1,\ldots,\sigma_n)=\sigma\langle k_1,\ldots,k_n\rangle \circ \bigl(\sigma_1\oplus\cdots \oplus \sigma_n\bigr)
\end{equation}
with (i) $\sigma_1\oplus\cdots \oplus \sigma_n$ the block sum induced by the $\sigma_j$ and (ii) $\sigma\langle k_1,\ldots,k_n\rangle$ the block permutation induced by $\sigma$ that permutes $n$ consecutive blocks of lengths $k_1,\ldots,k_n$.  Using Proposition \ref{prop:monoid}(1), one can check that $\As$-algebras are precisely monoids in $\M$.\dqed
\end{example}

\begin{example}[Commutative operad]\label{ex:operad-com}
There is a $1$-colored operad $\Com$ in $\M$, called the \index{commutative operad}\emph{commutative operad}, with entries \[\Com(n) = \tensorunit\] for $n \geq 0$, operadic composition induced by the isomorphism $\tensorunit \otimes \tensorunit \cong \tensorunit$, and unit the identity morphism.  It follows from Proposition \ref{prop:monoid}(2) that $\Com$-algebras are precisely commutative monoids in $\M$.\dqed
\end{example}

\begin{example}[Diagrams as operads]\label{ex:diagram-as-operad}
Suppose $\C$ is a small category with object set $\colorc$, and $F : \C \to \M$ is a $\C$-diagram in $\M$.  Then $F$ yields an $\colorc$-colored \index{diagram!as an operad}operad $\F$ in $\M$ with entries
\[\F\duc = \begin{cases} \coprodover{F\C(c,d)}\tensorunit & \text{ if $\uc=c\in \colorc$},\\ \varnothing & \text{ if $|\uc|\not=1$,}\end{cases}\] for $(\uc;d) \in \Profcc$, where $F\C(c,d)$ is the set of morphisms $Ff \in \M(Fc,Fd)$ for $f \in \C(c,d)$.  Since it is concentrated in unary entries, its equivariant structure is trivial.  Its colored units come from the identity morphisms in $\C$.  Its operadic composition $\gamma$ arises from the fact that $F$ is a functor.\dqed
\end{example}

\begin{example}[Operad for diagrams]\label{ex:operad-diag}
Suppose $\C$ is a small category with object set $\colorc$.  There is a $\colorc$-colored operad\index{colored operad!for diagrams}\index{diagram!operad for} $\Cdiag$ in $\M$ with entries
\[\Cdiag\duc = \begin{cases} \coprodover{\C(c,d)}\tensorunit & \text{ if $\uc=c \in \colorc$},\\ \varnothing & \text{ if $|\uc|\not=1$}\end{cases}\]
for $(\uc;d) \in \Profcc$.  Its equivariant structure is trivial.  Its colored units come from the identity morphisms in $\C$.  Its operadic composition $\gamma$ is induced by the categorical composition in $\C$.  One can check that $\Cdiag$-algebras are precisely $\C$-diagrams in $\M$.\dqed
\end{example}

\begin{example}[Operad for diagrams of monoids]\label{ex:diag-monoid-operad}
This example is a combination of Examples \ref{ex:operad-as} and \ref{ex:operad-diag}.  Suppose $\C$ is a small category with object set $\colorc$.  There is a $\colorc$-colored \index{colored operad!for diagrams of monoids}\index{monoid!operad for diagrams of}operad $\Ocm$ in $\M$ with entries \[\Ocm\duc = \coprodover{\Sigma_n \times \prod\limits_{j=1}^n \C(c_j,d)}\tensorunit \forspace \duc=\dconecn \in \Profcc.\]  A coproduct summand corresponding to an element $(\sigma,\uf) \in \Sigma_n \times \prod_j \C(c_j,d)$ is denoted by $\tensorunit_{(\sigma,\uf)}$.  We will describe the operad structure on $\Ocm$ in terms of the subscripts.

Its equivariant structure sends $\tensorunit_{(\sigma,\uf)}$ to $\tensorunit_{(\sigma\tau,\uf\tau)}$ for $\tau \in \Sigma_{|\uc|}$.  Its $c$-colored unit corresponds to $\tensorunit_{(\id_1,\Id_c)}$.  Its operadic composition 
\[\nicexy{\Ocm\duc \otimes \bigotimes\limits_{j=1}^n \Ocm\cjubj \ar[r]^-{\gamma} & \Ocm\dub}\]
corresponds to
\[\nicexy{\Bigl((\sigma,\uf);\bigl\{(\tau_j,\ug_j)\bigr\}_{j=1}^n\Bigr) \ar@{|->}[r] & \Bigl(\sigma(\tau_1,\ldots,\tau_n),\bigl(f_1\ug_1,\ldots,f_n\ug_n\bigr) \Bigr)}\] where \[f_j\ug_j=\bigl(f_jg_{j1},\ldots,f_jg_{jk_j}\bigr) \in \prod\limits_{i=1}^{k_j} \C(b_{ji},d) \ifspace \ug_j=\bigl(g_{j1},\ldots,g_{jk_j}\bigr) \in \prod\limits_{i=1}^{k_j} \C(b_{ji},c_j)\] and \[\sigma(\tau_1,\ldots,\tau_n) = \underbrace{\sigma\langle k_1,\ldots,k_n\rangle}_{\text{block permutation}} \circ~ \underbrace{(\tau_1\oplus\cdots\oplus\tau_n)}_{\text{block sum}} \in \Sigma_{k_1+\cdots+k_n}\]
as in \eqref{as-comp}.  

There is a canonical isomorphism \[\nicexy{\algmocm \ar[r]^-{\cong} & \Monmc}\] defined as follows.  Each $\Ocm$-algebra $(X,\lambda)$ has a restricted structure morphism \[\nicexy@C+.7cm{X_{\uc} \ar[r]^-{\lambda_{(\sigma,\uf)}} \ar[d]_-{(\sigma,\uf)~\mathrm{inclusion}} & X_d\\ \coprod\limits_{\Sigma_n\times\prod\limits_{j=1}^n \C(c_j,d)} X_{\uc} \ar[r]^-{\cong} & \Ocm\duc \otimes X_{\uc} \ar[u]_-{\lambda}}\] for each $(\sigma,\uf) \in \Sigma_n \times \prod_{j=1}^n \C(c_j,d)$.  For a morphism $f : c \to d \in \C$, there is a restricted structure morphism \[\nicexy@C+.5cm{X_c \ar[r]^-{\lambda_{(\id_1,f)}} & X_d}\in \M.\]  The associativity and unity axioms of $(X,\lambda)$ imply that this is a $\C$-diagram in $\M$.  

For each $c \in \C$, the restricted structure morphisms \[\nicexy@C+1.2cm{X_c\otimes X_c \ar[r]^-{\lambda_{(\id_2,\{\Id_c,\Id_c\})}} & X_c} \andspace \nicexy@C+.4cm{\tensorunit \ar[r]^-{\lambda_{(\id_0,*)}} & X_c}\] give $X_c$ the structure of a monoid in $\M$, once again by the associativity and unity axioms of $(X,\lambda)$.  One can check that this gives a $\C$-diagram of monoids in $\M$; i.e., the morphisms $\lambda_{(\id_1,f)}$ are compatible with the entrywise monoid structures.  In summary, $\Ocm$ is the $\colorc$-colored operad whose algebras are $\C$-diagrams of monoids in $\M$.  This identification is also given in \cite{bsw} Theorem 4.26.\dqed
\end{example}

\begin{example}[Operad for diagrams of commutative monoids]\label{ex:diag-com-operad}
This example is a combination of Examples \ref{ex:operad-com} and \ref{ex:operad-diag} and is a slight modification of Example \ref{ex:diag-monoid-operad}.  Suppose $\C$ is a small category with object set $\colorc$.  There is a $\colorc$-colored \index{commutative monoid!operad for diagrams of}\index{colored operad!for diagrams of commutative monoids}operad \label{notation:comc}$\Comc$ in $\M$ with entries \[\Comc\duc = \coprodover{\prod\limits_{j=1}^n\C(c_j,d)} \tensorunit \forspace \duc=\dconecn \in \Profcc.\]  Its operad structure is defined as in Example \ref{ex:diag-monoid-operad} by ignoring the first component.  

Moreover, with almost the same argument as in Example \ref{ex:diag-monoid-operad}, one can check that there is a canonical isomorphism \[\nicexy{\algmcomc \ar[r]^-{\cong} & \Commc}.\]  To see that the monoid multiplication \[\nicexy@C+1.2cm{X_c\otimes X_c \ar[r]^-{\mu_c~=~\lambda_{\{\Id_c,\Id_c\}}} & X_c} \in \M\] is commutative, observe that the pair $\{\Id_c,\Id_c\}$ is fixed by the permutation $(1~2)$.  So the equivariance axiom \eqref{operad-algebra-eq} implies that $\mu_c$ is commutative.  In summary, $\Comc$ is the $\colorc$-colored operad whose algebras are $\C$-diagrams of commutative monoids in $\M$.\dqed  
\end{example}

\chapter{Constructions on Operads}\label{ch:operad-construction}

In this chapter, we discuss several important constructions and properties of colored operads.  In Section \ref{sec:change-operad} to Section \ref{sec:change-base}, we discuss the category of algebras over a colored operad under a change of operads and a change of base categories.  In Section \ref{sec:localization-operad} and Section \ref{sec:algebra-local} we study localizations of colored operads, analogous to localizations of categories, and algebras over localized operads.  The material in the last two sections about localizations of operads is new.  Localizations of operads are needed later when we discuss the time-slice axiom in prefactorization algebras.

As in the previous chapter, $(\M,\otimes,\tensorunit)$ is a cocomplete symmetric monoidal closed category with an initial object $\varnothing$.  

\section{Change-of-Operad Adjunctions}\label{sec:change-operad}

In this section, we consider the category of algebras over an operad under an operad morphism.  Instead of restricting ourselves to operads with the same color set, we will need to consider morphisms between operads with different color sets.  So we first consider operads under a change of colors.

\begin{definition}\label{def:operad-change-color}
Suppose $(\O,\gammao)$ is a $\colorc$-colored operad in $\M$ in the sense of Definition \ref{def:operad-tree}, and $f : \colorb \to \colorc$ is a map of non-empty sets.
\begin{enumerate}
\item Define the object\label{notation:fstaro} $\fstaro \in \M^{\Profbb}$ by \[(\fstaro)\duc = \O\fdufc\] for $\bigl(\uc=(c_1,\ldots,c_m);d\bigr) \in \Profbb$, where $f\uc=(fc_1,\ldots,fc_m)$.
\item For each $\colorb$-colored tree $T$, define $fT$ as the $\colorc$-colored tree obtained from $T$ by applying $f$ to its $\colorb$-coloring.
\item For each $\colorb$-colored tree $T$, define the morphism $\gammafstaro_{T}$ by the commutative diagram
\[\nicexy{(\fstaro)[T] \ar[r]^-{\gammafstaro_T} \ar[d]_-{\Id} & (\fstaro)\bigl(\profoft\bigr)\\
\bigtensorover{v\in T} \O\finoutv = \O[fT] \ar[r]^-{\gammao_{fT}} & \O\bigl[\Prof(fT)\bigr]. \ar[u]_-{\Id}}\]
\end{enumerate}\end{definition}

\begin{proposition}\label{prop:operad-change-color}
Suppose $(\O,\gammao)$ is a $\colorc$-colored operad in $\M$, and $f : \colorb \to \colorc$ is a map of non-empty sets.  Then $\bigl(\fstaro,\gammafstaro\bigr)$ is \index{colored operad!change-of-color}a $\colorb$-colored operad in $\M$.
\end{proposition}

\begin{proof}
Observe that:
\begin{enumerate} \item $f\Cor_{(\uc;d)} = \Cor_{(f\uc;fd)}$ for each $(\uc;d) \in \Profbb$.
\item $f\bigl(T(H_v)_{v\in T}\bigr) = (fT)(fH_u)_{u \in fT}$ for each tree substitution $T(H_v)$ in $\uTreeb$.\end{enumerate}
Since $\gammafstaro_T = \gammao_{fT}$, the assertion follows from Definition \ref{def:operad-tree}.
\end{proof}

\begin{definition}\label{def:general-operad-map}
Suppose $\O$ is a $\colorc$-colored operad in $\M$, and $\P$ is a $\colord$-colored operad in $\M$.  An \index{operad morphism}\index{colored operad!morphism}\emph{operad morphism} $f : \O \to \P$ is a pair $(f_0,f_1)$ consisting of
\begin{itemize}\item a map $f_0 : \colorc \to \colord$ and
\item a morphism $f_1 : \O \to \fstar_0\P$ of $\colorc$-colored operads.\end{itemize}
The category of all colored operads in $\M$ is denoted by $\Operadm$.  We will sometimes abbreviate both $f_0$ and $f_1$ to $f$.
\end{definition}

Unpacking the definition we can express an operad morphism more explicitly as follows.

\begin{proposition}\label{prop:operad-map}
Suppose $(\O,\gammao)$ is a $\colorc$-colored operad, and $(\P,\gammap)$ is a $\colord$-colored operad in $\M$.  Then an operad morphism $f : \O \to \P$ consist of precisely
\begin{itemize}\item a map $f_0 : \colorc \to \colord$ and 
\item a morphism \[f_1 : \O\duc \to \P\fzerodufzeroc \in \M\] for each $(\uc;d) \in \Profcc$
\end{itemize}
such that the diagram
\[\nicexy{\O[T]\ar[d]_-{\gammao_T} \ar[r]^-{\bigtensorover{v\in T} f_1} & \P[f_0T] \ar[d]^-{\gammap_{f_0T}}\\ \O\bigl[\profoft] \ar[r]^-{f_1} & \P\bigl[\Prof(f_0T)\bigr]}\]
is commutative for each $\colorc$-colored tree $T$.
\end{proposition}

\begin{definition}\label{def:pullback-algebra}
Suppose $f : \O \to \P$ is an operad morphism with $\O$ a $\colorc$-colored operad and $\P$ a $\colord$-colored operad in $\M$.
\begin{enumerate}
\item For $X \in \Mtod$, define the object $\fstar X \in \Mtoc$ by \[(\fstar X)_c = X_{fc} \forspace c \in \colorc.\]
\item For a $\P$-algebra $(X,\theta)$, define the morphism $\theta^{\fstar X}$ as the composition in the diagram
\[\nicexy{\O\duc \otimes (\fstar X)_{\uc} \ar[r]^-{\theta^{\fstar X}} \ar[d]_-{(f,\Id)} & (\fstar X)_d\\ \P\fdufc \otimes X_{f\uc} \ar[r]^-{\theta} & X_{fd} \ar[u]_-{\Id}}\]
for each $(\uc;d) \in \Profcc$.
\end{enumerate}\end{definition}

A direct inspection of Definition \ref{def:operad-algebra-generating} yields the following result.

\begin{proposition}\label{prop:change-operad-pullback}
In the context of Definition \ref{def:pullback-algebra}:
\begin{enumerate}\item $\bigl(\fstar X, \theta^{\fstar X}\bigr)$ is an $\O$-algebra.
\item $\fstar$ defines \index{colored operadic algebra!pullback}functors \[\fstar : \Mtod \to \Mtoc \andspace \fstar : \algmp \to \algmo.\]
\end{enumerate}\end{proposition}

Before we discuss the adjunction associated to an operad morphism, let us first consider the following special case on underlying objects.

\begin{lemma}\label{lem:change-colored-object}
Suppose $f : \colorc \to \colord$ is a map of non-empty sets.  Then there is an adjunction
\[\nicexy{\Mtoc \ar@<2pt>[r]^-{f_!} & \Mtod \ar@<2pt>[l]^-{\fstar}}\]
with left adjoint $f_!$.\end{lemma}

\begin{proof} For $Y \in \Mtoc$, define an object $f_!Y \in \Mtod$ by \[(f_!Y)_d = \coprodover{c \in \finverse(d)} Y_c \forspace d \in \colord.\]  One checks directly that this defines a functor $f_!$ that is a left adjoint of $\fstar$.\end{proof}

\begin{theorem}\label{thm:change-operad}
Suppose $f : \O \to \P$ is an operad morphism with $\O$ a $\colorc$-colored operad and $\P$ a $\colord$-colored operad in $\M$.  Then there is an adjunction
\[\nicexy{\algmo \ar@<2pt>[r]^-{f_!} & \algmp \ar@<2pt>[l]^-{\fstar}},\]
called the \index{adjunction!change-of-operad}\index{change-of-operad adjunction}\emph{change-of-operad adjunction}, with left adjoint $f_!$.
\end{theorem}

\begin{proof}
Consider the solid-arrow diagram
\[\nicexy@C+.4cm{\algmo \ar@<2pt>@{.>}[r]^-{f_!} \ar@<1pt>[d]^-{U} & \algmp \ar@<2pt>[l]^-{\fstar} \ar@<1pt>[d]^-{U}\\ \Mtoc \ar@<3pt>[u]^-{\O\comp -}\ar@<2pt>[r]^-{f_!} & \Mtod\ar@<2pt>[l]^-{\fstar} \ar@<3pt>[u]^-{\P\comp  -}}\]
with the bottom adjunction from Lemma \ref{lem:change-colored-object} and the vertical adjunctions from  Proposition \ref{prop:algebra-bicomplete}.  There is an equality \[U\fstar = \fstar U : \algmp \to \Mtoc.\] Since the bottom horizontal functor $\fstar$ admits a left adjoint $f_!$ and since $\algmp$ is cocomplete, the Adjoint Lifting Theorem \cite{bor2} (Theorem 4.5.6) implies that the top horizontal functor $\fstar$ also admits a left adjoint.
\end{proof}

\begin{example}[Free-Forgetful Adjunction]For a $\colorc$-colored operad $\O$, the natural morphism $i : \I \to \O$, where $\I$ is the $\colorc$-colored unit operad in \eqref{unit-operad}, is an operad morphism.  In this case, the change-of-operad adjunction $i_! \dashv i^*$ is the \index{free-forgetful adjunction}\index{adjunction!free-forgetful}free-forgetful adjunction \[\nicexy{\Mtoc \ar@<2pt>[r]^-{\O\comp -} & \algmo \ar@<2pt>[l]^-{U}}\] in Proposition \ref{prop:algebra-bicomplete}.\dqed\end{example}

\begin{example}\label{ex:as-to-com}
There is an operad \index{associative operad!to commutative operad}\index{commutative operad!from associative operad}morphism $f : \As \to \Com$ from the associative operad in Example \ref{ex:operad-as} to the commutative operad in Example \ref{ex:operad-com}, given entrywise by the morphism \[\As(n) = \coprodover{\sigma \in \Sigma_n} \tensorunit \to \tensorunit = \Com(n)\] whose restriction to every copy of $\tensorunit$ in $\As(n)$ is the identity morphism.  In the change-of-operad adjunction \[\nicexy{\Monm=\algm(\As) \ar@<2pt>[r]^-{f_!} & \algm(\Com) = \Comm \ar@<2pt>[l]^-{\fstar}}\] the right adjoint $\fstar$ forgets about the commutativity of a commutative monoid.  The left adjoint $f_!$ sends a monoid to the commutative monoid generated by it.  For instance, if $\M=\Vectk$ and if $A$ is a monoid in $\M$ (i.e., a $\fieldk$-algebra), then $f_!A$ is the quotient $\fieldk$-algebra of $A$ by the ideal generated by all the \index{commutator}commutators $[a,b]=ab-ba$ with $a,b \in A$.\dqed
\end{example}

\begin{example}[Change-of-Monoids]\label{ex:change-of-monoid}
Suppose $f : A \to B$ is a morphism of monoids in $\M$.  Regarding $A$ and $B$ as $1$-colored operads concentrated in unary entries as in Example \ref{ex:monoid-unary-operad}, we can think of $f$ as an operad morphism.  With $A$ regarded as an operad, $A$-algebras are precisely $A$-modules\index{module!change-of-monoid}\index{adjunction!change-of-monoid} in the sense of Example \ref{ex:monoid-module}, and similarly for $B$-algebras.  In the change-of-operad adjunction \[\nicexy{\algm(A) \ar@<2pt>[r]^-{f_!} & \algm(B) \ar@<2pt>[l]^-{\fstar}}\] the right adjoint $\fstar$ is the restriction of the structure morphism to $A$.  If $\M=\Vectk$, then the left adjoint $f_!$ sends an $A$-module $(X,\theta)$ to the $B$-module $B \otimes_A X$, where the right $A$-action on $B$ is induced by $f$.\dqed
\end{example}

\begin{example}[Left Kan Extensions]\label{ex:change-of-diagram}
Suppose $F : \C \to \D$ is a functor between small categories with $\Obc=\colorc$ and $\Obd=\colord$.  Recall from Example \ref{ex:operad-diag} that there is a $\colorc$-colored operad $\Cdiag$ whose algebras are precisely $\C$-diagrams in $\M$.  There is an operad morphism \[\Fdiag : \Cdiag \to \Ddiag\] whose function on color sets $\colorc \to \colord$ is the object function of the functor $F$.  For a pair of objects $c,d \in \colorc$, the morphism \[\Cdiag\dc = \coprodover{f \in \C(c,d)} \tensorunit \to \coprodover{g \in \D(Fc,Fd)} \tensorunit=\Ddiag\FdFc\] identifies the copy of $\tensorunit$ in $\Cdiag\dc$ corresponding to $f \in \C(c,d)$ with the copy of $\tensorunit$ in $\Ddiag\FdFc$ corresponding to $Ff \in \D(Fc,Fd)$.  In the change-of-operad adjunction \[\nicexy{\Fun(\C,\M)=\algm(\Cdiag) \ar@<2pt>[r]^-{\Fdiag_!} & \algm(\Ddiag)=\Fun(\D,\M) \ar@<2pt>[l]^-{\Fdiagstar}}\] the right adjoint $\Fdiagstar = \Fun(F,\M)$ sends a $\D$-diagram $G : \D \to \M$ to the $\C$-diagram $GF : \C \to \M$.  For a $\C$-diagram $H : \C \to \M$, $\Fdiag_!H$ is the left \index{Kan extension}Kan extension of $H$ along $F$ in Theorem \ref{thm:left-kan-exists}.\dqed
\end{example}

\section{Model Category Structures}\label{sec:homotopical-algebra}

\subsection{Model Categories}

Before we discuss model category structures on $\algmo$ for an operad $\O$, let us first review some basic concepts of model categories, which were originally defined by\index{Quillen, D.} Quillen \cite{quillen}.  The reader is referred to the references \cite{hirschhorn,hovey,may-ponto,schwede-shipley} for more details.

The formulation of a model category below is due to \cite{may-ponto}.  Suppose $f : A \to B$ and $g : C \to D$ are morphisms in a category $\M$.  We write \index{left lifting property}\index{right lifting property}\label{notation:fboxslashg}$f \boxslash g$ if for each solid-arrow commutative diagram \[\nicexy{A \ar[d]_-{f} \ar[r] & C \ar[d]^-{g}\\ B \ar[r] \ar@{.>}[ur] & D}\] in $\M$, a dotted arrow exists that makes the entire diagram commutative.  For a class $\cala$ of morphisms in $\M$, define the classes of morphisms\label{notation:boxslasha}
\[\begin{split}
^{\boxslash}\!\cala &= \bigl\{f \in \M ~\vert~ f \boxslash a \text{ for all } a \in \cala\bigr\},\\
\cala^{\boxslash} &= \bigl\{g \in \M ~\vert~ a \boxslash g \text{ for all } a \in \cala\bigr\}.
\end{split}\]

A pair $(\call,\calr)$ of classes of morphisms in $\M$ \index{functorially factors}\emph{functorially factors} $\M$ if each morphism $h$ in $\M$ has a functorial factorization $h=gf$ such that $f \in \call$ and $g \in \calr$.  A \index{weak factorization system}\emph{weak factorization system} in a category $\M$ is a pair $(\call,\calr)$ of classes of morphisms in $\M$ such that (i) $(\call,\calr)$ functorially factors $\M$, (ii) $\call = \> {}^{\boxslash}\!\calr$, and (iii) $\calr = \call^{\boxslash}$.

A \index{model category}\emph{model category} is a complete and cocomplete category $\M$ equipped with three classes of morphisms\label{notation:wcf} $(\calw, \calc,\calf)$, called \index{weak equivalence}\emph{weak equivalences}, \index{cofibration}\emph{cofibrations}, and \index{fibration}\emph{fibrations}, such that:
\begin{itemize} \item $\calw$ has the \index{2of3@$2$-out-of-$3$ property}$2$-out-of-$3$ property.  In other words, for any morphisms $f$ and $g$ in $\M$ such that the composition $gf$ is defined, if any two of the three morphisms $f$, $g$, and $gf$ are in $\calw$, then so is the third.
\item $(\calc, \calf \cap \calw)$ and $(\calc \cap \calw,\calf)$ are weak factorization systems.\end{itemize}
For a model category $(\M,\calw,\calc,\calf)$, its \index{homotopy category}\index{category!homotopy}\emph{homotopy category}\label{notation:homotopy-cat} $\Ho(\M)$ is a \index{localization}$\calw$-localization of $\M$ as in Definition \ref{def:localization-cat}.  For a model category $\M$, its homotopy category always exists.

A model category $\M$ is:
\begin{enumerate}\item \emph{left proper}\index{left proper} if weak equivalences are closed under pushouts along cofibrations.
\item \index{cofibrantly generated model category}\emph{cofibrantly generated} if (i) it is equipped with two sets $\cali$ and $\calj$ of morphisms that permit the small object argument \cite{hirschhorn} (Definition 10.5.15), (ii) $\calf = \calj^{\boxslash}$, and (iii) $\calf \cap \calw = \cali^{\boxslash}$.
\item a \index{monoidal model category}\emph{monoidal model category} \cite{schwede-shipley} (Definition 3.1) if it is also a symmetric monoidal closed category that satisfies the following \index{pushout product axiom}\emph{pushout product axiom}:
\begin{quote} Given cofibrations $f : A \to B$ and $g : C \to D$, the pushout product $f \square g$ in the diagram \[\nicexy{A \otimes C \ar@{}[dr]|-{\mathrm{pushout}} \ar[r]^-{\Id_A \otimes g} \ar[d]_-{f \otimes \Id_C} & A \otimes D \ar[d] \ar[ddr]^-{f \otimes \Id_D}\\ B \otimes C \ar[r] \ar[drr]_-{\Id_B \otimes g} & Z \ar[dr]|-{f \square g} &\\ && B \otimes D}\] is a cofibration, which is also a weak equivalence if either $f$ or $g$ is also a weak equivalence.  Here \[Z = B \otimes C \coprodover{A \otimes C} A \otimes D\] is the object of the pushout square.\end{quote}
\end{enumerate}
In \cite{hovey} (Definition 4.2.6), a monoidal model category has an extra condition about the monoidal unit, which we do not need in this work.
 
\begin{example}\label{ex:model-cate}
Here are some basic examples of cofibrantly generated monoidal model categories.
\begin{enumerate}\item $\Top$ \index{topological space!model structure}is a cofibrantly generated monoidal model category \cite{hovey} (Section 2.4) in which a weak equivalence is a weak homotopy equivalence, i.e., a map that induces isomorphisms on all homotopy groups for all choices of base points in the domain.  A fibration is a Serre fibration. 
\item The category $\Sset$ of \index{simplicial set!model structure}simplicial sets is a left proper, cofibrantly generated monoidal model category \cite{hovey} (Chapter 3) in which a weak equivalence is a map whose geometric realization is a weak homotopy equivalence.  A cofibration is an injection.
\item For a field $\fieldk$, the category $\Chaink$ is a \index{chain complex!model structure}left proper, cofibrantly generated monoidal  model category \cite{quillen} with quasi-isomorphisms as weak equivalences, dimension-wise injections as cofibrations, and dimension-wise surjections as fibrations.  The homotopy category of $\Chaink$ is the derived category of chain complexes of $\fieldk$-vector spaces. 
\item The category $\Cat$ of small \index{small category!model structure}categories is a left proper, cofibrantly generated monoidal model category \cite{rezk}, called the \index{folk model structure}\emph{folk model structure}.  A weak equivalence is an equivalence of categories, i.e., a functor that is full, faithful, and essentially surjective.  A cofibration is a functor that is injective on objects.\dqed
\end{enumerate}\end{example}
 
\begin{example}\label{ex:cofgen-diagram}
For a cofibrantly generated model category $\M$ and a small category $\D$, the category $\M^{\D}$ of $\D$-diagrams in $\M$ inherits from $\M$ a cofibrantly generated model category structure \cite{hirschhorn} (11.6.1) with fibrations and weak equivalences defined entrywise in $\M$.  For instance, if $\D = \Sigmacopc$, then the \index{symmetric sequence!model structure}category \[\symseqcm = \M^{\Sigmacopc}\] of $\colorc$-colored symmetric sequences in $\M$ is a model category with weak equivalences and fibrations defined entrywise in $\M$.\dqed\end{example}

In a model category, an\index{acyclic fibration}\index{acyclic cofibration} \emph{acyclic (co)fibration} is a morphism that is both a (co)fibration and a weak equivalence. An object $Z$ is \index{fibrant}\emph{fibrant} if the unique morphism from $Z$ to the terminal object is a fibration.    A \index{fibrant replacement}\emph{fibrant replacement} of an object $X$ is a weak equivalence $X\to Z$ such that $Z$ is fibrant.  There is a functorial fibrant replacement $R$ given by applying the functorial factorization of the weak factorization system $(\calc\cap\calw,\calf)$ to the unique morphism to the terminal object.  An object $Y$ is \index{cofibrant}\emph{cofibrant} if the unique morphism from the initial object to $Y$ is a cofibration.  A \index{cofibrant replacement}\emph{cofibrant replacement} of an object $X$ is a weak equivalence $Y\to X$ such that $Y$ is cofibrant.  There is a functorial cofibrant replacement $Q$ given by applying the functorial factorization of the weak factorization system $(\calc,\calf\cap\calw)$ to the unique morphism from the initial object.

Suppose $F : \M \adjoint \N : U$ is an adjunction between model categories with left adjoint $F$.  Then $(F,U)$ is called a \index{Quillen adjunction}\emph{Quillen adjunction} if $U$ preserves fibrations and acyclic fibrations.  The \index{total left derived functor}\emph{total left derived functor}\label{notation:left-derived} \[\Lder F : \Ho(\M) \to \Ho(\N)\] is defined as the composition \[\nicexy{\Ho(\M) \ar[r]^-{\Ho(Q)} & \Ho(\M) \ar[r]^-{\Ho(F)} & \Ho(\N)}\] in which $Q$ is the functorial cofibrant replacement in $\M$.  The \index{total right derived functor}\emph{total right derived functor}\label{notation:right-derived} \[\R U : \Ho(\N) \to \Ho(\M)\] is defined as the composition \[\nicexy{\Ho(\N) \ar[r]^-{\Ho(R)} & \Ho(\N) \ar[r]^-{\Ho(U)} & \Ho(\M)}\]  in which $R$ is the functorial fibrant replacement in $\N$.  For a Quillen adjunction $(F,U)$, there is a \emph{derived adjunction}\index{derived adjunction}\index{adjunction!derived} \[\nicexy{\Ho(\M) \ar@<2pt>[r]^-{\Lder F} & \Ho(\N) \ar@<2pt>[l]^-{\R U}}\] between the homotopy categories with left adjoint $\Lder F$.

A Quillen adjunction $(F,U)$ is called a \index{Quillen equivalence}\emph{Quillen equivalence} if for each morphism $f : FX \to Y$ with $X \in \M$ cofibrant and $Y \in \N$ fibrant, $f$ is a weak equivalence in $\N$ if and only if its adjoint $X \to UY$ is a weak equivalence in $\M$.  For a Quillen equivalence, the derived adjunction is an adjoint equivalence between the homotopy categories.  

\begin{interpretation} The total left derived functor of $F$ is first a cofibrant replacement in the domain of $F$ and then $F$ itself.  The total right derived functor of $U$ is first a fibrant replacement in the domain of $U$ and then $U$ itself.  For a Quillen equivalence, the two model categories become adjoint equivalent via the derived adjunction after their weak equivalences are inverted.  We say that their homotopy theories are equivalent.\dqed
\end{interpretation}

\begin{example} The adjunction \[\nicexy{\Sset \ar@<2pt>[r]^-{|-|} & \Top \ar@<2pt>[l]^-{\Sing}}\] involving the geometric realization \index{geometric realization}functor and the singular simplicial set \index{singular simplicial set}functor is a Quillen equivalence.\dqed\end{example}

\subsection{Model Structure on Algebra Categories}

For more in-depth discussion of model category structure on the category of algebras over a colored operad, the reader may consult \cite{batanin-berger,berger-moerdijk-resolution,fresse-book,white-yau}.

\begin{definition}\label{def:admissibility}
Suppose $\M$ is a monoidal model category, and $\O$ is a $\colorc$-colored operad in $\M$.  
\begin{enumerate}
\item $\O$ is \emph{admissible}\index{admissible}\index{colored operad!admissibility} if $\algmo$ admits a model category structure in which a morphism $f=\{f_c\}_{c\in\colorc}\in\Mtoc$ is:
\begin{itemize}
\item a weak equivalence if and only if $f_c$ is a weak equivalence in $\M$ for each $c \in \colorc$;
\item a fibration if and only if $f_c$ is a fibration in $\M$ for each $c \in \colorc$;
\item a cofibration if and only if $f\boxslash g$ for all morphisms $g \in \algmo$ that are both weak equivalences and fibrations.
\end{itemize}
\item $\O$ is \emph{well-pointed}\index{well-pointed}\index{colored operad!well-pointed} if the $c$-colored unit $\operadunit_c : \tensorunit \to \O\cc$ is a cofibration for each $c \in \C$.
\item Suppose $\M$ is also cofibrantly generated.  The operad $\O$ is called \emph{$\Sigma$-cofibrant}\index{sigmacofibrant@$\Sigma$-cofibrant}\index{colored operad!sigmacofibrant@$\Sigma$-cofibrant} if its underlying $\colorc$-colored symmetric sequence is a cofibrant object in $\symseqcm$.
\item A morphism $f : \O \to \P$ of $\colorc$-colored operads in $\M$ is a \emph{weak equivalence}\index{weak equivalence}\index{colored operad!weak equivalence} if each entry of $f$ is a weak equivalence in $\M$.
\end{enumerate}
\end{definition}

\begin{example}\label{ex:all-admissible}
In the model categories $\Sset$, $\Chaink$ where $\fieldk$ has characteristic $0$, $\Cat$, and $\Top$, every colored operad is admissible.  The proofs for $\Sset$ and $\Chaink$ are in \cite{white-yau} (Section 8).  The method of proof for the $\Sset$ case also works for $\Cat$.  For $\Top$ and many other model categories, the admissibility of all colored operads is proved in \cite{batanin-berger,berger-moerdijk-resolution}.  Furthermore, in $\Sset$, $\Chaink$, and $\Cat$, every colored operad is well-pointed.  In $\Chaink$ every colored operad is $\Sigma$-cofibrant, which is a consequence of \index{Maschke's Theorem}Maschke's Theorem.\dqed\end{example}

The following comparison result is \cite{berger-moerdijk-resolution} Theorem 4.1 in the general colored case and \cite{berger-moerdijk-axiomatic} Theorem 4.4 in the one-colored case.

\begin{theorem}\label{thm:operad-comparison}
Suppose $\M$ is a monoidal model category, and $f : \O \to \P$ is a morphism between admissible $\colorc$-colored operads in $\M$.
\begin{enumerate}
\item The \index{change-of-operad adjunction}\index{adjunction!change-of-operad}change-of-operad adjunction $f_! \dashv \fstar$ in Theorem \ref{thm:change-operad} is a Quillen adjunction.
\item Suppose in addition that:
\begin{enumerate}\item $\M$ is left proper and cofibrantly generated with $\tensorunit$ cofibrant.
\item $f$ is a weak equivalence between well-pointed and $\Sigma$-cofibrant $\colorc$-colored operads.  
\end{enumerate}
Then the change-of-operad adjunction is a Quillen equivalence.
\end{enumerate}\end{theorem}

\begin{example}\label{ex:chain-operad-comparison}
If $f : \O \to \P$ is a weak equivalence of $\colorc$-colored operads in $\Chaink$, where $\fieldk$ has characteristic $0$, then the change-of-operad adjunction $f_! \dashv \fstar$ is a Quillen equivalence.\dqed\end{example}

\begin{remark}In \cite{white-yau-halt} there are general results extending the homotopical change-of-operad adjunction in Theorem \ref{thm:operad-comparison} to situations where $\O$ and $\P$ are colored operads in different monoidal model categories.  We will not need those results in this book, so we refer the interested reader to \cite{white-yau-halt}.\dqed\end{remark}

\section{Changing the Base Categories}\label{sec:change-base}

We will later need to consider operads transferred from one category to another category.  The following result is a special case of \cite{bluemonster} Theorem 12.11 and Corollary 12.13.

\begin{theorem}\label{thm:operad-transfer}
Suppose $F : \M \to \N$ is a symmetric monoidal functor between symmetric monoidal closed categories.  
\begin{enumerate}
\item $F$ prolongs to a \index{change-of-category functor}\index{colored operad!change-of-category}functor \[F_* : \Operadcm \to \Operadcn\] for every non-empty set $\colorc$.
\item Suppose $F$ admits a right adjoint $G$ that is also a symmetric monoidal functor.  Then the prolonged functors \[\nicexy{\Operadcm \ar@<2pt>[r]^-{F_*} & \Operadcn \ar@<2pt>[l]^-{G_*}}\] form an adjunction.
\end{enumerate}\end{theorem}

\begin{proof}
Let us describe the operad structure of $\Fstaro$ for a $\colorc$-colored operad $(\O,\gamma)$ in $\M$. For a pair $(\uc;d) \in \Profcc$, we define \[(\Fstaro)\duc = F\O\duc \in \N.\]  For each $\colorc$-colored tree $T \in \uTreec\duc$, we define the operadic structure morphism $\gamma^{\Fstaro}_T$ as the composition in the diagram
\[\nicexy{(\Fstaro)[T] \ar[r]^-{\gamma^{\Fstaro}_T} \ar[d]_-{\Id} & (\Fstaro)\duc\\
\bigtensorover{v\in T} F\O(v) \ar[r]^-{F_2} & F\bigl(\O[T]\bigr) \ar[u]_-{F\gamma_T}}\]
in which $\O(v)=\O\inoutv$.  The bottom horizontal morphism is an iteration of the monoidal structure of $F$.  One can now check that $(\Fstaro,\gamma^{\Fstaro})$ is a $\colorc$-colored operad in $\N$ in the sense of Definition \ref{def:operad-tree}.
\end{proof}

\begin{notation} In the setting of Theorem \ref{thm:operad-transfer}, for a $\colorc$-colored operad $\O$ in $\M$, we will often write the image $\Fstaro$ as\label{notation:oton} $\Oton$.\end{notation}

\begin{example}[Set operads to enriched operads]\label{ex:operad-set-m}
There is an adjunction of symmetric monoidal functors\index{set operad}\index{colored operad!enriched} \[\nicexy{\Set \ar@<2pt>[r]^-{\coprodover{(-)} \tensorunit} & \M\ar@<2pt>[l]^-{\M(\tensorunit,-)}}\] in which the left adjoint is \index{strong symmetric monoidal functor}strong symmetric monoidal.  For a $\colorc$-colored operad $\O$ in $\Set$, its image in $\M$ has entries \[\Otom\duc = \coprodover{\O\duc} \tensorunit\] for $(\uc;d) \in \Profcc$.\dqed\end{example}

\begin{example}[Change-of-rings]\label{ex:changeofring}
For a map $f : A \to B$ of associative and commutative rings, there is an adjunction of symmetric monoidal \index{adjunction!change-of-ring}functors \[\nicexy{\Chain_A \ar@<2pt>[r]^-{f_!} & \Chain_B\ar@<2pt>[l]^-{\Res}}\] with the left adjoint $f_!=-\otimes_A B$ and the right adjoint induced by restriction of scalars along $f$. For a $\colorc$-colored operad $\O$ in $\Chain_A$, its image $\O^B$ in $\Chain_B$ has entries \[\O^{B}\duc = \O\duc \otimes_A B\] for $(\uc;d) \in \Profcc$.\dqed\end{example}

\begin{example}Both the geometric realization functor \[|-| : \Sset \to \Top\] and its right adjoint, the singular simplicial set functor, are symmetric monoidal \cite{hovey} (Proposition 4.2.17).\dqed\end{example}

\section{Localizations of Operads}\label{sec:localization-operad}

In this section, we define the operad analogue of localizations of categories in Section \ref{sec:localization}.  For a category $\C$ and a set $S$ of morphisms in $\C$, recall that the $S$-localization $\Csinv$ is the category obtained from $\C$ by adjoining formal inverses $\finverse$ for $f \in S$.  For an operad $\O$ in $\Set$ and a set $S$ of unary elements, we will show in this section that there is an analogous $S$-localization $\Osinv$ in which formal inverses $\sinv$ for $s\in S$ are added.  The importance of this construction is what it does on algebras.  In Section \ref{sec:algebra-local} we will show that $\Osinv$-algebras are precisely the $\O$-algebras in which the structure morphisms corresponding to elements in $S$ are isomorphisms.

\begin{motivation} We will need localized operads later when we discuss the time-slice axiom in prefactorization algebras from an operad viewpoint.  The time-slice axiom is an invertibility condition that says that certain structure morphisms are isomorphisms.  Applied to the colored operad for prefactorization algebras, we will see that algebras over the localized operad have the desired invertible structure morphisms.\dqed
\end{motivation}

As before $\colorc$ is an arbitrary but fixed non-empty set.  We will use Definition \ref{def:operad-generating} of a $\colorc$-colored operad below.

\begin{definition}\label{def:unary-element}
Suppose $(\O,\gamma,\operadunit)$ is a $\colorc$-colored operad in $\Set$.  
\begin{enumerate}\item Elements in $\O\dc$ for $c,d\in \colorc$ are called \index{unary element}\index{colored operad!unary element}\emph{unary elements}.
\item A unary element $x \in \O\dc$ is said to be \index{invertible}\emph{invertible} if there exists a unary element $y \in \O\cd$, called an \emph{inverse of $x$}, such that 
\[\begin{split} 
\gamma(y;x) &= \operadunit_c \forspace \nicexy{\O\cd\times\O\dc\ar[r]^-{\gamma} & \O\cc},\\
\gamma(x;y) &= \operadunit_d \forspace \nicexy{\O\dc\times\O\cd\ar[r]^-{\gamma} & \O\dd}.\end{split}\]
Since an inverse of a unary element $x$ is unique if it exists, we will write it as $\xinv$.
\end{enumerate}
\end{definition}

The next definition is the operad version of a localization of a category in Definition \ref{def:localization-cat}.

\begin{definition}\label{def:operad-localization}
Suppose $\O$ is a $\colorc$-colored operad in $\Set$, and $S$ is a set of unary elements in $\O$.  An \index{colored operad!localization}\index{localization of an operad}\emph{$S$-localization of $\O$}, if it exists, is a pair $(\Osinv,\ell)$ consisting of
\begin{itemize}\item a $\colorc$-colored operad $\Osinv$ in $\Set$ and
\item a morphism of $\colorc$-colored operads $\ell : \O \to \Osinv$ 
\end{itemize}
that satisfies the following two properties.
\begin{enumerate}
\item $\ell(s)$ is invertible for each $s \in S$.
\item $(\Osinv,\ell)$ is initial with respect to the previous property: If $f : \O \to \P$ is an operad morphism such that $f(s)$ is invertible for each $s \in S$, then there exists a unique operad morphism \[f' : \Osinv \to \P \stspace f= f'\ell.\]
\begin{equation}\label{operad-localization-triangle}
\nicexy{& \O \ar[r]^-{\ell} \ar[d]_-{\forall\, f} & \Osinv \ar@{.>}[dl]^-{\exists !\, f'}\\ 
f(S) ~\mathrm{invertible} & \P &}
\end{equation}
\end{enumerate}
In this setting, $\ell$ is called the \index{localization morphism}\emph{$S$-localization morphism}.
\end{definition}

\begin{remark} In Definition \ref{def:operad-localization}:
\begin{enumerate}\item By the universal property \eqref{operad-localization-triangle}, an $S$-localization of $\O$, if it exists, is unique up to a unique isomorphism.  
\item We may assume that $S$ is closed under operadic composition.  Indeed, if $x$ and $y$ are invertible elements, then $z=\gamma(y;x)$, if it is defined, is also invertible with inverse $\gamma(\xinv;\yinv)$.  So if $S_*$ denotes the closure of $S$ under operadic composition, then the properties defining $\Osinv$ and $\O[\Sinv_*]$ are equivalent.\dqed
\end{enumerate}\end{remark}

The next observation is the operad version of Theorem \ref{thm:localization-cat}.  Its proof  is an adaptation of the proof of Theorem \ref{thm:localization-cat} by replacing linear graphs with trees.

\begin{theorem}\label{thm:operad-localization-exists}
Suppose $\O$ is a $\colorc$-colored operad in $\Set$, and $S$ is a set of unary elements in $\O$.  Then an $S$-localization of $\O$ exists.
\end{theorem}

\begin{proof}
Without loss of generality, we may assume that $S$ is closed under the operadic composition $\gamma$ in $\O$.  Choose a set $\Sinv$ that is disjoint from $\O$ and consists of symbols $\xinv$ for $x \in S$.  For $c,d\in \colorc$, the subset of $\Sinv$ consisting of $\xinv$ with $x\in S\cap\O\cd$ is denoted by $\Sinv\dc$.  We define an $S$-localization $\O'$ of $\O$ as follows.

For $(\uc;d) \in \Profcc$, $\O'\duc$ is the set of equivalence classes of pairs $(T,\phi)$ in which:
\begin{itemize}
\item $T \in \uTreec\duc$.
\item $\phi : \Vt(T) \to \O \sqcup \Sinv$ is a function that satisfies the following conditions:
\begin{itemize}
\item $\phi(v) \in \O\duc$ if $\profofv = \duc$ with $|\uc|\not=1$.
\item $\phi(v) \in \O\dc\sqcup\Sinv\dc$ if $\profofv=\dc$ with $c,d \in \colorc$.
\item If $u$ and $v$ are adjacent vertices in $T$, then one of $\phi(u)$ and $\phi(v)$ is in $\O$ with the other in $\Sinv$.
\end{itemize}\end{itemize}
Intuitively, the function $\phi$ decorates the vertices in $T$ by elements in $\O \sqcup \Sinv$ with the correct profiles such that adjacent vertices cannot be both decorated by $\O$ or both by $\Sinv$.

The equivalence relation $\sim$ on such pairs $(T,\phi)$ is generated by the following four types of identifications.
\begin{enumerate}
\item Suppose $u$ and $v$ are adjacent unary vertices in $T$ such that \[\phi(u) \in S\cap\O\cd \andspace \phi(v) = \inv{\phi(u)} \in \Sinv\dc.\]  Then we identify \[(T,\phi) \sim (T',\phi')\] in which:
\begin{itemize}\item Without the $\colorc$-coloring, $T'=T(\uparrow_u,\uparrow_v)$ with the exceptional edge $\uparrow_u$ (resp., $\uparrow_v$) substituted into $u$ (resp., $v$).  The $\colorc$-coloring of $T'$ is inherited from that of $T$.
\item $\phi'$ is the restriction of $\phi$ to $\Vt(T') = \Vt(T) \setminus \{u,v\}$.
\end{itemize}
\item Suppose $v \in \Vt(T)$ with $\profofv =\cc$ for some $c\in\colorc$ and that \[\phi(v)=\operadunit_c\in \O\cc.\]
\begin{enumerate}
\item If $v$ has no adjacent vertices, then $T=\Cor_{(c;c)}$ with unique vertex $v$.  In this case, we identify \[(T,\phi) \sim (\uparrow_c,\varnothing)\] in which $\varnothing$ is the trivial function with domain $\Vt(\uparrow_c)=\varnothing$.
\item If $v$ has only one adjacent vertex $u$, then one of the two flags in $v$ is a leg in $T$, while the other flag in $v$ is part of an internal edge adjacent to $u$.  In this case, we identify \[(T,\phi) \sim \bigl(T(\uparrow_v),\phi'\bigr)\] with 
\begin{itemize}\item $\uparrow_v\,=~\uparrow_c$ substituted into $v$;
\item $\phi'$ the restriction of $\phi$ to $\Vt\bigl(T(\uparrow_v)\bigr) = \Vt(T) \setminus\{v\}$.
\end{itemize}
\item If $v$ has two adjacent vertices $u$ and $w$, then one flag in $v$ is part of an internal edge $e=\{e_{\pm}\}$ adjacent to $u$, and the other flag in $v$ is part of an internal edge $f=\{f_{\pm}\}$ adjacent to $w$.  Moreover, both $u$ and $w$ are unary vertices such that \[\phi(u)=\xinv \andspace \phi(w)=\yinv \in \Sinv\] for some $x,y \in S$.  Switching the names $u$ and $w$ if necessary, we may assume that $v=\{e_+,f_-\}$, so the relevant part of $(T,\phi)$ looks like:
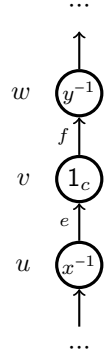
\begin{center}\begin{tikzpicture}
\node[plain] (w) {\footnotesize{$\yinv$}}; \node[left=.2cm of w] () {$w$};\node[above=.5cm of w,empty] (top) {$\cdots$};
\node[below=.5cm of w, plain] (v) {$\operadunit_c$}; \node[left=.2cm of v] () {$v$};
\node[below=.5cm of v, plain] (u) {\footnotesize{$\xinv$}}; \node[left=.2cm of u] () {$u$};
\node[below=.5cm of u, empty] (bot) {$\cdots$};
\draw[arrow] (bot) to (u); \draw[arrow] (u) to node{\scriptsize{$e$}} (v);
\draw[arrow] (v) to node{\scriptsize{$f$}} (w); \draw[arrow] (w) to (top);
\end{tikzpicture}\end{center}
Suppose $T'$ is the tree obtained from $T$ by (i) removing the four flags $\{e_{\pm},f_{\pm}\}$ and (ii) redefining $\bigl\{\out(w),\inp(u)\bigr\}$ as a single unary vertex $t$ with \[\inp(t)=\inp(u) \andspace \out(t)=\out(w).\]  We identify $(T,\phi) \sim (T',\phi')$ in which \[\phi' : \Vt(T') = \{t\} \sqcup \Vt(T) \setminus\{u,v,w\} \to \O \sqcup \Sinv\] is the restriction of $\phi$ away from $t$ and \[\phi'(t)=\inv{\gamma(x;y)} \in \Sinv.\]
\end{enumerate}
\item Suppose $v \in \Vt(T)$ with $\profofv =\cc$ for some $c\in\colorc$ and that \[\phi(v)=\inv{\operadunit_c}\in\Sinv\cc.\]  This can only happen if $\operadunit_c \in S$.
\begin{enumerate}
\item If $v$ has no adjacent vertices, then $T=\Cor_{(c;c)}$ with unique vertex $v$, and we identify $(T,\phi) \sim (\uparrow_c,\varnothing)$ as in Case (2)(a) above.
\item If $v$ has only one adjacent vertex $u$, then $(T,\phi) \sim \bigl(T(\uparrow_v),\phi'\bigr)$ as in Case (2)(b) above.
\item If $v$ has two adjacent vertices $u$ and $w$, then \[\phi(u)=x \andspace \phi(w)=y \in \O.\]  Proceeding as in Case (2)(c) above, the relevant part of $(T,\phi)$ now looks like:
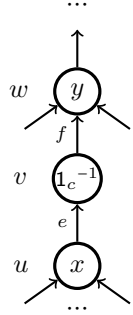
\begin{center}\begin{tikzpicture}
\node[plain] (w) {$y$}; \node[left=.2cm of w] () {$w$};\node[above=.5cm of w,empty] (top) {$\cdots$};
\node[below=.5cm of w, plain] (v) {\footnotesize{$\inv{\operadunit_c}$}}; \node[left=.2cm of v] () {$v$};
\node[below=.5cm of v, plain] (u) {$x$}; \node[below=.1cm of u] () {$\cdots$};
\node[left=.2cm of u] () {$u$};
\draw[inputleg] (w) to +(-.7cm,-.5cm); \draw[inputleg] (w) to +(.7cm,-.5cm);
\draw[inputleg] (u) to +(-.7cm,-.5cm); \draw[inputleg] (u) to +(.7cm,-.5cm);
\draw[arrow] (u) to node{\scriptsize{$e$}} (v);
\draw[arrow] (v) to node{\scriptsize{$f$}} (w); \draw[arrow] (w) to (top);
\end{tikzpicture}\end{center}
Suppose $T'$ is the tree obtained from $T$ by (i) removing the four flags $\{e_{\pm},f_{\pm}\}$ and (ii) redefining a single vertex \[t=\Bigl\{w\setminus\{f_+\},u\setminus\{e_-\}\Bigr\}\] with \[\out(t)=\out(w) \andspace \inp(t) = \inp(w) \compi \inp(u).\]  Here we assume $f_+$ is the $i$th input of $w$, and $\compi$ was defined in Definition \ref{def:compi}.  We identify $(T,\phi) \sim (T',\phi')$ in which \[\phi' : \Vt(T') = \{t\} \sqcup \Vt(T) \setminus\{u,v,w\} \to \O \sqcup \Sinv\] is the restriction of $\phi$ away from $t$ and \[\phi'(t)=y \compi x \in \O,\] which was defined in \eqref{compi-def}
\end{enumerate}
\item Suppose $v$ is a vertex in $T$ with $\phi(v) \in \O\bua$ and $\sigma \in \Sigma_{|\ua|}$.  Suppose $T^{\sigma}$ is the tree that is the same as $T$ except that its ordering at $v$ is the composition $\zeta_v\sigma$, where $\zeta_v$ is the ordering at $v$ in $T$.  Then we identify \[(T,\phi) \sim (T^{\sigma},\phi^{\sigma})\] in which \[\phi^{\sigma}(u) =\begin{cases}\phi(u) & \text{ if $u\not=v$},\\ \phi(v)\sigma & \text{ if $u=v$}. \end{cases}\]
\end{enumerate}
The equivalence class of $(T,\phi)$ is denoted by $[(T,\phi)]$.

We will use Definition \ref{def:operad-compi} of a $\colorc$-colored operad in the rest of this proof.  Next we define the $\colorc$-colored operad structure on $\O'$.  For $c \in \colorc$ the $c$-colored unit is the equivalence class of $(\uparrow_c,\varnothing)$.  The equivariant structure is induced by reordering the inputs \[(T,\phi)\sigma=(T\sigma,\phi),\] where $T\sigma$ is the same as $T$ except that its ordering is $\zeta\sigma$ with $\zeta$ the ordering of $T$.  This equivariant structure is well-defined in the sense that it respects the equivalence relation $\sim$ that defines $\O'$.  

For the $\compi$-composition in \eqref{operadic-compi}, suppose that $(T,\phi)$ represents an equivalence class in $\O'\duc$ with $|\uc|=n\geq 1$ and that $(T',\phi')$ represents an equivalence class in $\O'\ciub$.  First define
\begin{equation}\label{tree-compi}
T\compi T' = \graft\bigl(T; \uparrow_{c_1},\ldots,\uparrow_{c_{i-1}},T',\uparrow_{c_{i+1}},\ldots,\uparrow_{c_n}\bigr) \in \uTreec\sbinom{d}{\uc\compi\ub},
\end{equation}
which is a grafting as in Definition \ref{def:grafting}.
\begin{itemize}
\item If $(T',\phi') = (\uparrow_{c_i},\varnothing)$, then $T \compi \uparrow_{c_i} = T$, and we define
\[[(T,\phi)] \compi [(\uparrow_{c_i},\varnothing)] = [(T,\phi)].\]
\item If $(T,\phi) = (\uparrow_d,\varnothing)$, then we similarly define
\[[(\uparrow_d,\varnothing)] \compi [(T',\phi')] = [(T',\phi')].\]
\item If neither $T$ nor $T'$ is an exceptional edge, then $T\compi T'$ has exactly one internal edge $e=\{e_{\pm}\}$ that is neither an internal edge in $T$ nor in $T'$.  If $e$ is oriented from the vertex $u$ to the vertex $v$, then $u \in \Vt(T')$ and $v \in \Vt(T)$.  So a portion of $T\compi T'$ looks like:
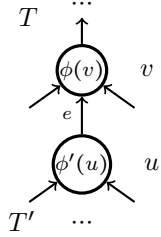
\begin{center}\begin{tikzpicture}
\node[plain] (v) {\footnotesize{$\phi(v)$}}; \node[above left=.3cm of v] () {$T$}; 
\node[right=.3cm of v] () {$v$}; \node[above=.2cm of v, empty](top){$\cdots$};
\node[below=.5cm of v, plain] (u) {\footnotesize{$\phi'(u)$}}; \node[below left=.3cm of u] () {$T'$};
\node[right=.3cm of u] () {$u$}; \node[below=.1cm of u, empty](bot){$\cdots$};
\draw[outputleg] (v) to +(0,.7cm);
\draw[inputleg] (v) to +(-.7cm,-.5cm); \draw[inputleg] (v) to +(.7cm,-.5cm);
\draw[inputleg] (u) to +(-.7cm,-.5cm); \draw[inputleg] (u) to +(.7cm,-.5cm);
\draw[arrow] (u) to node{\scriptsize{$e$}} (v);
\end{tikzpicture}\end{center}
\begin{itemize}
\item If one of $\phi'(u)$ and $\phi(v)$ is in $\Sinv$ with the other in $\O$, then we define
\begin{equation}\label{osinv-compi}
[(T,\phi)] \compi [(T',\phi')] = \bigl[(T\compi T',\phi\compi\phi')\bigr]
\end{equation}
in which $\phi\compi\phi'$ is induced by $\phi$ and $\phi'$ via the decomposition \[\Vt\bigl(T\compi T'\bigr) = \Vt(T) \sqcup \Vt(T').\]
\item If both $\phi'(u)$ and $\phi(v)$ are in $\O$, we first define $G \in \uTreec\binom{d}{\uc\compi\ub}$ as the tree obtained from $T\compi T'$ by (i) removing the two flags in $e$ and (ii) redefining a vertex
\[t = \bigl\{v\setminus\{e_+\},u\setminus\{e_-\}\bigr\}\] with \[\out(t)=\out(v) \andspace \inp(t)=\inp(v)\comp_j\inp(u).\]  Here we assume $e_+$ is the $j$th input of $v$.  We define \begin{equation}\label{oprime-compi}
[(T,\phi)] \compi [(T',\phi')] = [(G,\varphi)]
\end{equation} 
in which \[\nicexy{\Vt(G) = \{t\} \sqcup \bigl[\Vt(T) \setminus \{v\}\bigr] \sqcup \bigl[\Vt(T') \setminus \{u\}\bigr] \ar[r]^-{\varphi} & \O\sqcup\Sinv}\] is the restrictions of $\phi$ and $\phi'$ away from $t$ and \[\varphi(t) = \phi(v) \comp_j \phi'(u) \in \O.\]
\item If both $\phi'(u)=\xinv$ and $\phi(v)=\yinv$ are in $\Sinv$, then both $u$ and $v$ are unary vertices.  Using the same tree $G$ as in the previous case, we define
\[[(T,\phi)] \compi [(T',\phi')] = [(G,\pi)]\] in which $\pi$ is the restrictions of $\phi$ and $\phi'$ away from $t$ and \[\pi(t) = \inv{\gamma(x;y)} \in \Sinv.\]
\end{itemize}\end{itemize}
The $\colorc$-colored operad axioms of $\O$ and the unity and associativity of tree substitution (as in Corollary \ref{cor:treesub-assunity}) imply that the $\compi$-composition above is indeed well-defined, i.e., independent of the choices of the representatives $(T,\phi)$ and $(T',\phi')$ in their equivalence classes.  Furthermore, $\O'$ satisfies the axioms in Definition \ref{def:operad-compi} because of the existence of the $\colorc$-colored tree operad in Example \ref{ex:tree-operad}, so $\O'$ is a $\colorc$-colored operad in $\Set$.

Now we define a morphism \[\ell : \O \to \O' \in \M^{\Profcc}\] by setting 
\begin{equation}\label{ellofx}
\ell(x)=\bigl[\bigl(\Cor_x,\phi_x\bigr)\bigr]
\end{equation} 
for $x \in \O\duc$ in which:
\begin{itemize}\item $\Cor_x$ is the corolla $\Cor_{(\uc;d)}$ in Example \ref{ex:cd-corolla}.
\item $\phi_x$ sends the unique vertex in $\Cor_x$ to $x$.\end{itemize}
This defines a morphism $\ell : \O \to \O'$ of $\colorc$-colored operads.  Indeed, $\ell$ respects the $c$-colored units and the equivariant structures by the identifications of type (2)(a) and type (4) above.  It respects the $\compi$-composition by the definition \eqref{oprime-compi}.

By the identification of type (1) above, for each $s \in S \cap \O\cd$, its image \[\ell(s) = \bigl[\bigl(\Lin_{(d,c)},\phi_s\bigr)\bigr]\] is an invertible unary element with inverse \[\bigl[\bigl(\Lin_{(c,d)},\phi_{\sinv}\bigr)\bigr]\] in which $\phi_{\sinv}$ sends the unique vertex to $\sinv$.  Recall from Example \ref{ex:linear-graph} that $\Lin_{(c,d)} = \Cor_{(c;d)}$ is the linear graph with one vertex and profile $(c;d)$.

Finally, to prove the universal property \eqref{operad-localization-triangle}, suppose  $f : \O \to \P$ is an operad morphism with $\P$ a $\colord$-colored operad in $\Set$ such that $f(s)$ is invertible for each $s \in S$.  The requirement that the diagram \eqref{operad-localization-triangle} be commutative forces us to make the following definition of $f' : \O' \to \P$:
\begin{itemize}\item We define $f'=f : \colorc \to \colord$ on colors.  
\item For each $[(T,\phi)] \in \O'$, we define \[f'[(T,\phi)] = \gammap_{fT}\bigl(f'\phi(v)\bigr)_{v\in T}\]
in which:
\begin{enumerate}\item $fT$ is the $\colord$-colored tree obtained from $T$ by applying $f : \colorc \to \colord$ to its $\colorc$-coloring.
\item For each $v \in T$, \[f'\phi(v)=\begin{cases}f\phi(v) & \text{ if $\phi(v) \in \O$},\\ \inv{(fx)} & \text{ if $\phi(v)=\xinv \in \Sinv$ for some $x \in S$.}\end{cases}\]
\end{enumerate}
\end{itemize}
The operad axioms of $\P$ imply that (i) $f'$ is entrywise well-defined, i.e., independent of the choice of a representative $(T,\phi)$ in its equivalence class, and that (ii) it is an operad morphism.  The diagram \eqref{operad-localization-triangle} is commutative by construction.  As we mentioned above, the uniqueness of $f'$ is guaranteed by the commutativity of the diagram \eqref{operad-localization-triangle}.  Therefore, we have shown that $\O'$ is an $S$-localization of $\O$.
\end{proof}

\section{Algebras over Localized Operads}\label{sec:algebra-local}

In this section, we consider algebras over a localization of a colored operad.  For a $\colorc$-colored operad $\O$ in $\Set$, recall that $\Otom$ is the image of $\O$ in $\M$ via the change-of-category functor \[(-)^{\M} : \Operadcset \to \Operadcm\] induced by the strong symmetric monoidal functor $\coprod_{(-)} \tensorunit : \Set \to \M$.  This is an instance of Theorem \ref{thm:operad-transfer}.  First we consider what it means to be an algebra over a $\Set$-operad in $\M$.

\begin{theorem}\label{thm:algebra-set-operad}
Suppose $(\O,\gamma,\operadunit)$ is a $\colorc$-colored operad in $\Set$.  Then an $\Otom$-algebra is precisely a \index{algebra!over a set operad}\index{set operad!algebra}pair $(X,\theta)$ consisting of 
\begin{itemize}\item a $\colorc$-colored object $X \in \Mtoc$ and
\item a morphism \[\nicexy{X_{\uc} \ar[r]^-{\theta_p} & X_d \in \M}\] for each $p \in \O\duc$ with $(\uc;d)\in \Profcc$\end{itemize}
that satisfies the following axioms.
\begin{description}
\item[Associativity]
For $\bigl(\uc = (c_1, \ldots , c_n);d\bigr) \in \Profcc$ with $n \geq 1$, $\ub_j \in \Profc$ for $1 \leq j \leq n$, $\ub = (\ub_1,\ldots,\ub_n)$, $p \in \O\duc$, and $q_j \in \O\cjubj$, the associativity diagram\index{associativity!colored operadic algebra}
\begin{equation}\label{setoperad-ass}
\nicexy{X_{\ub}\ar@{=}[d] \ar[r]^-{\theta_{\gamma(p;q_1,\ldots,q_n)}} & X_d\\  X_{\ub_1}\otimes\cdots\otimes X_{\ub_n} \ar[r]^-{\bigtensorover{j} \theta_{q_j}} & X_{c_1} \otimes\cdots\otimes X_{c_n} \ar[u]_-{\theta_p}}
\end{equation}
in $\M$ is commutative.
\item[Unity] For each $c \in \colorc$, $\theta_{\operadunit_c}=\Id_{X_c}$.
\item[Equivariance]
For each $(\uc;d) \in \Profcc$ and each permutation $\sigma \in \Sigma_{|\uc|}$, the equivariance diagram\index{equivariance!colored operadic algebra}
\begin{equation}\label{setoperad-eq}
\nicexy{X_{\uc}\ar[d]_-{\theta_p} \ar[r]^-{\inv{\sigma}}_-{\cong}& X_{\uc\sigma}\ar[d]^-{\theta_{p\sigma}}\\ X_d \ar@{=}[r] & X_d}
\end{equation}
in $\M$ is commutative.
\end{description}
A \index{colored operadic algebra!morphism}\emph{morphism of $\O$-algebras} $f : (X,\theta) \to (Y,\xi)$ is a morphism $f : X \to Y$ of $\colorc$-colored objects in $\M$ such that the diagram
\begin{equation}\label{setoperadalg-morphism}
\nicexy{X_{\uc}\ar[d]_-{\theta_p}\ar[r]^-{\otimes f}& Y_{\uc}\ar[d]^-{\xi_p}\\ X_d \ar[r]^-{f}& Y_d}
\end{equation}
in $\M$ is commutative for all $(\uc;d) \in \Profcc$ and $p \in \O\duc$.
\end{theorem}

\begin{proof}
Each entry of $\O$ is a coproduct \[\Otom\duc = \coprodover{\O\duc} \tensorunit.\]  If $(X,\theta)$ is an $\Otom$-algebra in the sense of Definition \ref{def:operad-algebra-generating}, then for each $p \in \O\duc$ it has an induced $\O$-action structure morphism 
\[\nicexy@C+.3cm{X_{\uc} \ar[d]_-{\cong} \ar[rr]^-{\theta_p} && X_d\\ \tensorunit \otimes X_{\uc} \ar[r]^-{p}_-{\mathrm{inclusion}} & \coprodover{\O\duc} \bigl(\tensorunit \otimes X_{\uc}\bigr) \ar[r]^-{\cong} & \Otom\duc \otimes X_{\uc} \ar[u]_-{\theta}}\]
in which the bottom left horizontal morphism is the coproduct summand inclusion corresponding to $p$.  In terms of these structure morphisms $\theta_p$, the axioms stated above are simply those in Definition \ref{def:operad-algebra-generating}.  The converse also holds because a morphism \[\Otom\duc \otimes X_{\uc} \to X_d\] is unique determined by the morphisms $\theta_p$ as $p$ runs through $\O\duc$.
\end{proof}

Recall the change-of-operad adjunction in Theorem \ref{thm:change-operad}.  In the next result, we consider the change-of-operad adjunction induced by $\ellm$, which is the image of a localization morphism $\ell$ as in Definition \ref{def:operad-localization} under the change-of-category functor $(-)^{\M}$.

\begin{theorem}\label{thm:localization-algebra}
Suppose $\O$ is a $\colorc$-colored operad in $\Set$, and $S$ is a set of unary elements in $\O$.  Consider the change-of-operad \index{algebra!of localized operad}\index{colored operadic algebra!of localized operad}adjunction \[\nicexy{\algmotom \ar@<2pt>[r]^-{\ellm_!} & \algmosinvtom \ar@<2pt>[l]^-{(\ellm)^*}}\] induced by the image in $\M$ of the $S$-localization morphism $\ell : \O \to \Osinv$.
\begin{enumerate}\item The right adjoint $(\ellm)^*$ is full and faithful.
\item The counit of the adjunction \[\epsilon : \ellm_!(\ellm)^* \iso \Id_{\algmosinvtom}\] is a natural isomorphism.\end{enumerate}
\end{theorem}

\begin{proof}
By Theorem \ref{thm:algebra-set-operad} an $\Otom$-algebra morphism is a morphism of the underlying colored objects that respects the structure morphisms $\theta_p$, in the sense of \eqref{setoperadalg-morphism}, as $p$ runs through all of $\O$, and similarly for an $\Osinvtom$-algebra morphism.  It follows that the right adjoint $\ellmstar$ is faithful.

To see that $\ellmstar$ is full, it is enough to prove the following statement.
\begin{quote} Given $\Osinvm$-algebras $(X,\theta^X)$ and $(Y,\theta^Y)$ and a morphism $f : X \to Y$ of the underlying colored objects that respects the structure morphisms $\theta_{\ell(p)}$ for $p\in\O$, then $f$ is a morphism of $\Osinvm$-algebras.
\end{quote}
We prove this statement by the following series of reductions.  
\begin{enumerate}\item By the equivariance axiom \eqref{setoperad-eq} of $\Osinvm$-algebras, if $f$ respects $\theta_q$ for some $q \in \Osinv$, then it also respects $\theta_{q\sigma}$ for all permutations $\sigma$ for which $q\sigma$ is defined.
\item By Theorem \ref{thm:grafting-generate}, the definition \eqref{osinv-compi} of $\compi$ in $\Osinv$, the associativity \eqref{setoperad-ass} of $\Osinvm$-algebras, and the previous step, if $f$ respects $\theta_{[(\Cor_x,\phi_x)]}$ for all $x \in \O\sqcup\Sinv$, then $f$ is a morphism of $\Osinvm$-algebras.  Here $\Cor_x$ is the corolla with the same profile as $x$, and $\phi_x$ sends the unique vertex in $\Cor_x$ to $x$.  The reader is reminded that the operadic composition $\gamma$ in any colored operad can be written in terms of the various $\compi$-compositions as in \eqref{compi-to-gamma}.
\item By the definition of $\ell$ in \eqref{ellofx}, we are assuming that $f$ respects the structure morphism $\theta_{[(\Cor_x,\phi_x)]}$ for all $x \in \O$.  The associativity and unity of an $\Osinvm$-algebra imply that, for each $s\in S$, the structure morphism \[\theta_{\ell(s)}=\theta_{[(\Cor_s,\phi_s)]}\] is an isomorphism with inverse $\theta_{[(\Cor_{\sinv},\phi_{\sinv})]}$.  Since $f$ respects $\theta_{\ell(s)}$ for all $s \in S$, it also respects $\theta_{[(\Cor_{\sinv},\phi_{\sinv})]}$.  
\end{enumerate}
This finishes the proof of the first assertion.  The second assertion about the counit is a consequence of the first assertion and \cite{maclane} IV.3 Theorem 1.
\end{proof}

\begin{theorem}\label{osinvm-algebra}
In the setting of Theorem \ref{thm:localization-algebra}, an $\Otom$-algebra $(X,\theta)$ is in the image of the right adjoint $\ellmstar$ if and only if the structure morphisms $\theta_s$ are isomorphisms for all $s \in S$.
\end{theorem}

\begin{proof}
We already noted in the previous proof that, for an $\Osinvm$-algebra, the structure morphism  $\theta_{\ell(s)}$ is an isomorphism for each $s \in S$.  So for an $\Otom$-algebra in the image of $\ellmstar$, $\theta_s$ must be an isomorphism for each $s \in S$.  

For the converse, observe that by Theorem \ref{thm:grafting-generate} and the axioms in Theorem \ref{thm:algebra-set-operad}, the structure morphisms \[\bigl\{\theta_q : q \in \Osinv\bigr\}\] for an $\Osinvm$-algebra are uniquely determined by the subset \[\bigl\{\theta_{\ell(x)} : x \in \O\bigr\}.\]  So for an $\Otom$-algebra $(X,\theta)$ in which the structure morphisms $\theta_s$ are isomorphisms for all $s \in S$, we can first define 
\[\theta_{[(\Cor_x,\phi_x)]} = \theta_x \forspace x\in\O.\]  Then we use the associativity axiom \eqref{setoperad-ass} and the equivariance axiom \eqref{setoperad-eq} to define a general $\theta_q$ for $q \in \Osinv$.  The assumptions on $(X,\theta)$ ensure that these structure morphisms $\theta_q$ are well-defined and that they satisfy the axioms in Theorem \ref{thm:algebra-set-operad} for an $\Osinvm$-algebra.
\end{proof}

\begin{remark}\label{rk:osinvm-algebras}
By Theorem \ref{thm:localization-algebra} and Theorem \ref{osinvm-algebra}, we may regard the category $\algmosinvtom$ of $\Osinvm$-algebras as the full subcategory of $\algmotom$ consisting of the $\Otom$-algebras in which the structure morphisms $\theta_s$ are isomorphisms for all $s \in S$.\dqed\end{remark}

\chapter{Boardman-Vogt Construction of Operads}\label{ch:bv}

In this chapter we define the Boardman-Vogt construction of a colored operad in a symmetric monoidal category as an entrywise coend and study its naturality properties.  

\section{Overview}
The Boardman-Vogt construction was originally defined for a colored topological operad (without using the term operad) in \cite{boardman-vogt} and also in \cite{vogt}.  Our one-step formulation of the Boardman-Vogt construction in terms of a coend will be important when we apply it to the colored operads for algebraic quantum field theories and prefactorization algebras.  The very explicit nature of our coend definition will allow us to elucidate the structures in homotopy algebraic quantum field theories and homotopy prefactorization algebras.

As we will explain later, the Boardman-Vogt construction $\wo$ is a resolution of the original colored operad $\O$.  Its algebras are algebras over $\O$ up to coherent higher homotopies.  When $\O$ is a colored operad for algebraic quantum field theories or prefactorization algebras, $\wo$-algebras are homotopy algebraic quantum field theories or homotopy prefactorization algebras.  For instance, suppose $\O$ is the colored operad $\Cdiag$ for $\C$-diagrams.  If $X$ is an $\O$-algebra and if the composition $f\circ g$ is defined in $\O$, then $X_f \circ X_g$ is equal to $X_{f\circ g}$ by the associativity axiom of $\O$-algebras.  If $Y$ is a homotopy coherent $\C$-diagram, i.e., a $\wo$-algebra, then both $Y_f \circ Y_g$ and $Y_{f\circ g}$ are defined, but they are not equal in general.  Instead, there is another $\wo$-algebra structure morphism of $Y$ that is a homotopy from $Y_{f\circ g}$ to $Y_f \circ Y_g$.  There are other $\wo$-algebra structure morphisms that relate these homotopies, and so forth.  We will discuss homotopy coherent diagrams in Section \ref{sec:hcdiagram}.

The Boardman-Vogt construction of a colored operad is defined in Section \ref{sec:segment} and Section \ref{sec:coend-bv}.  An augmentation of the Boardman-Vogt construction over the original colored operad is defined in Section \ref{sec:augmentation-bv}.  The augmentation induces a change-of-operad adjunction, which allows us to go back and forth between algebras of the original colored operad and of the Boardman-Vogt construction.  In Section \ref{sec:morita} we construct an entrywise section of the augmentation and use it to show that, in familiar cases, the augmentation is a weak equivalence.  In particular, over $\Chaink$ the change-of-operad adjunction induced by the augmentation of a colored operad is always a Quillen equivalence.  Let us emphasize that, in order to define the Boardman-Vogt construction and to understand the structure of its algebras, a model structure on the base category and that the augmentation is a weak equivalence are not necessary.

In Section \ref{sec:filtration} we discuss a natural filtration of the Boardman-Vogt construction.  This filtration is not needed for applications to homotopy algebraic quantum field theories and homotopy prefactorization algebras, so the reader may skip this section safely.  One main point of this filtration is to show that, for one-colored operads, our one-step coend definition of the Boardman-Vogt construction is isomorphic to the sequential colimit definition by Berger and Moerdijk.  In \cite{berger-moerdijk-bv} the Boardman-Vogt construction of a one-colored operad was defined as the sequential colimit of an inductively defined sequence of morphisms, each being the pushout of some square involving the previous inductive step.  For the Boardman-Vogt construction of more general objects, including dioperad, properads, wheeled operads, and wheeled properads, the reader is referred to \cite{bvbook}.

Our coend definition of the Boardman-Vogt construction uses the language of trees from Chapter \ref{ch:tree}.  As before $(\M,\otimes,\tensorunit)$ is a cocomplete symmetric monoidal closed category with an initial object $\varnothing$, and $\colorc$ is an arbitrary non-empty set whose elements are called colors.

\section{Commutative Segments}\label{sec:segment}

To define the Boardman-Vogt construction of a colored operad, we will equip the internal edges in trees with a suitable length using the following concept from \cite{berger-moerdijk-bv} (Definition 4.1).  Recall the concept of a monoid in Section \ref{sec:monoids}.

\begin{definition}\label{def:segment}
A \index{segment}\emph{segment} in $\M$ is a tuple \label{notation:segment} $(J, \mu, 0, 1, \epsilon)$ in which:
\begin{itemize}\item $(J, \mu, 0)$ is a monoid in $\M$.
\item $1 : \tensorunit \to J$ is an \index{absorbing element}absorbing element.
\item $\epsilon : J \to \tensorunit$ is a \index{counit}counit.  
\end{itemize}
A \index{commutative segment}\emph{commutative segment} is a segment whose multiplication $\mu$ is commutative. \end{definition}

\begin{remark}
More explicitly, in a (commutative) segment, $J$ is a (commutative) monoid with multiplication $\mu : J \otimes J \to J$ and unit $0 : \tensorunit \to J$.  To say that $1 : \tensorunit \to J$ is an absorbing element means that the diagram
\[\nicexy{\tensorunit \otimes J \ar[d]_-{(\Id,\epsilon)} \ar[rr]^-{(1,\Id)} && J \otimes J \ar[d]_-{\mu} && J \otimes \tensorunit \ar[ll]_-{(\Id,1)} \ar[d]^-{(\epsilon,\Id)}\\
\tensorunit \otimes \tensorunit \ar[r]^-{\cong} & \tensorunit \ar[r]^-{1} & J & \tensorunit \ar[l]_-{1} & \tensorunit \otimes \tensorunit \ar[l]_-{\cong}}\]
is commutative.  The counit $\epsilon$ makes the diagrams
\[\nicexy{J \otimes J \ar[d]_-{\mu} \ar[r]^-{(\epsilon,\epsilon)} & \tensorunit\otimes\tensorunit \ar[d]^-{\cong}\\ J \ar[r]^-{\epsilon} & \tensorunit}\qquad
\nicexy{\tensorunit \ar[d]_-{1} \ar[r]^-{0}\ar[dr]^-{\Id} & J \ar[d]^-{\epsilon}\\ J \ar[r]^-{\epsilon} & \tensorunit}\]
commutative.  A commutative segment provides a concept of homotopy from the $0$-end $0 : \tensorunit \to J$ to the $1$-end $1 : \tensorunit \to J$.\dqed\end{remark}  

\begin{example}\label{ex:trivial-segment}
There is always a \index{trivial commutative segment}\emph{trivial commutative segment} $\tensorunit$ with $0,1,\epsilon=\Id_{\tensorunit}$ and $\mu : \tensorunit\otimes\tensorunit\cong\tensorunit$ the canonical isomorphism.\dqed
\end{example}

\begin{example}\label{ex:com-segment}
Here are some examples of non-trivial commutative segments.
\begin{enumerate}
\item In $\Top$ the \index{unit interval}\index{topological space!segment}unit interval $[0,1]$ equipped with the multiplication \[\mu(a,b) = \max\{a,b\}\] is a commutative segment.
\item In $\Cat$ the \index{small category!segment}category \[J = \bigl\{\nicexy@C-.5cm{0 \ar@{<->}[r]^-{\cong} & 1}\bigr\}\] with two objects $\{0,1\}$ and a unique isomorphism from $0$ to $1$ is a commutative segment with the multiplication induced by the maximum operation.
\item In $\Sset$ the simplicial interval, that is, the representable simplicial set\index{simplicial set!segment} $\Delta^1=\Delta(-,[1])$, is a commutative segment with the multiplication induced by the maximum operation.
\item In $\Chaink$ with $\fieldk$ a field of characteristic $0$, the \index{normalized chain complex}\index{chain complex!segment}normalized chain complex $J=N\Delta^1$ of $\Delta^1$ is a commutative segment whose structure is uniquely determined by that on the simplicial interval $\Delta^1$ and the monoidal structure of the normalized chain functor \cite{weibel} (8.3.6 page 265).  More explicitly, $J$ is a $2$-stage chain complex \[\nicexy{\cdots \ar[r] & 0 \ar[r] & \fieldk \ar[r]^-{(+,-)} & \fieldk \oplus \fieldk \ar[r] & 0 \ar[r] & \cdots}\] with $\fieldk$ in degree $1$ and $\fieldk \oplus \fieldk$ in degree $0$.  The morphisms $0,1 : \fieldk \to J$ correspond to the two copies of $\fieldk$ in degree $0$ in $J$, and the counit $\epsilon : J \to \fieldk$ is the identity morphism on each copy of $\fieldk$ in degree $0$.  The mapping cylinder of a chain complex $C$ (see, e.g., \cite{weibel} Exercise 1.5.3) is $J \otimes C$.  Two chain maps $f,g : C \to D$ are chain homotopic if and only if there is an extension $J \otimes C \to D$ whose restrictions to $C$ via $0$ and $1$ are $f$ and $g$, respectively.  We leave it to the reader to write down explicit formulas for the multiplication $\mu$ on $J$. 
\end{enumerate}
For the categories $\Top$, $\Sset$, $\Chaink$, and $\Cat$, unless otherwise specified, we will always use these commutative segments.\dqed
\end{example}

We will use the language of trees from Chapter \ref{ch:tree}.  In particular, recall that for a tree $T$, $|T|$ denotes the set of internal edges in $T$.  Also recall the exceptional edges in Example \ref{ex:colored-exedge} and the substitution category $\uTreec$ in Definition \ref{def:treesub-category}.

\begin{definition}\label{functor-J}
Suppose $(J,\mu,0,1,\epsilon)$ is a commutative segment in $\M$.  For each $(\uc;d) \in \Profcc$, define a functor\label{notation:functorj}\index{functor!induced by a segment} \[\J : \uTreecducop \to \M\] by the unordered monoidal product \[\J[T] = \bigotimes_{e \in |T|} J = J^{\otimes |T|}\] for $T \in \uTreecduc$.  For each morphism \[(H_v)_{v\in T} : T(H_v) \to T \in \uTreec\duc,\] the morphism \[\J[T] \to \J[T(H_v)] \in \M\] is induced by:
\begin{itemize}
\item $0 : \tensorunit \to J$ for each internal edge in each $H_v$, which must become an  internal edge in $T(H_v)$;
\item the multiplication $\mu : J \otimes J \to J$ if $H_v$ is an exceptional edge and if $v$ is adjacent to two other vertices in $T$;
\item the counit $\epsilon : J \to\tensorunit$ if $H_v$ is an exceptional edge and if $v$ is adjacent to only one other vertex in $T$;
\item the identity morphism of $\tensorunit$ if $H_v$ is an exceptional edge and if $v$ is not adjacent to any other vertices in $T$ (i.e., $T$ is a linear graph $\Lin_{(d,d)}$).
\end{itemize}\end{definition}

\begin{remark} The absorbing element $1 : \tensorunit \to J$ is not needed to define the functor $\J : \uTreecducop \to \M$.\dqed\end{remark}

Pick a commutative segment $(J,\mu,0,1,\epsilon)$ in $\M$.  The following observation will be needed to define the operad structure on the Boardman-Vogt construction.  We will use the morphism $1 : \tensorunit \to J$ of the commutative segment.

\begin{lemma}\label{lem:morphism-pi}
Suppose $T(H_v)_{v\in T}$ is a tree substitution of $\colorc$-colored trees, and $S$ is the set of internal edges in $T(H_v)_{v\in T}$ that are not in any of the $H_v$.  Then there is a morphism \[\nicexy{\bigtensorover{v\in T} \J[H_v] \ar[r]^-{\pi} & \J\bigl[T(H_v)_{v\in T}\bigr]
}\] of the form $\bigl(\bigotimes_S 1\bigr)\otimes \Id_{\bigotimes_{\sqcup_{v\in T} |H_v|}J}$ up to isomorphism.
\end{lemma}

\begin{proof}
Each internal edge in each $H_v$ becomes a unique internal edge in the tree substitution $T(H_v)_{v\in T}$, and there is a decomposition \[|T(H_v)| = S \sqcup \coprodover{v\in T}|H_v|.\]  The morphism $\pi$ is the composition
\[\nicexy{\bigtensorover{v\in T} \J[H_v] \ar[r]^-{\pi} \ar[d]_-{\cong} & \J\bigl[T(H_v)_{v\in T}\bigr]\\ \bigl(\bigtensorover{S} \tensorunit\bigr) \otimes \bigl(\bigtensorover{\coprodover{v\in T} |H_v|} J\bigr) \ar[r]^-{(\bigotimes_S 1,\Id)} & \bigtensorover{|T(H_v)|} J \ar@{=}[u]}\]
in which $1 : \tensorunit \to J$ is part of the commutative segment.
\end{proof}

\begin{interpretation} Intuitively, the morphism $\pi$ in Lemma \ref{lem:morphism-pi} assigns length $1$ to each new internal edge, i.e., those in $T(H_v)_{v\in T}$ that are not in any of the $H_v$.\dqed\end{interpretation}

\begin{example}\label{ex:segment-j} 
Consider the morphism \[(H_u,H_v,H_w) : K \to T \in \uTreec\] in Example \ref{ex:treesub}.  Counting the number of internal edges, we have
\[\begin{split} \J[T] &\cong J_c\otimes J_d, \qquad \J[K] \cong J_c\otimes J_f\otimes J_g,\\
\J[H_u] &= J_f, \qquad \J[H_v]=\tensorunit, \andspace \J[H_w]=J_g,\end{split}\]
in which we use $J_c$ to denote a copy of $J$ corresponding to a $c$-colored internal edge.  The morphism $\J[T] \to \J[K]$ is the composition in the following diagram.
\[\nicexy{\J[T]\cong J_c\otimes J_d \ar[d]_-{\cong} \ar[r]& J_c\otimes J_f\otimes J_g\cong \J[K]\\ J_c\otimes\tensorunit\otimes\tensorunit\otimes J_d \ar[r]^-{(\Id,0,0,\epsilon)} & J_c\otimes J_f\otimes J_g \otimes\tensorunit \ar[u]_-{\cong}}\]
Each of $H_u$ and $H_w$ has one internal edge.  This accounts for the morphisms $0 : \tensorunit \to J_f$ and $0 : \tensorunit \to J_g$.  The tree $H_v$ is the exceptional edge $\uparrow_d$, and $v$ is adjacent to only one vertex in $T$.  This accounts for the counit $\epsilon : J_d \to \tensorunit$.  

Since $K=T(H_u,H_v,H_w)$, the morphism $\pi$ in Lemma \ref{lem:morphism-pi} is \[\nicexy{\J[H_u]\otimes\J[H_v]\otimes\J[H_w] \cong \tensorunit \otimes J_f\otimes J_g \ar[r]^-{(1,\Id)} & J_c\otimes J_f \otimes J_g \cong \J[K]}\] with $1 : \tensorunit \to J_c$ corresponding to the $c$-colored internal edge in $K$.\dqed
\end{example}

\begin{example}
Suppose $L$ is the linear graph
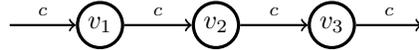
\begin{center}\begin{tikzpicture}
\matrix[row sep=.5cm, column sep=.9cm]{\node [plain] (1) {$v_1$}; & \node [plain] (2) {$v_2$}; & \node [plain] (3) {$v_3$};\\};
\draw [inputleg] (1) to node[swap]{\scriptsize{$c$}} +(-1.2cm,0);
\draw [arrow] (1) to node{\scriptsize{$c$}} (2);
\draw [arrow] (2) to node{\scriptsize{$c$}} (3);
\draw [outputleg] (3) to node{\scriptsize{$c$}} +(1.2cm,0);
\end{tikzpicture}\end{center}
in Example \ref{ex:linear-graph} with three vertices $\{v_1,v_2,v_3\}$, two internal edges, and each flag having color $c$.  Suppose $H_{v_i}=~\uparrow_c$ for each $i$, so there is a morphism \[\nicexy{L(H_{v_i})_{i=1}^3=~\uparrow_c \ar[r]^-{(H_{v_i})} & L}\] in $\uTreec\cc$.  The morphism $\J[L] \to \J[L(H_{v_i})_{i=1}^3]$ is given by either composition in the commutative diagram
\[\nicexy{\J[L]\cong J \otimes J \ar[d]_-{\mu} \ar[r]^-{(\epsilon,\epsilon)} & \tensorunit\otimes\tensorunit \ar[d]^-{\cong}\\ J \ar[r]^-{\epsilon} & \tensorunit= \J[L(H_{v_i})].}\]
The composition $\epsilon\mu$ corresponds to the factorization
\[L(H_{v_i})_{i=1}^3 = \Bigl(\bigl(L(H_{v_2})\bigr)(H_{v_1})\Bigr)(H_{v_3}),\]
while the other composition corresponds to the factorization
\[L(H_{v_i})_{i=1}^3 = \Bigl(\bigl(L(H_{v_1})\bigr)(H_{v_3})\Bigr)(H_{v_2}).\]
The morphism $\pi$ in Lemma \ref{lem:morphism-pi} is the isomorphism \[\J[H_{v_1}]\otimes \J[H_{v_2}]\otimes \J[H_{v_3}] = \tensorunit\otimes\tensorunit\otimes\tensorunit \iso \tensorunit = \J\bigl[L(H_{v_i})_{i=1}^3\bigr].\]\dqed
\end{example}

\section{Coend Definition of the BV Construction}\label{sec:coend-bv}

In this section, we define the Boardman-Vogt construction of a colored operad in a symmetric monoidal category as an entrywise coend.  Pick a commutative segment $(J,\mu,0,1,\epsilon)$ in $\M$.  Recall the concept of a coend in Definition \ref{def:coend}.

\begin{definition}\label{def:bv-operad}
Suppose $\O$ is a $\colorc$-colored operad in $\M$.  For each $(\uc;d) \in \Profcc$, define an object\index{Boardman-Vogt construction}\index{colored operad!Boardman-Vogt construction}
\begin{equation}\label{w-of-o}
\wo\duc= \int^{T \in \uTreec\duc} \J[T] \otimes \O[T] \in \M
\end{equation}
with $\J : \uTreecducop \to \M$ the functor in Definition \ref{functor-J} and $\O : \uTreec\duc \to \M$ the functor in Corollary \ref{cor:operad-functor-subcat}.  
\begin{itemize}
\item We call $\wo \in \M^{\Profcc}$ the \emph{Boardman-Vogt construction}, or \emph{BV construction}, of $\O$.
\item For $T \in \uTreec\duc$, we write\label{notation:omega-natural} \[\omega_T : \J[T] \otimes \O[T] \to \wo\duc\] for the natural morphism.
\end{itemize}
\end{definition}

\begin{interpretation} Intuitively, each entry of the Boardman-Vogt construction $\wo$ is made up of \index{decorated tree}decorated trees $\J[T]\otimes\O[T]$, with each internal edge decorated by the commutative segment $J$ and each vertex decorated by the entry of $\O$ with the same profile.  When $\O$ is a colored topological operad, one can check that the above definition of $\wo$ agrees with the original one in \cite{boardman-vogt,vogt} in terms of a quotient.  This is proved in \cite{bvbook} Example 3.4.7.\dqed\end{interpretation}

\begin{example}
In the setting of Example \ref{ex:treesub}, we have:
\[\begin{split}
\J[H_u] \otimes \O[H_u] &= J_f \otimes \O\sbinom{c}{a,b,f}\otimes \O\sbinom{f}{\varnothing},\\
\J[H_v] \otimes \O[H_v] &= \tensorunit\otimes\tensorunit,\\
\J[H_w] \otimes \O[H_w] &= J_g\otimes \O\sbinom{e}{c,g}\otimes \O\sbinom{g}{d},\\
\J[T] \otimes \O[T] &\cong J_c\otimes J_d \otimes \O\sbinom{e}{c,d}\otimes \O\sbinom{c}{a,b}\otimes \O\dd,\\
\J[K] \otimes \O[K] &\cong J_c\otimes J_f\otimes J_g \otimes \O\sbinom{e}{c,g}\otimes \O\sbinom{c}{a,b,f}\otimes \O\sbinom{f}{\varnothing}\otimes \O\sbinom{g}{d}.
\end{split}\]\dqed
\end{example}

\begin{example}\label{ex:jo-linear}
If $\uc = (c_0,\ldots,c_n)$ with $n\geq 1$ and if $\Lin_{\uc}$ is the linear graph in Example \ref{ex:linear-graph}, then we have
\[\J\bigl[\Lin_{\uc}\bigr] \otimes \O\bigl[\Lin_{\uc}\bigr] \cong \Bigl(\bigotimes\limits_{j=1}^{n-1} J_{c_j}\Bigr) \otimes \Bigl(\bigotimes\limits_{i=1}^n\O\sbinom{c_i}{c_{i-1}}\Bigr).\]\dqed
\end{example}

For a $\colorc$-colored operad $\O$ and a vertex $v$ in a $\colorc$-colored tree, recall our notation $\O(v) = \O\inoutv$ and $\uTreec(v) = \uTreec\inoutv$.

\begin{lemma}\label{lem:wo-of-t}
Suppose $\O$ is a $\colorc$-colored operad in $\M$, and $T$ is a $\colorc$-colored tree.  Then there is a canonical isomorphism
\[\wo[T] \cong \int^{\{H_v\}\in \prodover{v\in T} \uTreec(v)} \Bigl(\bigtensorover{v\in T} \J[H_v]\Bigr) \otimes \Bigl(\bigtensorover{v\in T} \O[H_v]\Bigr).\]
\end{lemma}

\begin{proof}
By definition there are canonical isomorphisms
\[\begin{split} \wo[T] &= \bigtensorover{v\in T} \wo(v)\\
&= \bigtensorover{v\in T} \left(\int^{H_v\in\uTreec(v)} \J[H_v] \otimes \O[H_v]\right)\\
&\cong\int^{\{H_v\}\in \prodover{v\in T} \uTreec(v)} \bigtensorover{v\in T} \Bigl(\J[H_v]\otimes \O[H_v]\Bigr)\\
&\cong \int^{\{H_v\}\in \prodover{v\in T} \uTreec(v)} \Bigl(\bigtensorover{v\in T} \J[H_v]\Bigr) \otimes \Bigl(\bigtensorover{v\in T} \O[H_v]\Bigr).\end{split}\]
The first isomorphism uses the naturality of coends.  The second isomorphism uses the symmetry in $\M$.
\end{proof}

Using Lemma \ref{lem:wo-of-t}, next we define the operad structure on the Boardman-Vogt construction.  We will use Definition \ref{def:operad-tree} of a $\colorc$-colored operad.

\begin{definition}\label{def:wo-operad-structure}
Suppose $\O$ is a $\colorc$-colored operad in $\M$.  For each pair $(\uc;d) \in \Profcc$ and each $\colorc$-colored tree $T$ with profile $(\uc;d)$, define the morphism \[\gamma_T : \wo[T] \to \wo\duc\] by insisting that the diagram\index{operadic structure morphism!Boardman-Vogt construction}
\begin{equation}\label{wo-operad-diagram}
\nicexy{\Bigl(\bigtensorover{v\in T} \J[H_v]\Bigr) \otimes \Bigl(\bigtensorover{v\in T} \O[H_v]\Bigr) \ar[d]_-{\{\omega_{H_v}\}_{v\in T}} \ar[r]^-{(\pi,\cong)} & \J\bigl[T(H_v)_{v\in T}\bigr] \otimes \O\bigl[T(H_v)_{v\in T}\bigr] \ar[d]^-{\omega_{T(H_v)_{v\in T}}}\\ \wo[T] \ar[r]^-{\gamma_T} & \wo\duc}
\end{equation}
be commutative for each $\{H_v\} \in \prod_{v\in T} \uTreec(v)$.  In the top horizontal morphism, $\pi$ is the morphism in Lemma \ref{lem:morphism-pi}, and the isomorphism is from Proposition \ref{prop:vertexdec-treesub}.  The left vertical natural morphism is from Lemma \ref{lem:wo-of-t}.
\end{definition}

\begin{lemma}\label{lem:gammat-welldefined}
The morphism $\gamma_T$ in \eqref{wo-operad-diagram} is well-defined.
\end{lemma}

\begin{proof}
For each vertex $v$ in $T$, suppose 
\[(D_{vu})_{u\in H_v} : H_v(D_{vu})_{u\in H_v} \to H_v\]
is a morphism in $\uTreec(v)$.  In the following diagram, we will abbreviate $H_v$ to $H$ and $D_{vu}$ to $D$, with $v$ and $u$ running through $\Vt(T)$ and $\Vt(H_v)$, respectively.   By Lemma \ref{lem:wo-of-t} it suffices to show that the outermost diagram in
\[\begin{footnotesize}\nicexy@R+.5cm@C-.5cm{\bigl(\bigtensorover{v} \J[H]\bigr) \otimes \bigl(\bigtensorover{v}\O[H(D)]\bigr) \ar[d]_{\tensorover{v} \J} \ar[dr]^-{\pi} \ar[rr]^-{\bigtensorover{v,u}\gammao_{D}} && \bigl(\bigtensorover{v} \J[H]\bigr) \otimes \bigl(\bigtensorover{v} \O[H]\bigr)\ar[d]^-{\pi}\\
\bigl(\bigtensorover{v} \J[H(D)]\bigr)\otimes \bigl(\bigtensorover{v} \O[H(D)]\bigr) \ar[d]_-{\pi} & \J[T(H)] \otimes \O[T(H)(D)] \ar[r]^-{\bigtensorover{v,u}\gammao_{D}} \ar[dl]_-{\J} & \J[T(H)] \otimes \O[T(H)] \ar[d]^-{\omega_{T(H)}}\\
\J[T(H)(D)] \otimes \O[T(H)(D)] \ar[rr]^-{\omega_{T(H)(D)}} && \wo\duc}\end{footnotesize}\]
is commutative, in which identity morphisms and isomorphisms are omitted.  The morphism \[\gammao_{D_{vu}} : \O[D_{vu}] \to \O(u)\] is the operadic structure morphism of $\O$ for $D_{vu}$ in \eqref{operadic-structure-map}.  The top left vertical morphism \[\J : \J[H_v] \to \J[H_v(D_{vu})]\] is the image under the functor $\J$ in Definition \ref{functor-J} of the morphism $(D_{vu})_{u\in H_v}$.  Similarly, the slanted morphism \[\J : \J[T(H_v)] \to \J\bigl[T(H_v)(D_{vu})\bigr]\] is the image under the functor $\J$ of the morphism \[(D_{vu})_{v,u} : T(H_v)(D_{vu}) \to T(H_v)\] in $\uTreec\duc$. The lower right trapezoid is commutative by the coend definition of $\wo\duc$.  The left triangle and the top trapezoid are commutative by inspection.
\end{proof}

\begin{interpretation}Intuitively, the morphism $\gamma_T$ in \eqref{wo-operad-diagram} is given by substituting decorated trees $\bigl\{\J[H_v]\otimes\O[H_v]\bigr\}_{v\in T}$ into $T$, with new internal edges (i.e., those in $T(H_v)_{v\in T}$ that are not in any of the $H_v$) given length $1$.\dqed\end{interpretation}

\begin{theorem}\label{thm:wo-operad}
Suppose $\O$ is a $\colorc$-colored operad in $\M$.  With the operadic structure morphisms $\gamma_T$ in Definition \ref{def:wo-operad-structure}, $\wo$ is a $\colorc$-colored operad.
\end{theorem}

\begin{proof}
For a corolla $\Corucd$ with unique vertex $v$, since \[\Corucd(H_v) = H_v,\] the top horizontal morphism in \eqref{wo-operad-diagram} is the identity morphism.  So $\gamma_{\Corucd}$ is the identity morphism.

To prove the associativity axiom \eqref{operad-tree-ass}, suppose $T(H_v)_{v\in T}$ is a tree substitution with $T \in \uTreec\duc$.  We want to show that the diagram
\begin{equation}\label{wo-gamma-associative}
\nicexy{\bigtensorover{v\in T}\wo[H_v] \ar[d]_-{\cong} \ar[r]^-{\bigtensorover{v}\gamma_{H_v}} & \bigtensorover{v\in T} \wo(v)=\wo[T] \ar[d]^-{\gamma_T}\\ \wo\bigl[T(H_v)_{v\in T}\bigr] \ar[r]^-{\gamma_{T(H_v)}} & \wo\duc}
\end{equation}
is commutative.  By Lemma \ref{lem:wo-of-t} and the naturality of coends, there are canonical isomorphisms
\[\begin{split}
\bigtensorover{v\in T}\wo[H_v]  
&\cong \bigtensorover{v\in T}\int^{\{D_{vu}\} \in \prodover{u\in H_v}\uTreec(u)} \Bigl(\bigtensorover{u\in H_v} \J[D_{vu}] \bigr) \otimes \Bigl(\bigtensorover{u\in H_v} \O[D_{vu}]\Bigr)\\
&\cong \int^{\{D_{vu}\}\in \prodover{v\in T,\,u\in H_v} \uTreec(u)} \Bigl(\bigtensorover{v,u} \J[D_{vu}] \bigr) \otimes \Bigl(\bigtensorover{v,u} \O[D_{vu}]\Bigr).\end{split}\]
In the following diagram, as in the proof of Lemma \ref{lem:gammat-welldefined}, we will abbreviate $H_v$ to $H$ and $D_{vu}$ to $D$, with $v$ and $u$ running through $\Vt(T)$ and $\Vt(H_v)$, respectively, and omit identity morphisms and isomorphisms.  To prove the commutativity of the diagram \eqref{wo-gamma-associative}, it is enough to show that the outermost diagram in
\[\nicexy{\bigl(\bigtensorover{v,u} \J[D]\bigr)\otimes \bigl(\bigtensorover{v,u} \O[D]\bigr) \ar[dd]_-{\pi} \ar[r]^-{\bigtensorover{v}\pi} & \bigl(\bigtensorover{v,u} \J[H(D)]\bigr)\otimes \bigl(\bigtensorover{v,u} \O[H(D)]\bigr)  \ar[d]^-{\pi}\\ & \J[T(H)(D)] \otimes \O[T(H)(D)] \ar[d]^-{\omega_{T(H)(D)}}\\ \J[T(H)(D)] \otimes \O[T(H)(D)] \ar[r]^-{\omega_{T(H)(D)}} \ar@(r,ul)[ur]^-{\Id} & \wo\duc}\]
is commutative.  The upper trapezoid is commutative by inspection, and the triangle is commutative by definition.
\end{proof}

\begin{remark}\label{rk:iterate-w}
By Theorem \ref{thm:wo-operad} the Boardman-Vogt construction can be iterated.  In other words, given a $\colorc$-colored operad $\O$, since $\wo$ is a $\colorc$-colored operad, it has a Boardman-Vogt construction $\W(\wo)$, which is a $\colorc$-colored operad.  Then $\W(\wo)$ has a Boardman-Vogt construction $\W(\W(\wo))$, which is again a $\colorc$-colored operad, and so forth.  However, we do not know of any applications of these iterated Boardman-Vogt constructions.\dqed
\end{remark}

We can also express the operad structure on the Boardman-Vogt construction $\wo$ in terms of the generating operations in Definition \ref{def:operad-generating}.  Using Theorem \ref{thm:operad-def2=def3} and its proof on $\wo$ and Definition \ref{def:wo-operad-structure}, we infer the following result.

\begin{proposition}\label{prop:bv-generating}
Suppose $\O$ is a $\colorc$-colored operad in $\M$.
\begin{enumerate}
\item For each $c \in \colorc$, the $c$-colored unit of $\wo$ is the composition
\[\nicexy{\tensorunit=\wo[\uparrow_c] \ar[r]^-{\cong} & \tensorunit \otimes \tensorunit=\J[\uparrow_c]\otimes\O[\uparrow_c] \ar[r]^-{\omega_{\uparrow_c}} & \wo\cc}\]
in which $\uparrow_c$ is the $c$-colored exceptional edge in Example \ref{ex:colored-exedge}.
\item For each pair $(\uc;d) \in \Profcc$ and permutation $\sigma \in \Sigma_{|\uc|}$, the equivariant structure of $\wo$ is uniquely determined by the commutative diagrams
\[\nicexy{\J[T]\otimes\O[T]\ar[d]_-{\omega_T} \ar[r]^-{\Id} & \J[T\sigma] \otimes\O[T\sigma] \ar[d]^-{\omega_{T\sigma}}\\ \wo\duc \ar[r]^-{\sigma} & \wo\ducsigma}\]
for $T \in \uTreec\duc$, where $T\sigma \in \uTreec\ducsigma$ is the same as $T$ except that its ordering is $\zeta_T\sigma$ with $\zeta_T$ the ordering of $T$.
\item For $\bigl(\uc=(c_1,\ldots,c_n);d) \in \Profcc$ with $n \geq 1$, $\ub_j \in \Profc$ for $1 \leq j \leq n$, and $\ub=(\ub_1,\ldots,\ub_n)$, the operadic composition $\gamma$ of $\wo$ is uniquely determined by the commutative diagrams
\[\begin{small}\nicexy{\J[T]\otimes\O[T] \otimes \bigotimes\limits_{j=1}^n \bigl(\J[T_j]\otimes\O[T_j]\bigr) \ar[dd]_-{\bigl(\omega_T,\bigotimes_j \omega_{T_j}\bigr)} \ar[r]^-{\mathrm{permute}}_-{\cong} & \Bigl(\J[T]\otimes \bigotimes\limits_{j=1}^n \J[T_j]\Bigr) \otimes\Bigl(\O[T] \otimes \bigotimes\limits_{j=1}^n\O[T_j]\Bigr) \ar[d]^-{(\pi,\cong)}\\ & \J[G]\otimes\O[G] \ar[d]^-{\omega_G}\\
\wo\duc\otimes\bigotimes\limits_{j=1}^n \wo\cjubj \ar[r]^-{\gamma} & \wo\dub}\end{small}\]
for $T \in \uTreecduc$, $T_j \in \uTreec\cjubj$ for $1 \leq j \leq n$, and \[G=\graft(T;T_1,\ldots,T_n) \in \uTreec\dub\] the grafting \eqref{def:grafting}.  Here $\pi=\bigotimes_S 1$ is the morphism in Lemma \ref{lem:morphism-pi} for the grafting $G$.
\end{enumerate}
\end{proposition}

\begin{remark}
In the last part of Proposition \ref{prop:bv-generating}, the morphism $\pi$ is $\bigotimes_S 1$, in which $1 : \tensorunit \to J$ is part of the commutative segment $J$.  The set $S$ is defined as the set of internal edges in the grafting $G$ that are neither in $T$ nor in any of the $T_j$.  For example, if neither $T$ nor any of the $T_j$ is an exceptional edge, then $S$ has exactly $n$ elements, one for each input of $T$.\dqed
\end{remark}

\section{Augmentation}\label{sec:augmentation-bv}

In this section, we observe that the Boardman-Vogt construction is augmented over the identity functor on the category of colored operads.  The augmentation induces a change-of-operad adjunction between the category of algebras over the Boardman-Vogt construction and the category of algebras over the original colored operad.  Furthermore, this adjunction is natural with respect to operad morphisms.  In the next section, we will see that in $\Chaink$ the change-of-operad adjunction induced by the augmentation is always a Quillen equivalence.

Recall the concept of an operad morphism in Definition \ref{def:general-operad-map}.  First we define what the Boardman-Vogt construction does to an operad morphism.

\begin{lemma}\label{lem:w-on-maps}
Suppose $f : (\O,\gammao) \to (\P,\gammap)$ is an operad morphism with $\O$ a $\colorc$-colored operad and $\P$ a $\colord$-colored operad.  Then there is an induced operad morphism\index{Boardman-Vogt construction!on operad morphism}\index{operad morphism!Boardman-Vogt construction} \[\wf : \wo \to \wofp\] that is $f : \colorc \to \colord$ on colors and is entrywise defined by the commutative diagram
\[\nicexy@C+.5cm{\J[T]\otimes\O[T] \ar[d]_-{\omega_T} \ar[r]^-{\bigl(\Id,\bigtensorover{v\in T} f\bigr)} & \J[fT] \otimes \P[fT] \ar[d]^-{\omega_{fT}}\\ \wo\duc \ar[r]^-{\wf} & \wofp\fdfuc}\]
for $(\uc;d) \in \Profcc$ and $T \in \uTreec\duc$, where $fT \in \uTreed\fdfuc$ is obtained from $T$ by applying $f$ to its $\colorc$-coloring.
\end{lemma}

\begin{proof} 
To see that the morphism $\wf$ is entrywise well-defined, it is enough to show that the outermost diagram in
\[\begin{small}\nicexy{\J[T]\otimes\O[T(H_v)] \ar[dr]_-{\bigtensorover{v,u}f} \ar[dd]_-{\J} \ar[r]^-{\bigtensorover{v}\gammao_{H_v}} & \J[T]\otimes\O[T] \ar[r]^-{\bigtensorover{v}f} & \J[fT]\otimes\P[fT] \ar[dd]^-{\omega_{fT}}\\
& \J[fT] \otimes \P\bigl[f(T(H_v))\bigr] \ar[ur]_-{\bigtensorover{v}\gammap_{fH_v}} \ar[d]_-{\J} &\\ \J[T(H_v)]\otimes \O[T(H_v)] \ar[r]^-{\bigtensorover{v,u}f} & \J\bigl[f(T(H_v))\bigr] \otimes \P\bigl[f(T(H_v))\bigr] \ar[r]^-{\omega_{f(T(H_v))}} & \wofp\fdfuc}\end{small}\]
is commutative for each $T\in \uTreec\duc$ and $\{H_v\} \in \prod_{v\in T} \uTreec(v)$, in which $v$ and $u$ run through $\Vt(T)$ and $\Vt(H_v)$, respectively.  The top triangle is commutative by Proposition \ref{prop:operad-map} because $f$ is an operad morphism.  The left trapezoid is commutative by inspection.  The right trapezoid is commutative by the coend definition of $\wofp\fdfuc$ because there is a morphism \[(fH_v) : f(T(H_v)) = (fT)(fH_v) \to fT\] in $\uTreed$.

To see that $\wf$ respects the operadic structure morphisms in Definition \ref{def:wo-operad-structure}, by Lemma \ref{lem:wo-of-t}, it is enough to show that the outermost diagram in
\[\nicexy{\bigl(\bigtensorover{v}\J[H_v]\bigr)\otimes \bigl(\bigtensorover{v} \O[H_v]\bigr) \ar[d]_-{\pi} \ar[r]^-{\bigtensorover{v,u}f} & \bigl(\bigtensorover{v}\J[fH_v]\bigr)\otimes \bigl(\bigtensorover{v} \P[fH_v]\bigr) \ar[d]^-{\pi}\\ 
\J[T(H_v)]\otimes \O[T(H_v)] \ar[d]_-{\bigtensorover{v,u}f} & \J\bigl[f(T(H_v))\bigr]\otimes \P\bigl[f(T(H_v))\bigr] \ar[d]^-{\omega_{f(T(H_v))}}\\
\J\bigl[f(T(H_v))\bigr]\otimes \P\bigl[f(T(H_v))\bigr] \ar[r]^-{\omega_{f(T(H_v))}} \ar@(r,ul)[ur]^-{\Id} & \wofp\fdfuc}\]
is commutative.  Both sub-diagrams are commutative by definition.
\end{proof}

\begin{interpretation} Intuitively, the morphism $\wf$ sends each decorated tree $\J[T] \otimes \O[T]$ to the decorated tree $\J[fT] \otimes \P[fT]$ by applying $f$ at each vertex.  The internal edges in $T$ and in $fT$ are canonically identified, so $\J[T]$ and $\J[fT]$ are the same.\dqed\end{interpretation}

Recall from Definition \ref{def:general-operad-map} that $\Operadm$ denotes the category of all colored operads in $\M$.

\begin{proposition}\label{prop:w-functor}
The Boardman-Vogt construction defines a functor\index{Boardman-Vogt construction!naturality} \[\W : \Operadm \to \Operadm\] that preserves color sets.
\end{proposition}

\begin{proof}
The assignment on objects is defined by Theorem \ref{thm:wo-operad}, and the assignment on morphisms is defined by Lemma \ref{lem:w-on-maps}.  The Boardman-Vogt construction of a $\colorc$-colored operad is a $\colorc$-colored operad.  The Boardman-Vogt construction preserves identity morphisms and composition of operad morphisms by the definition in Lemma \ref{lem:w-on-maps}.
\end{proof}

Next we define an augmentation of the Boardman-Vogt construction, which will allow us to compare the Boardman-Vogt construction with the original colored operad.

\begin{theorem}\label{thm:w-augmented}
There is a natural transformation\index{augmentation}\index{Boardman-Vogt construction!augmentation} \[\eta : \W \to \Id_{\Operadm}\] such that, for each $\colorc$-colored operad $(\O,\gammao)$ in $\M$, the operad morphism \[\eta : \wo \to \O\] fixes colors and is defined entrywise by the commutative diagrams
\[\nicexy{\J[T]\otimes\O[T] \ar[d]_-{\omega_T} \ar[r]^-{(\bigtensorover{|T|}\epsilon,\Id)} & \tensorunit[T]\otimes\O[T] \ar[r]^-{\cong} & \O[T] \ar[d]^-{\gammao_T}\\
\wo\duc \ar[rr]^-{\eta} && \O\duc}\]
for $(\uc;d) \in \Profcc$ and $T \in \uTreec\duc$.
\end{theorem}

\begin{proof}
To see that the morphism $\eta$ is entrywise well-defined, suppose \[(H_v)_{v\in T} : T(H_v) \to T\] is a morphism in $\uTreec\duc$.  It is enough to show that the outermost diagram in
\[\nicexy{\J[T]\otimes\O[T(H_v)] \ar[d]_-{\J} \ar[dr]^-{\cong\circ\bigotimes\epsilon} \ar[r]^-{\bigtensorover{v}\gammao_{H_v}} & \J[T] \otimes \O[T] \ar[r]^-{\bigtensorover{|T|}\epsilon} & \tensorunit[T]\otimes\O[T] \ar[d]^-{\cong}\\
\J[T(H_v)]\otimes\O[T(H_v)] \ar[d]_-{\bigtensorover{|T(H_v)|} \epsilon} & \O[T(H_v)] \ar@{=}[d] \ar[r]^-{\bigtensorover{v}\gammao_{H_v}} & \O[T] \ar[d]^-{\gammao_T}\\ 
\tensorunit[T(H_v)]\otimes \O[T(H_v)] \ar[r]^-{\cong} & \O[T(H_v)] \ar[r]^-{\gammao_{T(H_v)}} & \O\duc}\]
is commutative. The top trapezoid is commutativity by definition.  The left trapezoid is commutative by the fact that $\epsilon$ is the counit of the commutative segment $J$.  The lower right square is commutative by the associativity \eqref{operad-tree-ass} of the operadic structure morphism $\gammao$.

To see that $\eta$ is a morphism of $\colorc$-colored operads, we must show that the diagram \[\nicexy{\wo[T] \ar[d]_-{\gamma_T} \ar[r]^-{\bigtensorover{v} \eta} & \O[T] \ar[d]^-{\gammao_T}\\ \wo\duc \ar[r]^-{\eta} & \O\duc}\] is commutative for each $T \in \uTreec\duc$.  By Lemma \ref{lem:wo-of-t}, it is enough to show that the outermost diagram in 
\[\nicexy@C+.6cm{\Bigl(\bigtensorover{v\in T} \J[H_v]\Bigr) \otimes \Bigl(\bigtensorover{v\in T} \O[H_v]\Bigr) \ar[d]_-{\pi} \ar[r]^-{\bigtensorover{v}(\cong \circ \bigotimes\epsilon)} & \bigtensorover{v\in T} \O[H_v] \ar[dd]_-{\cong} \ar[r]^-{\bigtensorover{v}\gammao_{H_v}} & \bigtensorover{v\in T}\O(v) \ar@{=}[d]\\ \J[T(H_v)] \otimes \O[T(H_v)] \ar[d]_-{\bigotimes\epsilon} && \O[T] \ar[d]^-{\gammao_T}\\ 
\tensorunit[T(H_v)] \otimes \O[T(H_v)] \ar[r]^-{\cong} &
\O[T(H_v)] \ar[r]^-{\gammao_{T(H_v)}} & \O\duc}\]
is commutative.  The left sub-diagram is commutative by the fact that $\epsilon$ is the counit of $J$.  The right sub-diagram is commutative by the associativity of $\gammao$.

Finally, suppose $f : \O \to \P$ is an operad morphism with $(\O,\gammao)$ a $\colorc$-colored operad and $(\P,\gammap)$ a $\colord$-colored operad as in Lemma \ref{lem:w-on-maps}.  To show that the diagram
\begin{equation}\label{wf-eta}
\nicexy{\wo \ar[d]_-{\etao} \ar[r]^-{\wf} & \wofp \ar[d]^-{\etap}\\ \O\ar[r]^-{f} & \P}
\end{equation}
is commutative, it suffices to prove it in a typical $(\uc;d)$-entry.  By the coend definition of $\wo\duc$, it is enough to show that the outermost diagram in
\[\nicexy{\J[T]\otimes\O[T] \ar[d]_-{\cong\circ\bigotimes\epsilon} \ar[r]^-{\bigtensorover{v}f} & \J[fT]\otimes\P[fT] \ar[d]^-{\cong\circ\bigotimes\epsilon}\\ \O[T] \ar[d]_-{\gammao_T} \ar[r]^-{\bigtensorover{v}f} & \P[fT]\ar[d]^-{\gammap_{fT}}\\ \O\duc \ar[r]^-{f} & \P\fdfuc}\]
is commutative for each $T \in \uTreec\duc$.  The top square is commutative by naturality.  The bottom square is commutative by Proposition \ref{prop:operad-map}.
\end{proof}

\begin{definition}\label{def:bv-augmentation}
For each $\colorc$-colored operad $\O$ in $\M$, we call the morphism $\eta : \wo \to \O$ of $\colorc$-colored operads the \emph{augmentation} of the Boardman-Vogt construction.
\end{definition}

The following change-of-operad adjunction is a special case of Theorem \ref{thm:change-operad}. 

\begin{corollary}\label{cor:augmentation-adjunction}
For each $\colorc$-colored operad $\O$ in $\M$, the augmentation $\eta : \wo \to \O$ induces an adjunction\index{change-of-operad adjunction!induced by augmentation} \[\nicexy{\algmwo \ar@<2pt>[r]^-{\eta_!} & \algmo \ar@<2pt>[l]^-{\eta^*}}\]
with left adjoint $\eta_!$.
\end{corollary}

\begin{interpretation}This change-of-operad adjunction says that each $\O$-algebra pulls back to a $\wo$-algebra via the augmentation $\eta : \wo \to \O$.  Conversely, the left adjoint $\eta_!$ rectifies each $\wo$-algebra to an $\O$-algebra.  Looking ahead, when $\O$ is a colored operad for algebraic quantum field theories or prefactorization algebras, the change of operad adjunction will allow us to go back and forth between algebraic quantum field theories (resp., prefactorization algebras) and homotopy algebraic quantum field theories (resp., homotopy prefactorization algebras).\dqed\end{interpretation}

When applied to the commutative diagram \eqref{wf-eta} above, the change-of-operad adjunction in Theorem \ref{thm:change-operad} yields the following result.

\begin{corollary}\label{cor:wf-eta-adjunction}
Suppose $f : \O \to \P$ is an operad morphism in $\M$.  Then there is a diagram of change-of-operad adjunctions
\[\nicexy@C+.4cm@R+.3cm{\algmwo \ar@<2pt>[r]^-{(\wf)_!} \ar@<-2pt>[d]_-{\etao_!} 
& \algmwp \ar@<2pt>[l]^-{(\wf)^*} \ar@<-2pt>[d]_-{\etap_!} \\
\algmo \ar@<2pt>[r]^-{f_!} \ar@<-2pt>[u]_-{\etaostar}  
& \algmp \ar@<2pt>[l]^-{\fstar} \ar@<-2pt>[u]_-{\etapstar}}\]
in which \[f_! \circ \etao_! = \etap_! \circ (\wf)_! \andspace \etaostar\circ \fstar = (\wf)^*\circ \etapstar.\]
\end{corollary}

\begin{remark} The equality \[f_! \circ \etao_! = \etap_! \circ (\wf)_!\] says that the left adjoint diagram is commutative.  Similarly, the equality \[\etaostar\circ \fstar = (\wf)^*\circ \etapstar\] says that the right adjoint diagram is commutative.\dqed\end{remark}

\section{Homotopy Morita Equivalence}\label{sec:morita}

In this section, we construct an entrywise section of the augmentation, called the standard section, that preserves some of the operad structure, but is not an operad morphism in general.  Using the standard section, we will observe that in familiar model categories such as $\Top$, $\Sset$, and $\Chaink$, the augmentation is a weak equivalence from the Boardman-Vogt construction to the original colored operad.  Moreover, in $\Chaink$ the augmentation is always a homotopy Morita equivalence; i.e., the change-of-operad adjunction induced by the augmentation is a Quillen equivalence.

\begin{definition}\label{def:bv-coaugmentation}
Suppose $\O$ is a $\colorc$-colored operad in $\M$.  The \index{standard section}\emph{standard section}\label{notation:stsection} is the morphism \[\xi : \O \to \wo \in \M^{\Profcc}\] defined entrywise as the composition
\[\nicexy@C+.3cm{\O\duc \ar[r]^-{\cong} & \J\bigl[\Corucd\bigr] \otimes \O\bigl[\Corucd\bigr] \ar[r]^-{\omega_{\Corucd}} & \wo\duc}\in\M\] for $(\uc;d) \in \Profcc$, where $\Corucd$ is the $(\uc;d)$-corolla in Example \ref{ex:cd-corolla}.
\end{definition}

First we observe that the standard section is an entrywise right inverse of the augmentation.

\begin{proposition}\label{prop:zeta-eta}
Suppose $\O$ is a $\colorc$-colored operad in $\M$.  Then the diagram
\[\nicexy{\O\ar[r]^-{\xi} \ar[dr]_-{\Id} & \wo \ar[d]^-{\eta}\\ & \O}\]
in $\M^{\Profcc}$ is commutative.
\end{proposition}

\begin{proof}
For each pair $(\uc;d) \in \Profcc$, the $(\uc;d)$-entry of the composition $\eta\circ\xi$ is the top-right composition in the commutative diagram
\[\nicexy@C+.3cm{\O\duc \ar[dr]_-{\Id} \ar[r]^-{\cong} & \J\bigl[\Corucd\bigr] \otimes \O\bigl[\Corucd\bigr] \ar[r]^-{\omega_{\Corucd}} \ar[d]_-{\cong} & \wo\duc \ar[d]^-{\eta}\\
& \O\bigl[\Corucd\bigr]\ar[r]^-{\gammao_{\Corucd}} & \O\duc.}\] 
We finish the proof by noting that $\gammao_{\Corucd}$ is the identity morphism on $\O\duc$ by the unity axiom in Definition \ref{def:operad-tree}.\end{proof}

One might hope that the standard section is a morphism of colored operads, but we will see that this is not the case in general.  However, the standard section does preserve some of the operad structure.  

\begin{proposition}\label{prop:zeta-preserves}
Suppose $\O$ is a $\colorc$-colored operad in $\M$.  Then the standard section \[\xi : \O \to \wo \in \M^{\Profcc}\] in Definition \ref{def:bv-coaugmentation} preserves the equivariant structure and the colored units.
\end{proposition}

\begin{proof}
As we explained in the proof of Theorem \ref{thm:operad-def2=def3}, the equivariant structure comes from the operadic structure morphisms $\gamma_{\Corucd\tau}$, where $\Corucd\tau$ is the permuted corolla in Example \ref{ex:cd-permuted-corolla}.  Similarly, the colored units are the operadic structure morphisms $\gamma_{\uparrow_c}$ for the exceptional edges in Example \ref{ex:colored-exedge}. 

To show that the standard section preserves these operadic structure morphisms, consider more generally a $\colorc$-colored tree $T$ with profile $(\uc;d)$.  The standard section preserves the operadic structure morphism for $T$ if and only if the outermost diagram in 
\begin{equation}\label{zeta-preserves-diagram}
\nicexy@C-1.3cm@R+.4cm{\O[T] \ar[dd]_-{\gammao_T} \ar[dr]^-{\cong} \ar[rr]^-{\cong} && \bigtensorover{v\in T} \bigl(\J[\Cor_v]\otimes \O[\Cor_v]\bigr) \ar[dr]^-{\pi} \ar[rr]^-{\bigtensorover{v\in T}\omega_{\Cor_v}} && \wo[T] \ar[dd]^-{\gamma_T}\\
& \J[\Corucd]\otimes\O[T] \ar@{}[ur]|-{(*)} \ar[rr]^-{\J} \ar[dr]^-{\gammao_T} && \J[T]\otimes \O[T] \ar[dr]^-{\omega_T} \ar@{}[ur]|-{(c)} &\\
\O\duc \ar@{}[ur]|-{(a)} \ar[rr]^-{\cong} && \J[\Corucd] \otimes \O[\Corucd] \ar@{}[ur]|-{(b)} \ar[rr]^-{\omega_{\Corucd}} && \wo\duc}
\end{equation}
is commutative, in which $\Cor_v$ is the corolla with the same profile as $v$.  The sub-diagram (a) is commutative by definition.  The sub-diagram (b) is commutative by the coend definition of $\wo\duc$ because \[(T) : T=\Corucd(T) \to \Corucd\] is a morphism in $\uTreec\duc$.  The sub-diagram (c) is commutative by the definition \eqref{wo-operad-diagram} of $\gamma_T$ in $\wo$.  

In the sub-diagram $(*)$, the morphism $\pi$ is defined in Lemma \ref{lem:morphism-pi} and is isomorphic to $\bigotimes_{|T|}1$ with $1 : \tensorunit \to J$ a part of the commutative segment $J$ and $|T|$ the set of internal edges in $T$.  The morphism $\J$ is isomorphic to $\bigotimes_{|T|}0$, where $0 : \tensorunit \to J$ is also a part of the commutative segment.  If $T$ is either a permuted corolla or an exceptional edge, then the set $|T|$ is empty.  In this case, both $\pi$ and $\J$ are the identity morphism of $\tensorunit$, so $(*)$ is also commutative.
\end{proof}

\begin{remark}One can see from the diagram \eqref{zeta-preserves-diagram} that the standard section does not preserve the operadic structure morphism $\gamma_T$ in general.  Indeed, in the sub-diagram $(*)$, the morphisms $\pi = \bigotimes_{|T|} 1$ and $\J=\bigotimes_{|T|}0$ are different for most $T$.  Intuitively, the morphism $\pi$ assigns length $1$ to every internal edge in $T$, while $\J$ assigns length $0$ to every internal edge in $T$.\dqed\end{remark}

In the rest of this section, we will compare the categories of algebras over a colored operad and over its Boardman-Vogt construction.  Recall from Definition \ref{def:admissibility} the concept of a weak equivalence between $\colorc$-colored operads.  Next we observe that in familiar cases, the augmentation is a weak equivalence.

\begin{proposition}\label{prop:wo-weq-o}
Suppose $\M$ is $\Top$, $\Sset$, $\Chaink$, or $\Cat$ with the model category structure in Example \ref{ex:model-cate} and with the commutative segment in Example \ref{ex:com-segment}, and $\O$ is a $\colorc$-colored operad in $\M$.  Then the augmentation $\eta : \wo \to \O$ is a weak equivalence.
\end{proposition}

\begin{proof}
Let us consider the case $\M=\Top$ with $J=[0,1]$; the other cases are proved similarly.  By Proposition \ref{prop:zeta-eta}, we already know that \[\eta \circ \xi = \Id_{\O} \in \M^{\Profcc}.\]   It remains to show that \[\xi \circ \eta : \wo \to \wo\] is homotopic to the identity morphism.  For each $p \in [0,1]$, define $H_p$ by the commutative diagrams
\[\nicexy@C+.5cm{\{t_i\}_{i=1}^{|T|} \times \O[T] \ar[r]^-{\Id} \ar[d] & \bigl\{\min(p,t_i)\bigr\}_{i=1}^{|T|} \times \O[T] \ar[d]\\
[0,1]^{\times |T|} \times \O[T] \ar[d]_-{\omega_T} & [0,1]^{\times |T|}\times \O[T] \ar[d]^-{\omega_T}\\
\wo\duc \ar[r]^-{H_p} & \wo\duc}\] for $T \in \uTreec\duc$ and $t_i\in [0,1]$ for $1 \leq i \leq |T|$.  In other words, replace every internal edge length by its minimum with $p$.  Then $H_1$ is the identity morphism, and $H_0=\xi\circ\eta$ by the coend definition of $\wo\duc$.  So $\{H_p\}_{p\in [0,1]}$ defines a homotopy from $\xi\circ\eta$ to the identity morphism.
\end{proof}

\begin{remark} A statement and a proof similar to Proposition \ref{prop:wo-weq-o} for $\Top$ were first given in \cite{boardman-vogt,vogt}.\dqed\end{remark}

In abstract algebra, two unital associative rings are said to be \index{Morita equivalence}\emph{Morita equivalent} if their categories of left modules are equivalent.  Using the category of algebras, a similar concept of Morita equivalence also makes sense for colored operads.  Moreover, in the presence of a model category structure in the base category, it makes sense to consider a homotopy version of a Morita equivalence.

\begin{definition}\label{def:morita}
Suppose $f : \O \to \P$ is an operad morphism in a monoidal model category $\M$ with $\O$ and $\P$ admissible.  Then we say that $f$ is a \emph{homotopy Morita equivalence}\index{homotopy Morita equivalence} if the change-of-operad adjunction \[\nicexy{\algmo \ar@<2pt>[r]^-{f_!} & \algmp \ar@<2pt>[l]^-{\fstar}}\] is a \index{Quillen equivalence}Quillen equivalence.
\end{definition}

\begin{remark}
Suppose given an operad morphism $f : \O \to \P$ between admissible colored operads, such as the augmentation $\eta : \wo \to \O$ of a colored operad $\O$ in $\M=\Top$, $\Sset$, $\Chaink$, or $\Cat$.  Then the change-of-operad adjunction $f_! \dashv \fstar$ is already a Quillen adjunction.  Indeed, fibrations and acyclic fibrations in the algebra categories are defined entrywise in $\M$, so they are preserved by the right adjoint $\fstar$.  Therefore, the concept of a homotopy Morita equivalence is well-defined.\dqed
\end{remark}

Combining Proposition \ref{prop:wo-weq-o} and Example \ref{ex:chain-operad-comparison}, we obtain the following result that says that the augmentation of each colored operad over $\Chaink$ is a homotopy Morita equivalence. 

\begin{corollary}\label{cor:wo-o-chaink}
Suppose $\O$ is a $\colorc$-colored operad in $\M=\Chaink$, where $\fieldk$ is a field of characteristic zero.  Then the augmentation\index{augmentation!is homotopy Morita equivalence} $\eta : \wo \to \O$ is a homotopy Morita equivalence.  In other words, the change-of-operad adjunction
\[\nicexy{\algmwo \ar@<2pt>[r]^-{\eta_!} & \algmo \ar@<2pt>[l]^-{\eta^*}}\]
induced by the augmentation $\eta : \wo\to\O$ is a Quillen equivalence.
\end{corollary}

Since Quillen equivalences have the $2$-out-of-$3$ property, combining Corollary \ref{cor:wf-eta-adjunction} and Corollary \ref{cor:wo-o-chaink}, we infer that the Boardman-Vogt construction preserves homotopy Morita equivalences over $\Chaink$.

\begin{corollary}\label{cor:w-preserves-morita}
Suppose $f : \O \to \P$ is a homotopy Morita equivalence in $\M=\Chaink$, where $\fieldk$ is a field of characteristic zero.  Then $\wf : \wo \to \wofp$ is also a homotopy Morita equivalence.
\end{corollary}

\begin{remark}Although Corollary \ref{cor:wo-o-chaink} and Corollary \ref{cor:w-preserves-morita} are only stated for $\Chaink$, this is sufficient for most applications to (homotopy) algebraic quantum field theories and (homotopy) prefactorization algebras, which are often considered over $\Chaink$.\dqed\end{remark}

\section{Filtration}\label{sec:filtration}

In this section, we discuss a natural filtration of the Boardman-Vogt construction.  None of this is needed for applications to algebraic quantum field theories and prefactorization algebras.  The rest of this book is independent of this section, so the reader may skip this section safely.

In the coend definition of the Boardman-Vogt construction in Definition \ref{def:bv-operad}, we used the substitution category $\uTreec\duc$ in Definition \ref{def:treesub-category}.  To obtain a natural filtration of the Boardman-Vogt construction, we will use smaller substitution categories.

\begin{definition}\label{def:treen}
For each pair $(\uc;d) \in \Profcc$ and each $n \geq 0$, define the \emph{$n$th substitution category}\index{substitution category} $\uTreec_n\duc$ as the full subcategory of the substitution category $\uTreec\duc$ consisting of $\colorc$-colored trees with profile $\duc$ and with at most $n$ internal edges.
\end{definition}

\begin{example}
If $\uc\not= d$, then $\uTreec_0\duc$ contains only permuted corollas with profile $\duc$.  If   $\uc=d$, then $\uTreec_0\dd$ contains only the linear graph $\Lin_{(d,d)}$ and the $d$-colored exceptional edge $\uparrow_d$.\dqed
\end{example}

\begin{definition}\label{def:wn}
Suppose $\O$ is a $\colorc$-colored operad in $\M$, and $n \geq 0$.  Define the object $\wno \in \M^{\Profcc}$ entrywise as the coend \[\wno\duc=\int^{T\in\uTreecnduc} \J[T]\otimes\O[T]\] for $(\uc;d) \in \Profcc$.  Here \[\J : \uTreecnducop \to \M \andspace \O : \uTreecnduc \to \M\] are the restrictions of the functors in Definition \ref{functor-J} and Corollary \ref{cor:operad-functor-subcat}, respectively.  
\begin{itemize}
\item We call $\wno \in \M^{\Profcc}$ the\index{filtration}\index{Boardman-Vogt construction!filtration} \emph{$n$th filtration of the Boardman-Vogt construction} of $\O$.
\item For $T \in \uTreecnduc$, we write \[\omega_T : \J[T] \otimes \O[T] \to \wno\duc\] for the natural morphism.
\end{itemize}
\end{definition}

\begin{proposition}\label{prop:w-filtration}
Suppose $\O$ is a $\colorc$-colored operad in $\M$.  Then there is a natural diagram
\[\narrowxy{\O\cong\W_0\O\ar[r]^-{\iota_1} & \W_1\O \ar[r]^-{\iota_2} & \W_2\O\ar[r]^-{\iota_3} & \cdots \ar[r] & \colimover{n\geq 1}\,\wno \cong \wo}\]
in $\M^{\Profcc}$, in which $\iota_n$ is defined entrywise by the subcategory inclusion \[\uTreec_{n-1}\duc \subset \uTreecnduc.\]
\end{proposition}

\begin{proof}
The morphism $\O \to \W_0\O$ in $\M^{\Profcc}$ defined entrywise as the composition
\[\nicexy{\O\duc \ar[r]^-{\cong} & \J\bigl[\Corucd\bigr] \otimes \O\bigl[\Corucd\bigr] \ar[r]^-{\omega_{\Corucd}} & \W_0\O\duc}\] and the morphism $\W_0\O\to\O$ in $\M^{\Profcc}$ defined entrywise by the commutative diagrams
\[\nicexy{\J[T]\otimes\O[T]\ar[d]_-{\omega_T} \ar[r]^-{\cong} & \O[T] \ar[d]^-{\gammao_T}\\
\W_0\O\duc\ar[r] & \O}\]
for $T \in \uTreec_0\duc$ are mutual inverses by the coend definition of $\W_0\O\duc$.  The last isomorphism follows from the isomorphism \[\colimover{n\geq 1} ~\uTreecnduc \iso \uTreecduc\] of categories for each $(\uc;d) \in \Profcc$.
\end{proof}

To understand the above filtration better, we will decompose each morphism $\iota_n$ further as a pushout.  To define such a pushout, we will need the following definitions.  Recall the exceptional edges in Example \ref{ex:colored-exedge} and the permuted corollas in Example \ref{ex:cd-permuted-corolla}.  The intuitive meaning of the concepts in the following definition is explained in Remark \ref{rk:utreeceqn}.

\begin{definition}\label{def:weqn}
Suppose $\O$ is a $\colorc$-colored operad in $\M$, and $n \geq 1$.
\begin{enumerate}
\item For each $(\uc;d) \in \Profcc$, define \label{notation:utreeceqn}$\uTreeceqnduc$ as the subcategory of $\uTreecnduc$ consisting of $\colorc$-colored trees with profile $\duc$ and with exactly $n$ internal edges, in which a morphism $(H_v)_{v\in T} : K\to T$ must have $H_v$ a permuted corolla for each $v \in \Vt(T)$.
\item Define $\weqno \in \M^{\Profcc}$ entrywise as the coend\label{notation:weqno}
\[\weqno\duc= \int^{T\in\uTreeceqnduc} \J[T]\otimes\O[T]\]
for $(\uc;d) \in \Profcc$, in which \[\J : \uTreeceqnducop \to \M \andspace \O : \uTreeceqnduc \to \M\] are the restrictions of the functors in Definition \ref{functor-J} and Corollary \ref{cor:operad-functor-subcat}, respectively.
\item In a $\colorc$-colored tree $T$, a \emph{tunnel} is a vertex $v$ with $|\inp(v)|=1$ such that the input and the output have the same color.  For a tunnel $v$ whose input has color $c$, we will write $\uparrow_v$ for the exceptional edge $\uparrow_c$.  The set of all tunnels in $T$ is denoted by\label{notation:tunnels} $\Tun(T)$.  Define the object $\Ominus[T]$ and the morphism \[\nicexy{\Ominus[T]=\colimover{\varnothing\not=S\subseteq\Tun(T)} \O\bigl[T(\uparrow_v)_{v\in S}\bigr] \ar[r]^-{\alpha_T} & \O[T]} \in \M\]
in which the colimit is indexed by the category of non-empty subsets of $\Tun(T)$, where a morphism $S\to S'$ is a subset inclusion $S'\subseteq S$.  The morphisms that define the colimit and the morphism $\alpha_T$ are induced by the functor $\O$ in Corollary \ref{cor:operad-functor-subcat}.
\item For a $\colorc$-colored tree $T \in \uTreecnduc$, define the \emph{decomposition category}\label{notation:doft} $\Doft$ in which an object is a morphism \[(H_v)_{v\in K} : T=K(H_v)_{v\in K} \to K \inspace \uTreecnduc\] such that $\Vt(H_v)\not=\varnothing$ for all $v \in \Vt(K)$ and that at least one $H_u$ has $|H_u|\geq 1$.  A morphism \[(G_u)_{u\in K'} : \bigl(\nicexy@C+.3cm{T \ar[r]^-{(H_u')_{u\in K'}} & K'}\bigr) \to \bigl(\nicexy@C+.3cm{T \ar[r]^-{(H_v)_{v\in K}} & K}\bigr) \inspace \Doft\] is a morphism \[\nicexy@C+.5cm{K=K'(G_u)_{u\in K'} \ar[r]^-{(G_u)_{u\in K'}} & K'} \inspace\uTreecnduc\] such that $\Vt(G_u)\not=\varnothing$ for all $u \in K'$ and that the diagram
\[\nicexy{T=K(H_v)_{v\in K} \ar[r]^-{(H_v)_{v\in K}} \ar[d]_-{\Id} & K=K'(G_u)_{u\in K'} \ar[d]^-{(G_u)_{u\in K'}}\\  T=K'(H_u')_{u\in K'} \ar[r]^-{(H_u')_{u\in K'}} & K'}\]
in $\uTreecnduc$ is commutative.  Identity morphisms and composition are induced by those in $\uTreecnduc$.
\item For a $\colorc$-colored tree $T \in \uTreecnduc$, define the object $\Jminus[T]$ and the morphism \[\nicexy{\Jminus[T] = \colimover{(H_v)_{v\in K} \in \Doft} \J[K] \ar[r]^-{\beta_T} & \J[T]} \in \M\] induced by the functor $\J$ in Definition \ref{functor-J}.
\item For $T \in \uTreecnduc$, define the morphism $\delta_T$ as the pushout product $\beta_T\square \alpha_T$ in the diagram
\[\nicexy{\Jminus[T]\otimes \Ominus[T]\ar[r]^-{(\Id,\alpha_T)} \ar[d]_-{(\beta_T,\Id)} \ar@{}[dr]|-{\mathrm{pushout}} & \Jminus[T]\otimes \O[T]\ar[d] \ar[ddr]^-{(\beta_T,\Id)} &\\
\J[T]\otimes \Ominus[T] \ar[r] \ar[drr]_-{(\Id,\alpha_T)} & \jominus[T] \ar[dr]|-{\delta_T} &\\
&& \J[T]\otimes \O[T]}\]
in $\M$ in which $\jominus[T]$ is defined as the pushout of the square.
\item Define the object $\W'_{n-1}\O\duc$ and the morphism $\rho$
\[\nicexy{\W'_{n-1}\O\duc = \colimover{T\in \uTreeceqnduc}~ \jominus[T] \ar[r]^-{\rho} & \W_{n-1}\O\duc}\in \M\] in which the colimit is defined using the equivariant structure of $\O$.  The morphism $\rho$ is defined by the commutative diagrams
\[\begin{footnotesize}\nicexy@C-.3cm{&& \J[K]\otimes \O[T] \ar[d]_-{(\mathrm{natural},\Id)} \ar@/^1.5pc/[ddddr]^-{(\Id,\bigtensorover{v}\gammao_{H_v})} &\\
&& \Jminus[T]\otimes \O[T] \ar[d] & \\
\J[T]\otimes\O[T(\uparrow_s)]\ar[dd]_-{(\J,\Id)} \ar[r]^-{(\Id,\mathrm{nat.})} & \J[T]\otimes\Ominus[T] \ar[r] & \jominus[T] \ar[d] &\\
&& \W'_{n-1}\O\duc \ar[d]_-{\rho} &\\
\J[T(\uparrow_s)]\otimes\O[T(\uparrow_s)] \ar[rr]^-{\omega_{T(\uparrow_s)}} && \W_{n-1}\O\duc & \J[K]\otimes\O[K]\ar[l]_-{\omega_K}}\end{footnotesize}\]
for 
\begin{itemize}\item $T \in \uTreeceqnduc$, 
\item $\varnothing\not= S \subseteq \Tun(T)$ with $s \in S$, and 
\item objects $(H_v)_{v\in K} : T \to K$ in $\Doft$.
\end{itemize}
\item Define the morphism $\delta$ by the commutative diagrams \[\nicexy{\jominus[T] \ar[d]_-{\mathrm{natural}} \ar[r]^-{\delta_T} & \J[T] \otimes \O[T] \ar[d]^-{\omega_T}\\
\W'_{n-1}\O\duc \ar[r]^-{\delta} & \weqno\duc}\]
for $T \in \uTreeceqnduc$.
\end{enumerate}
\end{definition}

\begin{interpretation}\label{rk:utreeceqn}
Let us explain the intuitive meaning of the concepts in the previous definition.
\begin{itemize}\item In the category $\uTreeceqnduc$, a morphism is only allowed to change the ordering at each vertex.  In particular, there is no effect on the set of internal edges.  
\item $\weqno\duc$ is the coend of the decorated trees $\J[T]\otimes\O[T]$ over the category $\uTreeceqnduc$.  The $\J$ variable is unaffected by the morphisms in $\uTreeceqnduc$ because they do not change the set of internal edges.
\item $\Ominus[T]$ is the sub-object of the vertex-decorated tree $\O[T]=\bigotimes_{v\in T} \O(v)$ in which at least one tunnel is decorated by the corresponding colored unit of $\O$. 
\item $\Jminus[T]$ is the sub-object of the internal edge-decorated tree $\J[T] = J^{\otimes |T|}$ in which at least one internal edge is assigned length $0 : \tensorunit \to J$.  
\item The pushout $\jominus[T]$ is the sub-object of the decorated tree $\J[T]\otimes\O[T]$ such that 
\begin{itemize}\item at least one tunnel is decorated by the corresponding colored unit of $\O$,
\item or at least one internal edge is assigned length $0$, 
\item or both.
\end{itemize}  
\item $\W'_{n-1}\O\duc$ is the colimit of these sub-objects over the category $\uTreeceqnduc$.  The morphism \[\delta : \W'_{n-1}\O\duc\to\weqno\duc\] is the sum of the sub-object inclusions over the category $\uTreeceqnduc$.
\item The morphism \[\rho : \W'_{n-1}\O\duc \to \W_{n-1}\O\duc\] reduces the number of internal edges in each decorated tree in its domain using
\begin{itemize}\item the functor $\J$ for a morphism $T(\uparrow_s)_{s \in S} \to T$;
\item  the functor $\O$ for an object $(H_v)_{v\in K} : T \to K$ in $\Doft$.\dqed
\end{itemize}\end{itemize}\end{interpretation}

The main categorical property of the filtration in Proposition \ref{prop:w-filtration} is the following observation.

\begin{theorem}\label{thm:wn-pushout}
Suppose $\O$ is a $\colorc$-colored operad in $\M$, $n \geq 1$, and $(\uc;d) \in \Profcc$.  Then there is a pushout \[\nicexy{\W'_{n-1}\O\duc \ar[r]^-{\delta} \ar[d]_-{\rho} & \weqno\duc \ar[d]\\ \W_{n-1}\O\duc \ar[r]^-{\iota_n} & \W_n\O\duc}\] in $\M$, in which the right vertical morphism is induced by the subcategory inclusion $\uTreeceqnduc \subset \uTreecnduc$.
\end{theorem}  

\begin{proof}
The commutativity of the square follows from the definition of $\W_n\O\duc$ as a coend over the category $\uTreecnduc$.  To see that it has the universal property of a pushout, first note that a $\colorc$-colored tree $T \in \uTreecnduc$ is either in $\uTreeceqnduc$ or in $\uTreec_{n-1}\duc$, but not both.  Suppose given a commutative solid-arrow diagram
\[\nicexy{\W'_{n-1}\O\duc \ar[r]^-{\delta} \ar[d]_-{\rho} & \weqno\duc\ar[d] \ar[ddr]^-{B}&\\
\W_{n-1}\O\duc \ar[r]^-{\iota_n} \ar[drr]_-{A} & \W_n\O\duc \ar@{.>}[dr]|-{\chi} &\\ && Y}\] in $\M$ for some object $Y$ and morphisms $A$ and $B$.  Then the only possible extension $\chi$ must be defined by (i) the commutative diagram
\[\nicexy{\J[T]\otimes\O[T]\ar[d]_-{\omega_T} \ar[r]^-{\omega_T} & \weqno\duc \ar[d]^-{B}\\
\wno\duc\ar[r]^-{\chi} & Y}\] if $T \in \uTreeceqnduc$, and (ii) the commutative diagram
\[\nicexy{\J[T]\otimes\O[T]\ar[d]_-{\omega_T} \ar[r]^-{\omega_T} & \W_{n-1}\O\duc \ar[d]^-{A}\\ \wno\duc\ar[r]^-{\chi} & Y}\] if $T \in \uTreec_{n-1}\duc$.  Using the coend definition of $\W_n\O\duc$, one checks that this candidate $\chi$ is indeed a morphism $\W_n\O\duc \to Y$ that uniquely extends both $A$ and $B$.
\end{proof}

\begin{remark}\label{rk:bm-bv}
Proposition \ref{prop:w-filtration} and Theorem \ref{thm:wn-pushout} together imply that, in the one-colored case, our coend definition of the Boardman-Vogt construction is isomorphic to the one given by \index{Berger@Berger, C.}Berger and \index{Moerdijk@Moerdijk, I.}Moerdijk \cite{berger-moerdijk-bv}.  The main difference is that in \cite{berger-moerdijk-bv} $\wo$ is entrywise defined as a sequential colimit as in the filtration in Proposition \ref{prop:w-filtration}, in which the morphisms $\iota_n$ are inductively defined using a pushout similar to the one in Theorem \ref{thm:wn-pushout}.  In contrast, our coend definition of $\wo$, which does not appear in \cite{berger-moerdijk-bv}, describes the Boardman-Vogt construction in one step.  The coend definition is crucial for our understanding of $\wo$-algebras, as in the Coherence Theorem \ref{thm:wo-algebra-coherence}.  It will also be important for our study of homotopy algebraic quantum field theories and homotopy prefactorization algebras.\dqed
\end{remark}

\begin{remark}
In nice enough situations (e.g., when $\M=\Chaink$ with $\fieldk$ a field of characteristic zero), one can use the pushouts in Theorem \ref{thm:wn-pushout} to prove that each entry of each $\iota_n$ is an acyclic cofibration, so the same is true for the standard section $\xi : \O \to \wo$.  Furthermore, the augmentation $\eta : \wo \to \O$ is a cofibrant replacement of $\O$ in the model category of $\colorc$-colored operads in $\M$.  In the one-colored case, these properties are proved in \cite{berger-moerdijk-bv} Sections 4 and 5.  In the general colored case and for even more general objects than colored operads, these properties are proved in \cite{bvbook} Chapters 3-7.  We refer the interested reader to these sources for more details.   For applications to (homotopy) algebraic quantum field theories and (homotopy) prefactorization algebras, we will not need to use these homotopical properties.\dqed
\end{remark}

\chapter{Algebras over the Boardman-Vogt Construction}\label{ch:w-algebras}

This chapter is about the structure of algebras over the Boardman-Vogt construction of a colored operad and some key examples.  The categorical setting is the same as before, so $(\M,\otimes,\tensorunit)$ is a cocomplete symmetric monoidal closed category with an initial object $\varnothing$. Unless otherwise specified, $\colorc$ is an arbitrary non-empty set.

\section{Overview}
Since we intend to apply the Boardman-Vogt construction to colored operads for algebraic quantum field theories and prefactorization algebras, it is crucial that we be able to describe explicitly the structure of algebras over the Boardman-Vogt construction.

In Section \ref{sec:algebra-bv} we prove a coherence theorem for algebras over the Boardman-Vogt construction, which describes such an algebra explicitly in terms of certain structure morphisms and four axioms.  Our coend definition of the Boardman-Vogt construction plays an important role here.  It allows us to phrase the structure morphisms and axioms explicitly and non-inductively in terms of trees and tree substitution.  We will use this coherence theorem many times in the rest of this book.  The remaining sections of this chapter contain key examples that will be relevant in the discussion of homotopy algebraic quantum field theories and homotopy prefactorization algebras.  

In Section \ref{sec:hcdiagram} we explain the structure in homotopy coherent diagrams, which are algebras over the Boardman-Vogt construction of the $\C$-diagram operad.  A homotopy coherent $\C$-diagram in $\M$ is a relaxed version of a $\C$-diagram in $\M$ in which functoriality is replaced by specified homotopies that are also structure morphisms.  An algebraic quantum field theory and a prefactorization algebra each has an underlying $\C$-diagram in $\M$.  Therefore, a homotopy algebraic quantum field theory and a homotopy prefactorization algebra each has an underlying homotopy coherent $\C$-diagram.  

In Section \ref{sec:hinverse-bv} we discuss homotopy inverses in homotopy coherent $\C$-diagrams.  In a $\C$-diagram $X$ in $\M$, if $f$ is an isomorphism in $\C$, then the structure morphism $X(f)$ is also invertible.  In a homotopy coherent $\C$-diagram, this invertibility is expressed homotopically with specified homotopies that are also structure morphisms.  This will be important when we discuss a homotopy version of the time-slice axiom in homotopy algebraic quantum field theories and homotopy prefactorization algebras.

In Section \ref{sec:ainfinity-algebra} we discuss $A_{\infty}$-algebras, which are algebras over the Boardman-Vogt construction of the associative operad $\As$.  They are monoids up to coherent higher homotopies.  An algebraic quantum field theory is, in particular, a diagram of monoids.  Therefore, in a homotopy algebraic quantum field theory, each entry is an $A_\infty$-algebra.

In Section \ref{sec:einfinity-algebra} we discuss $E_{\infty}$-algebras, which are algebras over the Boardman-Vogt construction of the commutative operad $\Com$.  They are commutative monoids up to coherent higher homotopies.  Commutative monoids appear in some entries of prefactorization algebras.  In a homotopy prefactorization algebra, certain entries are $E_\infty$-algebras.  

In Section \ref{sec:hcdiagram-ainfinity} we discuss homotopy coherent diagrams of $A_\infty$-algebras.  Every algebraic quantum field theory has an underlying $\C$-diagram of monoids.  So every homotopy algebraic quantum field theory has an underlying homotopy coherent $\C$-diagram  of $A_\infty$-algebras.  Roughly speaking, for a $\C$-diagram of monoids, there are two directions in which homotopy can happen, namely the diagram direction and the monoid direction.  A homotopy coherent $\C$-diagram of $A_\infty$-algebras combines both of these directions.  In particular, it has an underlying homotopy coherent $\C$-diagram in $\M$ as well as an underlying objectwise $A_\infty$-algebra structure.

In Section \ref{sec:hcdiagram-einfinity} we discuss homotopy coherent diagrams of $E_\infty$-algebras. We will see later that there are adjunctions comparing algebraic quantum field theories and prefactorization algebras, although they are usually not equal.  However, as we will see in Section  \ref{sec:diag-com-monoid}, there is one case where they coincide.  When this happens, both the category of algebraic quantum field theories and the category of prefactorization algebras are canonically isomorphic to the category of $\C$-diagrams of commutative monoids in $\M$.  Therefore, in this case homotopy algebraic quantum field theories and homotopy prefactorization algebras have the structure of homotopy coherent diagrams of $E_\infty$-algebras.

\section{Coherence Theorem}\label{sec:algebra-bv}

Recall from Definition \ref{def:operad-algebra-generating} the concept of an algebra over a colored operad.  In this section, we prove the following coherence result for algebras over the Boardman-Vogt construction of a colored operad.

\begin{theorem}\label{thm:wo-algebra-coherence}
Suppose $(\O,\gammao)$ is a $\colorc$-colored operad in $\M$.  Then a $\wo$-algebra \index{Boardman-Vogt construction!algebra}\index{algebra!over the BV construction}\index{Coherence Theorem!for algebras over the BV construction}is exactly a pair $(X,\lambda)$ consisting of 
\begin{itemize}\item a $\colorc$-colored object $X$ in $\M$ and
\item a structure morphism\label{structure morphism!for algebras over the BV construction}
\begin{equation}\label{wo-algebra-restricted}
\nicexy{\J[T]\otimes\O[T]\otimes X_{\uc} \ar[r]^-{\lambda_T} & X_d}\in \M
\end{equation}
for each $(\uc;d) \in \Profcc$ and $T \in \uTreec\duc$
\end{itemize}
that satisfies the following four conditions.
\begin{description}
\item[Associativity] For $\bigl(\uc=(c_1,\ldots,c_n);d\bigr) \in \Profcc$ with $n \geq 1$, $T \in \uTreecduc$, $T_j \in \uTreec\cjubj$ for $1 \leq j \leq n$, $\ub=(\ub_1,\ldots,\ub_n)$, and \[G=\graft(T;T_1,\ldots,T_n) \in \uTreec\dub\] the grafting \eqref{def:grafting}, the diagram
\begin{equation}\label{wo-alg-ass}\begin{footnotesize}
\nicexy@C-3.3cm{& \J[T]\otimes\O[T] \otimes \Bigl(\bigotimes\limits_{j=1}^n \J[T_j]\otimes\O[T_j]\Bigr)\otimes X_{\ub} \ar[dl]_-{\mathrm{permute}}^-{\cong} \ar[dr]^-{\mathrm{permute}}_-{\cong} &\\
\J[T]\otimes\O[T] \otimes \bigotimes\limits_{j=1}^n\Bigl( \J[T_j]\otimes\O[T_j]\otimes X_{\ub_j}\Bigr) \ar[d]_-{(\Id,\bigotimes_j \lambda_{T_j})} && \Bigl(\J[T]\otimes \bigotimes\limits_{j=1}^n \J[T_j]\Bigr) \otimes\Bigl(\O[T] \otimes \bigotimes\limits_{j=1}^n\O[T_j]\Bigr) \otimes X_{\ub} \ar[d]^-{(\pi,\cong,\Id)}\\ \J[T]\otimes\O[T]\otimes X_{\uc} \ar[dr]_-{\lambda_T} 
&& \J[G]\otimes \O[G] \otimes X_{\ub} \ar[dl]^-{\lambda_G}\\ & X_d &}
\end{footnotesize}
\end{equation}
is commutative.  Here $\pi=\bigotimes_S 1$ is the morphism in Lemma \ref{lem:morphism-pi} for the grafting $G$.
\item[Unity] For each $c \in \colorc$, the composition
\begin{equation}\label{wo-alg-unity}
\nicexy{X_c \ar[r]^-{\cong} & \J[\uparrow_c]\otimes \O[\uparrow_c]\otimes X_c \ar[r]^-{\lambda_{\uparrow_c}} & X_c}
\end{equation} 
is the identity morphism of $X_c$.
\item[Equivariance] For each $T \in \uTreec\duc$ and permutation $\sigma \in \Sigma_{|\uc|}$, the diagram 
\begin{equation}\label{wo-alg-eq}
\nicexy@C+.5cm{\J[T]\otimes\O[T]\otimes X_{\uc} \ar[d]_-{(\Id,\sigmainv)} \ar[r]^-{\lambda_T} & X_d \ar@{=}[d]\\
\J[T\sigma]\otimes \O[T\sigma] \otimes X_{\uc\sigma} \ar[r]^-{\lambda_{T\sigma}} & X_d}
\end{equation}
is commutative, in which $T\sigma \in \uTreec\ducsigma$ is the same as $T$ except that its ordering is $\zeta_T\sigma$ with $\zeta_T$ the ordering of $T$.  The permutation $\sigmainv : X_{\uc} \iso X_{\uc\sigma}$ permutes the factors in $X_{\uc}$.
\item[Wedge Condition] For $T \in \uTreec\duc$, $H_v \in \uTreec(v)$ for each $v\in \Vt(T)$, and $K=T(H_v)_{v\in T}$ the tree substitution, the diagram
\begin{equation}\label{wo-alg-wedge}
\nicexy@C+.7cm{\J[T] \otimes \O[K] \otimes X_{\uc} \ar[d]_-{(\J,\Id)} \ar[r]^-{\bigl(\Id,\bigtensorover{v} \gammao_{H_v},\Id\bigr)} & \J[T]\otimes\O[T]\otimes X_{\uc} \ar[d]^-{\lambda_T}\\
\J[K]\otimes\O[K]\otimes X_{\uc} \ar[r]^-{\lambda_K} & X_d}
\end{equation}
is commutative.
\end{description}
A morphism $f : (X,\lambda^X) \to (Y,\lambda^Y)$ of $\wo$-algebras is a morphism of the underlying $\colorc$-colored objects that respects the structure morphisms in \eqref{wo-algebra-restricted} in the obvious sense.
\end{theorem}

\begin{proof}
Given a $\wo$-algebra $(X,\lambda)$ in the sense of Definition \ref{def:operad-algebra-generating}, we define the structure morphism $\lambda_T$ as the composition
\begin{equation}\label{restricted-woalg-def}
\nicexy@C+.5cm{\J[T]\otimes\O[T]\otimes X_{\uc} \ar[r]^-{\lambda_T} \ar[d]_-{(\omega_T,\Id)} & X_d\ar@{=}[d]\\
\wo\duc\otimes X_{\uc} \ar[r]^-{\lambda} & X_d}
\end{equation}
for $T \in \uTreec\duc$.  The wedge condition \eqref{wo-alg-wedge} is satisfied by the coend definition of $\wo\duc$ because \[(H_v)_{v\in T} : K=T(H_v)_{v\in T} \to T\] is a morphism in $\uTreec\duc$.  Using Proposition \ref{prop:bv-generating}, we infer that the above associativity, unity, and equivariance conditions \eqref{wo-alg-ass}-\eqref{wo-alg-eq} follow from those in Definition \ref{def:operad-algebra-generating}.

Conversely, given a pair $(X,\lambda)$ as in the statement above, we define the morphism \[\nicexy{\wo\duc\otimes X_{\uc} \ar[r]^-{\lambda} & X_d}\in\M\] for $(\uc;d)\in \Profcc$ by insisting that the diagram \eqref{restricted-woalg-def} be commutative for all $T \in \uTreec\duc$.  The wedge condition \eqref{wo-alg-wedge} guarantees that this morphism $\lambda$ is entrywise well-defined.  The associativity, unity, and equivariance axioms in Definition \ref{def:operad-algebra-generating} now follow from the assumed associativity, unity, and equivariance conditions \eqref{wo-alg-ass}-\eqref{wo-alg-eq}.
\end{proof}

The following observation says that the colored units of $\O$ also act as the identity on a $\wo$-algebra.  We will use this result when we discuss homotopy inverses in homotopy algebraic quantum field theories and homotopy prefactorization algebras.  Recall the linear graphs in Example \ref{ex:linear-graph}.

\begin{corollary}\label{cor:wo-alg-unit}
Suppose $(\O,\gammao)$ is a $\colorc$-colored operad in $\M$, and $(X,\lambda)$ is a $\wo$-algebra.  Then for each $c \in \colorc$, the diagram
\[\nicexy@C+.5cm{\J[\Lin_{(c,c)}]\otimes \O[\uparrow_c] \otimes X_c \ar[r]^-{(\Id,\gammao_{\uparrow_c},\Id)} \ar[d]_{\Id} & \J[\Lin_{(c,c)}]\otimes \O[\Lin_{(c,c)}]\otimes X_c \ar[d]^-{\lambda_{\Lin_{(c,c)}}}\\
\tensorunit\otimes\tensorunit\otimes X_c \ar[r]^-{\cong} & X_c}\]
is commutative, in which $\Lin_{(c,c)}$ is the linear graph with one vertex and profile $(c,c)$.
\end{corollary}

\begin{proof}
The diagram
\[\nicexy@C+.5cm{\J[\Lin_{(c,c)}]\otimes \O[\uparrow_c] \otimes X_c \ar[r]^-{(\Id,\gammao_{\uparrow_c},\Id)} \ar[d]_-{\Id} & \J[\Lin_{(c,c)}]\otimes \O[\Lin_{(c,c)}]\otimes X_c \ar[d]^-{\lambda_{\Lin_{(c,c)}}}\\
\J[\uparrow_c]\otimes\O[\uparrow_c]\otimes X_c \ar[r]^-{\lambda_{\uparrow_c}} & X_c}\]
is commutative by the wedge condition \eqref{wo-alg-wedge} because \[\nicexy{\Lin_{(c,c)}(\uparrow_c)=~\uparrow_c \ar[r]^-{(\uparrow_c)} & \Lin_{(c,c)}}\] is a morphism in $\uTreec\cc$.  By the unity condition \eqref{wo-alg-unity}, the bottom horizontal morphism $\lambda_{\uparrow_c}$ is the isomorphism $\tensorunit\otimes\tensorunit \otimes X_c \cong X_c$.
\end{proof}

Recall from Corollary \ref{cor:augmentation-adjunction} that the augmentation $\eta : \wo \to \O$ induces a change-of-operad adjunction \[\eta_! : \algmwo \adjoint \algmo : \eta^*.\]  The next observation describes the structure morphisms of a $\wo$-algebra that is the pullback of an $\O$-algebra.

\begin{corollary}\label{cor:woalg-from-oalg}
Suppose $(\O,\gammao)$ is a $\colorc$-colored operad in $\M$, and $(X,\lambda)$ is an $\O$-algebra.  For $T \in \uTreec\duc$, the structure morphism $\lambda_T$ in \eqref{wo-algebra-restricted} for the $\wo$-algebra $\eta^*(X,\lambda)$ is the composition
\[\nicexy{\J[T]\otimes\O[T]\otimes X_{\uc} \ar[rr]^-{\lambda_T} \ar[d]_-{(\bigotimes_{|T|}\epsilon, \Id)} && X_d\\ \tensorunit[T] \otimes\O[T]\otimes X_{\uc} \ar[r]^-{\cong} & \O[T]\otimes X_{\uc} \ar[r]^-{(\gammao_T,\Id)} & \O\duc\otimes X_{\uc} \ar[u]_-{\lambda}}\]
in which $\epsilon : J \to \tensorunit$ is the counit of the commutative segment $J$.
\end{corollary}

\begin{proof}
By Definition \ref{def:pullback-algebra} and \eqref{restricted-woalg-def}, $\lambda_T$ is the composition
\[\nicexy{\J[T]\otimes\O[T]\otimes X_{\uc} \ar[r]^-{\lambda_T} \ar[d]_-{(\omega_T,\Id)}& X_d \\
 \wo\duc \otimes X_{\uc} \ar[r]^-{(\eta,\Id)} & \O\duc\otimes X_{\uc}. \ar[u]_-{\lambda}}\]  Now we observe that \[\eta \circ \omega_T = (\gammao_T)(\cong)\Bigl(\bigotimes_{|T|}\epsilon, \Id\Bigr)\] by the definition of the augmentation in Theorem \ref{thm:w-augmented}.
\end{proof}

\begin{interpretation}
When an $\O$-algebra is regarded as a $\wo$-algebra, the structure morphism $\lambda_T$ is given by first forgetting the length of internal edges using the counit $\epsilon$.  Then one composes the elements in $\O$ using the operadic structure morphism $\gammao_T$, and follows that by the $\O$-action structure morphism.\dqed\end{interpretation}

\section{Homotopy Coherent Diagrams}\label{sec:hcdiagram}

For the next several sections, we will discuss some relevant examples of algebras over the Boardman-Vogt construction. Suppose $\C$ is a small category with object set $\colorc$.  In this section, we discuss algebras over the Boardman-Vogt construction of the colored operad for $\C$-diagrams, called homotopy coherent $\C$-diagrams.  We will explain that these algebras are $\C$-diagrams up to a family of coherent homotopies.  Homotopy coherent diagrams of topological spaces have a long history; see, for example \cite{berger-moerdijk-resolution,cordier-porter,cordier-porter2,vogt73}.  

\begin{motivation}
The physical relevance of homotopy coherent diagrams is that the \index{isotony axiom}isotony axiom in quantum field theory, sometimes called the \index{locality axiom}locality axiom, is not always satisfied in relevant examples; see, for example, \cite{bdhs,bss17}.  Instead, one should expect a homotopy version of functoriality, as suggested in \cite{bs17} Section 5.  Homotopy theory has taught us that when certain properties hold only up to homotopy (for example, homotopy associativity), there is usually a whole family of higher structure that encodes the specific homotopies and their relations.  We will see in the following few sections that the Boardman-Vogt construction is very convenient for encoding such a family of higher structure.  Homotopy coherent diagrams are also closely related to a homotopy version of the time-slice axiom, as we will explain in Section \ref{sec:hinverse-bv}.\dqed  
\end{motivation}

Recall from Example \ref{ex:operad-diag} that there is a $\colorc$-colored operad $\Cdiag$ whose algebras are $\C$-diagrams in $\M$.

\begin{definition}\label{def:hcdiagram}
Objects in the category $\algm(\Wcdiag)$ are called \index{diagram!homotopy coherent}\index{homotopy coherent diagram}\emph{homotopy coherent $\C$-diagrams in $\M$}, where $\Wcdiag$ is the Boardman-Vogt construction of the $\colorc$-colored operad $\Cdiag$.
\end{definition}

When applied to the $\colorc$-colored operad $\Cdiag$, Corollary \ref{cor:augmentation-adjunction} and Corollary \ref{cor:wo-o-chaink} yield the following adjunction.

\begin{corollary}\label{cor:wcdiag-adjunction}
The augmentation\index{Boardman-Vogt construction!of diagram operad} $\eta : \Wcdiag \to \Cdiag$ induces an adjunction \[\nicexy{\algm(\Wcdiag) \ar@<2pt>[r]^-{\eta_!} & \algm(\Cdiag) \ar@<2pt>[l]^-{\eta^*}}\] that is a Quillen equivalence if $\M=\Chaink$ with $\fieldk$ a field of characteristic $0$.
\end{corollary}

\begin{interpretation} This adjunction says that each $\C$-diagram in $\M$ can be regarded as a homotopy coherent $\C$-diagram in $\M$ via the augmentation $\eta$.  The left adjoint $\eta_!$ rectifies each homotopy coherent $\C$-diagram in $\M$ to a $\C$-diagram in $\M$.\dqed\end{interpretation}

Recall the linear graphs $\Lin_{\uc}$ in Example \ref{ex:linear-graph} and the substitution category $\uLinearc\dc$ for linear graphs in Definition \ref{def:treesub-category}.  The objects in $\uLinearc\dc$ are linear graphs with input color $c$ and output color $d$.  Its morphisms are given by tree substitution, but only for linear graphs.  The following is the coherence theorem for homotopy coherent $\C$-diagrams.  

\begin{theorem}\label{thm:hcdiagram}
A homotopy coherent $\C$-diagram in $\M$ is exactly a pair\index{Coherence Theorem!for homotopy coherent diagrams} $(X,\lambda)$ consisting of 
\begin{itemize}\item a $\colorc$-colored object $X$ in $\M$ and
\item a structure morphism\index{structure morphism!for homotopy coherent diagrams}
\begin{equation}\label{hcdiagram-structure-map}
\nicexy{\J[\Lin_{\uc}]\otimes X_{c_0} \ar[r]^-{\lambda_{\uc}^{\uf}} & X_{c_n}}\in \M
\end{equation}
for 
\begin{itemize}\item each profile $\uc=(c_0,\ldots,c_n)\in\Profc$ with $n \geq 0$;
\item each sequence of composable $\C$-morphisms $\uf=(f_1,\ldots,f_n)$ with $f_j \in \C(c_{j-1},c_j)$ for $1\leq j \leq n$
\end{itemize}
\end{itemize}
that satisfies the following three conditions.
\begin{description}
\item[Associativity] Suppose $0 \leq n \leq p$, $\uc=(c_0,\ldots,c_n)$, and $\uc'=(c_n,\ldots,c_p)\in \Profc$.  Suppose $f_j \in \C(c_{j-1},c_j)$ for each $1\leq j \leq p$ with $\uf=(f_1,\ldots,f_n)$ and $\uf'=(f_{n+1},\ldots,f_p)$.  Then the diagram
\begin{equation}\label{hcdiagram-ass}
\nicexy{\J[\Lin_{\uc'}] \otimes \J[\Lin_{\uc}]\otimes X_{c_0} \ar[d]_-{(\Id, \lambda^{\uf}_{\uc})} \ar[r]^-{(\pi,\Id)} & \J[\Lin_{(c_0,\ldots,c_p)}]\otimes X_{c_0} \ar[d]^-{\lambda^{(\uf,\uf')}_{(c_0,\ldots,c_p)}} \\
\J[\Lin_{\uc'}]\otimes X_{c_n} \ar[r]^-{\lambda^{\uf'}_{\uc'}} & X_{c_p}}
\end{equation}
is commutative, in which $\Lin_{(c_0,\ldots,c_p)}$ is regarded as the grafting \eqref{def:grafting} of $\Lin_{\uc'}$ and $\Lin_{\uc}$ with $\pi$ the morphism in Lemma \ref{lem:morphism-pi}.
\item[Unity] For each $c \in \colorc$, the composition
\begin{equation}\label{hcdiagram-unity}
\nicexy{X_c \ar[r]^-{\cong} & \J[\Lin_{(c)}]\otimes X_c \ar[r]^-{\lambda^{\varnothing}_{(c)}} & X_c}
\end{equation} 
is the identity morphism of $X_c$, where $\Lin_{(c)} =~\uparrow_c$ is the $c$-colored exceptional edge.
\item[Wedge Condition] Suppose \[\begin{split}\uc&=(c_0,\ldots,c_n)\in \Profc \text{ with $n \geq 1$},\\ \ub_j&=\bigl(c_{j-1}=b^j_0,b^j_1,\ldots,b^j_{k_j}=c_j\bigr)\in \Profc \text{ with $k_j\geq 0$ for $1 \leq j \leq n$, and}\\ \ub&=(\ub_1,\ldots,\ub_n).\end{split}\]
Suppose \[\begin{split}f^j_i &\in \C(b^j_{i-1},b^j_i) \text{ for each $1\leq j \leq n$ and $1 \leq i \leq k_j$},\\ \uf^j& =(f^j_1,\ldots,f^j_{k_j}),\\ \uf& =(\uf^1,\ldots,\uf^n),\text{ and}\\ f^j& =f^j_{k_j}\comp \cdots \comp f^j_1 \in \C(c_{j-1},c_j).\end{split}\]  
Then the diagram
\begin{equation}\label{hcdiagram-wedge}
\nicexy@C+1.5cm{\J[\Lin_{\uc}] \otimes X_{c_0} \ar[d]_-{(\J,\Id)} \ar[r]^-{\lambda^{(f^1,\ldots,f^n)}_{\uc}} & X_{c_n} \ar@{=}[d]\\
\J[\Lin_{\ub}]\otimes X_{c_0} \ar[r]^-{\lambda^{\uf}_{\ub}} & X_{c_n}}
\end{equation}
is commutative, in which $\Lin_{\ub}$ is regarded as the tree substitution \[\Lin_{\ub}=\Lin_{\uc}\Bigl(\Lin_{\ub_j}\Bigr)_{j=1}^n.\]
\end{description}
\end{theorem}

\begin{proof}
This is the special case of the Coherence Theorem \ref{thm:wo-algebra-coherence} for the $\colorc$-colored operad $\O=\Cdiag$.  Indeed, recall from Example \ref{ex:operad-diag} that the $\colorc$-colored operad $\Cdiag$ is concentrated in unary entries:
\[\Cdiag\duc = \begin{cases} \coprodover{\C(c,d)}\tensorunit & \text{ if $\uc=c \in \colorc$},\\ \varnothing & \text{ if $|\uc|\not=1$}\end{cases}\]
for $(\uc;d) \in \Profcc$.  Its equivariant structure is trivial.  Its colored units and operadic composition  come from the identity morphisms and the categorical composition in $\C$.  Since $\Cdiag$ is concentrated in unary entries, if $T \in \uTreec\duc$ is not a linear graph, then \[\Cdiag[T] = \bigtensorover{v\in T} \Cdiag\inoutv =\varnothing.\]  So when $T$ is not a linear graph, the structure morphism \[\nicexy{\J[T]\otimes\Cdiag[T]\otimes X_{\uc} \ar[r]^-{\lambda_T} & X_d}\] in \eqref{wo-algebra-restricted} for a $\Wcdiag$-algebra is the trivial morphism $\varnothing \to X_d$.  In particular, the equivariance condition \eqref{wo-alg-eq} is trivial for $\Wcdiag$-algebras.

For $\uc=(c_0,\ldots,c_n) \in \Profc$ with $n\geq 0$, there is a natural isomorphism
\[\Cdiag[\Lin_{\uc}] = \bigotimes_{j=1}^n \Cdiag\sbinom{c_j}{c_{j-1}} = \bigotimes_{j=1}^n \biggl[\coprodover{\C(c_{j-1},c_j)} \tensorunit\biggr] \cong \coprodover{\prod_{j=1}^n \C(c_{j-1},c_j)} \tensorunit.\]
This implies that there is a natural isomorphism \[\J[\Lin_{\uc}] \otimes \Cdiag[\Lin_{\uc}] \otimes X_{c_0} \cong \coprodover{\prod_{j=1}^n \C(c_{j-1},c_j)} \J[\Lin_{\uc}]\otimes X_{c_0}.\]  So the structure morphism \[\nicexy{\J[\Lin_{\uc}] \otimes \Cdiag[\Lin_{\uc}] \otimes X_{c_0} \ar[r]^-{\lambda_{\Lin_{\uc}}} & X_{c_n}}\]
in \eqref{wo-algebra-restricted} is uniquely determined by the restrictions $\lambda^{\uf}_{\uc}$ as stated in \eqref{hcdiagram-structure-map}.  The associativity, unity, and wedge conditions \eqref{hcdiagram-ass}-\eqref{hcdiagram-wedge} above are exactly those in the Coherence Theorem \ref{thm:wo-algebra-coherence} for linear graphs.
\end{proof}

\begin{interpretation}
Intuitively, one should think of the structure morphism $\lambda^{\uf}_{\uc}$ in \eqref{hcdiagram-structure-map} as determined by the decorated linear graph
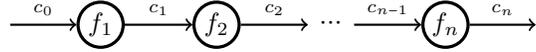
\begin{center}\begin{tikzpicture}
\matrix[row sep=.5cm, column sep=.9cm]{\node [plain] (1) {$f_1$}; & \node [plain] (2) {$f_2$}; & \node [empty] (3) {$\cdots$}; & \node [plain] (n) {$f_n$};\\};
\draw [inputleg] (1) to node[swap]{\scriptsize{$c_0$}} +(-1.2cm,0);
\draw [arrow] (1) to node{\scriptsize{$c_1$}} (2);
\draw [arrow] (2) to node{\scriptsize{$c_2$}} (3);
\draw [arrow] (3) to node{\scriptsize{$c_{n-1}$}} (n);
\draw [outputleg] (n) to node{\scriptsize{$c_n$}} +(1.2cm,0);
\end{tikzpicture}\end{center}
with $n$ vertices decorated by the $\C$-morphisms $f_j$.  The colors $c_j$ are the colors of the edges.  If $n=0$, then this is the $c_0$-colored exceptional edge $\uparrow_{c_0}$ with $\uc=(c_0)$ and $(f_j)=\varnothing$.\dqed
\end{interpretation}

Suppose $(X,\lambda)$ is a $\Wcdiag$-algebra, i.e., a homotopy coherent $\C$-diagram in $\M$.  In the next few examples, we will explain some of the structure on $X$ that suggests that it is a $\C$-diagram up to coherent higher homotopies.

\begin{example}[Assignment on morphisms]\label{ex1:hcdiagram}
For each morphism $f \in \C(c,d)$, the structure morphism in \eqref{hcdiagram-structure-map} yields the morphism \[\nicexy{X_c \ar[rr]^-{X_f} \ar[dr]_-{\cong} && X_d\\  & \J[\Lin_{(c,d)}]\otimes X_c \ar[ur]_-{\lambda^f_{(c,d)}} &}\] in $\M$.  If furthermore $f=\Id_c$, then $X_{\Id_c}$ is the identity morphism of $X_c$ by Corollary \ref{cor:wo-alg-unit}.  In $X$ is an actual $\C$-diagram, then it would preserve composition, i.e., $X_{fg} = X_f \circ X_g$ whenever $fg$ is defined.  For a homotopy coherent $\C$-diagram, we will see in the next example that $X_{(-)}$ preserves composition up to a specified homotopy.\dqed
\end{example}

\begin{example}[Homotopy preservation of composition]\label{ex2:hcdiagram}
Suppose $(f,g) \in \C(c,d)\times\C(b,c)$ is a pair of\index{homotopy functoriality} composable $\C$-morphisms.  Consider the diagram 
\[\nicexy@C+.5cm{\J[\Lin_{(b,d)}]\otimes X_b \ar[d]_-{(0,\Id)} \ar[dr]^-{\lambda^{fg}_{(b,d)}} & X_b \ar[l]_-{\cong} \ar[d]^-{X_{fg}} \\ 
\J[\Lin_{(b,c,d)}]\otimes X_b \ar[r]_-{\lambda^{(g,f)}_{(b,c,d)}} \ar@{}[ur]|(.35){(1)} & X_d\\
\tensorunit \otimes X_b \ar[u]^-{(1,\Id)} \ar@{}[r]|{(2)} &\\
\J[\Lin_{(c,d)}]\otimes\J[\Lin_{(b,c)}]\otimes X_b \ar[u]^-{\cong} \ar[r]^-{(\Id,\lambda^g_{(b,c)})} & \J[\Lin_{(c,d)}] \otimes X_c \ar[uu]^-{\lambda^f_{(c,d)}}\\
X_b \ar[u]^-{\cong} \ar[r]^-{X_g} & X_c \ar[u]^-{\cong} \ar@/_4pc/[uuu]_-{X_f}}\]
in $\M$, in which \[\J[\Lin_{(b,c,d)}]=J,\qquad \J[\Lin_{(b,d)}]=\J[\Lin_{(c,d)}]=\J[\Lin_{(b,c)}]=\tensorunit,\] and $0,1 : \tensorunit \to J$ are part of the commutative segment $J$.  This diagram is commutative:
\begin{itemize}\item The upper right triangle is the definition of $X_{fg}$.
\item The sub-diagram (1) is commutative by the wedge condition \eqref{hcdiagram-wedge} with $n=1$, $\uc=(b,d)$, $\ub_1=(b,c,d)$, and $\uf^1=(g,f)$.  
\item The sub-diagram (2) is commutative by the associativity condition \eqref{hcdiagram-ass} with $n=1$, $p=2$, $\uc=(b,c)$, $\uc'=(c,d)$, $\uf=(g)$, and $\uf'=(f)$.
\item The bottom rectangle is commutative by naturality and the definition of $X_g$.
\item The lower right stripe is the definition of $X_f$.
\end{itemize}

We will call the morphisms $0,1 : \tensorunit\to J$ the $0$-end and the $1$-end of $J$, respectively.  The above commutative diagram says that the structure morphism $\lambda^{(g,f)}_{(b,c,d)}$ is $X_{fg}$ at the $0$-end and the composition $X_f\circ X_g$ at the $1$-end.  So the structure morphism $\lambda^{(g,f)}_{(b,c,d)}$ is a homotopy from $X_{fg}$ to the composition $X_f\circ X_g$.  Therefore, a homotopy coherent $\C$-diagram preserves composition up to a specified homotopy.  It is important to observe that we are not just saying that the morphisms $X_{fg}$ and $X_f\circ X_g$ are homotopic.  Instead, a specific structure morphism $\lambda^{(g,f)}_{(b,c,d)}$ of a homotopy coherent $\C$-diagram acts as the homotopy.  There are higher homotopies for longer strings of composable $\C$-morphisms, as we will see in the next example.\dqed
\end{example}

\begin{example}[Homotopy preservation of triple composition]\label{ex3:hcdiagram}
Suppose given a triple of composable $\C$-morphisms \[(f,g,h) \in \C(c,d)\times\C(b,c)\times\C(a,b).\]  Consider the diagram
\[\nicexy@C+.6cm{J\otimes \tensorunit\otimes X_a \ar[r]^-{\cong} \ar[d]_-{(\Id,0,\Id)} & \J[\Lin_{(a,c,d)}]\otimes X_a \ar[d]^-{\lambda^{(gh,f)}_{(a,c,d)}}\\ 
\J[\Lin_{(a,b,c,d)}] \otimes X_a \ar[r]^-{\lambda^{(h,g,f)}_{(a,b,c,d)}} & X_d\\
J \otimes \tensorunit\otimes X_a \ar[u]^-{(\Id,1,\Id)} &\\
\J[\Lin_{(b,c,d)}] \otimes\J[\Lin_{(a,b)}]\otimes X_a \ar[u]^-{\cong} \ar[r]^-{(\Id,\lambda^h_{(a,b)})} & \J[\Lin_{(b,c,d)}]\otimes X_b \ar[uu]_-{\lambda^{(g,f)}_{(b,c,d)}}}\]
in $\M$, in which 
\[\begin{split}\J[\Lin_{(a,b,c,d)}] &\cong J \otimes J,\quad \J[\Lin_{(a,b)}]=\tensorunit,\\
\J[\Lin_{(a,c,d)}]&= \J[\Lin_{(b,c,d)}]=J.\end{split}\]
This diagram is commutative:
\begin{itemize}\item The top rectangle is commutative by the wedge condition \eqref{hcdiagram-wedge} with $n=2$, $\uc=(a,c,d)$, $\ub_1=(a,b,c)$, $\ub_2=(c,d)$, $\uf^1=(h,g)$, and $\uf^2=(f)$.
\item The bottom square is commutative by the associativity condition \eqref{hcdiagram-ass} with $n=1$, $p=3$, $\uc=(a,b)$, $\uc'=(b,c,d)$, $\uf=(h)$, and $\uf'=(g,f)$.
\end{itemize}
This commutative diagram says that the structure morphism $\lambda^{(h,g,f)}_{(a,b,c,d)}$ yields a higher homotopy from $\lambda^{(gh,f)}_{(a,c,d)}$ to the composition $\lambda^{(g,f)}_{(b,c,d)}\circ (\Id,\lambda^h_{(a,b)})$.  

Furthermore, as explained in Example \ref{ex2:hcdiagram}:
\begin{itemize} \item $\lambda^{(gh,f)}_{(a,c,d)}$ is a homotopy from $X_{fgh} : X_a \to X_d$ to $X_f\circ X_{gh} : X_a \to X_d$.
\item $\lambda^{(g,f)}_{(b,c,d)}$ is a homotopy from $X_{fg} : X_b \to X_d$ to $X_f\circ X_g : X_b \to X_d$.
\end{itemize}
Altogether the above commutative diagram expresses a specific homotopy from $X_{fgh}$ to $X_f\circ X_g\circ X_h$.\dqed
\end{example}

\begin{example}[Homotopy preservation of triple composition]\label{ex4:hcdiagram}
In Example \ref{ex3:hcdiagram} the commutative diagram only uses one copy of $J$ to express a higher homotopy between the homotopies $\lambda^{(gh,f)}_{(a,c,d)}$ and $\lambda^{(g,f)}_{(b,c,d)}\circ (\Id,\lambda^h_{(a,b)})$.  There is a similar commutative diagram
\[\nicexy@C+.6cm{\tensorunit \otimes J\otimes X_a \ar[r]^-{\cong} \ar[d]_-{(0,\Id)} & \J[\Lin_{(a,b,d)}]\otimes X_a \ar[d]^-{\lambda^{(h,fg)}_{(a,b,d)}}\\ 
\J[\Lin_{(a,b,c,d)}] \otimes X_a \ar[r]^-{\lambda^{(h,g,f)}_{(a,b,c,d)}} & X_d\\
\tensorunit\otimes J \otimes X_a \ar[u]^-{(1,\Id)} &\\
\J[\Lin_{(c,d)}]\otimes \J[\Lin_{(a,b,c)}] \otimes X_a \ar[u]^-{\cong} \ar[r]^-{(\Id,\lambda^{(h,g)}_{(a,b,c)})} & \J[\Lin_{(c,d)}]\otimes X_c \ar[uu]_-{\lambda^{f}_{(c,d)}}}\]
in $\M$ that makes use of the other copy of $J$ in $\J[\Lin_{(a,b,c,d)}]$.  Once again the top rectangle is commutative by the wedge condition \eqref{hcdiagram-wedge}, and the bottom square is commutative by the associativity condition \eqref{hcdiagram-ass}.  

This commutative diagram says that the structure morphism $\lambda^{(h,g,f)}_{(a,b,c,d)}$ yields:
\begin{itemize}\item a higher homotopy from $\lambda^{(h,fg)}_{(a,b,d)}$ to the composition $\lambda^{f}_{(c,d)}\circ (\Id,\lambda^{(h,g)}_{(a,b,c)})$;
\item another homotopy from $X_{fgh}$ to $X_f\circ X_g\circ X_h$.
\end{itemize}
For longer strings of composable $\C$-morphisms, there are similar commutative diagrams that express the structure morphisms $\lambda^{\uf}_{\uc}$ as a family of coherent higher homotopies.  The main point is that we are not trying to write down this infinite family of coherent homotopies from the ground up.  Instead, all of them are neatly packaged in the Boardman-Vogt construction $\Wcdiag$ of the $\colorc$-colored operad $\Cdiag$.\dqed
\end{example}

\section{Homotopy Inverses}\label{sec:hinverse-bv}

In this section, we discuss a homotopy coherent version of an inverse using the Boardman-Vogt construction.  

\begin{motivation} Physically homotopy inverses are homotopy manifestations of the time-slice axiom in both homotopy algebraic quantum field theories and homotopy prefactorization algebras.  The upshot of the time-slice axiom is that certain structure morphisms  in algebraic quantum field theories are supposed to be invertible, e.g., if they correspond to Cauchy morphisms between oriented, time-oriented, and globally hyperbolic Lorentzian manifolds.  The homotopy version of the time-slice axiom says that these structure morphisms are invertible up to specified homotopies.

As in Section \ref{sec:hcdiagram}, suppose $\C$ is a small category with object set $\colorc$.  If $X$ is a $\C$-diagram in $\M$ and if $f \in \C(c,d)$ is an isomorphism, then the morphism $X_f : X_c \to X_d$ in $\M$ is also an isomorphism with inverse $X_{\finverse}$, since \[\begin{split}X_f\circ X_{\finverse}&=X_{f\circ\finverse} = X_{\Id_d} = \Id_{X_d},\\ 
X_{\finverse}\circ X_f &= X_{\finverse \circ f} = X_{\Id_c} = \Id_{X_c}.\end{split}\]  If $X$ is a homotopy coherent $\C$-diagram, then we should replace the first equality in each line with a specified homotopy.  In other words, $X_f$ and $X_{\finverse}$ should be homotopy inverses of each other via specific structure morphisms.  We will explain this in the following result.\dqed\end{motivation}

We will reuse the notation in Example \ref{ex1:hcdiagram}.  

\begin{corollary}\label{cor:hinverse}
Suppose $(X,\lambda)$ is a homotopy coherent $\C$-diagram in $\M$, and $f \in \C(c,d)$ is an isomorphism with inverse $\finverse \in \C(d,c)$.  Then the morphisms \[X_f : X_c \to X_d \andspace X_{\finverse} : X_d \to X_c \in \M\] are \index{homotopy inverse}\index{homotopy coherent diagram!homotopy inverse}homotopy inverses of each other in the following sense.
\begin{enumerate}
\item $X_{\finverse}$ is a left homotopy inverse of $X_f$ in the sense that the diagram
\[\nicexy@C+.5cm{\J[\Lin_{(c,c)}]\otimes X_c \ar[d]_-{(0,\Id)} \ar[dr]^-{\lambda^{\finverse f}_{(c,c)}} & X_c \ar[l]_-{\cong} \ar[d]^-{\Id_{X_c}}\\ 
\J[\Lin_{(c,d,c)}] \otimes X_c \ar[r]_-{\lambda^{(f,\finverse)}_{(c,d,c)}} & X_c\\
\tensorunit \otimes X_c \ar[u]^-{(1,\Id)}&\\
\J[\Lin_{(d,c)}]\otimes\J[\Lin_{(c,d)}]\otimes X_c \ar[u]^-{\cong} \ar[r]^-{(\Id,\lambda^f_{(c,d)})} & \J[\Lin_{(d,c)}] \otimes X_d \ar[uu]^-{\lambda^{\finverse}_{(d,c)}}\\
X_c \ar[u]^-{\cong} \ar[r]^-{X_f} & X_d \ar[u]^-{\cong} \ar@/_4pc/[uuu]_-{X_{\finverse}}}\]
in $\M$ is commutative, in which \[\J[\Lin_{(c,d,c)}] = J\andspace \J[\Lin_{(c,c)}]=\J[\Lin_{(d,c)}]=\J[\Lin_{(c,d)}]=\tensorunit.\]  
\item $X_{\finverse}$ is a right homotopy inverse of $X_f$ in the sense that the diagram
\[\nicexy@C+.5cm{\J[\Lin_{(d,d)}]\otimes X_d \ar[d]_-{(0,\Id)} \ar[dr]^-{\lambda^{f\finverse}_{(d,d)}} & X_d \ar[l]_-{\cong} \ar[d]^-{\Id_{X_d}}\\ 
\J[\Lin_{(d,c,d)}] \otimes X_d \ar[r]_-{\lambda^{(\finverse,f)}_{(d,c,d)}} & X_d\\
\tensorunit \otimes X_d \ar[u]^-{(1,\Id)}&\\
\J[\Lin_{(c,d)}]\otimes\J[\Lin_{(d,c)}]\otimes X_d \ar[u]^-{\cong} \ar[r]^-{(\Id,\lambda^{\finverse}_{(d,c)})} & \J[\Lin_{(c,d)}] \otimes X_c \ar[uu]^-{\lambda^{f}_{(c,d)}}\\
X_d \ar[u]^-{\cong} \ar[r]^-{X_{\finverse}} & X_c \ar[u]^-{\cong} \ar@/_4pc/[uuu]_-{X_{f}}}\]
in $\M$ is commutative, in which $\J[\Lin_{(d,c,d)}] = J$.  
\end{enumerate}
\end{corollary}

\begin{proof}
The first assertion is the special case of Example \ref{ex2:hcdiagram} for the composable pair of $\C$-morphisms \[(\finverse,f) \in \C(d,c)\times\C(c,d),\] since by Example \ref{ex1:hcdiagram} $X_{\Id_c}$ is equal to $\Id_{X_c}$.  Similarly, the second assertion is the special case of Example \ref{ex2:hcdiagram} for the composable pair of $\C$-morphisms \[(f,\finverse) \in \C(c,d)\times\C(d,c).\] 
\end{proof}

\begin{interpretation}
In a homotopy coherent $\C$-diagram $(X,\lambda)$, the structure morphism $X_f$ for an invertible morphism $f$ in $\C$ has the structure morphism $X_{\finverse}$ as a two-sided  homotopy inverse.  Moreover, the two homotopies are the structure morphisms $\lambda^{(f,\finverse)}_{(c,d,c)}$ and $\lambda^{(\finverse,f)}_{(d,c,d)}$.  Therefore, a homotopy inverse and the homotopies are already encoded in the Boardman-Vogt construction $\Wcdiag$ of the colored operad $\Cdiag$.\dqed
\end{interpretation}

\section{$A_\infty$-Algebras}\label{sec:ainfinity-algebra}

In this section, we discuss a homotopy version of monoids, called \index{strongly homotopy associative algebra}strongly homotopy associative algebras or $A_{\infty}$-algebras, as algebras over the Boardman-Vogt construction of the associative operad.  

\begin{motivation} Recall from Example \ref{ex:operad-as} that the associative operad $\As$ is a $1$-colored operad in $\M$ whose category of algebras is canonically isomorphic to the category of monoids in $\M$ (Definition \ref{def:monoid}).  The physical relevance of $A_\infty$-algebras is that an algebraic quantum field theory is a diagram of monoids satisfying the causality axiom and possibly the time-slice axiom.  Strict associativity is not a homotopy invariant concept.  Instead, the work of \index{Stasheff@Stasheff,J.}Stasheff \cite{stasheff} taught us that a homotopy version of a monoid is an $A_\infty$-algebra.  Therefore, $A_\infty$-algebras will arise naturally in the study of homotopy algebraic quantum field theories.\dqed\end{motivation}

\begin{definition}\label{def:ainfinity-algebra}
Objects in the category $\algm(\Was)$ are called \index{ainfinityalgebra@$A_\infty$-algebra}\emph{$A_\infty$-algebras in $\M$}, where $\Was$ is the Boardman-Vogt construction of the associative operad $\As$.
\end{definition}

When applied to the associative operad, Corollary \ref{cor:augmentation-adjunction} and  Corollary \ref{cor:wo-o-chaink} yield the following adjunction.

\begin{corollary}\label{cor:was-adjunction}
The augmentation $\eta : \Was \to \As$ induces an adjunction \[\nicexy{\algm(\Was) \ar@<2pt>[r]^-{\eta_!} & \algm(\As) \ar@<2pt>[l]^-{\eta^*}}\] that is a Quillen equivalence if $\M=\Chaink$ with $\fieldk$ a field of characteristic zero.
\end{corollary}

\begin{interpretation} This adjunction says that each monoid in $\M$ can be regarded as an $A_\infty$-algebra in $\M$ via the augmentation $\eta$.  The left adjoint $\eta_!$ rectifies each $A_\infty$-algebra in $\M$ to a monoid in $\M$.\dqed\end{interpretation}

Since in this section we are discussing $1$-colored operads, we will be using $1$-colored trees.  The substitution category, as in Definition \ref{def:treesub-category}, of $1$-colored trees with $n$ inputs is denoted by $\uTree(n)$.  Its objects are $1$-colored trees with $n$ inputs, and its morphisms are given by tree substitution.  The following result is the coherence theorem for $A_\infty$-algebras.

\begin{theorem}\label{thm:ainfinity-algebra}
An $A_\infty$-algebra in $\M$ is exactly a pair\index{Coherence Theorem!for $A_\infty$-algebras} $(X,\lambda)$ consisting of
\begin{itemize}
\item an object $X \in \M$ and
\item a structure morphism\index{structure morphism!for $A_\infty$-algebras}
\begin{equation}\label{ainfinity-structure}
\nicexy@C+.8cm{\J[T] \otimes X^{\otimes n} \ar[r]^-{\lambda_T^{\{\sigma_v\}_{v\in T}}} & X}\in\M
\end{equation}
for 
\begin{itemize}\item each $T \in \uTree(n)$ with $n \geq 0$ and 
\item each $\{\sigma_v\}_{v\in T} \in \prod_{v\in T} \Sigma_{|\inp(v)|}$
\end{itemize}
\end{itemize}
that satisfies the following four conditions.
\begin{description}
\item[Associativity] Suppose $T \in \uTree(n)$ with $n \geq 1$, $T_j \in \uTree(k_j)$ with $k_j \geq 0$ and $1 \leq j \leq n$, $k=k_1+\cdots+k_n$, \[G=\graft(T;T_1,\ldots,T_n)\in \uTree(k)\] is the grafting, $\{\sigma_v\}\in \prod_{v\in T} \Sigma_{|\inp(v)|}$, and $\{\sigma^j_u\} \in \prod_{u\in T_j}\Sigma_{|\inp(u)|}$ for $1 \leq j \leq n$.  Then the diagram
\begin{equation}\label{ainfinity-ass}
\nicexy{\J[T] \otimes \Bigl(\bigotimes\limits_{j=1}^n \J[T_j]\Bigr) \otimes X^{\otimes k} \ar[d]_-{\mathrm{permute}}^-{\cong} \ar[r]^-{(\pi,\Id)} & \J[G] \otimes X^{\otimes k} \ar[dd]^-{\lambda_G^{\{\sigma_v,\sigma^j_u\}}}\\
\J[T] \otimes \bigotimes\limits_{j=1}^n \bigl(\J[T_j]\otimes X^{\otimes k_j}\bigr) \ar[d]_-{\bigl(\Id,\bigotimes_j \lambda_{T_j}^{\{\sigma^j_u\}_{u\in T_j}}\bigr)} &\\
\J[T] \otimes \bigotimes\limits_{j=1}^n X \ar[r]^-{\lambda_T^{\{\sigma_v\}_{v\in T}}} & X}
\end{equation}
is commutative.  Here $\pi=\bigotimes_S 1$ is the morphism in Lemma \ref{lem:morphism-pi} for the grafting $G$.  In the structure morphism $\lambda_G^{\{\sigma_v,\sigma^j_u\}}$, we have $v \in T$, $u \in T_j$, and $1 \leq j \leq n$.
\item[Unity]
The composition
\begin{equation}\label{ainfinity-unity}
\nicexy{X\ar[r]^-{\cong} & \J[\uparrow]\otimes X \ar[r]^-{\lambda^{\varnothing}_{\uparrow}} & X}
\end{equation} 
is the identity morphism of $X$, where $\uparrow$ is the $1$-colored exceptional edge.
\item[Equivariance]
For $T \in \uTree(n)$, $\sigma \in \Sigma_n$, and $\{\sigma_v\}\in \prod_{v\in T}\Sigma_{|\inp(v)|}$, the diagram 
\begin{equation}\label{ainfinity-eq}
\nicexy@C+1cm{\J[T]\otimes X^{\otimes n} \ar[d]_-{(\Id,\sigmainv)} \ar[r]^-{\lambda_T^{\{\sigma_v\}}} & X\ar@{=}[d] \\ \J[T\sigma]\otimes X^{\otimes n} \ar[r]^-{\lambda_{T\sigma}^{\{\sigma_v\}}} & X}
\end{equation}
is commutative, in which $T\sigma \in \uTree(n)$ is the same as $T$ except that its ordering is $\zeta_T\sigma$ with $\zeta_T$ the ordering of $T$.
\item[Wedge Condition]
Suppose $T \in \uTree(n)$, $H_v\in \uTree(|\inp(v)|)$ for $v \in \Vt(T)$, $K=T(H_v)_{v\in T}$ is the tree substitution, and $\sigma^v_u \in \Sigma_{|\inp(u)|}$ for each $v \in \Vt(T)$ and $u \in \Vt(H_v)$.  Then the diagram
\begin{equation}\label{ainfinity-wedge}
\nicexy@C+1cm{\J[T]\otimes X^{\otimes n} \ar[d]_-{(\J,\Id)} \ar[r]^-{\lambda_T^{\{\tau_v\}}} & X\ar@{=}[d]\\ \J[K]\otimes X^{\otimes n} \ar[r]^-{\lambda_K^{\{\sigma^v_u\}}} & X}
\end{equation}
is commutative.  For each $v \in \Vt(T)$, $\tau_v$ is defined as \[\tau_v = \gamma^{\As}_{H_v}\bigl(\{\sigma^v_u\}_{u\in H_v}\bigr) \in \Sigma_{|\inp(v)|}\]
in which \[\nicexy{\prodover{u\in H_v} \Sigma_{|\inp(u)|} = \As[H_v] \ar[r]^-{\gamma^{\As}_{H_v}} & \As(|\inp(v)|)=\Sigma_{|\inp(v)|}}\] is the operadic structure morphism for $H_v$ of the associative operad in $\Set$, as in \eqref{operadic-structure-map}.
\end{description}
\end{theorem}

\begin{proof}
This is the special case of the Coherence Theorem \ref{thm:wo-algebra-coherence} for the associative operad $\As$ in $\M$.  Indeed, recall that the entries of the associative operad in $\M$ are  \[\As(n) = \coprodover{\sigma\in \Sigma_n}\tensorunit\] for $n \geq 0$.  For a $1$-colored tree $T \in \uTree(n)$, there is a natural isomorphism \[\As[T] = \bigtensorover{v\in T} \As(|\inp(v)|) = \bigtensorover{v\in T} \Bigl[\coprodover{\sigma\in \Sigma_{|\inp(v)|}} \tensorunit\Bigr] \cong \coprodover{\{\sigma_v\}\in \prodover{v\in T}\Sigma_{|\inp(v)|}} \tensorunit.\]  It follows that there is a natural isomorphism \[\J[T]\otimes\As[T]\otimes X^{\otimes n} \cong \coprodover{\{\sigma_v\}\in \prodover{v\in T}\Sigma_{|\inp(v)|}} \J[T]\otimes X^{\otimes n}.\]  Therefore, the structure morphism $\lambda_T$ in \eqref{wo-algebra-restricted} is uniquely determined by the restrictions $\lambda^{\{\sigma_v\}_{v\in T}}_{T}$ as stated in \eqref{ainfinity-structure}.  The associativity, unity, equivariance, and wedge conditions \eqref{ainfinity-ass}-\eqref{ainfinity-wedge} above are exactly those in the Coherence Theorem \ref{thm:wo-algebra-coherence} for $1$-colored trees.
\end{proof}

Suppose $(X,\lambda)$ is an $A_\infty$-algebra in $\M$.  In the next few examples, we will explain some of the structure on $X$ that suggests that it is a monoid up to coherent higher homotopies.

\begin{example}[Multiplication]\label{ex1:ainfinity}
Suppose $\Cor_n$ is the $1$-colored corolla with $n$ legs; see Example \ref{ex:cd-corolla} where corollas were defined.  The structure morphism in \eqref{ainfinity-structure} yields the composition \[\nicexy{X^{\otimes n} \ar[rr]^-{\mu_n} \ar[dr]_-{\cong} && X\\ & \J[\Cor_n]\otimes X^{\otimes n} \ar[ur]_-{\lambda_{\Cor_n}^{\{\id_n\}}}}\] with $\J[\Cor_n]=\tensorunit$ and $\id_n \in \Sigma_{n}$ the identity permutation.  By Corollary \ref{cor:wo-alg-unit} \[\mu_1 : X \to X\] is the identity morphism on $X$, and so $\lambda_{\Cor_1}^{\{\id_1\}}$ is the isomorphism $\tensorunit \otimes X \cong X$.  If $X$ is a monoid, then we would expect \[\mu_2 : X\otimes X \to X\] to be strictly associative with $\mu_0 : \tensorunit \to X$ as a strict two-sided unit.  For an $A_\infty$-algebra, we expect homotopy associativity and a homotopy unit, as explained in the following examples.\dqed
\end{example}

\begin{example}[Left homotopy unit]\label{ex2:ainfinity}
Here we explain why $\mu_0 : \tensorunit \to X$ is a \index{homotopy unit}left homotopy unit of $\mu_2$.  Consider the grafting \[K= \graft\bigl(\Cor_2; \Cor_0,\uparrow\bigr) \in \uTree(1)\] which may be visualized as follows.
\begin{center}\begin{tikzpicture}
\node[plain] (v) {$v$}; \node[plain, below left=.5cm of v] (u) {$u$};
\draw[outputleg] (v) to +(0cm,.8cm); \draw[inputleg] (v) to +(.7cm,-.6cm);
\draw[arrow] (u) to (v);
\end{tikzpicture}\end{center}
Consider the diagram
\[\nicexy@C+.7cm{\J[\Cor_1]\otimes X \ar[d]_-{(0,\Id)} \ar[dr]^-{\lambda_{\Cor_1}^{\{\id_1\}}} & \tensorunit\otimes X \ar[l]_-{=} \ar[d]^-{\cong}\\
\J[K]\otimes X \ar[r]_-{\lambda_K^{\{\id_2,\id_0\}}} \ar@{}[ur]|(.3){(1)} & X\\
\tensorunit\otimes X \ar[u]^-{(1,\Id)} \ar@{}[r]|-{(2)} &\\
\J[\Cor_2] \otimes \bigl(\J[\Cor_0]\otimes X^{\otimes 0}\bigr)\otimes \bigl(\J[\uparrow]\otimes X\bigr) \ar[u]^-{\cong} \ar[r]^-{\bigl(\Id,\lambda_{\Cor_0}^{\{\id_0\}},\lambda_{\uparrow}^{\varnothing}\bigr)} & 
\J[\Cor_2]\otimes X \otimes X\ar[uu]^-{\lambda_{\Cor_2}^{\{\id_2\}}}\\
\tensorunit\otimes X \ar[u]^-{\cong} \ar[r]^-{(\mu_0,\Id)} & X\otimes X \ar[u]^-{\cong} \ar@/_4pc/[uuu]_-{\mu_2}}\]
in $\M$, in which \[\J[K]=J \andspace \J[\Cor_n]=\J[\uparrow]=X^{\otimes 0}=\tensorunit.\]

This diagram is commutative:
\begin{itemize}\item The top right triangle is commutative because $\lambda_{\Cor_1}^{\{\id_1\}}$ is the isomorphism $\tensorunit\otimes X\cong X$.
\item The triangle (1) is commutative by the wedge condition \eqref{ainfinity-wedge} for the tree substitution $K=\Cor_1(K)$.
\item The square (2) is commutative by the associativity condition \eqref{ainfinity-ass} using the grafting definition of $K$.
\item The bottom rectangle is commutative by the definition of $\mu_0$ and the unity condition \eqref{ainfinity-unity}.
\item The lower right stripe is the definition of $\mu_2$.
\end{itemize}
The above commutative diagram says that the structure morphism $\lambda_K^{\{\id_2,\id_0\}}$ is a homotopy from the isomorphism $\tensorunit\otimes X \cong X$ to the composition $\mu_2\circ(\mu_0,\Id)$.  So in an $A_\infty$-algebra, $\mu_0$ is a left homotopy unit of $\mu_2$.\dqed
\end{example}

\begin{example}[Right homotopy unit]\label{ex2.5:ainfinity}
Similarly,  consider the grafting \[G= \graft\bigl(\Cor_2;\uparrow,\Cor_0\bigr)\in \uTree(1)\] which may be visualized as follows.
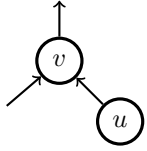
\begin{center}\begin{tikzpicture}
\node[plain] (v) {$v$}; \node[plain, below right=.5cm of v] (u) {$u$};
\draw[outputleg] (v) to +(0cm,.8cm); \draw[inputleg] (v) to +(-.7cm,-.6cm);
\draw[arrow] (u) to (v);
\end{tikzpicture}\end{center}
As above there is a commutative diagram
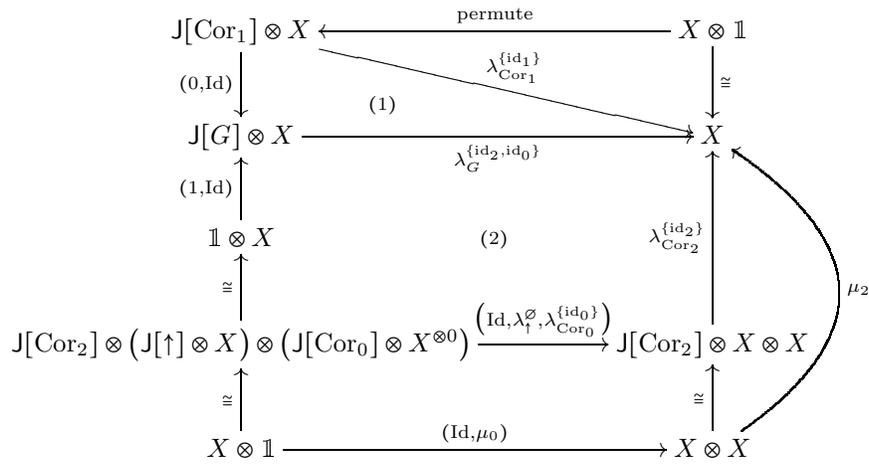
\[\nicexy@C+.7cm{\J[\Cor_1]\otimes X \ar[d]_-{(0,\Id)} \ar[dr]^-{\lambda_{\Cor_1}^{\{\id_1\}}} & X\otimes \tensorunit \ar[l]_-{\mathrm{permute}} \ar[d]^-{\cong}\\
\J[G]\otimes X \ar[r]_-{\lambda_G^{\{\id_2,\id_0\}}} \ar@{}[ur]|(.3){(1)} & X\\
\tensorunit\otimes X \ar[u]^-{(1,\Id)} \ar@{}[r]|-{(2)} &\\
\J[\Cor_2] \otimes \bigl(\J[\uparrow]\otimes X\bigr) \otimes \bigl(\J[\Cor_0]\otimes X^{\otimes 0}\bigr) \ar[u]^-{\cong} \ar[r]^-{\bigl(\Id,\lambda_{\uparrow}^{\varnothing},\lambda_{\Cor_0}^{\{\id_0\}}\bigr)} & 
\J[\Cor_2]\otimes X \otimes X\ar[uu]^-{\lambda_{\Cor_2}^{\{\id_2\}}}\\
X\otimes \tensorunit \ar[u]^-{\cong} \ar[r]^-{(\Id,\mu_0)} & X\otimes X \ar[u]^-{\cong} \ar@/_4pc/[uuu]_-{\mu_2}}\]
in $\M$.  The commutativity of this diagram says that the structure morphism $\lambda_G^{\{\id_2,\id_0\}}$ is a homotopy from the isomorphism $X\otimes\tensorunit\cong X$ to the composition $\mu_2\circ(\Id,\mu_0)$.  So in an $A_\infty$-algebra, $\mu_0$ is also a right homotopy unit of $\mu_2$.\dqed
\end{example}

\begin{example}[Homotopy associativity]\label{ex3:ainfinity}
Let us now observe that the structure morphism $\mu_2 : X^{\otimes 2} \to X$ is homotopy associative \index{homotopy associativity}in the following sense.  Consider the $1$-colored trees \[K=\graft\bigl(\Cor_2;\Cor_2,\uparrow\bigr) \andspace G=\graft\bigl(\Cor_2;\uparrow,\Cor_2\bigr)\in \uTree(3)\]
which may be visualized as follows.
\begin{center}\begin{tikzpicture}
\node[plain] (v) {$v$}; \node[plain, below left=.5cm of v] (u) {$u$}; 
\node[left=.5cm of v] {$K$}; \draw[arrow] (u) to (v);
\draw[outputleg] (v) to +(0cm,.8cm); \draw[inputleg] (v) to +(.7cm,-.6cm);
\draw[inputleg] (u) to +(-.7cm,-.6cm); \draw[inputleg] (u) to +(.7cm,-.6cm);
\node[plain, right=3cm of v] (v1) {$v$}; \node[plain, below right=.5cm of v1] (u1) {$u$}; 
\node[left=.5cm of v1] {$G$}; \draw[arrow] (u1) to (v1);
\draw[outputleg] (v1) to +(0cm,.8cm); \draw[inputleg] (v1) to +(-.7cm,-.6cm);
\draw[inputleg] (u1) to +(-.7cm,-.6cm); \draw[inputleg] (u1) to +(.7cm,-.6cm);
\end{tikzpicture}\end{center}
Consider the diagram
\[\nicexy{\tensorunit \otimes X^{\otimes 3} \ar[d]_-{(1,\Id)} & X^{\otimes 2}\otimes X \ar[l]_-{\cong} \ar[r]^-{(\mu_2,\Id)} & X^{\otimes 2} \ar[d]^-{\mu_2}\\
\J[K] \otimes X^{\otimes 3} \ar[rr]^-{\lambda_K^{\{\id_2,\id_2\}}} && X\\
\J[\Cor_3]\otimes X^{\otimes 3} \ar[u]^-{(0,\Id)} \ar[rr]^-{\lambda_{\Cor_3}^{\{\id_3\}}} \ar[d]_-{(0,\Id)} && X \ar@{=}[u] \ar@{=}[d]\\
\J[G] \otimes X^{\otimes 3} \ar[rr]^-{\lambda_G^{\{\id_2,\id_2\}}} && X\\
\tensorunit \otimes X^{\otimes 3} \ar[u]^-{(1,\Id)} & X \otimes X^{\otimes 2} \ar[l]_-{\cong} \ar[r]^-{(\Id,\mu_2)} & X^{\otimes 2} \ar[u]_-{\mu_2}}\]
in $\M$, in which \[\J[K]=\J[G]=J \andspace \J[\Cor_3]=\tensorunit.\]  

This diagram is commutative:
\begin{itemize}
\item The top rectangle is commutative by (i) the definition of $\mu_2$, (ii) the unity condition \eqref{ainfinity-unity}, and (iii) the associativity condition \eqref{ainfinity-ass} using the grafting definition of $K$.
\item The second rectangle from the top is commutative by the wedge condition \eqref{ainfinity-wedge} for the tree substitution $K=\Cor_3(K)$.
\item The third rectangle from the top is commutative by the wedge condition \eqref{ainfinity-wedge} for the tree substitution $G=\Cor_3(G)$.
\item The bottom rectangle is commutative by (i) the definition of $\mu_2$, (ii) the unity condition \eqref{ainfinity-unity}, and (iii) the associativity condition \eqref{ainfinity-ass} using the grafting definition of $G$.
\end{itemize}
The top half of the commutative diagram says that the structure morphism $\lambda_K^{\{\id_2,\id_2\}}$ is a homotopy from $\lambda_{\Cor_3}^{\{\id_3\}}$ to the composition $\mu_2\circ(\mu_2,\Id)$.  The bottom half of the commutative diagram says that the structure morphism $\lambda_G^{\{\id_2,\id_2\}}$ is a homotopy from $\lambda_{\Cor_3}^{\{\id_3\}}$ to the composition $\mu_2\circ(\Id,\mu_2)$.  The entire commutative diagram together exhibits a homotopy between the compositions $\mu_2\circ(\mu_2,\Id)$ and $\mu_2\circ(\Id,\mu_2)$.  So in an $A_\infty$-algebra, the morphism $\mu_2$ is homotopy associative.  

This is only the first layer of the higher homotopy associative structure in an $A_\infty$-algebra.  For example, similar to the discussion above, one can consider $1$-colored trees with more than one internal edges.  Any iterated composition of the various $\mu_k$'s as represented by a $1$-colored tree $T \in \uTree(n)$ is homotopic to the structure morphism $\lambda_{\Cor_n}^{\{\id_n\}}$ via the homotopy $\lambda_T^{\{\id_v\}}$, where $\id_v\in \Sigma_{|\inp(v)|}$ is the identity permutation for each $v \in \Vt(T)$.  The point is that we do not need to write these relations down one-by-one from the ground up.  Instead, all of the higher homotopy associative structure is neatly packed in the Boardman-Vogt construction $\Was$ of the associative operad.\dqed
\end{example}

\section{$E_\infty$-Algebras}\label{sec:einfinity-algebra}

In this section, we discuss a homotopy version of commutative monoids, called \index{strongly homotopy commutative algebra}strongly homotopy commutative algebras or $E_{\infty}$-algebras, as algebras over the Boardman-Vogt construction of the commutative operad.  

\begin{motivation}
Recall from Example \ref{ex:operad-com} that the commutative operad $\Com$ is a $1$-colored operad in $\M$ whose category of algebras is canonically isomorphic to the category of commutative monoids in $\M$ (Definition \ref{def:monoid}).  The physical relevance of $E_\infty$-algebras is that a prefactorization algebra includes a commutative monoid in its structure.  Strict commutativity is not a homotopy invariant concept.  A homotopy coherent version of a commutative monoid is an $E_\infty$-algebra.  Therefore, $E_\infty$-algebras will arise naturally in the study of homotopy prefactorization algebras.\dqed\end{motivation}

\begin{definition}\label{def:einfinity-algebra}
Objects in the category $\algm(\Wcom)$ are called\index{einfinityalgebra@$E_\infty$-algebra} \emph{$E_\infty$-algebras in $\M$}, where $\Wcom$ is the Boardman-Vogt construction of the commutative operad $\Com$.
\end{definition}

When applied to the commutative operad, Corollary \ref{cor:augmentation-adjunction} and Corollary \ref{cor:wo-o-chaink} yield the following adjunction.

\begin{corollary}\label{cor:wcom-adjunction}
The augmentation $\eta : \Wcom \to \Com$ induces an adjunction \[\nicexy{\algm(\Wcom) \ar@<2pt>[r]^-{\eta_!} & \algm(\Com) \ar@<2pt>[l]^-{\eta^*}}\] that is a Quillen equivalence if $\M=\Chaink$ with $\fieldk$ a field of characteristic zero
\end{corollary}

\begin{interpretation} This adjunction says that each commutative monoid in $\M$ can be regarded as an $E_\infty$-algebra in $\M$ via the augmentation $\eta$.  The left adjoint $\eta_!$ rectifies each $E_\infty$-algebra in $\M$ to a commutative monoid in $\M$.\dqed\end{interpretation}

When applied to the operad morphism $f : \As \to \Com$ in Example \ref{ex:as-to-com}, Corollary \ref{cor:wf-eta-adjunction} yields the following result.

\begin{corollary}\label{cor:was-wcom-adjunction}
There is a diagram of adjunctions
\[\nicexy@C+.4cm@R+.3cm{\algm(\Was) \ar@<2pt>[r]^-{(\wf)_!} \ar@<-2pt>[d]_-{\eta_!} 
& \algm(\Wcom) \ar@<2pt>[l]^-{(\wf)^*} \ar@<-2pt>[d]_-{\eta_!} \\
\algm(\As) \ar@<2pt>[r]^-{f_!} \ar@<-2pt>[u]_-{\eta^*}  
& \algm(\Com) \ar@<2pt>[l]^-{\fstar} \ar@<-2pt>[u]_-{\eta^*}}\]
in which \[f_! \circ \eta_! = \eta_! \circ (\wf)_! \andspace \eta^*\circ\fstar=(\wf)^*\circ \eta^*.\]
\end{corollary}

\begin{remark}The bottom adjunction, the left adjunction, and the right adjunction are the ones in Example \ref{ex:as-to-com}, Corollary \ref{cor:was-adjunction}, and Corollary \ref{cor:wcom-adjunction}, respectively.  In the top adjunction, the right adjoint $(\wf)^*$ sends each $E_\infty$-algebra to its underlying $A_\infty$-algebra.  The left adjoint $(\wf)_!$ sends each $A_\infty$-algebra to an $E_\infty$-algebra.\dqed\end{remark}

The following result is the coherence theorem for $E_\infty$-algebras.

\begin{theorem}\label{thm:einfinity-algebra}
An $E_\infty$-algebra in $\M$ is exactly a pair\index{Coherence Theorem!for $E_\infty$-algebra} $(X,\lambda)$ consisting of
\begin{itemize}
\item an object $X \in \M$ and
\item a structure morphism\index{structure morphism!for $E_\infty$-algebras}
\begin{equation}\label{einfinity-structure}
\nicexy{\J[T] \otimes X^{\otimes n} \ar[r]^-{\lambda_T} & X}
\end{equation}
for each $n \geq 0$ and $T \in \uTree(n)$
\end{itemize}
that satisfies the following four conditions.
\begin{description}
\item[Associativity] Suppose $T \in \uTree(n)$ with $n \geq 1$, $T_j \in \uTree(k_j)$ with $k_j \geq 0$ and $1 \leq j \leq n$, $k=k_1+\cdots+k_n$, and \[G=\graft(T;T_1,\ldots,T_n)\in \uTree(k)\] is the grafting.  Then the diagram
\begin{equation}\label{einfinity-ass}
\nicexy{\J[T] \otimes \Bigl(\bigotimes\limits_{j=1}^n \J[T_j]\Bigr) \otimes X^{\otimes k} \ar[d]_-{\mathrm{permute}}^-{\cong} \ar[r]^-{(\pi,\Id)} & \J[G] \otimes X^{\otimes k} \ar[dd]^-{\lambda_G}\\
\J[T] \otimes \bigotimes\limits_{j=1}^n \bigl(\J[T_j]\otimes X^{\otimes k_j}\bigr) \ar[d]_-{\bigl(\Id,\bigotimes_j \lambda_{T_j}\bigr)} &\\
\J[T] \otimes \bigotimes\limits_{j=1}^n X \ar[r]^-{\lambda_T} & X}
\end{equation}
is commutative.  Here $\pi=\bigotimes_S 1$ is the morphism in Lemma \ref{lem:morphism-pi} for the grafting $G$.
\item[Unity]
The composition
\begin{equation}\label{einfinity-unity}
\nicexy{X\ar[r]^-{\cong} & \J[\uparrow]\otimes X \ar[r]^-{\lambda_{\uparrow}} & X}
\end{equation} 
is the identity morphism of $X$, where $\uparrow$ is the $1$-colored exceptional edge.
\item[Equivariance]
For $T \in \uTree(n)$ and $\sigma \in \Sigma_n$, the diagram 
\begin{equation}\label{einfinity-eq}
\nicexy@C+1cm{\J[T]\otimes X^{\otimes n} \ar[d]_-{(\Id,\sigmainv)} \ar[r]^-{\lambda_T} & X\ar@{=}[d]\\ \J[T\sigma]\otimes X^{\otimes n} \ar[r]^-{\lambda_{T\sigma}} & X}
\end{equation}
is commutative, in which $T\sigma \in \uTree(n)$ is the same as $T$ except that its ordering is $\zeta_T\sigma$ with $\zeta_T$ the ordering of $T$.
\item[Wedge Condition]
Suppose $T \in \uTree(n)$, $H_v\in \uTree(|\inp(v)|)$ for $v \in \Vt(T)$, and $K=T(H_v)_{v\in T}$ is the tree substitution.  Then the diagram
\begin{equation}\label{einfinity-wedge}
\nicexy@C+1cm{\J[T]\otimes X^{\otimes n} \ar[d]_-{(\J,\Id)} \ar[r]^-{\lambda_T} & X\ar@{=}[d]\\
\J[K]\otimes X^{\otimes n} \ar[r]^-{\lambda_K} & X}
\end{equation}
is commutative.
\end{description}
\end{theorem}

\begin{proof}
This is the special case of the Coherence Theorem \ref{thm:wo-algebra-coherence} for the commutative operad $\Com$ in $\M$.  Indeed, recall that the entries of the commutative operad in $\M$ are  \[\Com(n) = \tensorunit \forspace n \geq 0.\]  For a $1$-colored tree $T \in \uTree(n)$, there is a natural isomorphism \[\Com[T] = \bigtensorover{v\in T} \Com(|\inp(v)|) = \bigtensorover{v\in T}\tensorunit \cong \tensorunit.\]  It follows that there is a natural isomorphism \[\J[T]\otimes\Com[T]\otimes X^{\otimes n} \cong \J[T]\otimes X^{\otimes n}.\]  Therefore, the structure morphism $\lambda_T$ in \eqref{wo-algebra-restricted} becomes the morphism $\lambda_T$ in \eqref{einfinity-structure}.  The associativity, unity, equivariance, and wedge conditions \eqref{einfinity-ass}-\eqref{einfinity-wedge} above are exactly those in the Coherence Theorem \ref{thm:wo-algebra-coherence} for $1$-colored trees.
\end{proof}

Suppose $(X,\lambda)$ is an $E_\infty$-algebra in $\M$.  In the next few examples, we will explain part of the structure on $X$.

\begin{example}[$A_\infty$-structure]\label{ex1:einfinity}
The  right adjoint $(\wf)^*$ in Corollary \ref{cor:was-wcom-adjunction} pulls $(X,\lambda)$ back to an $A_\infty$-algebra.  More explicitly, as an $A_\infty$-algebra, its structure morphisms $\lambda_T^{\{\sigma_v\}_{v\in T}}$ in \eqref{ainfinity-structure} are  equal to the structure morphism $\lambda_T$ in \eqref{einfinity-structure} for all choices of permutations $\{\sigma_v\} \in\prod_{v\in T} \Sigma_{|\inp(v)|}$.  In particular, the discussion in Example \ref{ex1:ainfinity} to Example \ref{ex3:ainfinity} also applies to an $E_\infty$-algebra.\dqed
\end{example}

\begin{example}[Strict commutativity]\label{ex2:einfinity}
Consider the tree substitution \[\Cor_n\sigma = \Cor_n\bigl(\Cor_n\sigma\bigr)\] for $n \geq 0$, $\sigma \in \Sigma_n$, $\Cor_n \in \uTree(n)$ the corolla with $n$ inputs, and $\Cor_n\sigma$ a permuted corolla.  Then the wedge condition \eqref{einfinity-wedge} yields the equality \[\lambda_{\Cor_n} = \lambda_{\Cor_n\sigma} : \tensorunit\otimes X^{\otimes n} \to X.\]  The equivariance condition \eqref{einfinity-eq} with $T=\Cor_n$ is the following commutative diagram.
\[\nicexy@C+1cm{X^{\otimes n} \cong \J[\Cor_n]\otimes X^{\otimes n} \ar[d]_-{(\Id,\sigmainv)} \ar[r]^-{\lambda_{\Cor_n}} & X\ar@{=}[d]\\ X^{\otimes n}\cong \J[\Cor_n\sigma]\otimes X^{\otimes n} \ar[r]^-{\lambda_{\Cor_n\sigma}} & X}\]
Since $\lambda_{\Cor_n} = \lambda_{\Cor_n\sigma}$, we conclude that the structure morphism $\lambda_{\Cor_n}$ is invariant under permutations of the $X$ factors in its domain.  For more general $1$-colored trees, the structure morphism is invariant under permutations of its domain factors up to a specified homotopy, as we will see in the next example.\dqed
\end{example}

\begin{example}[Homotopy commutativity]\label{ex3:einfinity}
Suppose $K \in \uTree(n)$ is not an exceptional edge, and $\sigma \in \Sigma_n$.  Recall that $|K|$ denotes the number of internal edges in $K$.  Consider the diagram
\[\nicexy@C+.6cm{\J[K] \otimes X^{\otimes n} \ar[r]^-{\lambda_K} & X\\
\J[\Cor_n] \otimes X^{\otimes n} \ar[u]^-{\bigl(0^{\otimes |K|}\circ \cong,\Id\bigr)} \ar[d]_-{\bigl(0^{\otimes |K|}\circ\cong, \Id\bigr)} \ar[r]^-{\lambda_{\Cor_n}} & X \ar@{=}[u] \ar@{=}[d]\\
\J[K\sigma]\otimes X^{\otimes n} \ar[d]_-{(\Id,\sigma)} \ar[r]^-{\lambda_{K\sigma}} & X\ar@{=}[d]\\ \J[K]\otimes X^{\otimes n} \ar[r]^-{\lambda_K} & X}\] 
in $\M$.  This diagram is commutative:
\begin{itemize}
\item The top square is commutative by the wedge condition \eqref{einfinity-wedge} for the tree substitution $K=\Cor_n(K)$, in which $0^{\otimes |K|}\circ \cong$ is the composition \[\nicexy{\J[\Cor_n]=\tensorunit \ar[r]^-{\cong} & \tensorunit^{\otimes |K|} \ar[r]^-{0^{\otimes |K|}} & J^{\otimes |K|} \cong \J[K]}\] with $0 : \tensorunit \to J$ a part of the commutative segment $J$.
\item The middle square is commutative for the same reason for the tree substitution $K\sigma = \Cor_n\bigl(K\sigma\bigr)$.
\item The bottom square is commutative by the equivariance condition \eqref{einfinity-eq}.
\end{itemize}
The entire \index{homotopy commutativity}commutative diagram together says that the structure morphism $\lambda_K$ is homotopic to the composition $\lambda_K \circ(\Id,\sigma)$.  So the structure morphism $\lambda_K$ is commutative in its domain $X$ factors up to a specified homotopy.\dqed
\end{example}

\section{Homotopy Coherent Diagrams of $A_\infty$-Algebras}\label{sec:hcdiagram-ainfinity}

In this section, we discuss a homotopy coherent version of a diagram of monoids using the Boardman-Vogt construction.  Fix a small category $\C$ with object set $\colorc$.

\begin{motivation} An algebraic quantum field theory on an orthogonal category $\Cbar$ is, first of all, a functor $\fraka : \C \to \Monm$ from $\C$ to monoids in $\M$.  Therefore, we should expect a homotopy algebraic quantum field theory to have the structure of a homotopy coherent $\C$-diagram of $A_\infty$-algebras.  This is a combination of the structures in Section \ref{sec:hcdiagram} and Section \ref{sec:ainfinity-algebra}, in the sense that it forgets to a homotopy coherent $\C$-diagram in $\M$ and that entrywise it is an $A_\infty$-algebra.  We saw in Example \ref{ex:diag-monoid-operad} that $\C$-diagrams of monoids in $\M$ are exactly algebras over the colored operad $\Ocm$.  Their homotopy coherent analogues should therefore be algebras over the Boardman-Vogt construction of $\Ocm$.\dqed
\end{motivation}

\begin{definition}\label{def:wocm-algebra}
Objects in the category $\algmwocm$ are called\index{homotopy coherent diagram!$A_\infty$-algebra} \emph{homotopy coherent $\C$-diagrams of $A_\infty$-algebras in $\M$}, where $\wocm$ is the Boardman-Vogt construction of the $\colorc$-colored operad $\Ocm$ in Example \ref{ex:diag-monoid-operad}.
\end{definition}

When applied to the colored operad $\Ocm$, Corollary \ref{cor:augmentation-adjunction} and Corollary \ref{cor:wo-o-chaink} yield the following adjunction.

\begin{corollary}\label{cor:wocm-adjunction}
The augmentation $\eta : \wocm \to \Ocm$ induces an adjunction \[\nicexy{\algm(\wocm) \ar@<2pt>[r]^-{\eta_!} & \algm(\Ocm) \cong \Monmc \ar@<2pt>[l]^-{\eta^*}}\] that is a Quillen equivalence if $\M=\Chaink$ with $\fieldk$ a field of characteristic zero
\end{corollary}

\begin{interpretation} Each $\C$-diagram of monoids in $\M$ can be regarded as a $\wocm$-algebra via the augmentation $\eta$.  The left adjoint $\eta_!$ rectifies each homotopy coherent $\C$-diagram of $A_\infty$-algebras in $\M$ to a $\C$-diagram of monoids in $\M$.\dqed\end{interpretation}

The colored operad $\Ocm$ for $\C$-diagrams of monoids in $\M$ is related to the associative operad $\As$ in Example \ref{ex:operad-as} and the $\C$-diagram operad $\Cdiag$ in Example \ref{ex:operad-diag} as follows.  We will denote the unique color in $\As$ by $*$.  A copy of $\tensorunit$ corresponding to an element $x$ will be denoted by $\tensorunit_x$.  The following observation is proved by a direct inspection.

\begin{lemma}\label{lem:ass-to-ocm}
Consider the $\colorc$-colored operad $\Ocm$ in Example \ref{ex:diag-monoid-operad}. 
\begin{enumerate}\item For each $c \in \colorc$, there is an operad morphism \[\nicexy{\As \ar[r]^-{\iota_c} & \Ocm}\] that sends $*$ to $c$ and is entrywise defined by the commutative diagrams \[\nicexy@C+.5cm{\tensorunit_{\sigma}\ar[d]_-{\mathrm{inclusion}} \ar[r]^-{=} & \tensorunit_{(\sigma,\{\Id_c\}_{j=1}^n)} \ar[d]^-{\mathrm{inclusion}}\\ \As(n)=\coprod\limits_{\sigma\in\Sigma_n}\tensorunit \ar[r]^-{\iota_c} & \coprod\limits_{\Sigma_n\times\prod\limits_{j=1}^n\C(c,c)} \tensorunit= \Ocm\ccc}\] for $n \geq 0$ and $\sigma \in \Sigma_n$.
\item There is a morphism of $\colorc$-colored operads \[\nicexy{\Cdiag \ar[r]^-{i} & \Ocm}\] that is entrywise defined by the commutative diagrams \[\nicexy@C+.5cm{\tensorunit_{f}\ar[d]_-{\mathrm{inclusion}} \ar[r]^-{=} & \tensorunit_{(\id_1,f)} \ar[d]^-{\mathrm{inclusion}}\\ \Cdiag\dc = \coprod\limits_{f\in \C(c,d)}\tensorunit \ar[r]^-{i} & \coprod\limits_{\Sigma_1\times\C(c,d)} \tensorunit = \Ocm\dc}\] for $c,d\in \colorc$ and $f \in \C(c,d)$.  In all other entries $\duc$, $i$ is the unique morphism from the initial object to $\Ocm\duc$.
\end{enumerate}
\end{lemma}

Applying Corollary \ref{cor:augmentation-adjunction} to the operad morphisms in Lemma \ref{lem:ass-to-ocm}, we obtain the following result.

\begin{corollary}\label{cor:ass-ocm-adjunction}
There is a diagram of change-of-operad adjunctions
\[\nicexy@C+.2cm@R+.3cm{\algmwas \ar@<2pt>[r]^-{(\W\iota_c)_!} \ar@<-2pt>[d]_-{\eta_!} 
& \algmwocm \ar@<2pt>[l]^-{\W\iota_c^*} \ar@<-2pt>[d]_-{\eta_!} \ar@<-2pt>[r]_-{\W i^*} & \algmwcdiag \ar@<-2pt>[l]_-{\W i_!} \ar@<-2pt>[d]_-{\eta_!} \\
\Monm\cong\algmas \ar@<2pt>[r]^-{(\iota_c)_!} \ar@<-2pt>[u]_-{\eta^*}  
& \algmocm \cong \Monmc \ar@<2pt>[l]^-{\iota_c^*} \ar@<-2pt>[u]_-{\eta^*} \ar@<-2pt>[r]_-{i^*} & \algmcdiag \cong \M^{\C} \ar@<-2pt>[l]_-{i_!} \ar@<-2pt>[u]_-{\eta^*}}\]
with commuting left adjoint diagrams and commuting right adjoint diagrams, where the left half is defined for each $c \in \colorc$.
\end{corollary}

\begin{interpretation} In the adjunction $(\iota_c)_! \dashv \iota_c^*$, the right adjoint $\iota_c^*$ remembers only the monoid at the $c$-colored entry.   In the adjunction $i_! \dashv i^*$, the right adjoint $i^*$ remembers only the underlying $\C$-diagram in $\M$.  The right adjoint $\W\iota_c^*$ remembers only the $A_\infty$-algebra at the $c$-colored entry, while $\W i^*$ remembers only the underlying homotopy coherent $\C$-diagram in $\M$.\dqed
\end{interpretation}

The following result is the coherence theorem for homotopy coherent $\C$-diagrams of $A_\infty$-algebras in $\M$.  If the base category is $\Set$, then we will denote $\Osubc^{\Set}$ by $\Osubc$.

\begin{theorem}\label{thm:wocm-coherence}
A $\wocm$-algebra is exactly a pair\index{Coherence Theorem!for hc diagrams of $A_\infty$-algebras} $(X,\lambda)$ consisting of 
\begin{itemize}\item a $\colorc$-colored object $X$ in $\M$ and
\item a structure morphism\index{structure morphism!for hc diagrams of $A_\infty$-algebras}
\begin{equation}\label{wocm-restricted}
\nicexy@C+1.5cm{\J[T]\otimes X_{\uc} \ar[r]^-{\lambda_T\left\{(\sigma^v,\uf^v)\right\}_{v\in T}} & X_d}\in \M
\end{equation}
for each $T \in \uTreec\duc$ with $(\uc;d) \in \Profcc$ and each \[\left\{(\sigma^v,\uf^v)\right\}_{v\in \Vt(T)} \in \prod_{v\in \Vt(T)} \biggl[\Sigma_{|\inp(v)|} \times \prod_{j=1}^{|\inp(v)|} \C\bigl(\inp(v)_j,\out(v)\bigr)\biggr] = \prod_{v\in \Vt(T)} \Osubc(v)\]
\end{itemize}
that satisfies the following four conditions.
\begin{description}
\item[Associativity] Suppose $\bigl(\uc=(c_1,\ldots,c_n);d\bigr) \in \Profcc$ with $n \geq 1$, $T \in \uTreecduc$, $T_j \in \uTreec\cjubj$ for $1 \leq j \leq n$, $\ub=(\ub_1,\ldots,\ub_n)$, \[G=\graft(T;T_1,\ldots,T_n) \in \uTreec\dub\] is the grafting \eqref{def:grafting}, $\left\{(\sigma^v,\uf^v)\right\}$ is as above, and \[\left\{(\sigma^u,\uf^u)\right\}_{u \in \Vt(T_j)} \in \prod_{u\in \Vt(T_j)} \biggl[ \Sigma_{|\inp(u)|} \times \prod_{k=1}^{|\inp(u)|} \C\bigl(\inp(u)_k,\out(u)\bigr)\biggr] =\prod_{u\in \Vt(T_j)} \Osubc(u)\] for each $1 \leq j \leq n$.  Then the diagram
\begin{equation}\label{wocm-ass}
\nicexy@R+.5cm@C-1.5cm{&\J[T]\otimes \Bigl(\bigotimes\limits_{j=1}^n \J[T_j]\Bigr)\otimes X_{\ub} \ar[dl]_-{\mathrm{permute}}^-{\cong} \ar[dr]^-{(\pi,\Id)} &\\
\J[T]\otimes \bigotimes\limits_{j=1}^n\bigl(\J[T_j]\otimes X_{\ub_j}\bigr) \ar[d]_-{\left(\Id,\bigtensorover{j} \lambda_{T_j}\left\{(\sigma^u,\uf^u)\right\}_{u\in T_j}\right)} && \J[G]\otimes X_{\ub} \ar[d]^-{\lambda_G\left\{(\sigma^w,\uf^w)\right\}_{w\in G}}\\
\J[T]\otimes X_{\uc} \ar[rr]^-{\lambda_T\left\{(\sigma^v,\uf^v)\right\}_{v\in T}} && X_d}
\end{equation}
is commutative.  Here $\pi=\bigotimes_S 1$ is the morphism in Lemma \ref{lem:morphism-pi} for the grafting $G$.
\item[Unity] For each $c \in \colorc$, the composition
\begin{equation}\label{wocm-unity}
\nicexy@C+.5cm{X_c \ar[r]^-{\cong} & \J[\uparrow_c]\otimes X_c \ar[r]^-{\lambda_{\uparrow_c}\{\varnothing\}} & X_c}
\end{equation} 
is the identity morphism of $X_c$.
\item[Equivariance] For each $T \in \uTreec\duc$, $\left\{(\sigma^v,\uf^v)\right\}$ as above, and permutation $\sigma \in \Sigma_{|\uc|}$, the diagram 
\begin{equation}\label{wocm-eq}
\nicexy@C+1.8cm{\J[T]\otimes X_{\uc} \ar[d]_-{(\Id,\sigmainv)} \ar[r]^-{\lambda_T\left\{(\sigma^v,\uf^v)\right\}_{v\in T}} & X_d\ar@{=}[d]\\
\J[T\sigma]\otimes X_{\uc\sigma} \ar[r]^-{\lambda_{T\sigma}\left\{(\sigma^v,\uf^v)\right\}_{v\in T\sigma}} & X_d}
\end{equation}
is commutative, in which $T\sigma \in \uTreec\ducsigma$ is the same as $T$ except that its ordering is $\zeta_T\sigma$ with $\zeta_T$ the ordering of $T$.  The permutation $\sigmainv : X_{\uc} \iso X_{\uc\sigma}$ permutes the factors in $X_{\uc}$.
\item[Wedge Condition] Suppose $T \in \uTreec\duc$, $H_v \in \uTreec(v)$ for each $v\in \Vt(T)$, $K=T(H_v)_{v\in T}$ is the tree substitution, and \[\left\{(\sigma^u,\uf^u)\right\}_{u\in \Vt(H_v)} \in \prod_{u\in \Vt(H_v)} \biggl[\Sigma_{|\inp(u)|} \times \prod_{j=1}^{|\inp(u)|} \C\bigl(\inp(u)_j,\out(u)\bigr)\biggr] =\prod_{u\in \Vt(H_v)} \Osubc(u)\] for each $v \in \Vt(T)$.  Then the diagram
\begin{equation}\label{wocm-wedge}
\nicexy@C+1.7cm{\J[T] \otimes X_{\uc} \ar[d]_-{(\J,\Id)} 
\ar[r]^-{\lambda_T\left\{(\tau^v,\ug^v)\right\}_{v\in T}} 
& X_d \ar@{=}[d]\\
\J[K]\otimes X_{\uc} \ar[r]^-{\lambda_K\left\{(\sigma^u,\uf^u)\right\}_{u\in K}} & X_d}
\end{equation}
is commutative.  Here for each $v\in \Vt(T)$, \[(\tau^v,\ug^v) = \gamma^{\Osubc}_{H_v}\Bigl(\{(\sigma^u,\uf^u)\}_{u\in H_v}\Bigr) \in \Osubc(v)=\Sigma_{|\inp(v)|} \times \prod_{j=1}^{|\inp(v)|} \C\bigl(\inp(v)_j,\out(v)\bigr)\] with \[\nicexy@C+.5cm{\Osubc[H_v] =\prod\limits_{u\in H_v} \Osubc(u) \ar[r]^-{\gamma^{\Osubc}_{H_v}} & \Osubc(v)}\]
the operadic structure morphism of $\Osubc$ for $H_v$ in \eqref{operadic-structure-map}.
\end{description}
A morphism $f : (X,\lambda^X) \to (Y,\lambda^Y)$ of $\wocm$-algebras is a morphism of the underlying $\colorc$-colored objects that respects the structure morphisms in \eqref{wocm-restricted} in the obvious sense.
\end{theorem}

\begin{proof}
This is the special case of the Coherence Theorem \ref{thm:wo-algebra-coherence} applied to the $\colorc$-colored operad $\Ocm$.  Indeed, since \[\Ocm\duc = \coprodover{\Sigma_n \times \prod\limits_{j=1}^n \C(c_j,d)}\tensorunit = \coprodover{\Osubc\duc}\tensorunit,\] for each $T \in \uTreec\duc$ there is a canonical isomorphism \[\Ocm[T] = \bigotimes_{v\in T} \Ocm(v) = \bigotimes_{v\in T}\Bigl(\coprodover{\Osubc(v)} \tensorunit\Bigr) \cong \coprodover{\prodover{v\in T}\Osubc(v)} \tensorunit.\]  This implies that there is a canonical isomorphism \[\J[T]\otimes \Ocm[T]\otimes X_{\uc} \cong \coprodover{\prodover{v\in T}\Osubc(v)} \J[T]\otimes X_{\uc}.\]  Therefore, the structure morphism $\lambda_T$ in \eqref{wo-algebra-restricted} is uniquely determined by the restricted structure morphisms $\lambda_T\{(\sigma^v,\uf^v)\}_{v\in T}$ in \eqref{wocm-restricted}.  The above associativity, unity, equivariance, and wedge conditions are those in Theorem \ref{thm:wo-algebra-coherence}.
\end{proof}

\begin{example}[Objectwise $A_\infty$-algebra]\label{ex:hcdiagmon-entrywise}
Suppose $(X,\lambda)$ is a $\wocm$-algebra, and $c \in \colorc$.  Under the right adjoint $\W\iota_c^*$ in Corollary \ref{cor:ass-ocm-adjunction}, we have that \[\W\iota_c^*(X,\lambda) \in \algmwas,\] i.e., an $A_\infty$-algebra.  Explicit, its underlying object is $X_c \in \M$.  For $T \in \uTree(n)$ and $\{\sigma_v\}_{v\in T} \in \prod_{v\in T} \Sigma_{|\inp(v)|}$, the $A_\infty$-algebra structure morphism \[\nicexy@C+.8cm{\J[T] \otimes X_c^{\otimes n} \ar[r]^-{\lambda_T^{\{\sigma_v\}_{v\in T}}} & X_c} \in \M\] in \eqref{ainfinity-structure} is the structure morphism \[\lambda_{T_c}\Bigl\{\bigl(\sigma_v,\{\Id_c\}_{j=1}^{|\inp(v)|}\bigr)\Bigr\}_{v\in T}\] in \eqref{wocm-restricted}, where $T_c \in \uTreec\ccc$ is the $c$-colored tree obtained from $T$ by replacing every edge color by $c$.\dqed
\end{example}

\begin{example}[Underlying homotopy coherent $\C$-diagram]\label{ex:hcdiagmon-hcdiag}
Suppose $(X,\lambda)$ is a $\wocm$-algebra.  Under the right adjoint $\W i^*$ in Corollary \ref{cor:ass-ocm-adjunction}, we have that \[\W i^*(X,\lambda) \in \algmwcdiag,\] i.e., a homotopy coherent $\C$-diagram in $\M$.  Explicitly, the homotopy coherent $\C$-diagram structure morphism 
\[\nicexy{\J[\Lin_{\uc}]\otimes X_{c_0} \ar[r]^-{\lambda_{\uc}^{\uf}} & X_{c_n}}\in \M\]
in \eqref{hcdiagram-structure-map} is the structure morphism \[\lambda_{\Lin_{\uc}}\Bigl\{\bigl(\id_1,f_j\bigr)\Bigr\}_{1\leq j \leq n}\] in \eqref{wocm-restricted}.\dqed
\end{example}

There is also a homotopy coherent compatibility between the homotopy coherent $\C$-diagram structure and the objectwise $A_\infty$-algebra structure in a $\wocm$-algebra.  We will explain it in details in Section \ref{sec:hcdiag-ainfinity} in the context that we actually care about, namely homotopy algebraic quantum field theories.

\section{Homotopy Coherent Diagrams of $E_\infty$-Algebras}\label{sec:hcdiagram-einfinity}

In this section, we discuss a homotopy coherent version of a diagram of commutative monoids using the Boardman-Vogt construction.  Fix a small category $\C$ with object set $\colorc$.

\begin{motivation} As we will see in Section \ref{sec:diag-com-monoid}, there is one situation where prefactorization algebras coincide with algebraic quantum field theories.  In this case, both categories are the categories of $\C$-diagrams of commutative monoids in $\M$.  Homotopy prefactorization algebras, which coincide with homotopy algebraic quantum field theories, are therefore homotopy coherent $\C$-diagrams of $E_\infty$-algebras in $\M$.\dqed
\end{motivation}

\begin{definition}\label{def:wcomc-algebra}
Objects in the category $\algmwcomc$ are called\index{homotopy coherent diagram!$E_\infty$-algebra} \emph{homotopy coherent $\C$-diagrams of $E_\infty$-algebras in $\M$}, where $\Wcomc$ is the Boardman-Vogt construction of the $\colorc$-colored operad $\Comc$ in Example \ref{ex:diag-com-operad}.
\end{definition}

When applied to the colored operad $\Comc$, Corollary \ref{cor:augmentation-adjunction} and Corollary \ref{cor:wo-o-chaink} yield the following adjunction.

\begin{corollary}\label{cor:wcomc-adjunction}
The augmentation $\eta : \Wcomc \to \Comc$ induces an adjunction \[\nicexy{\algmwcomc \ar@<2pt>[r]^-{\eta_!} & \algmcomc \cong \Commc \ar@<2pt>[l]^-{\eta^*}}\] that is a Quillen equivalence if $\M=\Chaink$ with $\fieldk$ a field of characteristic zero
\end{corollary}

\begin{interpretation} Each $\C$-diagram of commutative monoids in $\M$ can be regarded as a $\Wcomc$-algebra via the augmentation $\eta$.  The left adjoint $\eta_!$ rectifies each homotopy coherent $\C$-diagram of $E_\infty$-algebras in $\M$ to a $\C$-diagram of commutative monoids in $\M$.\dqed\end{interpretation}

The colored operad $\Comc$ for $\C$-diagrams of commutative monoids in $\M$ is related to the commutative operad $\Com$ in Example \ref{ex:operad-com} and the colored operad $\Ocm$ for $\C$-diagrams of monoids in $\M$ in Example \ref{ex:diag-monoid-operad} as follows.  We will denote the unique color in $\Com$ by $*$.   The following observation is proved by a direct inspection.

\begin{lemma}\label{lem:com-to-comc}
Consider the $\colorc$-colored operad $\Comc$ in Example \ref{ex:diag-com-operad}.
\begin{enumerate}\item For each $c \in \colorc$, there is an operad morphism \[\nicexy{\Com \ar[r]^-{\iota_c} & \Comc}\] that sends $*$ to $c$ and is entrywise defined by the commutative diagram \[\nicexy@C+.5cm{\tensorunit \ar[d]_-{=} \ar[r]^-{=} & \tensorunit_{\{\Id_c\}_{j=1}^n} \ar[d]^-{\mathrm{inclusion}}\\ \Com(n) \ar[r]^-{\iota_c} & \coprod\limits_{\prod\limits_{j=1}^n\C(c,c)} \tensorunit= \Comc\ccc}\] for $n \geq 0$.
\item There is a morphism of $\colorc$-colored operads \[\nicexy{\Ocm \ar[r]^-{p} & \Comc}\] that is entrywise defined by the commutative diagrams \[\nicexy@C+.5cm{\tensorunit_{(\sigma,\uf)}\ar[d]_-{\mathrm{inclusion}} \ar[r]^-{=} & \tensorunit_{\uf} \ar[d]^-{\mathrm{inclusion}}\\ \Ocm\duc= \coprod\limits_{\Sigma_n\times \prod\limits_{j=1}^n \C(c_j,d)}\tensorunit \ar[r]^-{p} & \coprod\limits_{\prod\limits_{j=1}^n \C(c_j,d)} \tensorunit = \Comc\duc}\] for $(\uc;d) \in \Profcc$ with $\uc=(c_1,\ldots,c_n)$, $\sigma\in \Sigma_n$, and $\uf \in \prod_{j=1}^n\C(c_j,d)$.
\end{enumerate}
\end{lemma}

Applying Corollary \ref{cor:augmentation-adjunction} to the operad morphisms in Lemma \ref{lem:com-to-comc}, we obtain the following result.

\begin{corollary}\label{cor:com-comc-adjunction}
There is a diagram of change-of-operad adjunctions
\[\begin{small}
\nicexy@R+.3cm{\algmwcom \ar@<2pt>[r]^-{(\W\iota_c)_!} \ar@<-2pt>[d]_-{\eta_!} 
& \algmwcomc \ar@<2pt>[l]^-{\W\iota_c^*} \ar@<-2pt>[d]_-{\eta_!} \ar@<-2pt>[r]_-{\W p^*} & \algmwocm \ar@<-2pt>[l]_-{\W p_!} \ar@<-2pt>[d]_-{\eta_!} \\
\Comm\cong\algmcom \ar@<2pt>[r]^-{(\iota_c)_!} \ar@<-2pt>[u]_-{\eta^*}  
& \algmcomc \cong \Commc \ar@<2pt>[l]^-{\iota_c^*} \ar@<-2pt>[u]_-{\eta^*} \ar@<-2pt>[r]_-{p^*} & \algmocm \cong \Monmc \ar@<-2pt>[l]_-{p_!} \ar@<-2pt>[u]_-{\eta^*}}\end{small}\]
with commuting left adjoint diagrams and commuting right adjoint diagrams, where the left half is defined for each $c \in \colorc$.
\end{corollary}

\begin{interpretation} In the adjunction $(\iota_c)_! \dashv \iota_c^*$, the right adjoint $\iota_c^*$ remembers only the commutative monoid at the $c$-colored entry.   In the adjunction $p_! \dashv p^*$, the right adjoint $p^*$ sends a $\C$-diagram of commutative monoids in $\M$ to its underlying $\C$-diagram of monoids in $\M$; i.e., it forgets about the commutativity.  The right adjoint $\W\iota_c^*$ remembers only the $E_\infty$-algebra at the $c$-colored entry.  The right adjoint $\W p^*$ sends a homotopy coherent $\C$-diagram of $E_\infty$-algebras in $\M$ to the underlying homotopy coherent $\C$-diagram of $A_\infty$-algebras in $\M$.  Combined with Corollary \ref{cor:ass-ocm-adjunction}, one can forget further down to the underlying homotopy coherent $\C$-diagram in $\M$.\dqed
\end{interpretation}

The following result is the coherence theorem for homotopy coherent $\C$-diagrams of $E_\infty$-algebras in $\M$.  If the base category is $\Set$, then we will write $\Comc$ as $\Comcset$.

\begin{theorem}\label{thm:wcomc-coherence}
A $\Wcomc$-algebra is exactly a pair\index{Coherence Theorem!for hc diagrams of $E_\infty$-algebras} $(X,\lambda)$ consisting of 
\begin{itemize}\item a $\colorc$-colored object $X$ in $\M$ and
\item a structure morphism\index{structure morphism!for hc diagrams of $E_\infty$-algebras}
\begin{equation}\label{wcomc-restricted}
\nicexy@C+1.2cm{\J[T]\otimes X_{\uc} \ar[r]^-{\lambda_T\left\{\uf^v\right\}_{v\in T}} & X_d}\in \M
\end{equation}
for each $T \in \uTreec\duc$ with $(\uc;d) \in \Profcc$ and each \[\left\{\uf^v\right\}_{v\in \Vt(T)} \in \prod_{v\in \Vt(T)} \prod_{j=1}^{|\inp(v)|} \C\bigl(\inp(v)_j,\out(v)\bigr) = \prod_{v\in \Vt(T)} \Comcset(v)\]
\end{itemize}
that satisfies the following four conditions.
\begin{description}
\item[Associativity] Suppose $\bigl(\uc=(c_1,\ldots,c_n);d\bigr) \in \Profcc$ with $n \geq 1$, $T \in \uTreecduc$, $T_j \in \uTreec\cjubj$ for $1 \leq j \leq n$, $\ub=(\ub_1,\ldots,\ub_n)$, \[G=\graft(T;T_1,\ldots,T_n) \in \uTreec\dub\] is the grafting \eqref{def:grafting}, $\{\uf^v\}$ is as above, and \[\left\{\uf^u\right\}_{u \in \Vt(T_j)} \in \prod_{u \in \Vt(T_j)} \prod_{k=1}^{|\inp(u)|} \C\bigl(\inp(u)_k,\out(u)\bigr) =\prod_{u\in \Vt(T_j)} \Comcset(u)\] for each $1 \leq j \leq n$.  Then the diagram
\begin{equation}\label{wcomc-ass}
\nicexy@R+.5cm@C-1.5cm{&\J[T]\otimes \Bigl(\bigotimes\limits_{j=1}^n \J[T_j]\Bigr)\otimes X_{\ub} \ar[dl]_-{\mathrm{permute}}^-{\cong} \ar[dr]^-{(\pi,\Id)} &\\
\J[T]\otimes \bigotimes\limits_{j=1}^n\bigl(\J[T_j]\otimes X_{\ub_j}\bigr) \ar[d]_-{\left(\Id,\bigtensorover{j} \lambda_{T_j}\left\{\uf^u\right\}_{u\in T_j}\right)} && \J[G]\otimes X_{\ub} \ar[d]^-{\lambda_G\left\{\uf^w\right\}_{w\in G}}\\
\J[T]\otimes X_{\uc} \ar[rr]^-{\lambda_T\left\{\uf^v\right\}_{v\in T}} && X_d}
\end{equation}
is commutative.  Here $\pi=\bigotimes_S 1$ is the morphism in Lemma \ref{lem:morphism-pi} for the grafting $G$.
\item[Unity] For each $c \in \colorc$, the composition
\begin{equation}\label{wcomc-unity}
\nicexy@C+.5cm{X_c \ar[r]^-{\cong} & \J[\uparrow_c]\otimes X_c \ar[r]^-{\lambda_{\uparrow_c}\{\varnothing\}} & X_c}
\end{equation} 
is the identity morphism of $X_c$.
\item[Equivariance] For each $T \in \uTreec\duc$, $\{\uf^v\}$ as above, and permutation $\sigma \in \Sigma_{|\uc|}$, the diagram 
\begin{equation}\label{wcomc-eq}
\nicexy@C+1.5cm{\J[T]\otimes X_{\uc} \ar[d]_-{(\Id,\sigmainv)} \ar[r]^-{\lambda_T\left\{\uf^v\right\}_{v\in T}} & X_d\ar@{=}[d]\\
\J[T\sigma]\otimes X_{\uc\sigma} \ar[r]^-{\lambda_{T\sigma}\left\{\uf^v\right\}_{v\in T\sigma}} & X_d}
\end{equation}
is commutative, in which $T\sigma \in \uTreec\ducsigma$ is the same as $T$ except that its ordering is $\zeta_T\sigma$ with $\zeta_T$ the ordering of $T$.  The permutation $\sigmainv : X_{\uc} \iso X_{\uc\sigma}$ permutes the factors in $X_{\uc}$.
\item[Wedge Condition] Suppose $T \in \uTreec\duc$, $H_v \in \uTreec(v)$ for each $v\in \Vt(T)$, $K=T(H_v)_{v\in T}$ is the tree substitution, and \[\left\{\uf^u\right\}_{u \in \Vt(H_v)} \in \prod_{u\in \Vt(H_v)} \prod_{j=1}^{|\inp(u)|} \C\bigl(\inp(u)_j,\out(u)\bigr) =\prod_{u\in \Vt(H_v)} \Comcset(u)\] for each $v \in \Vt(T)$.  Then the diagram
\begin{equation}\label{wcomc-wedge}
\nicexy@C+1.4cm{\J[T] \otimes X_{\uc} \ar[d]_-{(\J,\Id)} 
\ar[r]^-{\lambda_T\left\{\ug^v\right\}_{v\in T}} 
& X_d \ar@{=}[d]\\
\J[K]\otimes X_{\uc} \ar[r]^-{\lambda_K\left\{\uf^u\right\}_{u\in K}} & X_d}
\end{equation}
is commutative.  Here for each $v\in \Vt(T)$, \[\ug^v = \gamma^{\Comcset}_{H_v}\Bigl(\{\uf^u\}_{u\in H_v}\Bigr) \in \Comcset(v)= \prod_{j=1}^{|\inp(v)|} \C\bigl(\inp(v)_j,\out(v)\bigr)\] with \[\nicexy@C+.8cm{\Comcset[H_v] =\prod\limits_{u\in H_v} \Comcset(u) \ar[r]^-{\gamma^{\Comcset}_{H_v}} & \Comcset(v)}\]
the operadic structure morphism of $\Comcset$ for $H_v$ in \eqref{operadic-structure-map}.
\end{description}
A morphism $f : (X,\lambda^X) \to (Y,\lambda^Y)$ of $\Wcomc$-algebras is a morphism of the underlying $\colorc$-colored objects that respects the structure morphisms in \eqref{wcomc-restricted} in the obvious sense.
\end{theorem}

\begin{proof}
This is the special case of the Coherence Theorem \ref{thm:wo-algebra-coherence} applied to the $\colorc$-colored operad $\Comc$.  Indeed, since \[\Comc\duc = \coprodover{\prod\limits_{j=1}^n \C(c_j,d)}\tensorunit = \coprodover{\Comcset\duc}\tensorunit,\] for each $T \in \uTreec\duc$ there is a canonical isomorphism \[\Comc[T] = \bigotimes_{v\in T} \Comc(v) = \bigotimes_{v\in T}\Bigl(\coprodover{\Comcset(v)} \tensorunit\Bigr) \cong \coprodover{\prodover{v\in T}\Comcset(v)} \tensorunit.\]  This implies that there is a canonical isomorphism \[\J[T]\otimes \Comc[T]\otimes X_{\uc} \cong \coprodover{\prodover{v\in T}\Comcset(v)} \J[T]\otimes X_{\uc}.\]  Therefore, the structure morphism $\lambda_T$ in \eqref{wo-algebra-restricted} is uniquely determined by the restricted structure morphisms $\lambda_T\{\uf^v\}_{v\in T}$ in \eqref{wcomc-restricted}.  The above associativity, unity, equivariance, and wedge conditions are those in Theorem \ref{thm:wo-algebra-coherence}.
\end{proof}

\begin{example}[Objectwise $E_\infty$-algebra]\label{ex:hcdiagcom-entrywise}
Suppose $(X,\lambda)$ is a $\Wcomc$-algebra, and $c \in \colorc$.  Under the right adjoint $\W\iota_c^*$ in Corollary \ref{cor:com-comc-adjunction}, we have that \[\W\iota_c^*(X,\lambda) \in \algmwcom,\] i.e., an $E_\infty$-algebra.  Explicit, its underlying object is $X_c \in \M$.  For $T \in \uTree(n)$, the $E_\infty$-algebra structure morphism \[\nicexy@C+.5cm{\J[T] \otimes X_c^{\otimes n} \ar[r]^-{\lambda_T} & X_c} \in \M\] in \eqref{einfinity-structure} is the structure morphism \[\lambda_{T_c}\Bigl\{\{\Id_c\}_{j=1}^{|\inp(v)|}\Bigr\}_{v\in T}\] in \eqref{wcomc-restricted}, where $T_c \in \uTreec\ccc$ is the $c$-colored tree obtained from $T$ by replacing every edge color by $c$.\dqed
\end{example}

\begin{example}[Underlying homotopy coherent $\C$-diagram]\label{ex:hcdiagcom-hcdiag}
Suppose $(X,\lambda)$ is a $\Wcomc$-algebra.  Under the right adjoints 
\[\nicexy@C+.5cm{\algmwcomc \ar[r]^-{\W p^*} & \algmwocm \ar[r]^-{\W i^*} & \algmwcdiag}\]
in Corollary \ref{cor:com-comc-adjunction} and Corollary \ref{cor:ass-ocm-adjunction}, we have that \[(\W i^*)(\W p)^*(X,\lambda) \in \algmwcdiag,\] i.e., a homotopy coherent $\C$-diagram in $\M$.  Explicitly, the homotopy coherent $\C$-diagram structure morphism 
\[\nicexy{\J[\Lin_{\uc}]\otimes X_{c_0} \ar[r]^-{\lambda_{\uc}^{\uf}} & X_{c_n}}\in \M\]
in \eqref{hcdiagram-structure-map} is the structure morphism $\lambda_{\Lin_{\uc}}\bigl\{f_j\bigr\}_{1\leq j \leq n}$ in \eqref{wcomc-restricted}.\dqed
\end{example}

\part{AQFT and Prefactorization Algebras}\label{part:aqft-pfa}

\chapter{Algebraic Quantum Field Theories}\label{ch:aqft}

This chapter is about algebraic quantum field theories in the operadic framework of \cite{bsw}.  In Section \ref{sec:haag-kastler} we provide a brief description of the traditional Haag-Kastler approach to algebraic quantum field theories and how it may be generalized to an operadic framework.  In Section \ref{sec:aqft} we discuss orthogonal categories and algebraic quantum field theories defined on them.  In Section \ref{sec:aqft-operad} we discuss the colored operads in \cite{bsw} whose algebras are algebraic quantum field theories.  Many examples are discussed in Section \ref{sec:aqft-examples}, including diagrams of (commutative) monoids, chiral conformal, Euclidean, and locally covariant quantum field theories, various flavors of quantum gauge theories, and quantum field theories on spacetimes with timelike boundary.  In Section \ref{sec:aqft-homotopy} we study homotopical properties of the category of algebraic quantum field theories.

As in previous chapters, $(\M,\otimes,\tensorunit)$ is a fixed cocomplete symmetric monoidal closed category, such as $\Vectk$ and $\Chaink$, and $\colorc$ is a non-empty set.

\section{From Haag-Kastler Axioms to Operads}\label{sec:haag-kastler}

In this section, we provide a brief overview of the traditional approach to algebraic quantum field theories due to Haag and Kastler.  Then we review how the Haag-Kastler approach is modified to the operadic viewpoint in \cite{bsw}, which is what the rest of this chapter is about.

Haag and Kastler \cite{hk} defined an algebraic quantum field theory on a fixed Lorentzian spacetime $X$ as a rule $\fraka$ that assigns\index{Haag-Kastler axioms} 
\begin{itemize}\item to each suitable spacetime region $U \subseteq X$ a unital associative algebra $\fraka(U)$ and 
\item to each inclusion $f : U \subseteq V$ an injective algebra homomorphism $\fraka(f) : \fraka(U) \to \fraka(V)$.
\end{itemize}
The algebra $\fraka(U)$ is the algebra of quantum observables in $U$.  The homomorphism $\fraka(f)$ sends each observable in $U$ to an observable in the larger region $V$.  The condition that each $\fraka(f)$ be injective is called the \index{isotony axiom}\emph{isotony axiom}.  Moreover, it is assumed that the following axioms are satisfied.
\begin{description}
\item[Causality Axiom] If $U_1 \subseteq V$ and $U_2 \subseteq V$ \index{causality axiom}are causally disjoint regions in $V$, then each element in $\fraka(U_1)$ and each element in $\fraka(U_2)$ commute in $\fraka(V)$.
\item[Time-Slice Axiom] If $U \subseteq V$ contains a \index{Cauchy surface}Cauchy surface of $V$, then the homomorphism $\fraka(U) \iso \fraka(V)$ is an \index{time-slice axiom}isomorphism.
\end{description}
The causality axiom corresponds to the physical principle that effects do not travel faster than the speed of light.  The time-slice axiom says that observables in a small time interval determine all observables.

The Haag-Kastler approach is generalized in \cite{bfv} to the category of all oriented, time-oriented, and globally hyperbolic Lorentzian manifolds.  To obtain other flavors of quantum field theories, such as chiral conformal and Euclidean quantum field theories, the above framework is abstracted one step further in \cite{bsw} by replacing the category of spacetimes with an abstract small category $\C$ equipped with a set $\perp$ of pairs of morphisms $(g_1 : a \to c, g_2 : b \to c)$ with the same codomain.  Physically one interprets the objects in $\C$ as the spacetimes of interest and the morphisms as inclusions of smaller regions into larger regions.  A pair $(g_1,g_2) \in \perpen$ means that their domains $a$ and $b$ are suitably disjoint regions in the common codomain $c$.  The pair $\Cbar = (\C,\perp)$ is called an orthogonal category.  

The causality axiom is implemented using the set $\perp$ of orthogonality relations.  The time-slice axiom may be implemented by choosing a suitable set $S$ of morphisms in $\C$, corresponding to the Cauchy morphisms in the Lorentzian case.  In addition to the domain category, the target category can also be replaced by the category of monoids in a symmetric monoidal category $\M$, with $\M=\Vectk$ being the traditional case.  So now an algebraic quantum field theory on $\Cbar = (\C,\perp)$ is a functor $\fraka : \C \to \Monm$ that satisfies the causality axiom and, if a set $S$ of morphisms is given, the time-slice axiom. 

In Example \ref{ex:diag-monoid-operad} we saw that there is an $\Obc$-colored operad $\Ocm$ whose algebras are exactly $\C$-diagrams of monoids in $\M$.  With a bit more work, one can build the causality axiom into the colored operad.  So there is an $\Obc$-colored operad $\Ocbarm$ whose category of algebras is exactly the category of algebraic quantum field theories on $\Cbar=(\C,\perp)$.  To implement the time-slice axiom, one first replaces the small category $\C$ with its $S$-localization $\Csinv$ and the orthogonality relation $\perp$ by a suitable pushforward.  In Section \ref{sec:aqft-operad} we will discuss this operadic framework for algebraic quantum field theories.

One might wonder what happened to the isotony axiom, which requires that each homomorphism $\fraka(f)$ be injective.  Various models of \index{quantum gauge theory}quantum gauge theories actually do not satisfy the isotony axiom; see for example \cite{bbss,bss17,bdhs,bds14,bs17,dl,sdh}.  Since the operadic framework is general enough to include some flavors of quantum gauge theories, it is reasonable to drop the isotony axiom.

\section{AQFT as Functors}\label{sec:aqft}

In this section, we review the functor definition of algebraic quantum field theories as discussed in \cite{bsw}.  All the assertions in this section are from \cite{bsw}, where the reader may find more details.

The causality axiom says that certain elements from separated regions should commute.  The following concept of an orthogonality relation is used to formalize the idea of separated regions.  

\begin{definition}\label{def:orthogonal-category}
Suppose $\C$ is a small category.  An \emph{orthogonality relation}\index{orthogonality relation} on $\C$ is a subset\label{notation:perp} $\perp$ of pairs of morphisms in $\C$ such that if $(f,g) \in \perpen$, then $f$ and $g$ have the same codomain.  Furthermore, it is required to satisfy the following three axioms.
\begin{description}
\item[Symmetry] If $(f,g) \in \perpen$, then $(g,f) \in \perpen$.
\item[Post-Composition] If $(g_1,g_2) \in \perpen$, then $(fg_1,fg_2) \in \perpen$ for all composable morphisms $f$ in $\C$.
\item[Pre-Composition] If $(g_1,g_2) \in \perpen$, then $(g_1h_1,g_2h_2) \in \perpen$ for all composable morphisms $h_1$ and $h_2$ in $\C$. 
\end{description}
If\label{notation:fperpg} $(f,g) \in \perpen$, then we also write $f \perp g$ and say that $f$ and $g$ are\index{orthogonal morphisms} \emph{orthogonal}.  
\begin{itemize}
\item An\index{category!orthogonal} \emph{orthogonal category} is a small category equipped with an orthogonality relation.
\item An\index{functor!orthogonal} \emph{orthogonal functor}\label{notation:cperp} \[F : \Cbar = (\C,\perpc) \to (\D,\perpd) = \Dbar\] between orthogonal categories is a functor $F : \C \to \D$ that preserves the orthogonality relations in the sense that $(f,g) \in \perpenc$ implies $(Ff,Fg) \in \perpend$.
\item The category of orthogonal categories and orthogonal functors is denoted by\label{notation:orthcat} $\Orthcat$.
\end{itemize}
\end{definition}

Orthogonality relations can be pulled back and pushed forward via any functor.  The following observation is \cite{bsw} Lemma 4.29, which follows directly from Definition \ref{def:orthogonal-category}.

\begin{lemma}\label{lem:pullback-orthogonality}
Suppose $F : \C \to \D$ is a functor between small categories.
\begin{enumerate}
\item If $\perpc$ is an orthogonality relation on $\C$, then\label{notation:fpushforward}
\[F_*(\perpc) = \Bigl\{\bigl(fF(g_1)h_1,fF(g_2)h_2\bigr) : (g_1,g_2) \in \perpenc,\, f,h_1,h_2 \in \D\Bigr\}\]
is an orthogonality relation on $\D$, called the\index{pushforward orthogonality relation} pushforward of $\perpc$ along $F$.  Moreover, \[F : (\C,\perpc) \to \bigl(\D,F_*(\perpc)\bigr)\] is an orthogonal functor.
\item If $\perpd$ is an orthogonality relation on $\D$, then\label{notation:fpullback}
\[F^*(\perpd) = \Bigl\{(f_1,f_2) : \cod(f_1)=\cod(f_2),\,(Ff_1,Ff_2) \in \perpend \Bigr\}\]
is an orthogonality relation on $\C$, called the\index{pullback orthogonality relation} pullback of $\perpd$ along $F$.  Moreover, \[F : \bigl(\C,F^*(\perpd)\bigr) \to (\D,\perpd)\] is an orthogonal functor.
\end{enumerate}
\end{lemma}

Recall from Section \ref{sec:monoids} that $\Monm$ is the category of monoids in $\M$.  The isotony axiom in the Haag-Kastler setting \cite{hk} says that for each inclusion of regions there is a corresponding inclusion of algebras.  In the categorical setting, instead of algebra inclusions, we ask for a functor from the category of regions to the category of monoids.

\begin{definition}\label{def:aqft}
Suppose $\Cbar = (\C,\perp)$ is an orthogonal category, and $S$ is a set of morphisms in $\C$.
\begin{enumerate}
\item A functor $\fraka : \C \to \Monm$ satisfies the\index{causality axiom} \emph{causality axiom} if for each orthogonal pair $(g_1 : a \shortto c, g_2 : b \shortto c) \in \perpen$, the diagram
\begin{equation}\label{perp-com}
\nicexy@C+.7cm{\fraka(a) \otimes \fraka(b) \ar[d]_-{\bigl(\fraka(g_1),\fraka(g_2)\bigr)} \ar[r]^-{\bigl(\fraka(g_1),\fraka(g_2)\bigr)} & \fraka(c) \otimes \fraka(c) \ar[r]^-{\mathrm{permute}}_-{\cong} & \fraka(c) \otimes \frak(c) \ar[d]^-{\mu_c}\\ \fraka(c) \otimes \fraka(c) \ar[rr]^-{\mu_c} && \fraka(c)}
\end{equation}
in $\M$ is commutative, where $\mu_c$ is the monoid multiplication in $\fraka(c)$.
\item An \index{algebraic quantum field theory}\emph{algebraic quantum field theory on $\Cbar$} is a functor $\fraka : \C \to \Monm$ that satisfies the causality axiom.
\item The full subcategory of the diagram category $\Monm^{\C}$ whose objects are algebraic quantum field theories on $\Cbar$ is denoted by\label{notation:qftcbar} $\QFT(\Cbar)$.
\item A functor $\fraka : \C\to\Monm$ satisfies the\index{time-slice axiom!algebraic quantum field theory} \emph{time-slice axiom} with respect to $S$ if for each $s : a \to b \in S$, the morphism \[\fraka(s) : \fraka(a) \iso \fraka(b) \in \Monm\] is an isomorphism.
\item The full subcategory of $\QFT(\Cbar)$ consisting of algebraic quantum field theories on $\Cbar$ that satisfy the time-slice axiom with respect to $S$ is denoted by\label{notation:qftcbars} $\QFT(\Cbar,S)$.
\end{enumerate}
\end{definition}

\begin{interpretation}\label{int:orthogonal-category}
Physically the objects in the orthogonal category $\Cbar$ are the spacetime regions of interest.  The functor $\fraka$ assigns a monoid of quantum observes $\fraka(c)$ to each region $c$.  The orthogonality relation $\perp$ specifies the disjoint regions.  The causality axiom says that, if $a$ and $b$ are disjoint regions in $c$, then an observable from $a$ and an observable from $b$ commute in $c$. The set $S$ specifies the Cauchy morphisms.  The time-slice axiom says that Cauchy morphisms are sent to isomorphisms of monoids of quantum observables.  For an orthogonal category $\Cbar$, $\QFT(\Cbar)$ is the category of all the quantum field theories associated to the spacetime regions in $\C$.\dqed
\end{interpretation}

\begin{remark}In \cite{bsw} the causality axiom, the time-slice axiom, and an algebraic quantum field theory are called $\perp$-commutativity, $W$-constancy, and a $\perp$-commutative functor $\C \to \Monm$, respectively.  Moreover, $\QFT(\Cbar)$ is denoted by $\Mon_{\M}^{\Cbar}$ in \cite{bsw}.\dqed\end{remark}

Recall the concept of a localization of a category in Section \ref{sec:localization}.  The following result is \cite{bsw} Lemma 4.30, which uses the pushforward orthogonality relation in Lemma \ref{lem:pullback-orthogonality}.  

\begin{lemma}\label{lem:aqft-time-slice}
Suppose $\Cbar = (\C,\perp)$ is an orthogonal category, and $S$ is a set of morphisms in $\C$ with $S$-localization\index{localization of a category} $\ell : \C \to \Csinv$.  Suppose\label{notation:csinvbar} \[\Csinvbar = \Bigl(\Csinv, \ell_*(\perp)\Bigl)\] is the orthogonal category equipped with the pushforward of $\perp$ along $\ell$.  Then there is a canonical isomorphism \[\QFT(\Cbar,S) \cong \QFT(\Csinvbar).\]  Using this isomorphism, we will regard $\QFT(\Csinvbar)$ as a full subcategory of $\QFT(\Cbar)$.
\end{lemma}

\begin{proof}
Let us first indicate the correspondence between objects.  First, an object on the right side yields an object on the left side by pre-composition with the $S$-localization $\ell$.

On the other hand, suppose $\fraka : \C \to \Monm$ is a functor that satisfies the causality axiom and the time-slice axiom with respect to $S$.  Then by the universal property of the $S$-localization, there is a unique functor \[\frakb : \Csinv \to \Monm \stspace \fraka = \frakb \ell.\]  To see that $\frakb$ satisfies the causality axiom, recall that each orthogonal pair in the pushforward $\ell_*(\perp)$ has the form $\bigl(f\ell(g_1)h_1,f\ell(g_2)h_2\bigr)$ with
\begin{itemize}\item $(g_1 : a \shortto c) \perp (g_2 : b \shortto c)$ and 
\item $f : \ell c \shortto d, h_1 : x \shortto \ell a, h_2 : y \shortto \ell b\in \Csinv$.  
\end{itemize}
We want to know that the outermost diagram in 
\[\nicexy@C+1cm{\frakb x \otimes \frakb y \ar[d]_-{(\frakb h_1, \frakb h_2)} \ar[r]^-{(\frakb h_1, \frakb h_2)} & \frakb\ell a\otimes \frakb\ell b \ar[r]^-{(\frakb\ell g_1, \frakb\ell g_2)} & \frakb\ell c\otimes \frakb\ell c \ar[d]^-{(\frakb f, \frakb f)}\ar[dl]_-{\mathrm{permute}}^-{\cong}\\
\frakb\ell a \otimes \frakb\ell b \ar@{}[r]|-{(*)} \ar[d]_-{(\frakb\ell g_1, \frakb\ell g_2)} \ar[ur]^-{=} & \frakb\ell c \otimes \frakb\ell c \ar[d]_-{\mu_{\frakb\ell c}} \ar[dr]^-{(\frakb f,\frakb f)} & \frakb d \otimes \frakb d \ar[d]^-{\mathrm{permute}}_-{\cong}\\
\frakb\ell c \otimes \frakb\ell c \ar[d]_-{(\frakb f,\frakb f)} \ar[r]^-{\mu_{\frakb\ell c}} & \frakb\ell c \ar[dr]^-{\frakb f} & \frakb d \otimes \frakb d \ar[d]^-{\mu_{\frakb d}}\\
\frakb d \otimes \frakb d \ar[rr]^-{\mu_{\frakb d}} && \frakb d}\] 
is commutative.  The upper left triangle and the upper right triangle are commutative by definition and the symmetry in $\M$, respectively.  The bottom trapezoid is equal to the adjacent parallelogram, which is commutative because $\frakb f$ is a morphism of monoids in $\M$.  Since $\frakb\ell =\fraka$, the sub-diagram $(*)$ is the commutative diagram \eqref{perp-com}.

To see the correspondence between morphisms, we simply use the description in Theorem \ref{thm:localization-cat} of the morphisms in the localization.  
\end{proof}

\begin{interpretation} Lemma \ref{lem:aqft-time-slice} says that the time-slice axiom in algebraic quantum field theories can be implemented by replacing the orthogonal category with its localization along with the pushforward orthogonality relation.  Therefore, algebraic quantum field theories and those that satisfy the time-slice axiom can be studied in the same setting.\dqed\end{interpretation}

\section{AQFT as Operad Algebras}\label{sec:aqft-operad}

In this section, following \cite{bsw} we describe a colored operad whose algebras are algebraic quantum field theories on a given orthogonal category.  

\begin{motivation}From the previous section, an algebraic quantum field theory on an orthogonal category $(\C,\perp)$ is a functor $\fraka : \C \to \Monm$ that satisfies the causality axiom and, if a set $S$ of morphisms is given, the time-slice axiom with respect to $S$.   The time-slice axiom says that certain structure morphisms are invertible, which by Lemma \ref{lem:aqft-time-slice} can be implemented by using the $S$-localization of $\C$.  The functor $\fraka$ itself is a $\C$-diagram of monoids in $\M$, while the causality axiom is a form of commutativity.  As we saw in Examples \ref{ex:operad-com} and \ref{ex:diag-monoid-operad}, commutative monoids and diagrams of monoids can all be modeled using (colored) operads.  Therefore, it is natural to expect a colored operad whose algebras are algebraic quantum field theories.  Recall from Definition \ref{def:operad-generating} the description of a colored operad in terms of generating operations and generating axioms.\dqed\end{motivation}

\begin{definition}\label{def:aqft-operad}
Suppose $\Cbar = (\C,\perp)$ is an orthogonal category with $\Obc=\colorc$.  Define the following sets and functions.
\begin{description}
\item[Entries] Define the object\label{notation:ocbar} $\Ocbar \in \Set^{\Profcc}$\index{algebraic quantum field theory!operad}\index{colored operad!for AQFT} entrywise as the quotient set
\[\Ocbar\duc = \Bigl(\Sigma_{n} \times \prod\limits_{j=1}^n \C(c_j,d)\Bigr)\Big/\sim \forspace \duc=\dconecn \in \Profcc\] in which the equivalence relation $\sim$ is defined as follows.  For $(\sigma,\uf)$ and $(\sigma',\uf')$ in $\Sigma_{n} \times \prod_{j=1}^n \C(c_j,d)$, we define\label{notation:aqftsim} \[(\sigma,\uf) \sim (\sigma',\uf')\] if and only if the following two conditions hold:
\begin{itemize}
\item $\uf=\uf'$ in $\prod_{j=1}^n \C(c_j,d)$.
\item $\sigma\sigma'^{-1}$ factors as a product $\tau_1\cdots\tau_r$ of transpositions in $\Sigma_{n}$ such that, for each $1 \leq k \leq r$, the right permutation
\[\tau_k : \uf\sigmainv\tau_1\cdots \tau_{k-1} \to \uf\sigmainv\tau_1\ldots\tau_k\] is a transposition of two morphisms in $\C$ that are adjacent and orthogonal in $\uf\sigmainv\tau_1\cdots\tau_{k-1}$.
\end{itemize}
The equivalence class of $(\sigma,\uf)$ is denoted by\label{notation:sigmaf} $[\sigma,\uf]$.
\item[Equivariance] For $\tau \in \Sigma_{|\uc|}$, define the map
\[\nicexy{\Ocbar\duc \ar[r]^-{\tau} & \Ocbar\ductau}\] by $[\sigma,\uf]\tau = [\sigma\tau,\uf\tau]$.
\item[Colored Units] For $c \in \colorc$, the $c$-colored unit in $\Ocbar\cc$ is $[\id_1,\Id_c]$.
\item[Operadic Composition] For $(\uc;d) \in \Profcc$ with $|\uc|=n \geq 1$, $\ub_j=(b_{j1},\ldots,b_{jk_j}) \in \Profc$ for $1 \leq j \leq n$ with $|\ub_j|=k_j\geq 0$, and $\ub=(\ub_1,\ldots,\ub_n)$, define the map
\[\nicexy{\Ocbar\duc \times \prod\limits_{j=1}^n \Ocbar\cjubj \ar[r]^-{\gamma} & \Ocbar\dub}\] by
\[\gamma\Bigl([\sigma,\uf];\bigl\{[\tau_j,\ug_j]\bigr\}_{j=1}^n\Bigr) = \Bigl[\sigma(\tau_1,\ldots,\tau_n),\bigl(f_1\ug_1,\ldots,f_n\ug_n\bigr) \Bigr]\] where \[f_j\ug_j=\bigl(f_jg_{j1},\ldots,f_jg_{jk_j}\bigr) \in \prod\limits_{i=1}^{k_j} \C(b_{ji},d) \ifspace \ug_j=\bigl(g_{j1},\ldots,g_{jk_j}\bigr) \in \prod\limits_{i=1}^{k_j} \C(b_{ji},c_j)\] and \[\sigma(\tau_1,\ldots,\tau_n) = \underbrace{\sigma\langle k_1,\ldots,k_n\rangle}_{\text{block permutation}} \circ~ \underbrace{(\tau_1\oplus\cdots\oplus\tau_n)}_{\text{block sum}} \in \Sigma_{k_1+\cdots+k_n}\]
as in \eqref{as-comp}.
\end{description}
\end{definition}

\begin{interpretation} In the previous definition, one should think of $[\sigma,f]$ as a three-step operation:
\begin{enumerate}\item Apply the morphisms in $\uf$ to observables in $(c_1,\ldots,c_n)$.
\item Permute the result from the left by $\sigma$.
\item Multiply the observables in $d$.\end{enumerate}
The equivalence relation $\sim$ is generated by transpositions of adjacent orthogonal pairs.\dqed
\end{interpretation}

\begin{example}\label{ex:ocbar-unary}
The equivalence relation $\sim$ only has an effect when the sequence $\uf$ has length $>1$.  So for any colors $c,d \in \colorc$, there is a canonical bijection \[\nicexy{\C(c,d) \ar[r]^-{\cong} & \Sigma_1 \times \C(c,d) = \Ocbar\dc},\] sending $f \in \C(c,d)$ to $[\id_1,f] \in \Ocbar\dc$.\dqed
\end{example}

The following concept is the orthogonal version of an equivalence of categories in Definition \ref{def:equivalence-categories}.

\begin{definition}\label{def:orthogonal-equivalence}
An \index{equivalence!orthogonal}\index{orthogonal equivalence}\emph{orthogonal equivalence} is an orthogonal functor $F : \Cbar \to \Dbar$ such that
\begin{itemize}\item $F : \C \to \D$ is an equivalence of categories, and
\item $\perpc ~= F^*(\perpd)$.\end{itemize}
\end{definition}

Recall from Example \ref{ex:operad-set-m} (i) the strong symmetric monoidal functor $\Set \to \M$ that sends a set $S$ to $\coprod_S \tensorunit$ and (ii) the corresponding functor \[(-)^{\M} : \Operad(\Set) \to \Operad(\M)\] between categories of operads.  The following observations are the main categorical properties of the above construction.  They are from \cite{bsw} Proposition 4.11, Proposition 4.16, Theorem 4.27, Proposition 5.4, and Theorem 5.11.

\begin{theorem}\label{thm:ocbar-algebra}
Suppose $\Cbar = (\C,\perp)$ is an orthogonal category with $\Obc=\colorc$.
\begin{enumerate}
\item With the structure in Definition \ref{def:aqft-operad}, $\Ocbar$ is a $\colorc$-colored operad in $\Set$.
\item The construction $\O_{(-)}$ defines a functor \[\O_{(-)} : \Orthcat \to \Operad(\Set).\]
\item There is a canonical isomorphism 
\begin{equation}\label{aqft=operadalgebra}
\algm\bigl(\Ocbarm\bigr) \cong \QFT(\Cbar).
\end{equation}
\item For each set $S$ of morphisms in $\C$, the $S$-localization functor $\ell$ induces a change-of-operad adjunction
\begin{equation}\label{aqft-operad-timeslice}
\nicexy@C+.3cm{\QFT(\Cbar) \cong \algm\bigl(\Ocbarm\bigr) \ar@<2pt>[r]^-{(\Otom_\ell)_!} &  \algm\bigl(\Ocsinvbarm\bigr) \cong \QFT(\Cbar,S) \ar@<2pt>[l]^-{(\Otom_\ell)^*}}
\end{equation}
whose counit \[\epsilon : (\Otom_\ell)_!(\Otom_\ell)^* \iso \Id_{\algm\bigl(\Ocsinvbarm\bigr)}\] is a natural isomorphism.
\item Each orthogonal functor $F : \Cbar \to \Dbar$ induces a change-of-operad adjunction
\begin{equation}\label{aqft-operad-changecat}
\nicexy@C+.3cm{\QFT(\Cbar) \cong \algm\bigl(\Ocbarm\bigr) \ar@<2pt>[r]^-{(\Otom_F)_!} &  \algm\bigl(\Odbarm\bigr) \cong \QFT(\Dbar) \ar@<2pt>[l]^-{(\Otom_F)^*}}.
\end{equation}
\item If $F : \Cbar \to \Dbar$ is an orthogonal equivalence, then the change-of-operad adjunction in \eqref{aqft-operad-changecat} is an adjoint equivalence.
\end{enumerate}
\end{theorem}

\begin{proof}
For the first assertion, one checks directly that the structure morphisms for $\Ocbar$ are well-defined and that they satisfy the axioms in Definition \ref{def:operad-generating}.  The second assertion also follows from a direct inspection.

For the third assertion, let us describe the correspondence between objects.  From the left side, by Definition \ref{def:operad-algebra-generating} an $\Ocbarm$-algebra consists of a $\colorc$-colored object $X=\{X_c\}_{c\in \colorc}$ in $\M$ together with a structure morphism
\[\nicexy{\coprodover{\Ocbar\duc} X_{\uc} \cong \Ocbarm\duc \otimes X_{\uc} \ar[r]^-{\lambda} & X_d\in \M}\] for each $(\uc;d) \in \Profcc$ that satisfies the associativity, unity, and equivariance axioms.  The restriction of $\lambda$ to a copy of $X_{\uc}$ corresponding to an element $x \in \Ocbar\duc$, 
\[\nicexy@C+.4cm{X_{\uc} \ar[r]^-{x}_-{\text{summand}} \ar`u[rr]`[rr]^-{\lambda_x}[rr] & \Ocbarm\duc \otimes X_{\uc} \ar[r]^-{\lambda} & X_d},\]
will be denoted by $\lambda_x$.  Define a functor $\fraka_X : \C\to\M$ by setting \[\begin{split} \fraka_X(c)&=X_c \forspace c \in \colorc,\\
\fraka_X(f)&=\lambda_{[\id_1,f]} : X_c \to X_d \forspace f \in \C(c,d).\end{split}\]
One checks that $\fraka_X$ is well-defined. Moreover, it extends to a functor \[\fraka_X : \C\to\Monm\] such that, for each $c \in \colorc$, $\fraka_X(c)=X_c$ has monoid multiplication \[\lambda_{[\id_2,(\Id_c,\Id_c)]} : X_c \otimes X_c \to X_c\] and unit \[\lambda_{[\id_0,\varnothing]} : \tensorunit \to X_c.\]  That $\fraka_X$ satisfies the causality axiom is a consequence of the equivalence relation $\sim$ that defines each entry of $\Ocbar$.  So $\fraka_X$ is an algebraic quantum field theory on $\Cbar$.

For the converse, the key point is that the $\colorc$-colored operad $\Ocbar$ is generated by the elements
\begin{itemize}\item $\mu_c=[\id_2,(\Id_c,\Id_c)] \in \Ocbar\sbinom{c}{c,c}$, 
\item $[\id_1,f]\in \Ocbar\dc$, and
\item $1_c=[\id_0,\varnothing] \in \Ocbar\cempty$
\end{itemize}
for all $c,d\in \colorc$ and $f \in \C(c,d)$, and permutations.  Indeed, for each $m\geq 3$ and $c \in \C$, the element \[\mu_m=[\id_m,(\underbrace{\Id_c,\ldots,\Id_c}_{m})]\in \Ocbar\sbinom{c}{c,\ldots,c}\] is equal to \[\gamma\bigl(\mu_2; [\id_1,\Id_c],\mu_{m-1}\bigr),\] so by induction all the $\mu_m$ are generated by $\mu_2$ and the $c$-colored unit.  For $n \geq 2$ and $f_j \in \C(c_j,d)$ for $1 \leq j \leq n$, we have \[\bigl[\id_n,(f_1,\ldots,f_n)\bigr] = \gamma\Bigl(\mu_n; \bigl\{[\id_1,f_j]\bigr\}_{j=1}^n\Bigr) \in \Ocbar\dconecn.\]  So together with permutations the above elements generate all of $\Ocbar$.  Furthermore, one checks that all the generating relations among these generators are already reflected in the properties of an algebraic quantum field theory.  Therefore, using the previous paragraph and the axioms in Definition \ref{def:operad-algebra-generating}, an algebraic quantum field theory on $\Cbar$ determines an $\Ocbarm$-algebra.

The change-of-operad adjunction \eqref{aqft-operad-timeslice} in the fourth assertion is a consequence of Theorem \ref{thm:change-operad} applied to the morphism \[\Otom_\ell : \Ocbarm \to \Ocsinvbarm\] of $\colorc$-colored operads, the previous two assertions, and Lemma \ref{lem:aqft-time-slice}.  The counit is a natural isomorphism by \cite{maclane} VI.3 Theorem 1 because the right adjoint $(\Otom_\ell)^*$ is full and faithful, which in turn is true because on both sides a morphism is a natural assignment of a monoid morphism to each object in $\C$.

The change-of-operad adjunction \eqref{aqft-operad-changecat} in assertion (5) is a consequence of Theorem \ref{thm:change-operad} applied to the operad morphism \[\Otom_F : \Ocbarm \to \Odbarm\] and of assertions (2) and (3).

For assertion (6), since a left adjoint is unique up to a unique isomorphism, it is enough to show that the right adjoint $(\Otom_F)^*$ is an equivalence of categories, i.e., full, faithful, and essentially surjective.  By the isomorphism \eqref{aqft=operadalgebra} in assertion (3), it is enough to show that the functor \[\nicexy{\QFT(\Dbar) \ar[r]^-{F^*} & \QFT(\Cbar)}\] is an equivalence of categories.  For $\fraka \in \QFT(\Dbar)$, this functor is defined as \[F^*\fraka = \fraka F,\] i.e., pre-composition with $F : \C \to \D$.  Similarly, for $\fraka,\frakb \in \QFT(\Dbar)$, the function on morphisms \[\nicexy@R-.4cm{\QFT(\Dbar)(\fraka,\frakb) \ar@{=}[d] \ar[r]^-{F^*} & \QFT(\Cbar)(F^*\fraka,F^*\frakb) \ar@{=}[d] \\\Monm^{\D}(\fraka,\frakb) & \Monm^{\C}(\fraka F,\frakb F)}\] is given by pre-composition with $F$.  Using that $F : \C \to \D$ is an equivalence of categories, one checks that the function on morphisms $F^*$ is a bijection.  Therefore, the functor $F^*$ is full and faithful.

To see that the functor $F^*$ is essentially surjective, suppose $\fraka \in \QFT(\Cbar)$. We must show that there exist $\frakb \in \QFT(\Dbar)$ and an isomorphism $F^*\frakb \cong \fraka$.  Define a functor $\frakb : \D \to \Monm$ as follows.  
\begin{itemize}\item For each object $d \in \D$, since $F$ is an equivalence of categories, we can choose
\begin{itemize}\item an object $d' \in \C$ and
\item an isomorphism $\rho_d : d\iso Fd'$.  
\end{itemize}
We can furthermore insist that, if $d$ is in the image of $F$, then $d'$ is chosen from the $F$-pre-image of $d$ and that $\rho_d = \Id_d$.  Define \[\frakb(d) = \fraka(d') \in \Monm.\]
\item Suppose given a morphism $f \in \D(d_1,d_2)$.  In the previous step, we have chosen objects $d_1',d_2'\in \C$ and isomorphisms $d_1 \iso Fd_1'$ and $d_2 \iso Fd_2'$.  These choices yield bijections \[\nicexy{\C(d_1',d_2') \ar[r]^-{F}_-{\cong} & \D(Fd_1',Fd_2') \ar[r]^-{\cong} & \D(d_1,d_2)},\]
so $f$ has a unique pre-image $f' \in \C(d_1',d_2')$.  Define \[\frakb(f) = \fraka(f') : \frakb(d_1)=\fraka(d_1') \to \fraka(d_2') = \frakb(d_2) \in \Monm.\]
\end{itemize}
Using that $F$ is an orthogonal equivalence, one can check that this actually defines a functor $\frakb$ that satisfies the causality axiom, i.e., $\frakb \in \QFT(\Dbar)$.  Furthermore, by construction $\fraka$ and $F^*\frakb$ are naturally isomorphic as functors $\C \to \Monm$.  Therefore, $F^*$ is essentially surjective.
\end{proof}

\begin{interpretation}  Consider Theorem \ref{thm:ocbar-algebra}.
\begin{enumerate}
\item Via the isomorphism \eqref{aqft=operadalgebra}, the causality axiom of algebraic quantum field theories are built into the $\colorc$-colored operad $\Ocbar$ via the equivalence relation $\sim$ in Definition \ref{def:aqft-operad}.  In particular, from the operadic viewpoint, the causality axiom is not an extra property that a functor $\C \to \Monm$ may or may not satisfy.  Instead, every $\Ocbarm$-algebra already satisfies the causality axiom.  Using this isomorphism, we will identify algebraic quantum field theories on $\Cbar$ with $\Ocbarm$-algebras.
\item Similarly, via the isomorphism \[\algm\bigl(\Ocsinvbarm\bigr) \cong \QFT(\Cbar,S)\] the time-slice axiom with respect to $S$ is built into the $\colorc$-colored operad $\Ocsinvbar$.  So every $\Ocsinvbarm$-algebra already satisfies the time-slice axiom.  Using this isomorphism, we will identify algebraic quantum field theories on $\Cbar$ that satisfy the time-slice axiom with $\Ocsinvbarm$-algebras.
\item The right adjoint $(\Otom_\ell)^*$ in the change-of-operad adjunction \eqref{aqft-operad-timeslice} says that each algebraic quantum field theory on $\Cbar$ that satisfies the time-slice axiom is in particular an algebraic quantum field theory on $\Cbar$.  The left adjoint $(\Otom_\ell)_!$ assigns to each algebraic quantum field theory on $\Cbar$ another one that satisfies the time-slice axiom.
\item The change-of-operad adjunction in \eqref{aqft-operad-changecat} allows one to go back and forth between algebraic quantum field theories of different flavors, i.e., those on $\Cbar$ and those on $\Dbar$.\dqed
\end{enumerate}
\end{interpretation}

\begin{remark} In Theorem \ref{thm:ocbar-algebra} assertion (6), we observed that the change-of-operad adjunction associated to an orthogonal equivalence is an adjoint equivalence.  The proof given above uses (i) the canonical isomorphism \eqref{aqft=operadalgebra} and (ii) elementary facts about an orthogonal equivalence.  This line of argument is very similar to the well-known proof of Theorem \ref{thm:equivalence-categories} that characterizes equivalences of categories.  On the other hand, the proof of this adjoint equivalence given in \cite{bsw} Theorem 5.11 directly deals with the algebra categories $\algm(\Ocbarm)$ and $\algm(\Odbarm)$, and uses more sophisticated techniques.  Furthermore, Theorem \ref{thm:ocbar-algebra} assertion (6) has a homotopy version, given below in Theorem \ref{thm:aqft-model}.\dqed
\end{remark}

\section{Examples of AQFT}\label{sec:aqft-examples}

In this section we provide examples of orthogonal categories and algebraic quantum field theories.  The first two examples are the two extreme cases for the orthogonality relation.

\begin{example}[Diagrams of monoids]\label{ex:empty-causality}
Suppose $\C$ is a small category equipped with the empty orthogonality relation (i.e., $\perp~=\varnothing$).  Since the commutative diagram in the causality axiom \eqref{perp-com} never happens, an algebraic quantum field theory on the \index{minimal orthogonal category}\index{diagram of monoids!as AQFT}\index{algebraic quantum field theory!on minimal orthogonal category}\emph{minimal orthogonal category}\label{notation:cbarmin} \[\Cbarmin=(\C,\varnothing)\] is exactly a functor $\C \to \Monm$.  Therefore, there is an equality \[\QFT\bigl(\Cbarmin\bigr)=\Monm^{\C},\] the category of $\C$-diagrams of monoids in $\M$.\dqed
\end{example}

\begin{example}[Diagrams of commutative monoids]\label{ex:everything-causality}
Suppose $\C$ is a small category, and suppose $\perpmax$ is the set of all pairs of morphisms in $\C$ with the same codomain.  In particular, for each object $c \in \C$, we have $\Id_c \perpmax \Id_c$, so the causality axiom \eqref{perp-com} says that $\fraka(c)$ is a commutative monoid in $\M$.  For the \index{maximal orthogonal category}\index{diagram of commutative monoids!as AQFT}\index{algebraic quantum field theory!on maximal orthogonal category}\emph{maximal orthogonal category}\label{notation:cbarmax} \[\Cbarmax = (\C,\perpmax),\] each $\fraka \in \QFT(\Cbarmax)$ is in particular a $\C$-diagram of commutative monoids in $\M$.  Conversely, each $\C$-diagram of commutative monoids satisfies the causality axiom because the multiplication $\mu_c$ is commutative.  Therefore, in this case we have \[\QFT(\Cbarmax) = \Comm^{\C},\] the category of $\C$-diagrams of commutative monoids in $\M$.  We interpret this equality as saying that, when observables always commute, algebraic quantum field theories reduce to the classical case.\dqed
\end{example}

\begin{example}[Underlying diagrams of monoids]\label{ex:aqft-diagram}
For each orthogonal category $\Cbar = (\C,\perp)$, there are orthogonal functors \[\nicexy{\Cbarmin = (\C,\varnothing) \ar[r]^-{i_0} & \Cbar \ar[r]^-{i_1} & \Cbarmax = (\C,\perpmax)}\] whose underlying functors are the identity functors on $\C$.  By Theorem \ref{thm:ocbar-algebra} they induce the following two change-of-operad adjunctions.
\[\nicexy{\algm\bigl(\Ocbarminm\bigr) \ar[d]^-{\cong }\ar@<2pt>[r]^-{(\Otom_{i_0})_!} 
& \algm\bigl(\Ocbarm\bigr) \ar[d]^-{\cong} \ar@<2pt>[l]^-{(\Otom_{i_0})^*} \ar@<2pt>[r]^-{(\Otom_{i_1})_!} 
& \algm\bigl(\Ocbarmaxm\bigr) \ar[d]^-{\cong} \ar@<2pt>[l]^-{(\Otom_{i_1})^*}\\
\Monm^{\C}=\QFT\bigl(\Cbarmin\bigr) & \QFT(\Cbar) & \QFT\bigl(\Cbarmax\bigr) = \Comm^{\C}}\]
The right adjoint $(\Otom_{i_0})^*$ sends each algebraic quantum field theory on $\Cbar$ to its underlying $\C$-diagram of monoids.  The other right adjoint $(\Otom_{i_1})^*$ says that each $\C$-diagram of commutative monoids is in particular an algebraic quantum field theory on $\Cbar$.\dqed
\end{example}

The next three examples are about bounded lattices and (equivariant) topological spaces.

\begin{example}[Quantum field theories on bounded lattices]\label{ex:qft-lattice}
Suppose $(L,\leq)$ is a bounded lattice as in Example \ref{ex:lattice}, also regarded as a small category.  Two morphisms $g_1 : a \to c$ and $g_2 : b \to c$ in $L$ are orthogonal if and only if $a \wedge b = 0$, which is the least element in $L$.  \index{bounded lattice!AQFT}This defines an orthogonal category $(L,\perp)$ and algebraic quantum field theories on it.\dqed
\end{example}

\begin{example}[Quantum field theories on topological spaces]\label{ex:qft-space}
For each topological space $X$, recall from Example \ref{ex:openx} that $\Openx$ is a bounded lattice.  By Example \ref{ex:qft-lattice} there is an orthogonal \index{topological space!AQFT}category $\Openxbar$, where $U_1 \subset V$ and $U_2 \subset V$ are orthogonal if and only if $U_1$ and $U_2$ are disjoint.  Corresponding to the orthogonal category $\Openxbar$ is the category of algebraic quantum field theories on it.\dqed
\end{example}

\begin{example}[Quantum field theories on equivariant topological spaces]\label{ex:qft-eqspace}
Suppose $G$ is a group, and $X$ is a \index{equivariant topological space!AQFT}topological space in which $G$ acts on the left by homeomorphisms.  Suppose $\Openxg$ is the category in Example \ref{ex:eq-space}.  Define $\perp$ as the set of pairs of morphisms $\narrowxy{U_1 \ar[r]^-{i_1g_1} & V & U_2 \ar[l]_-{i_2g_2}}$ in $\Openxg$ of the form \[\nicexy{U_1 \ar[r]^-{g_1} & g_1U_1 \ar[r]^-{i_1} & V & g_2U_2 \ar[l]_-{i_2} & U_2 \ar[l]_-{g_2}}\] with $g_1,g_2 \in G$ and $i_1,i_2$ both inclusions such that $g_1U_1$ and $g_2U_2$ are disjoint.  This defines an orthogonal category $\Openxgbar$. If $G$ is the trivial group, then we recover the orthogonal category $\Openxbar$ in Example \ref{ex:qft-space}.  Corresponding to the orthogonal category $\Openxgbar$ is the category of algebraic quantum field theories on it.\dqed
\end{example}

Examples  \ref{ex:ccqft} to \ref{ex:causal-nets} below are from \cite{bsw} and are about quantum field theories defined on spacetimes without additional geometric structure.  The upshot is that the operadic framework in Section \ref{sec:aqft} includes many quantum field theories in the literature, including various flavors of chiral conformal, Euclidean, and locally covariant quantum field theories.  To specify a particular flavor of quantum field theories, we simply choose the right orthogonal category $\Cbar = (\C,\perp)$ and, if there is a version of the time-slice axiom, a suitable set of morphisms $S \subset \Mor(\C)$ to be localized.

\begin{example}[Chiral conformal quantum field theories]\label{ex:ccqft}
In the context of Example \ref{ex:man-cat}, there is an orthogonal category \[\Mandbar = (\Mand,\perp)\] in which $\Mand$ is the category of $d$-dimensional oriented manifolds with orientation-preserving open embeddings as morphisms.  Two morphisms $g_1 : X_1 \to X$ and $g_2 : X_2 \to X$ in $\Mand$ are orthogonal if and only if their images are disjoint subsets in $X$.  By Theorem \ref{thm:ocbar-algebra} there is a canonical isomorphism\index{chiral conformal quantum field theory} \[\algm\bigl(\Otom_{\Mandbar}\bigr) \cong \QFT\bigl(\Mandbar\bigr).\]  When $\M=\Vectk$ and $d=1$, the objects in $\QFT\bigl(\Mandbar\bigr)$ are coordinate-free chiral conformal nets of $\fieldk$-algebras that satisfy the commutativity axiom for observables localized in disjoint regions \cite{bdh}.\dqed
\end{example}

\begin{example}[Chiral conformal quantum field theories on discs]\label{ex:ccqft-int}
In the context of Example \ref{ex:disc-cat}, there is an orthogonal category \[\Discdbar = (\Discd,\perp)\] in which $\Discd$ is the full subcategory of $\Mand$ of $d$-dimensional oriented manifolds diffeomorphic to $\fieldr^d$.  The orthogonality relation is the pullback of that on $\Mand$ along the full subcategory inclusion \[j : \Discd\to \Mand.\]  In other words, two morphisms $g_1 : X_1 \to X$ and $g_2 : X_2 \to X$ in $\Discd$ are orthogonal if and only if their images are disjoint subsets in $X$.  By Theorem \ref{thm:ocbar-algebra} there is a canonical isomorphism \[\algm\bigl(\Otom_{\Discdbar}\bigr) \cong \QFT\bigl(\Discdbar\bigr).\]  When $\M=\Vectk$ and $d=1$, the objects in $\QFT\bigl(\Discdbar\bigr)$ are coordinate-free chiral conformal nets of $\fieldk$-algebras defined on intervals that satisfy the commutativity axiom for observables localized in disjoint intervals \cite{bdh}.

The full subcategory inclusion $j : \Discd \to \Mand$ induces an orthogonal functor by Lemma \ref{lem:pullback-orthogonality}.  So by Theorem \ref{thm:ocbar-algebra}, it induces a change-of-operad adjunction \[\nicexy@C+.3cm{\QFT\bigl(\Discdbar\bigr) \cong \algm\bigl(\Otom_{\Discdbar}\bigr) \ar@<2pt>[r]^-{(\Otom_j)_!} &  \algm\bigl(\Otom_{\Mandbar}\bigr) \cong \QFT\bigl(\Mandbar\bigr) \ar@<2pt>[l]^-{(\Otom_j)^*}}.\]  When $\M=\Vectk$ and $d=1$, this adjunction allows us to go back and forth between (i) coordinate-free chiral conformal nets of $\fieldk$-algebras defined on intervals that satisfy the commutativity axiom for observables localized in disjoint intervals and (ii) those defined on all $1$-dimensional oriented manifolds.\dqed
\end{example}

\begin{example}[Chiral conformal quantum field theories on a fixed manifold]\label{ex:ccqft-manifold}
For a fixed oriented manifold $X \in \Mand$, recall from Example \ref{ex:openx} the category $\Open(X)$ whose objects are open subsets of $X$ and whose morphisms are subset inclusions.  Denote the induced functor by \[\iota : \Open(X) \to \Mand,\] and equip $\Open(X)$ with the pullback orthogonality relation along $\iota$.  In other words, for open subsets $U_1,U_2 \subseteq V \subseteq X$, the inclusions $U_1 \subseteq V$ and $U_2 \subseteq V$ are orthogonal if and only if $U_1$ and $U_2$ are disjoint subsets of $V$.  This is a special case of Example \ref{ex:qft-space}.  By Theorem \ref{thm:ocbar-algebra} there is a canonical isomorphism \[\algm\bigl(\Otom_{\Openxbar}\bigr) \cong \QFT\bigl(\Openxbar\bigr).\]  When $\M=\Vectk$, $d=1$, and $X=S^1$, the objects in $\QFT\bigl(\Opensonebar\bigr)$ are chiral conformal nets of $\fieldk$-algebras on the circle \cite{kaw,rehren}.

The subcategory inclusion $\iota : \Open(X) \to \Mand$ induces an orthogonal functor by Lemma \ref{lem:pullback-orthogonality}.  So by Theorem \ref{thm:ocbar-algebra}, it induces a change-of-operad adjunction \[\nicexy@C+.5cm{\algm\bigl(\Otom_{\Openxbar}\bigr) \ar[d]_-{\cong} \ar@<2pt>[r]^-{(\Otom_\iota)_!} & \algm\bigl(\Otom_{\Mandbar}\bigr) \ar[d]^-{\cong}\ar@<2pt>[l]^-{(\Otom_\iota)^*}\\ \QFT\bigl(\Openxbar\bigr) & \QFT\bigl(\Mandbar\bigr)}.\]\dqed 
\end{example}

\begin{example}[Euclidean quantum field theories]\label{ex:euclidean-qft}
In the context of Example \ref{ex:riem-cat}, there is an orthogonal category \[\Riemdbar = (\Riemd,\perp)\] in which $\Riemd$ is the category with $d$-dimensional oriented Riemannian manifolds as objects and orientation-preserving isometric open embeddings as morphisms.  Two morphisms $g_1 : X_1 \to X$ and $g_2 : X_2 \to X$ in $\Riemd$ are orthogonal if and only if their images are disjoint subsets in $X$.  By Theorem \ref{thm:ocbar-algebra} there is a canonical isomorphism\index{Euclidean quantum field theory} \[\algm\bigl(\Otom_{\Riemdbar}\bigr) \cong \QFT\bigl(\Riemdbar\bigr).\]  When $\M=\Vectk$ the objects in $\QFT\bigl(\Riemdbar\bigr)$ are locally covariant versions of Euclidean quantum field theories that satisfy the commutativity axiom for observables localized in disjoint regions \cite{schlingemann}.  As in Example \ref{ex:ccqft-manifold}, we may also restrict to a fixed oriented Riemannian manifold $X$ and consider algebraic quantum field theories on $\Openxbar$ as objects in $\QFT\bigl(\Openxbar\bigr)$.\dqed
\end{example}

\begin{example}[Locally covariant quantum field theories]\label{ex:lcqft}
In the context of Example \ref{ex:loc-cat}, there is an orthogonal category \[\Locdbar = (\Locd,\perp)\] in which $\Locd$ is the category of $d$-dimensional oriented, time-oriented, and globally hyperbolic Lorentzian manifolds.  A morphism is an isometric embedding that preserves the orientations and time-orientations whose image is causally compatible and open.  Two morphisms $g_1 : X_1 \to X$ and $g_2 : X_2 \to X$ in $\Locd$ are orthogonal if and only if their images are causally disjoint subsets in $X$.  By Theorem \ref{thm:ocbar-algebra} there is a canonical isomorphism\index{locally covariant quantum field theory} \[\algm\bigl(\Otom_{\Locdbar}\bigr) \cong \QFT\bigl(\Locdbar\bigr).\]  When $\M=\Vectk$ the objects in $\QFT\bigl(\Locdbar\bigr)$ are casual locally covariant quantum field theories that do not necessarily satisfy the isotony axiom \cite{bfv,fewster,fv}.

To implement the time-slice axiom, recall that a morphism $f : X \to Y$ in $\Locd$ is a \emph{Cauchy morphism} if its image contains a Cauchy surface of $Y$.  The set of Cauchy morphisms is denoted by $S$.  By Theorem \ref{thm:ocbar-algebra} there is a change-of-operad adjunction \[\nicexy@C+.3cm{\QFT\bigl(\Locdbar\bigr) \cong \algm\bigl(\Otom_{\Locdbar}\bigr) \ar@<2pt>[r]^-{(\Otom_\ell)_!} & \algm\bigl(\Otom_{\Locdsinvbar}\bigr) \cong \QFT\bigl(\Locdbar,S\bigr) \ar@<2pt>[l]^-{(\Otom_\ell)^*}}.\]  When $\M$ is the category $\Vectk$, the objects on the right side are causal locally covariant quantum field theories satisfying the time-slice axiom but not necessarily the isotony axiom.\dqed
\end{example}

\begin{example}[Locally covariant quantum field theories on a fixed spacetime]\label{ex:causal-nets}
In the context of Example \ref{ex:gh-cat}, for each Lorentzian manifold $X \in \Locd$ consider the category $\Ghx$ of globally hyperbolic open subsets of $X$ with subset inclusions as morphisms.  As in Example \ref{ex:ccqft-manifold}, $\Ghx$ may be equipped with the pullback orthogonality relation along the subcategory inclusion $i : \Ghx \to \Locd$.   By Theorem \ref{thm:ocbar-algebra} there is a canonical isomorphism \[\algm\bigl(\Otom_{\Ghxbar}\bigr) \cong \QFT\bigl(\Ghxbar\bigr).\]  When $\M=\Vectk$ the objects in $\QFT\bigl(\Ghxbar\bigr)$ are locally covariant quantum field theories on $X$ that do not necessarily satisfy the time-slice axiom and the isotony axiom.

As in Example \ref{ex:lcqft}, to implement the time-slice axiom, suppose $S$ is the set of morphisms $U \subseteq V \subseteq X$ such that $U$ contains a Cauchy surface of $i(V)$.  By Theorem \ref{thm:ocbar-algebra} the $S$-localization functor \[\ell : \Ghx \to \Ghx[\Sinv]\] induces a change-of-operad adjunction \[\nicexy@C+.5cm{ \algm\bigl(\Otom_{\Ghxbar}\bigr)\ar[d]_-{\cong} \ar@<2pt>[r]^-{(\Otom_\ell)_!} & \algm\bigl(\Otom_{\Ghxsinvbar}\bigr) \ar[d]^-{\cong} \ar@<2pt>[l]^-{(\Otom_\ell)^*}\\ \QFT\bigl(\Ghxbar\bigr) & \QFT\bigl(\Ghxbar,S\bigr)}.\]  When $\M=\Vectk$ the objects on the right side are causal nets of $\fieldk$-algebras satisfying the time-slice axiom but not necessarily the isotony axiom \cite{hk}.\dqed
\end{example}

Examples \ref{ex:dynamical-gauge} to \ref{ex:qft-structured} below are from \cite{bs17} and are about quantum field theories defined on spacetimes with additional geometric structures such as principal bundles, connections, and spin structure.  A common feature is that the isotony axiom--which asks that each structure morphism $\fraka(f) : \fraka(X) \to \fraka(Y) \in \Monm$ be a monomorphism--is usually not satisfied.

\begin{example}[Dynamical quantum gauge theories on principal bundles]\label{ex:dynamical-gauge}
In the context of Example \ref{ex:bgloc-cat}, recall that for each Lie group $G$ there is a forgetful functor \[\pi : \Bgloc \to \Locd\] that forgets about the bundle structure, where $\Bgloc$ is the category of $d$-dimensional oriented, time-oriented, and globally hyperbolic Lorentzian manifolds equipped with a principal $G$-bundle.  Suppose:
\begin{itemize}\item $S_G=\pi^{-1}(S) \subset \Mor(\Bgloc)$ is the $\pi$-pre-image of the set $S$ of Cauchy morphisms in $\Locd$.
\item $\pi^*(\perp)$ is the pullback of the orthogonality relation $\perp$ in $\Locd$ in Example \ref{ex:lcqft}  along $\pi$.  
\end{itemize}
The forgetful functor $\pi$ and the universal property of localization induce a commutative diagram\index{dynamical quantum gauge theory}
\[\nicexy{\Bglocbar = (\Bgloc, \pi^*(\perp)) \ar[r]^-{\pi} \ar[d]_-{\ell} & (\Locd,\perp) = \Locdbar \ar[d]^-{\ell}\\
\Bglocsginvbar = \bigl(\Bglocsginv, \ell_*\pi^*(\perp)\bigr) \ar[r]^-{\pi'} & (\Locdsinv, \ell_*(\perp)) = \Locdsinvbar}\] in $\Orthcat$.  The right vertical morphism is the $S$-localization functor on $\Locd$, and $\ell_*(\perp)$ is the pushforward orthogonality relation along $\ell$.  The left vertical morphism is the $S_G$-localization functor on $\Bgloc$, and $\ell_*\pi^*(\perp)$ is the pushforward orthogonality relation of $\pi^*(\perp)$ along $\ell$. Since \[\ell\pi(S_G) \subseteq \ell(S)\] are all isomorphisms in $\Locdsinv$, by the universal property of $S_G$-localization, there is a unique functor \[\nicexy{\Bglocsginv \ar[r]^{\pi'} & \Locdsinv} \stspace \ell\pi = \pi'\ell.\]  A direct inspection shows that $\pi'$ is an orthogonal functor.

By Theorem \ref{thm:ocbar-algebra} the functor $\pi'$ induces a change-of-operad adjunction \[\nicexy@C+.5cm{\algm\bigl(\Otom_{\Bglocsginvbar}\bigr) \ar[d]_-{\cong} \ar@<2pt>[r]^-{(\Otom_{\pi'})_!} &  \algm\bigl(\Otom_{\Locdsinvbar}\bigr) \ar[d]^-{\cong} \ar@<2pt>[l]^-{(\Otom_{\pi'})^*}\\ \QFT\bigl(\Bglocbar,S_G\bigr) & \QFT\bigl(\Locdbar,S\bigr)}.\] When $\M=\Vectk$ objects on the left side include dynamical quantum gauge theories on principal $G$-bundles that do not necessarily satisfy the isotony axiom \cite{bdhs,bds14}.\dqed
\end{example}

\begin{example}[Charged matter quantum field theories on background gauge fields]\label{ex:charged-matter}
In the context of Example \ref{ex:bgconloc-cat}, recall that for each Lie group $G$ there is a forgetful functor \[\pi p : \Bgconloc \to \Locd\] that forgets about the bundle structure and the connection, where $\Bgconloc$ is the category of triples $(X,P,C)$ with $(X,P) \in \Bgloc$ and $C$ a connection on $P$.    Suppose:
\begin{itemize}\item $S_G=(\pi p)^{-1}(S) \subset \Mor(\Bgconloc)$ is the $(\pi p)$-pre-image of the set $S$ of Cauchy morphisms in $\Locd$.
\item $(\pi p)^*(\perp)$ is the pullback of the orthogonality relation $\perp$ in $\Locd$ in Example \ref{ex:lcqft}  along $\pi p$.  
\end{itemize}
Exactly as in Example \ref{ex:dynamical-gauge}, the forgetful functor $\pi p$ and the universal property of localization induce a commutative diagram\index{charged matter quantum field theory}
\[\begin{small}\nicexy{\Bgconlocbar = (\Bgconloc, (\pi p)^*(\perp)) \ar[r]^-{\pi p} \ar[d]_-{\ell} & (\Locd,\perp) = \Locdbar \ar[d]^-{\ell}\\
\Bgconlocsginvbar = \bigl(\Bgconlocsginv, \ell_*(\pi p)^*(\perp)\bigr) \ar[r]^-{\pi'} & (\Locdsinv, \ell_*(\perp)) = \Locdsinvbar}\end{small}\] in $\Orthcat$.

By Theorem \ref{thm:ocbar-algebra} the functor $\pi'$ induces a change-of-operad adjunction \[\nicexy@C+.5cm{\algm\bigl(\Otom_{\Bgconlocsginvbar}\bigr)\ar[d]_-{\cong} \ar@<2pt>[r]^-{(\Otom_{\pi'})_!} &  \algm\bigl(\Otom_{\Locdsinvbar}\bigr) \ar[d]^-{\cong} \ar@<2pt>[l]^-{(\Otom_{\pi'})^*}\\ \QFT\bigl(\Bgconlocbar,S_G\bigr) & \QFT\bigl(\Locdbar,S\bigr)}.\] When $\M=\Vectk$ objects on the left side include charged matter quantum field theories on background gauge fields that do not necessarily satisfy the isotony axiom \cite{sz,zahn}.\dqed
\end{example}

\begin{example}[Dirac and fermionic quantum field theories]\label{ex:dirac-qft}
In the context of Example \ref{ex:sloc-cat}, recall that there is a forgetful functor \[\pi : \Slocd \to \Locd,\] where $\Slocd$ is the category of $d$-dimensional oriented, time-oriented, and globally hyperbolic Lorentzian spin manifolds.   Suppose:
\begin{itemize}\item $S_{\pi} \subset \Mor(\Slocd)$ is the $\pi$-pre-image of the set $S$ of Cauchy morphisms in $\Locd$.
\item $\pi^*(\perp)$ is the pullback of the orthogonality relation $\perp$ in $\Locd$ in Example \ref{ex:lcqft} along $\pi$.  
\end{itemize}
Exactly, as in Example \ref{ex:dynamical-gauge}, the forgetful functor $\pi$ and the universal property of localization induce a commutative diagram
\[\nicexy{\Slocdbar = (\Slocd, \pi^*(\perp)) \ar[r]^-{\pi} \ar[d]_-{\ell} & (\Locd,\perp) = \Locdbar \ar[d]^-{\ell}\\
\Slocdsinvbar = \bigl(\Slocdsinv, \ell_*\pi^*(\perp)\bigr) \ar[r]^-{\pi'} & (\Locdsinv, \ell_*(\perp)) = \Locdsinvbar}\] in $\Orthcat$. 

By Theorem \ref{thm:ocbar-algebra} the functor $\pi'$ induces a change-of-operad adjunction \[\nicexy@C+.5cm{\algm\bigl(\Otom_{\Slocdsinvbar}\bigr)\ar[d]_-{\cong} \ar@<2pt>[r]^-{(\Otom_{\pi'})_!} &  \algm\bigl(\Otom_{\Locdsinvbar}\bigr) \ar[d]^-{\cong} \ar@<2pt>[l]^-{(\Otom_{\pi'})^*}\\ 
\QFT\bigl(\Slocdbar,S_{\pi}\bigr) & \QFT\bigl(\Locdbar,S\bigr)}.\] When $\M=\Vectk$ objects on the left side include Dirac quantum fields\index{Dirac quantum field theory} that do not necessarily satisfy the isotony axiom \cite{dhp,sanders,verch}.  Furthermore, when $\M$ is the symmetric monoidal category of $\fieldk$-supermodules, its monoids are $\fieldk$-superalgebras.  In this case, objects on the left side include fermionic quantum field theories\index{fermionic quantum field theory} that do not necessarily satisfy the isotony axiom \cite{bg11}.\dqed
\end{example}

\begin{example}[Quantum field theories on structured spacetimes]\label{ex:qft-structured}
Examples \ref{ex:dynamical-gauge}, \ref{ex:charged-matter}, and \ref{ex:dirac-qft} are subsumed by the following more general setting from \cite{bs17}.  Suppose given a functor $\pi : \Str \to \Locd$ between small categories.  One regards $\Str$ as the category of spacetimes with additional geometric structures with $\pi$ the forgetful functor that forgets about the additional structures.   Suppose:
\begin{itemize}\item $S_{\pi}\subset \Mor(\Str)$ is the $\pi$-pre-image of the set $S$ of Cauchy morphisms in $\Locd$.
\item $\pi^*(\perp)$ is the pullback of the orthogonality relation $\perp$ in $\Locd$ in Example \ref{ex:lcqft} along $\pi$.  
\end{itemize}
Exactly as in Example \ref{ex:dynamical-gauge}, the forgetful functor $\pi$ and the universal property of localization induce a commutative diagram\index{quantum field theory on structured spacetime}
\[\nicexy{\Strbar = (\Str, \pi^*(\perp)) \ar[r]^-{\pi} \ar[d]_-{\ell} & (\Locd,\perp) = \Locdbar \ar[d]^-{\ell}\\
\Strsinvbar = \bigl(\Strsinv, \ell_*\pi^*(\perp)\bigr) \ar[r]^-{\pi'} & (\Locdsinv, \ell_*(\perp)) = \Locdsinvbar}\] in $\Orthcat$.  

By Theorem \ref{thm:ocbar-algebra} the functor $\pi'$ induces a change-of-operad adjunction \[\nicexy@C+.3cm{\algm\bigl(\Otom_{\Strsinvbar}\bigr)\ar[d]_-{\cong} \ar@<2pt>[r]^-{(\Otom_{\pi'})_!} &  \algm\bigl(\Otom_{\Locdsinvbar}\bigr) \ar[d]^-{\cong} \ar@<2pt>[l]^-{(\Otom_{\pi'})^*}\\
\QFT\bigl(\Strbar,S_{\pi}\bigr) & \QFT\bigl(\Locdbar,S\bigr)}.\] Objects on the left side are quantum field theories on $\pi : \Str \to \Locd$ in the sense of \cite{bs17} that do not necessarily satisfy the isotony axiom.\dqed
\end{example}

The next example is from \cite{bds} and is about quantum field theories defined on spacetimes with timelike boundary.

\begin{example}[Algebraic quantum field theories on spacetime with timelike boundary]\label{ex:aqft-boundary}
Suppose $X$ is a spacetime with timelike boundary as in Example \ref{ex:regions}.  There is an orthogonal category \[\Regxbar = (\Regx,\perp)\] in which $\Regx$ is the category of regions in $X$.  
\begin{itemize}\item Two morphisms $g_1 : U_1 \to V$ and $g_2 : U_2 \to V$ in $\Regx$ are orthogonal if and only if $U_1$ and $U_2$ are causally disjoint in $V$.  
\item A \emph{Cauchy morphism} $i : U \to V$ in $\Regx$ is a morphism such that $D(U) = D(V)$, where $D(U)$ is the set of points $x \in X$ such that every inextensible piecewise smooth future directed causal curve from $x$ meets $U$.  
\item The set of all Cauchy morphisms in $\Regx$ is denoted by $S_X$.  
\end{itemize}
By Lemma \ref{lem:aqft-time-slice} and Theorem \ref{thm:ocbar-algebra}, there are canonical isomorphisms\index{quantum field theory on spacetime with timelike boundary} \[\algm\bigl(\Otom_{\Regxsinvbar}\bigr) \cong \QFT\bigl(\Regxsinvbar\bigr) \cong \QFT\bigl(\Regxbar,S_X\bigr).\]  Objects in $\QFT(\Regxbar,S_X)$ are exactly the algebraic quantum field theories on $X$ as in \cite{bds} Definition 3.1.

There is a full subcategory inclusion\label{notation:regxzero} \[j : \Regxzero \to \Regx\] in which $\Regxzero$ is the category of regions in the interior $X_0$ of $X$.  Suppose:
\begin{itemize}\item $S_{X_0}=j^{-1}(S_X) \subset \Mor(\Regxzero)$ is the $j$-pre-image of the set $S_X$ of Cauchy morphisms in $\Regx$.
\item $j^*(\perp)$ is the pullback of the orthogonality relation $\perp$ in $\Regx$ along $j$.  
\end{itemize}
Similar to Example \ref{ex:dynamical-gauge}, the full subcategory inclusion $j$ and the universal property of localization induce a commutative diagram
\[\begin{footnotesize}\nicexy@C-.3cm{\Regxzerobar = (\Regxzero, j^*(\perp)) \ar[r]^-{j} \ar[d]_-{\ell} & (\Regx,\perp) = \Regxbar \ar[d]^-{\ell}\\
\Regxzerosinvbar = \bigl(\Regxzerosinv, \ell_*j^*(\perp)\bigr) \ar[r]^-{j'} & (\Regxsinv, \ell_*(\perp)) = \Regxsinvbar}\end{footnotesize}\] in $\Orthcat$.

By Theorem \ref{thm:ocbar-algebra} the functor $j'$ induces a change-of-operad adjunction \[\nicexy@C+.3cm{\algm\bigl(\Otom_{\Regxzerosinvbar}\bigr) \ar[d]_-{\cong} \ar@<2pt>[r]^-{(\Otom_{j'})_!} &  \algm\bigl(\Otom_{\Regxsinvbar}\bigr) \ar[d]^-{\cong} \ar@<2pt>[l]^-{(\Otom_{j'})^*}\\ \QFT\bigl(\Regxzerobar,S_{X_0}\bigr) & \QFT\bigl(\Regxbar,S_X\bigr)}\] 
The right adjoint is the restriction functor, while the left adjoint is called the \index{universal extension functor}\emph{universal extension functor} in \cite{bds}.\dqed
\end{example}

\section{Homotopical Properties}\label{sec:aqft-homotopy}

In this section we study homotopical properties of the category $\QFT(\Cbar)$ of algebraic quantum field theories on an orthogonal category $\Cbar$.  For this to make sense, the base category $\M$ in this section is assumed to be a monoidal model category in which the colored operads under consideration are admissible in the sense of Definition \ref{def:admissibility}.  For example, one can take $\M$ to be $\Top$, $\Sset$, $\Cat$, or $\Chaink$ with $\fieldk$ a field of characteristic zero, in which all colored operads are admissible.

In Theorem \ref{thm:ocbar-algebra}(6) we noted that the change-of-operad adjunction
\[\nicexy@C+.3cm{\QFT(\Cbar) \cong \algm\bigl(\Ocbarm\bigr) \ar@<2pt>[r]^-{(\Otom_F)_!} &  \algm\bigl(\Odbarm\bigr) \cong \QFT(\Dbar) \ar@<2pt>[l]^-{(\Otom_F)^*}}\]
induced by an orthogonal equivalence $F : \Cbar \to \Dbar$ is an adjoint equivalence.  The following observation says that this is a Quillen equivalence as well.

\begin{theorem}\label{thm:aqft-model}
Suppose $F : \Cbar \to \Dbar$ is an orthogonal functor, and $\M$ is a monoidal model category in which the colored operads $\Ocbarm$ and $\Odbarm$ are admissible.
\begin{enumerate}\item The change-of-operad adjunction $(\Otom_F)_! \dashv (\Otom_F)^*$ is a Quillen adjunction.
\item If $F$ is an orthogonal equivalence, then the operad morphism \[\Otom_F : \Ocbarm \to \Odbarm\] is a \index{homotopy Morita equivalence}\index{algebraic quantum field theory!homotopy Morita equivalence}homotopy Morita equivalence; i.e., the change-of-operad adjunction is a Quillen equivalence.
\end{enumerate}
\end{theorem}

\begin{proof}
For assertion (1), in both model categories $\algmocbarm$ and $\algmodbarm$, fibrations and weak equivalences are defined entrywise in $\M$.  So by Definition \ref{def:pullback-algebra} the right adjoint $(\Otom_F)^*$ preserves fibrations and acyclic fibrations.

For assertion (2), by \cite{hovey} Corollary 1.3.16, it is enough to show that for each cofibrant object $X \in \algmocbarm$, the derived unit \[X \to (\Otom_F)^* R (\Otom_F)_! X\] is a weak equivalence, where $R$ is the functorial fibrant replacement in $\algmodbarm$.  The derived unit is the composition \[\nicexy@C+.4cm{X \ar[r]^-{\eta_X}_-{\cong} & (\Otom_F)^*(\Otom_F)_!X \ar[r]^-{(\Otom_F)^*r} & (\Otom_F)^* R (\Otom_F)_! X}\] with 
\begin{itemize}\item $\eta_X$ the unit of the change-of-operad adjunction and 
\item $r : (\Otom_F)_!X \to R(\Otom_F)_!X$ the fibrant replacement in $\algmodbarm$.  
\end{itemize}
Since the change-of-operad adjunction is an adjoint equivalence by Theorem \ref{thm:ocbar-algebra}(6), the unit and the counit are both natural isomorphisms.  So it remains to see that the morphism $(\Otom_F)^*r$ is a weak equivalence in $\algmocbarm$, i.e., an entrywise weak equivalence in $\M$.  Since $r$ is an entrywise weak equivalence, by the definition of the right adjoint $(\Otom_F)^*$ in Definition \ref{def:pullback-algebra}, $(\Otom_F)^*r$ is an entrywise weak equivalence in $\M$.
\end{proof}

\begin{interpretation} If two orthogonal categories are orthogonally equivalent (i.e., there is an orthogonal equivalence between them), then their categories of algebraic quantum field theories have equivalent homotopy theories.  In particular, these two categories of algebraic quantum field theories are equivalent both before and after inverting the weak equivalences.\dqed\end{interpretation}

\begin{example}[Chiral conformal, Euclidean, and locally covariant QFT]\label{ex:chiral-equivalent}
In the context of Example \ref{ex:man-cat} and Example \ref{ex:ccqft}, recall that $\Mand$ is a small category equivalent to the entire category of $d$-dimensional oriented manifolds with orientation-preserving open embeddings as morphisms.  Two different choices yield two equivalent orthogonal categories.  So by Theorem \ref{thm:ocbar-algebra}(6) and Theorem \ref{thm:aqft-model}(2) the change-of-operad adjunction between their categories of algebraic quantum field theories is both an adjoint equivalence and a Quillen equivalence.  The same can be said for Euclidean quantum field theories in Example \ref{ex:euclidean-qft} and locally covariant quantum field theories in Example \ref{ex:lcqft}.\dqed
\end{example}

\chapter{Homotopy Algebraic Quantum Field Theories}\label{ch:haqft}

In this chapter, we define homotopy algebraic quantum field theories on an orthogonal category $\Cbar$.  We observe that each of them has a homotopy coherent $\C$-diagram structure and a compatible objectwise $A_\infty$-algebra structure, and satisfies a homotopy coherent version of the causality axiom.  If a set of morphisms in $\C$ is chosen, then each homotopy algebraic quantum field theory also satisfies a homotopy coherent version of the time-slice axiom.

\section{Overview}
In Section \ref{sec:haqft-operad} we define homotopy algebraic quantum field theories over an orthogonal category $\Cbar$ as algebras over the Boardman-Vogt construction $\wocbarm$ of the colored operad $\Ocbarm$, which is the image in $\M$ of the colored operad $\Ocbar$ in Definition \ref{def:aqft-operad}.  Then we record some of their categorical properties.  This definition makes sense because, as we saw in \eqref{aqft=operadalgebra}, the category of algebraic quantum field theories on an orthogonal category $\Cbar$, as in Definition \ref{def:aqft}, is canonically isomorphic to the category of $\Ocbarm$-algebras.  Each colored operad $\O$ is equipped with an augmentation $\eta : \wo \to \O$ from its Boardman-Vogt construction.  In Section \ref{sec:morita} we saw that in favorable cases the Boardman-Vogt construction has the correct homotopy type in the sense that the augmentation $\eta : \wo \to \O$ is a weak equivalence and that the induced change-of-operad adjunction is a Quillen equivalence.  Furthermore, in Section \ref{sec:hcdiagram} to Section \ref{sec:einfinity-algebra} we observed that the Boardman-Vogt construction $\wo$ of a colored operad $\O$ encodes $\O$-algebras up to coherent higher homotopies.

In Section \ref{sec:example-haqft} we present a long list of examples of homotopy algebraic quantum field theories, using mostly the orthogonal categories in Section \ref{sec:aqft-examples}.  Among the examples are homotopy chiral conformal quantum field theories, homotopy Euclidean quantum field theories, homotopy locally covariant quantum field theories, and homotopy quantum field theories on spacetimes with additional geometric structure or timelike boundary.

Our main tool for understanding the structure in homotopy algebraic quantum field theories is the Coherence Theorem in Section \ref{sec:coherence-haqft}.  This coherence result describes a homotopy algebraic quantum field theory in terms of explicit structure morphisms indexed by trees and four generating axioms.  In the remaining sections in this chapter, we describe structure that exists on every homotopy algebraic quantum field theory using the Coherence Theorem.  

In Section \ref{sec:h-causality} we observe that each homotopy algebraic quantum field theory satisfies a homotopy coherent version of the causality axiom.  The causality axiom for an algebraic quantum field theory $\fraka \in \QFT(\Cbar)$ says that, for an orthogonal pair $(g_1 : a \shortto c ,g_2 : b \shortto c)$ in $\Cbar$, the images of $\fraka(a)$ and $\fraka(b)$ in $\fraka(c)$ commute.  The homotopy coherent version says that the diagram defining the causality axiom is homotopy commutative via specified homotopies that are also structure morphisms.

In Section \ref{sec:h-functoriality} we observe that every homotopy algebraic quantum field theory on $\Cbar$ has an underlying homotopy coherent $\C$-diagram structure.  This is the homotopy coherent version of the fact that each algebraic quantum field theory $\fraka : \Cbar \to \Monm$ can be composed with the forgetful functor to $\M$ to yield a $\C$-diagram in $\M$.  In Section \ref{sec:h-timeslice} we observe that this homotopy coherent $\C$-diagram structure satisfies a homotopy coherent version of the time-slice axiom.  The time-slice axiom for an algebraic quantum field theory says that certain structure morphisms are isomorphisms.  The homotopy coherent version of the time-slice axiom says that certain structure morphisms admit two-sided homotopy inverses via specified homotopies, where the homotopy inverses and the homotopies are also structure morphisms.

In Section \ref{sec:objective-ainfinity} we observe that each homotopy algebraic quantum field theory has an objectwise $A_\infty$-algebra structure.  This is the homotopy coherent version of the fact that, for each algebraic quantum field theory $\fraka : \Cbar \to \Monm$, each object $\fraka(c)$ is a monoid in $\M$ for $c\in \C$.  We saw in Section \ref{sec:ainfinity-algebra} that an $A_\infty$-algebra is a homotopy coherent version of a monoid.  Furthermore, in Section \ref{sec:hcdiag-ainfinity} we show that this objectwise $A_\infty$-algebra structure is compatible with the homotopy coherent $\C$-diagram structure via specified homotopies that are also structure morphisms.  This is the homotopy coherent version of the fact that an algebraic quantum field theory is, in particular, a diagram of monoids.  

An important point to keep in mind is that all of the above homotopy coherent structures, including the homotopies, are already encoded in the Boardman-Vogt construction $\wocbarm$.  This is of course the entire reason for using the Boardman-Vogt construction to define homotopy algebraic quantum field theories.  Our coend definition of the Boardman-Vogt construction plays a critical role here.  In fact, the Coherence Theorem \ref{thm:wo-algebra-coherence} for algebras over the Boardman-Vogt construction crucially depends on our coend definition of $\wo$.  A special case of this theorem is the Coherence Theorem for homotopy algebraic quantum field theories in Section \ref{sec:coherence-haqft}, from which the results in Sections \ref{sec:h-causality} to \ref{sec:hcdiag-ainfinity} follow.

Throughout this chapter $(\M,\otimes,\tensorunit)$ is a cocomplete symmetric monoidal closed category with an initial object $\varnothing$ and a commutative segment $(J,\mu,0,1,\epsilon)$ as in Definition \ref{def:segment}.

\section{Homotopy AQFT as Operad Algebras}\label{sec:haqft-operad}

In this section, we define homotopy algebraic quantum field theories using the Boardman-Vogt construction in Chapter \ref{ch:bv} and record their basic categorical properties.  

\begin{recollection}\label{rec:bv-operad}
For a $\colorc$-colored operad $\O$ in $\M$, recall that the Boardman-Vogt construction of $\O$ is a $\colorc$-colored operad $\wo$, which is entrywise defined as a coend \[\wo\duc = \int^{T \in \uTreec\duc} \J[T] \otimes\O[T] \in \M,\] where $\uTreec\duc$ is the substitution category of $\colorc$-colored trees with profile $\duc$ in Definition \ref{def:treesub-category}.  The functors \[\J : \uTreecducop \to \M \andspace \O : \uTreecduc \to \M\] are induced by $J$ and $\O$ and are defined in Definition \ref{functor-J} and Corollary \ref{cor:operad-functor-subcat}, respectively.  Geometrically $\J[T]\otimes\O[T]$ is the $\colorc$-colored tree $T$ whose internal edges are decorated by $J$ and whose vertices are decorated by $\O$.  Via the coend, the substitution category parametrizes the relations among such decorated trees.

The operad structure of the Boardman-Vogt construction, defined in Definition \ref{def:wo-operad-structure}, is induced by tree substitution.  It is equipped with a natural augmentation $\eta : \wo \to \O$ of $\colorc$-colored operads, defined in Theorem \ref{thm:w-augmented}.  Intuitively the augmentation forgets the lengths of the internal edges (i.e., the $\J$-component) and composes in the colored operad $\O$.

Since the colored operad $\Ocbar$, defined in Definition \ref{def:aqft-operad}, of an orthogonal category $\Cbar$ is defined over $\Set$, we will have to first transfer it to $\M$.  Recall from Example \ref{ex:operad-set-m} that the strong symmetric monoidal functor $\Set \to \M$, sending a set $S$ to the $S$-indexed coproduct $\coprod_S \tensorunit$, yields the change-of-category functor \[(-)^{\M} : \Operadcset \to \Operadcm.\]  The image of $\Ocbar$ in $\Operadc(\M)$ will be denoted by $\Ocbarm$.  Also recall from Definition \ref{def:operad-algebra-generating} the category of algebras over a colored operad.
\end{recollection}

\begin{definition}\label{def:haqft}
Suppose $\Cbar = (\C,\perp)$ is an orthogonal category with object set $\colorc$, and $\wocbarm\in \Operadcm$ is the Boardman-Vogt construction of $\Ocbarm \in \Operadcm$.  We define the category\label{notation:hqftcbar} \[\HQFT(\Cbar) = \algmwocbarm,\] whose objects are called \index{algebraic quantum field theory!homotopy}\index{homotopy algebraic quantum field theory}\emph{homotopy algebraic quantum field theories} on $\Cbar$.
\end{definition}

\begin{remark} In Definition \ref{def:haqft} we first transfer $\Ocbar \in \Operadcset$ to $\Ocbarm \in \Operadcm$, and then we apply $\W$ to $\Ocbarm$.  In particular, the Boardman-Vogt construction is \emph{not} apply to $\Ocbar$ because it depends on a choice of a commutative segment in $\M$.\dqed
\end{remark}

\begin{interpretation} One should think of the $\colorc$-colored operad $\wocbarm$ as made up of $\colorc$-colored trees whose internal edges are decorated by the commutative segment $J$ and whose vertices are decorated by elements in $\Ocbar$ with the correct profile.  A homotopy algebraic quantum field theory has structure morphisms indexed by these decorated $\colorc$-colored trees.  The precise statement is the Coherence Theorem \ref{thm:haqft-coherence} below.\dqed
\end{interpretation}

The following observation compares algebraic quantum field theories and homotopy algebraic quantum field theories.  It is a special case of Theorem \ref{thm:operad-comparison}(1), Corollary \ref{cor:augmentation-adjunction}, Corollary \ref{cor:wo-o-chaink}, and \eqref{aqft=operadalgebra}.  

\begin{corollary}\label{cor:haqft-aqft-adjunction}
Suppose $\Cbar = (\C,\perp)$ is an orthogonal category.  
\begin{enumerate}
\item The augmentation $\eta : \wocbarm \to \Ocbarm$ induces a change-of-operad adjunction 
\[\nicexy{\HQFT(\Cbar)=\algmwocbarm \ar@<2pt>[r]^-{\eta_!} & \algmocbarm\cong \QFT(\Cbar) \ar@<2pt>[l]^-{\eta^*}}.\] 
\item If $\M$ is a monoidal model category in which the colored operads $\Ocbarm$ and $\wocbarm$ are admissible, then the change-of-operad adjunction is a Quillen adjunction.
\item If $\M=\Chaink$ with $\fieldk$ a field of characteristic zero, then the change-of-operad adjunction is a Quillen equivalence.\index{homotopy Morita equivalence}\index{Quillen equivalence}
\end{enumerate}
\end{corollary}

\begin{interpretation} The right adjoint $\eta^*$ allows us to consider an algebraic quantum field theory on $\Cbar$ as a homotopy algebraic quantum field theory on $\Cbar$.  The left adjoint $\eta_!$ rectifies a homotopy algebraic quantum field theory to an algebraic quantum field theory.  Furthermore, if $\M$ is $\Chaink$, then the augmentation $\eta$ is a homotopy Morita equivalence.  In particular, the homotopy theory of homotopy algebraic quantum field theories is equivalent to the homotopy theory of algebraic quantum field theories over the same orthogonal category.  So there is no loss of homotopical information by considering homotopy algebraic quantum field theories compared to algebraic quantum field theories.\dqed\end{interpretation}

The next observation is about changing the orthogonal categories.  It is a consequence of Theorem \ref{thm:operad-comparison}(1), Theorem \ref{thm:w-augmented}, Theorem \ref{thm:aqft-model}, and Corollary \ref{cor:haqft-aqft-adjunction}.  The second assertion below uses the fact that Quillen equivalences have the $2$-out-of-$3$ property.

\begin{corollary}\label{cor:haqft-adjunction-diagram}
Suppose $F : \Cbar \to \Dbar$ is an orthogonal functor.
\begin{enumerate}\item There is an induced diagram of change-of-operad adjunctions
\[\nicexy@C+.7cm{\HQFT(\Cbar)=\algmwocbarm \ar@<-2pt>[d]_-{\eta_!} \ar@<2pt>[r]^-{(\W\Otom_F)_!} & 
\algmwodbarm= \HQFT(\Dbar) \ar@<-2pt>[d]_-{\eta_!} \ar@<2pt>[l]^-{(\W\Otom_F)^*}\\ \QFT(\Cbar) \cong \algmocbarm \ar@<2pt>[r]^-{(\Otom_F)_!} \ar@<-2pt>[u]_-{\eta^*} & \algmodbarm \cong \QFT(\Dbar) \ar@<2pt>[l]^-{(\Otom_F)^*} \ar@<-2pt>[u]_-{\eta^*}}\]
such that \[(\Otom_F)_!\eta_! = \eta_! (\W\Otom_F)_! \andspace  \eta^*(\Otom_F)^* = (\W\Otom_F)^*\eta^*.\]
\item If $\M$ is a monoidal model category in which the colored operads $\Ocbarm$, $\Odbarm$,  $\wocbarm$, and $\wodbarm$ are admissible, then all four change-of-operad adjunctions are Quillen adjunctions.
\item If $F$ is an orthogonal equivalence and if $\M=\Chaink$ with $\fieldk$ a field of characteristic zero, then all four change-of-operad adjunctions are Quillen equivalences.
\end{enumerate}
\end{corollary}

\begin{interpretation} The right adjoint $(\wom_F)^*$ sends each homotopy algebraic quantum field theory on $\Dbar$ to one on $\Cbar$.  The left adjoint $(\wom_F)_!$ sends each homotopy algebraic quantum field theory on $\Cbar$ to one on $\Dbar$.  The equality \[(\Otom_F)_!\eta_! = \eta_! (\W\Otom_F)_!\] means that the left adjoint diagram is commutative.  The equality \[\eta^*(\Otom_F)^* = (\W\Otom_F)^*\eta^*\] means that the right adjoint diagram is commutative.  Moreover, if $F$ is an orthogonal equivalence and if $\M$ is $\Chaink$, then all four operad morphisms in the commutative diagram \[\nicexy@C+.4cm{\wocbarm \ar[r]^-{\wom_F} \ar[d]_-{\eta} & \wodbarm \ar[d]^-{\eta}\\ \Ocbarm \ar[r]^-{\Otom_F} & \Odbarm}\] are homotopy Morita equivalences. In particular, the homotopy theory of homotopy algebraic quantum field theories on $\Cbar$ is equivalent to the homotopy theory of homotopy algebraic quantum field theories on $\Dbar$.\dqed
\end{interpretation}

\section{Examples of Homotopy AQFT}\label{sec:example-haqft}

In this section, we apply Corollary \ref{cor:haqft-aqft-adjunction} and Corollary \ref{cor:haqft-adjunction-diagram} to the orthogonal categories and orthogonal functors in Section \ref{sec:aqft-examples} to obtain examples of homotopy algebraic quantum field theories.  

\begin{example}[Homotopy coherent diagrams of $A_\infty$-algebras]\label{ex:hcdiag-ainfinity}
For each orthogonal category\index{homotopy coherent diagram!$A_\infty$-algebra}\index{homotopy coherent diagram!$E_\infty$-algebra}  $\Cbar = (\C,\perp)$, there are two orthogonal functors \[\nicexy{\Cbarmin = (\C,\varnothing) \ar[r]^-{i_0} & \Cbar \ar[r]^-{i_1} & \Cbarmax = (\C,\perpmax)}\] as in Example \ref{ex:aqft-diagram}.  By Corollary \ref{cor:haqft-adjunction-diagram} there is an induced diagram whose middle squares consist of change-of-operad adjunctions
\[\nicexy{\HQFT\bigl(\Cbarmin\bigr) \ar@{=}[d] & \HQFT(\Cbar) \ar@{=}[d] & \HQFT\bigl(\Cbarmax\bigr)\ar@{=}[d] \\
\algm\bigl(\W\Ocbarminm\bigr) \ar@<-2pt>[d]_-{\eta_!} \ar@<2pt>[r]^-{(\wom_{i_0})_!} 
& \algm\bigl(\W\Ocbarm\bigr) \ar@<-2pt>[d]_-{\eta_!} \ar@<2pt>[l]^-{(\wom_{i_0})^*} \ar@<2pt>[r]^-{(\wom_{i_1})_!} 
& \algm\bigl(\W\Ocbarmaxm\bigr) \ar@<-2pt>[d]_-{\eta_!}  \ar@<2pt>[l]^-{(\wom_{i_1})^*}\\
\algm\bigl(\Ocbarminm\bigr) \ar[d]^-{\cong}\ar@<2pt>[r]^-{(\Otom_{i_0})_!} \ar@<-2pt>[u]_-{\eta^*}
& \algm\bigl(\Ocbarm\bigr) \ar[d]^-{\cong} \ar@<2pt>[l]^-{(\Otom_{i_0})^*} \ar@<2pt>[r]^-{(\Otom_{i_1})_!} \ar@<-2pt>[u]_-{\eta^*} 
& \algm\bigl(\Ocbarmaxm\bigr) \ar[d]^-{\cong} \ar@<2pt>[l]^-{(\Otom_{i_1})^*} \ar@<-2pt>[u]_-{\eta^*}\\
\Monm^{\C}=\QFT\bigl(\Cbarmin\bigr) & \QFT(\Cbar) & \QFT\bigl(\Cbarmax\bigr) = \Comm^{\C}}\]
with commutative left/right adjoint diagrams.  Since $\Monm^{\C}$ is the category of $\C$-diagrams of monoids in $\M$, in view of Theorems \ref{thm:hcdiagram} and \ref{thm:ainfinity-algebra}, we interpret \[\HQFT\bigl(\Cbarmin\bigr) = \algm\bigl(\W\Ocbarminm\bigr)\] as the category of homotopy coherent $\C$-diagrams of $A_\infty$-algebras.  The right adjoint $(\wom_{i_0})^*$ sends each homotopy algebraic quantum field theory on $\Cbar$ to its underlying homotopy coherent $\C$-diagram of $A_\infty$-algebras.  We will explain this structure in more details in Section \ref{sec:hcdiag-ainfinity}.\dqed
\end{example}

\begin{example}[Homotopy chiral conformal quantum field theories]\label{ex:homotopy-chiral}
When applied to the orthogonal category $\Mandbar = (\Mand,\perp)$ in Example \ref{ex:ccqft}, Corollary \ref{cor:haqft-aqft-adjunction} gives a change-of-operad adjunction 
\[\nicexy{\HQFT\bigl(\Mandbar\bigr)=\algm\bigl(\wom_{\Mandbar}\bigr) \ar@<2pt>[r]^-{\eta_!} & \algm\bigl(\Otom_{\Mandbar}\bigr)\cong \QFT\bigl(\Mandbar\bigr) \ar@<2pt>[l]^-{\eta^*}}\] between chiral conformal quantum field theories and homotopy chiral conformal quantum field theories.\index{homotopy chiral conformal quantum field theory}\index{chiral conformal quantum field theory!homotopy}  Moreover, this adjunction is a Quillen equivalence when $\M=\Chaink$ with $\fieldk$ a field of characteristic zero.\dqed
\end{example}

\begin{example}[Homotopy chiral conformal QFT on discs]\label{ex:hccqft-int}
In the context of Example \ref{ex:ccqft-int}, there is an orthogonal functor \[j : \Discdbar \to \Mandbar.\]  By Corollary \ref{cor:haqft-adjunction-diagram}, there is an induced diagram of change-of-operad adjunctions
\[\nicexy@C+.2cm{\HQFT\bigl(\Discdbar\bigr)=\algm\bigl(\wom_{\Discdbar}\bigr) \ar@<-2pt>[d]_-{\eta_!} \ar@<2pt>[r]^-{(\W\Otom_j)_!} & 
\algm\bigl(\wom_{\Mandbar}\bigr)= \HQFT\bigl(\Mandbar\bigr) \ar@<-2pt>[d]_-{\eta_!} \ar@<2pt>[l]^-{(\W\Otom_j)^*}\\
\QFT\bigl(\Discdbar\bigr) \cong \algm\bigl(\Otom_{\Discdbar}\bigr) \ar@<2pt>[r]^-{(\Otom_j)_!} \ar@<-2pt>[u]_-{\eta^*}
&  \algm\bigl(\Otom_{\Mandbar}\bigr) \cong \QFT\bigl(\Mandbar\bigr) \ar@<2pt>[l]^-{(\Otom_j)^*} \ar@<-2pt>[u]_-{\eta^*}}\]
with commutative left/right adjoint diagrams.  When $d=1$ the vertical adjunction on the left goes between chiral conformal quantum field theories\index{homotopy chiral conformal quantum field theory!on discs} defined on intervals and their homotopy analogues.\dqed
\end{example}

\begin{example}[Homotopy chiral conformal QFT on a fixed manifold]\label{ex:hccqft}
In the context of Example \ref{ex:ccqft-manifold}, each oriented manifold $X \in \Mand$ induces an orthogonal functor \[\iota : \Openxbar \to \Mandbar.\]  By Corollary \ref{cor:haqft-adjunction-diagram}, there is an induced diagram of change-of-operad adjunctions
\[\nicexy@C+.2cm{\HQFT\bigl(\Openxbar\bigr)=\algm\bigl(\wom_{\Openxbar}\bigr) \ar@<-2pt>[d]_-{\eta_!} \ar@<2pt>[r]^-{(\W\Otom_\iota)_!} & 
\algm\bigl(\wom_{\Mandbar}\bigr)= \HQFT\bigl(\Mandbar\bigr) \ar@<-2pt>[d]_-{\eta_!} \ar@<2pt>[l]^-{(\W\Otom_\iota)^*}\\
\QFT\bigl(\Openxbar\bigr) \cong \algm\bigl(\Otom_{\Openxbar}\bigr) \ar@<2pt>[r]^-{(\Otom_\iota)_!} \ar@<-2pt>[u]_-{\eta^*}
&  \algm\bigl(\Otom_{\Mandbar}\bigr) \cong \QFT\bigl(\Mandbar\bigr) \ar@<2pt>[l]^-{(\Otom_\iota)^*} \ar@<-2pt>[u]_-{\eta^*}}\]
with commutative left/right adjoint diagrams.  The vertical adjunction on the left goes between chiral conformal quantum field theories\index{homotopy chiral conformal quantum field theory!on a manifold} defined on $X$ and their homotopy analogues.\dqed
\end{example}

\begin{example}[Homotopy Euclidean quantum field theories]\label{ex:homotopy-euclidean}
When applied to the orthogonal category $\Riemdbar = (\Riemd,\perp)$ in Example \ref{ex:euclidean-qft}, Corollary \ref{cor:haqft-aqft-adjunction} gives a change-of-operad adjunction 
\[\nicexy{\HQFT\bigl(\Riemdbar\bigr)=\algm\bigl(\wom_{\Riemdbar}\bigr) \ar@<2pt>[r]^-{\eta_!} & \algm\bigl(\Otom_{\Riemdbar}\bigr)\cong \QFT\bigl(\Riemdbar\bigr) \ar@<2pt>[l]^-{\eta^*}}\] between Euclidean quantum field theories and \index{Euclidean quantum field theory!homotopy}\index{homotopy Euclidean quantum field theory}homotopy Euclidean quantum field theories.  Moreover, this adjunction is a Quillen equivalence when $\M=\Chaink$ with $\fieldk$ a field of characteristic zero.\dqed
\end{example}

\begin{example}[Homotopy locally covariant quantum field theories]\label{ex:homotopy-lcqft}
When applied to the orthogonal category $\Locdbar = (\Locd,\perp)$ in Example \ref{ex:lcqft}, Corollary \ref{cor:haqft-aqft-adjunction} gives a change-of-operad adjunction 
\[\nicexy@R-.2cm{\algm\bigl(\wom_{\Locdsinvbar}\bigr) \ar@{=}[d] \ar@<2pt>[r]^-{\eta_!} & \algm\bigl(\Otom_{\Locdsinvbar}\bigr) \ar[d]^-{\cong} \ar@<2pt>[l]^-{\eta^*}\\
\HQFT\bigl(\Locdsinvbar\bigr) & \QFT\bigl(\Locdbar,S\bigr)}\] between locally covariant quantum field theories satisfying the time-slice axiom and \index{locally covariant quantum field theory!homotopy}\index{homotopy locally covariant quantum field theory}their homotopy analogues.  Moreover, this adjunction is a Quillen equivalence when $\M=\Chaink$ with $\fieldk$ a field of characteristic zero.\dqed
\end{example}

\begin{example}[Homotopy locally covariant QFT on a fixed spacetime]\label{ex:hcausal-nets}
In the context of Example \ref{ex:causal-nets}, there is an orthogonal functor  \[\ell : \Ghxbar \to \Ghxsinvbar,\] where $\Ghx$ is the category of globally hyperbolic open subsets of $X \in \Locd$ with subset inclusions as morphisms.   By Corollary \ref{cor:haqft-adjunction-diagram}, there is an induced diagram whose middle square consists of change-of-operad adjunctions
\[\nicexy@C+.7cm{\HQFT\bigl(\Ghxbar\bigr) \ar@{=}[d] & \HQFT\bigl(\Ghxsinvbar\bigr)\ar@{=}[d]\\ 
\algm\bigl(\wom_{\Ghxbar}\bigr) \ar@<-2pt>[d]_-{\eta_!} \ar@<2pt>[r]^-{(\W\Otom_{\ell})_!} & \algm\bigl(\wom_{\Ghxsinvbar}\bigr)\ar@<-2pt>[d]_-{\eta_!}\ar@<2pt>[l]^-{(\W\Otom_{\ell})^*}\\
\algm\bigl(\Otom_{\Ghxbar}\bigr) \ar[d]_-{\cong}\ar@<2pt>[r]^-{(\Otom_{\ell})_!} \ar@<-2pt>[u]_-{\eta^*} &  \algm\bigl(\Otom_{\Ghxsinvbar}\bigr) \ar[d]^-{\cong} \ar@<2pt>[l]^-{(\Otom_{\ell})^*} \ar@<-2pt>[u]_-{\eta^*}\\
\QFT\bigl(\Ghxbar\bigr) & \QFT\bigl(\Ghxbar,S\bigr)}\] 
with commutative left/right adjoint diagrams. When $\M$ is the category $\Chaink$, the vertical adjunction on the right goes between causal nets of $\fieldk$-algebras satisfying the time-slice axiom but not necessarily the isotony axiom and their \index{homotopy locally covariant quantum field theory!on a spacetime}homotopy analogues.\dqed
\end{example}

\begin{example}[Homotopy dynamical quantum gauge theories on principal bundles]\label{ex:hdynamical-gauge}
In the context of Example \ref{ex:dynamical-gauge}, there is an orthogonal functor \[\nicexy{\Bglocsginvbar \ar[r]^{\pi'} & \Locdsinvbar},\] where $\Bgloc$ is the category of $d$-dimensional oriented, time-oriented, and globally hyperbolic Lorentzian manifolds equipped with a principal $G$-bundle.   By Corollary \ref{cor:haqft-adjunction-diagram}, there is an induced diagram whose middle square consists of change-of-operad adjunctions\index{dynamical quantum gauge theory!homotopy}\index{homotopy dynamical quantum gauge theory}
\[\nicexy@C+.7cm{\HQFT\bigl(\Bglocsginvbar\bigr) \ar@{=}[d] & \HQFT\bigl(\Locdsinvbar\bigr)\ar@{=}[d]\\ 
\algm\bigl(\wom_{\Bglocsginvbar}\bigr) \ar@<-2pt>[d]_-{\eta_!} \ar@<2pt>[r]^-{(\W\Otom_{\pi'})_!} & \algm\bigl(\wom_{\Locdsinvbar}\bigr)\ar@<-2pt>[d]_-{\eta_!}\ar@<2pt>[l]^-{(\W\Otom_{\pi'})^*}\\
\algm\bigl(\Otom_{\Bglocsginvbar}\bigr) \ar[d]_-{\cong}\ar@<2pt>[r]^-{(\Otom_{\pi'})_!} \ar@<-2pt>[u]_-{\eta^*} &  \algm\bigl(\Otom_{\Locdsinvbar}\bigr) \ar[d]^-{\cong} \ar@<2pt>[l]^-{(\Otom_{\pi'})^*} \ar@<-2pt>[u]_-{\eta^*}\\
\QFT\bigl(\Bglocbar,S_G\bigr) & \QFT\bigl(\Locdbar,S\bigr)}\] 
with commutative left/right adjoint diagrams.  When $\M=\Chaink$ the vertical adjunction on the left goes between dynamical quantum gauge theories on principal $G$-bundles that do not necessarily satisfy the isotony axiom and their homotopy analogues.\dqed
\end{example}

\begin{example}[Homotopy charged matter QFT on background gauge fields]\label{ex:hcharged-matter}
In the context of Example \ref{ex:charged-matter}, there is an orthogonal functor \[\nicexy{\Bgconlocsginvbar \ar[r]^{\pi'} & \Locdsinvbar},\] where $\Bgconloc$ is the category of $d$-dimensional oriented, time-oriented, and globally hyperbolic Lorentzian manifolds equipped with a principal $G$-bundle and a connection.  By Corollary \ref{cor:haqft-adjunction-diagram}, there is an induced diagram whose middle square consists of change-of-operad \index{homotopy charged matter quantum field theory}\index{charged matter quantum field theory!homotopy}adjunctions
\[\nicexy@C+.7cm{\HQFT\bigl(\Bgconlocsginvbar\bigr) \ar@{=}[d] & \HQFT\bigl(\Locdsinvbar\bigr)\ar@{=}[d]\\ 
\algm\bigl(\wom_{\Bgconlocsginvbar}\bigr) \ar@<-2pt>[d]_-{\eta_!} \ar@<2pt>[r]^-{(\W\Otom_{\pi'})_!} & \algm\bigl(\wom_{\Locdsinvbar}\bigr) \ar@<-2pt>[d]_-{\eta_!}\ar@<2pt>[l]^-{(\W\Otom_{\pi'})^*}\\
\algm\bigl(\Otom_{\Bgconlocsginvbar}\bigr) \ar[d]_-{\cong}\ar@<2pt>[r]^-{(\Otom_{\pi'})_!} \ar@<-2pt>[u]_-{\eta^*} &  \algm\bigl(\Otom_{\Locdsinvbar}\bigr) \ar[d]^-{\cong} \ar@<2pt>[l]^-{(\Otom_{\pi'})^*} \ar@<-2pt>[u]_-{\eta^*}\\
\QFT\bigl(\Bgconlocbar,S_G\bigr) & \QFT\bigl(\Locdbar,S\bigr)}\] 
with commutative left/right adjoint diagrams.  When $\M=\Chaink$ the vertical adjunction on the left goes between charged matter quantum field theories on background gauge fields that do not necessarily satisfy the isotony axiom and their homotopy analogues.\dqed
\end{example}

\begin{example}[Homotopy Dirac quantum field theories]\label{ex:hdirac}
In the context of Example \ref{ex:dirac-qft}, there is an orthogonal functor \[\nicexy{\Slocdsinvbar \ar[r]^{\pi'} & \Locdsinvbar},\] where $\Slocd$ is the category of $d$-dimensional oriented, time-oriented, and globally hyperbolic Lorentzian spin manifolds.   By Corollary \ref{cor:haqft-adjunction-diagram}, there is an induced diagram whose middle square consists of change-of-operad \index{homotopy Dirac quantum field theory}\index{Dirac quantum field theory!homotopy}adjunctions
\[\nicexy@C+.7cm{\HQFT\bigl(\Slocdsinvbar\bigr) \ar@{=}[d] & \HQFT\bigl(\Locdsinvbar\bigr)\ar@{=}[d]\\ 
\algm\bigl(\wom_{\Slocdsinvbar}\bigr) \ar@<-2pt>[d]_-{\eta_!} \ar@<2pt>[r]^-{(\W\Otom_{\pi'})_!} & \algm\bigl(\wom_{\Locdsinvbar}\bigr)\ar@<-2pt>[d]_-{\eta_!}\ar@<2pt>[l]^-{(\W\Otom_{\pi'})^*}\\
\algm\bigl(\Otom_{\Slocdsinvbar}\bigr) \ar[d]_-{\cong}\ar@<2pt>[r]^-{(\Otom_{\pi'})_!} \ar@<-2pt>[u]_-{\eta^*} &  \algm\bigl(\Otom_{\Locdsinvbar}\bigr) \ar[d]^-{\cong} \ar@<2pt>[l]^-{(\Otom_{\pi'})^*} \ar@<-2pt>[u]_-{\eta^*}\\
\QFT\bigl(\Slocdbar,S_{\pi}\bigr) & \QFT\bigl(\Locdbar,S\bigr)}\] 
with commutative left/right adjoint diagrams.  When $\M=\Chaink$ the vertical adjunction on the left goes between Dirac quantum fields that do not necessarily satisfy the isotony axiom and their homotopy analogues.\dqed
\end{example}

\begin{example}[Homotopy QFT on structured spacetimes]\label{ex:hqft-structured}
Subsuming Examples \ref{ex:hdynamical-gauge} to \ref{ex:hdirac}, consider a functor $\pi : \Str \to \Locd$ between small categories in the context of Example \ref{ex:qft-structured}.  It induces an \index{homotopy quantum field theory on structured spacetime}\index{quantum field theory on structured spacetime!homotopy}orthogonal functor
\[\nicexy{\Strsinvbar = \bigl(\Strsinv, \ell_*\pi^*(\perp)\bigr) \ar[r]^-{\pi'} & (\Locdsinv, \ell_*(\perp)) = \Locdsinvbar}.\]  By Corollary \ref{cor:haqft-adjunction-diagram}, there is an induced diagram whose middle square consists of change-of-operad adjunctions
\[\nicexy@C+.7cm{\HQFT\bigl(\Strsinvbar\bigr) \ar@{=}[d] & \HQFT\bigl(\Locdsinvbar\bigr)\ar@{=}[d]\\ 
\algm\bigl(\wom_{\Strsinvbar}\bigr) \ar@<-2pt>[d]_-{\eta_!} \ar@<2pt>[r]^-{(\W\Otom_{\pi'})_!} & \algm\bigl(\wom_{\Locdsinvbar}\bigr)\ar@<-2pt>[d]_-{\eta_!}\ar@<2pt>[l]^-{(\W\Otom_{\pi'})^*}\\
\algm\bigl(\Otom_{\Strsinvbar}\bigr)\ar[d]_-{\cong} \ar@<2pt>[r]^-{(\Otom_{\pi'})_!} \ar@<-2pt>[u]_-{\eta^*} &  \algm\bigl(\Otom_{\Locdsinvbar}\bigr) \ar@<-2pt>[u]_-{\eta^*}\ar[d]^-{\cong} \ar@<2pt>[l]^-{(\Otom_{\pi'})^*}\\
\QFT\bigl(\Strbar,S_{\pi}\bigr) & \QFT\bigl(\Locdbar,S\bigr)}\] with commutative left/right adjoint diagrams.  The vertical adjunction on the left goes between quantum field theories on $\pi : \Str \to \Locd$ that do not necessarily satisfy the isotony axiom and their homotopy analogues.\dqed
\end{example}

\begin{example}[Homotopy AQFT on spacetime with timelike boundary]\label{ex:haqft-boundary}
When applied to the orthogonal category $\Regxbar = (\Regx,\perp)$ in Example \ref{ex:aqft-boundary}, Corollary \ref{cor:haqft-aqft-adjunction} gives a \index{homotopy quantum field theory on spacetime with timelike boundary}\index{quantum field theory on spacetime with timelike boundary!homotopy}change-of-operad adjunction 
\[\nicexy@R-.2cm{\algm\bigl(\wom_{\Regxsinvbar}\bigr) \ar@{=}[d] \ar@<2pt>[r]^-{\eta_!} & \algm\bigl(\Otom_{\Regxsinvbar}\bigr) \ar[d]^-{\cong} \ar@<2pt>[l]^-{\eta^*}\\
\HQFT\bigl(\Regxsinvbar\bigr) & \QFT\bigl(\Regxbar,S_X\bigr)}\] between algebraic quantum field theories on $X$ and their homotopy analogues.  Moreover, this adjunction is a Quillen equivalence when $\M=\Chaink$ with $\fieldk$ a field of characteristic zero.

Moreover, the full subcategory inclusion $j : \Regxzero \to \Regx$ induces an orthogonal functor
\[\nicexy{\Regxzerosinvbar \ar[r]^-{j'} & \Regxsinvbar},\] where $\Regxzero$ is the category of regions in the interior $X_0$ of $X$.  By Corollary \ref{cor:haqft-adjunction-diagram}, there is an induced diagram whose middle square consists of change-of-operad adjunctions
\[\nicexy@C+.7cm{\HQFT\bigl(\Regxzerosinvbar\bigr) \ar@{=}[d] & \HQFT\bigl(\Regxsinvbar\bigr)\ar@{=}[d]\\ 
\algm\bigl(\wom_{\Regxzerosinvbar}\bigr) \ar@<-2pt>[d]_-{\eta_!} \ar@<2pt>[r]^-{(\W\Otom_{j'})_!} & \algm\bigl(\wom_{\Regxsinvbar}\bigr)\ar@<-2pt>[d]_-{\eta_!}\ar@<2pt>[l]^-{(\W\Otom_{j'})^*}\\
\algm\bigl(\Otom_{\Regxzerosinvbar}\bigr)\ar[d]_-{\cong} \ar@<2pt>[r]^-{(\Otom_{j'})_!} \ar@<-2pt>[u]_-{\eta^*} &  \algm\bigl(\Otom_{\Regxsinvbar}\bigr) \ar@<-2pt>[u]_-{\eta^*}\ar[d]^-{\cong} \ar@<2pt>[l]^-{(\Otom_{j'})^*}\\
\QFT\bigl(\Regxzerobar,S_{X_0}\bigr) & \QFT\bigl(\Regxbar,S_X\bigr)}\] with commutative left/right adjoint diagrams.  The vertical adjunction on the left goes between algebraic quantum field theories on the interior $X_0$ and their homotopy analogues.\dqed
\end{example}

\section{Coherence Theorem}\label{sec:coherence-haqft}

For the rest of this chapter, we will study the structure of homotopy algebraic quantum field theories.  In Definition \ref{def:haqft} we defined a homotopy algebraic quantum field theory on an orthogonal category $\Cbar = (\C,\perp)$ as an algebra over the colored operad $\wocbarm\in \Operadcm$, which is the Boardman-Vogt construction of the colored operad $\Ocbarm \in \Operadcm$.  Recall that $\Ocbarm$ is the image under the change-of-category functor \[\Operadcset \to \Operadcm\] of the colored operad $\Ocbar$ in Definition \ref{def:aqft-operad}.    

The following coherence result describes homotopy algebraic quantum field theories in terms of generating structure morphisms and generating relations.  Recall from Notation \ref{not:x-sub-c} the shorthand \[X_{\uc} = X_{c_1} \otimes \cdots \otimes X_{c_m}\] for each $\colorc$-colored object $X$ and $\uc=(c_1,\ldots,c_m) \in \Profc$.  Also recall from Notation \ref{notation:a-of-v} that for $A \in \M^{\Profcc}$ and a vertex $v$ in a $\colorc$-colored tree, $A(v)$ is the shorthand for the entry $A\inoutv$.

\begin{theorem}\label{thm:haqft-coherence}
Suppose $\Cbar = (\C,\perp)$ is an orthogonal category with object set $\colorc$.  Then a \index{homotopy algebraic quantum field theory!coherence}\index{Coherence Theorem!for homotopy algebraic quantum field theories}homotopy algebraic quantum field theory on $\Cbar$ is exactly a pair $(X,\lambda)$ consisting of
\begin{itemize}\item a $\colorc$-colored object $X = \{X_c\}_{c\in \colorc}$ in $\M$ and
\item a structure morphism\index{structure morphism!for homotopy algebraic quantum field theory}
\begin{equation}\label{haqft-restricted}
\nicexy@C+.7cm{\J[T] \otimes X_{\uc} \ar[r]^{\lambda_T^{\{f^v\}}} & X_d \in \M}
\end{equation}
for 
\begin{itemize}\item each $T \in \uTreec\duc$ with $\duc \in \Profcc$ and 
\item each $\{f^v\} \in \prod_{v\in \Vt(T)} \Ocbar(v)$
\end{itemize}
\end{itemize}
that satisfies the following four conditions.
\begin{description}
\item[Associativity] For $\bigl(\uc=(c_1,\ldots,c_n);d\bigr) \in \Profcc$ with $n \geq 1$, $T \in \uTreecduc$, $T_j \in \uTreec\cjubj$ for $1 \leq j \leq n$, $\ub=(\ub_1,\ldots,\ub_n)$, \[G=\graft(T;T_1,\ldots,T_n) \in \uTreec\dub\] the grafting \eqref{def:grafting},  $\{f^v\} \in \prod_{v\in \Vt(T)} \Ocbar(v)$, and $\{f^u_j\} \in \prod_{u\in \Vt(T_j)} \Ocbar(u)$ for $1 \leq j \leq n$, the diagram
\begin{equation}\label{haqft-ass}
\nicexy{\J[T]\otimes \Bigl(\bigotimes\limits_{j=1}^n \J[T_j]\Bigr) \otimes X_{\ub} \ar[d]_-{\mathrm{permute}}^-{\cong} \ar[r]^-{\pi} & \J[G]\otimes X_{\ub} \ar[dd]^-{\lambda_G^{\{f^v\},\{f^u_j\}_{j=1}^n}} \\
\J[T]\otimes\bigotimes\limits_{j=1}^n\bigl( \J[T_j]\otimes X_{\ub_j}\bigr) \ar[d]_-{\bigl(\Id,\bigotimes_j \lambda_{T_j}^{\{f^u_j\}}\bigr)} &\\
\J[T]\otimes X_{\uc} \ar[r]^-{\lambda_T^{\{f^v\}}} & X_d}
\end{equation}
is commutative.  Here $\pi=\bigotimes_S 1$ is the morphism in Lemma \ref{lem:morphism-pi} for the grafting $G$.
\item[Unity] For each $c \in \colorc$, the composition
\begin{equation}\label{haqft-unity}
\nicexy{X_c \ar[r]^-{\cong} & \J[\uparrow_c]\otimes  X_c \ar[r]^-{\lambda_{\uparrow_c}^{\varnothing}} & X_c}
\end{equation} 
is the identity morphism of $X_c$.
\item[Equivariance] For each $T \in \uTreec\duc$, permutation $\sigma \in \Sigma_{|\uc|}$, and $\{f^v\} \in \prod_{v\in \Vt(T)} \Ocbar(v)$, the diagram 
\begin{equation}\label{haqft-eq}
\nicexy@C+.7cm{\J[T]\otimes X_{\uc} \ar[d]_-{(\Id,\sigmainv)} \ar[r]^-{\lambda_T^{\{f^v\}}} & X_d \ar@{=}[d]\\ \J[T\sigma]\otimes X_{\uc\sigma} \ar[r]^-{\lambda_{T\sigma}^{\{f^v\}}} & X_d}
\end{equation}
is commutative, in which $T\sigma \in \uTreec\ducsigma$ is the same as $T$ except that its ordering is $\zeta_T\sigma$ with $\zeta_T$ the ordering of $T$.  The permutation $\sigmainv : X_{\uc} \iso X_{\uc\sigma}$ permutes the factors in $X_{\uc}$.
\item[Wedge Condition] For $T \in \uTreec\duc$, $H_v \in \uTreec(v)$ for each $v\in \Vt(T)$,  $K=T(H_v)_{v\in T}$ the tree substitution, and $\{f^u_v\}\in\prod_{u\in \Vt(H_v)} \Ocbar(u)$ for each $v \in \Vt(T)$, the diagram
\begin{equation}\label{haqft-wedge}
\nicexy@C+.8cm{\J[T] \otimes X_{\uc} \ar[d]_-{(\J,\Id)} \ar[r]^-{\lambda_T^{\{h^v\}}} & X_d \ar@{=}[d] \\ \J[K]\otimes X_{\uc} \ar[r]^-{\lambda_K^{\{f^u_v\}_{u\in K}}} & X_d}
\end{equation}
is commutative, in which \[h^v = \gamma^{\Ocbar}_{H_v}\bigl(\{f^u_v\}_{u\in H_v}\bigr) \in \Ocbar(v)\] for each $v \in \Vt(T)$ with \[\gamma^{\Ocbar}_{H_v} : \Ocbar[H_v] \to \Ocbar(v) \in \Set\] the operadic structure morphism \eqref{operadic-structure-map} of $\Ocbar$ for $H_v$.
\end{description}
A morphism $f : (X,\lambda^X) \to (Y,\lambda^Y)$ of homotopy algebraic quantum field theories on $\Cbar$ is a morphism of the underlying $\colorc$-colored objects that respects the structure morphisms in \eqref{haqft-restricted} in the obvious sense.
\end{theorem}

\begin{proof}
This is the special case of the Coherence Theorem \ref{thm:wo-algebra-coherence} for the $\colorc$-colored operad $\Ocbarm$.  Indeed, recall that the $\colorc$-colored operad $\Ocbarm$ has entries
\[\Ocbarm\duc = \coprodover{\Ocbar\duc} \tensorunit\]
for $(\uc;d) \in \Profcc$.  For each $\colorc$-colored tree $T$, there is a natural isomorphism \[\Ocbarm[T] = \bigtensorover{v\in T}\, \Ocbarm\inoutv = \bigtensorover{v\in T} \, \Bigl[\coprodover{\Ocbar(v)} \tensorunit\Bigr] \cong \coprodover{\prodover{v\in T}\Ocbar(v)} \tensorunit.\]  It follows that there is a natural isomorphism \[\J[T]\otimes\Ocbarm[T] \otimes X_{\uc} \cong \coprodover{\prodover{v\in T}\Ocbar(v)} \J[T]\otimes X_{\uc}.\]  Therefore, the structure morphism $\lambda_T$ in \eqref{wo-algebra-restricted} is uniquely determined by the restrictions $\lambda^{\{f^v\}}_{T}$ as stated in \eqref{haqft-restricted}.  The associativity, unity, equivariance, and wedge conditions \eqref{haqft-ass}-\eqref{haqft-wedge} are exactly those in the Coherence Theorem \ref{thm:wo-algebra-coherence}.
\end{proof}

\section{Homotopy Causality Axiom}\label{sec:h-causality}

In this section, we explain that every homotopy algebraic quantum field theory satisfies a homotopy coherent version of the causality axiom.

\begin{motivation}
An algebraic quantum field theory $\fraka \in \QFT(\Cbar)$ on an orthogonal category $\Cbar=(\C,\perp)$ satisfies the causality axiom \eqref{perp-com}.  It says that for each orthogonal pair $(g_1 : a \shortto c, g_2 : b \shortto c) \in \perpen$, the diagram\label{notation:12permutation}
\[\nicexy@C+1cm{\fraka(a) \otimes \fraka(b) \ar[r]^-{\bigl(\fraka(g_1),\fraka(g_2)\bigr)} & \fraka(c) \otimes \fraka(c) \ar@<2pt>[r]^-{\mu_c(1~2)} \ar@<-2pt>[r]_-{\mu_c} & \fraka(c)}\] is commutative, where $(1~2)$ is the symmetry permutation on $\fraka(c)^{\otimes 2}$.  For a homotopy algebraic quantum field theory, we should expect this diagram to commute up to specified homotopies.\dqed
\end{motivation}

To explain the homotopy version of the causality axiom, we need the following notations.  As before, using the canonical bijection in Example \ref{ex:ocbar-unary}, for a morphism $f \in \C(c,d)$, we will abbreviate an element $[\id_1,f] \in \Ocbar\dc$ to just $f$.

\begin{assumption}\label{assumption:hcausality}
Suppose $\Cbar = (\C,\perp)$ is an orthogonal category with object set $\colorc$, and $(g_1 : a \shortto c, g_2 : b \shortto c)$ is an orthogonal pair in $\Cbar$.  
\begin{itemize}
\item Suppose $C = \Cor_{(c,c;c)} \in \uTreec\sbinom{c}{c,c}$, and $C_{ab} = \Cor_{(a,b;c)} \in \uTreec\cab$. 
\item Suppose $L_1 = \Lin_{(a,c)} \in \uTreec\ca$, and $L_2 = \Lin_{(b,c)} \in \uTreec\cb$.
\item Define the grafting $T = \graft\bigl(C; L_1,L_2\bigr) \in \uTreec\cab$, which we may visualize as follows.
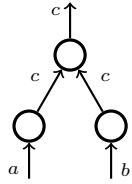
\begin{center}\begin{tikzpicture} 
\node (x) {}; \node [smallplain, above=.6cm of x] (w) {};
\node [smallplain, left=.2cm of x] (u) {}; \node [smallplain, right=.2cm of x] (v) {};
\draw [outputleg] (w) to node[near end]{\scriptsize{$c$}} +(0,.7cm);
\draw [inputleg] (u) to node[near end, swap]{\scriptsize{$a$}} +(0,-.7cm);
\draw [inputleg] (v) to node[near end]{\scriptsize{$b$}} +(0,-.7cm);
\draw [arrow] (u) to node{\scriptsize{$c$}} (w); 
\draw [arrow] (v) to node[swap]{\scriptsize{$c$}} (w);
\end{tikzpicture}
\end{center}
Note that $T$ is the $2$-level tree $T\bigl({(a),(b)};(c,c);c\bigr)$ in Example \ref{ex:twolevel-tree}, and it has two internal edges.
\item Denote by $\Id_c^2$ the element $\bigl[\id_2, \{\Id_c,\Id_c\}\bigr]\in \Ocbar\sbinom{c}{c,c}$.
\item Denote by $\tau$ the element $\bigl[(1~2), \{\Id_c,\Id_c\}\bigr]\in \Ocbar\sbinom{c}{c,c}$, where $(1~2)$ is the non-identity permutation in $\Sigma_2$.
\item Denote by $\ug$ the element $\bigl[\id_2,\{g_1,g_2\}\bigr] \in \Ocbar\cab$.
\end{itemize}
\end{assumption}

The following result is the homotopy coherent version of the causality axiom.  To simplify the notation, we will omit writing some of the identity morphisms below.

\begin{theorem}\label{thm:hcausality}
In the context of Assumption \ref{assumption:hcausality}, suppose $(X,\lambda)$ is a homotopy algebraic quantum field theory on $\Cbar$.  \index{causality axiom!homotopy}\index{homotopy causality axiom}Then the diagram
\[\begin{footnotesize}\nicexy@C+.4cm{\tensorunit^{\otimes 2} \otimes X_a \otimes X_b \ar[d]_-{1^{\otimes 2}} 
& \J[C] \otimes \bigl(\J[L_1]\otimes X_a\bigr) \otimes \bigl(\J[L_2]\otimes X_b\bigr) \ar[l]_-{\cong} \ar[r]^-{\bigl(\lambda_{L_1}^{g_1}, \lambda_{L_2}^{g_2}\bigr)} 
& \J[C] \otimes X_c^{\otimes 2} \ar[d]^-{\lambda_{C}^{\Id_c^2}}\\
J^{\otimes 2} \otimes X_a \otimes X_b \ar[r]^-{\cong} & \J[T] \otimes X_a \otimes X_b \ar[r]^-{\lambda_T^{\left\{\Id_c^2,g_1,g_2\right\}}} & X_c\\
\tensorunit^{\otimes 2} \otimes X_a \otimes X_b \ar[d]_-{0^{\otimes 2}} \ar[u]^-{0^{\otimes 2}} 
& \J[C_{ab}] \otimes X_a \otimes X_b \ar[l]_-{\cong} \ar[r]^-{\lambda_{C_{ab}}^{\ug}} \ar@{}[u]|-{(1)} \ar@{}[d]|-{(2)} & X_c \ar@{=}[u] \ar@{=}[d]\\
J^{\otimes 2} \otimes X_a \otimes X_b \ar[r]^-{\cong} & \J[T] \otimes X_a \otimes X_b \ar[r]^-{\lambda_T^{\left\{\tau,g_1,g_2\right\}}} & X_c\\
\tensorunit^{\otimes 2} \otimes X_a \otimes X_b \ar[u]^-{1^{\otimes 2}} 
& \J[C] \otimes \bigl(\J[L_1]\otimes X_a\bigr) \otimes \bigl(\J[L_2]\otimes X_b\bigr) \ar[l]_-{\cong} \ar[r]^-{\bigl(\lambda_{L_1}^{g_1}, \lambda_{L_2}^{g_2}\bigr)} 
& \J[C] \otimes X_c^{\otimes 2} \ar[u]_-{\lambda_{C}^{\tau}}}\end{footnotesize}\]
in $\M$ is commutative, where $0,1 : \tensorunit \to J$ are part of the commutative segment $J$.
\end{theorem}

\begin{proof}
This is a consequence of the Coherence Theorem \ref{thm:haqft-coherence} for homotopy algebraic quantum field theories.  Indeed, in the above diagram:
\begin{enumerate}
\item The top and bottom rectangles are commutative by the associativity condition \eqref{haqft-ass} and the grafting definition of $T$.
\item The rectangle (1) is commutative by the wedge condition \eqref{haqft-wedge} applied to the tree substitution \[T = C_{ab}(T) \in \uTreec\cab\] and by the equalities \[\gamma^{\Ocbar}_T\bigl(\Id^2_c; g_1,g_2\bigr)=\bigl[\id_2,\{g_1,g_2\}\bigr]=\ug\in \Ocbar\cab.\]
\item The rectangle (2) is commutative by the same wedge condition and the equalities \[\ug = \bigl[(1~2),\{g_1,g_2\}\bigr] = \gamma^{\Ocbar}_T\bigl(\tau;g_1,g_2\bigr) \in \Ocbar\cab,\] the first of which holds because $g_1$ and $g_2$ are orthogonal.
\end{enumerate}
\end{proof}

\begin{interpretation}
Theorem \ref{thm:hcausality} is a precise form of the statement that the diagram 
\[\nicexy@C+1cm{X_a \otimes X_b \ar[r]^-{\left(X_{g_1},X_{g_2}\right)} \ar[d]_-{\left(X_{g_1},X_{g_2}\right)}  & X_c \otimes X_c \ar[d]^-{\mu_2^c}\\
X_c \otimes X_c \ar[r]^-{\mu_2^c (1~2)} & X_c}\] is commutative up to homotopy.  Here $\mu_2^c$ is the binary multiplication in the $A_\infty$-algebra $X_c$ as in Example \ref{ex1:ainfinity}.  So Theorem \ref{thm:hcausality} says that each homotopy algebraic quantum field theory satisfies the causality axiom up to specified homotopies that are also structure morphisms.\dqed
\end{interpretation}

Theorem \ref{thm:hcausality} can be applied to all the orthogonal categories in Section \ref{sec:aqft-examples}.

\begin{example}[Homotopy causality in homotopy chiral conformal QFT]\label{ex:hcausality-chiral}
Applied \index{homotopy chiral conformal quantum field theory!homotopy causality axiom}to the orthogonal category $\Mandbar$ in Example \ref{ex:ccqft}, Theorem \ref{thm:hcausality} says that every $\wom_{\Mandbar}$-algebra satisfies the causality axiom up to specified homotopies.\dqed\end{example}

\begin{example}[Homotopy causality in homotopy chiral conformal QFT on discs]\label{ex:hcausality-chiral-interval}
Applied \index{homotopy chiral conformal quantum field theory on discs!homotopy causality axiom}to the orthogonal category $\Discdbar$ in Example \ref{ex:ccqft-int}, Theorem \ref{thm:hcausality} says that every $\wom_{\Discdbar}$-algebra satisfies the causality axiom up to specified homotopies.\dqed\end{example}

\begin{example}[Homotopy causality in homotopy chiral conformal QFT on a fixed manifold]\label{ex:hcausality-chiral-manfold}
Applied \index{homotopy chiral conformal quantum field theory on a manifold!homotopy causality axiom}to the orthogonal category $\Openxbar$ for $X \in \Mand$ in Example \ref{ex:ccqft-manifold}, Theorem \ref{thm:hcausality} says that every $\wom_{\Openxbar}$-algebra satisfies the causality axiom up to specified homotopies.\dqed\end{example}

\begin{example}[Homotopy causality in homotopy Euclidean QFT]\label{ex:hcausality-euclidean}
Applied \index{homotopy Euclidean quantum field theory!homotopy causality axiom}to the orthogonal category $\Riemdbar$ in Example \ref{ex:euclidean-qft}, Theorem \ref{thm:hcausality} says that every $\wom_{\Riemdbar}$-algebra satisfies the causality axiom up to specified homotopies.\dqed\end{example}

\begin{example}[Homotopy causality in homotopy locally covariant QFT]\label{ex:hcausality-lcqft}
Applied \index{homotopy locally covariant quantum field theory!homotopy causality axiom}to the orthogonal category $\Locdsinvbar$ in Example \ref{ex:lcqft},Theorem \ref{thm:hcausality} says that every $\wom_{\Locdsinvbar}$-algebra satisfies the causality axiom up to specified homotopies.\dqed\end{example}

\begin{example}[Homotopy causality in homotopy locally covariant QFT on a fixed spacetime]\label{ex:hcausality-lcqft-fixed}
Applied \index{homotopy locally covariant quantum field theory on a spacetime!homotopy causality axiom}to the orthogonal category $\Ghxsinvbar$ for $X \in \Locd$ in Example \ref{ex:causal-nets}, Theorem \ref{thm:hcausality} says that every $\wom_{\Ghxsinvbar}$-algebra satisfies the causality axiom up to specified homotopies.\dqed\end{example}

\begin{example}[Homotopy causality in homotopy dynamical quantum gauge theories on principal bundles]\label{ex:hcausality-dynamical}
Applied \index{homotopy dynamical quantum gauge theory!homotopy causality axiom}to the orthogonal category $\Bglocsginvbar$ in Example \ref{ex:dynamical-gauge}, Theorem \ref{thm:hcausality} says that every $\wom_{\Bglocsginvbar}$-algebra satisfies the causality axiom up to specified homotopies.\dqed\end{example}

\begin{example}[Homotopy causality in homotopy charged matter QFT on background gauge fields]\label{ex:hcausality-charged}
Applied \index{homotopy charged matter quantum field theory!homotopy causality axiom}to the orthogonal category $\Bgconlocsginvbar$ in Example \ref{ex:charged-matter}, Theorem \ref{thm:hcausality} says that every $\wom_{\Bgconlocsginvbar}$-algebra satisfies the causality axiom up to specified homotopies.\dqed\end{example}

\begin{example}[Homotopy causality in homotopy Dirac QFT]\label{ex:hcausality-dirac}
Applied \index{homotopy Dirac quantum field theory!homotopy causality axiom}to the orthogonal category $\Slocdsinvbar$ in Example \ref{ex:dirac-qft}, Theorem \ref{thm:hcausality} says that every $\wom_{\Slocdsinvbar}$-algebra satisfies the causality axiom up to specified homotopies.\dqed\end{example}

\begin{example}[Homotopy causality in homotopy QFT on structured spacetimes]\label{ex:hcausality-structured-spacetime}
Applied \index{homotopy quantum field theory on structured spacetime!homotopy causality axiom}to the orthogonal category $\Strsinvbar$ in Example \ref{ex:qft-structured}, Theorem \ref{thm:hcausality} says that every $\wom_{\Strsinvbar}$-algebra satisfies the causality axiom up to specified homotopies.\dqed\end{example}

\begin{example}[Homotopy causality in homotopy QFT on spacetime with timelike boundary]\label{ex:hcausality-timelike-boundary}
Applied \index{homotopy quantum field theory on spacetime with timelike boundary!homotopy causality axiom}to the orthogonal category $\Regxsinvbar$, where $X$ is a spacetime with timelike boundary, in Example \ref{ex:aqft-boundary}, Theorem \ref{thm:hcausality} says that every $\wom_{\Regxsinvbar}$-algebra satisfies the causality axiom up to specified homotopies.\dqed\end{example}

\section{Homotopy Coherent Diagrams}\label{sec:h-functoriality}

Using the Coherence Theorem \ref{thm:haqft-coherence}, for the next few sections we will explain the structure that exists in homotopy algebraic quantum field theories, i.e., in $\wocbarm$-algebras.   In this section, we explain the homotopy coherent diagram structure that exists on each homotopy algebraic quantum field theory.

\begin{motivation}
Recall from Example \ref{ex:operad-diag} that for each small category $\C$ with object set $\colorc$, there is a $\colorc$-colored operad $\Cdiag$ in $\M$ whose algebras are $\C$-diagrams in $\M$.  The algebras over the Boardman-Vogt construction $\Wcdiag$ are homotopy coherent $\C$-diagrams in $\M$, as we explained in Section \ref{sec:hcdiagram}.  An algebraic quantum field theory on an orthogonal category $\Cbar$ is, first of all, a functor $\C \to \Monm$.  Composing with the forgetful functor to $\M$, each algebraic quantum field theory yields an underlying $\C$-diagram in $\M$.  So we should expect a homotopy algebraic quantum field theory on $\Cbar$ to have the structure of a homotopy coherent $\C$-diagram in $\M$.\dqed 
\end{motivation}

For any colors $c,d \in \colorc$, recall from Example \ref{ex:ocbar-unary} that there is a canonical bijection \[\C(c,d) \cong \Ocbar\dc,\] sending $f \in \C(c,d)$ to $[\id_1,f]$.  This induces a canonical isomorphism \[\Ocbarm\dc =  \coprodover{\Ocbar\dc}\tensorunit \cong \coprodover{\C(c,d)} \tensorunit,\] which we will use below.

\begin{theorem}\label{thm:cdiag-ocbar}
Suppose $\Cbar = (\C,\perp)$ is an orthogonal category with object set $\colorc$.  
\begin{enumerate}
\item There is a morphism of $\colorc$-colored operads \[p : \Cdiag \to \Ocbarm\] in $\M$ defined entrywise as follows.
\begin{itemize}\item For $c,d \in \colorc$, the morphism
\[\nicexy{\Cdiag\dc = \coprodover{\C(c,d)} \tensorunit \ar[r]^-{p}_-{\cong} & \Ocbarm\dc} \in \M\]
sends the copy of $\tensorunit$ in $\Cdiag\dc$ corresponding to $f \in \C(c,d)$ to the copy of $\tensorunit$ in $\Ocbarm\dc$ corresponding to $[\id_1,f] \in \Ocbar\dc$.
\item If $|\uc|\not= 1$, then \[\nicexy{\Cdiag\duc = \varnothing \ar[r]^-{p} & \Ocbarm\duc} \in \M\] is the unique morphism from the initial object $\varnothing$.
\end{itemize}
\item The morphism $p$ induces a change-of-operad adjunction \[\nicexy@C+.4cm{\algm\bigl(\Wcdiag\bigr) \ar@<2pt>[r]^-{(\W p)_!} & \algm\bigl(\wocbarm\bigr) = \HQFT(\Cbar) \ar@<2pt>[l]^-{(\W p)^*}}\] between the category of \index{homotopy coherent diagram}homotopy coherent $\C$-diagrams in $\M$ and the category of homotopy algebraic quantum field theories on $\Cbar$.
\end{enumerate}
\end{theorem}

\begin{proof}
For assertion (1), note that the equivariant structure on $\Cdiag$ is trivial, since it is concentrated in unary entries, and its operadic composition is given by the categorical composition in $\C$.  A direct inspection of Definition \ref{def:aqft-operad} of $\Ocbar$ shows that $p$ is entrywise well-defined and respects the colored units and operadic composition.  Therefore, $p$ is a morphism of $\colorc$-colored operads.

Assertion (2) follows from assertion (1), the naturality of the Boardman-Vogt construction in Proposition \ref{prop:w-functor}, and Theorem \ref{thm:change-operad}.
\end{proof}

\begin{interpretation}
The right adjoint $(\W p)^*$ sends each homotopy algebraic quantum field theory on $\Cbar$ to its underlying $\Wcdiag$-algebra, i.e., homotopy coherent $\C$-diagram in $\M$.  More explicitly, suppose $(X,\lambda) \in \HQFT(\Cbar)$, and recall the Coherence Theorem \ref{thm:hcdiagram} for homotopy coherent $\C$-diagrams in $\M$.  Suppose given a profile $\uc=(c_0,\ldots,c_n) \in \Profc$ and a sequence of composable morphisms \[\uf=\{f_j\} \in \prod_{j=1}^n \C(c_{j-1},c_j) \cong \prod_{j=1}^n \Ocbar\sbinom{c_j}{c_{j-1}}.\] Then the structure morphism
\begin{equation}\label{haqft-hcdiagram}
\nicexy{\J[\Lin_{\uc}]\otimes \bigl[(\W p)^*X\bigr]_{c_0} = \J[\Lin_{\uc}]\otimes X_{c_0} \ar[r]^-{\lambda_{\uc}^{\uf}} & X_{c_n}=\bigl[(\W p)^*X\bigr]_{c_n}\in \M}
\end{equation}
in \eqref{hcdiagram-structure-map} of the homotopy coherent $\C$-diagram $(\W p)^*X$ is given by the structure morphism $\lambda_{\Lin_{\uc}}^{\{[\id_1,f_j]\}}$ in \eqref{haqft-restricted}.\dqed\end{interpretation}

Theorem \ref{thm:cdiag-ocbar} can be applied to all the orthogonal categories in Section \ref{sec:aqft-examples}.

\begin{example}[Homotopy coherent diagram in homotopy chiral conformal QFT]\label{ex:hcdiagram-chiral}
Applied \index{homotopy chiral conformal quantum field theory!homotopy coherent diagram}to the orthogonal category $\Mandbar$ in Example \ref{ex:ccqft}, we obtain the adjunction \[\nicexy@C+.4cm{\algm\bigl(\W(\Mand)^{\diag}\bigr) \ar@<2pt>[r]^-{(\W p)_!} & \algm\bigl(\wom_{\Mandbar}\bigr) = \HQFT(\Mandbar) \ar@<2pt>[l]^-{(\W p)^*}}\] between the category of homotopy coherent $\Mand$-diagrams in $\M$ and the category of homotopy chiral conformal quantum field theories.\dqed
\end{example}

\begin{example}[Homotopy coherent diagram in homotopy chiral conformal QFT on discs]\label{ex:hcdiagram-chiral-interval}
Applied \index{homotopy chiral conformal quantum field theory on discs!homotopy coherent diagram}to the orthogonal category $\Discdbar$ in Example \ref{ex:ccqft-int}, we obtain the adjunction \[\nicexy@C+.4cm{\algm\bigl(\W(\Discd)^{\diag}\bigr) \ar@<2pt>[r]^-{(\W p)_!} & \algm\bigl(\wom_{\Discdbar}\bigr) = \HQFT(\Discdbar) \ar@<2pt>[l]^-{(\W p)^*}}\] between the category of homotopy coherent $\Discd$-diagrams in $\M$ and the category of homotopy chiral conformal quantum field theories defined on oriented manifolds diffeomorphic to $\fieldr^d$.\dqed
\end{example}

\begin{example}[Homotopy coherent diagram in homotopy chiral conformal QFT on a fixed manifold]\label{ex:hcdiagram-chiral-manfold}
Applied \index{homotopy chiral conformal quantum field theory on a manifold!homotopy coherent diagram}to the orthogonal category $\Openxbar$ for $X \in \Mand$ in Example \ref{ex:ccqft-manifold}, we obtain the adjunction \[\nicexy@C+.4cm{\algm\bigl(\W(\Openx)^{\diag}\bigr) \ar@<2pt>[r]^-{(\W p)_!} & \algm\bigl(\wom_{\Openxbar}\bigr) = \HQFT(\Openxbar) \ar@<2pt>[l]^-{(\W p)^*}}\] between the category of homotopy coherent $\Openx$-diagrams in $\M$ and the category of homotopy chiral conformal quantum field theories defined on $X$.\dqed
\end{example}

\begin{example}[Homotopy coherent diagram in homotopy Euclidean QFT]\label{ex:hcdiagram-euclidean}
Applied \index{homotopy Euclidean quantum field theory!homotopy coherent diagram}to the orthogonal category $\Riemdbar$ in Example \ref{ex:euclidean-qft}, we obtain the adjunction \[\nicexy@C+.4cm{\algm\bigl(\W(\Riemd)^{\diag}\bigr) \ar@<2pt>[r]^-{(\W p)_!} & \algm\bigl(\wom_{\Riemdbar}\bigr) = \HQFT(\Riemdbar) \ar@<2pt>[l]^-{(\W p)^*}}\] between the category of homotopy coherent $\Riemd$-diagrams in $\M$ and the category of homotopy Euclidean quantum field theories.\dqed
\end{example}

\begin{example}[Homotopy coherent diagram in homotopy locally covariant QFT]\label{ex:hcdiagram-lcqft}
Applied \index{homotopy locally covariant quantum field theory!homotopy coherent diagram}to the orthogonal category $\Locdsinvbar$ in Example \ref{ex:lcqft}, we obtain the adjunction \[\nicexy@C+.4cm{\algm\bigl(\W(\Locdsinv)^{\diag}\bigr) \ar@<2pt>[r]^-{(\W p)_!} & \algm\bigl(\wom_{\Locdsinvbar}\bigr) = \HQFT(\Locdsinvbar) \ar@<2pt>[l]^-{(\W p)^*}}\] between the category of homotopy coherent $\Locdsinv$-diagrams in $\M$ and the category of homotopy locally covariant quantum field theories.\dqed
\end{example}

\begin{example}[Homotopy coherent diagram in homotopy locally covariant QFT on a fixed spacetime]\label{ex:hcdiagram-lcqft-fixed}
Applied \index{homotopy locally covariant quantum field theory on a spacetime!homotopy coherent diagram}to the orthogonal category $\Ghxsinvbar$ for $X \in \Locd$ in Example \ref{ex:causal-nets}, we obtain the adjunction \[\nicexy@C+.4cm{\algm\bigl(\W(\Ghxsinv)^{\diag}\bigr) \ar@<2pt>[r]^-{(\W p)_!} & \algm\bigl(\wom_{\Ghxsinvbar}\bigr) = \HQFT(\Ghxsinvbar) \ar@<2pt>[l]^-{(\W p)^*}}\] between the category of homotopy coherent $\Ghxsinv$-diagrams in $\M$ and the category of homotopy locally covariant quantum field theories defined on $X$.\dqed
\end{example}

\begin{example}[Homotopy coherent diagram in homotopy dynamical quantum gauge theories on principal bundles]\label{ex:hcdiagram-dynamical}
Applied \index{homotopy dynamical quantum gauge theory!homotopy coherent diagram}to the orthogonal category $\Bglocsginvbar$ in Example \ref{ex:dynamical-gauge}, we obtain the adjunction \[\nicexy@C+.4cm{\algm\bigl(\W(\Bglocsginv)^{\diag}\bigr) \ar@<2pt>[r]^-{(\W p)_!} & \algm\bigl(\wom_{\Bglocsginvbar}\bigr) = \HQFT(\Bglocsginvbar) \ar@<2pt>[l]^-{(\W p)^*}}\] between the category of homotopy coherent $\Bglocsginv$-diagrams in $\M$ and the category of homotopy dynamical quantum gauge theories on principal $G$-bundles.\dqed
\end{example}

\begin{example}[Homotopy coherent diagram in homotopy charged matter QFT on background gauge fields]\label{ex:hcdiagram-charged}
Applied \index{homotopy charged matter quantum field theory!homotopy coherent diagram}to the orthogonal category $\Bgconlocsginvbar$ in Example \ref{ex:charged-matter}, we obtain the adjunction \[\nicexy@C+.4cm@R-.2cm{\algm\bigl(\W(\Bgconlocsginv)^{\diag}\bigr) \ar@<2pt>[r]^-{(\W p)_!} & \algm\bigl(\wom_{\Bgconlocsginvbar}\bigr) \ar@<2pt>[l]^-{(\W p)^*}\ar@{=}[d]\\ & \HQFT(\Bgconlocsginvbar)}\] between the category of homotopy coherent $\Bgconlocsginv$-diagrams in $\M$ and the category of homotopy charged matter quantum field theories on background gauge fields.\dqed
\end{example}

\begin{example}[Homotopy coherent diagram in homotopy Dirac QFT]\label{ex:hcdiagram-dirac}
Applied \index{homotopy Dirac quantum field theory!homotopy coherent diagram}to the orthogonal category $\Slocdsinvbar$ in Example \ref{ex:dirac-qft}, we obtain the adjunction \[\nicexy@C+.4cm{\algm\bigl(\W(\Slocdsinv)^{\diag}\bigr) \ar@<2pt>[r]^-{(\W p)_!} & \algm\bigl(\wom_{\Slocdsinvbar}\bigr) = \HQFT(\Slocdsinvbar) \ar@<2pt>[l]^-{(\W p)^*}}\] between the category of homotopy coherent $\Slocdsinv$-diagrams in $\M$ and the category of homotopy Dirac quantum fields.\dqed
\end{example}

\begin{example}[Homotopy coherent diagram in homotopy QFT on structured spacetimes]\label{ex:hcdiagram-structured-spacetime}
Applied \index{homotopy quantum field theory on structured spacetime!homotopy coherent diagram}to the orthogonal category $\Strsinvbar$ in Example \ref{ex:qft-structured}, we obtain the adjunction \[\nicexy@C+.4cm{\algm\bigl(\W(\Strsinv)^{\diag}\bigr) \ar@<2pt>[r]^-{(\W p)_!} & \algm\bigl(\wom_{\Strsinvbar}\bigr) = \HQFT(\Strsinvbar) \ar@<2pt>[l]^-{(\W p)^*}}\] between the category of homotopy coherent $\Strsinv$-diagrams in $\M$ and the category of homotopy  quantum field theories on $\pi : \Str \to \Locd$.\dqed
\end{example}

\begin{example}[Homotopy coherent diagram in homotopy QFT on spacetime with timelike boundary]\label{ex:hcdiagram-timelike-boundary}
Applied \index{homotopy quantum field theory on spacetime with timelike boundary!homotopy coherent diagram}to the orthogonal category $\Regxsinvbar$, where $X$ is a spacetime with timelike boundary, in Example \ref{ex:aqft-boundary}, we obtain the adjunction \[\nicexy@C+.4cm@R-.2cm{\algm\bigl(\W(\Regxsinv)^{\diag}\bigr) \ar@<2pt>[r]^-{(\W p)_!} & \algm\bigl(\wom_{\Regxsinvbar}\bigr) \ar@<2pt>[l]^-{(\W p)^*} \ar@{=}[d]\\ & \HQFT(\Regxsinvbar) }\] between the category of homotopy coherent $\Regxsinv$-diagrams in $\M$ and the category of homotopy algebraic quantum field theories on $X$.\dqed
\end{example}

\section{Homotopy Time-Slice Axiom}\label{sec:h-timeslice}

In this section, we explain that for an orthogonal category $\Cbar = (\C,\perp)$ and a set $S$ of morphisms in $\C$, every homotopy algebraic quantum field theory on the orthogonal \index{localization of a category}category $\Csinvbar = \bigl(\Csinv, \ell_*(\perp)\bigr)$ satisfies a homotopy coherent version of the time-slice axiom.

\begin{motivation}
For a chosen set $S$ of morphisms in $\C$, recall from Definition \ref{def:aqft} that an algebraic quantum field theory $\fraka$ on $\Cbar$ satisfies the time-slice axiom with respect to $S$ if the structure morphism $\fraka(s) \in \Monm$ is an isomorphism for each morphism $s \in S$.  By Lemma \ref{lem:aqft-time-slice} we know that algebraic quantum field theories on $\Cbar$ that satisfy the time-slice axiom with respect to $S$ are exactly the algebraic quantum field theories on $\Csinvbar= \bigl(\Csinv, \ell_*(\perp)\bigr)$, with 
\begin{itemize}\item $\Csinv$ the $S$-localization of the category $\C$ and 
\item $\ell_*(\perp)$ the pushforward of the orthogonality relation $\perp$ in $\C$ along the $S$-localization $\ell : \C \to \Csinv$.
\end{itemize}
Therefore, we should expect each homotopy algebraic quantum field theory on $\Csinvbar$, i.e., $\wocsinvbarm$-algebra, to have homotopy inverses for the structure morphisms in $S$.  

In Theorem \ref{thm:cdiag-ocbar} we constructed a morphism of $\colorc$-colored operads \[p : \Csinvdiag \to \Ocsinvbarm\] together with an induced change-of-operad adjunction  \[\nicexy@C+.4cm{\algm\bigl(\Wcsinvdiag\bigr) \ar@<2pt>[r]^-{(\W p)_!} & \algm\bigl(\wocsinvbarm\bigr) = \HQFT(\Csinvbar) \ar@<2pt>[l]^-{(\W p)^*}}\] between the category of homotopy coherent $\Csinv$-diagrams in $\M$ and the category of homotopy algebraic quantum field theories on $\Csinvbar$.  The right adjoint $(\W p)^*$ leaves the underlying entries unchanged, so each $\wocsinvbarm$-algebra has an underlying homotopy coherent $\Csinv$-diagram in $\M$.\dqed
\end{motivation}

For the following result on homotopy time-slice, recall from Example \ref{ex:linear-graph} the notation $\Lin_{\uc}$ for a linear graph associated to a profile $\uc$ and from Definition \ref{def:segment} the morphisms $0,1 : \tensorunit \to J$ as part of the commutative segment $J$.  To simplify the notation, using the canonical bijection \[\nicexy{\Csinv(c,d) \ar[r]^-{\cong} & \Sigma_1 \times \Csinv(c,d) = \Ocsinvbar\dc}, \quad \nicexy{f  \ar@{|->}[r] &[\id_1,f]}\] in Example \ref{ex:ocbar-unary}, we will abbreviate an element $[\id_1,f] \in \Ocsinvbar\dc$ to $f$.  We will use the notation $\lambda_T^{\{f^v\}}$ in \eqref{haqft-restricted} for a structure morphism of a homotopy algebraic quantum field theory.

\begin{theorem}\label{thm:homotopy-timeslice}
Suppose $\Cbar = (\C,\perp)$ is an orthogonal category, and $S$ is a set of morphisms in $\C$.  Suppose $(X,\lambda)$ is a homotopy algebraic quantum field theory on $\Csinvbar=\bigl(\Csinv, \ell_*(\perp)\bigr)$, i.e., a $\wocsinvbarm$-algebra.  Suppose $f : c \to d$ is a morphism in $S$ with inverse $\finverse : d \to c \in \Csinv$.  Then the structure morphism\index{homotopy time-slice axiom}\index{time-slice axiom!homotopy}\index{homotopy algebraic quantum field theory!homotopy time-slice axiom} \[\nicexy@C+.5cm{\J[\Lin_{(d,c)}] \otimes X_d = \tensorunit \otimes X_d \ar[r]^-{\lambda_{\Lin_{(d,c)}}^{\{\finverse\}}} & X_c} \in \M\] is a two-sided \index{homotopy inverse}homotopy inverse of the structure morphism \[\nicexy@C+.5cm{\J[\Lin_{(c,d)}] \otimes X_c = \tensorunit \otimes X_c \ar[r]^-{\lambda_{\Lin_{(c,d)}}^{\{f\}}} & X_d} \in \M\] in the following sense.
\begin{enumerate}
\item $\lambda_{\Lin_{(d,c)}}^{\{\finverse\}}$ is a left homotopy inverse of $\lambda_{\Lin_{(c,d)}}^{\{f\}}$ in the sense that the diagram
\[\nicexy@C+.8cm{\tensorunit \otimes X_c = \J[\Lin_{(c,c)}]\otimes X_c \ar[d]_-{(0,\Id)} \ar[dr]^-{\lambda^{\{\finverse f\}}_{\Lin_{(c,c)}}} & X_c \ar[l]_-{\cong} \ar[d]^-{\Id_{X_c}}\\ 
J \otimes X_c = \J[\Lin_{(c,d,c)}] \otimes X_c \ar[r]_-{\lambda^{\{f,\finverse\}}_{\Lin_{(c,d,c)}}} & X_c\\ \tensorunit \otimes X_c \ar[u]^-{(1,\Id)}&\\
\J[\Lin_{(d,c)}]\otimes\J[\Lin_{(c,d)}]\otimes X_c \ar[u]^-{\cong} \ar[r]^-{\bigl(\Id,\lambda^{\{f\}}_{\Lin_{(c,d)}}\bigr)} & \J[\Lin_{(d,c)}] \otimes X_d \ar[uu]_-{\lambda^{\{\finverse\}}_{\Lin_{(d,c)}}}}\]
in $\M$ is commutative.
\item $\lambda_{\Lin_{(d,c)}}^{\{\finverse\}}$ is a right homotopy inverse of $\lambda_{\Lin_{(c,d)}}^{\{f\}}$ in the sense that the diagram
\[\nicexy@C+.8cm{\tensorunit \otimes X_d = \J[\Lin_{(d,d)}]\otimes X_d \ar[d]_-{(0,\Id)} \ar[dr]^-{\lambda^{\{f\finverse\}}_{\Lin_{(d,d)}}} & X_d \ar[l]_-{\cong} \ar[d]^-{\Id_{X_d}}\\ 
J \otimes X_d = \J[\Lin_{(d,c,d)}] \otimes X_d \ar[r]_-{\lambda^{\{\finverse,f\}}_{\Lin_{(d,c,d)}}} & X_d\\ \tensorunit \otimes X_d \ar[u]^-{(1,\Id)}&\\
\J[\Lin_{(c,d)}]\otimes\J[\Lin_{(d,c)}]\otimes X_d \ar[u]^-{\cong} \ar[r]^-{\bigl(\Id,\lambda^{\{\finverse\}}_{\Lin_{(d,c)}}\bigr)} & \J[\Lin_{(c,d)}] \otimes X_c \ar[uu]_-{\lambda^{\{f\}}_{\Lin_{(c,d)}}}}\]
in $\M$ is commutative.
\end{enumerate}
\end{theorem}

\begin{proof}
This is Corollary \ref{cor:hinverse} applied to the underlying homotopy coherent $\Csinv$-diagram in $\M$ of $(X,\lambda)$.
\end{proof}

\begin{interpretation}
Every homotopy algebraic quantum field theory on $\Csinvbar=\bigl(\Csinv, \ell_*(\perp)\bigr)$ satisfies a homotopy version of the time-slice axiom in the sense that each structure morphism $\lambda_{\Lin_{(c,d)}}^{\{f\}}$ for $f \in S$ has the structure morphism $\lambda_{\Lin_{(d,c)}}^{\{\finverse\}}$ as a two-sided homotopy inverse via specified homotopies.  Furthermore, the homotopies $\lambda^{\{f,\finverse\}}_{\Lin_{(c,d,c)}}$ and $\lambda^{\{\finverse,f\}}_{\Lin_{(d,c,d)}}$ are specified structure morphisms of the homotopy algebraic quantum field theory.  In other words, each two-sided homotopy inverse and the corresponding homotopies are already encoded in the Boardman-Vogt construction $\wocsinvbarm$ of $\Ocsinvbarm$.\dqed
\end{interpretation}

\begin{example}[Homotopy time-slice in homotopy locally covariant QFT]\label{ex:hts-hlcqft}
In the context of Example \ref{ex:hcdiagram-lcqft}, every\index{homotopy locally covariant quantum field theory!homotopy time-slice axiom} $\wom_{\Locdsinvbar}$-algebra satisfies the homotopy time-slice axiom in the sense of Theorem \ref{thm:homotopy-timeslice}.\dqed
\end{example}

\begin{example}[Homotopy time-slice in homotopy locally covariant QFT on a fixed spacetime]\label{ex:hts-hlcqft-fixed}
In the context of Example \ref{ex:hcdiagram-lcqft-fixed}, every\index{homotopy locally covariant quantum field theory on a spacetime!homotopy time-slice axiom}  $\wom_{\Ghxsinvbar}$-algebra satisfies the homotopy time-slice axiom in the sense of Theorem \ref{thm:homotopy-timeslice}.\dqed
\end{example}

\begin{example}[Homotopy time-slice in homotopy dynamical quantum gauge theories on principal bundles]\label{ex:hts-bundle}
In the context of Example \ref{ex:hcdiagram-dynamical}, every\index{homotopy dynamical quantum gauge theory!homotopy time-slice axiom}  $\wom_{\Bglocsginvbar}$-algebra satisfies the homotopy time-slice axiom in the sense of Theorem \ref{thm:homotopy-timeslice}.\dqed
\end{example}

\begin{example}[Homotopy time-slice in homotopy charged matter QFT on background gauge fields]\label{ex:hts-charged}
In the context of Example \ref{ex:hcdiagram-charged}, every\index{homotopy charged matter quantum field theory!homotopy time-slice axiom}  $\wom_{\Bgconlocsginvbar}$-algebra satisfies the homotopy time-slice axiom in the sense of Theorem \ref{thm:homotopy-timeslice}.\dqed
\end{example}

\begin{example}[Homotopy time-slice in homotopy Dirac quantum fields]\label{ex:hts-dirac}
In the context of Example \ref{ex:hcdiagram-dirac}, every\index{homotopy Dirac quantum field theory!homotopy time-slice axiom}  $\wom_{\Slocdsinvbar}$-algebra satisfies the homotopy time-slice axiom in the sense of Theorem \ref{thm:homotopy-timeslice}.\dqed
\end{example}

\begin{example}[Homotopy time-slice in homotopy QFT on structured spacetimes]\label{ex:hts-structured-spacetime}
In the context of Example \ref{ex:hcdiagram-structured-spacetime}, every\index{homotopy  quantum field theory on structured spacetime!homotopy time-slice axiom}  $\wom_{\Strsinvbar}$-algebra satisfies the homotopy time-slice axiom in the sense of Theorem \ref{thm:homotopy-timeslice}.\dqed
\end{example}

\begin{example}[Homotopy time-slice in homotopy QFT on spacetime with timelike boundary]\label{ex:hts-boundary}
In the context of Example \ref{ex:hcdiagram-timelike-boundary}, every\index{homotopy quantum field theory on spacetime with timelike boundary!homotopy time-slice axiom}  $\wom_{\Regxsinvbar}$-algebra satisfies the homotopy time-slice axiom in the sense of Theorem \ref{thm:homotopy-timeslice}.\dqed
\end{example}

\section{Objectwise $A_\infty$-Algebra}\label{sec:objective-ainfinity}

In this section, we explain that each homotopy algebraic quantum field theory on an orthogonal category is objectwise an $A_\infty$-algebra.

\begin{motivation}
Recall from Example \ref{ex:operad-as} that there is a $1$-colored operad $\As$ whose algebras are precisely monoids in $\M$.  In Section \ref{sec:ainfinity-algebra} we observed that algebras over the Boardman-Vogt construction $\Was$, called $A_\infty$-algebras, are in a precise sense monoids up to coherent higher homotopies.  An algebraic quantum field theory $\fraka$ on an orthogonal category $\Cbar$ is, first of all, a functor $\fraka : \C \to \Monm$.  So for each object $c \in \C$, $\fraka(c)$ is a monoid in $\M$.  Therefore, it is reasonable to expect that a homotopy algebraic quantum field theory on $\Cbar$ is objectwise an $A_\infty$-algebra in $\M$.\dqed
\end{motivation}

In the following result, we assume the unique color for the $1$-colored operad $\As$ is $*$.    Recall from Definition \ref{def:aqft-operad} the set $\Ocbar\duc$.  We will write $\tensorunit_x$ for a copy of $\tensorunit$ indexed by an element $x$.

\begin{theorem}\label{thm:haqft-entrywise-ainfinity}
Suppose $\Cbar = (\C,\perp)$ is an orthogonal category, and $c \in \C$.
\begin{enumerate}
\item There is an operad morphism \[j_c : \As \to \Ocbarm\] that sends $*$ to $c$ and is defined entrywise by the commutative diagrams \[\nicexy{\tensorunit_{\sigma} \ar[r]^-{=} \ar[d]_-{\sigma~ \mathrm{summand}} & \tensorunit_{[\sigma,\{\Id_c\}_{i=1}^n]} \ar[d]^-{[\sigma,\{\Id_c\}_{i=1}^n]~ \mathrm{summand}} \\
\As(n) = \coprodover{\sigma \in \Sigma_n} \tensorunit \ar[r]^-{j_c} & \coprodover{\Ocbar\ccc}\tensorunit =\Ocbarm\ccc}\]
for $n \geq 0$ and $\sigma \in \Sigma_n$, where $(c,\ldots,c)$ has $n$ copies of $c$.
\item The operad morphism $j_c$ induces a change-of-operad adjunction \[\nicexy@C+.4cm{\algm\bigl(\Was\bigr) \ar@<2pt>[r]^-{(\W j_c)_!} & \algm\bigl(\wocbarm\bigr) = \HQFT(\Cbar) \ar@<2pt>[l]^-{(\W j_c)^*}}\] between the category of\index{ainfinityalgebra@$A_\infty$-algebra} $A_\infty$-algebras in $\M$ and the category of\index{homotopy algebraic quantum field theory!objectwise $A_\infty$-algebra} homotopy algebraic quantum field theories on $\Cbar$.
\end{enumerate}
\end{theorem}

\begin{proof}
It follows directly from Example \ref{ex:operad-as} and Definition \ref{def:aqft-operad} that $j_c$ is a well-defined operad morphism from the $\{*\}$-colored operad $\As$ to the $\Obc$-colored operad $\Ocbarm$.  Assertion (2) follows from assertion (1), the naturality of the Boardman-Vogt construction in Proposition \ref{prop:w-functor}, and Theorem \ref{thm:change-operad}.
\end{proof}

\begin{interpretation}
For each $c \in \colorc = \Obc$, the right adjoint $(\W j_c)^*$ sends each homotopy algebraic quantum field theory $(X,\lambda)$ on the orthogonal category $\Cbar$ to the $A_\infty$-algebra $X_c$.  Explicitly, for each $n \geq 0$, $T \in \uTree^{\{*\}}(n)$, and $\{\sigma_v\} \in \prod_{v\in T} \Sigma_{|\inp(v)|}$, the structure morphism 
\[\nicexy@C+.7cm{\J[T] \otimes \bigl[(\W j_c)^*X\bigr]^{\otimes n} = \J[T] \otimes X_c^{\otimes n} \ar[r]^-{\lambda_T^{\{\sigma_v\}_{v\in T}}} & X_c = (\W j_c)^*X}\]
in \eqref{ainfinity-structure} of the $A_\infty$-algebra $(\W j_c)^*X$ is given by the structure morphism
\begin{equation}\label{entrywise-ainfinity}
\lambda_{T_c}^{\bigl\{[\sigma_v,\{\Id_c\}_{i=1}^{|\inp(v)|}]\bigr\}_{v\in T_c}} \withspace [\sigma_v,\{\Id_c\}_{i=1}^{|\inp(v)|}] \in \Ocbar\ccc
\end{equation} 
in \eqref{haqft-restricted}.  Here $T_c \in \uTreec\ccc$ is the $\colorc$-colored tree obtained from $T$ by switching each of its edge color from $*$ to $c$.\dqed
\end{interpretation}

\begin{remark}\label{rk:nonassociative-qft}
We speculate that the objectwise $A_\infty$-algebra structure in each homotopy algebraic quantum field theory may be related to non-associative quantum field theory \cite{dzh} and non-associative gauge theory \cite{majid,medeiros,okubo,ram}.\dqed  
\end{remark}

\begin{example}[Objectwise $A_\infty$-structure in homotopy chiral conformal QFT]\label{ex:ainfinity-hccqft}
In the context of Example \ref{ex:homotopy-chiral}, every\index{homotopy chiral conformal quantum field theory!objectwise $A_\infty$-algebra}  $\wom_{\Mandbar}$-algebra has an $A_\infty$-algebra structure in each color in the sense of Theorem \ref{thm:haqft-entrywise-ainfinity}.\dqed
\end{example}

\begin{example}[Objectwise $A_\infty$-structure in homotopy chiral conformal QFT on discs]\label{ex:ainfinity-hccqft-interval}
In the context of Example \ref{ex:hccqft-int}, every\index{homotopy chiral conformal quantum field theory on discs!objectwise $A_\infty$-algebra} $\wom_{\Discdbar}$-algebra has an $A_\infty$-algebra structure in each color in the sense of Theorem \ref{thm:haqft-entrywise-ainfinity}.\dqed
\end{example}

\begin{example}[Objectwise $A_\infty$-structure in homotopy chiral conformal QFT on a fixed manifold]\label{ex:ainfinity-hccqft-manifold}
In the context of Example \ref{ex:hccqft}, every\index{homotopy chiral conformal quantum field theory on a manifold!objectwise $A_\infty$-algebra} $\wom_{\Openxbar}$-algebra has an $A_\infty$-algebra structure in each color in the sense of Theorem \ref{thm:haqft-entrywise-ainfinity}.\dqed
\end{example}

\begin{example}[Objectwise $A_\infty$-structure in homotopy Euclidean QFT]\label{ex:ainfinity-euclidean}
In the context of Example \ref{ex:homotopy-euclidean}, every\index{homotopy Euclidean quantum field theory!objectwise $A_\infty$-algebra} $\wom_{\Riemdbar}$-algebra has an $A_\infty$-algebra structure in each color in the sense of Theorem \ref{thm:haqft-entrywise-ainfinity}.\dqed
\end{example}

\begin{example}[Objectwise $A_\infty$-structure in homotopy locally covariant QFT]\label{ex:ainfinity-hlcqft}
In the context of Example \ref{ex:homotopy-lcqft}, every\index{homotopy locally covariant quantum field theory!objectwise $A_\infty$-algebra} $\wom_{\Locdsinvbar}$-algebra has an $A_\infty$-algebra structure in each color in the sense of Theorem \ref{thm:haqft-entrywise-ainfinity}.\dqed
\end{example}

\begin{example}[Objectwise $A_\infty$-structure in homotopy locally covariant QFT on a fixed spacetime]\label{ex:ainfinity-hlcqft-fixed}
In the context of Example \ref{ex:hcausal-nets}, every\index{homotopy locally covariant quantum field theory on a spacetime!objectwise $A_\infty$-algebra}  $\wom_{\Ghxsinvbar}$-algebra has an $A_\infty$-algebra structure in each color in the sense of Theorem \ref{thm:haqft-entrywise-ainfinity}.\dqed
\end{example}

\begin{example}[Objectwise $A_\infty$-structure in homotopy dynamical quantum gauge theories on principal bundles]\label{ex:ainfinity-dynamical}
In the context of Example \ref{ex:hdynamical-gauge}, \index{homotopy dynamical quantum gauge theory!objectwise $A_\infty$-algebra} every\\ $\wom_{\Bglocsginvbar}$-algebra has an $A_\infty$-algebra structure in each color in the sense of Theorem \ref{thm:haqft-entrywise-ainfinity}.\dqed
\end{example}

\begin{example}[Objectwise $A_\infty$-structure in homotopy charged matter QFT on background gauge fields]\label{ex:ainfinity-charged}
In the context of Example \ref{ex:hcharged-matter}, every\index{homotopy charged matter quantum field theory!objectwise $A_\infty$-algebra}  $\wom_{\Bgconlocsginvbar}$-algebra has an $A_\infty$-algebra structure in each color in the sense of Theorem \ref{thm:haqft-entrywise-ainfinity}.\dqed
\end{example}

\begin{example}[Objectwise $A_\infty$-structure in homotopy Dirac QFT]\label{ex:ainfinity-dirac}
In the context of Example \ref{ex:hdirac}, every\index{homotopy Dirac quantum field theory!objectwise $A_\infty$-algebra}  $\wom_{\Slocdsinvbar}$-algebra has an $A_\infty$-algebra structure in each color in the sense of Theorem \ref{thm:haqft-entrywise-ainfinity}.\dqed
\end{example}

\begin{example}[Objectwise $A_\infty$-structure in homotopy QFT on structured spacetimes]\label{ex:ainfinity-structured}
In the context of Example \ref{ex:hqft-structured}, every\index{homotopy quantum field theory on structured spacetime!objectwise $A_\infty$-algebra} $\wom_{\Strsinvbar}$-algebra has an $A_\infty$-algebra structure in each color in the sense of Theorem \ref{thm:haqft-entrywise-ainfinity}.\dqed
\end{example}

\begin{example}[Objectwise $A_\infty$-structure in homotopy AQFT on spacetime with timelike boundary]\label{ex:ainfinity-boundary}
In the context of Example \ref{ex:haqft-boundary}, every\index{homotopy quantum field theory on spacetime with timelike boundary!objectwise $A_\infty$-algebra} $\wom_{\Regxsinvbar}$-algebra has an $A_\infty$-algebra structure in each color in the sense of Theorem \ref{thm:haqft-entrywise-ainfinity}.\dqed
\end{example}

\section{Homotopy Coherent Diagrams of $A_\infty$-Algebras}\label{sec:hcdiag-ainfinity}

In Section \ref{sec:h-functoriality} we explained that each homotopy algebraic quantum field theory on an orthogonal category $\Cbar$ has the structure of a homotopy coherent $\C$-diagram in $\M$.  Moreover, in Section \ref{sec:objective-ainfinity} we observed that each entry of a homotopy algebraic quantum field theory on $\Cbar$ has the structure of an $A_\infty$-algebra.  In this section, we explain how the homotopy coherent $\C$-diagram structure and the objectwise $A_\infty$-algebras in a homotopy algebraic quantum field theory are compatible with each other.

\begin{motivation}
For an orthogonal category $\Cbar$, an algebraic quantum field theory is, first of all, a functor $\fraka : \C \to \Monm$, i.e., a $\C$-diagram of monoids in $\M$.  For each morphism $g : c \to d$ in $\C$, its image $\fraka(g) : \fraka(c) \to \fraka(d)$ is a morphism of monoids in $\M$.  So it respects the multiplications and the units in the sense that the diagrams \[\nicexy{(\fraka(c))^{\otimes p} \ar[d]_-{\bigotimes \fraka(g)} \ar[r]^-{\mu} & \fraka(c) \ar[d]^-{\fraka(g)}\\ (\fraka(d))^{\otimes p} \ar[r]^-{\mu} & \fraka(d)}\qquad
\nicexy{\tensorunit \ar[r]^-{\operadunit_{\fraka(c)}} \ar[dr]_{\operadunit_{\fraka(d)}} & \fraka(c) \ar[d]^-{\fraka(g)} \\ & \fraka(d)}\] 
are commutative for all $p \geq 2$.  In other words, the $\C$-diagram structure and the monoid structure commute.  When these structures are replaced by a homotopy coherent $\C$-diagram and objectwise $A_\infty$-algebras, we should expect the structures to commute up to specified homotopies.  The homotopy coherent version is necessarily more involved because the structure morphisms in a homotopy coherent $\C$-diagram \eqref{haqft-hcdiagram} are indexed by linear graphs, while those in an $A_\infty$-algebra \eqref{entrywise-ainfinity} are indexed by trees.\dqed\end{motivation}

Recall the minimal orthogonal category $\Cbarmin=(\C,\varnothing)$ in Example \ref{ex:empty-causality}.

\begin{corollary}\label{cor:haqft-hcdiag-ainfinity}
Suppose $\C$ is a small category with object set $\colorc$.
\begin{enumerate}\item There is an equality \[\Ocbarminm=\Ocm \in \Operadcm,\] where:
\begin{itemize}\item $\Ocbarminm$ is the image in $\Operadcm$ of $\Ocbarmin$ as in Definition \ref{def:aqft-operad}.
\item $\Ocm$ is the $\colorc$-colored operad in Example \ref{ex:diag-monoid-operad}.
\end{itemize}
\item Suppose $\Cbar=(\C,\perp)$ is an orthogonal category.  Then the identity functor on $\C$ defines an orthogonal functor \[\nicexy{\Cbarmin=(\C,\varnothing) \ar[r]^-{i_0} & \Cbar}\] and  induces a diagram of\index{homotopy algebraic quantum field theory!homotopy coherent diagram of $A_\infty$-algebras}\index{homotopy coherent diagram!$A_\infty$-algebra} change-of-operad adjunctions
\[\nicexy@C+.7cm@R+.3cm{\algmwocm \ar@<2pt>[r]^-{(\wom_{i_0})_!} \ar@<-2pt>[d]_-{\eta_!} 
& \algmwocbarm = \HQFT(\Cbar) \ar@<2pt>[l]^-{(\wom_{i_0})^*} \ar@<-2pt>[d]_-{\eta_!} \\
\Monmc \cong \algmocm \ar@<2pt>[r]^-{(\Otom_{i_0})_!} \ar@<-2pt>[u]_-{\eta^*}  
& \algmocbarm \cong \QFT(\Cbar) \ar@<2pt>[l]^-{(\Otom_{i_0})^*} \ar@<-2pt>[u]_-{\eta^*}}\]
with commuting left adjoint diagram and commuting right adjoint diagram.
\end{enumerate}
\end{corollary}

\begin{proof} The first assertion follows from the definition of $\Ocm$ in Example \ref{ex:diag-monoid-operad} and Definition \ref{def:aqft-operad} of $\Ocbarminm$, where the equivalence relation $\sim$ is trivial for the empty orthogonality relation.  The second assertion follows from the first assertion, Example \ref{ex:operad-set-m}, Corollary \ref{cor:wf-eta-adjunction}, and Theorem \ref{thm:ocbar-algebra}(2).
\end{proof}

\begin{interpretation} Via the right adjoint $(\wom_{i_0})^*$, every homotopy algebraic quantum field theory $(X,\lambda)$ has an underlying $\wocm$-algebra, i.e., homotopy coherent $\C$-diagram of $A_\infty$-algebras as in Definition \ref{def:wocm-algebra}.  Therefore, by Corollary \ref{cor:ass-ocm-adjunction}, $(X,\lambda)$ has an underlying homotopy coherent $\C$-diagram in $\M$ and an objectwise $A_\infty$-algebra structure.  These are the structures in Theorem \ref{thm:cdiag-ocbar} and Theorem \ref{thm:haqft-entrywise-ainfinity}.\dqed
\end{interpretation}

To explain the compatibility between these two structures precisely, we need the following notations.  Recall our convention that, for a vertex $v$ in a $\colorc$-colored tree, we often abbreviate its profile $\profofv \in \Profcc$ to  $(v)$.  Using the canonical bijection in Example \ref{ex:ocbar-unary}, for a morphism $f \in \C(c,d)$, we will abbreviate an element $[\id_1,f] \in \Ocbar\dc$ to just $f$.

\begin{assumption}\label{assumption:hcdiag-ainfinity}
Suppose $\Cbar = (\C,\perp)$ is an orthogonal category with object set $\colorc$.  Suppose:
\begin{itemize}\item $T_c \in \uTree^{\{c\}}\ccc$ is a $c$-colored tree for some color $c \in \colorc$, where $(c,\ldots,c) \in \Profc$ has $p \geq 0$ copies of $c$.  
\item $T_d \in \uTree^{\{d\}}\ddd$ is the $d$-colored tree obtained from $T_c$ by replacing every edge color by $d$.  
\item $\uc=\bigl(c=c_0,c_1,\ldots,c_n=d\bigr) \in \Profc$ is a profile with $n \geq 1$.
\item $L = \Lin_{\uc} \in \uTreec\dc$ is the linear graph for the profile $\uc$.  
\end{itemize}
Define the graftings\index{grafting} \[\begin{split}T^1 &= \graft(L;T_c) \in \uTreec\dcc,\\
T^2 &= \graft\bigl(T_d;\underbrace{\Lin_{(c,d)},\ldots,\Lin_{(c,d)}}_{\text{$p$ copies}}\bigr) \in \uTreec\dcc.\end{split}\]

We may visualize $T^1$ (on the left) and $T^2$ (on the right) as follows.
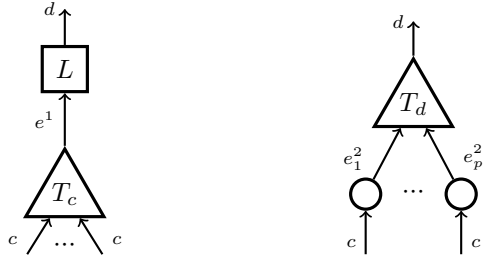
\begin{center}\begin{tikzpicture}
\node [rectplain] (L) {$L$}; \node [triangular, below=.7cm of L] (T) {$T_c$}; 
\node [below=.2cm of T] () {$\cdots$};
\draw [outputleg] (L) to node[at end]{\scriptsize{$d$}} +(0,.8cm);
\draw [arrow] (T) to node{\scriptsize{$e^1$}} (L);
\draw [inputleg] (T) to node[at end, swap]{\scriptsize{$c$}} +(-.5cm,-.8cm);
\draw [inputleg] (T) to node[at end]{\scriptsize{$c$}} +(.5cm,-.8cm);
\node [right=4cm of T] (dots) {$\cdots$};
\node [triangular, above=.7cm of dots] (Td) {$T_d$};
\node [smallplain, left=.15cm of dots] (L2) {}; 
\node [smallplain, right=.15cm of dots] (L3) {};
\draw [outputleg] (Td) to node[at end]{\scriptsize{$d$}} +(0,1.1cm);
\draw [arrow] (L2) to node[at start]{\scriptsize{$e^2_1$}} (Td);
\draw [arrow] (L3) to node[at start, swap]{\scriptsize{$e^2_p$}} (Td);
\draw [inputleg] (L2) to node[near end, swap]{\scriptsize{$c$}} +(0cm,-.8cm);
\draw [inputleg] (L3) to node[near end]{\scriptsize{$c$}} +(0cm,-.8cm);
\end{tikzpicture}
\end{center}
In $T^1$, the output flag of the $c$-colored tree $T_c$ is grafted with the input leg of $L$.  The $c$-colored internal edge in $T_1$ that extends from $T_c$ to $L$, which is created by the grafting, is denoted by $e^1$.  In $T_2$, the output flag of a copy of $\Lin_{(c,d)}$ is grafted with each of the $p$ input legs of $T_d$.  The $d$-colored internal edge that extends from the $k$th copy of $\Lin_{(c,d)}$ (from the left) to $T_d$ is denoted by $e^2_k$.  Note that if $T_c$ is a corolla with $p=0$, then 
\begin{equation}\label{T-pzero}
\begin{tikzpicture} 
\node [smallplain] (Tc) {}; \draw [outputleg] (Tc) to node[near end]{\scriptsize{$c$}} +(0,.6cm);
\node[left=.2cm of Tc] () {$T_c =$};
\node[right=.2cm of Tc] () {$\in \uTree^{\{c\}}\cempty$,};
\node [smallplain, right=5cm of Tc] (Td) {}; 
\draw [outputleg] (Td) to node[near end]{\scriptsize{$d$}} +(0,.6cm);
\node[left=.2cm of Td] () {$T^2 = T_d =$};
\node[right=.2cm of Td] () {$\in \uTree^{\{d\}}\dempty$.};
\end{tikzpicture}
\end{equation}
So in this case the grafting in $T^2$ is trivial because $T_d$ has no input legs.

Suppose \[\{g_j\} \in \prod_{j=1}^n \C(c_{j-1},c_j) \cong \prod_{j=1}^n \Ocbar\sbinom{c_j}{c_{j-1}}\] is a sequence of $n$ composable morphisms in $\C$.  For each vertex $v$ in $T_c$, suppose
\begin{equation}\label{ftov}
f_v = \bigl[\sigma_v, \{f_v^i\}_{i=1}^{|\inp(v)|}\bigr] \in \Ocbar(v) = \Ocbar\ccc
\end{equation}
in which $(c,\ldots,c)$ has $|\inp(v)|$ copies of $c$, $\sigma_v \in \Sigma_{|\inp(v)|}$, and each $f^i_v \in \C(c,c)$.  Define \[q_v = \bigl[\sigma_v, \{\Id_d\}_{i=1}^{|\inp(v)|}\bigr] \in \Ocbar\ddd,\] in which $(d,\ldots,d)$ has $|\inp(v)|$ copies of $d$.  Note that if 
\begin{equation}\label{gammat1}
\gamma^{\Ocbar}_{T_1}\Bigl(\{g_j\}_{j=1}^n, \{f_v\}_{v\in T_c}\Bigr) = \bigl[\sigma, \{h_i\}_{i=1}^p\bigr] \in \Ocbar\dcc
\end{equation}
for some permutation $\sigma \in \Sigma_p$ and morphisms $h_i \in\C(c,d)$, then we also have
\begin{equation}\label{gammat1t2}
\gamma^{\Ocbar}_{T_1}\Bigl(\{g_j\}_{j=1}^n, \{f_v\}_{v\in T_c}\Bigr) = 
\gamma^{\Ocbar}_{T_2} \Bigl(\{q_v\}_{v\in T_d}, \{h_i\}_{i=1}^p\Bigr),
\end{equation}
where by our notational convention $\{h_i\} \in \prod_{i=1}^p \Ocbar\dc$.  This equality follows from the fact that each $g_j$ is associated with the trivial permutation $\id_1 \in \Sigma_1$.
\end{assumption}

The following result is the homotopy coherent compatibility between the homotopy coherent $\C$-diagram structure and the objectwise $A_\infty$-algebra structure in a homotopy algebraic quantum field theory.  A copy of the morphism $1 : \tensorunit \to J$ corresponding to an internal edge $e$ will be denoted by $1_e$ below.  To simplify the notation, we will omit writing some of the identity morphisms below.

\begin{theorem}\label{thm:hcdiag-ainfinity}
Suppose $(X,\lambda)$ is a homotopy algebraic quantum field theory on $\Cbar$ in the setting of Assumption \ref{assumption:hcdiag-ainfinity}.  Then the diagram
\[\begin{footnotesize}\nicexy@C-.3cm{\tensorunit\otimes \J[L] \otimes \J[T_c]\otimes X_c^{\otimes p} \ar[d]_-{1_{e^1}} & \J[L] \otimes \J[T_c]\otimes X_c^{\otimes p} \ar[l]_-{\cong} \ar[rr]^-{\lambda_{T_c}^{\{f_v\}}} && \J[L]\otimes X_c \ar[d]^-{\lambda_L^{\{g_j\}}}\\
\J[T_1]\otimes X_c^{\otimes p} \ar[rrr]^-{\lambda_{T_1}^{\{g_j\}_{j=1}^n,\{f_v\}}} &&& X_d \ar@{=}[d]\\
\tensorunit^{\otimes |T_1|} \otimes X_c^{\otimes p} \cong \tensorunit^{\otimes |T_2|} \otimes X_c^{\otimes p} \ar[u]^-{0^{\otimes |T_1|}} \ar[d]_-{0^{\otimes |T_2|}}
& \J[\Cor_{(\uc;d)}] \otimes X_c^{\otimes p} \ar[l]_-{\cong} \ar[rr]^-{\lambda_{\Cor_{(\uc;d)}}^{[\sigma,\{h_i\}_{i=1}^p]}} && X_d \ar@{=}[d]\\
\J[T_2] \otimes X_c^{\otimes p} \ar[rrr]^-{\lambda_{T_2}^{\{q_v\},\{h_i\}_{i=1}^p}} &&& X_d\\
\J[T_d] \otimes \tensorunit^{\otimes p} \otimes X_c^{\otimes p}
\ar[u]^-{\bigotimes\limits_{i=1}^p 1_{e^2_i}}
& \J[T_d] \otimes \bigotimes\limits_{i=1}^p \bigl[\J[\Lin_{(c,d)}]\otimes X_c\bigr] \ar[l]_-{\cong} \ar[rr]^-{\bigtensorover{i} \lambda_{\Lin_{(c,d)}}^{\{h_i\}}}
&& \J[T_d] \otimes X_d^{\otimes p} \ar[u]_-{\lambda_{T_d}^{\{q_v\}}}}\end{footnotesize}\]
is commutative, where the morphisms $0,1 : \tensorunit \to J$ are part of the commutative segment $J$.
\end{theorem}

\begin{proof}
This is a consequence of the Coherence Theorem \ref{thm:haqft-coherence} for homotopy algebraic quantum field theories.  Indeed, in the above diagram from top to bottom:
\begin{itemize}
\item The top rectangle is commutative by the associativity condition \eqref{haqft-ass} and the grafting definition of $T_1$.
\item The second rectangle is commutative by \eqref{gammat1} and the wedge condition \eqref{haqft-wedge} applied to the tree substitution $T_1 = \Cor_{(\uc;d)}(T_1)$.
\item The third rectangle is commutative by \eqref{gammat1t2} and the wedge condition \eqref{haqft-wedge} applied to the tree substitution $T_2 = \Cor_{(\uc;d)}(T_2)$.
\item The bottom rectangle  is commutative by the associativity condition \eqref{haqft-ass} and the grafting definition of $T_2$.
\end{itemize}
\end{proof}

Theorem \ref{thm:hcdiag-ainfinity} applies to all the homotopy algebraic quantum field theories in Section \ref{sec:example-haqft}.  In the following examples, we will explain some special cases of Theorem \ref{thm:hcdiag-ainfinity}.

\begin{example}[Homotopy compatibility]\label{ex:h-compatibility}
In \eqref{ftov} suppose each $f_v^i = \Id_c$.  Then  \[\begin{split}f_v &= \bigl[\sigma_v, \{\Id_c\}_{i=1}^{|\inp(v)|}\bigr] \in\Ocbar\ccc \forspace v \in \Vt(T_c),\\ h_i &= g_n\circ\cdots\circ g_1 \in \C(c,d)\cong \Ocbar\dc \forspace 1 \leq i \leq p.\end{split}\]
In the commutative diagram in Theorem \ref{thm:hcdiag-ainfinity}:
\begin{itemize}
\item The structure morphism \[\nicexy@C+.3cm{\J[T_c]\otimes X_c^{\otimes p} \ar[r]^-{\lambda_{T_c}^{\{f_v\}}} & X_c}\] is part of the $A_\infty$-algebra $X_c$ as explained in \eqref{entrywise-ainfinity}.
\item Similarly, the structure morphism \[\nicexy@C+.3cm{\J[T_d] \otimes X_d^{\otimes p} \ar[r]^-{\lambda_{T_d}^{\{q_v\}}} & X_d}\] is part of the $A_\infty$-algebra $X_d$.
\item The structure morphisms \[\nicexy{\J[L]\otimes X_c \ar[r]^-{\lambda_L^{\{g_j\}}} & X_d}
\andspace \nicexy@C+.3cm{\J[\Lin_{(c,d)}] \otimes X_c \ar[r]^-{\lambda_{\Lin_{(c,d)}}^{\{h_i\}}} & X_d}\]
for $1 \leq i \leq p$ are part of the homotopy coherent $\C$-diagram structure of $(X,\lambda)$ as explained in \eqref{haqft-hcdiagram}.
\end{itemize}
Therefore, the entire commutative diagram in Theorem \ref{thm:hcdiag-ainfinity} is expressing an up-to-homotopy compatibility between the homotopy coherent $\C$-diagram structure and the objectwise $A_\infty$-algebra structure via specified homotopies in each homotopy algebraic quantum field theory.\dqed 
\end{example}

\begin{example}[Homotopy preservation of homotopy units]\label{ex:homotopy-pres-hunits}
In the context of Example \ref{ex:h-compatibility}, suppose further that $T_c=\Cor_{(\varnothing;c)}$ with $p=0$ as in \eqref{T-pzero} and that $n=1$, so $L = \Lin_{(c,d)}$.  Then the commutative diagram in Theorem \ref{thm:hcdiag-ainfinity} is a precise version of the statement that, for $g \in \C(c,d)$, the diagram \[\nicexy{\tensorunit \ar[r]^-{\mu_0^c} \ar[dr]_-{\mu_0^d} & X_c \ar[d]^-{X_{g}}\\ & X_d}\]
is commutative up to homotopy.  Here $X_{g}$ is the notation in Example \ref{ex1:hcdiagram} for a homotopy coherent $\C$-diagram.  Similarly, $\mu_0^c$ and $\mu_0^d$ are the two-sided homotopy units\index{homotopy unit} in the $A_\infty$-algebras $X_c$ and $X_d$, respectively, as explained in Examples \ref{ex2:ainfinity} and \ref{ex2.5:ainfinity}.  Therefore, in this case Theorem \ref{thm:hcdiag-ainfinity} says that, in each homotopy algebraic quantum field theory, the homotopy coherent $\C$-diagram structure preserves the two-sided homotopy units in the objectwise $A_\infty$-algebras up to specified homotopies.\dqed
\end{example}

\begin{example}[Homotopy preservation of multiplication]\label{ex:homotopy-pres-mult}
In the context of Example \ref{ex:h-compatibility}, suppose further that \[T_c=\Cor_{(c,\ldots,c;c)}\] is a $c$-colored corolla with $p$ inputs and that $n=1$, so $L = \Lin_{(c,d)}$.  Then the commutative diagram in Theorem \ref{thm:hcdiag-ainfinity} is a precise version of the statement that, for $g \in \C(c,d)$, the diagram \[\nicexy{X_c^{\otimes p} \ar[r]^-{\mu_p^c} \ar[d]_-{\bigtensorover{i} X_{g}} & X_c \ar[d]^-{X_{g}}\\ X_d^{\otimes p} \ar[r]^-{\mu_p^d} & X_d}\] is commutative up to homotopy.  Here $\mu_p^c$ and $\mu_p^d$ are the multiplications in the $A_\infty$-algebras $X_c$ and $X_d$, respectively, as in Example \ref{ex1:ainfinity}.  Therefore, in this case Theorem \ref{thm:hcdiag-ainfinity} says that, in each homotopy algebraic quantum field theory, the homotopy coherent $\C$-diagram structure preserves the multiplications in the objectwise $A_\infty$-algebras up to specified homotopies.\dqed
\end{example}

\chapter{Prefactorization Algebras}\label{ch:pfa}

In this chapter, we define prefactorization algebras on a configured category as algebras over a suitable colored operad and observe their basic structure.  In Section \ref{sec:costello} we briefly review prefactorization algebras on a topological space in the original sense of Costello-Gwilliam \cite{cg}.  Configured categories are abstractions of the category $\Openx$ for a topological space $X$ and are defined in Section \ref{sec:configured-cat}.  The colored operad and prefactorization algebras associated to a configured category are defined in Section \ref{sec:pfa-operad}.  The coherence theorems for prefactorization algebras, with or without the time-slice axiom, are also recorded in that section.

In Section \ref{sec:pfa-pointed-diagram} it is shown that every prefactorization algebra has an underlying pointed diagram.  In Section \ref{sec:pfa-com-monoid} we observe that some entries of a prefactorization algebra are equipped with the structure of a commutative monoid.  This applies, in particular, to the $0$-entry of each prefactorization algebra on the configured category of a bounded lattice with least element $0$.  In Section \ref{sec:pfa-module-com}, we show that, for each prefactorization algebra $Y$ on the configured category of a bounded lattice, the commutative monoid $Y_0$ acts on every other entry, and the underlying diagram is a diagram of left $Y_0$-modules.

In Section \ref{sec:diag-com-monoid} we show that every diagram of commutative monoids can be realized as a prefactorization algebra.  In Sections \ref{sec:pfa-pointed-diagram} and \ref{sec:diag-com-monoid} we also give evidence that prefactorization algebras and algebraic quantum field theories are closely related.  A detailed study of their relationship is the subject of Chapter \ref{ch:comparing}.  In Section \ref{sec:config-equivalence} we show that equivalences of configured categories yield equivalent and Quillen equivalent categories of prefactorization algebras.

Throughout this chapter, $(\M,\otimes,\tensorunit)$ is a fixed cocomplete symmetric monoidal closed category, such as $\Vectk$ and $\Chaink$.

\section{Costello-Gwilliam Prefactorization Algebras}\label{sec:costello}

Prefactorization algebras and their variants in the sense of Costello-Gwilliam \cite{cg} provide a mathematical framework for quantum field theories that is analogous to deformation quantization in quantum mechanics.  In \cite{cg} 3.1.1 a \emph{prefactorization algebra}\index{prefactorization algebra!Costello-Gwilliam} on a topological space $X$ valued in $\M$ is defined as a functor \[\scF : \Openx \to \M\] that functorially assigns to each open subset $U \subset X$ an object $\scF(U) \in \M$.  If $U_1,\ldots,U_n \subset V \in \Openx$ are pairwise disjoint open subsets of $V$, then $\scF$ is also equipped with a structure morphism \[\nicexy@C+.7cm{\scF(U_1) \otimes \cdots \otimes \scF(U_n) \ar[r]^-{\scF^V_{U_1,\ldots,U_n}} & \scF(V)} \in \M.\]  These structure morphisms are required to satisfy some natural associativity, unity, and equivariance axioms.  

In particular, if $\varnothing_X \subset X$ denotes the empty subset, then $\scF(\varnothing_X)$ is equipped with an associative and commutative multiplication \[\nicexy@C+.5cm{\scF(\varnothing_X) \otimes \scF(\varnothing_X) \ar[r]^-{\scF^{\varnothing_X}_{\varnothing_X,\varnothing_X}} & \scF(\varnothing_X)}.\]  If this multiplicative structure is equipped with a two-sided unit, making $\scF(\varnothing_X)$ into a commutative monoid in $\M$, then $\scF$ is called a \emph{unital prefactorization algebra on $X$}.

Physically $X$ is the spacetime of interest.  A prefactorization algebra $\scF$ assigns to each open subset $U \subset X$ an object $\scF(U)$ of quantum observables.  For an inclusion $U \subset V$ of open subsets of $X$, the structure morphism \[\nicexy@C+.5cm{\scF(U) \ar[r]^-{\scF^V_U} & \scF(V)} \in \M\] sends observables in $U$ to observables in $V$.  If the open subsets $U_1,\ldots,U_n \subset V$ are suitably disjoint, then we can combine the observables in the $U_i$'s via the structure morphism $\scF^V_{U_1,\ldots,U_n}$.  

There is also a $G$-equivariant analogue of prefactorization algebras when the topological space $X$ is equipped with an action by a group $G$.  In this case, the category $\Openx$ is replaced by its $G$-equivariant analogue $\Openxg$ in Example \ref{ex:eq-space}.   As defined in \cite{cg} 3.7.1.1, a \emph{$G$-equivariant prefactorization algebra on $X$}\index{prefactorization algebra!equivariant}\index{equivariant prefactorization algebra} is a prefactorization algebra on $X$ defined by a functor \[\scF : \Openxg \to \M,\] so now there are structure isomorphisms \[\nicexy@C+.5cm{\scF(U) \ar[r]^-{\scF(g)}_-{\cong} & \scF(gU)} \in \M\] for open subsets $U \subset X$ and elements $g \in G$.  If $U_1,\ldots,U_n \subset V$ are pairwise disjoint open subsets of $V \in \Openx$ and if $g \in G$, then it is required that the diagram \[\nicexy@C+1.3cm{\scF(U_1) \otimes \cdots \otimes \scF(U_n) \ar[d]_-{\scF^V_{U_1,\ldots,U_n}} \ar[r]^-{\left(\scF(g),\ldots,\scF(g)\right)} & \scF(gU_1) \otimes \cdots \otimes \scF(gU_n) \ar[d]^-{\scF^{gV}_{gU_1,\ldots,gU_n}}\\ 
\scF(V) \ar[r]^-{\scF(g)} & \scF(gV)}\]
in $\M$ be commutative.

Since prefactorization algebras on a topological space and algebraic quantum field theories on an orthogonal category are both mathematical frameworks for quantum field theories, one might wonder what the difference is.  An algebraic quantum field theory on an orthogonal category is entrywise a monoid in $\M$, so observables in the same object of observables can always be multiplied.  On the other hand, a prefactorization algebra on $X$ is entrywise an object in $\M$.  In particular, observables in a prefactorization algebra on $X$ cannot be multiplied unless they come from pairwise disjoint open subsets.  Despite this difference, in Sections \ref{sec:pfa-pointed-diagram}, Section \ref{sec:diag-com-monoid}, and Chapter \ref{ch:comparing}, we will see that these two mathematical approaches to quantum field theories--algebraic quantum field theories and prefactorization algebras--are actually closely related.

To facilitate the comparison between prefactorization algebras and algebraic quantum field theories, we will take a more categorical approach to the former.  A Costello-Gwilliam prefactorization algebra on a topological space $X$ is defined as a functor $\Openx \to \M$ with some extra structure and properties.  As we pointed out in Example \ref{ex:openx}, the category $\Openx$ is a bounded lattice, i.e., a lattice with both a least element and a greatest element.  We take the abstraction one step further.  In order to specify the structure morphism $\scF^V_{U_1,\ldots,U_n}$, we need two things: 
\begin{itemize}\item Each $U_i$ is equipped with a morphism $U_i \to V$.
\item The $U_i$'s are pairwise disjoint in a suitable sense.
\end{itemize}
We will achieve this below by a new concept called a configured category.  Basically, we simply incorporate the finite families of morphisms $\{U_i \to V\}_{i=1}^n$ into the data of our category and impose some natural axioms as suggested by $\Openx$.  This is analogous to an orthogonal category in Definition \ref{def:orthogonal-category}, where a concept of disjointedness is built into the data of the category via a set $\perp$ of pairs of morphisms with a common codomain.

\section{Configured Categories}\label{sec:configured-cat}

In this section, we define configured categories, from which we will later define prefactorization algebras, and provide some key examples.

\begin{definition}\label{def:configcat}
A \emph{configured category}\index{configured category}\index{category!configured} $\Chat = (\C,\Configc)$ is a pair consisting of
\begin{itemize}\item a small category $\C$ and 
\item a set $\Configc$ in which each element, called a \index{configuration}\emph{configuration}, is a pair $(d;\{f_i\})$ with 
\begin{itemize}\item $d \in \C$ and 
\item $\{f_i\}$ a finite, possibly empty, sequence $\bigl\{f_i : c_i \to d\bigr\}_{i=1}^n$ of morphisms in $\C$ with codomain $d$.\end{itemize}
\end{itemize}
It is required that the following four axioms hold.
\begin{description}
\item[Symmetry] If $\bigl(d;\{f_i\}_{i=1}^n\bigr) \in \Configc$ and if $\sigma \in \Sigma_n$, then $\bigl(d;\{f_{\sigma(i)}\}_{i=1}^n\bigr) \in \Configc$.
\item[Subset] If $(d;\{f_i\}) \in \Configc$ and $\{f_j'\}$ is a possibly empty subsequence of $\{f_i\}$, then $(d;\{f_j'\}) \in \Configc$.
\item[Inclusivity] If $f : c \to d$ is a morphism in $\C$, then $(d;\{f\}) \in \Configc$.
\item[Composition] If $\bigl(d;\bigl\{f_i : c_i \to d\bigl\}_{i=1}^n\bigr) \in \Configc$ with $n \geq 1$ and if for each $1 \leq i \leq n$, $\bigl(c_i;\bigl\{g_{ij} : b_{ij} \to c_i\bigl\}_{j=1}^{k_i}\bigr) \in \Configc$, then the \emph{composition}
\begin{equation}\label{configured-composition}
\Bigl(d;\bigl\{f_ig_{ij} : b_{ij} \to d\bigr\}_{1\leq i \leq n,\, 1 \leq j \leq k_i}\Bigr)
\end{equation}
is also a configuration.
\end{description}
\end{definition}

\begin{example} The composition \eqref{configured-composition} of $\bigl(d;\{f_1,f_2\}\bigr)$, $\bigl(c_1;\{g_{11},g_{12}\}\bigr)$, and $\bigl(c_2;\{g_{21},g_{22},g_{23}\}\bigr)$ is \[\Bigl(d; \{f_1g_{11}, f_1g_{12}, f_2g_{21}, f_2g_{22}, f_2g_{23}\}\Bigr).\]  If we replace $\bigl(c_2;\{g_{21},g_{22},g_{23}\}\bigr)$ with $(c_2;\varnothing)$, then the composition becomes \[\Bigl(d; \{f_1g_{11}, f_1g_{12}\}\Bigr).\] \dqed
\end{example}

\begin{interpretation}\label{int:configured-category}
From a physical perspective, one should think of the objects in a configured category $(\C,\Configc)$ as the spacetime regions of interest, e.g., oriented manifolds of a fixed dimension.  A morphism $f : c \to d$ in $\C$ should be thought of as an inclusion of the spacetime region $c$ into a bigger spacetime region $d$.   A configuration $\bigl(d;\{f_i : c_i \to d\}_{i=1}^n\bigr)$ is expressing the idea that the spacetime regions $c_1,\ldots,c_n$ are pairwise disjoint in $d$. 
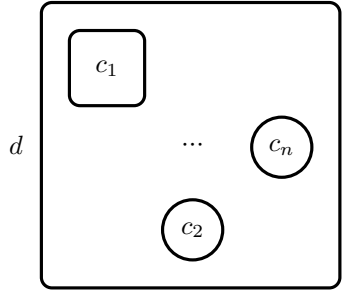
\begin{center}\begin{tikzpicture}
\node (dots) {$\cdots$}; \node[bigplain, rounded corners, above left=.5cm of dots] (c1) {$c_1$};
\node [xplain, below=.5cm of dots] (c2) {$c_2$}; \node [xplain, right=.5cm of dots] (cn) {$c_n$};
\node[draw=black,inner sep=10pt,very thick,rounded corners,
fit=(c1) (c2) (cn)] (d) {}; \node [left=.1cm of d] () {$d$};
\end{tikzpicture}\end{center}

All four axioms in Definition \ref{def:configcat} are physically motivated by this picture.  Indeed, the symmetry axiom is just about relabeling the pairwise disjoint regions.  The subset and inclusivity axioms are immediate.  The composition \eqref{configured-composition} says that, if the spacetime regions $c_i$'s are pairwise disjoint in $d$, and if the spacetime regions $b_{ij}$'s are pairwise disjoint in $c_i$ for each $i$, then the entire collection $\{b_{ij}\}_{i,j}$ is pairwise disjoint in $d$.
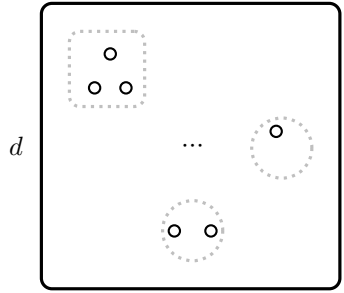
\begin{center}\begin{tikzpicture}
\node (dots) {$\cdots$}; 
\node[bigplain, rounded corners, lightgray, dotted, above left=.5cm of dots] (c1) {};
\node [above left=1cm of dots] (c1o) {}; \node [xsplain, above=.03cm of c1o] () {};
\node [xsplain, below left=.03cm of c1o] () {}; \node [xsplain, below right=.03cm of c1o] () {};
\node [xplain, lightgray, dotted,below=.5cm of dots] (c2) {}; 
\node [below=.8cm of dots] (c2o) {}; \node [xsplain, left=.03cm of c2o] () {};
\node [xsplain, right=.03cm of c2o] () {};
\node [xplain, lightgray, dotted,right=.5cm of dots] (cn) {};
\node [right=.5cm of dots] (cno) {}; \node [xsplain, above right=.05cm of cno] (){};
\node[draw=black,inner sep=10pt,very thick,rounded corners,
fit=(c1) (c2) (cn)] (d) {}; \node [left=.1cm of d] () {$d$};
\end{tikzpicture}\end{center}
In this picture, the three small discs inside $c_1$ are the pairwise disjoint regions $b_{11}$, $b_{12}$, and $b_{13}$, and the two small discs inside $c_2$ are the disjoint regions $b_{21}$ and $b_{22}$.  The only small disc inside $c_n$ is $b_{n1}$.\dqed
\end{interpretation}

\begin{definition}\label{def:config-functor}
Suppose $(\C,\Configc)$ is a configured category.
\begin{enumerate}
\item For objects $c_1,\ldots,c_n,d \in \C$, the set of configurations $\bigl(d;\bigl\{c_i \to d\bigr\}_{i=1}^n\bigr)$ is denoted by $\Configc\duc$, where $\uc=(c_1,\ldots,c_n)$. 
\item A \emph{configured functor}\index{configured functor}\index{functor!configured} \[F : (\C,\Configc) \to (\D,\Configd)\] between two configured categories is a functor $F : \C \to\D$ that preserves the configurations, i.e., \[(Fd;\{Ff_i\}) \in \Configd \ifspace (d; \{f_i\}) \in \Configc.\]
\item The category of configured categories and configured functors is denoted by\label{notation:configcat} $\Configcat$.
\end{enumerate}
\end{definition}

\begin{lemma}\label{lem:configcat-basic} 
Suppose $(\C,\Configc)$ is a configured category.
\begin{enumerate}\item For each object $d \in \C$, $(d;\varnothing)$ belongs to $\Configc\dempty$.
\item The composition \eqref{configured-composition} belongs to $\Configc\dub$, where $\ub_i=(b_{i1},\ldots,b_{ik_i})$ for $1 \leq i \leq n$ and $\ub = (\ub_1,\ldots,\ub_n)$.
\end{enumerate}
\end{lemma}

\begin{proof} For the first assertion, first note that $(d;\{\Id_d\}) \in \Configc$ by the inclusivity axiom.  So by the subset axiom, we infer that $(d;\emptyset) \in \Configc$.  The second assertion follows directly from the definition.
\end{proof}

\begin{notation}\label{not:configcat-notation} Suppose $(\C,\Configc)$ is a configured category.
\begin{enumerate}
\item We call\label{notation:emptyconfiguration} $(d;\varnothing) \in \Configc\dempty$ the \emph{empty configuration at $d$}.\index{empty configuration}
\item To simplify the presentation, for a configuration $\bigl(d;\{f_i\}_{i=1}^n\bigr) \in \Configc\duc$, we will often omit $d$, which is the common codomain of the morphisms $f_i$ for $1\leq i \leq n$, and simply write\label{notation:nconfig} $\{f_i\}_{i=1}^n$ or $\{f_i\}$.
\item If $\{f_i\}_{i=1}^n \in \Configc$, then we call it an \emph{$n$-ary configuration}.\index{configuration!$n$-ary}
\end{enumerate}
\end{notation}

Some examples of configured categories follow.  Many more examples will be given in Section \ref{sec:relation-orthcat}, where we will show that every orthogonal category yields a configured category.

\begin{example}[Minimal configured category]\label{ex:min-configuration}
Suppose $\C$ is a small category.  Define $\Configcmin$ to be the set consisting of
\begin{itemize}
\item $(d;\varnothing)$ for all objects $d \in \C$ and
\item $(d;\{f\})$ for all morphisms $f \in \C(c,d)$ with $c,d \in \C$.
\end{itemize}
Then\label{notation:chatmin} \[\Chatmin = (\C,\Configcmin)\] is a configured category, called the \index{minimal configured category}\emph{minimal configured category on $\C$}.\dqed
\end{example}

\begin{example}[Maximal configured category]\label{ex:max-configuration}
Suppose $\C$ is a small category.  Define $\Configcmax$ to be the set of all pairs $(d;\{f_i\})$ with $d \in \C$ and $\{f_i\}$ any possibly empty finite sequence of morphisms in $\C$ with codomain $d$.  Then\label{notation:chatmax} \[\Chatmax = (\C,\Configcmax)\] is a configured category, called the \index{maximal configured category}\emph{maximal configured category on $\C$}.  For objects $c_1,\ldots,c_n,d \in \C$ with $n \geq 0$ and $\uc=(c_1,\ldots,c_n)$, by our notational convention above, we have \[\Configcmax\duc = \begin{cases}\prod\limits_{i=1}^n \C(c_i,d) & \text{ if $n \geq 1$},\\ 
\{(d;\emptyset)\} & \text{ if $n=0$}.\end{cases}\]
If $(\C,\Configc)$ is a configured category, then there are configured functors \[\nicexy{\bigl(\C,\Configcmin\bigr) \ar[r]^-{i_0} & \bigl(\C,\Configc\bigr) \ar[r]^-{i_1} & \bigl(\C,\Configcmax\bigr)}\] in which $i_0$ and $i_1$ are both the identity functors on $\C$.\dqed
\end{example}

\begin{example}[Configured categories of bounded lattices]\label{ex:boundedlatter-config}
Suppose $(L,\leq)$ is a \index{bounded lattice!configured category}bounded lattice with least element $0$, considered as a small category as in Example \ref{ex:lattice}.  For $c_1,\ldots,c_n,d \in L$, suppose $\uc=(c_1,\ldots,c_n)$.  Define the set
\[\Configl\duc= \begin{cases} \prod\limits_{i=1}^n L(c_i,d) & \text{ if each $c_i \leq d$ and $c_p \wedge c_q=0$ for all $1 \leq p\not=q \leq n$},\\ \varnothing & \text{ otherwise}.\end{cases}\]
In the first case, $\Configl\duc$ is a one-element set because each morphism set $L(c_i,d)$ with $c_i \leq d$ is a one-element set if $n \geq 1$, while an empty product is also a one-element set if $n=0$. This forms a configured category\label{notation:lhat} \[\Lhat =\bigl(L,\Configl\bigr).\]  To check the composition axiom in \eqref{configured-composition}, the key point is that if $a \wedge b = 0$ (i.e., if $a$ and $b$ have the least element $0$ as their greatest lower bound), then $0$ is the only lower bound of $a$ and $b$.  The other three axioms are immediate from the definition.  

Note that if $(d,\{c_i \leq d\})$ is a configuration, then we can add any finite number of copies of $0 \leq d$ to the finite sequence $\{c_i \leq d\}$ to yield another configuration.  In particular, for each $d \in L$, $\bigl(d;\{0\leq d\}_{i=1}^n\bigr)$ is a configuration for each $n \geq 0$.\dqed
\end{example}

\begin{example}[Configured categories of topological spaces]\label{ex:open-configuration}
Suppose $X$ is a topological space.  Recall from Example \ref{ex:openx} the bounded lattice $(\Openx,\subset)$.  By Example \ref{ex:boundedlatter-config} this yields a \index{topological space!configured category}configured category \[\Openxhat =\bigl(\Openx,\Configx\bigr).\] More explicitly, the category $\Openx$ has open subsets of $X$ as objects and subset inclusions as morphisms.  For open subsets $U_1,\ldots,U_n,V \subset X$, suppose $\uU = (U_1,\ldots,U_n)$. Then \[\Configx\VuU = \begin{cases} \prod\limits_{i=1}^n\Openx(U_i,V) & \text{ if the $U_i$'s are pairwise disjoint subsets of $V$},\\ \varnothing & \text{ otherwise}.\end{cases}\]
As in the previous example, in the first case, $\Configx\VuU$ is a one-element set.  Moreover, for each open subset $V \subset X$, $\bigl(V;\{\emptyset_X \subset V\}_{i=1}^n\bigr)$ is a configuration for each $n \geq 0$, where $\emptyset_X$ is the empty subset of $X$.\dqed
\end{example}

\begin{example}[Configured categories of equivariant topological spaces]\label{ex:eq-space-configuration}
Suppose $G$ is a group, and $X$ is a topological space in which $G$ acts on the left by homeomorphisms.  Suppose $\Openxg$ is the category in Example \ref{ex:eq-space}.  For open subsets $U_1,\ldots,U_n,V \subset X$, suppose $\uU = (U_1,\ldots,U_n)$. Define the set $\Configxg\VuU$ as consisting of finite sequences \[(g_1,\ldots,g_n)\in \prod_{i=1}^n \Openxg(U_i,V)\] such that:
\begin{itemize}\item The $g_iU_i$'s are pairwise disjoint subsets of $V$.
\item Each $g_i \in G$ is regarded as the composition \[\nicexy@C+.4cm{U_i \ar[r]^-{g_i} & g_iU_i \ar[r]^-{\text{inclusion}} & V}\] in $\Openxg$.  
\end{itemize}
This defines a\index{equivariant topological space!configured category} configured category \[\Openxghat =\bigl(\Openxg,\Configxg\bigr).\]  If $G$ is the trivial group, then we recover the configured category $\Openxhat$ in Example \ref{ex:open-configuration}.\dqed
\end{example}

\section{Prefactorization Algebras as Operad Algebras}\label{sec:pfa-operad}

In this section, we define prefactorization algebras as algebras over some colored operads associated with configured categories.  We record the coherence theorems for prefactorization algebras, with or without the time-slice axiom.  We recover the prefactorization algebras of Costello-Gwilliam \cite{cg} when the configured category is $\Openxhat$ for a topological space $X$.  We also recover their equivariant prefactorization algebras when the configured category is $\Openxghat$ for a topological space $X$ equipped with an action by a group $G$.  

\begin{motivation} Given a configured category $\Chat = (\C,\Configc)$, for the moment let us think of its objects as the spacetime regions of interest as in Interpretation \ref{int:configured-category}.  From the prefactorization algebra perspective, a quantum field theory $\scF$ on $\Chat$ is an assignment that associates to each spacetime region $c \in \C$ an object $\scF(c)$, say a chain complex, of quantum observables on $c$.  If $c_1, \ldots, c_n$ are suitably  disjoint spacetime regions in $d$, then we should be able to combine the observables in the form of a map \[\scF(c_1) \otimes \cdots \otimes \scF(c_n) \to \scF(d).\]  These multiplication maps should satisfy some natural conditions with respect to sub-regions.  The following colored operad is designed to modeled this structure.\dqed
\end{motivation}

\begin{definition}\label{def:ochat-opread}
Suppose $\Chat = (\C,\Config)$ is a configured category with object set $\colorc$.  Define the following sets and functions.
\begin{description}
\item[Entries] Define the object\label{notation:ochat} $\Ochat \in \Set^{\Profcc}$ \index{set operad!configured category}\index{prefactorization algebra!operad}entrywise as \[\Ochat\duc = \Config\duc \forspace \duc \in \Profcc.\]
\item[Equivariance] For $\sigma \in \Sigma_{|\uc|}$, define the map \[\nicexy{\Ochat\duc \ar[r]^-{\sigma} & \Ochat\ducsigma} \byspace \{f_i\}\sigma = \{f_{\sigma(i)}\}\] for $\{f_i\} \in\Config\duc$.
\item[Colored Units] For each $c\in \colorc$, the $c$-colored unit in $\Ochat\cc$ is $\{\Id_c\}$.
\item[Operadic Composition] For $(\uc;d) \in \Profcc$ with $|\uc|=n \geq 1$, $\ub_i=(b_{i1},\ldots,b_{ik_i}) \in \Profc$ for $1 \leq i \leq n$, and $\ub=(\ub_1,\ldots,\ub_n)$, define the map \[\nicexy{\Ochat\duc \times \prod\limits_{i=1}^n \Ochat\ciubi \ar[r]^-{\gamma} & \Ocbar\dub}\] as the composition \[\gamma\Bigl(\{f_i\}_{i=1}^n; \{g_{1j}\}_{j=1}^{k_1}, \ldots, \{g_{nj}\}_{j=1}^{k_n}\Bigr) 
= \bigl\{f_ig_{ij}\bigr\}_{1\leq i \leq n,\, 1 \leq j \leq k_i}\] in \eqref{configured-composition} for $\{f_i\} \in \Ochat\duc$ and $\{g_{ij}\}_{j=1}^{k_i} \in \Ochat\ciubi$ with $1 \leq i \leq n$.
\end{description}
\end{definition}

\begin{lemma}\label{lem:ochat-operad}
In the setting of Definition \ref{def:ochat-opread}:
\begin{enumerate}\item $\Ochat$ is a $\colorc$-colored operad in $\Set$.
\item This construction defines a functor \[\O_{(-)} : \Configcat \to \Operad(\Set)\] from the category of configured categories to the category of colored operads in $\Set$.
\end{enumerate}
\end{lemma}

\begin{proof}
One checks directly that $\Ochat$ satisfies the axioms in Definition \ref{def:operad-generating}, so it is a $\colorc$-colored operad in $\Set$.  The naturality of this construction is also checked by a direct inspection.
\end{proof}

Recall from Example \ref{ex:operad-set-m} the strong symmetric monoidal functor $\Set \to \M$, which sends each set $S$ to the coproduct $\coprod_S \tensorunit$, and the induced change-of-category functor \[(-)^{\M} : \Operadcset \to\Operadcm\] between the categories of $\colorc$-colored operads.  We will consider the image in $\M$ of $\Ochat$, denoted by\label{notation:ochatm} $\Ochatm$.  Also recall from Definition \ref{def:operad-localization} the $S$-localization $\Osinv$ of a $\colorc$-colored operad $\O$ in $\Set$ for a set $S$ of unary elements in $\O$.

\begin{definition}\label{def:pfa}
Suppose $\Chat = (\C,\Config)$ is a configured category.  
\begin{enumerate}\item Define the category\label{notation:pfachat} \[\PFA(\Chat) = \algmochatm,\] whose objects are called \index{prefactorization algebra!on a configured category}\emph{prefactorization algebras on $\Chat$}.
\item Suppose $S$ is a set of morphisms in $\C$, regarded as a subset of $\Config$ by the inclusivity axiom.  Define the category\label{notation:pfachats} \[\PFA(\Chat,S) = \algmochatsinvm,\] whose objects are called \index{prefactorization algebra!time-slice axiom}\index{time-slice axiom!prefactorization algebra}\emph{prefactorization algebras on $\Chat$ satisfying the time-slice axiom with respect to $S$}.
\end{enumerate}
\end{definition}

\begin{remark} In the previous definition, a morphism $s : c \to d \in S$ is regarded as a configuration $\{s\} \in \Config\dc = \Ochat\dc$, hence also a unary element in $\Ochat$.  So the $S$-localization $\Ochatsinv$ exists by Theorem \ref{thm:operad-localization-exists}.\dqed
\end{remark}

The following coherence theorem is a special case of Theorem \ref{thm:algebra-set-operad}.  It explains precisely what a prefactorization algebra is.

\begin{theorem}\label{thm:ochat-algebra}
Suppose $\Chat = (\C,\Config)$ is a configured category with object set $\colorc$.  Then an $\Ochatm$-algebra is precisely a pair\index{prefactorization algebra!coherence}\index{Coherence Theorem!for prefactorization algebras} $(X,\lambda)$ consisting of
\begin{itemize}\item a $\colorc$-colored object $X=\{X_c\}_{c\in \colorc}$ in $\M$ and
\item a structure morphism\index{structure morphism!for prefactorization algebras}  
\begin{equation}\label{pfa-structure-morphism}
\nicexy@C+.5cm{\bigotimes\limits_{i=1}^n X_{c_i} \ar[r]^-{\lambda\{f_i\}_{i=1}^n} & X_d} \in \M
\end{equation}
for 
\begin{itemize}\item each $\duc = \dconecn \in \Profcc$ and
\item each configuration $\{f_i\}_{i=1}^n \in \Config\duc$
\end{itemize}
\end{itemize}
that satisfies the following associativity, unity, and equivariance axioms.
\begin{description}
\item[Associativity] For $(\uc;d) \in \Profcc$ with $|\uc|=n\geq 1$, $\ub_i=(b_{i1},\ldots,b_{ik_i}) \in \Profc$ for $1 \leq i \leq n$, $\ub=(\ub_1,\ldots,\ub_n)$, configurations $\{f_i\} \in \Config\duc$, and $\{g_{ij}\} \in \Config\ciubi$ for $1 \leq i \leq n$, the associativity diagram
\begin{equation}\label{pfa-ass}
\nicexy@C+.8cm{\bigotimes\limits_{i=1}^n \bigotimes\limits_{j=1}^{k_i} X_{b_{ij}} \ar[d]_-{\bigotimes\limits_{i=1}^n \lambda\{g_{ij}\}_{j=1}^{k_i}} \ar[r]^-{\lambda\{f_ig_{ij}\}_{i,j}} & X_d \ar@{=}[d]\\ \bigotimes\limits_{i=1}^n X_{c_i} \ar[r]^-{\lambda\{f_i\}} & X_d}
\end{equation}
in $\M$ is commutative.
\item[Unity] For each $c\in \colorc$, $\lambda\{\Id_c\}$ is equal to $\Id_{X_c}$.
\item[Equivariance] For each configuration $\{f_i\} \in \Config\duc$ with $|\uc|=n$ and $\sigma \in \Sigma_n$, the equivariance diagram
\begin{equation}\label{pfa-eq}
\nicexy{\bigotimes\limits_{i=1}^n X_{c_i} \ar[dr]_-{\lambda\{f_i\}} \ar[rr]^-{\sigmainv} 
&& \bigotimes\limits_{i=1}^n X_{c_{\sigma(i)}} \ar[dl]^-{\lambda\{f_{\sigma(i)}\}}\\ & X_d &}
\end{equation}
in $\M$ is commutative.
\end{description}
A morphism of $\Ochatm$-algebras $\varphi : (X,\lambda^X) \to (Y,\lambda^Y)$ is a morphism $\varphi : X \to Y$ of $\colorc$-colored objects in $\M$ that respects the structure morphisms in \eqref{pfa-structure-morphism} in the sense that the diagram
\begin{equation}\label{ochatalg-morphism}
\nicexy@C+.5cm{\bigotimes\limits_{i=1}^n X_{c_i} \ar[d]_-{\lambda^X\{f_i\}} \ar[r]^-{\bigotimes\limits_{i=1}^n \varphi_{c_i}} & \bigotimes\limits_{i=1}^n Y_{c_i} \ar[d]^-{\lambda^Y\{f_i\}}\\ X_d \ar[r]^-{\varphi_d} & Y_d}
\end{equation}
in $\M$ is commutative for each configuration $\{f_i\} \in \Config\duc$ with $|\uc|=n$. 
\end{theorem}

\begin{example}[Costello-Gwilliam prefactorization algebras]\label{ex:cg-pfa}
Consider the configured category $\Openxhat$ in Example \ref{ex:open-configuration} for a topological space $X$.  A prefactorization algebra\index{prefactorization algebra!Costello-Gwilliam} on $\Openxhat$, i.e., an $\Otom_{\Openxhat}$-algebra in Theorem \ref{thm:ochat-algebra}, is precisely a \emph{unital prefactorization algebra on $X$} in the sense of \cite{cg} 3.1.1.1 and 3.1.2.3.\dqed
\end{example}

\begin{example}[Costello-Gwilliam equivariant prefactorization algebras]\label{ex:cg-eqpfa}
Consider the configured category $\Openxghat$ in Example \ref{ex:eq-space-configuration} for a topological space $X$ equipped with a left action by a group $G$.  A \index{prefactorization algebra!equivariant}\index{equivariant prefactorization algebra}prefactorization algebra on $\Openxghat$, i.e., an $\Otom_{\Openxghat}$-algebra in Theorem \ref{thm:ochat-algebra}, is precisely a \emph{unital $G$-equivariant prefactorization algebra on $X$} in the sense of \cite{cg} 3.7.1.1.\dqed
\end{example}

The following coherence theorem is a special case of Theorems \ref{thm:localization-algebra} and \ref{osinvm-algebra}.  It explains precisely what a \index{Coherence Theorem!for prefactorization algebras with time-slice}prefactorization algebra satisfying the time-slice axiom is.

\begin{theorem}\label{thm:pfa-timeslice}
Suppose $\Chat = (\C,\Config)$ is a configured category, and $S$ is a set of morphisms in $\C$.
\begin{enumerate}\item The $S$-localization\index{localization of an operad} morphism $\ell : \Ochat \to \Ochatsinv$ induces the change-of-operad adjunction \[\nicexy{\PFA(\Chat)=\algmochatm \ar@<2pt>[r]^-{\ellm_!} & \algmochatsinvm=\PFA(\Chat,S) \ar@<2pt>[l]^-{(\ellm)^*}}\] whose right adjoint $(\ellm)^*$ is full and faithful and whose counit \[\epsilon : \ellm_!(\ellm)^* \iso \Id_{\algmochatsinvm}\] is a natural isomorphism.
\item Via the right adjoint $(\ellm)^*$, $\Ochatsinvm$-algebras are equivalent to $\Ochatm$-algebras whose structure morphisms $\lambda\{s\}$ are isomorphisms for all $s \in S$.
\end{enumerate}
\end{theorem}

\begin{interpretation} A prefactorization algebra on a configured category $\Chat$ satisfies the time-slice axiom with respect to $S$ precisely when the structure morphisms $\lambda\{s\}$ are invertible for all $s \in S$.  This is the exact analogue of the time-slice axiom for algebraic quantum field theories in Definition \ref{def:aqft}, which may also be implemented by replacing the orthogonal category with its $S$-localization as in Lemma \ref{lem:aqft-time-slice}.  For prefactorization algebras, the time-slice axiom may be implemented by replacing the colored operad $\Ochat$ with its $S$-localization $\Ochatsinv$.\dqed
\end{interpretation}

The following result compares prefactorization algebras, with or without the time-slice axiom, on different configured categories.

\begin{corollary}\label{cor:pfa-adjunction}
Suppose $F : \Chat = (\C,\Configc) \to (\D,\Configd)=\Dhat$ is a configured functor, and $S$ is a set of morphisms in $\D$.  Define \[\Szero = F^{-1}(S) = \Bigl\{g \in \Morc : Fg \in S \Bigr\}\] to be the $F$-pre-image of $S$.  Then there is an induced diagram of change-of-operad adjunctions
\[\nicexy@C+.5cm{\PFA(\Chat)=\algmochatm \ar@<-2pt>[d]_-{\ellm_!} \ar@<2pt>[r]^-{(\Otom_F)_!} 
& \algmodhatm=\PFA(\Dhat) \ar@<2pt>[l]^-{(\Otom_F)^*} \ar@<-2pt>[d]_-{\ellm_!}\\
\PFA(\Chat,\Szero)=\algmochatszeroinvm \ar@<-2pt>[u]_-{(\ellm)^*} \ar@<2pt>[r]^-{(\Otom_{F'})_!} 
& \algmodhatsinvm=\PFA(\Dhat,S) \ar@<2pt>[l]^-{(\Otom_{F'})^*} \ar@<-2pt>[u]_-{(\ellm)^*}}\] 
in which \[(\Otom_{F'})_! \ellm_!= \ellm_!(\Otom_F)_! \andspace 
(\ellm)^*(\Otom_{F'})^* = (\Otom_F)^*(\ellm)^*.\]
\end{corollary}

\begin{proof}
Consider the solid-arrow diagram \[\nicexy{\Ochat \ar[r]^-{\O_F} \ar[d]_-{\ell} & \Odhat \ar[d]^-{\ell}\\ \Ochatszeroinv \ar@{.>}[r]^-{\O_{F'}} & \Odhatsinv}\] of colored operads in $\Set$, where \[\ell : \Ochat \to \Ochatszeroinv \andspace \ell : \Odhat \to \Odhatsinv\] are the $\Szero$-localization of $\Ochat$ and the $S$-localization of $\Odhat$, respectively.  Since every unary element in \[\ell\O_F(\Szero) \subset \ell(S)\] is invertible in $\Odhatsinv$, by the universal property of $\Szero$-localization, there is a unique operad morphism $\O_{F'}$ that makes the entire diagram commutative.  This diagram becomes a commutative diagram of colored operads in $\M$ once we apply the change-of-category functor $(-)^{\M}$.  The desired diagram of change-of-operad adjunctions is obtained by applying Theorem \ref{thm:change-operad}.
\end{proof}

\begin{example}[Costello-Gwilliam locally constant prefactorization algebras]\label{ex:locally-constant-pfa}
In the\index{prefactorization algebra!locally constant}\index{locally constant prefactorization algebra} configured category $\Openr$, suppose $S$ is the set of inclusions of open intervals.  By Theorem \ref{thm:pfa-timeslice} a prefactorization algebra on $\Openrhat$ satisfying the time-slice axiom with respect to $S$ is equivalent to a prefactorization algebra on $\Openrhat$ whose structure morphisms $\lambda\{s\}$ are isomorphisms for all $s \in S$.  These are precisely the \emph{locally constant} unital prefactorization algebras on $\fieldr$ in \cite{cg} 3.2.0.1.\dqed
\end{example}

\section{Pointed Diagram Structure}\label{sec:pfa-pointed-diagram}

In the following few sections, we will provide more examples of prefactorization algebras.  Along the way, we provide evidence that prefactorization algebras are closely related to algebraic quantum field theories, a relationship that will be made precise in Chapter \ref{ch:comparing}.  In this section, we observe that every prefactorization algebra has an underlying pointed diagram, which itself can be realized as a prefactorization algebra on the minimal configured category.  There is a free-forgetful adjunction between the category of prefactorization algebras on the minimal configured category and the category of algebraic quantum field theories on the minimal orthogonal category.

First we need the following definition.

\begin{definition}\label{def:pointed-digram}
Suppose $\C$ is a small category.  
\begin{enumerate}\item A $\C$-diagram $\scF : \C \to \M$ is\index{pointed diagram} \emph{pointed} if it is equipped with a $c$-colored unit \[\operadunit_c : \tensorunit \to \scF(c) \in \M\] for each object $c \in \C$ such that the diagram \[\nicexy{\tensorunit \ar@{=}[r] \ar[d]_-{\operadunit_c} & \tensorunit \ar[d]^-{\operadunit_d} &\\ \scF(c) \ar[r]^-{\scF(f)} & \scF(d)}\] is commutative for each morphism $f : c \to d \in \C$.  
\item A natural transformation between two pointed $\C$-diagrams is \emph{pointed} if it preserves the colored units.
\item The category of pointed $\C$-diagrams in $\M$ and pointed natural transformations is denoted by\label{notation:mcstar} $\Mcstar$.
\end{enumerate}
\end{definition}

\begin{example} By forgetting the multiplicative structure, every $\C$-diagram of monoids in $\M$ has an underlying pointed $\C$-diagram, where the colored units are the units of the monoids.\dqed\end{example}

\begin{proposition}\label{prop:pointed-diagram}
Suppose $\C$ is a small category.  Then there is a canonical isomorphism\index{colored operad!for pointed diagrams} \[\algm\bigl(\Otom_{\Chatmin}\bigr) = \PFA(\Chatmin) \cong \Mcstar\] between the category of prefactorization algebras on the minimal configured category $\Chatmin = (\C,\Configcmin)$ in Example \ref{ex:min-configuration} and the category of pointed $\C$-diagrams in $\M$.  
\end{proposition}

\begin{proof}
Both a pointed $\C$-diagram in $\M$ and a prefactorization algebra on $\Chatmin$ assign to each object $c \in \C$ an object $\scF(c) \in \M$.  To see that pointed $\C$-diagrams in $\M$ are precisely the prefactorization algebras on $\Chatmin$, we use the Coherence Theorem \ref{thm:ochat-algebra}.  There are only two kinds of configurations in $\Configcmin$:
\begin{itemize}\item $(c;\varnothing)$ for all objects $c \in \C$ and
\item $(d;\{f\})$ for all morphisms $f \in \C(c,d)$ with $c,d \in \C$.
\end{itemize}
If $\scF$ is a prefactorization algebra on $\Chatmin$, then its only structure morphisms \eqref{pfa-structure-morphism} are \[\lambda\bigl\{(c;\varnothing)\bigr\} = \operadunit_c : \tensorunit \to \scF(c) \in \M\] for objects $c \in \C$ and \[\lambda\bigl\{(d;\{f\})\bigr\} = \scF(f) : \scF(c) \to \scF(d) \in \M\] for morphisms $f \in \C(c,d)$.

The equivariance condition \eqref{pfa-eq} is trivial, and the unity condition says that \[\scF(\Id_c)=\Id_{\scF(c)} \forspace c \in \C.\]  The associativity condition \eqref{pfa-ass} must have $n=1$.  If $k_1=0$, then the associativity condition is the diagram in Definition \ref{def:pointed-digram} that defines pointed $\C$-diagrams.  If $k_1=1$, then the associativity condition is the commutative diagram \[\nicexy@C+.5cm{\scF(b) \ar[d]_-{\scF(g)} \ar[r]^-{\scF(fg)} & \scF(d) \ar@{=}[d]\\ \scF(c) \ar[r]^-{\scF(f)} & \scF(d)}\]
for all objects $b,c,d \in \C$ and composable morphisms $(f,g) \in \C(c,d)\times\C(b,c)$.  Therefore, a prefactorization algebra on $\Chatmin$ is precisely a pointed $\C$-diagram in $\M$.  

Similarly, to see the correspondence between morphisms, we use \eqref{ochatalg-morphism} in the Coherence Theorem \ref{thm:ochat-algebra}.  If $n=0$ then \eqref{ochatalg-morphism} is the preservation of colored units.  If $n=1$ then \eqref{ochatalg-morphism} is the commutative square that defines a natural transformation between two $\C$-diagrams in $\M$.
\end{proof}

Recall from Definition \ref{def:ochat-opread} the colored operad $\Ochat$ for a configured category $\Chat$.  The following observation is a consequence of Corollary \ref{cor:pfa-adjunction} and Proposition \ref{prop:pointed-diagram}.

\begin{corollary}\label{cor:pfa-underlying-diagram}
Suppose $\Chat = (\C,\Config)$ is a configured category.  Then the configured functor \[\nicexy{\Chatmin = (\C,\Configcmin) \ar[r]^-{i_0} & (\C,\Config)=\Chat},\] whose underlying functor is the identity functor on $\C$, induces a change-of-operad \index{prefactorization algebra!pointed diagram}adjunction \[\nicexy@C+.5cm{\Mcstar \cong \PFA(\Chatmin) = \algm\bigl(\Otom_{\Chatmin}\bigr) \ar@<2pt>[r]^-{(\Otom_{i_0})_!} & \algmochatm = \PFA(\Chat) \ar@<2pt>[l]^-{(\Otom_{i_0})^*}}\] between the category of pointed $\C$-diagrams in $\M$ and the category of prefactorization algebras on $\Chat$.
\end{corollary}

\begin{interpretation} Each prefactorization algebra $(X,\lambda)$ on a configured category $\Chat=(\C,\Config)$ has an underlying pointed $\C$-diagram in $\M$.  For a morphism $f : c \to d$ in $\C$, the corresponding morphism is the structure morphism \[\nicexy{X_c \ar[r]^-{\lambda\{f\}} & X_d}\in \M\] with $\{f\} \in \Config\dc$.  For each object $c \in \C$, the $c$-colored unit is the structure morphism \[\nicexy@C+.5cm{\tensorunit \ar[r]^-{\lambda\{(c;\varnothing)\}} & X_c}\in \M\] with $(c;\varnothing) \in \Config\cempty$.\dqed
\end{interpretation}

Recall from Definition \ref{def:aqft-operad} the colored operad $\Ocbar$ for an orthogonal category $\Cbar$.

\begin{example}\label{ex:cbarmin-chatmin}
In Example \ref{ex:empty-causality} we noted that the category $\QFT(\Cbarmin)$ of algebraic quantum field theories on $\Cbarmin=(\C,\varnothing)$, where $\varnothing$ is the empty orthogonality relation, is the category $\Monm^{\C}$ of $\C$-diagrams in $\Monm$.  There is a forgetful functor \[\algm\bigl(\Otom_{\Cbarmin}\bigr)\cong\QFT(\Cbarmin)=\Monm^{\C}\to \Mcstar \cong \PFA(\Chatmin)=\algm\bigl(\Otom_{\Chatmin}\bigr)\] from the category of $\C$-diagrams of monoids in $\M$ to the category of pointed $\C$-diagrams in $\M$ that forgets about the multiplicative structure.  This relationship between algebraic quantum field theories and prefactorization algebras is conceptual rather than random, as we now explain.\dqed
\end{example}

\begin{proposition}\label{lem:ochatmin-to-ocbarmin}
Suppose $\C$ is a small category with object set $\colorc$.  Then there is a\index{comparison morphism}\index{prefactorization algebra!comparison morphism}\index{algebraic quantum field theory!comparison morphism} morphism \[\nicexy{\O_{\Chatmin} \ar[r]^-{\deltamin} & \O_{\Cbarmin}}\in\Operadcset\] that is entrywise defined as follows.
\begin{itemize}
\item $\deltamin$ is a canonical bijection on each $0$-ary entry $\cempty$ for $c\in \colorc$ and each unary entry $\dc$ for $c,d\in \colorc$.
\item $\deltamin$ is the unique morphism from the empty set to $\O_{\Cbarmin}\duc$ if $|\uc| \geq 2$.
\end{itemize}
\end{proposition}

\begin{proof}
Recall that $\Configcmin$ only has $0$-ary configurations $(c;\varnothing) \in \Configcmin\cempty$ for $c \in \colorc$ and unary configurations $(d;\{f\}) \in \Configcmin\dc$ for $f \in \C(c,d)$.  So there are  canonical bijections on the $0$-ary entries \[\nicexy{\O_{\Chatmin}\cempty = \Configcmin\cempty = \bigl\{(c;\varnothing)\bigr\} \ar[r]^-{\deltamin}_-{\cong}& \Sigma_0 \times * = \O_{\Cbarmin}\cempty}\] for $c\in \colorc$ and on the unary entries \[\nicexy{\O_{\Chatmin}\dc =\Configcmin\dc = \C(c,d) \ar[r]^-{\deltamin}_-{\cong} & \Sigma_1 \times \C(c,d) = \O_{\Cbarmin}\dc}\] for $c,d\in \colorc$.  For $\uc = (c_1,\ldots,c_n)$ with $n \geq 2$, the morphism $\deltamin$ is defined as the unique morphism \[\nicexy{\O_{\Chatmin}\duc = \Configcmin\duc = \varnothing \ar[r]^-{\deltamin} & \Sigma_n \times \prod\limits_{j=1}^n \C(c_j,d) = \O_{\Cbarmin}\duc}.\]  A direct inspection shows that $\deltamin$ is a well-defined morphism of $\C$-colored operads in $\Set$.  Indeed, there is no equivariance relation to check because $\O_{\Chatmin}$ is concentrated in $0$-ary and unary entries.  The preservation by $\deltamin$ of colored units and operadic composition follows from the fact that on both sides these structures are given by identity morphisms and composition in $\C$.
\end{proof}

Recall the change-of-category functor \[(-)^{\M} : \Operadcset \to \Operadcm\] in Example \ref{ex:operad-set-m}.  The following result is a consequence of Theorem \ref{thm:change-operad} and Proposition \ref{lem:ochatmin-to-ocbarmin}.

\begin{corollary}\label{cor:cbarmin-chatmin}
Suppose $\C$ is a small category with object set $\colorc$.  Then the \index{diagram of monoids!as AQFT}morphism \[\nicexy{\Otom_{\Chatmin} \ar[r]^-{\deltaminm} & \Otom_{\Cbarmin}}\] of $\colorc$-colored operads induces a change-of-operad adjunction \[\nicexy{\Mcstar \cong \PFA(\Chatmin)=\algm\bigl(\Otom_{\Chatmin}\bigr) \ar@<2pt>[r]^-{\deltaminmst} & \algm\bigl(\Otom_{\Cbarmin}\bigr)\cong\QFT(\Cbarmin)=\Monm^{\C} \ar@<2pt>[l]^-{\deltaminmstar}}\] whose right adjoint $\deltaminmstar$ is the forgetful functor in Example \ref{ex:cbarmin-chatmin}.
\end{corollary}

\begin{interpretation}
In the minimal case, there is a morphism $\deltaminm$ from the $\colorc$-colored operad $\Otom_{\Chatmin}$ defining prefactorization algebras ($=$ pointed $\C$-diagrams in $\M$) to the $\colorc$-colored operad $\Otom_{\Cbarmin}$ for algebraic quantum field theories ($=$ $\C$-diagrams of monoids in $\M$).  This morphism of $\colorc$-colored operads induces a free-forgetful adjunction between the algebra categories.\dqed
\end{interpretation}

\section{Commutative Monoid Structure}\label{sec:pfa-com-monoid}

In this section, we observe that some entries of a prefactorization algebra are equipped with the structure of a commutative monoid.  In particular, this applies to the empty subset for prefactorization algebras on a topological space.  Recall from Example \ref{ex:operad-com} the commutative operad $\Com$, which is a $1$-colored operad whose algebras are commutative monoids in $\M$.  We will denote its unique color by $*$. For the following result, the example to keep in mind is the empty subset $\varnothing_X \subset X$ in a topological space $X$.

\begin{proposition}\label{prop:com-to-ochat}
Suppose $\Chat = (\C,\Config)$ is a \index{commutative monoid}\index{commutative operad}\index{prefactorization algebra!commutative monoid}configured category with object set $\colorc$, and $c \in \colorc$ such that \[\{\Id_c\}_{i=1}^n \in \Config\ccc\] for all $n$.  
\begin{enumerate}\item Then there is an operad morphism \[\nicexy{\Com \ar[r]^-{\iota_c} & \Ochatm}\] that sends $*$ to $c \in \colorc$ and that is entrywise defined by the summand inclusion \[\nicexy@C+.7cm{\Com(n) = \tensorunit \ar[r]^-{\{\Id_c\}_{i=1}^n}_-{\mathrm{summand}} & \coprodover{\Config\ccc}\tensorunit = \Ochatm\ccc} \forspace n \geq 0.\]
\item There is an induced change-of-operad adjunction \[\nicexy{\Comm = \algm(\Com) \ar@<2pt>[r]^-{(\iota_c)_!} & \algmochatm = \PFA(\Chat) \ar@<2pt>[l]^-{\iota_c^*}}.\]
\end{enumerate}
\end{proposition}

\begin{proof}
For the first assertion, one checks directly that this is a well-defined operad morphism.  The second assertion follows from the first assertion and Theorem \ref{thm:change-operad}.
\end{proof}

\begin{interpretation}
With $c \in \colorc$ as in Proposition \ref{prop:com-to-ochat}, if $(Y,\lambda)$ is a prefactorization algebra on $\Chat$, then $Y_c$ is equipped with the structure of a commutative monoid.  More explicitly, in the context of the Coherence Theorem \ref{thm:ochat-algebra}, the monoid multiplication in $Y_c$ is the structure morphism \[\nicexy@C+.7cm{Y_c \otimes Y_c \ar[r]^-{\lambda\{\Id_c,\Id_c\}} & Y_c} \in \M\] with $\{\Id_c,\Id_c\} \in \Config\sbinom{c}{c,c}$, and its unit is the structure morphism \[\nicexy@C+.7cm{\tensorunit \ar[r]^-{\lambda\{(c;\varnothing)\}} & Y_c} \in \M\] with $(c;\varnothing) \in \Config\cempty$.\dqed
\end{interpretation}

\begin{example}[Commutative monoid structure in prefactorization algebras on bounded lattices]\label{ex:com-lattice-pfa}
Consider the configured category $\Lhat = (L,\Configl)$ for a \index{bounded lattice!prefactorization algebra}bounded lattice $(L,\leq)$ in Example \ref{ex:boundedlatter-config}.  The least element $0 \in L$ has the property that \[\{\Id_{0}\}_{i=1}^n \in \Configl\sbinom{0}{0,\ldots,0} \forspace n\geq 0\] because $0 \leq d$ for all $d \in L$ and $0 \wedge 0 = 0$.  Therefore, by Proposition \ref{prop:com-to-ochat}, if $Y$ is a prefactorization algebra on $\Lhat$, then $Y_{0}$ is equipped with a commutative monoid structure whose multiplication is the structure morphism \[\nicexy@C+.7cm{Y_0 \otimes Y_0 \ar[r]^-{\lambda\{\Id_0,\Id_0\}} & Y_0} \in \M\] with $\{\Id_0,\Id_0\} \in \Configl\sbinom{0}{0,0}$ and whose unit is the structure morphism \[\nicexy@C+.7cm{\tensorunit \ar[r]^-{\lambda\{(0;\varnothing)\}} & Y_0} \in \M\] with $(0;\varnothing) \in \Configl\sbinom{0}{\varnothing}$.\dqed
\end{example}

\begin{example}[Commutative monoid structure in Costello-Gwilliam prefactorization algebras]\label{ex:com-pfa}
Consider the empty subset\index{prefactorization algebra!Costello-Gwilliam} $\varnothing_X \subset X$ in the configured category $\Openxhat$ in Example \ref{ex:open-configuration} for a topological space $X$.  This is a special case of Example \ref{ex:com-lattice-pfa} with $L=\Openx$ and least element $0=\varnothing_X$.  Therefore, by Proposition \ref{prop:com-to-ochat}, if $Y$ is a prefactorization algebra on $\Openxhat$ (i.e., a Costello-Gwilliam unital prefactorization algebra on $X$), then $Y_{\varnothing_X}$ is equipped with a commutative monoid structure.\dqed
\end{example}

\begin{example}[Commutative monoid structure in Costello-Gwilliam equivariant prefactorization algebras]\label{ex:com-eq-pfa}
Consider the configured category $\Openxghat$ for a topological \index{equivariant prefactorization algebra}\index{prefactorization algebra!equivariant}space $X$ with a left action by a group $G$ in Example \ref{ex:eq-space-configuration}.  The empty subset $\varnothing_X \subset X$ has the property that \[\bigl\{\Id_{\varnothing_X}\bigr\}_{i=1}^n \in \Configxg\sbinom{\varnothing_X}{\varnothing_X,\ldots,\varnothing_X} \forspace n\geq 0.\] Therefore, by Proposition \ref{prop:com-to-ochat}, if $Y$ is a prefactorization algebra on $\Openxghat$ (i.e., a Costello-Gwilliam unital $G$-equivariant prefactorization algebra on $X$), then $Y_{\varnothing_X}$ is equipped with a commutative monoid structure.\dqed \end{example}

\section{Diagrams of Modules over a Commutative Monoid}\label{sec:pfa-module-com}

In Example \ref{ex:com-lattice-pfa} we observed that, for each prefactorization algebra $Y$ on the configured category of a bounded lattice, the entry $Y_0$ is equipped with the structure of a commutative monoid.  In this section, we first observe that every other entry of $Y$ is equipped with the structure of a left $Y_0$-module in the sense of Definition \ref{def:module-monoid}.  Then we show that these left $Y_0$-modules are compatible with the diagram structure in Corollary \ref{cor:pfa-underlying-diagram}.  As in Example \ref{ex:lattice}, we will regard a lattice $(L,\leq)$ also as a category, where a morphism $c \to d$ exists if and only if $c \leq d$.

\begin{proposition}\label{prop:pfa-module-com}
Suppose $\Lhat = (L,\Configl)$ is the configured category of a bounded lattice $(L,\leq)$ with least element $0 \in L$ as in Example \ref{ex:boundedlatter-config}, and $(Y,\lambda)$ is a prefactorization algebra on $\Lhat$.  Then for each element $d \in L$, the entry $Y_d$ is equipped with the structure of a\index{commutative monoid!module}\index{prefactorization algebra!module} left $Y_0$-module via the structure morphism \[\nicexy@C+.7cm{Y_0 \otimes Y_d \ar[r]^-{\lambda\{0_d,\Id_d\}} & Y_d} \in \M\] with 
\begin{itemize}\item $0_d : 0\to d \in L$ the unique morphism and 
\item $\{0_d,\Id_d\} \in \Configl\sbinom{d}{0,d}$.
\end{itemize}
\end{proposition}

\begin{proof}
This is a consequence of the Coherence Theorem \ref{thm:ochat-algebra}.  To see that the required associativity diagram \[\nicexy@C+1cm{Y_0\otimes Y_0 \otimes Y_d \ar[d]_-{\bigl(\lambda\{\Id_0,\Id_0\},\Id\bigr)} \ar[r]^-{\bigl(\Id, \lambda\{0_d,\Id_d\}\bigr)} & Y_0 \otimes Y_d\ar[d]^-{\lambda\{0_d,\Id_d\}}\\ Y_0 \otimes Y_d \ar[r]^-{\lambda\{0_d,\Id_d\}} & Y_d}\] of a left $Y_0$-module is commutative, we apply the associativity condition \eqref{pfa-ass} to the equalities
\[\begin{split}
&\gamma\Bigl(\{0_d,\Id_d\}; \{\Id_0,\Id_0\}, \{\Id_d\}\Bigr)\\ &= \{0_d,0_d,\Id_d\}\\
&= \gamma\Bigl(\{0_d,\Id_d\}; \{\Id_0\}, \{0_d,\Id_d\}\Bigr) \in \Configl\sbinom{d}{0,0,d}.
\end{split}\]
Therefore, both composites in the previous diagram are equal to the structure morphism \[\nicexy@C+1cm{Y_0 \otimes Y_0 \otimes Y_d \ar[r]^-{\lambda\{0_d,0_d,\Id_d\}} & Y_d}\in \M.\]

Similarly, to see that the required unity diagram \[\nicexy@C+1cm{\tensorunit \otimes Y_d \ar[r]^-{\bigl(\lambda\{(0;\varnothing)\},\Id\bigr)} \ar[d]_-{\cong} & Y_0 \otimes Y_d \ar[d]^-{\lambda\{0_d,\Id_d\}}\\ Y_d \ar@{=}[r] & Y_d}\] of a left $Y_0$-module is commutative, first note that \[\Id_{Y_d} = \lambda\{\Id_d\}\] by the unity condition in Theorem \ref{thm:ochat-algebra}.  Therefore, the associativity condition \eqref{pfa-ass} applied to the equality \[\gamma\Bigl(\{0_d,\Id_d\}; \{(0;\varnothing)\}, \{\Id_d\}\Bigr) = \{\Id_d\} \in \Configl\dd\] yields the desired unity diagram.
\end{proof}

\begin{motivation}In Proposition \ref{prop:pfa-module-com} we observed that, for a bounded lattice $(L,\leq)$ with least element $0$ and for a prefactorization algebra $(Y,\lambda)$ on the configured category $\Lhat$, the entry $Y_0$ is equipped with the structure of a commutative monoid, and every other entry $Y_d$ is equipped with the structure of a left $Y_0$-module.  These left $Y_0$-module structures should be compatible with the $L$-diagram structure.  In the next result, we will consider the underlying $L$-diagram structure instead of pointed $L$-diagram.\dqed
\end{motivation}

\begin{corollary}\label{cor:lattice-diagram-modules}
Suppose $\Lhat = (L,\Configl)$ is the configured category of a bounded lattice $(L,\leq)$ with least element $0 \in L$ as in Example \ref{ex:boundedlatter-config}, and $(Y,\lambda)$ is a prefactorization algebra on $\Lhat$.  Then the underlying $L$-diagram in $\M$ of $(Y,\lambda)$ in Corollary \ref{cor:pfa-underlying-diagram} becomes an $L$-diagram of left $Y_0$-modules when equipped with the structure morphisms in Proposition \ref{prop:pfa-module-com}.
\end{corollary}

\begin{proof}
Suppose $g : c \to d$ in $L$, i.e., $c \leq d$.  We must show that the diagram
 \[\nicexy@C+1cm{Y_0 \otimes Y_c \ar[d]_-{\bigl(\Id,\lambda\{g\}\bigr)} \ar[r]^-{\lambda\{0_c,\Id_c\}} & Y_c \ar[d]^-{\lambda\{g\}}\\ Y_0 \otimes Y_d \ar[r]^-{\lambda\{0_d,\Id_d\}} & Y_d}\] 
in $\M$ is commutative, where $0_c : 0 \to c \in L$.  There are equalities
\[\begin{split} &\gamma\Bigl(\{0_d,\Id_d\}; \{\Id_0\}, \{g\}\Bigr)\\ &= \{0_d,g\}\\
&= \gamma\Bigl(\{g\}; \{0_c,\Id_c\}\Bigr) \in \Configl\sbinom{d}{0,c}.
\end{split}\]
Moreover, the unity condition in Theorem \ref{thm:ochat-algebra} implies that \[\lambda\{\Id_0\} = \Id_{Y_0}.\]  So the associativity condition \eqref{pfa-ass} applied to the above equalities implies that both composites in the above diagram are equal to the structure morphism \[\nicexy@C+.7cm{Y_0 \otimes Y_c \ar[r]^-{\lambda\{0_d,g\}} & Y_d}\]
in $\M$.
\end{proof}

\begin{interpretation} For each prefactorization algebra $Y$ on the configured category $\Lhat$ of a bounded lattice $L$:
\begin{itemize}\item $Y_0$ is equipped with the structure of a commutative monoid.
\item Every other entry $Y_d$ is equipped with the structure of a left $Y_0$-module.
\item The underlying $L$-diagram of $Y$ is an $L$-diagram of left $Y_0$-modules.\dqed
\end{itemize}
\end{interpretation}

\begin{example}[Costello-Gwilliam prefactorization algebras]\label{ex:pfa-mod-com-top}
For the configured category $\Openxhat$ of a topological \index{prefactorization algebra!Costello-Gwilliam}space $X$ and for a prefactorization algebra $(Y,\lambda)$ on $\Openxhat$, the entry $Y_{\varnothing_X}$ is a commutative monoid.  Furthermore, for each open subset $U \subset X$, the entry $Y_U$ is equipped with the structure of a left $Y_{\varnothing_X}$-module.   Furthermore, the underlying $\Openx$-diagram in $\M$ of $Y$ is an $\Openx$-diagram of left $Y_{\varnothing_X}$-modules.\dqed
\end{example}

\begin{example}[Costello-Gwilliam equivariant prefactorization algebras]\label{ex:pfa-mod-com-eqtop}
Suppose $X$ is a topological \index{equivariant prefactorization algebra}\index{prefactorization algebra!equivariant}space with a left action by a group $G$.  Consider the configured category $\Openxghat$ in Example \ref{ex:eq-space-configuration}.  There is a configured functor \[\nicexy{\Openxhat \ar[r]^-{\iota} & \Openxghat}\] that is the identity assignment on objects (i.e., open subsets of $X$) and morphisms (i.e., inclusions of open subsets).  By Lemma \ref{lem:ochat-operad} and Example \ref{ex:operad-set-m}, it induces a morphism of $\colorc$-colored operads \[\nicexy{\Otom_{\Openxhat} \ar[r]^-{\Otom_{\iota}} & \Otom_{\Openxghat}}\] where $\colorc = \Ob\bigl(\Openx\bigr)$.  By Theorem \ref{thm:change-operad} there is a change-of-operad adjunction
\[\nicexy{\PFA\bigl(\Openxhat\bigr) = \algm\bigl(\Otom_{\Openxhat}\bigr) \ar@<2pt>[r]^-{(\Otom_{\iota})_!} & \algm\bigl(\Otom_{\Openxghat}\bigr) =\PFA\bigl(\Openxghat\bigr) \ar@<2pt>[l]^-{(\Otom_{\iota})^*}},\] in which the right adjoint $(\Otom_{\iota})^*$ forgets about the structure isomorphisms \[\lambda\{g\} : Y_{U} \iso Y_{gU}\] for $g \in G$ and $U \in \Openx$. Therefore, by Example \ref{ex:pfa-mod-com-top} for each prefactorization algebra $(Y,\lambda)$ on $\Openxghat$, $Y_{\varnothing_X}$ is a commutative monoid, and every other entry $Y_U$ for $U \in \Openx$ is equipped with the structure of a left $Y_{\varnothing_X}$-module.  Furthermore, the underlying $\Openx$-diagram in $\M$ of $Y$ is an $\Openx$-diagram of left $Y_{\varnothing_X}$-modules.\dqed
\end{example}

\section{Diagrams of Commutative Monoids}\label{sec:diag-com-monoid}

In this section, we observe that diagrams of commutative monoids can be realized as prefactorization algebras on the maximal configured category $\Chatmax = (\C,\Configcmax)$ in Example \ref{ex:max-configuration}.   Furthermore, these diagrams of commutative monoids coincide with algebraic quantum field theories on the maximal orthogonal category $\Cbarmax = (\C,\perpmax)$ in Example \ref{ex:everything-causality}.  Taking $\C$ to be the category of (complex) $n$-manifolds, we recover prefactorization algebras on (complex) $n$-manifolds in the sense of Costello-Gwilliam.

\begin{proposition}\label{prop:deltamax}
Suppose $\C$ is a small category with object set $\colorc$.  Then there is a canonical \index{comparison morphism}\index{prefactorization algebra!comparison}\index{algebraic quantum field theory!comparison}isomorphism \[\nicexy{\O_{\Chatmax} \ar[r]^-{\deltamax}_-{\cong} & \O_{\Cbarmax}}\] of $\colorc$-colored operads in $\Set$.
\end{proposition}

\begin{proof}
By definition every pair of morphisms in $\C$ with the same codomain are orthogonal in $\Cbarmax$.  Therefore, two elements are equal \[[\sigma,\uf]=[\sigma',\uf'] \in \O_{\Cbarmax}\duc\] if and only if \[\uf=\uf' \in \prod_{i=1}^{|\uc|} \C(c_i,d).\]
\begin{itemize}
\item On $0$-ary entries there is a canonical bijection \[\nicexy{\O_{\Chatmax}\cempty = \Configcmax\cempty= \bigl\{(c;\varnothing)\bigr\} \ar[r]^-{\cong} & \Sigma_0\times * =\O_{\Cbarmax}\cempty}\] for $c \in \colorc$.
\item For $n$-ary entries with $n \geq 1$, there is a canonical bijection \[\nicexy{\O_{\Chatmax}\duc = \Configcmax\duc = \prod\limits_{i=1}^n \C(c_i,d) \ar[r]^-{\cong} & \O_{\Cbarmax}\duc}\] for $(\uc;d) \in \Profcc$ with $\uc = (c_1,\ldots,c_n)$. 
\end{itemize}
The required morphism $\deltamax$ is defined as these canonical isomorphisms.  Moreover, $\deltamax$ is a well-defined morphism of $\colorc$-colored operads because on both sides the structures are defined using the identity morphisms and the categorical composition in $\C$.
\end{proof}

\begin{example}[Prefactorization algebras are AQFT in the classical case]\label{ex:deltamax}
Applying the change-of-category functor \[(-)^{\M} : \Operadcset \to \Operadcm\] to $\deltamax$, we obtain a canonical isomorphism \[\nicexy{\Otom_{\Chatmax} \ar[r]^-{\deltamaxm}_-{\cong} & \Otom_{\Cbarmax}}\] of $\colorc$-colored operads in $\M$.  Therefore, the induced functor on algebra \index{prefactorization algebra!diagram of commutative monoids}\index{algebraic quantum field theory!diagram of commutative monoids}\index{diagram of commutative monoids}categories \[\nicexy{\PFA(\Chatmax)=\algm\bigl(\Otom_{\Chatmax}\bigr) & \algm\bigl(\Otom_{\Cbarmax}\bigr)\cong\QFT(\Cbarmax)=\Comm^{\C} \ar[l]_-{\deltamaxmstar}^-{\cong}}\] is also an isomorphism, where the equality $\QFT(\Cbarmax)=\Comm^{\C}$ is from Example \ref{ex:everything-causality}. In other words, in the maximal case (i.e., with $\Configcmax$ and $\perpmax$), prefactorization algebras coincide with algebraic quantum field theories, which in turn are precisely $\C$-diagrams of commutative monoids in $\M$.  Physically we interpret this isomorphism as saying that the two mathematical approaches to quantum field theory both reduce to the classical case where observables form commutative monoids.\dqed
\end{example}

\begin{example}[Prefactorization algebras as symmetric monoidal functors]\label{ex2:deltamax}
If the small category $\C$ has all small coproducts, then there is another nice description of $\C$-diagrams of commutative monoids in $\M$.  Indeed, by Proposition \ref{prop:finite-coprod} the category $\Comm^{\C}$ is canonically isomorphism to the category $\SMFun(\C,\M)$ of symmetric monoidal functors, where $\C$ is regarded as a\index{symmetric monoidal functor}\index{prefactorization algebra!symmetric monoidal functor}\index{algebraic quantum field theory!symmetric monoidal functor} symmetric monoidal category under coproducts.  So there are canonical isomorphisms \[\PFA(\Chatmax) \cong \QFT(\Cbarmax) = \Comm^{\C} \cong \SMFun(\C,\M)\] from the category of prefactorization algebras on $\Chatmax$ to the category of symmetric monoidal functors $\C \to \M$.\dqed
\end{example}

\begin{example}[Costello-Gwilliam prefactorization algebras on manifolds]\label{ex:pfa-manifolds}
Suppose $\Embn$ is a small category equivalent to the category of smooth $n$-manifolds with open embeddings as morphisms.  Symmetric monoidal functors $\Embn \to \M$ are called\index{prefactorization algebra!on manifolds} \emph{prefactorization algebras on $n$-manifolds with values in $\M$} in \cite{cg} Definition 6.3.0.2.  By Example \ref{ex2:deltamax} the category of such symmetric monoidal functors is isomorphic to the category of prefactorization algebras on the maximal configured category $\Embnhatmax$ and the category of algebraic quantum field theories on the maximal orthogonal category $\Embnbarmax$.\dqed
\end{example}

\begin{example}[Costello-Gwilliam prefactorization algebras on complex manifolds]\label{ex:pfa-cpmanifolds}
Suppose $\Holn$ is a small category equivalent to the category of complex $n$-manifolds with open holomorphic embeddings as morphisms.  Symmetric monoidal functors $\Holn \to \M$ are called \index{prefactorization algebra!on complex manifolds}\emph{prefactorization algebras on complex $n$-manifolds with values in $\M$} in \cite{cg} Definition 6.3.2.2.  By Example \ref{ex2:deltamax} the category of such symmetric monoidal functors is isomorphic to the category of prefactorization algebras on the maximal configured category $\Holnhatmax$ and the category of algebraic quantum field theories on the maximal orthogonal category $\Holnbarmax$.\dqed
\end{example}

\section{Configured and Homotopy Morita Equivalences}\label{sec:config-equivalence}

In Theorem \ref{thm:ocbar-algebra}(6) we observed that the equivalence type of the category $\QFT(\Cbar)$ of algebraic quantum field theories on $\Cbar$ is an invariant of the equivalence type of the orthogonal category $\Cbar$.  In this section, we prove a prefactorization algebra analogue of this result as well as a homotopical version.  First we need the following configured analogue of an orthogonal equivalence.

\begin{definition}\label{def:configured-equivalence}
A \index{configured equivalence}\index{equivalence!configured}\emph{configured equivalence} \[F : (\C,\Configc) \to (\D,\Configd)\] between configured categories is an equivalence $F : \C \to \D$ of categories such that \[(d;\{f_i\}) \in \Configc \iffspace (Fd;\{Ff_i\}) \in \Configd.\]
\end{definition}

\begin{example}\label{ex:config-eq-cat}
In the context of Example \ref{ex:pfa-manifolds}, different choices of a small category equivalent to the category of smooth $n$-manifolds yield configured categories connected by configured equivalences.  The same is true in the context of Example \ref{ex:pfa-cpmanifolds} for complex $n$-manifolds.\dqed\end{example}

\begin{theorem}\label{thm:config-equivalence}
Suppose $F : \Chat \to \Dhat$ is a configured equivalence.  Then the change-of-operad adjunction \[\nicexy@C+.5cm{\PFA(\Chat)=\algmochatm \ar@<2pt>[r]^-{(\Otom_F)_!} 
& \algmodhatm =\PFA(\Dhat) \ar@<2pt>[l]^-{(\Otom_F)^*}}\] is an adjoint equivalence.
\end{theorem}

\begin{proof}
By Theorem \ref{thm:equivalence-categories} it is enough to show that the right adjoint $(\Otom_F)^*$ is an equivalence of categories, i.e., full, faithful, and essentially surjective.  Since $F : \C \to \D$ is an equivalence of categories, for each object $d \in \D$, we can choose
\begin{itemize}\item an object $c_d \in \C$ and 
\item an isomorphism $h_d : Fc_d \iso d$ in $\D$.  
\end{itemize}
We can further insist that, if $d=Fc$ for some $c \in \C$, then
\begin{itemize}\item $c_{Fc}$ is chosen from within the $F$-pre-image of $Fc$, i.e., $Fc_{Fc}=Fc$, and 
\item $h_{Fc}$ is $\Id_{Fc}$. 
\end{itemize}
By the inclusivity axiom, each $\{h_d\}$ is a configuration in $\Dhat$.    We now check the three required properties of $(\Otom_F)^*$.  To simplify the presentation, we will write $(\Otom_F)^*(X,\lambda^X)$ as $X^*$ for each $\Odhatm$-algebra $(X,\lambda^X)$ and similarly for morphisms.  

To see that $(\Otom_F)^*$ is faithful, suppose \[\phi,\psi : (X,\lambda^X) \to (Y,\lambda^Y)\] are two morphisms of $\Odhatm$-algebras such that \[\phi^* = \psi^* : X^*\to Y^*\in \algmochatm.\] We must show that $\phi=\psi$ in $\algmodhatm$.  It is sufficient to prove this equality color-wise, so suppose $d \in \D$.  By the associativity and the unity conditions in the Coherence Theorem \ref{thm:ochat-algebra}, the structure morphism \[\nicexy@C+.5cm{X_{Fc_d} \ar[r]^-{\lambda^X\{h_d\}} & X_d} \in \M\] is invertible with inverse $\lambda^X\{h_d^{-1}\}$.  Since $\phi \in \algmodhatm$, the diagram \[\nicexy@C+.7cm{X^*_{c_d} = X_{Fc_d} \ar[r]^-{\lambda^X\{h_d\}}_-{\cong} \ar[d]_-{\phi^*_{c_d}~=}^-{\phi_{Fc_d}} & X_d \ar[d]^-{\phi_d}\\
Y^*_{c_d} = Y_{Fc_d} \ar[r]^-{\lambda^Y\{h_d\}}_-{\cong} & Y_d}\] is a special case of \eqref{ochatalg-morphism}, so it is commutative.  By the invertibility of $\lambda^X\{h_d\}$, we infer the equality \[\phi_d = \lambda^Y\{h_d\} \circ \phi^*_{c_d} \circ \lambda^X\{h_d^{-1}\}.\]  The same is true with $\psi \in \algmodhatm$ in place of $\phi$, so \[\psi_d = \lambda^Y\{h_d\} \circ \psi^*_{c_d} \circ \lambda^X\{h_d^{-1}\}.\] Since $\phi^*_{c_d} = \psi^*_{c_d}$ by assumption, we conclude that $\phi_d = \psi_d$.  This proves that the right adjoint $(\Otom_F)^*$ is faithful.

To see that $(\Otom_F)^*$ is full, suppose \[\varphi : X^* \to Y^* \in \algmochatm\] for some $\Odhatm$-algebras $(X,\lambda^X)$ and $(Y,\lambda^Y)$.  We must show that \[\varphi = \phi^* \forsomespace \phi : X \to Y \in \algmodhatm.\]  We define such a morphism $\phi$ entrywise as the composition
\begin{equation}\label{phid-def}
\nicexy@C+.7cm{X^*_{c_d} = X_{Fc_d} \ar[d]_-{\varphi_{c_d}} & X_d \ar[l]_-{\lambda^X\{h_d^{-1}\}}^-{\cong} \ar[d]^-{\phi_d}\\ Y^*_{c_d} = Y_{Fc_d} \ar[r]^-{\lambda^Y\{h_d\}}_-{\cong} & Y_d}
\end{equation}
for $d \in \D$, where the object $c_d \in \C$ and the isomorphism $h_d : Fc_d \iso d \in\D$ are as in the first paragraph.  

To show that $\phi$ is a morphism of $\Odhatm$-algebras, suppose $\{f_i\} \in \Configd\dud$ with $\ud=(d_1,\ldots,d_n)$ and each $f_i \in \D(d_i,d)$.  We must show that the diagram \[\nicexy{\bigotimes\limits_{i=1}^n X_{d_i} \ar[d]_-{\lambda^X\{f_i\}} \ar[r]^-{\bigotimes_i \phi_{d_i}} & \bigotimes\limits_{i=1}^n Y_{d_i} \ar[d]^-{\lambda^Y\{f_i\}}\\ X_d \ar[r]^-{\phi_d} & Y_d}\] in $\M$ is commutative.  For each $1 \leq i \leq n$, the composition \[\nicexy@C+.5cm{Fc_{d_i} \ar[d]_-{h_{d_i}}^-{\cong} \ar[r]^-{h_d^{-1}f_ih_{d_i}} & Fc_d\\ d_i \ar[r]^-{f_i} & d \ar[u]^-{\cong}_-{h_d^{-1}}}\] in $\D$ has a unique $F$-pre-image $g_i \in \C(c_{d_i},c_d)$ because $F$ is full and faithful.  Moreover, if $f_i$ is the identity morphism of $d$, then $g_i$ is the identity morphism of $c_d$.  By the inclusivity axiom and the composition axiom in Definition \ref{def:configcat}, there is a configuration 
\begin{equation}\label{fg=hfh}
\{Fg_i\}_{i=1}^n = \bigl\{h_d^{-1}f_ih_{d_i}\bigr\}_{i=1}^n \in \Configd\sbinom{Fc_d}{Fc_{d_1},\ldots,Fc_{d_n}}.
\end{equation}  
Since $F$ is a configured equivalence, this implies that there is a configuration 
\begin{equation}\label{gi-configuration}
\{g_i\}_{i=1}^n \in \Configc\sbinom{c_d}{c_{d_1},\ldots,c_{d_n}}.
\end{equation} 

The diagram
\[\nicexy@C+.3cm@R+.3cm{\bigotimes\limits_{i=1}^n X_{d_i}\ar[ddd]_-{\lambda^X\{f_i\}} \ar[rrr]^-{\bigotimes_i \phi_{d_i}} \ar[dr]^-{\bigotimes_i \lambda^X\{h_{d_i}^{-1}\}}
&&& \bigotimes\limits_{i=1}^n Y_{d_i} \ar[ddd]^-{\lambda^Y\{f_i\}}\\
& \bigotimes\limits_{i=1}^n X^*_{c_{d_i}} \ar[d]_-{\lambda^{X^*}\{g_i\}} \ar[r]^-{\bigotimes_i \varphi_{c_{d_i}}} & \bigotimes\limits_{i=1}^n Y^*_{c_{d_i}} \ar[d]^-{\lambda^{Y^*}\{g_i\}} \ar[ur]^-{\bigotimes_i \lambda^Y\{h_{d_i}\}} &\\
& X^*_{c_d} \ar[r]^-{\varphi_{c_d}} & Y^*_{c_d} \ar[dr]^-{\lambda^Y\{h_d\}} &\\
X_d \ar[rrr]^-{\phi_d} \ar[ur]^-{\lambda^X\{h_{d}^{-1}\}} &&& Y_d}\] in $\M$ is commutative:
\begin{itemize}
\item The top and bottom trapezoids are commutative by the definition of $\phi$ in \eqref{phid-def}.  
\item The left and right trapezoids are commutative by the associativity condition \eqref{pfa-ass} for $\lambda^X$ and $\lambda^Y$ and the equality in \eqref{fg=hfh}.
\item The middle square is commutative by \eqref{ochatalg-morphism} and \eqref{gi-configuration} because $\varphi$ is a morphism of $\Ochatm$-algebras.
\end{itemize}
So $\phi$ is a morphism of $\Odhatm$-algebras.

To show that $\phi^* = \varphi$, first observe that for each object $d \in \D$, there are equalities 
\[\phi^*_{c_d} = \phi_{Fc_d} = \varphi_{c_{Fc_d}}= \varphi_{c_d}\] by the definition of $\phi$.  Now suppose $c\in \C$.  We must show that $\phi^*_c=\varphi_c$.  We just proved that
\begin{equation}\label{phicfc}
\phi^*_{c_{Fc}} = \varphi_{c_{Fc}},
\end{equation} 
since $Fc \in \D$.  Note that $Fc_{Fc} = Fc$  by our choices of the objects $c_?$ in the first paragraph.  Since $F$ is full and faithful, there exists a unique isomorphism
\begin{equation}\label{rciso}
r_c : c \iso c_{Fc}\in \C \stspace Fr_c = \Id_{Fc}.
\end{equation}  
Since $\varphi$ is a morphism of $\Ochatm$-algebras, the diagram 
\[\nicexy@C+.4cm{X^*_c \ar[d]_-{\lambda^{X^*}\{r_c\}}^-{\cong} \ar[r]^-{\varphi_c} & Y^*_c \ar[d]^-{\lambda^{Y^*}\{r_c\}}_-{\cong} \\ X^*_{c_{Fc}} \ar[r]^-{\varphi_{c_{Fc}}} & Y^*_{c_{Fc}}}\]
in $\M$ is commutative.  So there is an equality
\begin{equation}\label{varphic-decomp}
\varphi_c = \lambda^{Y^*}\{r_c^{-1}\} \circ \varphi_{c_{Fc}} \circ \lambda^{X^*}\{r_c\}.
\end{equation}  
Similarly, since $\phi^*$ is a morphism of $\Ochatm$-algebras, there is an equality
\begin{equation}\label{phistarc-decomp}
\phi^*_c = \lambda^{Y^*}\{r_c^{-1}\} \circ \phi^*_{c_{Fc}} \circ \lambda^{X^*}\{r_c\}.
\end{equation}  
The desired equality $\phi^*_c = \varphi_c$ now follows from \eqref{phicfc}, \eqref{varphic-decomp}, and \eqref{phistarc-decomp}.  Therefore, the right adjoint $(\Otom_F)^*$ is full.

Finally, to prove that $(\Otom_F)^*$ is essentially surjective, suppose $(Z,\lambda^Z)$ is an  $\Ochatm$-algebra.  We must show that there exist 
\begin{itemize}\item $(W,\lambda^W) \in \algmodhatm$ and 
\item an isomorphism $Z \iso W^*$ of $\Ochatm$-algebras.  
\end{itemize}
First we define the entries of $W$ as \[W_d = Z_{c_d} \forspace d \in \D.\]  For each $c \in \C$, there is a canonical isomorphism \[\nicexy@C+.5cm{Z_c \ar[r]^-{\lambda^Z\{r_c\}}_-{\cong} & Z_{c_{Fc}} = W_{Fc} = W^*_c}\in \C\] with $r_c \in \C(c,c_{Fc})$ the isomorphism in \eqref{rciso}.

To define the $\Odhatm$-algebra structure morphism of $W$, suppose as above that $\{f_i\} \in \Configd\dud$ with each $f_i \in \D(d_i,d)$.  In \eqref{gi-configuration} we observed that there exists a unique configuration \[\{g_i\} \in \Configc \stspace \{Fg_i\}=\{h_d^{-1}f_ih_{d_i}\} \in \Configd.\] We now define the structure morphism $\lambda^W\{f_i\}$ as in the commutative diagram 
\begin{equation}\label{lambdaw-def}
\nicexy@C+.5cm{\bigotimes\limits_{i=1}^n W_{d_i} \ar@{=}[d] \ar[r]^-{\lambda^W\{f_i\}} & W_d\\ \bigotimes\limits_{i=1}^n Z_{c_{d_i}} \ar[r]^-{\lambda^Z\{g_i\}} & Z_{c_d} \ar@{=}[u]}
\end{equation}
in $\M$.  Since $(Z,\lambda^Z)$ satisfies the associativity, unity, and equivariance conditions in the Coherence Theorem \ref{thm:ochat-algebra}, it follows that $(W,\lambda^W)$ is an $\Odhatm$-algebra.  It remains to show that $\bigl\{\lambda^Z\{r_c\}\bigr\}_{c\in \C}$ is a morphism of $\Ochatm$-algebras.

Suppose \[\{p_j\} \in \Configc\sbinom{c}{c_1,\ldots,c_m}\] is a configuration with each $p_j \in \C(c_j,c)$, so \[\{Fp_j\} \in \Configd\sbinom{Fc}{Fc_1,\ldots,Fc_m}.\]  We must show that the diagram
\begin{equation}\label{lambdaz-square}
\nicexy@C+1cm{\bigotimes\limits_{j=1}^m Z_{c_j} \ar[d]_-{\lambda^Z\{p_j\}} \ar[r]^-{\bigotimes_j \lambda^Z\{r_{c_j}\}} & \bigotimes\limits_{j=1}^m W^*_{c_j} = \bigotimes\limits_{j=1}^m Z_{c_{Fc_j}} \ar[d]_-{\lambda^{W^*}\{p_j\}}^-{=~\lambda^W\{Fp_j\}}\\
Z_c \ar[r]^-{\lambda^Z\{r_c\}} & W^*_c=Z_{c_{Fc}}}
\end{equation}
in $\M$ is commutative.  Since $\{Fp_j\} \in \Configd$, as in \eqref{fg=hfh} there exists a unique configuration 
\begin{equation}\label{kj-configuration}
\{k_j\} \in \Configc\sbinom{c_{Fc}}{c_{Fc_1},\ldots,c_{Fc_m}} \stspace \{Fk_j\} = \{Fp_j\} \in \Configd\sbinom{Fc}{Fc_1,\ldots,Fc_m}
\end{equation} 
since \[h^{-1}_{Fc} =\Id_{Fc} \andspace h_{Fc_j}=\Id_{Fc_j}.\]  By the definition \eqref{lambdaw-def} we have \[\lambda^W\{Fp_j\} = \lambda^Z\{k_j\}.\]  Therefore, by the associativity condition \eqref{pfa-ass} for $(Z,\lambda^Z)$, to show the commutativity of the square \eqref{lambdaz-square}, it is enough to prove that the diagram \[\nicexy{c_j \ar[d]_-{p_j} \ar[r]^-{r_{c_j}} & c_{Fc_j} \ar[d]^-{k_j}\\ c \ar[r]^-{r_c} & c_{Fc}}\] in $\C$ is commutative for each $1 \leq j \leq m$.  Since $F$ is faithful, it is enough to show that the $F$-image diagram \[\nicexy{Fc_j \ar[d]_-{Fp_j} \ar[r]^-{Fr_{c_j}} & Fc_{Fc_j}=Fc_j \ar[d]^-{Fk_j}\\ Fc \ar[r]^-{Fr_c} & Fc_{Fc}=Fc}\] in $\D$ is commutative.  But since $Fr_c$ and each $Fr_{c_j}$ are the identity morphisms by \eqref{rciso} and since $Fp_j=Fk_j$ by \eqref{kj-configuration}, we conclude that the above square is commutative.  Therefore, the right adjoint $(\Otom_F)^*$ is essentially surjective.
\end{proof}

\begin{theorem}\label{thm:config-morita}
Suppose $F : \Chat \to \Dhat$ is a configured  functor, and $\M$ is a monoidal model category in which the colored operads $\Ochatm$ and $\Odhatm$ are admissible.
\begin{enumerate}\item The change-of-operad adjunction \[\nicexy@C+.5cm{\PFA(\Chat)=\algmochatm \ar@<2pt>[r]^-{(\Otom_F)_!} 
& \algmodhatm =\PFA(\Dhat) \ar@<2pt>[l]^-{(\Otom_F)^*}}\] is a Quillen adjunction.
\item If $F$ is a configured equivalence, then the operad morphism \[\Otom_F : \Ochatm \to \Odhatm\] is a\index{homotopy Morita equivalence}\index{Quillen equivalence} homotopy Morita equivalence; i.e., the change-of-operad adjunction is a Quillen equivalence.
\end{enumerate}
\end{theorem}

\begin{proof}
The proof is the same as that of Theorem \ref{thm:aqft-model}.  Here we use Theorem \ref{thm:config-equivalence} to infer that the unit of the change-of-operad adjunction is a natural isomorphism.
\end{proof}

\begin{interpretation} If two configured categories are connected by a configured equivalence, then their categories of prefactorization algebras have equivalent homotopy theories.  In particular, these two categories of prefactorization algebras are equivalent both before and after inverting the weak equivalences.\dqed\end{interpretation}

\begin{example}[Costello-Gwilliam prefactorization algebras on (complex) $n$-manifolds]\label{ex2:config-eq-cat}
In the context of Example \ref{ex:pfa-manifolds}, different\index{prefactorization algebra!on manifolds}\index{prefactorization algebra!on complex manifolds} choices of a small category equivalent to the category of smooth $n$-manifolds yield configured categories connected by configured equivalences.  By Theorem \ref{thm:config-equivalence} and Theorem \ref{thm:config-morita}, for any two different choices their categories of prefactorization algebras are connected by a change-of-operad adjunction that is both an adjoint equivalence and a Quillen equivalence.  The same is true in the context of Example \ref{ex:pfa-cpmanifolds} for complex $n$-manifolds.\dqed\end{example}

\chapter{Homotopy Prefactorization Algebras}\label{ch:hpa}

In this chapter, we define homotopy prefactorization algebras on a configured category and study their structure.  

\section{Overview}
In Section \ref{sec:hpa-operad} we define homotopy prefactorization algebras on a configured category $\Chat$ as algebras over the Boardman-Vogt construction $\wochatm$ of $\Ochatm$.  This definition makes sense because prefactorization algebras on $\Chat$ are defined as algebras over the colored operad $\Ochatm$.  The Boardman-Vogt construction comes with an augmentation $\eta : \wochatm \to \Ochatm$, which induces a change-of-operad adjunction between the category of prefactorization algebras and the category of homotopy prefactorization algebras.  In favorable cases, this change-of-operad adjunction is a Quillen equivalence.

Examples of homotopy prefactorization algebras are given in Section \ref{sec:ex-hpa}.  They include homotopy coherent versions of prefactorization algebras satisfying the time-slice axiom, Costello-Gwilliam (equivariant) prefactorization algebras, and prefactorization algebras on (complex) manifolds.  Moreover, homotopy coherent diagrams of $E_\infty$-algebras are  homotopy prefactorization algebras on the maximal configured category.

In Section \ref{sec:coherence-hpa} we record the coherence theorem for homotopy prefactorization algebras.  This is a special case of the Coherence Theorem \ref{thm:wo-algebra-coherence} for algebras over the Boardman-Vogt construction.  This coherence theorem is our main tool for understanding the structure on homotopy prefactorization algebras.

In Section \ref{sec:hpa-hcdiag} we observe that every homotopy prefactorization algebra on a configured category $\Chat$ has an underlying homotopy coherent pointed $\C$-diagram.  This is the homotopy coherent version of the fact that every prefactorization algebra on $\Chat$ has an underlying pointed $\C$-diagram.  Compared to a homotopy coherent $\C$-diagram, a homotopy coherent pointed $\C$-diagram has additional structure morphisms parametrized by truncated linear graphs in Example \ref{ex:truncated-linear-graph}.

In Section \ref{sec:hpa-hinverse} we observe that, when a set of morphisms $S$ in $\C$ is chosen, $\wochatsinvm$-algebras are homotopy coherent versions of prefactorization algebras on $\Chat$ satisfying the time-slice axiom with respect to $S$.  In particular, the structure morphism corresponding to each morphism $f \in S$ is invertible up to specified homotopies that are also structure morphisms.  This is the homotopical analogue of Theorem \ref{thm:pfa-timeslice}, which says that prefactorization algebras satisfying the time-slice axiom with respect to $S$ are equivalent to prefactorization algebras whose structure morphisms corresponding to all $f \in S$ are invertible.

In Section \ref{sec:hpa-einfinity} we show that some entries of a homotopy prefactorization algebra on a configured category $\Chat$ are $E_\infty$-algebras.  For example, this is the case if $\Chat$ comes from a bounded lattice $L$ with least element $0$.  In this case, for each homotopy prefactorization algebra on $\Lhat$, the $0$-entry is equipped with the structure of an $E_\infty$-algebra.  This is the homotopical analogue of the fact that, for each prefactorization algebra on $\Lhat$, the $0$-entry is equipped with the structure of a commutative monoid.  It follows that each homotopy Costello-Gwilliam (equivariant) prefactorization algebra has an $E_\infty$-algebra at the entry corresponding to the empty subset.

In Section \ref{sec:hpa-einfinity-module} we show that for a bounded lattice $L$ with least element $0$ and for a homotopy prefactorization algebra $Y$ on $\Lhat$, every other entry of $Y$ admits the structure of an $E_\infty$-module over the $E_\infty$-algebra $Y_0$.  This is the homotopy coherent version of the fact that, for a prefactorization algebra $Y$ on $\Lhat$, each entry $Y_d$ is a left $Y_0$-module.  In particular, this applies to homotopy Costello-Gwilliam (equivariant) prefactorization algebras.  

In Section \ref{sec:hpa-hcdiag-mod} we show that the objectwise $E_\infty$-module structure in Section \ref{sec:hpa-einfinity-module} is homotopically compatible with the homotopy coherent diagram structure in Section \ref{sec:hpa-hcdiag}.  This is the homotopical analogue of the fact that, for a prefactorization algebra $Y$ on $\Lhat$, the underlying $L$-diagram of $Y$ is actually an $L$-diagram of left $Y_0$-modules.

Finally, in Section \ref{sec:hpa-hcdiag-einfinity} we observe that every homotopy coherent $\C$-diagram of $E_\infty$-algebras can be realized as a homotopy prefactorization algebra on the maximal configured category on $\C$.  This implies that homotopy Costello-Gwilliam prefactorization algebras on (complex) manifolds are homotopy coherent diagrams of $E_\infty$-algebras.  This is the homotopical analogue of the definition of Costello-Gwilliam prefactorization algebras on (complex) manifolds as symmetric monoidal functors.

Throughout this chapter $(\M,\otimes,\tensorunit)$ is a cocomplete symmetric monoidal closed category with an initial object $\varnothing$ and a commutative segment $(J,\mu,0,1,\epsilon)$ as in Definition \ref{def:segment}.  For a small category $\C$, its object set will be denoted by $\colorc$.

\section{Homotopy Prefactorization Algebras as Operad Algebras}\label{sec:hpa-operad}

In this section, we define homotopy prefactorization algebras on a configured category using the Boardman-Vogt construction in Chapter \ref{ch:bv} and record their basic categorical properties.  

\begin{recollection}\label{rec:hpa-bv-operad}
For a $\colorc$-colored operad $\O$ in $\M$, its Boardman-Vogt construction $\wo$ is the  $\colorc$-colored operad with entries \[\wo\duc = \int^{T \in \uTreec\duc} \J[T] \otimes\O[T] \in \M,\] where $\uTreec\duc$ is the substitution category of $\colorc$-colored trees with profile $\duc$ in Definition \ref{def:treesub-category}.  The functors \[\J : \uTreecducop \to \M \andspace \O : \uTreecduc \to \M\] are induced by $J$ and $\O$ and are defined in Definition \ref{functor-J} and Corollary \ref{cor:operad-functor-subcat}, respectively.   The operad structure on $\wo$, defined in Definition \ref{def:wo-operad-structure}, is induced by tree substitution.  It is equipped with a natural augmentation $\eta : \wo \to \O$ of $\colorc$-colored operads, defined in Theorem \ref{thm:w-augmented}.  One should think of the augmentation as forgetting the lengths of the internal edges (i.e., the $\J$-component) and composing in the colored operad $\O$.

The strong symmetric monoidal functor $\Set \to \M$ in Example \ref{ex:operad-set-m}, sending a set $S$ to the $S$-indexed coproduct $\coprod_S \tensorunit$, yields the change-of-category functor \[(-)^{\M} : \Operadcset \to \Operadcm.\]  For the colored operad $\Ochat$ in Definition \ref{def:ochat-opread} for a configured category $\Chat = (\C,\Config)$, its image in $\Operadc(\M)$ will be denoted by $\Ochatm$.  Its entries are \[\Ochatm\duc = \coprod_{\Config\duc}\tensorunit \forspace \duc \in\Profcc.\] Also recall from Definition \ref{def:operad-algebra-generating} the category of algebras over a colored operad.
\end{recollection}

\begin{definition}\label{def:hpa}
Suppose $\Chat = (\C,\Config)$ is a configured category with object set $\colorc$, and $\wochatm\in \Operadcm$ is the Boardman-Vogt construction of $\Ochatm \in \Operadcm$.  
\begin{enumerate}\item We define the category\label{notation:hpfachat} \[\HPFA(\Chat) = \algmwochatm,\] whose objects are called \index{homotopy prefactorization algebra}\index{prefactorization algebra!homotopy}\emph{homotopy prefactorization algebras} on $\Chat$.
\item Suppose $S$ is a set of morphisms in $\C$.  We define the category\label{notation:hpfachats} \[\HPFA(\Chat,S) = \algmwochatsinvm,\] whose objects are called \index{homotopy time-slice axiom!homotopy prefactorization algebra}\emph{homotopy prefactorization algebras on $\Chat$ satisfying the homotopy time-slice axiom with respect to $S$}.
\end{enumerate}
\end{definition}

\begin{interpretation} The $\colorc$-colored operad $\wochatm$ is made up of $\colorc$-colored trees whose internal edges are decorated by the commutative segment $J$ and whose vertices are decorated by elements in the colored operad $\Ochat$ (i.e., configurations in $\Chat$) with the correct profile.  A homotopy prefactorization algebra has structure morphisms indexed by these decorated $\colorc$-colored trees.  The precise statement is the Coherence Theorem \ref{thm:hpa-coherence} below.\dqed
\end{interpretation}

The following observation compares prefactorization algebras and homotopy prefactorization algebras.  It is a special case of Theorem \ref{thm:operad-comparison}(1), Corollary \ref{cor:augmentation-adjunction}, and Corollary \ref{cor:wo-o-chaink}.  

\begin{corollary}\label{cor:hpa-pfa-adjunction}
Suppose $\Chat = (\C,\Config)$ is a configured category.  
\begin{enumerate}
\item The augmentation $\eta : \wochatm \to \Ochatm$ induces a change-of-operad adjunction 
\[\nicexy{\HPFA(\Chat)=\algmwochatm \ar@<2pt>[r]^-{\eta_!} & \algmochatm= \PFA(\Chat) \ar@<2pt>[l]^-{\eta^*}}.\] 
\item If $\M$ is a monoidal model category in which the colored operads $\Ochatm$ and $\wochatm$ are admissible, then the change-of-operad adjunction is a Quillen adjunction.
\item If $\M=\Chaink$ with $\fieldk$ a field of characteristic zero, then the change-of-operad adjunction is a \index{Quillen equivalence}\index{homotopy Morita equivalence}Quillen equivalence.
\end{enumerate}
\end{corollary}

\begin{interpretation} The right adjoint $\eta^*$ allows us to consider a prefactorization algebra on $\Chat$ as a homotopy prefactorization algebra on $\Chat$.  The left adjoint $\eta_!$ rectifies a homotopy prefactorization algebra to a prefactorization algebra.  Furthermore, if $\M$ is $\Chaink$, then the augmentation $\eta$ is a homotopy Morita equivalence.  In particular, the homotopy theory of homotopy prefactorization algebras is equivalent to the homotopy theory of prefactorization algebras over the same configured category.  So there is no loss of homotopical information by considering homotopy prefactorization algebras.\dqed\end{interpretation}

The next observation is about changing the configured categories.  It is a consequence of Theorem \ref{thm:operad-comparison}(1), Theorem \ref{thm:w-augmented}, Theorem \ref{thm:config-morita}, and Corollary \ref{cor:hpa-pfa-adjunction}.  The second assertion below uses the fact that Quillen equivalences have the $2$-out-of-$3$ property.

\begin{corollary}\label{cor:hpa-adjunction-diagram}
Suppose $F : \Chat \to \Dhat$ is a configured functor.
\begin{enumerate}\item There is an induced diagram of change-of-operad adjunctions
\[\nicexy@C+.7cm{\HPFA(\Chat)=\algmwochatm \ar@<-2pt>[d]_-{\eta_!} \ar@<2pt>[r]^-{(\W\Otom_F)_!} & 
\algmwodhatm= \HPFA(\Dhat) \ar@<-2pt>[d]_-{\eta_!} \ar@<2pt>[l]^-{(\W\Otom_F)^*}\\ \PFA(\Chat) =\algmochatm \ar@<2pt>[r]^-{(\Otom_F)_!} \ar@<-2pt>[u]_-{\eta^*} & \algmodhatm =\PFA(\Dhat) \ar@<2pt>[l]^-{(\Otom_F)^*} \ar@<-2pt>[u]_-{\eta^*}}\]
such that \[(\Otom_F)_!\eta_! = \eta_! (\W\Otom_F)_! \andspace  \eta^*(\Otom_F)^* = (\W\Otom_F)^*\eta^*.\]
\item If $\M$ is a monoidal model category in which the colored operads $\Ochatm$, $\Odhatm$,  $\wochatm$, and $\wodhatm$ are admissible, then all four change-of-operad adjunctions are Quillen adjunctions.
\item If $F$ is a configured equivalence and if $\M=\Chaink$ with $\fieldk$ a field of characteristic zero, then all four change-of-operad adjunctions are Quillen equivalences.
\end{enumerate}
\end{corollary}

\begin{interpretation} The right adjoint $(\wom_F)^*$ sends each homotopy prefactorization algebra on $\Dhat$ to one on $\Chat$.  The left adjoint $(\wom_F)_!$ sends each homotopy prefactorization algebra on $\Chat$ to one on $\Dhat$.  The equality \[(\Otom_F)_!\eta_! = \eta_! (\W\Otom_F)_!\] means that the left adjoint diagram is commutative.  The equality \[\eta^*(\Otom_F)^* = (\W\Otom_F)^*\eta^*\] means that the right adjoint diagram is commutative.  Moreover, if $F$ is a configured equivalence and if $\M$ is $\Chaink$, then all four operad morphisms in the commutative diagram \[\nicexy@C+.4cm{\wochatm \ar[r]^-{\wom_F} \ar[d]_-{\eta} & \wodhatm \ar[d]^-{\eta}\\ \Ochatm \ar[r]^-{\Otom_F} & \Odhatm}\] are homotopy Morita equivalences. In particular, the homotopy theory of homotopy prefactorization algebras on $\Chat$ is equivalent to the homotopy theory of homotopy prefactorization algebras on $\Dhat$.\dqed
\end{interpretation}

\section{Examples}\label{sec:ex-hpa}

In this section, we list some examples of homotopy prefactorization algebras.

\begin{example}[Homotopy coherent pointed diagrams]\label{ex:hcpdiag}
For\index{homotopy coherent pointed diagram} a small category $\C$, consider the minimal configured category $\Chatmin = (\C,\Configcmin)$ on $\C$ in Example \ref{ex:min-configuration}.  By Corollary \ref{cor:hpa-pfa-adjunction} and Proposition \ref{prop:pointed-diagram}, the augmentation \[\eta : \wom_{\Chatmin} \to \Otom_{\Chatmin}\] induces a change-of-operad adjunction 
\[\nicexy{\HPFA(\Chatmin)=\algm\bigl(\wom_{\Chatmin}\bigr) \ar@<2pt>[r]^-{\eta_!} & \algm\bigl(\Otom_{\Chatmin}\bigr)= \PFA(\Chatmin)\cong\Mcstar \ar@<2pt>[l]^-{\eta^*}}\] between the category of homotopy prefactorization algebras on $\Chatmin$ and the category of pointed $\C$-diagrams in $\M$.  Therefore, $\wom_{\Chatmin}$-algebras are homotopy coherent versions of pointed $\C$-diagrams in $\M$.  We will study them in details in Section \ref{sec:hpa-hcdiag}.  Furthermore, if $\M=\Chaink$ with $\fieldk$ a field of characteristic zero, then this adjunction is a Quillen equivalence.\dqed
\end{example}

\begin{example}[Underlying homotopy coherent pointed diagrams]\label{ex:underlying-hcpdiag}
Suppose $\Chat = (\C,\Config)$ is a configured category.  In Corollary \ref{cor:pfa-underlying-diagram} we saw that the identity functor on $\C$ induces a configured functor \[\nicexy{\Chatmin = (\C,\Configcmin) \ar[r]^-{i_0} & (\C,\Config)=\Chat}.\] By Corollary \ref{cor:hpa-adjunction-diagram} there is an induced diagram of change-of-operad adjunctions
\[\nicexy@C+.7cm{\HPFA(\Chatmin)=\algm\bigl(\wom_{\Chatmin}\bigr) \ar@<-2pt>[d]_-{\eta_!} \ar@<2pt>[r]^-{(\W\Otom_{i_0})_!} & 
\algmwochatm= \HPFA(\Chat) \ar@<-2pt>[d]_-{\eta_!} \ar@<2pt>[l]^-{(\W\Otom_{i_0})^*}\\ \Mcstar \cong \PFA(\Chatmin) =\algm\bigl(\Otom_{\Chatmin}\bigr) \ar@<2pt>[r]^-{(\Otom_{i_0})_!} \ar@<-2pt>[u]_-{\eta^*} & \algmochatm =\PFA(\Chat) \ar@<2pt>[l]^-{(\Otom_{i_0})^*} \ar@<-2pt>[u]_-{\eta^*}}\]
with commuting left adjoint diagram and commuting right adjoint diagram.  In particular, via the right adjoin $(\W\Otom_{i_0})^*$, every homotopy prefactorization algebra on $\Chat$ has an underlying $\wochatminm$-algebra, i.e., homotopy coherent pointed $\C$-diagram in $\M$.\dqed
\end{example}

\begin{example}[Homotopy prefactorization algebras with time-slice]\label{ex:hpatimeslice}
Suppose $\Chat = (\C,\Config)$ is a configured category, and $S$ is a set of morphisms in $\C$.  By the naturality of the Boardman-Vogt construction in Theorem \ref{thm:w-augmented} and the change-of-category functor $(-)^{\M}$, the $S$-localization morphism $\ell : \Ochat \to \Ochatsinv$ yields a commutative diagram \[\nicexy@C+.4cm{\wochatm \ar[d]_-{\eta} \ar[r]^-{\wellm} & \wochatsinvm \ar[d]^-{\eta}\\ \Ochatm \ar[r]^-{\ellm} & \Ochatsinvm}\] of $\colorc$-colored operads in $\M$, where $\colorc=\Obc$.  By Corollary \ref{cor:hpa-adjunction-diagram} there is an induced diagram of change-of-operad adjunctions
\[\nicexy@C+.7cm{\HPFA(\Chat)=\algmwochatm \ar@<-2pt>[d]_-{\eta_!} \ar@<2pt>[r]^-{\wellmst} & 
\algm\bigl(\wochatsinvm\bigr) = \HPFA(\Chat,S) \ar@<-2pt>[d]_-{\eta_!} \ar@<2pt>[l]^-{\wellmstar}\\ \PFA(\Chat) =\algmochatm \ar@<2pt>[r]^-{\ellmst} \ar@<-2pt>[u]_-{\eta^*} & \algmochatsinvm =\PFA(\Chat,S) \ar@<2pt>[l]^-{\ellmstar} \ar@<-2pt>[u]_-{\eta^*}}\]
with commuting left adjoint diagram and commuting right adjoint diagram.  Objects in $\algm\bigl(\wochatsinvm\bigr)$ are homotopy prefactorization algebras on $\Chat$ satisfying the homotopy time-slice axiom with respect to $S$ in Definition \ref{def:hpa}.

In Theorem \ref{thm:pfa-timeslice} we noted that $\Ochatsinvm$-algebras are equivalent to prefactorization algebras on $\Chat$ whose structure morphisms $\lambda\{s\}$ are isomorphisms for all $s \in S$.  Therefore, in each $\wochatsinvm$-algebra, the structure morphisms $\lambda\{s\}$ should be invertible up to coherent homotopies.  We will explain this in details in Section \ref{sec:hpa-hinverse}.\dqed
\end{example}

\begin{example}[Homotopy prefactorization algebras on bounded lattices]\label{ex:hpa-lattice}
For a\index{bounded lattice} bounded lattice $(L,\leq)$ with least element $0$, consider the configured category $\Lhat = (L,\Configl)$ in Example \ref{ex:boundedlatter-config}.   By Corollary \ref{cor:hpa-pfa-adjunction} the augmentation \[\eta : \wom_{\Lhat} \to \Otom_{\Lhat}\] induces a change-of-operad adjunction 
\[\nicexy{\HPFA(\Lhat)=\algm\bigl(\wom_{\Lhat}\bigr) \ar@<2pt>[r]^-{\eta_!} & \algm\bigl(\Otom_{\Lhat}\bigr)= \PFA(\Lhat)\ar@<2pt>[l]^-{\eta^*}}\] between the category of homotopy prefactorization algebras on $\Lhat$ and the category of prefactorization algebras on $\Lhat$.  Furthermore, if $\M=\Chaink$ with $\fieldk$ a field of characteristic zero, then this adjunction is a Quillen equivalence.

For a prefactorization algebra $(Y,\lambda)$ on $\Lhat$, we saw in Example \ref{ex:com-lattice-pfa} that $Y_0$ is equipped with the structure of a commutative monoid.  Furthermore, in Corollary \ref{cor:lattice-diagram-modules} we observed that the underlying $L$-diagram of $Y$ is an $L$-diagram of left $Y_0$-modules.  Therefore, for a homotopy prefactorization algebra $(Y,\lambda)$ on $\Lhat$:
\begin{itemize}\item $Y_0$ is equipped with the structure of an $E_\infty$-algebra.
\item Every other entry $Y_d$ with $d \in L$ is an $E_\infty$-module over $Y_0$.
\item These $E_\infty$-modules over $Y_0$ are compatible with the homotopy coherent $L$-diagram structure.
\end{itemize}
We will study these structures in details in Section \ref{sec:hpa-einfinity} to Section \ref{sec:hpa-hcdiag-mod}.\dqed
\end{example}

\begin{example}[Homotopy Costello-Gwilliam prefactorization algebras]\label{ex:hcgpfa}
For a topological\index{prefactorization algebra!Costello-Gwilliam} space $X$, consider the configured category $\Openxhat$ in Example \ref{ex:open-configuration}.  Since $\Openxhat$ is an example of $\Lhat$ for the bounded lattice $\Openx$, everything in Example \ref{ex:hpa-lattice} applies to $\Openxhat$.  In particular, for a $\wom_{\Openxhat}$-algebra $(Y,\lambda)$:
\begin{itemize}\item $Y$ has an underlying homotopy coherent pointed $\Openx$-diagram in $\M$ by Example \ref{ex:hcpdiag}.
\item $Y_{\varnothing_X}$ is equipped with the structure of an $E_\infty$-algebra, where $\varnothing_X \subset X$ is the empty subset.
\item Every other entry $Y_U$ with $U \in \Openx$ is an $E_\infty$-module over $Y_{\varnothing_X}$.
\item These $E_\infty$-modules over $Y_{\varnothing_X}$ are compatible with the homotopy coherent $\Openx$-diagram structure.\dqed
\end{itemize}
\end{example}

\begin{example}[Homotopy Costello-Gwilliam equivariant prefactorization algebras]\label{ex:hcgeqpfa}
For\index{equivariant prefactorization algebra}\index{prefactorization algebra!equivariant} a topological space $X$ equipped with a left action by a group $G$, consider the configured category $\Openxghat$ in Example \ref{ex:eq-space-configuration}.  In Example \ref{ex:pfa-mod-com-eqtop} we saw that there is a configured functor \[\nicexy{\Openxhat \ar[r]^-{\iota} & \Openxghat}.\]  By Corollary \ref{cor:hpa-adjunction-diagram} there is an induced diagram of change-of-operad adjunctions
\[\nicexy@C+.8cm{\HPFA(\Openxhat) \ar@{=}[d] & \HPFA(\Openxghat) \ar@{=}[d]\\ 
\algm\bigl(\wom_{\Openxhat}\bigr) \ar@<-2pt>[d]_-{\eta_!} \ar@<2pt>[r]^-{(\W\Otom_{\iota})_!} & \algm\bigl(\wom_{\Openxghat}\bigr)\ar@<-2pt>[d]_-{\eta_!} \ar@<2pt>[l]^-{(\W\Otom_{\iota})^*}\\ 
\algm\bigl(\Otom_{\Openxhat}\bigr) \ar@<2pt>[r]^-{(\Otom_{\iota})_!} \ar@<-2pt>[u]_-{\eta^*} & \algm\bigl(\Otom_{\Openxghat}\bigr) \ar@<2pt>[l]^-{(\Otom_{\iota})^*} \ar@<-2pt>[u]_-{\eta^*}\\
\PFA(\Openxhat) \ar@{=}[u] & \PFA(\Openxghat) \ar@{=}[u]}\]
with commuting left adjoint diagram and commuting right adjoint diagram.   In particular, via the right adjoin $(\W\Otom_{\iota})^*$, every homotopy prefactorization algebra on $\Openxghat$ also has the structure stated in Example \ref{ex:hcgpfa}.\dqed
\end{example}

\begin{example}[Homotopy coherent diagrams of $E_\infty$-algebras]\label{ex:hcdiag-einfinity}
For\index{homotopy coherent diagram!$E_\infty$-algebra} a small category $\C$, consider the maximal configured category $\Chatmax = (\C,\Configcmax)$ on $\C$ in Example \ref{ex:max-configuration}.  By Corollary \ref{cor:hpa-pfa-adjunction} and Example \ref{ex:deltamax}, the augmentation \[\eta : \wom_{\Chatmax} \to \Otom_{\Chatmax}\] induces a change-of-operad adjunction 
\[\nicexy{\HPFA(\Chatmax)=\algm\bigl(\wom_{\Chatmax}\bigr) \ar@<2pt>[r]^-{\eta_!} & \algm\bigl(\Otom_{\Chatmax}\bigr)= \PFA(\Chatmax)\cong\Comm^{\C} \ar@<2pt>[l]^-{\eta^*}}\] between the category of homotopy prefactorization algebras on $\Chatmax$ and the category of $\C$-diagrams of commutative monoids in $\M$.  Therefore, $\wom_{\Chatmax}$-algebras should be homotopy coherent $\C$-diagrams of $E_\infty$-algebras.  We will explain this in details in Section \ref{sec:hpa-hcdiag-einfinity}.  Furthermore, if $\M=\Chaink$ with $\fieldk$ a field of characteristic zero, then this adjunction is a Quillen equivalence.\dqed
\end{example}

\begin{example}[Homotopy Costello-Gwilliam prefactorization algebras on manifolds]\label{ex:hpfa-manifolds}
Suppose\index{homotopy prefactorization algebra!on manifolds} $\Embn$ is a small category equivalent to the category of smooth $n$-manifolds with open embeddings as morphisms.  Recall from Example \ref{ex:pfa-manifolds} that symmetric monoidal functors $\Embn \to \M$ are called prefactorization algebras on $n$-manifolds with values in $\M$.  The category of such objects is isomorphic to the category $\PFA(\Embnhatmax)$ of prefactorization algebras on $\Embnhatmax$.  By Example \ref{ex:hcdiag-einfinity} this category is related to the category of homotopy prefactorization algebras on $\Embnhatmax$, i.e., $\wom_{\Embnhatmax}$-algebras, via the change-of-operad adjunction.  In Section \ref{sec:hpa-hcdiag-einfinity} we will see that homotopy prefactorization algebras on $\Embnhatmax$ are homotopy coherent $\Embn$-diagrams of $E_\infty$-algebras.\dqed
\end{example}

\begin{example}[Homotopy Costello-Gwilliam prefactorization algebras on complex manifolds]\label{ex:hpfa-cpmanifolds}
Suppose\index{homotopy prefactorization algebra!on complex manifolds} $\Holn$ is a small category equivalent to the category of complex $n$-manifolds with open holomorphic embeddings as morphisms.  Recall from Example \ref{ex:pfa-cpmanifolds} that symmetric monoidal functors $\Holn \to \M$ are called prefactorization algebras on complex $n$-manifolds with values in $\M$.  The category of such objects is isomorphic to the category $\PFA(\Holnhatmax)$ of prefactorization algebras on $\Holnhatmax$.  By Example \ref{ex:hcdiag-einfinity} this category is related to the category of homotopy prefactorization algebras on $\Holnhatmax$, i.e., $\wom_{\Holnhatmax}$-algebras, via the change-of-operad adjunction.  In Section \ref{sec:hpa-hcdiag-einfinity} we will see that homotopy prefactorization algebras on $\Holnhatmax$ are homotopy coherent $\Holn$-diagrams of $E_\infty$-algebras.\dqed
\end{example}

\section{Coherence Theorem}\label{sec:coherence-hpa}

In this section, we record the following coherence theorem for homotopy prefactorization algebras.  In the remaining sections of this chapter, we will use this coherence theorem to study the structure of homotopy prefactorization algebras.  For a vertex $v$ in a colored tree, recall our convention of writing $(v)$ for its profile.  So if $\profofv = \duc$, then $\Config(v) = \Config\duc$ for a configured category $(\C,\Config)$ as in Definition \ref{def:config-functor}.

\begin{theorem}\label{thm:hpa-coherence}
Suppose\index{Coherence Theorem!for homotopy prefactorization algebras}\index{homotopy prefactorization algebra!coherence} $\Chat = (\C,\Config)$ is a configured category with object set $\colorc$.  Then a $\wochatm$-algebra is precisely a pair $(X,\lambda)$ consisting of
\begin{itemize}\item a $\colorc$-colored object $X$ in $\M$ and
\item a structure morphism\index{structure morphism!for homotopy prefactorization algebras}
\begin{equation}\label{wochatm-restricted}
\nicexy@C+1.3cm{\J[T]\otimes X_{\uc} \ar[r]^-{\lambda_T\left\{\uf^v\right\}_{v\in T}} & X_d}\in \M
\end{equation}
for 
\begin{itemize}\item each $T \in \uTreec\duc$ with $(\uc;d) \in \Profcc$ and 
\item each $\left\{\uf^v\right\}_{v\in T} \in \prod_{v\in T} \Config(v)$
\end{itemize}
\end{itemize}
that satisfies the following four conditions.
\begin{description}
\item[Associativity] Suppose $\bigl(\uc=(c_1,\ldots,c_n);d\bigr) \in \Profcc$ with $n \geq 1$, $T \in \uTreecduc$, $T_j \in \uTreec\cjubj$ for $1 \leq j \leq n$, $\ub=(\ub_1,\ldots,\ub_n)$,  \[G=\graft(T;T_1,\ldots,T_n) \in \uTreec\dub\] is the grafting \eqref{def:grafting}, $\left\{\uf^v\right\}$ is as above, and $\left\{\uf^u\right\} \in \prod_{u\in T_j} \Config(u)$ for each $1 \leq j \leq n$.  Then the diagram
\begin{equation}\label{wochatm-ass}
\nicexy@R+.4cm@C+.8cm{\J[T]\otimes \Bigl(\bigotimes\limits_{j=1}^n \J[T_j]\Bigr)\otimes X_{\ub} \ar[d]_-{\mathrm{permute}}^-{\cong} \ar[r]^-{(\pi,\Id)} & \J[G]\otimes X_{\ub} \ar[dd]^-{\lambda_G\left\{\uf^w\right\}_{w\in G}}\\
\J[T]\otimes \bigotimes\limits_{j=1}^n\Bigl(\J[T_j]\otimes X_{\ub_j}\Bigr) \ar[d]_-{\left(\Id,\bigtensorover{j} \lambda_{T_j} \left\{\uf^u\right\}_{u\in T_j}\right)} & \\ 
\J[T]\otimes X_{\uc} \ar[r]^-{\lambda_T\left\{\uf^v\right\}_{v\in T}} & X_d}
\end{equation}
is commutative.  Here $\pi=\bigotimes_S 1$ is the morphism in Lemma \ref{lem:morphism-pi} for the grafting $G$.
\item[Unity] For each $c \in \colorc$, the composition
\begin{equation}\label{wochatm-unity}
\nicexy@C+.5cm{X_c \ar[r]^-{\cong} & \J[\uparrow_c]\otimes X_c \ar[r]^-{\lambda_{\uparrow_c}\{\varnothing\}} & X_c}
\end{equation} 
is the identity morphism of $X_c$.
\item[Equivariance] For each $T \in \uTreec\duc$, $\left\{\uf^v\right\}$ as above, and permutation $\sigma \in \Sigma_{|\uc|}$, the diagram 
\begin{equation}\label{wochatm-eq}
\nicexy@C+1.4cm{\J[T]\otimes X_{\uc} \ar[d]_-{(\Id,\sigmainv)} \ar[r]^-{\lambda_T\left\{\uf^v\right\}_{v\in T}} & X_d \ar@{=}[d]\\
\J[T\sigma]\otimes X_{\uc\sigma} \ar[r]^-{\lambda_{T\sigma}\left\{\uf^v\right\}_{v\in T\sigma}} & X_d}
\end{equation}
is commutative, in which $T\sigma \in \uTreec\ducsigma$ is the same as $T$ except that its ordering is $\zeta_T\sigma$ with $\zeta_T$ the ordering of $T$.  The permutation $\sigmainv : X_{\uc} \iso X_{\uc\sigma}$ permutes the factors in $X_{\uc}$.
\item[Wedge Condition] Suppose $T \in \uTreec\duc$, $H_v \in \uTreec(v)$ for each $v\in \Vt(T)$, $K=T(H_v)_{v\in T}$ is the tree substitution, and $\left\{\uf^u\right\} \in \prod_{u\in H_v} \Config(u)$ for each $v \in \Vt(T)$.  Then the diagram
\begin{equation}\label{wochatm-wedge}
\nicexy@C+1.5cm{\J[T] \otimes X_{\uc} \ar[d]_-{(\J,\Id)} \ar[r]^-{\lambda_T\left\{\ug^v\right\}_{v\in T}} & X_d \ar@{=}[d]\\ \J[K]\otimes X_{\uc} \ar[r]^-{\lambda_K\left\{\uf^w\right\}_{w\in K}} & X_d}
\end{equation}
is commutative.  Here for each $v \in \Vt(T)$, \[\ug^v = \gamma^{\Ochat}_{H_v}\Bigl(\left\{\uf^u\right\}_{u\in H_v}\Bigr) \in \Ochat(v)=\Config(v)\] with \[\nicexy@C+.5cm{\Ochat[H_v] = \prodover{u\in H_v} \Ochat(u) \ar[r]^-{\gamma^{\Ochat}_{H_v}} & \Ochat(v)}\] the operadic structure morphism of $\Ochat$ for $H_v$ in \eqref{operadic-structure-map}.
\end{description}
A morphism $f : (X,\lambda^X) \to (Y,\lambda^Y)$ of $\wochatm$-algebras is a morphism of the underlying $\colorc$-colored objects that respects the structure morphisms in \eqref{wochatm-restricted} in the obvious sense.
\end{theorem}

\begin{proof}
This is the special case of the Coherence Theorem \ref{thm:wo-algebra-coherence} applied to the $\colorc$-colored operad $\Ochatm$.  Indeed, since \[\Ochatm\duc = \coprodover{\Ochat\duc}\tensorunit = \coprodover{\Config\duc}\tensorunit,\] for each $T \in \uTreec\duc$ there is a canonical isomorphism \[\Ochatm[T] = \bigotimes_{v\in T} \Ochatm(v) = \bigotimes_{v\in T}\Bigl(\coprodover{\Config(v)} \tensorunit\Bigr) \cong \coprodover{\prodover{v\in T}\Config(v)} \tensorunit.\]  This implies that there is a canonical isomorphism \[\J[T]\otimes \Ochatm[T]\otimes X_{\uc} \cong \coprodover{\prodover{v\in T}\Config(v)} \J[T]\otimes X_{\uc}.\]  Therefore, the structure morphism $\lambda_T$ in \eqref{wo-algebra-restricted} is uniquely determined by the restricted structure morphisms $\lambda_T\left\{\uf^v\right\}_{v\in T}$ in \eqref{wochatm-restricted}.  The above associativity, unity, equivariance, and wedge conditions are those in the Coherence Theorem \ref{thm:wo-algebra-coherence}.
\end{proof}

\begin{interpretation} In a homotopy prefactorization algebra on a configured category $\Chat$, the structure morphism $\lambda_T\left\{\uf^v\right\}_{v\in T}$ is specified by (i) first choosing a $\colorc$-colored tree $T \in \uTreec\duc$ and (ii) then choosing a configuration $\uf^v \in \Config(v)$ for each vertex $v$ in $T$.  In other words, the structure morphisms are parametrized by $\colorc$-colored trees whose internal edges are decorated by the commutative segment $J$ and whose vertices are decorated by configurations in $\Chat$ with the correct profile.\dqed
\end{interpretation}

\begin{example}[Homotopy prefactorization algebras on bounded lattices]\label{ex:hpacoherence-lattice}
For a\index{bounded lattice} bounded lattice $(L,\leq)$ with least element $0$, consider the configured category $\Lhat = (L,\Configl)$ in Example \ref{ex:boundedlatter-config}.   For any two elements $c,d \in L$, there is a morphism $c \to d$ in $L$, which is necessarily unique, if and only if $c \leq d$.  Each subset $\Configl\duc$ of configurations is either empty or a one-element set.  If $v \in \Vt(T)$ has profile $\baoneam$, then \[\uf^v \in \Configl(v)=\Configl\baoneam\] if and only if 
\begin{itemize}\item $a_i \leq b$ in $L$ for each $1 \leq i \leq m$ and
\item $a_i \wedge a_j = 0$ for all $1 \leq i \not= j \leq m$.
\end{itemize}
In this case, \[\uf^v=\bigl\{a_i\to b\bigr\}_{i=1}^m\] is the unique element in $\prod_{i=1}^m L(a_i,b)$.\dqed
\end{example}

\begin{example}[Homotopy Costello-Gwilliam prefactorization algebras]\label{ex:hcgpfa-coherence}
For\index{prefactorization algebra!Costello-Gwilliam} a topological space $X$, consider the configured category $\Openxhat$ in Example \ref{ex:open-configuration}.  Since $\Openxhat$ is an example of $\Lhat$ for the bounded lattice $\Openx$, everything in Example \ref{ex:hpacoherence-lattice} applies to $\Openxhat$.  Suppose $T$ is an $\Openx$-colored tree, and $v \in \Vt(T)$ has profile $\vuoneum$ with $U_1,\ldots,U_m,V\in \Openx$.  Then \[\uf^v \in \Configx(v)=\Configx\vuoneum\] if and only if $\{U_i\}_{i=1}^m$ are pairwise disjoint subsets of $V$.  In this case, \[\uf^v=\bigl\{U_i\subset V\bigr\}_{i=1}^m\] is the unique element in $\prod_{i=1}^m \Openx(U_i,V)$.\dqed
\end{example}

\begin{example}[Homotopy Costello-Gwilliam equivariant prefactorization algebras]\label{ex:hcgeqpfa-coherence}
For\index{equivariant prefactorization algebra}\index{prefactorization algebra!equivariant} a topological space $X$ equipped with a left action by a group $G$, consider the configured category $\Openxghat$ in Example \ref{ex:eq-space-configuration}.  The objects in $\Openxg$ are the objects in $\Openx$, i.e., open subsets of $X$, but there are more morphisms in $\Openxg$ than in $\Openx$.  Suppose $T$ is an $\Openx$-colored tree, and $v \in \Vt(T)$ has profile $\vuoneum$ with $U_1,\ldots,U_m,V\in \Openx$.  Then \[\uf^v \in \Configxg(v)=\Configxg\vuoneum\] if and only if $\uf^v$ has the form \[\Bigl\{\nicexy@C+.3cm{U_i \ar[r]^-{g_i} & g_iU_i \ar[r]^-{\mathrm{inclusion}} & V}\Bigr\}_{i=1}^m \in \prod_{i=1}^m \Openxg(U_i,V)\] for some $g_1,\ldots,g_m \in G$ such that $\{g_iU_i\}_{i=1}^m$ are pairwise disjoint subsets of $V$.\dqed
\end{example}

\section{Homotopy Coherent Pointed Diagrams}\label{sec:hpa-hcdiag}

Using the Coherence Theorem \ref{thm:hpa-coherence}, for the next few sections we will explain the structure that exists in homotopy prefactorization algebras, i.e., in $\wochatm$-algebras.   In this section, we explain the homotopy coherent pointed diagram structure that exists on each homotopy prefactorization algebra.

\begin{definition}\label{def:hcpt-diagram}
Objects in the category $\algmwochatminm$ are called \index{homotopy coherent pointed diagram}\emph{homotopy coherent pointed $\C$-diagrams in $\M$}.
\end{definition}

\begin{interpretation} The $\colorc$-colored operad $\Ochatminm$ is the operad for pointed $\C$-diagrams in $\M$, so algebras over its Boardman-Vogt construction are homotopy coherent pointed $\C$-diagrams in $\M$.  We saw in Example \ref{ex:underlying-hcpdiag} that the right adjoint in the change-of-operad adjunction \[\nicexy@C+.7cm{\HPFA(\Chatmin)= \algmwochatminm \ar@<2pt>[r]^-{(\W\Otom_{i_0})_!} & 
\algmwochatm= \HPFA(\Chat) \ar@<2pt>[l]^-{(\W\Otom_{i_0})^*}}\] sends each homotopy prefactorization algebra on $\Chat$ to its underlying homotopy coherent pointed $\C$-diagram in $\M$.\dqed
\end{interpretation}

To understand homotopy coherent pointed $\C$-diagrams in $\M$, first we make explicit the colored operad $\Ochatminm$.  The next result is a consequence of the definition of $\Chatmin$ in Example \ref{ex:min-configuration} and of Definition \ref{def:ochat-opread}.

\begin{lemma}\label{lem:ochatminm}
Suppose\index{pointed diagram!operad}\index{colored operad!for pointed diagrams} $\C$ is a small category with object set $\colorc$.  Then the $\colorc$-colored operad $\Ochatminm$ has entries \[\Ochatminm\duc = \begin{cases}\tensorunit & \text{ if $\uc=\varnothing$},\\ \coprodover{\C(c,d)} \tensorunit & \text{ if $\uc=c\in \colorc$},\\
\varnothing & \text{ if $|\uc|\geq 2$}\end{cases}\] for $(\uc;d) \in \Profcc$.
\end{lemma}

\begin{motivation} The following result is the coherence theorem for homotopy coherent pointed diagrams.  A pointed $\C$-diagram in $\M$ consists of a $\C$-diagram in $\M$ and compatible colored units for objects in $\C$.  Therefore, a homotopy coherent pointed $\C$-diagram in $\M$ should contain a homotopy coherent $\C$-diagram in $\M$ along with homotopically compatible homotopy colored units.\dqed
\end{motivation}

We will refer to (i) the Coherence Theorem \ref{thm:hcdiagram} for homotopy coherent $\C$-diagrams in $\M$, (ii) the linear graphs $\Lin_?$ in Example \ref{ex:linear-graph}, and (iii) the truncated linear graphs $\lin_{?}$ in Example \ref{ex:truncated-linear-graph}.

\begin{theorem}\label{thm:hcptdiagram}
A \index{Coherence Theorem!for homotopy coherent pointed diagrams}\index{homotopy coherent pointed diagram!coherence}homotopy coherent pointed $\C$-diagram in $\M$ is exactly a triple $(X,\lambda,\theta)$ consisting of 
\begin{itemize}\item a homotopy coherent $\C$-diagram $(X,\lambda)$ in $\M$ and
\item a structure morphism\index{structure morphism!for homotopy coherent pointed diagrams}
\begin{equation}\label{hcptdiagram-structure-map}
\nicexy{\J[\lin_{\uc}]\ar[r]^-{\theta_{\uc}^{\uf}} & X_{c_n}}\in \M
\end{equation}
for 
\begin{itemize}\item each profile $\uc=(c_1,\ldots,c_n)\in\Profc$ with $n \geq 1$;
\item each sequence of composable $\C$-morphisms $\uf=(f_2,\ldots,f_n)$ with $f_j \in \C(c_{j-1},c_j)$ for $2\leq j \leq n$
\end{itemize}
\end{itemize}
that satisfies the following two conditions.
\begin{description}
\item[Associativity] Suppose $1 \leq n \leq p$, $\uc=(c_1,\ldots,c_n)$, and $\uc'=(c_n,\ldots,c_p)\in \Profc$.  Suppose $f_j \in \C(c_{j-1},c_j)$ for each $2\leq j \leq p$ with $\uf=(f_2,\ldots,f_n)$ and $\uf'=(f_{n+1},\ldots,f_p)$.  Then the diagram
\begin{equation}\label{hcptdiagram-ass}
\nicexy@C+.5cm@R+.4cm{\J[\Lin_{\uc'}] \otimes \J[\lin_{\uc}] \ar[d]_-{(\Id, \theta^{\uf}_{\uc})} \ar[r]^-{\pi} & \J\left[\lin_{(c_1,\ldots,c_p)}\right] \ar[d]^-{\theta^{(\uf,\uf')}_{(c_1,\ldots,c_p)}} \\
\J[\Lin_{\uc'}]\otimes X_{c_n} \ar[r]^-{\lambda^{\uf'}_{\uc'}} & X_{c_p}}
\end{equation}
is commutative.  Here the truncated linear graph $\lin_{(c_1,\ldots,c_p)}$ is regarded as the grafting \eqref{def:grafting} of the linear graph $\Lin_{\uc'}$ and the truncated linear graph $\lin_{\uc}$ with $\pi$ the morphism in Lemma \ref{lem:morphism-pi}.
\item[Wedge Condition] Suppose $\uc=(c_1,\ldots,c_n)\in \Profc$ with $n \geq 1$, 
\[\ub_j=\begin{cases}\bigl(b^1_1,\ldots,b^1_{k_1}=c_1\bigr) & \text{ with $k_1\geq 1$ for $j=1$},\\ \bigl(c_{j-1}=b^j_0,b^j_1,\ldots,b^j_{k_j}=c_j\bigr) & \text{ with $k_j\geq 0$ for $2 \leq j \leq n$}\end{cases}\] 
in $\Profc$, and $\ub=(\ub_1,\ldots,\ub_n)$.  Suppose $f^j_i \in \C(b^j_{i-1},b^j_i)$ for each $1\leq j \leq n$ and $1 \leq i \leq k_j$ except for $f^1_1$, 
\[\uf^j=\begin{cases} (f^1_2,\ldots,f^1_{k_1}) & \text{ if $j=1$},\\
(f^j_1,\ldots,f^j_{k_j}) & \text{ if $2\leq j \leq n$},\end{cases}\] 
$\uf=(\uf^1,\ldots,\uf^n)$, and 
\[f^j=\begin{cases} f^1_{k_1} \comp \cdots \comp f^1_2 \in \C(b^1_1,c_1) & \text{ if $j=1$},\\
f^j_{k_j}\comp \cdots \comp f^j_1 \in \C(c_{j-1},c_j) & \text{ if $2 \leq j\leq n$}.\end{cases}\]  
Then the diagram
\begin{equation}\label{hcptdiagram-wedge}
\nicexy@C+1.5cm{\J[\lin_{\uc}] \ar[d]_-{\J} \ar[r]^-{\theta^{(f^1,\ldots,f^n)}_{\uc}} & X_{c_n} \ar@{=}[d]\\
\J[\lin_{\ub}] \ar[r]^-{\theta^{\uf}_{\ub}} & X_{c_n}}
\end{equation}
is commutative.  Here the truncated linear graph $\lin_{\ub}$ is regarded as the tree substitution \[\lin_{\ub}=\lin_{\uc}\Bigl(\lin_{\ub_1}, \Lin_{\ub_2},\ldots, \Lin_{\ub_n}\Bigr).\]
\end{description}
\end{theorem}

\begin{proof}
This is the special case of the Coherence Theorem \ref{thm:wo-algebra-coherence} for the $\colorc$-colored operad $\Ochatminm$.  Indeed, by Lemma \ref{lem:ochatminm} the equivariant structure on $\Ochatminm$ is trivial.  Since $\Ochatminm$ is concentrated in $0$-ary and unary entries, if $T \in \uTreec\duc$ is neither a linear graph nor a truncated linear graph, then \[\Ochatminm[T] = \bigtensorover{v\in T} \Ochatminm\inoutv =\varnothing.\]  In this case, the structure morphism \[\nicexy{\J[T]\otimes\Ochatminm[T]\otimes X_{\uc} \ar[r]^-{\lambda_T} & X_d}\] in \eqref{wo-algebra-restricted} for a $\wochatminm$-algebra is the trivial morphism $\varnothing \to X_d$.  In particular, the equivariance condition \eqref{wo-alg-eq} is trivial for $\wochatminm$-algebras.  For linear graphs $T$, we have exactly the structure of a homotopy coherent $\C$-diagram in $\M$ as in Theorem \ref{thm:hcdiagram}.

For $\uc=(c_1,\ldots,c_n) \in \Profc$ with $n\geq 1$, we have natural isomorphisms
\[\Ochatminm[\lin_{\uc}] = \Ochatminm\sbinom{c_1}{\varnothing} \otimes \biggl[\bigotimes_{j=2}^n \Ochatminm\sbinom{c_j}{c_{j-1}}\biggr] \cong \bigotimes_{j=2}^n \biggl[\coprodover{\C(c_{j-1},c_j)} \tensorunit\biggr] \cong \coprodover{\prod_{j=2}^n \C(c_{j-1},c_j)} \tensorunit.\]
This implies that there is a natural isomorphism \[\J[\lin_{\uc}] \otimes \Ochatminm[\lin_{\uc}] \cong \coprodover{\prod_{j=2}^n \C(c_{j-1},c_j)} \J[\lin_{\uc}].\]  So the structure morphism \[\nicexy@C+.5cm{\J[\lin_{\uc}] \otimes \Ochatminm[\lin_{\uc}] \ar[r]^-{\lambda_{\lin_{\uc}}} & X_{c_n}}\]
in \eqref{wo-algebra-restricted} is uniquely determined by the restrictions $\theta^{\uf}_{\uc}$ in \eqref{hcptdiagram-structure-map}.  The associativity and wedge conditions are exactly those in the Coherence Theorem \ref{thm:wo-algebra-coherence} for (truncated) linear graphs.
\end{proof}

\begin{interpretation} A homotopy coherent pointed $\C$-diagram in $\M$ has a homotopy coherent $\C$-diagram in $\M$ and additional structure morphisms $\theta^{\uf}_{\uc}$ for truncated linear graphs.  One should think of the structure morphism $\theta^{\uf}_{\uc}$ in \eqref{hcptdiagram-structure-map} as determined by the decorated truncated linear graph
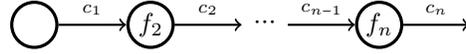
\begin{center}\begin{tikzpicture}
\matrix[row sep=.5cm, column sep=.9cm]{\node [plain] (1) {}; & \node [plain] (2) {$f_2$}; & \node [empty] (3) {$\cdots$}; & \node [plain] (n) {$f_n$};\\};
\draw [arrow] (1) to node{\scriptsize{$c_1$}} (2);
\draw [arrow] (2) to node{\scriptsize{$c_2$}} (3);
\draw [arrow] (3) to node{\scriptsize{$c_{n-1}$}} (n);
\draw [outputleg] (n) to node{\scriptsize{$c_n$}} +(1.2cm,0);
\end{tikzpicture}\end{center}
with all but the first vertices decorated by the $\C$-morphisms $f_j \in \C(c_{j-1},c_j)$.  Note that if $n=1$, then $(f_j)=\varnothing$.\dqed
\end{interpretation}

\begin{example}[Homotopy colored units]\label{ex:h-colored-units}
Suppose\index{homotopy unit} $(X,\lambda,\theta)$ is a homotopy coherent pointed $\C$-diagram in $\M$.  It has structure morphisms \[\nicexy{X_c \ar[rr]^-{X_f} \ar[dr]_-{\cong} && X_d\\  & \J[\Lin_{(c,d)}]\otimes X_c \ar[ur]_-{\lambda^f_{(c,d)}} &}\] for $f \in \C(c,d)$ as in Example \ref{ex1:hcdiagram} and \[\nicexy{\tensorunit = \J[\lin_{(c)}] \ar[r]^-{\theta^{\varnothing}_{(c)}} & X_c}\] for $c \in \C$.  If $X$ is actually a pointed $\C$-diagram in $\M$ as in Definition \ref{def:pointed-digram}, then the diagram \[\nicexy{\tensorunit \ar[d]_-{\theta^{\varnothing}_{(c)}} \ar@{=}[r] & \tensorunit \ar[d]^-{\theta^{\varnothing}_{(d)}}\\ X_c \ar[r]^-{X_f} & X_d}\]
is commutative.  For a homotopy coherent pointed $\C$-diagram in $\M$, this diagram is homotopy commutative in the following sense.

The diagram \[\nicexy@C+.5cm{\tensorunit= \J[\lin_{(d)}] \ar[d]_-{0} \ar[r]^-{\theta^{\varnothing}_{(d)}} & X_d \ar@{=}[d]\\ J=\J[\lin_{(c,d)}] \ar[r]^-{\theta^{f}_{(c,d)}} & X_d\\ \tensorunit \ar[u]^-{1} \ar[d]_-{\cong} &\\ 
\J[\Lin_{(c,d)}] \otimes \J[\lin_{(c)}] \ar[r]^-{(\Id,\theta^{\varnothing}_{(c)})} & \J[\Lin_{(c,d)}] \otimes X_c \ar[uu]_-{\lambda^{f}_{(c,d)}}}\] is commutative, where $0,1 : \tensorunit \to J$ are part of the commutative segment $J$.
\begin{itemize}\item The top rectangle is commutative by the wedge condition \eqref{hcptdiagram-wedge} for the tree substitution \[\lin_{(c,d)} = \lin_{(d)}\bigl(\lin_{(c,d)}\bigr).\]  Visually $\lin_{(c,d)}$ and $\lin_{(d)}$ are the truncated linear graphs
\begin{center}\begin{tikzpicture}
\node [plain] (1) {}; \node [plain, right=.9cm of 1] (2) {$f$};  \node [plain, right=3cm of 2] (3){};
\draw [arrow] (1) to node{\scriptsize{$c$}} (2);
\draw [outputleg] (2) to node{\scriptsize{$d$}} +(1.2cm,0);
\draw [outputleg] (3) to node{\scriptsize{$d$}} +(1.2cm,0);
\end{tikzpicture}\end{center}
and similarly for $\lin_{(c)}$.
\item The bottom square is commutative by the associativity condition \eqref{hcptdiagram-ass} with $\uc=(c)$ and $\uc'=(c,d)$.
\end{itemize}
In other words, $\theta^{\varnothing}_{(c)}$ is a homotopy colored unit that is preserved by  $\lambda^f_{(c,d)}$ up to the homotopy $\theta^f_{(c,d)}$ that is also a structure morphism.\dqed
\end{example}

\section{Homotopy Time-Slice Axiom}\label{sec:hpa-hinverse}

In this section, we explain a \index{homotopy time-slice axiom!homotopy prefactorization algebra}\index{homotopy prefactorization algebra!homotopy time-slice axiom}homotopy coherent version of the time-slice axiom in homotopy prefactorization algebras.

\begin{motivation} 
Suppose $\Chat = (\C,\Config)$ is a configured category, and $S$ is a set of morphisms in $\C$.  In Example \ref{ex:hpatimeslice} we saw that there is a change-of-operad adjunction
\[\nicexy@C+.7cm{\HPFA(\Chat)=\algmwochatm \ar@<2pt>[r]^-{\wellmst} & 
\algm\bigl(\wochatsinvm\bigr) =\HPFA(\Chat,S) \ar@<2pt>[l]^-{\wellmstar}}.\] The objects on the right side are homotopy prefactorization algebras on $\Chat$ satisfying the homotopy time-slice axiom with respect to $S$.  In view of Theorem \ref{thm:pfa-timeslice}, for each $s \in S$, we therefore expect the structure morphism $\lambda\{s\}$ to be invertible up to coherent homotopies that are also structure morphisms.  We will explain this in the following examples.\dqed
\end{motivation}

\begin{example}[Left homotopy inverses]\label{ex:hap-hinverse}
Suppose\index{homotopy inverse} $(X,\lambda)$ is a homotopy prefactorization algebra on $\Chat$ satisfying the homotopy time-slice axiom with respect to $S$, i.e., a $\wochatsinvm$-algebra.  Since $\Ochatsinvm$ has entries \[\Ochatsinvm\duc = \coprodover{\Ochatsinv\duc} \tensorunit \forspace \duc \in \Profcc,\] there is a canonical isomorphism \[\Ochatsinvm[T] = \bigotimes_{v\in T} \Ochatsinvm(v) = \bigotimes_{v\in T} \biggl[\coprodover{\Ochatsinv(v)} \tensorunit\biggr] \cong \coprodover{\prodover{v\in T} \Ochatsinv(v)} \tensorunit\] for each $\colorc$-colored tree $T \in \uTree\duc$.  Therefore, there is a canonical isomorphism \[\J[T] \otimes \Ochatsinvm[T] \otimes X_{\uc} \cong \coprodover{\prodover{v\in T} \Ochatsinv(v)} \J[T]\otimes X_{\uc}.\]  It follows that the structure morphisms $\lambda_T$ in \eqref{wo-algebra-restricted} is uniquely determined by the restricted structure morphisms \[\nicexy@C+1cm{\J[T] \otimes X_{\uc} \ar[r]^-{\lambda_T\left\{\uf^v\right\}_{v\in T}} & X_d}\in \M\] for $T \in \uTree\duc$ and $\left\{\uf^v\right\}_{v\in T} \in \prodover{v\in T} \Ochatsinv(v)$.

Suppose $f : c \to d$ is a morphism in $S$, so \[f \in \Ochatsinv\dc \andspace \finverse \in \Ochatsinv\cd.\]  Note that \[\gamma^{\Ochatsinv}_{\Lin_{(c,d,c)}}\bigl(f,\finverse\bigr) = \operadunit_c \in \Ochatsinv\cc,\] the $c$-colored unit in the colored operad $\Ochatsinv$.  Here \[\nicexy@C+1.2cm{\Ochatsinv[\Lin_{(c,d,c)}] \cong \Ochatsinv\dc \times \Ochatsinv\cd \ar[r]^-{\gamma^{\Ochatsinv}_{\Lin_{(c,d,c)}}} & \Ochatsinv\cc}\] is the operadic structure morphism \eqref{operadic-structure-map} of $\Ochatsinv$ for the linear graph $\Lin_{(c,d,c)}$.  

The diagram \[\nicexy@C+1.5cm{\tensorunit\otimes X_c = \J[\Lin_{(c,c)}] \otimes X_c \ar[d]_-{(0,\Id)} \ar[r]^-{\lambda_{\Lin_{(c,c)}}\{\operadunit_c\}} & X_c \ar@{=}[d]\\
J \otimes X_c = \J[\Lin_{(c,d,c)}] \otimes X_c \ar[r]^-{\lambda_{\Lin_{(c,d,c)}}\{f,\finverse\}} & X_c\\ \tensorunit \otimes X_c \ar[u]^-{(1,\Id)} \ar[d]_-{\cong} &\\
\J[\Lin_{(d,c)}] \otimes \J[\Lin_{(c,d)}] \otimes X_c \ar[r]^-{\left(\Id,\lambda_{\Lin_{(c,d)}}\{f\}\right)} & \J[\Lin_{(d,c)}] \otimes X_d \ar[uu]_-{\lambda_{\Lin_{(d,c)}}\{\finverse\}}}\]
is commutative:
\begin{itemize}\item The top rectangle is commutative by the wedge condition \eqref{wo-alg-wedge} for the tree substitution \[\Lin_{(c,d,c)} = \Lin_{(c,c)}\bigl(\Lin_{(c,d,c)}\bigr).\]
\item The bottom square is commutative by the associativity condition \eqref{wo-alg-ass} for the grafting \[\Lin_{(c,d,c)} = \graft\bigl(\Lin_{(d,c)}; \Lin_{(c,d)}\bigr).\]
\end{itemize}
The structure morphism $\lambda_{\Lin_{(c,c)}}\{\operadunit_c\}$ is isomorphic to the identity morphism on $X_c$ by Corollary \ref{cor:wo-alg-unit} .  Therefore, the above commutative diagram says that the structure morphism $\lambda_{\Lin_{(d,c)}}\{\finverse\}$ is a left homotopy inverse of the structure morphism $\lambda_{\Lin_{(c,d)}}\{f\}$ via the homotopy $\lambda_{\Lin_{(c,d,c)}}\{f,\finverse\}$ that is also a structure morphism.\dqed
\end{example}

\begin{example}[Right homotopy inverses]\label{ex2:hap-hinverse}
Similarly, the diagram \[\nicexy@C+1.5cm{\tensorunit\otimes X_d = \J[\Lin_{(d,d)}] \otimes X_d \ar[d]_-{(0,\Id)} \ar[r]^-{\lambda_{\Lin_{(d,d)}}\{\operadunit_d\}} & X_d \ar@{=}[d]\\
J \otimes X_d = \J[\Lin_{(d,c,d)}] \otimes X_d \ar[r]^-{\lambda_{\Lin_{(d,c,d)}}\{\finverse,f\}} & X_d\\ \tensorunit \otimes X_d \ar[u]^-{(1,\Id)} \ar[d]_-{\cong} &\\
\J[\Lin_{(c,d)}] \otimes \J[\Lin_{(d,c)}] \otimes X_d \ar[r]^-{\left(\Id,\lambda_{\Lin_{(d,c)}}\{\finverse\}\right)} & \J[\Lin_{(c,d)}] \otimes X_c \ar[uu]_-{\lambda_{\Lin_{(c,d)}}\{f\}}}\]
is commutative, and $\lambda_{\Lin_{(d,d)}}\{\operadunit_d\}$ is isomorphic to the identity morphism on $X_d$.  Therefore, the commutative diagram says that the structure morphism $\lambda_{\Lin_{(d,c)}}\{\finverse\}$ is a right homotopy inverse of the structure morphism $\lambda_{\Lin_{(c,d)}}\{f\}$ via the homotopy $\lambda_{\Lin_{(d,c,d)}}\{\finverse,f\}$ that is also a structure morphism.\dqed
\end{example}

\section{$E_\infty$-Algebra Structure}\label{sec:hpa-einfinity}

In this section, we explain that some entries in a homotopy prefactorization algebra are $E_\infty$-algebras as in Definition \ref{def:einfinity-algebra}.

\begin{motivation} Suppose $\Chat = (\C,\Config)$ is a configured category with object set $\colorc$, and $c \in \colorc$ such that $\{\Id_c\}_{i=1}^n \in \Config\ccc$ for all $n$.  For each prefactorization algebra $(Y,\lambda)$ on $\Chat$, we saw in Proposition \ref{prop:com-to-ochat} that its $c$-colored entry $Y_c$ is equipped with the structure of a commutative monoid.  For a homotopy prefactorization algebra on $\Chat$, we expect the entry $Y_c$ to be an $E_\infty$-algebra.\dqed
\end{motivation}

The following result is a consequence of Corollary \ref{cor:wf-eta-adjunction} and Proposition \ref{prop:com-to-ochat}.

\begin{corollary}\label{cor:hpa-einfinity}
Suppose\index{homotopy prefactorization algebra!$E_\infty$-algebra}\index{einfinityalgebra@$E_\infty$-algebra} $\Chat = (\C,\Config)$ is a configured category with object set $\colorc$, and $c \in \colorc$ such that \[\{\Id_c\}_{i=1}^n \in \Config\ccc\] for all $n$.  Then the operad morphism \[\nicexy{\Com \ar[r]^-{\iota_c} & \Ochatm}\] in Proposition \ref{prop:com-to-ochat} induces a diagram of change-of-operad adjunctions
\[\nicexy{\algmwcom \ar@<2pt>[r]^-{(\W\iota_c)_!} \ar@<-2pt>[d]_-{\eta_!} & \algmwochatm = \HPFA(\Chat) \ar@<-2pt>[d]_-{\eta_!} \ar@<2pt>[l]^-{\W\iota_c^*}\\
\Comm = \algm(\Com) \ar@<2pt>[r]^-{(\iota_c)_!} \ar@<-2pt>[u]_-{\eta^*} & \algmochatm = \PFA(\Chat) \ar@<2pt>[l]^-{\iota_c^*} \ar@<-2pt>[u]_-{\eta^*}}\]
with commuting left adjoint diagram and commuting right adjoint diagram.
\end{corollary}

\begin{interpretation}\label{int:hpa-einfinity-algebra} The right adjoint $\W\iota_c^*$ sends each homotopy prefactorization algebra $(Y,\lambda)$ on $\Chat$ to the $E_\infty$-algebra \[\W\iota_c^*(Y,\lambda) \in \algmwcom.\]  The underlying object is the entry $Y_c\in\M$.  For each one-colored tree $T \in \Tree(n)$, the $E_\infty$-algebra structure morphism \[\nicexy{\J[T]\otimes Y_c^{\otimes n}\ar[r]^-{\lambda_T} & Y_c}\] in \eqref{einfinity-structure} is the structure morphism \[\lambda_{T_c}\Bigl\{\{\Id_c\}_{i=1}^{|\inp(v)|}\Bigr\}_{v\in T_c}\] in \eqref{wochatm-restricted}.  Here $T_c \in \uTreec\ccc$ is the $c$-colored tree obtained from $T$ by replacing every edge color by $c$.  For each vertex $v \in T_c$, \[\{\Id_c\}_{i=1}^{|\inp(v)|} \in \Config\ccc\] is a configuration by assumption.\dqed
\end{interpretation}

\begin{example}[Homotopy prefactorization algebras on bounded lattices]\label{ex:hpa-einfinity-lattice} 
For a\index{bounded lattice} bounded lattice $(L,\leq)$ with least element $0$, consider the configured category $\Lhat = (L,\Configl)$ in Example \ref{ex:boundedlatter-config}.  The least element $0 \in L$ has the property that \[\{\Id_0\}_{i=1}^n \in \Configl\zerozerozero \forspace n\geq 0.\]  If $(Y,\lambda)$ is a homotopy prefactorization algebra on $\Lhat$, i.e., a $\wolhatm$-algebra, then the entry $Y_0 \in \M$ is equipped with the structure of an $E_\infty$-algebra by Corollary \ref{cor:hpa-einfinity}.\dqed
\end{example}

\begin{example}[Homotopy Costello-Gwilliam prefactorization algebras]\label{ex:hpa-einfinity-cg}
For a\index{homotopy prefactorization algebra!Costello-Gwilliam} topological space $X$, consider the configured category \[\Openxhat =\bigl(\Openx,\Configx\bigr)\] in Example \ref{ex:open-configuration}.  The category $\Openx$ is a bounded lattice with least element $\varnothing_X \subset X$, the empty subset of $X$.  The configured category $\Openxhat$ has the form $\Lhat$ in Example \ref{ex:boundedlatter-config}.  Therefore, as in Example \ref{ex:hpa-einfinity-lattice}, if $(Y,\lambda)$ is a homotopy prefactorization algebra on $\Openxhat$, i.e., a $\wom_{\Openxhat}$-algebra, then the entry $Y_{\varnothing_X} \in \M$ is equipped with the structure of an $E_\infty$-algebra by Corollary \ref{cor:hpa-einfinity}.\dqed
\end{example}

\begin{example}[Homotopy Costello-Gwilliam equivariant prefactorization algebras]\label{ex:hpa-einfinty-cgeq}
Suppose\index{homotopy equivariant prefactorization algebra} $G$ is a group, and $X$ is a topological space in which $G$ acts on the left by homeomorphisms.  Consider the configured category \[\Openxghat =\bigl(\Openxg,\Configxg\bigr)\] in Example \ref{ex:eq-space-configuration}.  The empty subset $\varnothing_X \subset X$ has the property that \[\bigl\{\Id_{\varnothing_X}\bigr\}_{i=1}^n \in \Configxg\sbinom{\varnothing_X}{\varnothing_X,\ldots,\varnothing_X} \forspace n\geq 0.\] If $(Y,\lambda)$ is a homotopy prefactorization algebra on $\Openxghat$, i.e., a $\wom_{\Openxghat}$-algebra, then the entry $Y_{\varnothing_X} \in \M$ is equipped with the structure of an $E_\infty$-algebra by Corollary \ref{cor:hpa-einfinity}.\dqed
\end{example}

\section{Objectwise $E_\infty$-Module}\label{sec:hpa-einfinity-module}

Suppose $(L,\leq)$ is an arbitrary but fixed bounded lattice with least element $0$.  Consider the configured category $\Lhat = (L,\Configl)$ in Example \ref{ex:boundedlatter-config}.  We saw in Examle \ref{ex:hpa-einfinity-lattice} that for each homotopy prefactorization algebra on $\Lhat$, the $0$-entry is an $E_\infty$-algebra.  In this section, we explain that every other entry has the structure of an $E_\infty$-module over the $E_\infty$-algebra at the $0$-entry.

\begin{motivation} For a monoid $A$ in $\M$, recall from Definition \ref{def:module-monoid} that a left $A$-module is an object $X$ equipped with a left action $m : A \otimes X \to X$ that satisfies the associativity and unity axioms.  These two axioms can  be read off from those of a monoid in Proposition \ref{prop:monoid} by replacing the last $A$-entry by $X$.  We can define modules over an $A_\infty$-algebra or an $E_\infty$-algebra in the same way, by replacing the  last entry with the module in each structure morphism and each axiom.  For our current objective of understanding homotopy prefactorization algebras, we will need the following concept of an $E_\infty$-module over an $E_\infty$-algebra.  To define $E_\infty$-modules, we will need two colors, one color $0$ for the $E_\infty$-algebra and one color $d$ for the $E_\infty$-modules on which it acts.\dqed
\end{motivation}

Recall the concept of a directed path in Definition \ref{def:directed-path}.  We first define the colored trees that parametrize the structure morphisms of an $E_\infty$-module.  In the next definition, $0$ and $d$ are two distinct symbols, not necessarily elements in a bounded lattice.  In practice, $0$ is the least element in a bounded lattice $L$, and $0 \not= d \in L$.

\begin{definition}\label{def:painted-tree}
A \index{tree!$\zerod$-}\emph{$\zerod$-tree} is a $\zerod$-colored tree \[T \in \uTreezerod\dzerozerod\] in which $(0,\ldots,0)$ is a possibly empty profile of copies of $0$.  It is required that the following conditions be satisfied. 
\begin{itemize}\item $T$ has at least one input, the last of which and the output are colored by $d$.  All other inputs of $T$ are colored by $0$.
\item Each $v \in \Vt(T)$ has at least one input.
\item If $T$ does not have any vertices, then $T$ is the $d$-colored exceptional edge $\uparrow_d$.
\item If $T$ has a non-empty set of vertices, consider the unique directed path $P^d_T$ in $T$ whose initial vertex contains the last input of $T$ and whose terminal vertex contains the output of $T$. 
\begin{itemize}\item Each vertex $v \in P^d_T$ has profile $\dzerozerod$, where $(0,\ldots,0)$ has $|\inp(v)|-1$ copies of $0$.
\item Each vertex $v\not\in P^d_T$ has profile $\zerozerozero$, where $(0,\ldots,0)$ has $|\inp(v)|$ copies of $0$.
\end{itemize}
\end{itemize}
\end{definition}

\begin{interpretation} In a $\zerod$-tree $T$, the directed path $P^d_T$ from the last input to the output is $d$-colored, where we abbreviate the singleton $\{d\}$ to $d$.  All other edges in $T$ are colored by $0$.  For the vertex $v$ that contains the last input of $T$, the last input of $v$ is also the last input of $T$, both of which are $d$-colored.\dqed
\end{interpretation}

\begin{example} Every $d$-colored linear graph $\Lin_{(d,d,\ldots,d)}$ as in Example \ref{ex:linear-graph} is a $\zerod$-tree.  On the other hand, a truncated linear graph as in Example \ref{ex:truncated-linear-graph} cannot be a $\zerod$-tree because it does not have any inputs.\dqed\end{example}

\begin{example} The $(\uc;d)$-corolla in Example \ref{ex:cd-corolla} is a $\zerod$-tree if and only if $\uc$ has the form $(0,\ldots,0,d)$, where $(0,\ldots,0)$ is a possibly empty profile of copies of $0$.\dqed\end{example}

\begin{example} The $2$-level tree $T\bigl(\{\ub_j\};\uc;d\bigr)$ in Example \ref{ex:twolevel-tree} is a $\zerod$-tree if and only if the following conditions hold:
\begin{itemize}
\item Every $k_j \geq 1$.
\item $c_i=0$ for $1 \leq i \leq m-1$, and $c_m=d$.
\item $b_{j,l}=0$ for all $1 \leq j \leq m$ and $1 \leq l \leq k_j$, except for $b_{m,k_m}=d$.
\end{itemize}
For example, the $2$-level tree on the right
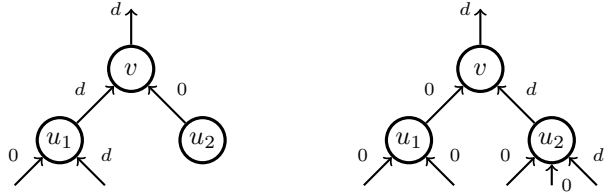
\begin{center}\begin{tikzpicture}
\node [plain] (v) {$v$}; \node [plain, below left=.7cm of v] (u1) {$u_1$}; 
\node [plain, below right=.7cm of v] (u2) {$u_2$};
\draw [outputleg] (v) to node[at end]{\scriptsize{$d$}} +(0,.8cm);
\draw [arrow] (u1) to node{\scriptsize{$d$}} (v);
\draw [arrow] (u2) to node[swap]{\scriptsize{$0$}} (v);
\draw [inputleg] (u1) to node[swap]{\scriptsize{$0$}} +(-.6cm,-.6cm);
\draw [inputleg] (u1) to node{\scriptsize{$d$}} +(.6cm,-.6cm);
\node [plain, right=4cm of v] (v2) {$v$}; \node [plain, below left=.7cm of v2] (u21) {$u_1$}; 
\node [plain, below right=.7cm of v2] (u22) {$u_2$};
\draw [outputleg] (v2) to node[at end]{\scriptsize{$d$}} +(0,.8cm);
\draw [arrow] (u21) to node{\scriptsize{$0$}} (v2);
\draw [arrow] (u22) to node[swap]{\scriptsize{$d$}} (v2);
\draw [inputleg] (u21) to node[swap]{\scriptsize{$0$}} +(-.6cm,-.6cm);
\draw [inputleg] (u21) to node{\scriptsize{$0$}} +(.6cm,-.6cm);
\draw [inputleg] (u22) to node[swap]{\scriptsize{$0$}} +(-.6cm,-.6cm);
\draw [inputleg] (u22) to node[at end]{\scriptsize{$0$}} +(0cm,-.6cm);
\draw [inputleg] (u22) to node{\scriptsize{$d$}} +(.6cm,-.6cm);
\end{tikzpicture}\end{center}
is a $\zerod$-tree, but the one on the left is not a $\zerod$-tree because $u_2$ does not have any inputs.\dqed
\end{example}

\begin{example} Suppose $T$ is a $\zerod$-tree with $n \geq 1$ inputs, and $T_n$ is also a $\zerod$-tree.  For each $1 \leq i \leq n-1$, suppose $T_i \in \uTreezero$ in which every vertex has at least one input.  Then the grafting \[G=\graft\bigl(T;T_1,\ldots,T_n)\] is also a $\zerod$-tree.  Indeed, the last input of $T_n$ becomes the last input of the grafting $G$, and the output of $T$ becomes the output of $G$.  The output of $T_n$ and the last input of $T$ are both $d$-colored.  The $d$-colored edges in $T$ and $T_n$ yield the required $d$-colored directed path from the last input of the grafting to its output.  All other edges are colored by $0$.\dqed
\end{example}

\begin{example}
Suppose $T$ is a $\zerod$-tree with $n$ inputs, and $H_v \in \uTreezerod(v)$ is a $\zerod$-tree if $v\in P^d_T$.  Otherwise, $H_v\in\uTreezero(v)$ in which every vertex has at least one input.  Then the tree substitution $K=T(H_v)_{v\in T}$ is also a $\zerod$-tree.  Indeed, the unique $d$-colored directed paths in $H_v$ for $v \in P^d_T$ together form the unique $d$-colored directed path in $K$ from its last input to the output.  All other edges in $K$ are $0$-colored.\dqed
\end{example}

The following definition of an $E_\infty$-module is modeled after the Coherence Theorem \ref{thm:einfinity-algebra} for $E_\infty$-algebras.

\begin{definition}\label{def:einfinity-module}
Suppose $(A,\lambda^A)$ is an $E_\infty$-algebra.  An\index{einfinitymodule@$E_\infty$-module} \emph{$E_\infty$-module over $(A,\lambda)$} is a pair $(X,\lambda^X)$ consisting of
\begin{itemize}
\item an object $X \in \M$ and
\item a structure morphism\index{structure morphism!for $E_\infty$-modules}
\begin{equation}\label{einfinity-module-structure}
\nicexy@C+.5cm{\J[T] \otimes A^{\otimes n-1} \otimes X \ar[r]^-{\lambda^X_T} & X}\in \M
\end{equation}
for each $\zerod$-tree $T$ with $n \geq 1$ inputs
\end{itemize}
that satisfies the following four conditions.
\begin{description}
\item[Associativity] Suppose $T$ and $T_n$ are $\zerod$-trees in which $T$ has $n \geq 1$ inputs, and $T_j\in\uTreezero$ for $1 \leq j \leq n-1$ in which every vertex has at least one input.  Suppose $T_j$ has $k_j$ inputs for $1\leq j \leq n$, and $k=k_1+\cdots+k_n$.  Suppose $G=\graft(T;T_1,\ldots,T_n)$ is the grafting.  Then the diagram
\begin{equation}\label{einfinity-module-ass}
\nicexy@C-.7cm@R+.3cm{\J[T] \otimes \Bigl(\bigotimes\limits_{j=1}^n \J[T_j]\Bigr) \otimes A^{\otimes k-1} \otimes X \ar[d]_-{\mathrm{permute}}^-{\cong} \ar[r]^-{(\pi,\Id)} & \J[G] \otimes A^{\otimes k-1} \otimes X \ar[dd]^-{\lambda^X_G}\\
\J[T] \otimes \Bigl[\bigotimes\limits_{j=1}^{n-1} \bigl(\J[T_j]\otimes A^{\otimes k_j}\bigr)\Bigr] \otimes \bigl(\J[T_n] \otimes A^{\otimes k_n-1} \otimes X\bigr) \ar[d]_-{\bigl(\Id,\bigotimes\limits_{j=1}^{n-1} \lambda^A_{T_j}, \lambda^X_{T_n}\bigr)} &\\
\J[T] \otimes A^{\otimes n-1} \otimes X \ar[r]^-{\lambda^X_T} & X}
\end{equation}
is commutative.  Here $\pi=\bigotimes_S 1$ is the morphism in Lemma \ref{lem:morphism-pi} for the grafting $G$.
\item[Unity]
The composition
\begin{equation}\label{einfinity-module-unity}
\nicexy{X\ar[r]^-{\cong} & \J[\uparrow_d]\otimes X \ar[r]^-{\lambda^X_{\uparrow}} & X}
\end{equation} 
is the identity morphism of $X$, where $\uparrow_d$ is the $d$-colored exceptional edge.
\item[Equivariance]
For a $\zerod$-tree $T$ with $n\geq 1$ inputs and $\sigma \in \Sigma_{n-1}$, the diagram 
\begin{equation}\label{einfinity-module-eq}
\nicexy@C+1cm{\J[T]\otimes A^{\otimes n-1} \otimes X \ar[d]_-{(\Id,\sigmainv,\Id)} \ar[r]^-{\lambda^X_T} & X\ar@{=}[d]\\ \J[T\sigma]\otimes A^{\otimes n-1} \otimes X \ar[r]^-{\lambda^X_{T\sigma}} & X}
\end{equation}
is commutative.  Here $T\sigma$ is the $\zerod$-tree obtained from $T$ by replacing its ordering $\zeta_T$ by $\zeta_T\circ (\sigma \oplus \id_1)$.
\item[Wedge Condition]
Suppose $T$ is a $\zerod$-tree with $n$ inputs, and $H_v \in \uTreezerod(v)$ is a $\zerod$-tree if $v\in P^d_T$.  For $v \in \Vt(T)$ with $v \not\in P^d_T$, $H_v\in\uTreezero(v)$ in which every vertex has at least one input.  Suppose $K=T(H_v)_{v\in T}$ is the tree substitution.  Then the diagram
\begin{equation}\label{einfinity-module-wedge}
\nicexy@C+1cm{\J[T]\otimes A^{\otimes n-1} \otimes X \ar[d]_-{(\J,\Id)} \ar[r]^-{\lambda^X_T} & X\ar@{=}[d]\\ \J[K]\otimes A^{\otimes n-1} \otimes X \ar[r]^-{\lambda^X_K} & X}
\end{equation}
is commutative.
\end{description}
\end{definition}

In a bounded lattice $(L,\leq)$ with least element $0$, regarded as a small category, the unique morphism $0 \to d$ is denoted by $0_d$ for $d \in L$.  In particular, $0_0 = \Id_0$.  The next result is the main observation in this section.

\begin{corollary}\label{cor:hpa-lattice-emodule}
Suppose $(L,\leq)$ is a \index{bounded lattice}bounded lattice with least element $0$, and $(Y,\lambda)$ is a homotopy prefactorization algebra on $\Lhat$, i.e., a $\wolhatm$-algebra.  For each $d \in L$, the entry $Y_d$ is an $E_\infty$-module over the $E_\infty$-algebra $Y_0$ when equipped with the structure morphisms
\[\nicexy@C+1cm{\J[T] \otimes Y_0^{\otimes n-1} \otimes Y_d \ar[r]^-{\lambda_T\left\{\uf^v\right\}_{v\in T}} & Y_d} \in \M\]
in \eqref{wochatm-restricted} for $\zerod$-trees $T$ with $n\geq 1$ inputs.  Here for $v \in \Vt(T)$,
\begin{equation}\label{ufsupv}
\uf^v = \begin{cases} \bigl\{0_d,\ldots,0_d,\Id_d\bigr\} \in \Configl\dzerozerod & \text{ if $v \in P^d_T$},\\ \{\Id_0\}_{i=1}^{|\inp(v)|} \in \Configl\zerozerozero & \text{ if $v \not\in P^d_T$.}\end{cases}
\end{equation}
\end{corollary}

\begin{proof}
This is a special case of the Coherence Theorem \ref{thm:hpa-coherence} applied to the configured category $\Lhat$.  To check that the $E_\infty$-module associativity \eqref{einfinity-module-ass} is a special case of the associativity condition \eqref{wochatm-ass}, we use the fact, which we explained in Interpretation \ref{int:hpa-einfinity-algebra}, that the $E_\infty$-algebra structure morphisms of $Y_0$ are the structure morphisms \[\lambda_{T_0}\Bigl\{\{\Id_0\}_{i=1}^{|\inp(v)|}\Bigr\}_{v\in T_0}\] with $T_0 \in \uTreezero$.  

To check that the $E_\infty$-module wedge condition \eqref{einfinity-module-wedge} is a special case of the wedge condition \eqref{wochatm-wedge}, we use the fact that, for $v \in P^d_T$, $H_v\in\uTreezerod(v)$ is a $\zerod$-tree.   For $u \in \Vt(H_v)$, $\uf^u \in \Configl(u)$ is defined as in \eqref{ufsupv}.   So we have \[\gamma^{\Olhat}_{H_v}\Bigl(\left\{\uf^u\right\}_{u\in H_v}\Bigr) = \bigl\{0_d,\ldots,0_d,\Id_d\bigr\} \in \Configl\dzerozerod = \Configl(v).\]  On the other hand, for $v \in \Vt(T)$ with $v \not\in P^d_T$, we have that $H_v \in \uTreezero(v)$ and that \[\gamma^{\Olhat}_{H_v}\Bigl(\bigl\{\{\Id_0\}_{i=1}^{|\inp(u)|}\bigr\}_{u\in H_v}\Bigr) = \{\Id_0\}_{i=1}^{|\inp(v)|} \in \Configl\zerozerozero = \Configl(v).\] 
\end{proof}

\begin{example}[Homotopy Costello-Gwilliam prefactorization algebras]\label{ex:hpa-emod-cg}
For \index{homotopy prefactorization algebra!Costello-Gwilliam}a topological space $X$, consider the configured category \[\Openxhat =\bigl(\Openx,\Configx\bigr)\] in Example \ref{ex:open-configuration}.  The category $\Openx$ is a bounded lattice with least element $\varnothing_X \subset X$, the empty subset of $X$.  The configured category $\Openxhat$ has the form $\Lhat$ in Example \ref{ex:boundedlatter-config}.  Therefore, by Corollary \ref{cor:hpa-lattice-emodule}, if $(Y,\lambda)$ is a homotopy prefactorization algebra on $\Openxhat$, i.e., a $\wom_{\Openxhat}$-algebra, then each entry $Y_U$ with $U \in \Openx$ is equipped with the structure of an $E_\infty$-module over the $E_\infty$-algebra $Y_{\varnothing_X}$ in Example \ref{ex:hpa-einfinity-cg}.\dqed
\end{example}

\begin{example}[Homotopy Costello-Gwilliam equivariant prefactorization algebras]\label{ex:hpa-emod-cgeq}
Suppose\index{homotopy equivariant prefactorization algebra} $G$ is a group, and $X$ is a topological space in which $G$ acts on the left by homeomorphisms.  Consider the configured category \[\Openxghat =\bigl(\Openxg,\Configxg\bigr)\] in Example \ref{ex:eq-space-configuration}.  By the change-of-operad adjunction \[\nicexy@C+.8cm{\algm\bigl(\wom_{\Openxhat}\bigr) \ar@<2pt>[r]^-{(\W\Otom_{\iota})_!} & \algm\bigl(\wom_{\Openxghat}\bigr) \ar@<2pt>[l]^-{(\W\Otom_{\iota})^*}}\] in Example \ref{ex:hcgeqpfa}, every homotopy prefactorization algebra $(Y,\lambda)$ on $\Openxghat$ has an underlying homotopy prefactorization algebra on $\Openxhat$.  Therefore, by Example \ref{ex:hpa-emod-cg}, each entry $Y_U$ with $U \in \Openx$ is equipped with the structure of an $E_\infty$-module over the $E_\infty$-algebra $Y_{\varnothing_X}$.\dqed
\end{example}

\section{Homotopy Coherent Diagrams of $E_\infty$-Modules}\label{sec:hpa-hcdiag-mod}

Suppose $(L,\leq)$ is an arbitrary but fixed bounded lattice with least element $0$.  Consider the configured category $\Lhat = (L,\Configl)$ in Example \ref{ex:boundedlatter-config}.  In this section, we explain that, for each homotopy prefactorization algebra on $\Lhat$, the\index{einfinitymodule@$E_\infty$-module!homotopy coherent diagram}\index{homotopy coherent diagram!$E_\infty$-module} objectwise $E_\infty$-module structure over the $E_\infty$-algebra at the $0$-entry in Section \ref{sec:hpa-einfinity-module} is compatible with the homotopy coherent diagram structure in Section \ref{sec:hpa-hcdiag}.

\begin{motivation} In Corollary \ref{cor:lattice-diagram-modules} we saw that, for a prefactorization algebra $(Y,\lambda)$ on $\Lhat$, the objectwise left $Y_0$-module structure is compatible with the $L$-diagram structure.  For a homotopy prefactorization algebra $(Y,\lambda)$ on $\Lhat$, the entry $Y_0$ is equipped with the structure of an $E_\infty$-algebra by Example \ref{ex:hpa-einfinity-lattice}.  Furthermore, by Corollary \ref{cor:hpa-lattice-emodule}, every other entry $Y_d$ with $d\in L$ is equipped with the structure of an $E_\infty$-module over the $E_\infty$-algebra $Y_0$.   We expect the homotopy coherent $L$-diagram structure in $Y$ in Section \ref{sec:hpa-hcdiag} to be homotopically compatible with the entrywise $E_\infty$-module structure.  To explain precisely how they are compatible, we need the following notation.\dqed
\end{motivation}

\begin{assumption}\label{assumption:hpa-hcdiag-mod}
Suppose 
\begin{itemize}\item $\uc = \bigl(c=c_0,\ldots,c_m=d\bigr) \in \profl$ with $m \geq 1$ and $c=c_0 \leq \cdots \leq c_m=d$ in $L$.
\item $L_{\uc}=\Lin_{\uc} \in \Linear^L\dc$ is the corresponding linear graph in Example \ref{ex:linear-graph}. 
\item $L_{cd} = \Lin_{(c,d)} \in \Linear^L \dc$.
\item $g_i \in L(c_{i-1},c_i)$ is the unique element for $1 \leq i \leq m$, and $g = g_m \cdots g_1 \in L(c,d)$.
\item $T_d \in \uTreezerod\dzerozerod$ is a $\zerod$-tree as in Definition \ref{def:painted-tree}, in which $(0,\ldots,0)$ has $n-1$ copies of $0$ for some $n \geq 1$.
\item $T_c\in \uTreezeroc\czerozeroc$ is the $\zeroc$-tree obtained from $T_d$ by replacing every $d$-colored edge by a $c$-colored edge.  
\end{itemize}
Define the $\zerocd$-colored trees
\[T^1 = \graft\Bigl(L_{\uc};T_c\Bigr), \quad 
T^2 = \graft\Bigl(T_d; \underbrace{\uparrow_0,\ldots,\uparrow_0}_{n-1~ \mathrm{copies}},L_{cd}\Bigr), \andspace C = \Cor_{\bigl((0,\ldots,0,c);d\bigr)}\]
in $\uTreezerocd\dzerozeroc$.  Here $\uparrow_0$ is the $\{0\}$-colored exceptional edge in Example \ref{ex:colored-exedge}, and $\Cor_?$ is the corolla in Example \ref{ex:cd-corolla}, with $\graft$ the grafting in Definition \ref{def:grafting}.  

We can visualize $T^1$ (on the left) and $T^2$ (on the right) as follows.
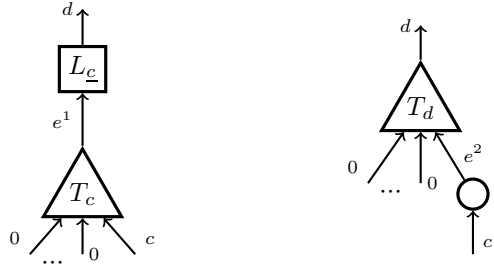
\begin{center}\begin{tikzpicture}
\node [rectplain] (L) {$L_{\uc}$}; \node [triangular, below=.7cm of L] (T) {$T_c$}; 
\node [below=.5cm of T] (b1) {}; \node [left=.1mm of b1] (d1) {$\cdots$};
\draw [outputleg] (L) to node[at end]{\scriptsize{$d$}} +(0,.8cm);
\draw [arrow] (T) to node{\scriptsize{$e^1$}} (L);
\draw [inputleg] (T) to node[at end, swap]{\scriptsize{$0$}} +(-.7cm,-.8cm);
\draw [inputleg] (T) to node[at end, inner sep=2pt]{\scriptsize{$0$}} +(0cm,-.8cm);
\draw [inputleg] (T) to node[at end]{\scriptsize{$c$}} +(.7cm,-.8cm);
\node [right=4cm of T] (dots) {};
\node [triangular, above=.7cm of dots] (Td) {$T_d$}; 
\node [smallplain, right=.35cm of dots] (L3) {}; \node [left=.1mm of dots] (d2) {$\cdots$};
\draw [outputleg] (Td) to node[at end]{\scriptsize{$d$}} +(0,1.1cm);
\draw [arrow] (L3) to node[near start, swap, inner sep=2pt]{\scriptsize{$e^2$}} (Td);
\draw [inputleg] (Td) to node[at end, swap]{\scriptsize{$0$}} +(-.7cm,-1cm);
\draw [inputleg] (Td) to node[at end, inner sep=2pt]{\scriptsize{$0$}} +(0cm,-1cm);
\draw [inputleg] (L3) to node[near end]{\scriptsize{$c$}} +(0cm,-.8cm);
\end{tikzpicture}
\end{center}
In $T^1$ the $c$-colored internal edge connecting $T_c$ to $L_{\uc}$ is denoted by $e^1$.  In $T^2$ the $d$-colored internal edge connecting the linear graph $L_{cd}$ to $T_d$ is denoted by $e^2$.

For $v \in \Vt(T_d)$, the configuration $\uf^v \in \Configl$ in \eqref{ufsupv} will be denoted by $\uf^v_d$.  The corresponding $c$-colored version, with $d$ replaced by $c$ everywhere, is denoted by $\uf^v_c$.
\end{assumption}

The following is the main result of this section.  A copy of the morphism $1 : \tensorunit \to J$ indexed by an internal edge $e$ will be denoted by $1_e$.  To simplify the notation, we will omit writing some of the identity morphisms below. We will use the notation in the Coherence Theorem \ref{thm:hpa-coherence} for homotopy prefactorization algebras.

\begin{theorem}\label{thm:hpadiag-einfinity-module}
Suppose $(Y,\lambda)$ is a homotopy prefactorization algebra on $\Lhat$, i.e., a $\wolhatm$-algebra.  Under Assumption \ref{assumption:hpa-hcdiag-mod}, the diagram
\[\nicexy@C+1cm{\J[L_{\uc}] \otimes \J[T_c] \otimes Y_0^{\otimes n-1} \otimes Y_c \ar[r]^-{\lambda_{T_c}\{\uf^v_c\}_{v\in T_c}} \ar[d]_-{(1_{e^1})(\cong)} & \J[L_{\uc}] \otimes Y_c \ar[d]^-{\lambda_{L_{\uc}}\{g_i\}_{i=1}^m}\\
\J[T^1] \otimes Y_0^{\otimes n-1} \otimes Y_c \ar[r]^-{\lambda_{T_1}\left\{\{\uf^v_c\}_{v\in T_c},\, \{g_i\}_{i=1}^n\right\}} & Y_d \ar@{=}[d]\\
\J[C]\otimes Y_0^{\otimes n-1} \otimes Y_c \ar[r]^-{\lambda_C\left\{0_d,\ldots,0_d,g\right\}}  \ar[u]^-{(0^{\otimes |T^1|})(\cong)} \ar[d]_-{(0^{\otimes |T^2|})(\cong)} & Y_d \ar@{=}[d]\\
\J[T^2] \otimes Y_0^{\otimes n-1} \otimes Y_c  \ar[r]^-{\lambda_{T^2}\left\{g,\,\{\uf^v_d\}_{v\in T_d}\right\}} & Y_d\\
\J[T_d] \otimes Y_0^{\otimes n-1} \otimes \J[L_{cd}] \otimes Y_c \ar[r]^-{\lambda_{L_{cd}}\{g\}}  \ar[u]^-{(1_{e^2})(\cong)} & \J[T_d] \otimes Y_0^{\otimes n-1} \otimes Y_d  \ar[u]_-{\lambda_{T_d}\{\uf^v_d\}_{v\in T_d}}}\]
is commutative, where $0,1 : \tensorunit \to J$ are part of the commutative segment $J$.
\end{theorem}

\begin{proof}
This follows from the Coherence Theorem \ref{thm:hpa-coherence}.  Indeed, in the above diagram from top to bottom:
\begin{itemize}\item The first rectangle is commutative by the associativity condition \eqref{wochatm-ass} and the grafting definition of $T^1$.
\item The second rectangle is commutative by the wedge condition \eqref{wochatm-wedge} applied to the tree substitution $T^1 = C(T^1)$.
\item The third rectangle is commutative by the same wedge condition applied to the tree substitution $T^2 = C(T^2)$.
\item The bottom rectangle is commutative by the associativity condition \eqref{wochatm-ass}, the grafting definition of $T^2$, and the unity condition \eqref{wochatm-unity}.
\end{itemize}
In the second and the third rectangles, one observes that the set $\Configl\dzerozeroc$ contains only the configuration \[\bigl\{\underbrace{0_d,\ldots, 0_d}_{n-1~\mathrm{copies}}, g\bigr\} = \gamma^{\Olhat}_{T^1}\Bigl(\{\uf^v_c\}_{v\in T_c},\, \{g_i\}_{i=1}^n\Bigr) = \gamma^{\Olhat}_{T^2}\Bigl(g,\,\{\uf^v_d\}_{v\in T_d}\Bigr).\] Along the left side of the diagram, the top and the bottom isomorphisms are of the form $? \cong \tensorunit \otimes ?$.  The middle two isomorphisms are of the form $? \cong \tensorunit^{\otimes |T^r|}\otimes ?$ with $r=1,2$.
\end{proof}

\begin{interpretation} In Theorem \ref{thm:hpadiag-einfinity-module}, the structure morphisms \[\lambda_{T_c}\left\{\uf^v_c\right\}_{v\in T_c} \andspace \lambda_{T_d}\left\{\uf^v_d\right\}_{v\in T_d}\] are $E_\infty$-module structure morphisms of $Y_c$ and $Y_d$, respectively, as in Corollary \ref{cor:hpa-lattice-emodule}.  The structure morphisms \[\lambda_{L_{\uc}}\{g_i\}_{i=1}^m \andspace \lambda_{L_{cd}}\{g\}\] are part of the underlying homotopy coherent $L$-diagram of $(Y,\lambda)$.  Therefore, the commutative diagram says that the homotopy coherent $L$-diagram structure of $(Y,\lambda)$ commutes with the objectwise $E_\infty$-module structure over the $E_\infty$-algebra $Y_0$ up to specified homotopies that are also structure morphisms.\dqed
\end{interpretation}

\begin{example}\label{ex1:hcgpfa-hcdiag-emod}
In Assumption \ref{assumption:hpa-hcdiag-mod}, suppose $m=1$, so $\uc=(c,d)$ and $L_{\uc}=L_{cd}=\Lin_{(c,d)}$.  In this case, the commutative diagram in Theorem \ref{thm:hpadiag-einfinity-module} says that the diagram \[\nicexy@C+1.2cm{\J[L_{cd}]\otimes \J[T_c] \otimes Y_0^{\otimes n-1} \otimes Y_c \ar[r]^-{\lambda_{T_c}\{\uf^v_c\}_{v\in T_c}} \ar[d]_-{\cong}
& \J[L_{cd}] \otimes Y_c \ar[dd]^-{\lambda_{L_{cd}}\{g\}}\\
\J[T_d] \otimes Y_0^{\otimes n-1} \otimes \J[L_{cd}] \otimes Y_c \ar[d]_-{\lambda_{L_{cd}}\{g\}}  & \\ \J[T_d] \otimes Y_0^{\otimes n-1} \otimes Y_d  \ar[r]^-{\lambda_{T_d}\{\uf^v_d\}_{v\in T_d}} & Y_d}\] is commutative up to specified homotopies that are also structure morphisms.  This is the homotopy coherent analogue of the commutative diagram \[\nicexy@C+1cm{Y_0 \otimes Y_c \ar[d]_-{\bigl(\Id,\lambda\{g\}\bigr)} \ar[r]^-{\lambda\{0_c,\Id_c\}} & Y_c \ar[d]^-{\lambda\{g\}}\\
Y_0 \otimes Y_d \ar[r]^-{\lambda\{0_d,\Id_d\}} & Y_d}\] in Corollary \ref{cor:lattice-diagram-modules} for a prefactorization algebra $(Y,\lambda)$ on $\Lhat$.\dqed
\end{example}

\begin{example}[Homotopy Costello-Gwilliam prefactorization algebras]\label{ex2:hcgpfa-hcdiag-emod}
For\index{homotopy prefactorization algebra!Costello-Gwilliam} a topological space $X$, the configured category \[\Openxhat =\bigl(\Openx,\Configx\bigr)\] in Example \ref{ex:open-configuration} has the form $\Lhat$ for the bounded lattice $\Openx$  with least element $\varnothing_X \subset X$.  Therefore, Theorem \ref{thm:hpadiag-einfinity-module} applies to every homotopy prefactorization algebra $(Y,\lambda)$ on $\Openxhat$.  So the homotopy coherent $\Openx$-diagram structure in $Y$ is homotopically compatible with the objectwise $E_\infty$-module structure over the $E_\infty$-algebra $Y_{\varnothing_X}$.\dqed
\end{example}

\begin{example}[Homotopy Costello-Gwilliam equivariant prefactorization algebras]\label{ex3:hcgpfa-hcdiag-emod}
Suppose\index{homotopy equivariant prefactorization algebra} $G$ is a group, and $X$ is a topological space in which $G$ acts on the left by homeomorphisms.  The homotopy prefactorization algebras on the configured category \[\Openxghat =\bigl(\Openxg,\Configxg\bigr)\] in Example \ref{ex:eq-space-configuration} are related to those on the configured category $\Openxhat$ via the change-of-operad adjunction \[\nicexy@C+1cm@R-.3cm{\algm\bigl(\wom_{\Openxhat}\bigr) \ar@{=}[d] \ar@<2pt>[r]^-{(\W\Otom_{\iota})_!} & \algm\bigl(\wom_{\Openxghat}\bigr) \ar@{=}[d] \ar@<2pt>[l]^-{(\W\Otom_{\iota})^*}\\ \HPFA\bigl(\Openxhat\bigr) & \HPFA\bigl(\Openxghat\bigr)}\] in Example \ref{ex:hcgeqpfa}.  In particular, every homotopy prefactorization algebra $(Y,\lambda)$ on $\Openxghat$ has an underlying homotopy prefactorization algebra on $\Openxhat$.  Therefore, by Example \ref{ex2:hcgpfa-hcdiag-emod}, for each homotopy prefactorization algebra $(Y,\lambda)$ on $\Openxghat$, the homotopy coherent $\Openx$-diagram structure in $Y$ is homotopically compatible with the objectwise $E_\infty$-module structure over the $E_\infty$-algebra $Y_{\varnothing_X}$.\dqed
\end{example}

\section{Homotopy Coherent Diagrams of $E_\infty$-Algebras}\label{sec:hpa-hcdiag-einfinity}

Suppose $\C$ is a small category.  In this section, we explain that homotopy prefactorization algebras on the maximal configured category $\Chatmax = (\C,\Configcmax)$ in Example \ref{ex:max-configuration} are\index{homotopy coherent diagram!$E_\infty$-algebra}\index{einfinityalgebra@$E_\infty$-algebra!homotopy coherent diagram} homotopy coherent $\C$-diagrams of $E_\infty$-algebras.  Recall the colored operads $\Comc$ in Example \ref{ex:diag-com-operad} and $\Ochat$ in Definition \ref{def:ochat-opread}.

\begin{lemma}\label{lem:hcdiag-einf-algebra}
Suppose $\C$ is a small category with object set $\colorc$.  There is an equality \[\Ochatmaxm = \Comc\] of $\colorc$-colored operads in $\M$.
\end{lemma}

\begin{proof}
Both $\colorc$-colored operads have entries \[\Ochatmaxm\duc= \coprodover{\Configcmax\duc} \tensorunit = \coprodover{\prod\limits_{j=1}^n \C(c_j,d)} \tensorunit = \Comc\duc\] for $\duc=\dconecn \in \Profcc$.  From Example \ref{ex:diag-com-operad} and Definition \ref{def:ochat-opread}, their operad structures also coincide.
\end{proof}

Recall from Definition \ref{def:wcomc-algebra} that $\Wcomc$-algebras are called homotopy coherent $\C$-diagrams of $E_\infty$-algebras in $\M$.

\begin{corollary}\label{cor:hcdiag-einf-algebra}
There is an equality \[\algmwochatmaxm = \algmwcomc\] between the category of homotopy prefactorization algebras on $\Chatmax$, i.e., $\wochatmaxm$-algebras and the category of homotopy coherent $\C$-diagrams of $E_\infty$-algebras in $\M$.
\end{corollary}

\begin{proof}
We first apply the Boardman-Vogt construction in Theorem \ref{thm:wo-operad} to the equality of colored operads in Lemma \ref{lem:hcdiag-einf-algebra} and then take the category of algebras.
\end{proof}

\begin{example}[Homotopy Costello-Gwilliam prefactorization algebras on manifolds]\label{ex:hpfa-manifolds-einf}
Suppose\index{homotopy prefactorization algebra!on manifolds} $\Embn$ is a small category equivalent to the category of smooth $n$-manifolds with open embeddings as morphisms.  Recall from Example \ref{ex:pfa-manifolds} that symmetric monoidal functors $\Embn \to \M$ are called prefactorization algebras on $n$-manifolds with values in $\M$.  The category of such objects is isomorphic to the category $\PFA(\Embnhatmax)$ of prefactorization algebras on $\Embnhatmax$.  By Corollary \ref{cor:hcdiag-einf-algebra} there is an equality \[\algm\bigl(\wom_{\Embnhatmax}\bigr) = \algm\bigl(\W\Com^{\Embn}\bigr)\] between the category of homotopy prefactorization algebras on the maximal configured category $\Embnhatmax$, i.e., $\wom_{\Embnhatmax}$-algebras and the category of homotopy coherent $\Embn$-diagrams of $E_\infty$-algebras in $\M$.\dqed
\end{example}

\begin{example}[Homotopy Costello-Gwilliam prefactorization algebras on complex manifolds]\label{ex:hpfa-cpmanifolds-einf}
Suppose\index{homotopy prefactorization algebra!on complex manifolds} $\Holn$ is a small category equivalent to the category of complex $n$-manifolds with open holomorphic embeddings as morphisms.  Recall from Example \ref{ex:pfa-cpmanifolds} that symmetric monoidal functors $\Holn \to \M$ are called prefactorization algebras on complex $n$-manifolds with values in $\M$.  The category of such objects is isomorphic to the category $\PFA(\Holnhatmax)$ of prefactorization algebras on $\Holnhatmax$.  By Corollary \ref{cor:hcdiag-einf-algebra} there is an equality \[\algm\bigl(\wom_{\Holnhatmax}\bigr) = \algm\bigl(\W\Com^{\Holn}\bigr)\] between the category of homotopy prefactorization algebras on the maximal configured category $\Holnhatmax$, i.e., $\wom_{\Holnhatmax}$-algebras and the category of homotopy coherent $\Holn$-diagrams of $E_\infty$-algebras in $\M$.\dqed
\end{example}

\chapter{Comparing Prefactorization Algebras and AQFT}\label{ch:comparing}

In this chapter, we compare (homotopy) prefactorization algebras and (homotopy) algebraic quantum field theories.

Recall from Chapter \ref{ch:aqft} and Chapter \ref{ch:haqft} that algebraic quantum field theories and their homotopy analogues are defined as algebras over the colored operads $\Ocbarm$ and $\wocbarm$ for an orthogonal category $\Cbar$.  In Chapter \ref{ch:pfa} and Chapter \ref{ch:hpa}, prefactorization algebras and their homotopy analogues are defined as algebras over the colored operads $\Ochatm$ and $\wochatm$ for a configured category $\Chat$.  To compare these objects, in Section \ref{sec:relation-orthcat} we first observe that every orthogonal category yields a configured category in which the configurations are the finite sequences of pairwise orthogonal morphisms.

In Section \ref{sec:confcat-to-orthcat} we observe that we can also go backward, from configured categories to orthogonal categories, by restricting to binary configurations.  More formally, the category of orthogonal categories  embeds in the category of configured categories as a full reflective subcategory.  We will show by examples that this is not an adjoint equivalence, so the two categories are genuinely different.

In Section \ref{sec:adjunction-pfa-aqft} we show that, for each configured category, there is a comparison morphism from the colored operad defining prefactorization algebras to the colored operad defining algebraic quantum field theories.  This comparison morphism is well-behaved with respect to configured functors and the time-slice axiom.  As a consequence, we have various comparison adjunctions between (homotopy) prefactorization algebras and (homotopy) algebraic quantum field theories.

In Section \ref{sec:example-operadic-comparison} we illustrate the comparison adjunctions with many examples.  In Section \ref{sec:pfa-from-aqft} we identify precisely the prefactorization algebras that come from algebraic quantum field theories.

As in previous chapters, $(\M,\otimes,\tensorunit)$ is a cocomplete symmetric monoidal closed category, such as $\Chaink$, with a commutative segment $(J,\mu,0,1,\epsilon)$ as in Definition \ref{def:segment}.  For a small category $\C$, its object set is denoted by $\colorc$.

\section{Orthogonal Categories as Configured Categories}\label{sec:relation-orthcat}

In this section, we show that orthogonal categories as in Definition \ref{def:orthogonal-category} yield configured categories as in Definition \ref{def:configcat}.

\begin{definition}\label{def:orthcat-to-confcat}
Suppose $\Cbar = (\C,\perpc)$ is an orthogonal category.  For objects $c_1,\ldots,c_n,d\in \C$ with $n \geq 0$ and $\uc = (c_1,\ldots,c_n)$, define $\Configc\duc$ as the set of pairs $\bigl(d;\{f_i\}_{i=1}^n\bigr)$ such that
\begin{itemize}\item $\{f_i\}_{i=1}^n \in \prod_{i=1}^n \C(c_i,d)$ and
\item if $1 \leq i \not=j \leq n$, then $f_i \perpc f_j$.
\end{itemize}
\end{definition}

\begin{interpretation} Configurations in $\Configc$ are those $\{f_i\} \in \prod\C(c_i,d)$ such that the $f_i$'s are pairwise orthogonal.  As in Notation \ref{not:configcat-notation} we will usually write $(d;\{f_i\})$ as $\{f_i\}$.\dqed\end{interpretation}

Recall from Definitions \ref{def:orthogonal-category} and \ref{def:config-functor} that $\Orthcat$ is the category of orthogonal categories and that $\Configcat$ is the category of configured categories.

\begin{proposition}\label{prop:orthcat-to-confcat}
Suppose $\Cbar = (\C,\perpc)$ is an orthogonal category.  
\begin{enumerate}
\item With $\Configc$ as in Definition \ref{def:orthcat-to-confcat}, $\Psi\Cbar = (\C,\Configc)$\label{notation:Psi} is\index{orthogonal category!to configured category} a configured category.  
\item This construction defines a functor \[\Psi : \Orthcat \to \Configcat\] that leaves the underlying categories and functors unchanged.
\item $\Psi$ sends each orthogonal equivalence to a configured equivalence.
\end{enumerate}
\end{proposition}

\begin{proof}
For the first assertion, the subset axiom and the inclusivity axiom follow directly from the definition of $\Configc$.  The symmetry axiom follows from that of the orthogonality relation $\perpc$.  The composition axiom follows from the fact that the orthogonality relation is closed under both post-compositions and pre-compositions.  Indeed, using the notation in \eqref{configured-composition}, we need to show that the $f_ig_{ij}$'s are pairwise orthogonal.
\begin{itemize} \item If $1 \leq j \not=j' \leq k_i$ for some $1\leq i \leq n$, then \[f_ig_{ij} \perpc f_ig_{ij'}\] by the post-composition axiom of $\perpc$ because $g_{ij} \perpc g_{ij'}$ by assumption.
\item If $1 \leq i\not=i' \leq n$, $1 \leq j \leq k_i$, and $1 \leq j' \leq k_{i'}$, then \[f_ig_{ij} \perpc f_{i'}g_{i'j'}\] by the pre-composition axiom of $\perpc$ because $f_i \perpc f_{i'}$ by assumption.
\end{itemize}
Therefore, $\Chat$ is a configured category.  

For the functoriality of this construction, observe that an orthogonal functor $F$ sends a configuration $\{f_i\}$ to $\{Ff_i\}$, where the $Ff_i$'s are pairwise orthogonal because the $f_i$'s are.  The last assertion follows immediately from the definition.
\end{proof}

In particular, every orthogonal category in Section \ref{sec:aqft-examples} is sent by the functor $\Psi$ to a configured category.

\begin{example}[Minimal and maximal orthogonal categories]\label{ex:Psi-minmax}
For\index{minimal orthogonal category}\index{maximal orthogonal category} each small category $\C$, there are equalities \[\Psi\Cbarmin = (\C,\Configcmin) = \Chatmin \andspace \Psi\Cbarmax = (\C,\Configcmax) = \Chatmax.\] Here $\Cbarmin$ and $\Cbarmax$ are the minimal and maximal orthogonal categories in Examples \ref{ex:empty-causality} and \ref{ex:everything-causality}.  On the other hand, $\Chatmin$ and $\Chatmax$ are the minimal and maximal configured categories in Examples \ref{ex:min-configuration} and \ref{ex:max-configuration}.\dqed
\end{example}

\begin{example}[Bounded lattices]\label{ex:Psi-lattice}
Suppose\index{bounded lattice} $(L,\leq)$ is a bounded lattice with least element $0$ as in Example \ref{ex:lattice}.  There is an equality \[\Psi(L,\perp) = (L,\Configl) = \Lhat\] with $(L,\perp)$ the orthogonal category in Example \ref{ex:qft-lattice} and $\Lhat$ the configured category in Example \ref{ex:boundedlatter-config}.\dqed
\end{example}

\begin{example}[Topological spaces]\label{ex:Psi-space}
For each topological space\index{topological space} $X$, there is an equality \[\Psi\Openxbar = \Openxhat\] with $\Openxbar$ the orthogonal category in Example \ref{ex:qft-space} and $\Openxhat$ the configured category in Example \ref{ex:open-configuration}.\dqed
\end{example}

\begin{example}[Equivariant topological spaces]\label{ex:Psi-eqspace}
Suppose $G$ is a group, and $X$ is a topological space\index{equivariant topological space} in which $G$ acts on the left by homeomorphisms.  There is an equality \[\Psi\Openxgbar = \Openxghat\] with $\Openxgbar$ the orthogonal category in Example \ref{ex:qft-eqspace} and $\Openxghat$ the configured category in Example \ref{ex:eq-space-configuration}.\dqed
\end{example}

\begin{example}[Oriented manifolds]\label{ex:Psi-man}
Recall from Example \ref{ex:ccqft} the orthogonal category $(\Mand,\perp)$, where $\Mand$ is the category of $d$-dimensional \index{oriented manifold}oriented manifolds with orientation-preserving open embeddings as morphisms in Example \ref{ex:man-cat}.  Two morphisms $g_1 : X_1 \to X$ and $g_2 : X_2 \to X$ in $\Mand$ are orthogonal if and only if their images are disjoint subsets in $X$.  In the configured category \[\Psi(\Mand,\perp) = (\Mand,\Config),\] a finite sequence of morphisms $\{g_i : X_i \shortto X\}_{i=1}^n$ is a configuration if and only if the images $g_iX_i$ are pairwise disjoint in $X$.\dqed
\end{example}

\begin{example}[Discs]\label{ex:Psi-disc}
Recall from Example \ref{ex:ccqft-int} the orthogonal category $(\Discd,\perp)$, where $\Discd$ is the full subcategory of $\Mand$ of $d$-dimensional oriented manifolds diffeomorphic to $\fieldr^d$ in Example \ref{ex:disc-cat}.  In the configured category \[\Psi(\Discd,\perp) = (\Discd,\Config),\] a finite sequence of morphisms $\{g_i : X_i \shortto X\}_{i=1}^n$ is a configuration if and only if their images $g_iX_i$ are pairwise disjoint in $X$.\dqed
\end{example}

\begin{example}[Oriented Riemannian manifolds]\label{ex:Psi-riem}
Recall from Example \ref{ex:euclidean-qft} the orthogonal category\index{Riemannian manifold} $(\Riemd,\perp)$, where $\Riemd$ is the category with $d$-dimensional oriented Riemannian manifolds as objects and orientation-preserving isometric open embeddings as morphisms in Example \ref{ex:riem-cat}.  In the configured category \[\Psi(\Riemd,\perp) = (\Riemd,\Config),\] a finite sequence of morphisms $\{g_i : X_i \shortto X\}_{i=1}^n$ is a configuration if and only if their images $g_iX_i$ are pairwise disjoint in $X$.\dqed
\end{example}

\begin{example}[Lorentzian manifolds]\label{ex:Psi-loc}
Consider the orthogonal category\index{Lorentzian manifold} $(\Locd,\perp)$ in Example \ref{ex:lcqft}, where $\Locd$ is the category of $d$-dimensional oriented, time-oriented, and globally hyperbolic Lorentzian manifolds in Example \ref{ex:loc-cat}.  In the configured category \[\Psi(\Locd,\perp) = (\Locd,\Config),\] a finite sequence of morphisms $\{g_i : X_i \shortto X\}_{i=1}^n$ is a configuration if and only if their images $g_iX_i$ are pairwise causally disjoint in $X$.\dqed
\end{example}

\begin{example}[A fixed spacetime]\label{ex:Psi-ghx}
Recall from Example \ref{ex:causal-nets} the orthogonal category $(\Ghx,\perp)$ with $X \in \Locd$ and $\Ghx$ the category of\index{globally hyperbolic open subset} globally hyperbolic open subsets of $X$ with subset inclusions as morphisms in Example \ref{ex:gh-cat}.  In the configured category \[\Psi(\Ghx,\perp) = (\Ghx,\Config),\] a finite sequence of morphisms $\{g_i : U_i \shortto V\}_{i=1}^n$ is a configuration if and only if the $U_i$'s are pairwise causally disjoint in $V$.\dqed
\end{example}

\begin{example}[Lorentzian manifolds with bundles]\label{ex:Psi-bgloc}
Consider the orthogonal category $(\Bgloc, \pi^*(\perp))$ in Example \ref{ex:dynamical-gauge}, where $\Bgloc$ is the category of $d$-dimensional oriented, time-oriented, and globally hyperbolic Lorentzian manifolds equipped with a principal $G$-bundle in Example \ref{ex:bgloc-cat}.  The functor \[\pi : \Bgloc \to \Locd\] forgets about the bundle structure, with $\pi^*(\perp)$ the pullback of the orthogonality relation $\perp$ in $\Locd$.  In the configured category \[\Psi\bigl(\Bgloc,\pi^*(\perp)\bigr) = (\Bgloc,\Config),\] a finite sequence of morphisms $\bigl\{g_i : (X_i,P_i) \shortto (X,P)\bigr\}_{i=1}^n$ is a configuration if and only if the images $g_iX_i$ are pairwise causally disjoint in $X$.\dqed
\end{example}

\begin{example}[Lorentzian manifolds with bundles and connections]\label{ex:Psi-bgconloc}
Consider the orthogonal category $(\Bgconloc, (\pi p)^*(\perp))$ in Example \ref{ex:charged-matter}, where $\Bgconloc$ is the category of triples $(X,P,C)$ with $(X,P) \in \Bgloc$ and $C$ a connection on $P$ in Example \ref{ex:bgconloc-cat}.  The functor \[\pi p : \Bgconloc \to \Locd\] forgets about the bundle structure and the connection, with $(\pi p)^*(\perp)$ the pullback of the orthogonality relation $\perp$ in $\Locd$.  In the configured category \[\Psi\bigl(\Bgconloc,(\pi p)^*(\perp)\bigr) = (\Bgconloc,\Config),\] a finite sequence of morphisms $\bigl\{g_i : (X_i,P_i,C_i) \shortto (X,P,C)\bigr\}_{i=1}^n$ is a configuration if and only if the images $g_iX_i$ are pairwise causally disjoint in $X$.\dqed
\end{example}

\begin{example}[Lorentzian spin manifolds]\label{ex:Psi-slocd}
Consider the orthogonal category\index{spin manifold} $(\Slocd, \pi^*(\perp))$ in Example \ref{ex:dirac-qft}, where $\Slocd$ is the category of $d$-dimensional oriented, time-oriented, and globally hyperbolic Lorentzian spin manifolds in Example \ref{ex:sloc-cat}.  The functor \[\pi : \Slocd \to \Locd\] forgets about the spin structure, with $\pi^*(\perp)$ the pullback of the orthogonality relation $\perp$ in $\Locd$.  In the configured category \[\Psi\bigl(\Slocd,\pi^*(\perp)\bigr) = (\Slocd,\Config),\] a finite sequence of morphisms $\bigl\{g_i : (X_i,P_i,\psi_i) \shortto (X,P,\psi)\bigr\}_{i=1}^n$ is a configuration if and only if the images $g_iX_i$ are pairwise causally disjoint in $X$.\dqed
\end{example}

\begin{example}[Structured spacetimes]\label{ex:Psi-str}
Consider\index{spacetime!structured} the orthogonal category $(\Str, \pi^*(\perp))$ in Example \ref{ex:qft-structured}, with \[\pi : \Str \to \Locd\] a functor between small categories and $\pi^*(\perp)$ the pullback of the orthogonality relation $\perp$ in $\Locd$.  In the configured category \[\Psi\bigl(\Str,\pi^*(\perp)\bigr) = (\Str,\Config),\] a finite sequence of morphisms $\{g_i : X_i \shortto X\}_{i=1}^n$ is a configuration if and only if the images $(\pi g_i)(\pi X_i)$ are pairwise causally disjoint in $\pi X$.\dqed
\end{example}

\begin{example}[Spacetime with timelike boundary]\label{ex:Psi-boundary}
Suppose $X$ is a\index{spacetime!with timelike boundary} spacetime with timelike boundary, and $\Regx$ is the category of regions in $X$ in Example \ref{ex:regions}.  Consider the orthogonal category $(\Regx,\perp)$ in Example \ref{ex:aqft-boundary}.  In the configured category \[\Psi(\Regx,\perp) = (\Regx,\Config),\] a finite sequence of morphisms $\{g_i : U_i \shortto V\}_{i=1}^n$ is a configuration if and only if the $U_i$'s are pairwise causally disjoint in $V$.\dqed
\end{example}

\section{Configured Categories to Orthogonal Categories}\label{sec:confcat-to-orthcat}

In this section, we observe that each configured category yields an orthogonal category in which an orthogonal pair is exactly a binary configuration.  Moreover, the category of orthogonal categories embeds as a full reflective subcategory of the category of configured categories.

\begin{definition}\label{def:confcat-to-orthcat}
Suppose $\Chat = (\C,\Configc)$ is a configured category.  Define $\perpc$ as the set of pairs $\{g_1,g_2\}$ in $\Configc$.
\end{definition}

\begin{proposition}\label{prop:confcat-to-orthcat}
Suppose $\Chat = (\C,\Configc)$ is a configured category.  
\begin{enumerate}
\item With $\perpc$ as in Definition \ref{def:confcat-to-orthcat}, $\Phi\Chat = (\C,\perpc)$\label{notation:Phi} is an orthogonal category.\index{configured category!to orthogonal category}
\item This construction defines a functor \[\Phi : \Configcat\to\Orthcat\] that leaves the underlying categories and functors unchanged.
\item $\Phi$ sends each configured equivalence to an orthogonal equivalence.
\end{enumerate}
\end{proposition}

\begin{proof}
The symmetry of $\perpc$ follows from that of $\Configc$.  By the inclusivity axiom of $\Configc$, each morphism $f$ in $\C$ yields a configuration $\{f\}$.  So the composition axiom of $\Configc$ implies both the post-composition axiom and the pre-composition axiom of $\perpc$.  For the second assertion, observe that a configured functor preserves all the configurations, in particular the binary configurations $\{g_1,g_2\}$.  The last assertion follows immediately from the definition.
\end{proof}

\begin{definition}\label{def:phi-of-chat}
For a configured category $\Chat$, we call $\Phi\Chat$ the \emph{associated orthogonal category}.\index{associated orthogonal category}\index{orthogonal category!associated}
\end{definition}

The next observation says that the category $\Orthcat$ of orthogonal categories embeds in the category $\Configcat$ of configured categories via the functor $\Psi$ in Proposition \ref{prop:orthcat-to-confcat} as a full reflective subcategory.

\begin{theorem}\label{thm:confcat-orthcat-adjunction}
There is an\index{reflective subcategory} adjunction \[\nicexy{\Configcat \ar@<2pt>[r]^-{\Phi} & \Orthcat \ar@<2pt>[l]^{\Psi}}\] with left adjoint $\Phi$ such that the counit $\Phi\Psi \to \Id_{\Orthcat}$ is the identity natural transformation.  In particular, every orthogonal category is the $\Phi$-image of some configured category.
\end{theorem}

\begin{proof}
Since both functors $\Phi$ and $\Psi$ leave the underlying categories and functors unchanged, to establish the adjunction, it suffices to prove the following statement.  Suppose $\Chat = (\C,\Configc)$ is a configured category and $\Dbar = (\D,\perpd)$ is an orthogonal category.  For each functor $F : \C \to \D$, the following statements are equivalent:
\begin{enumerate}
\item $F$ sends each binary configuration in $\Chat$ to an orthogonal pair in $\Dbar$.
\item The image of each configuration in $\Chat$ under $F$ is pairwise orthogonal in $\Dbar$.
\end{enumerate}
To see that (2) implies (1), simply note that each binary configuration is a configuration.  To see that (1) implies (2), suppose $\{g_i\}_{i=1}^n$ is a configuration in $\Chat$ with $1 \leq i\not=j \leq n$.  By the subset axiom of a configured category, $\{g_i,g_j\}$ is also a configuration in $\Chat$.  So by (1) their images $\{Fg_i,Fg_j\}$ are orthogonal in $\Dbar$.

The equality \[\Phi\Psi = \Id_{\Orthcat}\] follows directly from the definition of the functors $\Phi$ and $\Psi$.
\end{proof}

In Theorem \ref{thm:confcat-orthcat-adjunction} we observed that the counit $\Phi\Psi \to \Id_{\Orthcat}$ is the identity natural transformation, so each orthogonal category $\Cbar$ is equal to $\Phi\Psi\Cbar$.  In particular, this is true for all the orthogonal categories in Section \ref{sec:aqft-examples}.  Below are some examples.

\begin{example}[Empty orthogonality and minimal configuration]\label{ex:empty-minconfig}
For each small category $\C$, there are equalities \[\Psi(\C,\emptyset) = (\C,\Configcmin) \andspace \Phi(\C,\Configcmin) = (\C,\emptyset).\] Here $\emptyset$ is the empty orthogonality relation in Example \ref{ex:empty-causality}, and $(\C,\Configcmin)$ is the\index{minimal configured category} minimal configured category on $\C$ in Example \ref{ex:min-configuration}.\dqed
\end{example}

\begin{example}[Maximal orthogonality and maximal configuration]\label{ex:max-orthconfig}
For each small category $\C$, there are equalities \[\Psi(\C,\perpmax) = (\C,\Configcmax) \andspace \Phi(\C,\Configcmax) = (\C,\perpmax).\] Here $\perpmax$ is the orthogonality relation in Example \ref{ex:everything-causality}, and $(\C,\Configcmax)$ is the\index{maximal configured category} maximal configured category on $\C$ in Example \ref{ex:max-configuration}.\dqed
\end{example}

\begin{example}[Orthogonality and configuration of bounded lattices]\label{ex:lattice-orthconfig}
For\index{bounded lattice} each bounded lattice $(L,\leq)$, there are equalities \[\Psi(L,\perp) = (L,\Configl) \andspace \Phi(L,\Configl) = (L,\perp).\] Here $\perp$ is the orthogonality relation in Example \ref{ex:qft-lattice}, and $(L,\Configl)$ is the configured category on $L$ in Example \ref{ex:boundedlatter-config}.\dqed
\end{example}

\begin{example}[Orthogonality and configuration of topological spaces]\label{ex:top-orthconfig}
For each\index{topological space} topological space $X$, there are equalities \[\Psi\bigl(\Openxbar\bigr) = \Openxhat \andspace \Phi\bigl(\Openxhat\bigr) = \Openxbar.\] Here $\Openxbar$ is the orthogonal category in Example \ref{ex:qft-space}, and $\Openxhat$ is the configured category in Example \ref{ex:open-configuration}.\dqed
\end{example}

\begin{example}[Orthogonality and configuration of equivariant topological spaces]\label{ex:eqtop-orthconfig}
For each\index{equivariant topological space} topological space $X$ with a left action by a group $G$, there are equalities \[\Psi\bigl(\Openxgbar\bigr) = \Openxghat \andspace \Phi\bigl(\Openxghat\bigr) = \Openxgbar.\] Here $\Openxgbar$ is the orthogonal category in Example \ref{ex:qft-eqspace}, and $\Openxghat$ is the configured category in Example \ref{ex:eq-space-configuration}.\dqed
\end{example}

The next two examples show that the unit of the adjunction $\Id_{\Configcat} \to \Psi\Phi$ is not a natural isomorphism.  Therefore, the adjunction $\Phi \dashv \Psi$ is not an adjoint equivalence.

\begin{example}[$\Psi$ is not essentially surjective]\label{ex:psi-not-esssur}
Consider the category $\C$ with four objects $\Obc = \{a,b_1,b_2,b_3\}$ and only three non-identity morphisms $f_i : b_i \to a$ for $i=1,2,3$.  We may visualize the category $\C$ as follows.
\[\nicexy{& b_2 \ar[d]^-{f_2} &\\ b_1\ar[r]^-{f_1} & a & b_3\ar[l]_-{f_3}}\]
For any two objects $c,d$ in $\C$, define the sets:
\[\begin{split} \Configc\cempty &= *,\\ 
\Configc\dc &= \C(c,d),\\
\Configc\sbinom{a}{c,d} &= \C(c,a) \times \C(d,a) \quad\text{ if $c\not= d$ and $c,d \in \{b_1,b_2,b_3\}$}\end{split}\]  All other sets $\Configc\duc$ are empty.  Then $\Chat = (\C,\Configc)$ is a configured category.

However, the configured category $\Chat$ is not in the essential image of the functor $\Psi$.  Indeed, if it is in the essential image of $\Psi$, then $\Configc$ is as in Definition \ref{def:orthcat-to-confcat} for some orthogonality relation $\perpc$ on $\C$.  By the definition of $\Configc$, we have $f_i \perpc f_j$ for $1 \leq i\not=j \leq 3$.  But then $\{f_1,f_2,f_3\}$ is also pairwise orthogonal, so it forms a configuration, which contradicts the definition of $\Configc$.\dqed
\end{example}

\begin{example}[$\Phi$ is not an embedding]\label{ex:phi-not-embedding}
Suppose $\C$ is the category in Example \ref{ex:psi-not-esssur}.  Define $\Configczero$ to be the same as $\Configc$ except that \[\Configczero\sbinom{a}{x,y,z} = \C(x,a) \times \C(y,a) \times \C(z,a) \ifspace \{x,y,z\} = \{b_1,b_2,b_3\}\] as sets.  Then $(\C,\Configczero)$ is also a configured category, which is not isomorphic to $(\C,\Configc)$ since the latter has no triple configurations.  However, the images $\Phi(\C,\Configc)$ and $\Phi(\C,\Configczero)$ are equal as orthogonal categories because $\Configc$ and $\Configczero$ have the same binary configurations.  So the functor $\Phi$ is not an embedding.\dqed
\end{example}

\section{Comparison Adjunctions}\label{sec:adjunction-pfa-aqft}

In this section, we show that for each configured category, there is a comparison morphism from the colored operad for prefactorization algebras to the colored operad for algebraic quantum field theories.  This comparison morphism induces a comparison adjunction between the category of (homotopy) prefactorization algebras and the category of (homotopy) algebraic quantum field theories on the associated orthogonal category.  The comparison morphism is also compatible with changing the configured category and with the time-slice axiom.

Recall from Definition \ref{def:aqft-operad} the colored operad $\Ocbar$ for an orthogonal category $\Cbar$ and from Definition \ref{def:ochat-opread} the colored operad $\Ochat$ for a configured category $\Chat$.  Also recall from Example \ref{ex:operad-set-m} the change-of-category functor \[(-)^{\M} : \Operadcset \to \Operadcm.\] We will use the augmentation $\eta : \W \to \Id$ of the Boardman-Vogt construction in Theorem \ref{thm:w-augmented}.

\begin{theorem}\label{thm:compare-pfa-aqft}
Suppose $\Chat = (\C,\Config)$ is a configured category with object set $\colorc$, and $\Phi\Chat = \Cbar$ is the associated orthogonal category.
\begin{enumerate}\item There is a morphism\index{comparison morphism} \[\nicexy{\Ochat \ar[r]^-{\delta} & \Ocbar}\] of $\colorc$-operads in $\Set$ that is entrywise defined as \[\nicexy{\Ochat\duc \ar[r]^-{\delta} & \Ocbar\duc},\qquad \nicexy{\{f_i\}_{i=1}^n \ar@{|->}[r] & [\id_n,\{f_i\}_{i=1}^n}]\] for $\duc=\dconecn \in \Profcc$ and $\{f_i\}_{i=1}^n \in \Ochat\duc = \Config\duc$.
\item There is a commutative diagram\index{homotopy comparison morphism} \[\nicexy@C+.4cm{\wochatm \ar[d]_-{\eta} \ar[r]^-{\W\deltam} & \wocbarm \ar[d]^-{\eta} \\ \Ochatm \ar[r]^-{\deltam} & \Ocbarm}\] of $\colorc$-colored operads in $\M$, in which $\eta : \wochatm \to \Ochatm$ and $\eta : \wocbarm \to \Ocbarm$ are the augmentations of $\Ochatm$ and $\Ocbarm$.
\item There is a diagram of change-of-operad adjunctions\index{comparison adjunction}\index{homotopy comparison adjunction}
\[\nicexy@C+.7cm@R+.3cm{\HPFA(\Chat)=\algmwochatm \ar@<2pt>[r]^-{\W\deltam_!} \ar@<-2pt>[d]_-{\eta_!} & \algmwocbarm= \HQFT(\Cbar) \ar@<2pt>[l]^-{(\W\deltam)^*} \ar@<-2pt>[d]_-{\eta_!} \\ \PFA(\Chat)=\algmochatm \ar@<2pt>[r]^-{\deltam_!} \ar@<-2pt>[u]_-{\eta^*}  & \algmocbarm \cong \QFT(\Cbar) \ar@<2pt>[l]^-{(\deltam)^*} \ar@<-2pt>[u]_-{\eta^*}}\]
with commuting left adjoint diagram and commuting right adjoint diagram.
\end{enumerate}
\end{theorem}

\begin{proof}
For the first assertion, first observe that the morphism $\delta$ is entrywise a well-defined function.  That $\delta$ preserves the colored units and the operadic compositions follows immediately from the definition.  To see that $\delta$ preserves the equivariant structures, suppose given a configuration $\{f_i\}_{i=1}^n \in \Config\duc$ and a permutation $\sigma \in \Sigma_n$.  We must show that the middle equality in \[\delta\Bigl(\{f_i\}_{i=1}^n\,\sigma\Bigr) = \bigl[\id_n, \{f_{\sigma(i)}\}_{i=1}^n\bigr] = \bigl[\sigma, \{f_{\sigma(i)}\}_{i=1}^n\bigr] = \Bigl(\delta\{f_i\}_{i=1}^n\Bigr)\sigma\] holds in $\Ocbar\duc$.  For any $1 \leq i \not=j \leq n$, by the subset axiom of a configured category, we have that $\{f_i,f_j\} \in \Config$.  It follows that $f_i \perp f_j$ in $(\C,\perp) = \Cbar$, since $\perp$ is defined as the set of binary configurations.  In other words, the $f_i$'s are pairwise orthogonal.  So the middle equality above holds by the definition of the equivalence relation $\sim$ that defines $\Ocbar$.

The second assertion follows from the first assertion, the change-of-category functor $(-)^{\M}$, and the naturality of the Boardman-Vogt construction.  The third assertion follows from the second assertion and Corollary \ref{cor:wf-eta-adjunction}.
\end{proof}

\begin{definition}\label{def:operadic-comparison}
In the setting of Theorem \ref{thm:compare-pfa-aqft}:
\begin{enumerate}\item The morphism\label{notation:compmorphism} \[\delta : \Ochat \to \Ocbar\] of $\colorc$-colored operads and its image $\deltam$ in $\M$ are called the \emph{comparison morphisms}.  
\item The morphism \[\W\deltam : \wochatm \to \wocbarm\] of $\colorc$-colored operads in $\M$ is called the \emph{homotopy comparison morphism}.  
\item The adjunction $\deltam_! \dashv (\deltam)^*$ is called the \emph{comparison adjunction}.
\item The adjunction $\W\deltam_! \dashv (\W\deltam)^*$ is called the \emph{homotopy comparison adjunction}. 
\end{enumerate}
\end{definition}

\begin{interpretation} The comparison adjunction \[\nicexy@C+.5cm{\PFA(\Chat)=\algmochatm \ar@<2pt>[r]^-{\deltam_!} & \algmocbarm \cong \QFT(\Cbar) \ar@<2pt>[l]^-{(\deltam)^*}}\] compares prefactorization algebras on $\Chat$ with algebraic quantum field theories on the associated orthogonal category $\Cbar = \Phi\Chat$.  The homotopy comparison adjunction \[\nicexy@C+.7cm{\HPFA(\Chat)=\algmwochatm \ar@<2pt>[r]^-{\W\deltam_!} & \algmwocbarm= \HQFT(\Cbar) \ar@<2pt>[l]^-{(\W\deltam)^*}}\] compares homotopy prefactorization algebras on $\Chat$ with homotopy algebraic quantum field theories on the associated orthogonal category $\Cbar$.\dqed  
\end{interpretation}

The following observation is the relative version of Theorem \ref{thm:compare-pfa-aqft}.

\begin{corollary}\label{cor:compare-pfa-aqft-relative}
Suppose $F : \Chat \to \Dhat$ is a configured functor with  $\Phi\Chat = \Cbar$ and $\Phi\Dhat = \Dbar$.
\begin{enumerate}\item There is a commutative diagram \[\nicexy@C+.3cm{\Ochat \ar[r]^-{\O_F} \ar[d]_-{\delta} & \Odhat \ar[d]^-{\delta}\\ \Ocbar \ar[r]^-{\O_{\Phi F}} & \Odbar}\] of colored operads in $\Set$ with both morphisms $\delta$ as in Theorem \ref{thm:compare-pfa-aqft}.
\item There is a commutative cube \[\nicexy@C+.6cm{\wochatm \ar[dd]_-{\eta} \ar[dr]^{\W\deltam} \ar[rr]^-{\W\Otom_F} && \wodhatm \ar'[d][dd]_-{\eta} \ar[dr]^-{\W\deltam} &\\
& \wocbarm \ar[dd]^(.7){\eta} \ar[rr]^(.3){\W\Otom_{\Phi F}} && \wodbarm \ar[dd]^-{\eta}\\
\Ochatm \ar[dr]^-{\deltam} \ar'[r][rr]^-{\Otom_F} && \Odhatm \ar[dr]^-{\deltam} &\\
& \Ocbarm \ar[rr]^-{\Otom_{\Phi F}} && \Odbarm}\] of colored operads in $\M$, in which every morphism $\eta$ is an augmentation.
\item There is a diagram of change-of-operad adjunctions \[\nicexy{\algmwochatm \ar[dd]_-{\eta_!} \ar[dr]^{\W\deltam_!} \ar[rr]^-{(\W\Otom_F)_!} && \algmwodhatm \ar'[d][dd]_-{\eta_!} \ar[dr]^-{\W\deltam_!} &\\
& \algmwocbarm \ar[dd]^(.7){\eta_!} \ar[rr]^(.3){(\W\Otom_{\Phi F})_!} && \algmwodbarm \ar[dd]^-{\eta_!}\\
\algmochatm \ar[dr]^-{\deltam_!} \ar'[r][rr]^-{(\Otom_F)_!} && \algmodhatm \ar[dr]^-{\deltam_!} &\\ & \algmocbarm \ar[rr]^-{(\Otom_{\Phi F})_!} && \algmodbarm}\] 
in which only the left adjoints are displayed.  In each face, the left adjoint diagram and the right adjoint diagram are commutative.
\end{enumerate}
\end{corollary}

\begin{proof} The first assertion follows from Theorem \ref{thm:compare-pfa-aqft}(1) and the fact that $\Phi$ leaves the underlying functor $F$ unchanged.  The second assertion follows from the first assertion, the change-of-category functor $(-)^{\M}$, and the naturality of the Boardman-Vogt construction in Theorem \ref{thm:w-augmented}.  The last assertion follows from assertion (2) and Theorem \ref{thm:change-operad}.
\end{proof}

\begin{interpretation} In the commutative cube in Corollary \ref{cor:compare-pfa-aqft-relative}(3), the left and the right faces are the diagrams in Theorem \ref{thm:compare-pfa-aqft}(3) for $\Chat$ and $\Dhat$, respectively.  The front face is the diagram in Corollary \ref{cor:haqft-adjunction-diagram}, and the back face is the diagram in Corollary \ref{cor:hpa-adjunction-diagram}.  The bottom face compares prefactorization algebras on $\Chat$ and $\Dhat$ and algebraic quantum field theories on $\Cbar$ and $\Dbar$.  The top face compares homotopy prefactorization algebras on $\Chat$ and $\Dhat$ and homotopy algebraic quantum field theories on $\Cbar$ and $\Dbar$.\dqed
\end{interpretation}

Next we consider the situation when the\index{time-slice axiom}\index{homotopy time-slice axiom} time-slice axiom is present.  Recall the time-slice axiom from Definition \ref{def:aqft} for algebraic quantum field theories and from Definition \ref{def:pfa} for prefactorization algebras.  Also recall localization of a category $\ell : \C \to \Csinv$ from Definition \ref{def:localization-cat} and localization of a colored operad $\ell : \O \to \Osinv$ from Definition \ref{def:operad-localization}.

\begin{corollary}\label{cor:pfa-aqft-timeslice}
Suppose $\Chat = (\C,\Config)$ is a configured category with object set $\colorc$, and $\Phi\Chat = \Cbar$ is the associated orthogonal category.  Suppose $S$ is a set of morphisms in $\C$.
\begin{enumerate}
\item There exists a unique morphism \[\delta_S : \Ochatsinv \to \Ocsinvbar\] of $\colorc$-colored operads such that the diagram \[\nicexy@C+.3cm{\Ochat \ar[r]^-{\ell} \ar[d]_-{\delta} & \Ochatsinv \ar[d]^-{\delta_S}\\ \Ocbar \ar[r]^-{\O_\ell} & \Ocsinvbar}\] of $\colorc$-colored operads in $\Set$ is commutative, in which the morphism $\delta$ is from Theorem \ref{thm:compare-pfa-aqft}(1).
\item There is a commutative cube \[\nicexy@C+.3cm{\wochatm \ar[dd]_-{\eta} \ar[dr]^{\W\deltam} \ar[rr]^-{\W\ellm} && \wochatsinvm \ar'[d][dd]_-{\eta} \ar[dr]^-{\W\deltam_S} &\\ & \wocbarm \ar[dd]^(.7){\eta} \ar[rr]^(.3){\W\Otom_{\ell}} && \wocsinvbarm \ar[dd]^-{\eta}\\
\Ochatm \ar[dr]^-{\deltam} \ar'[r][rr]^-{\ellm} && \Ochatsinvm \ar[dr]^-{\deltam_S} &\\
& \Ocbarm \ar[rr]^-{\Otom_{\ell}} && \Ocsinvbarm}\] of $\colorc$-colored operads in $\M$, in which every morphism $\eta$ is an augmentation.
\item There is a diagram of change-of-operad adjunctions \[\nicexy@C-.6cm{\algmwochatm \ar[dd]_-{\eta_!} \ar[dr]^{\W\deltam_!} \ar[rr]^-{\W\ellm_!} && \algmwochatsinvm \ar'[d][dd]_-{\eta_!} \ar[dr]^-{(\W\deltam_S)_!} &\\
& \algmwocbarm \ar[dd]^(.7){\eta_!} \ar[rr]^(.3){(\W\Otom_{\ell})_!} && \algmwocsinvbarm \ar[dd]^-{\eta_!}\\
\algmochatm \ar[dr]^-{\deltam_!} \ar'[r][rr]^-{\ellm_!} && \algmochatsinvm \ar[dr]^-{(\deltam_S)_!} &\\ & \algmocbarm \ar[rr]^-{(\Otom_{\ell})_!} && \algmocsinvbarm}\] in which only the left adjoints are displayed.  In each face, the left adjoint diagram and the right adjoint diagram are commutative.
\end{enumerate}
\end{corollary}

\begin{proof}
For the first assertion, consider the solid-arrow diagram \[\nicexy@C+.3cm{\Ochat \ar[r]^-{\ell} \ar[d]_-{\delta} & \Ochatsinv \ar@{.>}[d]^-{\delta_S}\\ \Ocbar \ar[r]^-{\O_\ell} & \Ocsinvbar}\] of $\colorc$-colored operads in $\Set$.  For each morphism $f \in S$, regarded as a unary element in the $\colorc$-colored operad $\Ochat$, the unary element \[\O_\ell\delta(\{f\}) = [\id_1,f]\] is invertible in $\Ocsinvbar$ with inverse $[\id_1,\finverse]$.  This is true because the morphism $f \in \Csinv$ is invertible.  Therefore, by the universal property of an $S$-localization of $\Ochat$, there exists a unique morphism $\delta_S$ that makes the diagram commutative.

The second assertion follows from the first assertion, the change-of-category functor $(-)^{\M}$, and the naturality of the Boardman-Vogt construction in Theorem \ref{thm:w-augmented}.  The last assertion follows from assertion (2) and Theorem \ref{thm:change-operad}.
\end{proof}

\begin{interpretation} The adjunction \[\nicexy{\PFA(\Chat,S)=\algmochatsinvm \ar@<2pt>[r]^-{(\deltam_S)_!} & \algmocsinvbarm\cong \QFT(\Cbar,S) \ar@<2pt>[l]^-{(\deltam_S)^*}}\] compares (i) prefactorization algebras on the configured category $\Chat$ satisfying the time-slice axiom with respect to $S$ and (ii) algebraic quantum field theories on the associated orthogonal category $\Cbar$ satisfying the time-slice axiom with respect to $S$. The adjunction \[\nicexy{\HPFA(\Chat,S)=\algmwochatsinvm \ar@<2pt>[r]^-{(\W\deltam_S)_!} & \algmwocsinvbarm= \HQFT(\Csinvbar) \ar@<2pt>[l]^-{(\W\deltam_S)^*}}\] compares (i) homotopy prefactorization algebras on $\Chat$ satisfying the homotopy time-slice axiom with respect to $S$ and (ii) homotopy algebraic quantum field theories on $\Csinvbar$.\dqed
\end{interpretation}

\section{Examples of Comparison}\label{sec:example-operadic-comparison}

In this section, we provide a long list of examples that illustrate the comparison adjunctions in Section \ref{sec:adjunction-pfa-aqft} between (homotopy) prefactorization algebras and (homotopy) algebraic quantum field theories.

\begin{example}[The minimal case]\label{ex:comparison-minimal}
Suppose $\C$ is a small category.  In Example \ref{ex:empty-minconfig} we noted that \[\Phi(\C,\Configcmin) = \Cbarmin,\] with $\Chatmin=(\C,\Configcmin)$ the minimal configured category on $\C$ in Example \ref{ex:min-configuration} and $\Cbarmin = (\C,\emptyset)$  the orthogonal category with the empty orthogonality relation in Example \ref{ex:empty-causality}.  The comparison morphism \[\nicexy{\Ochatminm \ar[r]^-{\deltaminm} & \Ocbarminm = \Ocm
}\] was described in Proposition \ref{lem:ochatmin-to-ocbarmin}, in which the equality comes from Example \ref{ex:diag-monoid-operad}.  In Proposition \ref{prop:pointed-diagram} we noted that $\Ochatminm$ is the $\colorc$-colored operad whose algebras are pointed $\C$-diagrams in $\M$. In Corollary \ref{cor:cbarmin-chatmin} we observed that the comparison adjunction is the free-forgetful \index{diagram of monoids}\index{pointed diagram}adjunction \[\nicexy{\Mcstar \cong \PFA(\Chatmin)=\algm\bigl(\Otom_{\Chatmin}\bigr) \ar@<2pt>[r]^-{\deltaminmst} & \algm\bigl(\Otom_{\Cbarmin}\bigr)\cong\QFT(\Cbarmin)=\Monm^{\C} \ar@<2pt>[l]^-{\deltaminmstar}}\] between pointed $\C$-diagrams in $\M$ and $\C$-diagrams of monoids in $\M$. 

Furthermore, the homotopy comparison \index{homotopy pointed diagram}\index{homotopy coherent diagram!$A_\infty$-algebra}adjunction \[\nicexy{\HPFA(\Chatmin) =\algmwochatminm \ar@<2pt>[r]^-{(\W\deltaminm)_!} & \algmwocm = \algmwocbarminm = \HQFT(\Cbarmin) \ar@<2pt>[l]^-{(\W\deltaminm)^*}}\] is the free-forgetful adjunction between the category of homotopy coherent pointed $\C$-diagrams in $\M$ in Definition \ref{def:hcpt-diagram} and the category of homotopy coherent $\C$-diagrams of $A_\infty$-algebras in $\M$ in Definition \ref{def:wocm-algebra}.\dqed
\end{example}

\begin{example}[The classical case]\label{ex:comparison-classical-case}
Suppose $\C$ is a small category.  In Example \ref{ex:max-orthconfig} we noted that $\Phi\Chatmax = \Cbarmax$.  The isomorphism \[\nicexy{\O_{\Chatmax} \ar[r]^-{\deltamax}_-{\cong} & \O_{\Cbarmax}}\] in Proposition \ref{prop:deltamax} coincides with the comparison morphism $\delta$ in Theorem \ref{thm:compare-pfa-aqft}.  In Example \ref{ex:deltamax} we noted that the induced\index{diagram of commutative monoids} functor $(\deltam)^* = \deltamaxmstar$ is an isomorphism \[\nicexy{\PFA(\Chatmax)=\algm\bigl(\Otom_{\Chatmax}\bigr) & \algm\bigl(\Otom_{\Cbarmax}\bigr)\cong\QFT(\Cbarmax)=\Comm^{\C} \ar[l]_-{\deltamaxmstar}^-{\cong}}.\] In this case, the category of prefactorization algebras and the category of algebraic quantum field theories are both isomorphic to the category $\Commc$ of $\C$-diagrams of commutative monoids in $\M$.  We interpreted this situation as saying that the two mathematical approaches to quantum field theory coincide in the classical case.  

Furthermore, since the comparison morphism \[\nicexy{\Comc=\Ochatmaxm \ar[r]^-{\deltam}_-{\cong} & \Ocbarmaxm}\] is an isomorphism of $\colorc$-colored operads with the first equality from Lemma \ref{lem:hcdiag-einf-algebra}, the homotopy comparison\index{homotopy coherent diagram!$E_\infty$-algebra} morphism \[\nicexy{\Wcomc= \wochatmaxm \ar[r]^-{\W\deltam}_-{\cong} & \wocbarmaxm}\] is also an isomorphism of $\colorc$-colored operads.  Therefore, the induced functor is an isomorphism \[\begin{small}\nicexy@C+.2cm{\HPFA(\Chatmax) = \algmwcomc = \algmwochatmaxm & \algmwocbarmaxm = \HQFT\bigl(\Cbarmax\bigr) \ar[l]_-{(\W\deltam)^*}^-{\cong}}.\end{small}\]  In this case, both the category of homotopy prefactorization algebras and the category of homotopy algebraic quantum field theories are isomorphic to the category of homotopy coherent $\C$-diagrams of $E_\infty$-algebras in $\M$ in Definition \ref{def:wcomc-algebra}.\dqed
\end{example}

\begin{example}[Homotopy Costello-Gwilliam PFA and QFT on manifolds]\label{ex:comparison-hpfa-manifolds}
For\index{homotopy prefactorization algebra!on manifolds}\index{homotopy algebraic quantum field theory!on manifolds} any small category $\Embn$ equivalent to the category of smooth $n$-manifolds with open embeddings as morphisms, we observed in Example \ref{ex:hpfa-manifolds} that the category of prefactorization algebras on $n$-manifolds with values in $\M$ in the sense of Costello-Gwilliam is isomorphic to the categories \[\Comm^{\Embn}= \PFA(\Embnhatmax) \cong \QFT(\Embnbarmax).\]  By Example \ref{ex:comparison-classical-case} there is also an isomorphism  \[\nicexy@R-.3cm@C+.7cm{\algm\bigl(\W\Com^{\Embn}\bigr) = \algm\bigl(\wom_{\Embnhatmax}\bigr) \ar@{=}[d] & \algm\bigl(\wom_{\Embnbarmax}\bigr) \ar@{=}[d] \ar[l]_-{(\W\deltam)^*}^-{\cong}\\
\HPFA(\Embnhatmax) & \HQFT\bigl(\Embnbarmax\bigr)}.\]
So the category of homotopy prefactorization algebras on $n$-manifolds and the category of homotopy algebraic quantum field theories on $\Embnbarmax$ are both isomorphic to the category of homotopy coherent $\Embn$-diagrams of $E_\infty$-algebras in $\M$.\dqed
\end{example}

\begin{example}[Homotopy Costello-Gwilliam PFA and QFT on complex manifolds]\label{ex:comparison-hpfa-cpmanifolds}
Example \ref{ex:comparison-hpfa-manifolds} also holds if $\Embn$ is replaced by any small category $\Holn$ equivalent\index{homotopy prefactorization algebra!on complex manifolds}\index{homotopy algebraic quantum field theory!on complex manifolds} to the category of complex $n$-manifolds with open holomorphic embeddings as morphisms.  The category of prefactorization algebras on complex $n$-manifolds with values in $\M$ in the sense of Costello-Gwilliam is isomorphic to the categories \[\Comm^{\Holn}= \PFA(\Holnhatmax) \cong \QFT(\Holnbarmax).\]  By Example \ref{ex:comparison-classical-case} there is also an isomorphism  \[\nicexy@R-.3cm@C+.7cm{\algm\bigl(\W\Com^{\Holn}\bigr) = \algm\bigl(\wom_{\Holnhatmax}\bigr) \ar@{=}[d] & \algm\bigl(\wom_{\Holnbarmax}\bigr) \ar@{=}[d] \ar[l]_-{(\W\deltam)^*}^-{\cong}\\
\HPFA(\Holnhatmax) & \HQFT\bigl(\Holnbarmax\bigr)}.\]
In other words, the category of homotopy prefactorization algebras on complex $n$-manifolds and the category of homotopy algebraic quantum field theories on $\Holnbarmax$ are both isomorphic to the category of homotopy coherent $\Holn$-diagrams of $E_\infty$-algebras in $\M$.\dqed
\end{example}

\begin{example}[Homotopy PFA and QFT on bounded lattices and topological spaces]\label{ex:compare-pfaqft-lattice}
Suppose\index{bounded lattice}\index{homotopy prefactorization algebra!on bounded lattices}\index{homotopy algebraic quantum field theory!on bounded lattices} $(L,\leq)$ is a bounded lattice as in Example \ref{ex:lattice}.  Consider the configured category $\Lhat$ in Example \ref{ex:boundedlatter-config}.  Then \[\Phi\Lhat = \Lbar = (L,\perp)\] is the orthogonal category in Example \ref{ex:qft-lattice}.   By Theorem \ref{thm:compare-pfa-aqft} there is a diagram of change-of-operad adjunctions
\[\nicexy@C+.7cm@R+.3cm{\HPFA(\Lhat)=\algm\bigl(\wom_{\Lhat}\bigr) \ar@<2pt>[r]^-{\W\deltam_!} \ar@<-2pt>[d]_-{\eta_!} & \algm\bigl(\wom_{\Lbar}\bigr) = \HQFT(\Lbar) \ar@<2pt>[l]^-{(\W\deltam)^*} \ar@<-2pt>[d]_-{\eta_!} \\ \PFA(\Lhat)=\algm\bigl(\Otom_{\Lhat}\bigr) \ar@<2pt>[r]^-{\deltam_!} \ar@<-2pt>[u]_-{\eta^*}  & \algm\bigl(\Otom_{\Lbar}\bigr) \cong \QFT(\Lbar) \ar@<2pt>[l]^-{(\deltam)^*} \ar@<-2pt>[u]_-{\eta^*}}\]
with commuting left/right adjoint diagrams that compares (homotopy) prefactorization algebras on $\Lhat$ and (homotopy) algebraic quantum field theories on $\Lbar$.  In particular, this works when $\Lhat$ is the configured category\index{topological space}\index{homotopy prefactorization algebra!on topological spaces}\index{homotopy algebraic quantum field theory!on topological spaces} $\Openxhat$ in Example \ref{ex:open-configuration} for some topological space $X$, and $\Lbar$ is the orthogonal category $\Openxbar$ in Example \ref{ex:qft-space}.\dqed
\end{example}

\begin{example}[Homotopy Costello-Gwilliam equivariant PFA and QFT]\label{ex:compare-pfaqft-eqspace}
Suppose\index{equivariant topological space}\index{homotopy equivariant prefactorization algebra}\index{homotopy algebraic quantum field theory!on equivariant topological spaces}   $G$ is a group, and $X$ is a topological space in which $G$ acts on the left by homeomorphisms.  Consider the configured category $\Openxghat$ in Example \ref{ex:eq-space-configuration}.  Then \[\Phi\Openxghat = \Openxgbar\] is the orthogonal category in Example \ref{ex:qft-eqspace}.   By Theorem \ref{thm:compare-pfa-aqft} there is a diagram of change-of-operad adjunctions
\[\nicexy@C+.7cm{\HPFA(\Openxghat) \ar@{=}[d] & \HQFT(\Openxgbar) \ar@{=}[d]\\ \algm\bigl(\wom_{\Openxghat}\bigr) \ar@<2pt>[r]^-{\W\deltam_!} \ar@<-2pt>[d]_-{\eta_!} & \algm\bigl(\wom_{\Openxgbar}\bigr) \ar@<2pt>[l]^-{(\W\deltam)^*} \ar@<-2pt>[d]_-{\eta_!} \\ \algm\bigl(\Otom_{\Openxghat}\bigr) \ar@<2pt>[r]^-{\deltam_!} \ar@<-2pt>[u]_-{\eta^*}  & \algm\bigl(\Otom_{\Openxgbar}\bigr) \ar[d]^-{\cong} \ar@<2pt>[l]^-{(\deltam)^*} \ar@<-2pt>[u]_-{\eta^*}\\ \PFA(\Openxghat) \ar@{=}[u] & \QFT(\Openxgbar)}\]
with commuting left/right adjoint diagrams that compares (homotopy) $G$-equivariant prefactorization algebras on $X$ and (homotopy) algebraic quantum field theories on $\Openxgbar$.\dqed
\end{example}

\begin{example}[Homotopy chiral conformal PFA and QFT]\label{ex:compare-pfaqft-chiral}
Consider\index{homotopy chiral conformal prefactorization algebra}\index{homotopy chiral conformal quantum field theory} the category $\Mand$ of $d$-dimensional oriented manifolds with orientation-preserving open embeddings as morphisms in Example \ref{ex:man-cat}.  Define $\Config$ to be the set of finite sequences of morphisms $\{g_i : X_i \shortto X\}_{i=1}^n$ in $\Mand$ such that the images $g_iX_i$ are pairwise disjoint subsets of $X$.  Then \[\Mandhat = (\Mand,\Config)\] is the configured category in Example \ref{ex:Psi-man} such that \[\Phi\Mandhat = \Mandbar\] is the orthogonal category in Example \ref{ex:ccqft}.  Algebras over the colored operads $\Otom_{\Mandhat}$ and $\wom_{\Mandhat}$ are called \emph{chiral conformal prefactorization algebras} and \emph{homotopy chiral conformal prefactorization algebras}, respectively.  By Theorem \ref{thm:compare-pfa-aqft} there is a diagram of change-of-operad adjunctions
\[\nicexy@C+.5cm{\HPFA(\Mandhat) = \algm\bigl(\wom_{\Mandhat}\bigr) \ar@<2pt>[r]^-{\W\deltam_!} \ar@<-2pt>[d]_-{\eta_!} & \algm\bigl(\wom_{\Mandbar}\bigr)= \HQFT(\Mandbar) \ar@<2pt>[l]^-{(\W\deltam)^*} \ar@<-2pt>[d]_-{\eta_!} \\ 
\PFA(\Mandhat)  = \algm\bigl(\Otom_{\Mandhat}\bigr) \ar@<2pt>[r]^-{\deltam_!} \ar@<-2pt>[u]_-{\eta^*}  & \algm\bigl(\Otom_{\Mandbar}\bigr) \cong \QFT(\Mandbar) \ar@<2pt>[l]^-{(\deltam)^*} \ar@<-2pt>[u]_-{\eta^*}}\]
with commuting left/right adjoint diagrams that compares (homotopy) chiral conformal prefactorization algebras and (homotopy) chiral conformal quantum field theories.\dqed
\end{example}

\begin{example}[Homotopy chiral conformal PFA and QFT on discs]\label{ex:compare-pfaqft-discs}
Consider\index{homotopy chiral conformal prefactorization algebra on discs}\index{homotopy chiral conformal quantum field theory on discs}  the full subcategory $\Discd$ of $\Mand$ consisting of $d$-dimensional oriented manifolds diffeomorphic to $\fieldr^d$.  Define $\Config$ to be the set of finite sequences of morphisms $\{g_i : X_i \shortto X\}_{i=1}^n$ in $\Discd$ such that the images $g_iX_i$ are pairwise disjoint subsets of $X$.  Then \[\Discdhat = (\Discd,\Config)\] is the configured category in Example \ref{ex:Psi-disc} such that \[\Phi\Discdhat = \Discdbar\] is the orthogonal category in Example \ref{ex:ccqft-int}.  Algebras over the colored operads $\Otom_{\Discdhat}$ and $\wom_{\Discdhat}$ are called \emph{chiral conformal prefactorization algebras on discs} and \emph{homotopy chiral conformal prefactorization algebras on discs}, respectively.  By Theorem \ref{thm:compare-pfa-aqft} there is a diagram of change-of-operad adjunctions
\[\nicexy@C+.5cm{\HPFA(\Discdhat) = \algm\bigl(\wom_{\Discdhat}\bigr) \ar@<2pt>[r]^-{\W\deltam_!} \ar@<-2pt>[d]_-{\eta_!} & \algm\bigl(\wom_{\Discdbar}\bigr)= \HQFT(\Discdbar) \ar@<2pt>[l]^-{(\W\deltam)^*} \ar@<-2pt>[d]_-{\eta_!} \\ 
\PFA(\Discdhat)  = \algm\bigl(\Otom_{\Discdhat}\bigr) \ar@<2pt>[r]^-{\deltam_!} \ar@<-2pt>[u]_-{\eta^*}  & \algm\bigl(\Otom_{\Discdbar}\bigr) \cong \QFT(\Discdbar) \ar@<2pt>[l]^-{(\deltam)^*} \ar@<-2pt>[u]_-{\eta^*}}\]
with commuting left/right adjoint diagrams that compares (homotopy) chiral conformal prefactorization algebras on discs and (homotopy) chiral conformal quantum field theories on discs.\dqed
\end{example}

\begin{example}[Homotopy Euclidean PFA and QFT]\label{ex:compare-pfaqft-euclidean}
Recall\index{homotopy Euclidean prefactorization algebra}\index{homotopy Euclidean quantum field theory} from Example \ref{ex:riem-cat} the category $\Riemd$ with $d$-dimensional oriented Riemannian manifolds as objects and orientation-preserving isometric open embeddings as morphisms.   Define $\Config$ to be the set of finite sequences of morphisms $\{g_i : X_i \shortto X\}_{i=1}^n$ in $\Riemd$ such that the images $g_iX_i$ are pairwise disjoint subsets of $X$.   Then \[\Riemdhat = (\Riemd,\Config)\] is the configured category in Example \ref{ex:Psi-riem} such that \[\Phi\Riemdhat = \Riemdbar\] is the orthogonal category in Example \ref{ex:euclidean-qft}.  Algebras over the colored operads $\Otom_{\Riemdhat}$ and $\wom_{\Riemdhat}$ are called \emph{Euclidean prefactorization algebras} and \emph{homotopy Euclidean prefactorization algebras}, respectively.  By Theorem \ref{thm:compare-pfa-aqft} there is a diagram of change-of-operad adjunctions
\[\nicexy@C+.5cm{\HPFA(\Riemdhat) = \algm\bigl(\wom_{\Riemdhat}\bigr) \ar@<2pt>[r]^-{\W\deltam_!} \ar@<-2pt>[d]_-{\eta_!} & \algm\bigl(\wom_{\Riemdbar}\bigr)= \HQFT(\Riemdbar) \ar@<2pt>[l]^-{(\W\deltam)^*} \ar@<-2pt>[d]_-{\eta_!} \\ 
\PFA(\Riemdhat)  = \algm\bigl(\Otom_{\Riemdhat}\bigr) \ar@<2pt>[r]^-{\deltam_!} \ar@<-2pt>[u]_-{\eta^*}  & \algm\bigl(\Otom_{\Riemdbar}\bigr) \cong \QFT(\Riemdbar) \ar@<2pt>[l]^-{(\deltam)^*} \ar@<-2pt>[u]_-{\eta^*}}\]
with commuting left/right adjoint diagrams that compares (homotopy) Euclidean prefactorization algebras and (homotopy) Euclidean quantum field theories.\dqed
\end{example}

\begin{example}[Homotopy locally covariant PFA and QFT]\label{ex:compare-pfaqft-lc}
Recall\index{homotopy locally covariant prefactorization algebra}\index{homotopy locally covariant quantum field theory} from Example \ref{ex:loc-cat} the category $\Locd$ of $d$-dimensional oriented, time-oriented, and globally hyperbolic Lorentzian manifolds.  A morphism is an isometric embedding that preserves the orientations and time-orientations whose image is causally compatible and open.  Define $\Config$ to be the set of finite sequences of morphisms $\{g_i : X_i \shortto X\}_{i=1}^n$ in $\Locd$ such that the images $g_iX_i$ are pairwise causally disjoint subsets of $X$.   Then \[\Locdhat = (\Locd,\Config)\] is the configured category in Example \ref{ex:Psi-loc} such that \[\Phi\Locdhat = \Locdbar\] is the orthogonal category in Example \ref{ex:lcqft}.  Algebras over the colored operads $\Otom_{\Locdhat}$ and $\wom_{\Locdhat}$ are called \emph{locally covariant prefactorization algebras} and \emph{homotopy locally covariant prefactorization algebras}, respectively.  By Theorem \ref{thm:compare-pfa-aqft} there is a diagram of change-of-operad adjunctions
\[\nicexy@C+.5cm{\HPFA(\Locdhat) = \algm\bigl(\wom_{\Locdhat}\bigr) \ar@<2pt>[r]^-{\W\deltam_!} \ar@<-2pt>[d]_-{\eta_!} & \algm\bigl(\wom_{\Locdbar}\bigr)= \HQFT(\Locdbar) \ar@<2pt>[l]^-{(\W\deltam)^*} \ar@<-2pt>[d]_-{\eta_!} \\ 
\PFA(\Locdhat)  = \algm\bigl(\Otom_{\Locdhat}\bigr) \ar@<2pt>[r]^-{\deltam_!} \ar@<-2pt>[u]_-{\eta^*}  & \algm\bigl(\Otom_{\Locdbar}\bigr) \cong \QFT(\Locdbar) \ar@<2pt>[l]^-{(\deltam)^*} \ar@<-2pt>[u]_-{\eta^*}}\]
with commuting left/right adjoint diagrams that compares (homotopy) locally covariant prefactorization algebras and (homotopy) locally covariant quantum field theories.

To incorporate the time-slice axiom, suppose $S$ is the set of Cauchy morphisms in $\Locd$.  By Corollary \ref{cor:pfa-aqft-timeslice} there is a diagram of change-of-operad adjunctions
\[\begin{footnotesize}\nicexy@C+.2cm{\HPFA(\Locdhat,S) = \algm\bigl(\W\O_{\Locdhat}[\Sinv]^{\M}\bigr) \ar@<2pt>[r]^-{(\W\deltam_S)_!} \ar@<-2pt>[d]_-{\eta_!} & \algm\bigl(\wom_{\Locdsinvbar}\bigr)= \HQFT(\Locdsinvbar) \ar@<2pt>[l]^-{(\W\deltam_S)^*} \ar@<-2pt>[d]_-{\eta_!} \\ 
\PFA(\Locdhat,S)  = \algm\bigl(\O_{\Locdhat}[\Sinv]^{\M}\bigr) \ar@<2pt>[r]^-{(\deltam_S)_!} \ar@<-2pt>[u]_-{\eta^*}  & \algm\bigl(\Otom_{\Locdsinvbar}\bigr) \cong \QFT(\Locdbar,S) \ar@<2pt>[l]^-{(\deltam_S)^*} \ar@<-2pt>[u]_-{\eta^*}}\end{footnotesize}\]
with commuting left/right adjoint diagrams that compares (homotopy) locally covariant prefactorization algebras and (homotopy) locally covariant quantum field theories satisfying the time-slice axiom.\dqed
\end{example}

\begin{example}[Homotopy locally covariant PFA and QFT on a fixed spacetime]\label{ex:compare-pfaqft-ghx}
For\index{homotopy locally covariant prefactorization algebra on a spacetime}\index{homotopy locally covariant quantum field theory on a spacetime}   each Lorentzian manifold $X \in \Locd$, recall from Example \ref{ex:gh-cat} the category $\Ghx$ of globally hyperbolic open subsets of $X$ with subset inclusions as morphisms.   Define $\Config$ to be the set of finite sequences of morphisms $\{g_i : U_i \shortto V\}_{i=1}^n$ in $\Ghx$ such that the $U_i$'s are pairwise causally disjoint subsets of $V$.   Then \[\Ghxhat = (\Ghx,\Config)\] is the configured category in Example \ref{ex:Psi-ghx} such that \[\Phi\Ghxhat = \Ghxbar\] is the orthogonal category in Example \ref{ex:causal-nets}.  Algebras over the colored operads $\Otom_{\Ghxhat}$ and $\wom_{\Ghxhat}$ are called \emph{locally covariant prefactorization algebras on $X$} and \emph{homotopy locally covariant prefactorization algebras on $X$}, respectively.  By Theorem \ref{thm:compare-pfa-aqft} there is a diagram of change-of-operad adjunctions
\[\nicexy@C+.5cm{\HPFA(\Ghxhat) = \algm\bigl(\wom_{\Ghxhat}\bigr) \ar@<2pt>[r]^-{\W\deltam_!} \ar@<-2pt>[d]_-{\eta_!} & \algm\bigl(\wom_{\Ghxbar}\bigr)= \HQFT(\Ghxbar) \ar@<2pt>[l]^-{(\W\deltam)^*} \ar@<-2pt>[d]_-{\eta_!} \\ 
\PFA(\Ghxhat)  = \algm\bigl(\Otom_{\Ghxhat}\bigr) \ar@<2pt>[r]^-{\deltam_!} \ar@<-2pt>[u]_-{\eta^*}  & \algm\bigl(\Otom_{\Ghxbar}\bigr) \cong \QFT(\Ghxbar) \ar@<2pt>[l]^-{(\deltam)^*} \ar@<-2pt>[u]_-{\eta^*}}\]
with commuting left/right adjoint diagrams that compares (homotopy) locally covariant prefactorization algebras on $X$ and (homotopy) locally covariant quantum field theories on $X$.

To incorporate the time-slice axiom, suppose $S$ is the set of morphisms $U \subseteq V$ in $\Ghx$ such that $U$ contains a Cauchy surface of $i(V)$, where $i : \Ghx \to \Locd$ is the subcategory inclusion.  By Corollary \ref{cor:pfa-aqft-timeslice} there is a diagram of change-of-operad adjunctions
\[\nicexy@C+.5cm{\HPFA(\Ghxhat,S) \ar@{=}[d] & \HQFT(\Ghxsinvbar) \ar@{=}[d]\\ 
\algm\bigl(\W\O_{\Ghxhat}[\Sinv]^{\M}\bigr) \ar@<2pt>[r]^-{(\W\deltam_S)_!} \ar@<-2pt>[d]_-{\eta_!} & \algm\bigl(\wom_{\Ghxsinvbar}\bigr)\ar@<2pt>[l]^-{(\W\deltam_S)^*} \ar@<-2pt>[d]_-{\eta_!} \\ 
\algm\bigl(\O_{\Ghxhat}[\Sinv]^{\M}\bigr) \ar@<2pt>[r]^-{(\deltam_S)_!} \ar@<-2pt>[u]_-{\eta^*}  & \algm\bigl(\Otom_{\Ghxsinvbar}\bigr) \ar[d]^-{\cong} \ar@<2pt>[l]^-{(\deltam_S)^*} \ar@<-2pt>[u]_-{\eta^*}\\ \PFA(\Ghxhat,S) \ar@{=}[u] & \QFT(\Ghxbar,S)}\]
with commuting left/right adjoint diagrams that compares (homotopy) locally covariant prefactorization algebras on $X$ and (homotopy) locally covariant quantum field theories on $X$ satisfying the time-slice axiom.\dqed
\end{example}

\begin{example}[Homotopy PFA and QFT on structured spacetimes]\label{ex:compare-pfaqft-str}
As\index{homotopy prefactorization algebra on structured spacetime}\index{homotopy quantum field theory on structured spacetime} in Example \ref{ex:qft-structured}, suppose \[\pi : \Str \to \Locd\] is a functor between small categories, and $S_{\pi} \subset \Mor(\Str)$ is the $\pi$-pre-image of the set $S$ of Cauchy morphisms in $\Locd$.  Define $\Config$ to be the set of finite sequences of morphisms $\{g_i : X_i \shortto X\}_{i=1}^n$ in $\Str$ such that the images $(\pi g_i)(\pi X_i)$ are pairwise causally disjoint subsets of $\pi X$.   Then \[\Strhat = (\Str,\Config)\] is the configured category in Example \ref{ex:Psi-str} such that \[\Phi\Strhat = \Strbar\] is the orthogonal category in Example \ref{ex:qft-structured}.  Algebras over the colored operads $\O_{\Strhat}[\Sinv_{\pi}]^{\M}$ and $\W\O_{\Strhat}[\Sinv_{\pi}]^{\M}$ are called \emph{prefactorization algebras on $\pi$} and \emph{homotopy prefactorization algebras on $\pi$}, respectively.  

By Corollary \ref{cor:pfa-aqft-timeslice} there is a diagram of change-of-operad adjunctions
\[\begin{small}\nicexy@C+.2cm{\HPFA(\Strhat,S_{\pi}) = \algm\bigl(\W\O_{\Strhat}[\Sinv_{\pi}]^{\M}\bigr) \ar@<2pt>[r]^-{(\W\deltam_{S_{\pi}})_!} \ar@<-2pt>[d]_-{\eta_!} & \algm\bigl(\wom_{\Strsinvbar}\bigr)= \HQFT(\Strsinvbar) \ar@<2pt>[l]^-{(\W\deltam_{S_{\pi}})^*} \ar@<-2pt>[d]_-{\eta_!} \\ 
\PFA(\Strhat,S_{\pi})  = \algm\bigl(\O_{\Strhat}[\Sinv_{\pi}]^{\M}\bigr) \ar@<2pt>[r]^-{(\deltam_{S_{\pi}})_!} \ar@<-2pt>[u]_-{\eta^*}  & \algm\bigl(\Otom_{\Strsinvbar}\bigr) \cong \QFT(\Strbar,S_{\pi}) \ar@<2pt>[l]^-{(\deltam_{S_{\pi}})^*} \ar@<-2pt>[u]_-{\eta^*}}\end{small}\]
with commuting left/right adjoint diagrams that compares (homotopy) prefactorization algebras on $\pi$ and (homotopy) quantum field theories on $\pi$.  For example, this applies to the following functors.
\begin{itemize}\item The forgetful functor \[\pi : \Bgloc \to \Locd\] in Example \ref{ex:dynamical-gauge}, where $\Bgloc$ is the category of $d$-dimensional oriented, time-oriented, and globally hyperbolic Lorentzian manifolds equipped with a principal $G$-bundle. 
\item The forgetful functor \[\pi p : \Bgconloc \to \Locd\] in Example \ref{ex:charged-matter},  where $\Bgconloc$ is the category of triples $(X,P,C)$ with $(X,P) \in \Bgloc$ and $C$ a connection on $P$.    
\item The forgetful functor \[\pi : \Slocd \to \Locd\] in Example \ref{ex:dirac-qft}, where $\Slocd$ is the category of $d$-dimensional oriented, time-oriented, and globally hyperbolic Lorentzian spin manifolds.\dqed
\end{itemize}
\end{example}

\begin{example}[Homotopy PFA and QFT on spacetime with timelike boundary]\label{ex:compare-pfaqft-boundary}
Consider\index{homotopy prefactorization algebra on spacetime with timelike boundary}\index{homotopy quantum field theory on spacetime with timelike boundary}   the category $\Regx$ in Example \ref{ex:regions} for a spacetime  $X$ with timelike boundary.  Define $\Config$ to be the set of finite sequences of morphisms $\{g_i : U_i \shortto V\}_{i=1}^n$ in $\Regx$ such that the $U_i$'s are pairwise causally disjoint subsets of $V$.   Then \[\Regxhat = (\Regx,\Config)\] is the configured category in Example \ref{ex:Psi-boundary} such that \[\Phi\Regxhat = \Regxbar\] is the orthogonal category in Example \ref{ex:aqft-boundary}. Suppose $S_X \subset \Mor(\Regx)$ is the set of Cauchy morphisms as in Example \ref{ex:aqft-boundary}. Algebras over the colored operads $\O_{\Regxhat}[\Sinv_{X}]^{\M}$ and $\W\O_{\Regxhat}[\Sinv_{X}]^{\M}$ are called \emph{prefactorization algebras on $X$} and \emph{homotopy prefactorization algebras on $X$}, respectively. 

By Corollary \ref{cor:pfa-aqft-timeslice} there is a diagram of change-of-operad adjunctions
\[\nicexy@C+.5cm{\HPFA(\Regxhat,S_{X}) \ar@{=}[d] & \HQFT(\Regxsinvbar) \ar@{=}[d]\\
\algm\bigl(\W\O_{\Regxhat}[\Sinv_{X}]^{\M}\bigr) \ar@<2pt>[r]^-{(\W\deltam_{S_{X}})_!} \ar@<-2pt>[d]_-{\eta_!} & \algm\bigl(\wom_{\Regxsinvbar}\bigr) \ar@<2pt>[l]^-{(\W\deltam_{S_{X}})^*} \ar@<-2pt>[d]_-{\eta_!} \\ 
\algm\bigl(\O_{\Regxhat}[\Sinv_{X}]^{\M}\bigr) \ar@<2pt>[r]^-{(\deltam_{S_{X}})_!} \ar@<-2pt>[u]_-{\eta^*}  & \algm\bigl(\Otom_{\Regxsinvbar}\bigr) \ar[d]^-{\cong} \ar@<2pt>[l]^-{(\deltam_{S_{X}})^*} \ar@<-2pt>[u]_-{\eta^*}\\
\PFA(\Regxhat,S_{X}) \ar@{=}[u] & \QFT(\Regxbar,S_{X})}\]
with commuting left/right adjoint diagrams that compares (homotopy) prefactorization algebras on $X$ and (homotopy) algebraic quantum field theories on $X$.\dqed
\end{example}

\section{Prefactorization Algebras from AQFT}\label{sec:pfa-from-aqft}

In this section, we identify the essential image of the right adjoint \[\nicexy@C+.5cm{\PFA(\Chat)=\algmochatm & \algmocbarm \cong \QFT(\Cbar) \ar@<2pt>[l]_-{(\deltam)^*}}\] in the comparison adjunction in Theorem \ref{thm:compare-pfa-aqft}.  In other words, we characterize the prefactorization algebras that come from algebraic quantum field theories.  We will use the notation in the Coherence Theorem \ref{thm:ochat-algebra} for prefactorization algebras.

\begin{theorem}\label{thm:deltamstar-image}
Suppose $\Chat = (\C,\Config)$ is a configured category with object set $\colorc$, and $\Phi\Chat = \Cbar$ is the associated orthogonal category.  Suppose $(X,\lambda) \in \algmochatm$.  Then the following two statements are equivalent.\index{prefactorization algebra!from AQFT}\index{algebraic quantum field theory!to prefactorization algebra}
\begin{enumerate}\item There exist $B \in \algmocbarm$ and an isomorphism \[(X,\lambda) \cong (\deltam)^*B.\]
\item For each $c \in \colorc$, the object $X_c \in \M$ can be equipped with a monoid structure $(X_c,\mu_c,\operadunit_c)$ such that the following three conditions are satisfied.
\begin{enumerate}\item The structure morphism \[\nicexy@C+.7cm{\tensorunit \ar[r]^-{\lambda\{(d;\varnothing)\}} & X_d}\in \M\] is the monoid unit $\operadunit_d$ of $X_d$ for each $d \in \colorc$.
\item For each morphism $f : c \to d$ in $\C$, the structure morphism \[\nicexy@C+.7cm{X_c \ar[r]^-{\lambda\{f\}} & X_d}\in \M\] respects the monoid structure.
\item For each configuration $\{f_i : c_i \shortto d\}_{i=1}^n \in \Config\duc$ with $n \geq 2$, the diagram \[\nicexy@C+.8cm{\bigotimes\limits_{i=1}^n X_{c_i} \ar[d]_-{\bigotimes\limits_{i=1}^n \lambda\{f_i\}} \ar[r]^-{\lambda\{f_i\}_{i=1}^n} & X_d \ar@{=}[d]\\ \bigotimes\limits_{i=1}^n X_d \ar[r]^-{\mu_d} & X_d}\] is commutative, in which $\mu_d$ is the $(n-1)$-fold iterate of the monoid multiplication on $X_d$.
\end{enumerate}
\end{enumerate}
\end{theorem}

\begin{proof}
The implication (1) $\Longrightarrow$ (2) follows from the definition of the operad morphism $\delta : \Ochat \to \Ocbar$ and the fact that each $\Ocbarm$-algebra, i.e., algebraic quantum field theory on $\Cbar$, is a functor $\C \to \Monm$.  Every $\Ochatm$-algebra of the form $(\deltam)^*B$ for some $B \in \algmocbarm$ satisfies the three conditions in (2).  Therefore, so does any $\Ochatm$-algebra in the essential image of $(\deltam)^*$.

For (2) $\Longrightarrow$ (1), suppose $(X,\lambda)$ satisfies the conditions in (2).  Define a functor $B : \C \to \Monm$ by setting \[\begin{split} &B(c) = (X_c,\mu_c,\operadunit_c) \forspace c \in \colorc,\\
&\nicexy@C+.7cm{B(c) \ar[r]^-{B(f)\,=\, \lambda\{f\}} & B(d)} \forspace f \in \C(c,d).\end{split}\] The functoriality of $B$ follows from the inclusivity axiom of a configured category and the associativity condition \eqref{pfa-ass} of $(X,\lambda)$.  

To check that $B$ satisfies the causality axiom \eqref{perp-com}, suppose given an orthogonal pair $(f : a \shortto c) \perp (g : b \shortto c)$.  By the definition of $\Phi$, this means that $\{f,g\}$ is a configuration.  We must show that the outermost diagram in \[\nicexy@R+.6cm@C+1cm{X_a \otimes X_b \ar[dd]_-{(\lambda\{f\},\lambda\{g\})} \ar[ddr]_-{\lambda\{f,g\}} \ar[dr]^-{\mathrm{permute}} \ar[r]^-{(\lambda\{f\},\lambda\{g\})} & X_c^{\otimes 2} \ar[r]^-{\mathrm{permute}} & X_c^{\otimes 2} \ar[dd]^-{\mu_c}\\ & X_b \otimes X_a \ar[ur]^-{(\lambda\{g\},\lambda\{f\})} \ar[d]^-{\lambda\{g,f\}} & \\ X_c^{\otimes 2} \ar[r]^-{\mu_c} & X_c \ar@{=}[r] & X_c}\] is commutative.
\begin{itemize}\item The top triangle is commutative by the naturality of the symmetry isomorphism in $\M$.
\item The left triangle and the right trapezoid are commutative by assumption (2)(c), since $\{f,g\}$ and $\{g,f\}$ are both configurations.
\item The middle triangle is commutative by the equivariance condition \eqref{pfa-eq} of $(X,\lambda)$.
\end{itemize}
Therefore, $B$ is an algebraic quantum field theory on $\Cbar$.  By the assumed conditions (2)(a)-(2)(c), we also have that $(X,\lambda) = (\deltam)^*B$.
\end{proof}

\begin{interpretation} Physically, Theorem \ref{thm:deltamstar-image} tells us which prefactorization algebras on $\Chat$ arise from algebraic quantum field theories on the associated orthogonal category $\Cbar$.  We will see below examples of both kinds, i.e., prefactorization algebras that arise from algebraic quantum field theories and those that do not.\dqed
\end{interpretation}

Recall from Example \ref{ex:open-configuration} the configured category $\Openxhat$ for a topological space $X$.

\begin{example}[Costello-Gwilliam associative prefactorization algebras on $\fieldr$]\label{ex:cgpfa-real}
Here\index{prefactorization algebra!on $\fieldr$} we provide examples of prefactorization algebras that do not come from algebraic quantum field theories.

Suppose $(A,\mu,\varepsilon)$ is a monoid in $\M$.  Define a prefactorization algebra $(A^{\fact},\lambda)$ on the configured category $\Openrhat$ as follows.  For the empty subset of $\fieldr$, we define $A^{\fact}_{\varnothing} = \tensorunit$.  For each open interval $(a,b) \subset \fieldr$, we define $A^{\fact}_{(a,b)} = A$.  For a finite disjoint union of open intervals $V = \coprod_{j=1}^n (a_j,b_j)$ with $b_j \leq a_{j+1}$ for $1 \leq j \leq n-1$, we define \[A^{\fact}_V = A^{\fact}_{(a_1,b_1)} \otimes \cdots \otimes A^{\fact}_{(a_n,b_n)} = A^{\otimes n}.\]  For a general disjoint union of open intervals $U = \coprod_{i \in I} (a_i,b_i)$, we define \[A^{\fact}_U = \colimover{J \subset I} ~ A^{\fact}_{\coprod_{j\in J} (a_j,b_j)} \in \M\] with the colimit indexed by the partially ordered set of finite subsets $J \subset I$ under inclusion.  If $J \subset J'$ are finite subsets of $I$, then the morphism \[A^{\fact}_{\coprod_{j\in J} (a_j,b_j)} \to A^{\fact}_{\coprod_{j'\in J'} (a_{j'},b_{j'})}\] is induced by the unit $\varepsilon : \tensorunit \to A$ for each element in $J' \setminus J$.

Consider a configuration $\bigl\{f_i : U_i \subset V\bigr\}_{i=1}^n$ in $\Openrhat$; i.e., the $U_i$'s are pairwise disjoint open subsets in $V$.  The structure morphism \[\nicexy@C+.8cm{\bigotimes\limits_{i=1}^n A^{\fact}_{U_i} \ar[r]^-{\lambda\{f_i\}_{i=1}^n} & A^{\fact}_V}\in \M\] in \eqref{pfa-structure-morphism} is defined by the colimits involved, the equivariance condition \eqref{pfa-eq}, and the following special cases on open intervals.
\begin{itemize}\item If $n=0$, then \[\lambda\{\varnothing\} : A^{\fact}_{\varnothing}=\tensorunit \to A = A^{\fact}_{(a,b)}\] is the unit $\varepsilon : \tensorunit \to A$.
\item If $n= 1$ and if $f : (a,b) \subset (c,d)$, then \[\lambda\{f\} : A^{\fact}_{(a,b)}=A \to A=A^{\fact}_{(c,d)}\] is the identity morphism.
\item If $n \geq 2$ and if $f_i : (a_i,b_i) \subset (c,d)$ are pairwise disjoint in $(c,d)$ for $1 \leq i \leq n$ with $b_i \leq a_{i+1}$ for $1 \leq i \leq n-1$, then \[\nicexy@C+.8cm{\bigotimes\limits_{i=1}^n A^{\fact}_{(a_i,b_i)} = \bigotimes\limits_{i=1}^n A \ar[r]^-{\lambda\{f_i\}_{i=1}^n} & A = A^{\fact}_{(c,d)}}\] is the $(n-1)$-fold iterate of the multiplication $\mu$.
\end{itemize}

For instance, consider the inclusions \[f_1 : (2,3) \subset (1,8) \andspace f_2 : (4,6) \subset (1,8).\] 
\begin{itemize}\item The structure morphisms \[\nicexy{A^{\fact}_{(2,3)} = A \ar[r]^-{\lambda\{f_1\}} & A = A^{\fact}_{(1,8)}} \andspace \nicexy{A^{\fact}_{(4,6)} = A \ar[r]^-{\lambda\{f_2\}} & A =A^{\fact}_{(1,8)}}\] are both the identity morphism.   
\item The structure morphism \[\nicexy@C+1cm{A^{\fact}_{(2,3)} \otimes A^{\fact}_{(4,6)} =A^{\otimes 2} \ar[r]^-{\lambda\{f_1,f_2\}} & A = A^{\fact}_{(1,8)}}\] is the multiplication $\mu$.  So for open intervals already in the correct order in $\fieldr$, the structure morphism is just the multiplication.  
\item On the other hand, for the configuration $\{f_2,f_1\}$, the structure morphism \[\nicexy@C+1cm{A^{\fact}_{(4,6)} \otimes A^{\fact}_{(2,3)} = A^{\otimes 2} \ar[r]^-{\lambda\{f_2,f_1\}} & A}\] is the opposite multiplication $\mu \circ (1~2)$; i.e., permute the two domain factors before multiplying. So for open intervals not in the correct order in $\fieldr$, we must first permute the domain factors back to the correct order before multiplying.
\end{itemize}  

One can check that $A^{\fact}$ is actually a prefactorization algebra on $\Openrhat$ using the Coherence Theorem \ref{thm:ochat-algebra}.  This prefactorization algebra is a key example in \cite{cg} Section 3.1.1.  Moreover, by Theorem \ref{thm:deltamstar-image}, $A^{\fact}$ is \emph{not} in the essential image of the right adjoint $(\deltam)^*$ because it does not satisfy condition (2)(c) there.  Indeed, if it satisfies condition (2)(c), then the structure morphism $\lambda\{f_2,f_1\}$ in the previous paragraph would just be the multiplication $\mu$, which is not true.\dqed
\end{example}

\begin{example}[Costello-Gwilliam commutative prefactorization algebras on $\fieldr$]\label{ex:cgpfa-real-commutative}
Suppose $(A,\mu,\varepsilon)$ is a commutative monoid.  Then the  prefactorization algebra $(A^{\fact},\lambda)$ on the configured category $\Openrhat$ in Example \ref{ex:cgpfa-real} is in the image of the right adjoint $(\deltam)^*$.  In other words, it arises from an algebraic quantum field theory on the associated orthogonal category $\Openrbar$.  Indeed, since $A$ is a commutative monoid, each entry $A^{\fact}_U$ inherits from $A$ the structure of a commutative monoid.  So this construction defines a functor \[A^{\fact}_? : \Openr \to \Comm,\] and the causality axiom \eqref{perp-com} is satisfied.  In other words, $A^{\fact}_?$ is an algebraic quantum field theory on $\Openrbar$.  Applying the right adjoint $(\deltam)^*$, it becomes the prefactorization algebra $A^{\fact}$ on $\Openrhat$.\dqed
\end{example}

\begin{example}[Costello-Gwilliam symmetric prefactorization algebras]\label{ex:cgpfa-sym}
Here we provide examples of prefactorization algebras that come from algebraic quantum field theories.  Suppose\index{prefactorization algebra!symmetric}  \[F : \Openr \to \M\] is any functor, and suppose \[\Com : \M \to \Comm\] is the free commutative monoid functor, which is left adjoint to the forgetful functor.  Then their composition \[\Com \circ F : \Openr \to \Comm\] defines an algebraic quantum field theory on $\Openrbar$.  Applying the right adjoint, it becomes a prefactorization algebra on $\Openrhat$.  A prefactorization algebra of the form $\Com \circ F$ is another example from \cite{cg} Section 3.1.1.\dqed
\end{example}

\begin{example}[The right adjoint is not injective]
In this example, we illustrate that the right adjoint $(\deltam)^*$, from algebraic quantum field theories to prefactorization algebras, is in general not injective on objects.  Suppose $X$ is an indiscrete topological space; i.e., \[\Openx = \{\varnothing \subset X\}\] is a category with only two objects and one non-identity morphism.  Suppose $(A,\mu,\varepsilon)$ is a monoid in $\M$.  Define a functor \[F^A : \Openx \to\Monm\] by setting \[\begin{split}F^A(\varnothing) &= (\tensorunit, \tensorunit \otimes \tensorunit \cong \tensorunit,\Id_{\tensorunit}),\\ F^A(X) &= (A,\mu,\varepsilon),\\
F^A(\varnothing \subset X) &= \varepsilon : \tensorunit \to A.\end{split}\]  
In the orthogonal category $\Openxbar$, there are only four orthogonal pairs: \[\varnothing \subset \varnothing \supset \varnothing,\quad \varnothing \subset X \supset \varnothing,\quad \varnothing \subset X \supset X,\andspace X \subset X \supset \varnothing.\]  It follows that the functor $F^A$ satisfies the causality axiom \eqref{perp-com} because the multiplication $\mu$ on $A$ is not involved.  So $F^A$ is an algebraic quantum field theory on $\Openxbar$.

In the prefactorization algebra $(\deltam)^*F^A$ on $\Openxhat$, the only structure morphisms are of the forms \[\tensorunit^{\otimes n} \cong \tensorunit,\quad \nicexy{\tensorunit^{\otimes n} \cong \tensorunit \ar[r]^-{\varepsilon} & A}, \andspace \nicexy{\overbrace{\tensorunit \otimes \cdots \otimes A \otimes \cdots \otimes \tensorunit}^{\mathrm{only}~\mathrm{one}~ A} \ar[r]^-{\cong} & A}.\]  They correspond to the unique configurations in \[\Configx\sbinom{\varnothing}{\varnothing,\ldots,\varnothing},\quad \Configx\sbinom{X}{\varnothing,\ldots,\varnothing},\andspace \Configx\sbinom{X}{\varnothing,\ldots,X,\ldots,\varnothing}.\] In particular, the multiplication $\mu$ is not needed to specify this prefactorization algebra.  Therefore, if we change $(A,\mu,\varepsilon)$ to the monoid $A^{\op} = \bigl(A, \mu \circ (1~2),\varepsilon\bigr)$ with the opposite multiplication, then there is an equality \[(\deltam)^*F^A = (\deltam)^*F^{A^{\op}}\] of prefactorization algebras on $\Openxhat$.  However, the algebraic quantum field theories $F^A$ and $F^{A^{\op}}$ are different because their values at $X$ are different.\dqed
\end{example}

\chapter*{List of Notations}

\newcommand{\where}[1]{\> \> \pageref{#1} \> \>}
\newcommand{\blob}{\> \> \> \> \hspace{2em}}

\begin{tabbing}
\textbf{Notation} \= \hspace{1.5cm}\= \textbf{Page}\= \hspace{.5cm}\=\textbf{Description} \\
$\Obc$ \where{notation:objects-category} class of objects in a category $\C$\\
$\C(a,b)$ \where{notation:morphism-set}  set of morphisms from $a$ to $b$ in $\C$\\
$\Id_a$ \where{notation:identity-morphism} identity morphism of $a$\\
$\Mor(\C)$ \where{notation:morphism-C} class of all morphisms in $\C$\\
$\Cop$ \where{notation:opposite-category} opposite category of $\C$\\
$\cong$ \where{notation:iso} isomorphism\\
$\Fun(\C,\D)$, $\D^{\C}$ \where{ex:functor-cat} category of functors from $\C$ to $\D$\\
$a \downarrow \C$ \where{ex:undercat} under category\\
$\Set$ \where{ex:set} category of sets\\
$\fieldk$ \where{notation:fieldk} a field\\
$\Vectk$ \where{ex:vectk} category of $\fieldk$-vector spaces\\
$\Chaink$ \where{ex:chaink} category of chain complexes of $\fieldk$-vector spaces\\
$\Top$ \where{ex:top} category of compactly generated weak\\
\blob Hausdorff spaces\\
$\Delta$ \where{ex:simplex-cat} simplex category\\ 
$\Sset$ \where{ex:simplicial-cat} category of simplicial sets\\
$\Cat$ \where{ex:cat} category of small categories\\
$(S,\leq)$ \where{ex:lattice} partially ordered set\\
$a \vee b$ \where{notation:lub} least upper bound\\
$a \wedge b$ \where{notation:glb} greatest lower bound\\
$\Openx$ \where{ex:openx} category of open subsets in $X$\\
$\Openxg$ \where{ex:eq-space} category of open subsets in $X$ with a $G$-action\\
$\Mand$ \where{ex:man-cat} $d$-dimensional oriented manifolds\\
$\fieldr$ \where{notation:fieldr} field of real numbers\\
$\Discd$ \where{ex:disc-cat} oriented manifolds diffeomorphic to $\fieldr^d$\\
$\Riemd$ \where{ex:riem-cat} $d$-dimensional oriented Riemannian manifolds\\
$\Locd$ \where{ex:loc-cat}  $d$-dimensional Lorentzian manifolds\\
$\Ghx$ \where{ex:gh-cat} globally hyperbolic open subsets of $X$\\
$\Bgloc$ \where{ex:bgloc-cat} Lorentzian manifolds with principal $G$-bundles\\
$\Bgconloc$ \where{ex:bgconloc-cat} Lorentzian manifolds with principal $G$-bundles\\
\blob and connections\\
$\Slocd$ \where{ex:sloc-cat} $d$-dimensional Lorentzian spin manifolds\\
$\Regx$ \where{ex:regions} causally convex open subsets in $X$\\
$\limit\, F$ \where{notation:limit} limit of $F$\\
$\colim\, F$ \where{notation:colim} colimt of $F$\\
$\emptyset$ \where{notation:initialobject} initial object\\
$\prod$, $\coprod$ \where{notation:product} product, coproduct\\
$\int^{c\in \C} F(c,c)$ \where{def:coend} coend of $F : \Cop \times \C \to \M$ \\
$F \dashv G$ \where{notation:adjoints} $F$ is left adjoint to $G$\\
$\Lan_F H$ \where{def:left-kan} left Kan extension of $H$ along $F$\\
$\otimes$ \where{def:monoidal-category} monoidal product\\
$\tensorunit$ \where{def:monoidal-category} monoidal unit\\
$\xi_{X,Y}$ \where{def:symmetric-monoidal-category} symmetry isomorphism\\
$\Homm$ \where{notation:internal-hom} internal hom in $\M$\\
$\MFun(\M,\N)$ \where{ex:monoidal-functor-cat} category of monoidal functors\\
$\SMFun(\M,\N)$ \where{ex:monoidal-functor-cat} category of symmetric monoidal functors\\
$\Monm$ \where{notation:monm} category of monoids in $\M$\\
$\Comm$ \where{notation:comm} category of commutative monoids in $\M$\\
$(\Csinv,\ell)$ \where{def:localization-cat} $S$-localization of $\C$\\
$\Flag(G)$ \where{notation:flag} set of flags in a graph \\
$\lambda_G$ \where{notation:flag} partition of a graph \\
$G_0$ \where{notation:flag} exceptional cell of a graph $G$\\
$\iota_G$ \where{notation:flag} involution of a graph \\
$\pi_G$ \where{notation:flag}  free involution on $\iota_G$-fixed points in $G_0$\\
$|v|$ \where{notation:vertex} cardinality of a vertex $v$ \\
$\Vt(G)$ \where{notation:vertex} set of vertices in a graph\\
$\Leg(G)$ \where{notation:leg} set of legs in a graph \\
$\Ed(G)$ \where{notation:edge} set of edges in a graph \\
$\bigcircle$ \where{notation:exceptional-loop} exceptional loop \\
$\edge$ \where{notation:exceptional-loop} exceptional edge \\
$\varnothing$ \where{notation:empty-graph} empty graph \\
$\kappa : \Flag(G) \shortto \colorc$ \where{notation:coloring} $\colorc$-coloring of a graph\\
$\delta : \Flag(G) \shortto \{\pm 1\}$ \where{notation:direction} direction of a graph\\
$\zeta_v$ \where{notation:ordering} ordering at a vertex $v$\\
$\zeta_T$ \where{notation:ordering} ordering of $T$\\
$\uc$, $(c_1,\ldots,c_n)$ \where{notation:profile} a $\colorc$-profile\\
$\Profc$ \where{notation:profc} set of $\colorc$-profiles\\
$(\uc;d)$, $\duc$ \where{notation:duc} an element in $\Profcc$\\
$\profofz$, $\inout{z}$ \where{notation:profz} profile of $z$\\
$\Treec\duc$ \where{notation:treecduc} isomorphism classes of trees with profile $\duc$\\
$\Linearc\dc$ \where{def:linear-graph} isomorphism classes of linear graphs with profile $\dc$\\
$\uparrow_c$ \where{ex:colored-exedge} $c$-colored exceptional edge\\
$\Lin_{\uc}$ \where{ex:linear-graph} linear graph for profile $\uc$\\
$\lin_{(c_1,\ldots,c_n)}$ \where{ex:truncated-linear-graph} truncated linear graph for profile $(c_1,\ldots,c_n)$\\
$\Cor_{(\uc;d)}$ \where{ex:cd-corolla} $(\uc;d)$-corolla\\
$\Cor_{(\uc;d)}\tau$ \where{ex:cd-permuted-corolla} permuted corolla\\
$T\left(\{\ub_j\};\uc;d\right)$ \where{ex:twolevel-tree} $2$-level tree\\
$T(H_v)_{v\in \Vt(T)}$ \where{def:tree-substitution} tree substitution\\
$\uTreec$, $\uTreec\duc$ \where{def:treesub-category} substitution categories\\
$\uLinearc$, $\uLinearc\dc$ \where{notation:ulinear} substitution categories for linear graphs\\
$\graft(\cdots)$ \where{notation:grafting} grafting\\
$\Sigma_n$ \where{notation:sigman} symmetric group on $n$ letters\\
$\id_n$ \where{notation:sigman} identity permutation in $\Sigma_n$\\
$\sigma\ua$ \where{notation:left-permutation} left permutation of a profile $\ua$\\
$\Sigmac$ \where{notation:sigmac} groupoid of $\colorc$-profiles with left permutations\\
$\Sigmacop$ \where{notation:sigmac} groupoid of $\colorc$-profiles with right permutations\\
$\ua\sigma$ \where{notation:right-permutation} right permutation of a profile $\ua$\\
$\symseqcm$ \where{notation:symseqcm} $\colorc$-colored symmetric sequences in $\M$\\
$X(\uc;d)$, $X\duc$ \where{notation:symseqcm} an entry of a $\colorc$-colored symmetric sequence\\
$\M^{\colorc}$ \where{notation:mtoc} $\colorc$-colored objects in $\M$\\
$X \circ Y$ \where{circle-product} $\colorc$-colored circle product\\
$\Operadcm$ \where{def:operad} $\colorc$-colored operads in $\M$\\
$\O(n)$ \where{notation:oofn} $n$th entry of a $1$-colored operad $\O$\\
$\gamma$ \where{operadic-composition} operadic composition\\
$\operadunit_c$ \where{c-colored-unit} $c$-colored unit\\
$\sigma\langle k_1,\ldots,k_n\rangle$ \where{operadic-eq-1} block permutation\\
$\tau_1 \oplus \cdots \oplus \tau_n$ \where{operadic-eq-2} block sum\\
$\compi$ \where{def:compi} $\compi$-composition\\
$A[T]$ \where{notation:vertex-dec} $A$-decoration of $T$\\
$\gamma_T$ \where{operadic-structure-map} operadic structure morphism\\
$X_{\uc}$ \where{not:x-sub-c} $X_{c_1} \otimes \cdots \otimes X_{c_m}$\\
$\algmo$ \where{def:operad-algebra} category of $\O$-algebras in $\M$\\
$\End(X)$ \where{ex:endomorphism-operad} endomorphism operad\\
$\Treeopc$ \where{ex:tree-operad} $\colorc$-colored tree operad\\
$\As$ \where{ex:operad-as} associative operad\\
$\Com$ \where{ex:operad-com} commutative operad\\
$\Cdiag$ \where{ex:operad-diag} operad for $\C$-diagrams\\
$\Ocm$ \where{ex:diag-monoid-operad} operad for $\C$-diagrams of monoids\\
$\Comc$ \where{notation:comc} operad for $\C$-diagrams of commutative monoids\\
$\fstaro$ \where{notation:fstaro} pullback of $\O$ along $f$\\
$(-)_!$ \where{thm:change-operad} induced left adjoint\\
$(-)^*$ \where{thm:change-operad} induced right adjoint\\
$f \boxslash g$ \where{notation:fboxslashg} $f$ has the left lifting property with respect to $g$\\
$^{\boxslash}\!\cala$ \where{notation:boxslasha} morphisms with the LLP with respect to $\cala$\\
$\cala^{\boxslash}$ \where{notation:boxslasha} morphisms with the RLP with respect to $\cala$\\
$(\calw, \calc,\calf)$ \where{notation:wcf} weak equivalences, cofibrations, and fibrations\\
$\Ho(\M)$ \where{notation:homotopy-cat} homotopy category of $\M$\\
$\Lder F$ \where{notation:left-derived} total left derived functor of $F$\\
$\R U$ \where{notation:right-derived} total right derived functor of $U$\\
$(\Osinv,\ell)$ \where{def:operad-localization} $S$-localization of $\O$\\
$(-)^{\M}$ \where{notation:oton} change-of-category functor\\
$(J, \mu, 0, 1, \epsilon)$ \where{def:segment} segment\\
$\J$ \where{notation:functorj} functor $\uTreecducop \to \M$ defined by $J$\\
$\wo$ \where{w-of-o} Boardman-Vogt construction of $\O$\\
$\omega_T$ \where{notation:omega-natural} natural morphism $\J[T] \otimes \O[T] \to \wo\duc$\\
$\eta : \wo \shortto \O$ \where{thm:w-augmented} augmentation of the BV construction\\
$\xi : \O \shortto \wo$ \where{notation:stsection} standard section\\
$\uTreec_n\duc$ \where{def:treen} $n$th substitution category\\
$\wno$ \where{def:wn} $n$th filtration of the Boardman-Vogt construction\\
$\uTreeceqnduc$ \where{notation:utreeceqn} substitution category with permuted corollas\\
\blob as morphisms\\
$\weqno$ \where{notation:weqno} coend defined using $\uTreeceqnduc$\\
$\Tun(T)$ \where{notation:tunnels} set of tunnels in $T$\\
$\Doft$ \where{notation:doft} decomposition category\\
$\lambda_T$ \where{wo-algebra-restricted} structure morphism of a $\wo$-algebra\\
$T\sigma$ \where{wo-alg-eq} $T$ with ordering permuted by $\sigma$\\
$\lambda_{\uc}^{\uf}$ \where{hcdiagram-structure-map} structure morphism for a homotopy\\ \blob coherent diagram\\
$\lambda_T^{\{\sigma_v\}_{v\in T}}$ \where{ainfinity-structure} structure morphism for an $A_\infty$-algebra\\
$\lambda_T\left\{(\sigma^v,\uf^v)\right\}_{v\in T}$ \where{wocm-restricted} structure morphism for a homotopy coherent\\ \blob diagram of $A_\infty$-algebras\\
$\lambda_T\left\{\uf^v\right\}_{v\in T}$ \where{wcomc-restricted} structure morphism for a homotopy coherent\\ \blob diagram of $E_\infty$-algebras\\
$\perp$ \where{notation:perp} orthogonality relation\\
$f \perp g$, $(f,g) \in \perpen$ \where{notation:fperpg} $f$ and $g$ are orthogonal\\
$\Cbar$, $(\C,\perp)$ \where{notation:cperp} orthogonal category\\
$\Orthcat$ \where{notation:orthcat} category of orthogonal categories\\
$F_*(\perp)$ \where{notation:fpushforward} pushforward of $\perp$ along $F$\\
$F^*(\perp)$ \where{notation:fpullback} pullback of $\perp$ along $F$\\
$\QFT(\Cbar)$ \where{notation:qftcbar} algebraic quantum field theories on $\Cbar$\\
$\QFT(\Cbar,S)$ \where{notation:qftcbars} algebraic quantum field theories with time-slice\\
$\Csinvbar$ \where{notation:csinvbar} orthogonal category of $\Csinv$\\
$\Ocbar$ \where{notation:ocbar} $\Set$-operad associated to $\Cbar$\\
$\sim$ \where{notation:aqftsim} equivalence relation defining $\Ocbar$\\
$[\sigma,\uf]$ \where{notation:sigmaf} equivalence class of $(\sigma,\uf)$\\
$\Ocbarm$ \where{aqft=operadalgebra} operad for algebraic quantum field theories on $\Cbar$\\
$\Cbarmin$ \where{notation:cbarmin} minimal orthogonal category on $\C$\\
$\Cbarmax$ \where{notation:cbarmax} maximal orthogonal category on $\C$\\
$\Regxzero$ \where{notation:regxzero} regions in the interior of $X$\\
$\wocbarm$ \where{def:haqft} Boardman-Vogt construction of $\Ocbarm$\\
$\HQFT(\Cbar)$ \where{notation:hqftcbar} homotopy algebraic quantum field theories on $\Cbar$\\
$\lambda_T^{\{f^v\}}$ \where{haqft-restricted} structure morphism of homotopy AQFT\\
$(1~2)$ \where{notation:12permutation} non-identity permutation in $\Sigma_2$\\
$\Chat$, $(\C,\Config)$ \where{def:configcat} a configured category\\
$\Config$ \where{def:configcat} set of configurations\\
$\Configcat$ \where{notation:configcat} category of configured categories\\
$(d;\varnothing)$ \where{notation:emptyconfiguration} empty configuration at $d$\\
$\{f_i\}_{i=1}^n$ \where{notation:nconfig} an $n$-ary configuration\\
$\Chatmin$, $(\C,\Configcmin)$ \where{notation:chatmin} minimal configured category on $\C$\\
$\Chatmax$, $(\C,\Configcmax)$ \where{notation:chatmax} maximal configured category on $\C$\\
$\Lhat$, $(L,\Configl)$ \where{notation:lhat} configured category on a bounded lattice $L$\\
$\Ochat$ \where{notation:ochat} $\Set$-operad of $\Chat$\\
$\Ochatm$ \where{notation:ochatm} operad for prefactorization algebras on $\Chat$\\
$\PFA(\Chat)$ \where{notation:pfachat} prefactorization algebras on $\Chat$\\
$\PFA(\Chat,S)$ \where{notation:pfachats} prefactorization algebras with time-slice\\
$\lambda\{f_i\}_{i=1}^n$ \where{pfa-structure-morphism} structure morphism for a prefactorization algebra\\
$\Mcstar$ \where{notation:mcstar} pointed $\C$-diagrams in $\M$\\
$\Embn$ \where{ex:pfa-manifolds} category of smooth $n$-manifolds\\
$\Holn$ \where{ex:pfa-cpmanifolds} category of complex $n$-manifolds\\
$\wochatm$ \where{def:hpa} Boardman-Vogt construction of $\Ochatm$\\
$\HPFA(\Chat)$ \where{notation:hpfachat} homotopy prefactorization algebras on $\Chat$\\
$\HPFA(\Chat,S)$ \where{notation:hpfachats} homotopy prefactorization algebras with time-slice\\
$\lambda_T\left\{\uf^v\right\}_{v\in T}$ \where{wochatm-restricted} structure morphism of a homotopy PFA\\
$\theta_{\uc}^{\uf}$ \where{hcptdiagram-structure-map} structure morphism of a homotopy\\ \blob coherent pointed $\C$-diagram\\
$\Psi$ \where{notation:Psi} functor $\Orthcat \to \Configcat$\\
$\Phi$ \where{notation:Phi} functor $\Configcat \to \Orthcat$\\
$\delta$, $\deltam$ \where{notation:compmorphism} comparison morphisms\\
$\W\deltam$ \where{notation:compmorphism} homotopy comparison morphism\\

\end{tabbing}


\printindex

\end{document}
